\documentclass[12pt]{article}
\usepackage{graphicx}
\usepackage{amsmath,amsthm,amssymb,mathrsfs,mathtools}
\usepackage{caption}
\usepackage[OT1]{fontenc} 
\usepackage{tabularx}
\usepackage[margin=1in]{geometry}
\usepackage{kpfonts}
\usepackage{inconsolata}
\usepackage{color}
\usepackage{authblk}
\usepackage{enumitem}
\usepackage{bm}
\usepackage[colorlinks,breaklinks=true]{hyperref}
\usepackage{tikz}
\usetikzlibrary{shapes.geometric, arrows}
\usepackage[nameinlink,noabbrev,sort,capitalise]{cleveref}
\usepackage{nameref}
\usepackage{thmtools}
\usepackage{xcolor}
\AtBeginDocument{%
	\hypersetup{%
    linkcolor={red!50!black},%
    citecolor={blue!50!black},%
    urlcolor={blue!80!black}%
}%
}%
\renewcommand{\footnotesize}{\fontsize{9pt}{11pt}\selectfont}
\usepackage[page]{appendix}

\usepackage[pagewise]{lineno}
\usepackage{titlesec}
\usepackage{titletoc}

\titlespacing\section{0pt}{6pt plus 2pt minus 2pt}{6pt plus 2pt minus 2pt}
\titlespacing\subsection{0pt}{4pt plus 2pt minus 2pt}{4pt plus 2pt minus 2pt}
\titlespacing\subsubsection{0pt}{2pt plus 2pt minus 2pt}{2pt plus 2pt minus 2pt}
\titlespacing*{\paragraph}{0pt}{1.25ex plus 1ex minus .2ex}{0.5em}

\usepackage[american]{babel}
\usepackage[style=authoryear-comp,giveninits=true,backend=biber,safeinputenc]{biblatex}
\allowdisplaybreaks
\newtheorem{thm}{Theorem}
\crefname{thm}{Theorem}{Theorems}

\newtheorem{lem}{Lemma}
\numberwithin{lem}{section}
\crefname{lem}{Lemma}{Lemmas}
\newtheorem{prop}{Proposition}
\newtheorem{definition}{Definition}
\newtheorem{axiom}{Axiom}
\crefname{axiom}{Axiom}{Axioms}
\newtheorem{cor}{Corollary}%[thm]

\newtheorem{remark}{Remark}
\newtheorem{innercustomthm}{Theorem}
\newenvironment{customthm}[1]
  {\renewcommand\theinnercustomthm{#1}\innercustomthm}
  {\endinnercustomthm}
\newtheoremstyle{example}{}{}{}{}{\bfseries}{\smallskip}{\newline}{}
\theoremstyle{example}
\newtheorem{eg}{Example}
\crefname{eg}{Example}{Examples}

\newcommand{\overbar}[1]{\mkern 1.5mu\overline{\mkern-1.5mu#1\mkern-1.5mu}\mkern 1.5mu}
\renewcommand{\d}{\mathrm{d}}
\newcommand{\defn}{:=}

\newcommand{\supp}{\mathrm{supp}}
\newcommand{\diag}{\mathrm{diag}}
\newcommand{\intr}{\mathrm{int}}
\renewcommand{\H}{\mathrm{Hess}}
\newcommand{\E}{\mathbb{E}}
\newcommand{\R}{\mathbb{R}}
\newcommand{\C}{\mathcal{C}}
\newcommand{\K}{\mathcal{K}}

\newcommand{\Ex}{\mathcal{R}}
\newcommand{\Se}{\mathcal{E}}

\newcommand{\dom}{\mathrm{dom}}
\newcommand{\ri}{\mathrm{relint}}

\newcommand{\awb}{\textcolor{blue}}

\newcommand{\rp}[1]{\langle #1 \rangle}

\usepackage{setspace}
\usepackage{indentfirst}
\newcommand{\dred}{\textcolor{red!50!black}}
\newcommand{\ddd}{\mathrm{d}}%  differential d  (upright)
\newcommand{\dd}{\,\ddd}% differential closing an integral
\newcommand{\Ec}{\Gamma}
\newcommand{\EC}{\mathcal{G}}%{{\mathpalette\makebbGamma\relax}}

\title{The Cost of Optimally Acquired Information\thanks{This paper supersedes the working paper \textcite{bloedel-zhong-2020}. We thank Nageeb Ali, Doug Bernheim, Simon Board, Andrew Caplin, Mark Dean, Henrique de Oliveira, Tommaso Denti, John Geanakoplos, Marina Halac, Ben H\'{e}bert, Ian Jewitt, Emir Kamenica, Elliot Lipnowski, Alessandro Lizzeri, Jay Lu, Paul Milgrom, Stephen Morris, Luciano Pomatto, Doron Ravid, Ilya Segal, Colin Stewart, Philipp Strack, Balazs Szentes, Omer Tamuz, Can Urgun, Leeat Yariv, and numerous conference and seminar audiences for helpful comments. Bloedel thanks the B.F. Haley and E.S. Shaw
Fellowship for Economics for its financial support and Caltech for its hospitality while part of this work was completed.}}
\author{\vspace{-1em}Alexander W. Bloedel\thanks{{Department of Economics, UCLA. Email: \url{abloedel@econ.ucla.edu}}} \hspace{2em} Weijie Zhong\thanks{Graduate School of Business, Stanford University. Email: \url{weijie.zhong@stanford.edu}}}
\date{\vspace{-1em}\today}

\begin{document}

\maketitle
%\vspace{-1.1cm}
\begin{abstract}
This paper introduces a framework for modeling the cost of information acquisition based on the principle of cost-minimization. We study the reduced-form \emph{indirect cost} of information generated by the sequential minimization of a primitive \emph{direct cost} function. Indirect cost functions: (i) are characterized by a novel recursive property, \emph{sequential learning-proofness}; (ii) provide an optimization foundation for the popular class of ``uniformly posterior separable'' costs; and (iii) can often be tractably calculated from their underlying direct costs. We apply the framework by identifying fundamental modeling tradeoffs in the rational inattention literature and two new indirect cost functions that balance these tradeoffs.
%This paper introduces a framework for modeling the cost of information acquisition based on the core tenet of cost-minimization. We study the ``reduced-form'' \emph{indirect cost} of information that arises from the sequential minimization of a ``primitive'' \emph{direct cost} function. Indirect cost functions: (i) are characterized by a simple recursive property, \emph{sequential learning-proofness}; (ii) provide an optimization foundation for the widely studied class of “uniformly posterior separable” costs; and (iii) can be tractably calculated from their underlying direct costs via incrementally informative ``diffusion'' signals. We apply the framework by identifying novel modeling tradeoffs in the rational inattention literature and two new indirect cost functions---\emph{Total Information} and the \emph{Minimal Likelihood Ratio (MLR) cost}---that balance these tradeoffs.
\end{abstract}

\setlength\abovedisplayskip{9pt}
\setlength\belowdisplayskip{9pt}

\section{Introduction}

Information is a valuable but costly resource. There is a unified paradigm for modeling the value of information based on the extent to which it facilitates decision-making \parencite{blackwell-experiment51}. There is less consensus on how to model its cost. In this paper, we introduce a framework for modeling the cost of information based on the core tenet of production theory: that outputs are produced at minimal cost by combining inputs optimally.

Our framework features a Bayesian decision-maker (DM) who learns about an uncertain state by acquiring costly information in the form of Blackwell experiments (i.e., signals correlated with the state). The DM's ``primitive'' information acquisition technology is described by an arbitrary \emph{direct cost} function over experiments. Given any ``target'' experiment, the DM produces it as cheaply as possible by optimizing over all sequential information acquisition strategies that generate at least as much information as the target. We define the DM's \emph{indirect cost} function as the minimum expected cost of producing target experiments in this manner. The indirect cost function then represents the DM's ``reduced-form'' cost of acquiring information in any downstream decision problem.

%We call the DM's minimal expected cost of producing target experiments his \emph{indirect cost} function. The indirect cost function then represents the DM's ``reduced-form'' cost of acquiring information in any downstream decision problem (as in the rational inattention literature, e.g., \cite{sims_JME2003,matejka_mckay_AER2015,mackowiak2023rational}).

We propose this framework as a unified way to capture two key features of real-world information acquisition. First, across a wide range of settings, it is both feasible and optimal for the DM to acquire information piece-by-piece in a sequential fashion. For example, in a standard \emph{statistical sampling problem} \parencite{wald-ams1945}, a firm learns about the demand for a new product by sequentially sampling consumers (e.g., via surveys, A/B tests, or RCTs), subject to a physical or pecuniary direct cost that depends on the sample's size and features.\footnote{E.g., multi-stage RCTs for pharmaceuticals \parencite{FDA2018D3124} and sequential A/B tests of tech products \parencite{johari2022always}. Similarly, scientific research and industrial R\&D often involve multiple adaptively designed stages of experimentation. 
%For instance, \textcite{FDA2018D3124} encourages the use of multi-stage RCTs in the context of clinical trials for pharmaceutical products. Meanwhile, sequential A/B testing is common in the tech industry (e.g., \cite{johari2022always}).
} In a typical \emph{encoding problem} \parencite{shannon-1948}, an online consumer chooses between products by sequentially querying their attributes (e.g., on a price-comparison website), incurring a cognitive or computational direct cost for each query. In a generic \emph{perception task} \parencite{ratcliff1978theory}, a lab subject faced with a visual stimulus gradually contemplates how to classify it, paying a direct cost of delay or cognitive effort while he thinks.\footnote{%Sequential learning is often relevant even when information acquisition may initially appear to be one-shot. 
Oftentimes, even apparently ``static'' information acquisition strategies are actually sequential. In statistical sampling, ``non-sequential'' (i.e., fixed sample size) procedures still take time to implement and can be viewed as non-contingent sequential procedures. In perception tasks, subjects' response times are short but non-zero (e.g., seconds).
%Sequential learning is also common in many other settings, e.g., scientific research and industrial R\&D often involve multiple adaptively designed stages of experimentation. Even in settings where information acquisition may initially appear to be one-shot, there is usually some degree of sequentiality. In statistical sampling problems, non-sequential (i.e., fixed sample size) procedures still take time to implement and can be interpreted as non-contingent sequential procedures. In perception tasks, lab subjects' response times are short (e.g., on the order of seconds) but still nonzero.
}

Second, the cost of information is highly context-specific. In the above examples, to paraphrase \textcite[p. 161]{sims-handbook2010}, the physical or pecuniary costs of generating new information through statistical sampling may bear no relation to the cognitive or computational costs of processing freely available information in encoding and perception tasks.

Although these two features are ubiquitous, extant theories of costly information acquisition capture at most one of them. 
%Despite the ubiquity of these two features, extant theories of information cost capture at most one of them. 
On the one hand, a classic approach is to study \emph{sequential} learning with \emph{specific direct costs}, as in the literatures on sequential sampling in statistics \parencite{wald-ams1945,wald-ecma1947,abg-ecma1949}, optimal encoding in information theory \parencite{shannon-1948,huffman1952method}, and drift-diffusion models of perception in psychology and neuroscience \parencite{ratcliff2008diffusion,fehr2011neuroeconomic,fudenberg2018speed}. These frameworks explicitly model the DM's production procedure but, by %virtue of adopting 
focusing on specific ``units for information,'' are ``useful but only on very limited problems'' \parencite[p. 120]{arrow-empirica1996}. On the other hand, the modern flexible information acquisition (``rational inattention'') literature eschews the underlying production procedure and instead 
%abstract away from the underlying production procedure and instead 
studies \emph{one-shot} learning with \emph{reduced-form} cost functions \parencite{sims_JME2003,matejka_mckay_AER2015}. %as in the flexible information acquisition (or ``rational inattention'') literature \parencite{sims_JME2003,mackowiak2023rational}. 
This approach justifies various reduced-form cost functions via context-specific axioms \parencite{hebert2021neighborhood,caplin2022rationally,denti2022experimental,pomatto2023cost} and their implications for choice behavior \parencite{caplin_dean_aer2015,ddmo-te2017,denti2022posterior,dean2023experimental}. But, by design, 
%eschewing the underlying production procedure, 
it does not address where these cost functions ``come from'' or whether they are ``rationalizable'' via sequential optimization.

Our framework bridges these two paradigms. By allowing for arbitrary direct costs, it enables one to study both the \emph{context-free} implications of sequential optimization and various \emph{context-specific} cost functions. By requiring indirect costs to arise from sequential optimization, it imposes discipline on the notion of ``reduced-form'' information cost.

\paragraph{The Indirect Cost of Information.} Our first contribution is to characterize the full class of indirect costs. We do so via a novel recursive property that we call \emph{sequential learning-proofness} (\nameref{axiom:slp}). 
%full class of indirect costs in terms of a novel recursive property that we call \emph{sequential learning-proofness} (\nameref{axiom:slp}). 
A cost function is \nameref{axiom:slp} if the cost of acquiring any target experiment in one shot is weakly lower than the expected cost of decomposing it into two steps. \nameref{axiom:slp} represents a mild ``internal consistency'' or ``robustness'' requirement for reduced-form models of information cost: if the DM's cost function were \emph{not} \nameref{axiom:slp}, then he could optimize away some of its features using a simple two-step strategy. We show that a cost function is the indirect cost generated by \emph{some} underlying direct cost  \emph{if and only if} it is \nameref{axiom:slp} (\cref{prop:1}). Thus, \nameref{axiom:slp} fully characterizes the ``context-free'' implications of sequential optimization.

To provide a more concrete characterization, we then show (\cref{thm:UPS}) that a cost function is \nameref{axiom:slp} and  \emph{\nameref{defi:ll}}---a mild notion of ``local differentiability''---\emph{if and only if} it is \emph{uniformly posterior separable} (\nameref{defi:ups}). That is, letting $\Theta$ denote the set of states, there is some convex ``potential function'' $H : \Delta(\Theta) \to (-\infty, +\infty]$ such that the cost function is given by 
\begin{align}
    C_\text{ups}^H(\pi) :=\E_{\pi}\left[H(q) -H(p) \right] \tag{UPS}
\end{align}
%for some convex \emph{potential function} $H : \Delta(\Theta) \to (-\infty, +\infty]$,
for every distribution $\pi \in \Delta(\Delta(\Theta))$ of Bayesian posteriors $q \in \Delta(\Theta)$ induced by some experiment and prior belief  $p \in \Delta(\Theta)$. The class of \nameref{defi:ups} cost functions (introduced by \cite{caplin2022rationally}) includes most specifications studied in the rational inattention literature, including \hyperref[MI]{Mutual Information} \parencite{sims_JME2003,matejka_mckay_AER2015} and the more general family of neighborhood-based costs \parencite{hebert2021neighborhood}. \cref{thm:UPS} offers a novel optimality- and tractability-based foundation for the \nameref{defi:ups} model.

\begin{figure}[t]
\vspace{-1em}
    \centering\tikzset{every picture/.style={line width=0.75pt}} %set default line width to 0.75pt        

\begin{tikzpicture}[x=0.75pt,y=0.75pt,yscale=-1,xscale=1]
%uncomment if require: \path (0,312); %set diagram left start at 0, and has height of 312

%Rounded Rect [id:dp7010069609141982] 
\draw  [dash pattern={on 4.5pt off 4.5pt}] (350.06,75.62) .. controls (350.06,67.43) and (356.7,60.8) .. (364.88,60.8) -- (467.85,60.8) .. controls (476.03,60.8) and (482.67,67.43) .. (482.67,75.62) -- (482.67,255.02) .. controls (482.67,263.2) and (476.03,269.83) .. (467.85,269.83) -- (364.88,269.83) .. controls (356.7,269.83) and (350.06,263.2) .. (350.06,255.02) -- cycle ;
%Rounded Rect [id:dp6641215903765686] 
\draw  [fill={rgb, 255:red, 0; green, 0; blue, 0 }  ,fill opacity=0.07 ] (353.67,100.58) .. controls (353.67,95.24) and (358,90.92) .. (363.33,90.92) -- (470.66,90.92) .. controls (476,90.92) and (480.32,95.24) .. (480.32,100.58) -- (480.32,256.28) .. controls (480.32,261.61) and (476,265.94) .. (470.66,265.94) -- (363.33,265.94) .. controls (358,265.94) and (353.67,261.61) .. (353.67,256.28) -- cycle ;
%Rounded Rect [id:dp8352279423990964] 
\draw  [fill={rgb, 255:red, 0; green, 0; blue, 0 }  ,fill opacity=0.16 ] (356.91,196.41) .. controls (356.91,193.17) and (359.53,190.55) .. (362.76,190.55) -- (471.22,190.55) .. controls (474.45,190.55) and (477.07,193.17) .. (477.07,196.41) -- (477.07,257.41) .. controls (477.07,260.65) and (474.45,263.27) .. (471.22,263.27) -- (362.76,263.27) .. controls (359.53,263.27) and (356.91,260.65) .. (356.91,257.41) -- cycle ;
%Rounded Rect [id:dp48514262051231527] 
\draw  [fill={rgb, 255:red, 0; green, 0; blue, 0 }  ,fill opacity=0.16 ] (135.83,196.14) .. controls (135.83,192.91) and (138.45,190.29) .. (141.69,190.29) -- (250.15,190.29) .. controls (253.38,190.29) and (256,192.91) .. (256,196.14) -- (256,257.15) .. controls (256,260.38) and (253.38,263) .. (250.15,263) -- (141.69,263) .. controls (138.45,263) and (135.83,260.38) .. (135.83,257.15) -- cycle ;
%Rounded Rect [id:dp8643621312401525] 
\draw   (132.67,69.16) .. controls (132.67,63.83) and (137,59.5) .. (142.33,59.5) -- (249.66,59.5) .. controls (255,59.5) and (259.32,63.83) .. (259.32,69.16) -- (259.32,259.94) .. controls (259.32,265.28) and (255,269.61) .. (249.66,269.61) -- (142.33,269.61) .. controls (137,269.61) and (132.67,265.28) .. (132.67,259.94) -- cycle ;

% Text Node
\draw (157,42) node [anchor=north west][inner sep=0.75pt]  [font=\normalsize] [align=left] {\textit{Direct Cost}};
% Text Node
\draw (373,72) node [anchor=north west][inner sep=0.75pt]  [font=\normalsize] [align=left] {\textit{Indirect Cost}};
% Text Node
\draw (361.5,175) node [anchor=north west][inner sep=0.75pt]  [font=\footnotesize] [align=left] {\textit{Regular Indirect Cost}};
% Text Node
\draw (388.81,94.28) node [anchor=north west][inner sep=0.75pt]  [font=\small]  {$\boldsymbol{\Leftrightarrow }$  \textbf{SLP}};
% Text Node
\draw (388.81,193.2) node [anchor=north west][inner sep=0.75pt]  [font=\small]  {$\boldsymbol{\Leftrightarrow}$ \textbf{UPS}};
% Text Node
\draw (165.15,193.2) node [anchor=north west][inner sep=0.75pt]  [font=\small]  {$\boldsymbol{\Leftrightarrow }$ \textbf{FLIEs}};
% Text Node
\draw (189.07,129) node [anchor=north west][inner sep=0.75pt]  [font=\large]  {${\displaystyle C}$};
% Text Node
\draw (396.55,128.15) node [anchor=north west][inner sep=0.75pt]  [font=\large]  {$\Phi ( C)$};
% Text Node
\draw (189.19,223.5) node [anchor=north west][inner sep=0.75pt]  [font=\large]  {${\displaystyle C}$};
% Text Node
\draw (398,219.19) node [anchor=north west][inner sep=0.75pt]  [font=\large]  {$C_\text{ups}^{H}$};
% Text Node
\draw (150,248) node [anchor=north west][inner sep=0.75pt]  [font=\scriptsize] [align=left] {with kernel $\H H$};
% Text Node
\draw (371,248) node [anchor=north west][inner sep=0.75pt]  [font=\scriptsize] [align=left] {with kernel $\H H$};
% Text Node
\draw (384,151) node [anchor=north west][inner sep=0.75pt]  [font=\scriptsize] [align=left] {with kernel $k_{C}$};
%%
% Text Node
\draw (162,151) node [anchor=north west][inner sep=0.75pt]  [font=\scriptsize] [align=left] {with kernel $k_{C}$};
% Text Node
%\draw (270.03,104.47) node [anchor=north west][inner sep=0.75pt]  [font=\footnotesize] [align=center] {\begin{minipage}[lt]{48.98pt}\setlength\topsep{0pt}
%\begin{center}
%\textit{Sequential}\\\textit{Optimization}
%\end{center}
%\end{minipage}};
\draw (270.03,105) node [anchor=north west][inner sep=0.75pt]  [font=\footnotesize] [align=center] 
{\textit{Sequential}\\\textit{Optimization}};
% Text Node
\draw (291.18,147.58) node [anchor=north west][inner sep=0.75pt]  [font=\huge]  {$\mathrm{\Phi} $};
% Text Node
%\draw (272.78,198) node [anchor=north west][inner sep=0.75pt]  [font=\footnotesize] [align=center] {\textit{Incremental}\\\textit{Learning}};
% Connection
\draw [line width=1.5]    (207.07,137.25) -- (389.55,137.25) ;
\draw [shift={(393.55,137.25)}, rotate = 180] [fill={rgb, 255:red, 0; green, 0; blue, 0 }  ][line width=0.08]  [draw opacity=0] (11.61,-5.58) -- (0,0) -- (11.61,5.58) -- cycle    ;
% Connection
\draw [line width=0.75]    (216.19,229.06) -- (385.15,229.25)(216.18,232.06) -- (385.15,232.25) ;
\draw [shift={(394.15,230.76)}, rotate = 180] [fill={rgb, 255:red, 0; green, 0; blue, 0 }  ][line width=0.08]  [draw opacity=0] (8.93,-4.29) -- (0,0) -- (8.93,4.29) -- cycle    ;
\draw [shift={(207.19,230.55)}, rotate = 0] [fill={rgb, 255:red, 0; green, 0; blue, 0 }  ][line width=0.08]  [draw opacity=0] (8.93,-4.29) -- (0,0) -- (8.93,4.29) -- cycle    ;

\end{tikzpicture}
    \caption{Indirect cost and the sequential learning map (\hyperref[prop:1]{Theorems 1}--\ref{thm:flie}).}
    \label{fig:1}
\end{figure}
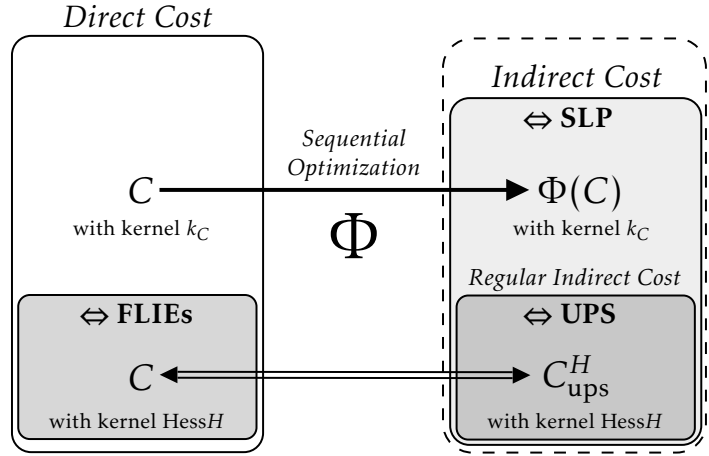

\paragraph{The Sequential Learning Map.} Our second contribution is to characterize the \emph{sequential learning map}, $\Phi$, that transforms each direct cost $C$ into its corresponding indirect cost $\Phi(C)$ (see \cref{fig:1}). This map determines how properties of a given direct cost function are transformed under optimization. Conversely, the \emph{pre-image of} this map determines the ``primitive'' economic assumptions that are implicitly imposed on the underlying direct cost when one adopts a particular functional form for the indirect cost.

Central to our characterization is an object that we call the \emph{kernel} of a cost function, which is a matrix-valued function that generalizes the standard notion of a Hessian. The kernel summarizes the cost of ``incremental evidence,'' i.e., experiments that shift posterior beliefs only locally (analogous to  continuous-time diffusion signals). Our key observation is that the kernel of any direct cost is \emph{invariant} under the sequential learning map. That is, the cost of incremental evidence cannot be reduced through optimization.

We proceed in two steps. First, we develop general lower and upper bounds on the sequential learning map. For any direct cost $C$, the indirect cost $\Phi(C)$ is (i) \emph{locally} bounded below by the kernel of $C$ and (ii) \emph{globally} bounded above by the \nameref{defi:ups} cost obtained by integrating the kernel of $C$ (\cref{thm:qk}). Economically, the \nameref{defi:ups} upper bound represents the expected cost of the \emph{incremental learning} strategy that only acquires incremental evidence.

Second, we show that the upper bound is tight \emph{if and only if} the indirect cost $\Phi(C)$ is \nameref{defi:ll}/\nameref{defi:ups}. 
%
%In other words, the subset of \nameref{defi:ll}/\nameref{defi:ups} indirect costs comprises precisely the indirect costs that are generated from direct costs for which incremental learning is an optimal strategy. 
%That is, we obtain an exact characterization of the sequential learning map for the co-domain of \nameref{defi:ll}/\nameref{defi:ups} indirect costs. 
Specifically, a direct cost $C$ generates the indirect cost $\Phi(C) = C_\text{ups}^H$ \emph{if and only if}: (i) the kernel of $C$ equals the Hessian of the potential function $H$, and (ii) $C$ \emph{favors learning via incremental evidence} (\nameref{axiom:flie}), i.e., weakly exceeds the expected cost of incremental learning (\cref{thm:flie}). This result yields an exact characterization of the sequential learning map for the co-domain of \nameref{defi:ll}/\nameref{defi:ups} indirect costs, and demonstrates that such indirect costs arise precisely when incremental learning is an optimal strategy.

\cref{thm:flie} helps delineate the range of applications in which  \nameref{defi:ll}/\nameref{defi:ups} indirect costs are economically reasonable. 
%, depending on whether it is natural to assume the direct cost \nameref{axiom:flie}.
%as it is natural for the direct cost to \hyperref[axiom:flie]{FLIE} in some applications but not in others. 
It also offers a tractable method for calculating \nameref{defi:ll}/\nameref{defi:ups} from their direct costs and vice versa, which we illustrate via several examples.

%\cref{thm:qk,thm:flie} offers a tractable tool for linking properties of the direct cost to its indirect cost. Specifically, it allows us to examine whether a property imposed on direct cost is carried over to the indirect cost, and if not, how does optimization transform the property. Conversely, it delineates the primitive property on the direct cost that is required to justify a property on the indirect cost. \par

%The characterizations we developed are illustrated in \cref{fig:1}. Then, we leveraged the framework to study several natural axioms of information cost. Our third contribution is to establish---and resolve---a trilemma of information cost.

\paragraph{Information Cost Trilemmas.} Our third contribution is to apply our framework by characterizing the implications of sequential optimization in specific economic contexts. This exercise serves two purposes: to pinpoint specific indirect cost functions for use in applications, and to elucidate modeling tradeoffs in the rational inattention literature. 

To these ends, we study how our notion of indirect/\nameref{axiom:slp} cost interacts with two axioms that the literature has advocated for imposing on reduced-form cost functions. The first axiom, \hyperref[axiom:prior:invariant]{Prior Invariance}, requires the cost of any given Blackwell experiment to be independent of the DM's prior beliefs. This property is natural when modeling physical or pecuniary information costs (e.g., statistical sampling or R\&D).\footnote{Various authors have advocated for \hyperref[axiom:prior:invariant]{Prior Invariance} on these and related grounds. See, for instance,  \textcite{woodford-reference2012,gentzkow_kamenica_costly,mensch-cardinal-information,denti2022experimental}.} 
The second axiom, \dred{Constant Marginal Cost} (\hyperref[axiom:CMC:0]{CMC}), posits that the cost of running two independent experiments together equals the sum of their individual costs. \textcite{pomatto2023cost} propose this property as a non-parametric way of modeling costs that are ``linear in sample size,'' which is a familiar and natural assumption in statistical sampling problems.

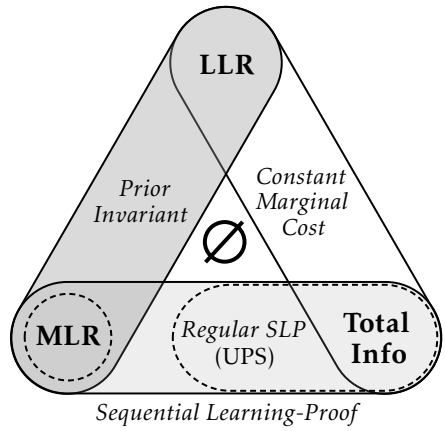
\begin{figure}[t]
\vspace{-1em}
 \centering
    \tikzset{every picture/.style={line width=0.75pt}} %set default line width to 0.75pt        

\begin{tikzpicture}[x=0.75pt,y=0.75pt,yscale=-1,xscale=1]

%Rounded Rect [id:dp5207642762729766] 
\draw   (471.29,48.25) .. controls (484.75,40.47) and (501.96,45.09) .. (509.74,58.55) -- (590.01,197.58) .. controls (597.78,211.04) and (593.17,228.26) .. (579.71,236.03) -- (579.71,236.03) .. controls (566.25,243.8) and (549.03,239.19) .. (541.26,225.73) -- (460.99,86.69) .. controls (453.22,73.23) and (457.83,56.02) .. (471.29,48.25) -- cycle ;
%Rounded Rect [id:dp2973989522023799] 
\draw  [fill={rgb, 255:red, 155; green, 155; blue, 155 }  ,fill opacity=0.37 ] (499.37,48.25) .. controls (512.83,56.02) and (517.45,73.23) .. (509.68,86.69) -- (429.4,225.73) .. controls (421.63,239.19) and (404.42,243.8) .. (390.96,236.03) -- (390.96,236.03) .. controls (377.5,228.26) and (372.88,211.04) .. (380.66,197.58) -- (460.93,58.55) .. controls (468.7,45.09) and (485.91,40.47) .. (499.37,48.25) -- cycle ;
%Rounded Rect [id:dp9118838359553011] 
\draw  [fill={rgb, 255:red, 155; green, 155; blue, 155 }  ,fill opacity=0.17 ] (377.08,211.47) .. controls (377.08,195.93) and (389.68,183.33) .. (405.23,183.33) -- (565.77,183.33) .. controls (581.31,183.33) and (593.92,195.93) .. (593.92,211.47) -- (593.92,211.47) .. controls (593.92,227.01) and (581.31,239.61) .. (565.77,239.61) -- (405.23,239.61) .. controls (389.68,239.61) and (377.08,227.01) .. (377.08,211.47) -- cycle ;
%Shape: Circle [id:dp008476840302349076] 
\draw  [line width=1.5]  (474.67,163.1) .. controls (474.67,157.58) and (479.14,153.1) .. (484.66,153.1) .. controls (490.19,153.1) and (494.66,157.58) .. (494.66,163.1) .. controls (494.66,168.62) and (490.19,173.1) .. (484.66,173.1) .. controls (479.14,173.1) and (474.67,168.62) .. (474.67,163.1) -- cycle ;
%Straight Lines [id:da749981325917144] 
\draw [line width=1.5]    (493.72,152.98) -- (475.02,173.33) ;

%Rounded Rect [id:dp36856389872110806] 
\draw  [fill={rgb, 255:red, 0; green, 0; blue, 0 }  ,fill opacity=0 ][dash pattern={on 2.25pt off 1.5pt}] (458.5,211.52) .. controls (458.5,196.74) and (472.48,184.77) .. (487.25,184.77) -- (565.15,184.77) .. controls (579.92,184.77) and (591.9,196.74) .. (591.9,211.52) -- (591.9,211.52) .. controls (591.9,226.29) and (579.92,238.27) .. (565.15,238.27) -- (487.25,238.27) .. controls (472.48,238.27) and (458.5,226.29) .. (458.5,211.52) -- cycle ;
%Shape: Circle [id:dp9658065600881709] 
\draw  [dash pattern={on 2.25pt off 1.5pt}][line width=0.75]  (383.8,211.2) .. controls (383.8,199.13) and (393.58,189.35) .. (405.65,189.35) .. controls (417.72,189.35) and (427.5,199.13) .. (427.5,211.2) .. controls (427.5,223.26) and (417.72,233.05) .. (405.65,233.05) .. controls (393.58,233.05) and (383.8,223.26) .. (383.8,211.2) -- cycle ;

% Text Node
\draw (494.3,208.5) node  [font=\footnotesize,color={rgb, 255:red, 0; green, 0; blue, 0 }  ,opacity=1 ] [align=left] {\textit{Regular SLP}};
% Text Node
\draw (478,214.5) node [anchor=north west][inner sep=0.75pt]  [font=\footnotesize] [align=left] {(UPS)};
% Text Node
\draw (561.5,203.5) node  [font=\small,color={rgb, 255:red, 0; green, 0; blue, 0 }  ,opacity=1 ] [align=left] {\textbf{Total}};
% Text Node
\draw (562.5,219) node  [font=\small,color={rgb, 255:red, 0; green, 0; blue, 0 }  ,opacity=1 ] [align=left] {\textbf{Info}};
% Text Node
\draw (470.3,67) node [anchor=north west][inner sep=0.75pt]  [font=\small] [align=left] {\textbf{LLR}};
% Text Node
\draw (388,204) node [anchor=north west][inner sep=0.75pt]  [font=\small] [align=left] {\textbf{MLR}};
% Text Node
\draw (499,124) node [anchor=north west][inner sep=0.75pt]  [font=\footnotesize] [align=left] {\textit{Constant}};
% Text Node
\draw (499,136.5) node [anchor=north west][inner sep=0.75pt]  [font=\footnotesize] [align=left] {\textit{Marginal}};
% Text Node
\draw (512,150) node [anchor=north west][inner sep=0.75pt]  [font=\footnotesize] [align=left] {\textit{Cost}};
% Text Node
\draw (430,130) node [anchor=north west][inner sep=0.75pt]  [font=\footnotesize] [align=left] {\textit{Prior}};
% Text Node
\draw (417,143) node [anchor=north west][inner sep=0.75pt]  [font=\footnotesize] [align=left] {\textit{Invariant}};
% Text Node
\draw (418,243) node [anchor=north west][inner sep=0.75pt]  [font=\footnotesize] [align=left] {\textit{Sequential Learning-Proof}};

\end{tikzpicture}
    \vspace{-2.5em}
\iffalse
\hspace{-2em}\centering
\begin{minipage}{.5\textwidth}
    \centering
    \include{Figure/trilemma}
    \vspace{-2em}
    \end{minipage}
    \begin{minipage}{.5\textwidth}
    \centering
    \include{Figure/trilemma_new}
    \vspace{-2.5em}
    \end{minipage}
\fi
\iffalse
    \centering
    \include{Figure/trilemma}
    \vspace{-2.5em}
\fi
    \captionof{figure}{The information cost trilemma (\cref{thm:trilemma}).}
    \label{fig:trilemma}
\end{figure}

We offer two characterization results. First, we establish an \emph{information cost trilemma} (\cref{thm:trilemma}) among the three natural properties of \nameref{axiom:slp}, \hyperref[axiom:prior:invariant]{Prior Invariance}, and \hyperref[axiom:CMC:0]{CMC}. In particular, an information cost function can satisfy any two of these properties, but no nonzero cost function can satisfy all three of them (see \cref{fig:trilemma}). \textcite{pomatto2023cost} have shown that the (essentially) unique \hyperref[axiom:prior:invariant]{Prior Invariant} and \hyperref[axiom:CMC:0]{CMC} cost function is the \emph{Log-Likelihood Ratio} (\ref{eqn:LLR}) cost.\footnote{Formally, \textcite{pomatto2023cost} also impose a mild ``\hyperref[axiom:DL]{Dilution Linearity}'' axiom that is implied by \nameref{axiom:slp}.} We show that the unique \nameref{axiom:slp} and \hyperref[axiom:CMC:0]{CMC} cost function is the \emph{\nameref{defi:TI}} cost, a novel \nameref{defi:ups} cost defined via the potential function 
 \begin{align}
        H_\text{TI}(q):= \sum_{\theta,\theta'\in\Theta}\gamma_{\theta,\theta'} q (\theta)  \log\left(\frac{q (\theta) }{q (\theta') }\right) ,\tag{TI}
    \end{align}
where the coefficients $\gamma_{\theta,\theta'} \geq0$ control the cost of distinguishing between pairs of states (and can be arbitrary). %\footnote{\nameref{defi:TI} has other desirable properties, e.g., it can also be represented as an expectation (under the DM's prior) over state-contingent LLR costs and viewed as a (finite-state) generalization of the \emph{Fisher Information cost} (\cite{hebert2021neighborhood}).} 
We interpret \nameref{defi:TI} as being the natural reduced-form cost function in applications where \hyperref[axiom:CMC:0]{CMC} is a desirable assumption, such as statistical sampling problems. 
%\footnote{As we discuss in \cref{ssec:trilemma}, another lesson from this exercise is that \hyperref[axiom:CMC:0]{CMC} is generally not preserved under the sequential learning map---e.g., the indirect LLR cost does not satisfy it, and for \nameref{defi:TI} it is perhaps best interpreted as an ``emergent'' property of the optimization process.} 
To show that the remaining two-way intersection is nonempty, we construct the \nameref{axiom:slp} and \hyperref[axiom:prior:invariant]{Prior Invariant} \emph{Minimal Likelihood Ratio (\nameref{defi:MLR})} cost, defined as
\begin{align}
        C_\text{MLR}(\pi):=\E_{\pi}\left[ 1-\min_{\theta\in\supp(p)} \frac{q(\theta)}{p (\theta)}\right]\tag{MLR}
    \end{align}
for every distribution $\pi \in \Delta(\Delta(\Theta))$ over posteriors $q$ induced by some experiment and prior belief $p$. As we demonstrate, the \nameref{defi:MLR} cost---which is \emph{not} \nameref{defi:ll}/\nameref{defi:ups}---arises as the indirect cost in a canonical model of continuous-time Poisson sampling (\cref{eg:Poisson:0}). %We argue that the \nameref{defi:MLR} cost is useful in certain applications, such as modeling costly monitoring in games. 

The main tension in the trilemma is between \nameref{axiom:slp} and \hyperref[axiom:prior:invariant]{Prior Invariance}. For instance, no \hyperref[axiom:prior:invariant]{Prior Invariant} costs commonly studied in the literature (e.g., the \ref{eqn:LLR} cost) are \nameref{axiom:slp}. Moreover, no \nameref{axiom:slp} cost that is \nameref{defi:ll}/\nameref{defi:ups} can be \hyperref[axiom:prior:invariant]{Prior Invariant}. Our framework suggests that the tension between these two properties is natural: since \nameref{axiom:slp} costs are derived from \emph{expected} cost-minimization, they ``should'' \emph{endogenously} depend on prior beliefs. It also suggests that a natural way to alleviate this tension is to interpret \hyperref[axiom:prior:invariant]{Prior Invariance} as a ``primitive'' property of \emph{direct} costs, rather than as a ``reduced-form'' property of \emph{indirect} costs. Following this logic, we introduce the novel class of \emph{Sequentially Prior Invariant} (\nameref{defi:spi}) cost functions: indirect costs that are generated by \hyperref[axiom:prior:invariant]{Prior Invariant} direct costs.

\begin{figure}[t]
\vspace{-1em}
\centering
    \tikzset{every picture/.style={line width=0.75pt}} %set default line width to 0.75pt        

\begin{tikzpicture}[x=0.75pt,y=0.75pt,yscale=-1,xscale=1]

%Rounded Rect [id:dp06144288179576274] 
\draw  [fill={rgb, 255:red, 0; green, 0; blue, 0 }  ,fill opacity=0.06 ] (188.31,101.43) .. controls (190.58,78.88) and (210.71,62.45) .. (233.25,64.72) -- (365.47,78.06) .. controls (388.01,80.33) and (404.45,100.45) .. (402.17,123) -- (402.17,123) .. controls (399.9,145.54) and (379.78,161.98) .. (357.23,159.7) -- (225.02,146.37) .. controls (202.47,144.09) and (186.04,123.97) .. (188.31,101.43) -- cycle ;
%Rounded Rect [id:dp8995130464231654] 
%%%% WHITE PORTION / CMC RECTANGLE %%%%
\draw  [fill={rgb, 255:red, 255; green, 255; blue, 255 }  ,fill opacity=0.16 ] (188.37,109.79) .. controls (186.14,87.23) and (202.62,67.15) .. (225.18,64.93) -- (356.7,51.96) .. controls (379.25,49.74) and (399.33,66.22) .. (401.56,88.77) -- (401.56,88.77) .. controls (403.78,111.32) and (387.3,131.4) .. (364.75,133.63) -- (233.23,146.59) .. controls (210.67,148.82) and (190.59,132.34) .. (188.37,109.79) -- cycle ;
%Shape: Ellipse [id:dp1836857530192304] 
\draw   (228.85,146.73) .. controls (206.21,146.67) and (187.9,128.27) .. (187.95,105.63) .. controls (188.01,82.99) and (206.41,64.67) .. (229.05,64.73) .. controls (251.69,64.79) and (270,83.19) .. (269.95,105.83) .. controls (269.89,128.47) and (251.49,146.78) .. (228.85,146.73) -- cycle ;
%Rounded Rect [id:dp03053233705742342] 
\draw  [fill={rgb, 255:red, 0; green, 0; blue, 0 }  ,fill opacity=0.16 ] (56.05,105.31) .. controls (56.1,82.64) and (74.52,64.32) .. (97.18,64.37) -- (229.02,64.7) .. controls (251.68,64.75) and (270,83.17) .. (269.95,105.83) -- (269.95,105.83) .. controls (269.89,128.49) and (251.48,146.81) .. (228.82,146.76) -- (96.98,146.44) .. controls (74.32,146.38) and (55.99,127.97) .. (56.05,105.31) -- cycle ;
%Shape: Ellipse [id:dp8792760119811185] 
\draw  [dash pattern={on 2.25pt off 1.5pt}] (96.78,145.41) .. controls (74.56,145.35) and (56.6,127.33) .. (56.65,105.16) .. controls (56.71,82.99) and (74.76,65.06) .. (96.98,65.11) .. controls (119.2,65.17) and (137.16,83.19) .. (137.11,105.36) .. controls (137.05,127.53) and (119,145.46) .. (96.78,145.41) -- cycle ;

% Text Node
\draw (149.26,99) node [anchor=north west][inner sep=0.75pt]  [font=\normalsize] [align=left] {\textit{SPI}};
% Text Node
\draw (327.65,59.2) node [anchor=north west][inner sep=0.75pt]  [font=\normalsize] [align=left] {\textit{CMC}};
% Text Node
\draw (327.65,141.35) node [anchor=north west][inner sep=0.75pt]  [font=\normalsize] [align=left] {\textit{UPS}};
% Text Node
\draw (285.8,99) node [anchor=north west][inner sep=0.75pt]  [font=\small] [align=left] {Total Info};
% Text Node
\draw (208,99) node [anchor=north west][inner sep=0.75pt]  [font=\normalsize] [align=left] {\textbf{Wald}};
% Text Node
\draw (78.43,92) node [anchor=north west][inner sep=0.75pt]  [font=\footnotesize] [align=left] {\textit{SLP} \&};
% Text Node
\draw (57,104) node [anchor=north west][inner sep=0.75pt]  [font=\footnotesize] [align=left] {\textit{Prior Invariant}};
% Text Node
%\draw (69.93,111.58) node [anchor=north west][inner sep=0.75pt]  [font=\footnotesize] [align=left] {\textit{Invariant}};

\end{tikzpicture}
    \vspace{-2.5em}
    \caption{{\small Resolving the information cost trilemma  (\cref{thm:wald}).}}
    \label{fig:wald-thm}
\end{figure}
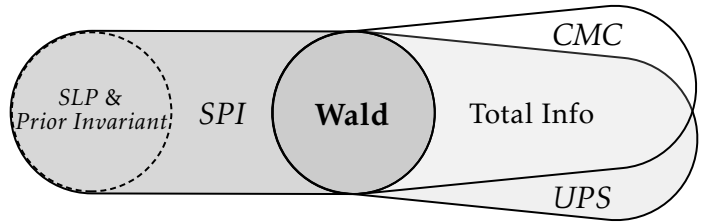

Our second characterization result (\cref{thm:wald}) uses the notion of \nameref{defi:spi} indirect cost to resolve the information cost trilemma. Central to this result is the \emph{\ref{eqn:MS} cost} function of  \textcite{morris-strack-sampling}, which is 
%defined for a binary state space $\Theta  = \{0,1\}$ as
% %
% %Given a binary state space $\Theta  = \{0,1\}$ and letting $q \in [0,1]$ denote the probability of $\theta=1$, \textcite{morris-strack-sampling} define the \emph{\ref{eqn:MS} cost} function as
%  \begin{align}
%      C_{\text{Wald}}:=C_\text{ups}^{H^*} \quad\text{where } \ H^*(q):=(2q(1)-1)\log\left( \frac{q(1)}{1-q(1)} \right), \label{eqn:MS}\tag{Wald}
%  \end{align}
% i.e.,
the special case of \nameref{defi:TI} for binary state spaces (i.e., $\Theta =\{0,1\}$) and with symmetric coefficients (i.e., $\gamma_{0,1} = \gamma_{1,0}$). We establish a three-way equivalence: a cost function is \nameref{defi:spi} and \hyperref[axiom:CMC:0]{CMC} \emph{if and only if} it is \nameref{defi:spi} and \nameref{defi:ll}/\nameref{defi:ups} \emph{if and only if} it is proportional to the \ref{eqn:MS} cost (see \cref{fig:wald-thm}). Therefore, from a positive perspective, the \ref{eqn:MS} cost resolves the trilemma and---equally importantly---demonstrates that the \nameref{defi:ups} model can be justified via optimization of a \hyperref[axiom:prior:invariant]{Prior Invariant} direct cost. However, there is a caveat: the \ref{eqn:MS} cost is the \emph{unique} cost function with either of these virtues and, being defined only for binary-state settings, it is very special. 

\cref{thm:wald} thus identifies a new \emph{modeler's trilemma} among the three modeling desiderata of realism (\nameref{defi:spi}), tractability (\hyperref[defi:ll]{Regularity}/\nameref{defi:ups}), and scalability (to general state spaces). That is, a cost function can satisfy any two of these properties, but cannot satisfy all three.

%, which we refer to as the \emph{modeler's trilemma}. Every modeler desires three things: realism (\nameref{defi:spi}), tractability (Regularity), and generality (general state space). Pick any two; you can't have all three. 

\paragraph{Roadmap.} We review %the remainder of 
the related literature below. \cref{section:model} presents the framework. \cref{sec:C,sec:phi} characterize the class of indirect cost functions and the sequential learning map, respectively. \cref{section:applications} develops the information cost trilemma and other applications. \cref{section:discussion} discusses extensions and open questions. Proofs of our main results are in \cref{app:main}. Auxiliary results and additional proofs are in \dred{Online Appendices} \ref{app:omitted-results}--\ref{app:additional-proofs}.

\subsection{Related Literature}

This paper sits at the intersection of two literatures. First, as summarized above, our sequential learning framework offers a new perspective on the reduced-form cost functions studied in the flexible information acquisition literature.\footnote{See \textcite{mackowiak2023rational} for a recent survey of this literature.} We elaborate on these connections throughout the paper. Second, as we explain here, our work directly relates to several papers that, in effect, study special cases of our framework by analyzing the indirect costs generated by specific direct costs and specific forms of sequential learning. 

%This paper sits at the intersection of two literatures. First, as noted above, our framework offers a new perspective on the reduced-form cost functions studied in the flexible information acquisition literature.\footnote{See \textcite{mackowiak2023rational} for a recent survey of this literature.} We discuss these connections throughout the paper. Second, there are several papers that, in effect, study special cases of our framework. 

The first example of an ``indirect cost'' is due to \textcite{shannon-1948}, who introduces \hyperref[MI]{Mutual Information} and shows that it \emph{approximates} the indirect cost generated by a particular direct cost under which only partitional experiments are feasible. In \cref{ssec:illustrative}, we use our framework to characterize all direct costs that \emph{exactly} generate \hyperref[MI]{Mutual Information}.

%The first example of an indirect cost is due to \textcite{shannon-1948}, who shows that \hyperref[MI]{Mutual Information} \emph{approximates} the indirect cost generated by a specific direct cost. In \cref{ssec:illustrative}, we instead characterize all direct costs that \emph{exactly} generate \hyperref[MI]{Mutual Information}. 

We build on \textcite{morris-strack-sampling}, who show that \nameref{defi:ups} cost functions represent the expected cost of sequentially sampling continuous-time diffusion signals (see \cref{eg:Diffusion:0} in \cref{ssec:leading-examples}). However, \textcite{morris-strack-sampling} rely on two simplifying assumptions: (i) their DM samples from an \emph{exogenous signal process}, choosing only when to stop, and (ii) there are \emph{only two states}. 
%\footnote{When there are more than two states, \textcite{morris-strack-sampling} show that, because the signal process is exogenous, only a ``small set'' of target experiments can be implemented by some stopping strategy.} 
Concurrent to our work, \textcite[Proposition 7]{hebert2023rational} derive a related result for many-state settings where the DM has access to richer signal processes, but nonetheless has an \emph{exogenous preference} for diffusion signals.\footnote{\textcite{hebert2023rational} assume that the DM has a ``preference for gradual learning,'' which is stronger than \nameref{axiom:flie} for ``smooth \hyperref[eqn:PS]{Posterior Separable}'' direct costs and is not well-defined for other direct costs. 
An earlier draft of \textcite{hebert2023rational}, concurrent to \textcite{morris-strack-sampling}, assumed that \emph{only} diffusion signals are feasible.} Both of these results can be viewed as special cases of the ``sufficiency'' direction of our \cref{thm:flie}, which shows that the indirect cost is \nameref{defi:ups} \emph{if} the direct cost \nameref{axiom:flie} (i.e., diffusion sampling is optimal). Meanwhile, the ``necessity'' direction of our \cref{thm:flie} establishes a conceptually and technically novel converse: the indirect cost is \nameref{defi:ups} \emph{only if} the direct cost \nameref{axiom:flie}. More broadly, the novelty of our approach stems from allowing the DM to \emph{flexibly choose} the signal process, imposing \emph{no (a priori) restrictions} on the direct cost, considering the \emph{full class} of indirect costs, and using the \emph{new notion} of \nameref{axiom:slp} to characterize this class. 

There is also a literature that builds on our working paper \parencite{bloedel-zhong-2020}. Several papers adopt \nameref{axiom:slp} as an axiom on reduced-form cost functions in various applications (e.g., \cite{muller2023rational,wong2023dynamic,li2022selling}). \textcite[Section 5]{hebert2023rational} and \textcite{miao2024dynamic} apply \nameref{defi:TI} in dynamic decision problems. %Subsequent to our working paper, 
\textcite[Section 2]{denti2022experimental} revisit the special case of our framework with \hyperref[axiom:prior:invariant]{Prior Invariant} direct costs and develop an extension of our finding (implied by \cref{thm:wald}) that no nonzero, \emph{full-domain} \nameref{defi:ups} cost is \nameref{defi:spi}.

\section{Model}\label{section:model}

\subsection{Primitives}

A Bayesian decision-maker (DM) can acquire information about an unknown \emph{state} $\theta \in \Theta$, where $\Theta$ is a finite set with $|\Theta|\geq 2$. The DM's \emph{beliefs} are denoted by $p, q  \in \Delta(\Theta)$, 
%Formally, we identify $\Delta(\Theta)$ with the simplex in $\R^\Theta$ and endow it with the subspace topology.\footnote{That is, we identify $\Delta(\Theta)$ with the simplex $\left\{x \in \mathbb{R}^\Theta_+ : \sum_{\theta \in \Theta} x_\theta = 1\right\}$, and deem a set $W \subseteq \Delta(\Theta)$ if and only if $W = W' \cap \Delta(\Theta)$ for some open set $W' \subseteq \mathbb{R}^\Theta$. Note that, in this topology, $\Delta(\Theta)$ is open in itself.} 
where $p$ denotes a generic \emph{prior} belief and $q$ denotes a generic \emph{posterior} belief. We endow $\Delta(\Theta)$ with the subspace topology and denote by $\Delta^\circ(\Theta)$ the subset of full-support beliefs.\footnote{More generally, for any Polish (resp., compact metrizable) space $X$, we denote by $\Delta(X)$ the set of Borel probability measures on $X$ equipped with the weak$^*$ topology; this renders $\Delta(X)$ itself a Polish (resp., compact metrizable) space. For any $Y\subseteq X$, we denote by $\Delta(Y)$ the subset of probability measures supported on $Y$, i.e., $\Delta(Y)=\{\pi\in \Delta(X) \mid\supp(\pi)\subseteq Y\}$.
%For the special case of $X = \Theta$, this is equivalent to viewing $\Delta(\Theta)$ as the simplex in $\R^{|\Theta|}$ equipped with the subspace topology, whereby $X\subseteq \Delta(\Theta)$ is deemed open if and only if $X = \widehat{X} \cap \Delta(\Theta)$ for some $\widehat{X} \subseteq \mathbb{R}^{|\Theta|}$ that is open in the Euclidean topology on $\R^{|\Theta|}$.\label{fn:subspace} 
%Note that, under this convention, both $\Delta(\Theta)$ and $\Delta^\circ(\Theta) = \{p \in \Delta(\Theta) \mid p(\theta) >0 \ \forall \theta \in \Theta\}$ are open in $\Delta(\Theta)$.
}

The DM acquires information via \emph{experiments}. Each experiment $\sigma = (S, (\sigma_\theta)_{\theta \in \Theta})$ specifies of a Polish space $S$ of signal realizations and, for each state $\theta \in \Theta$, a conditional distribution $\sigma_\theta \in \Delta(S)$ over signals. Every experiment and prior belief $p$ induce a \emph{random posterior} $\pi \in \Delta(\Delta(\Theta))$ describing the distribution over the DM's signal-contingent Bayesian posteriors $q$, where Bayes' rule requires that $p = \E_\pi [ q ]$. Conversely, every random posterior $\pi$ can be generated by some experiment starting from the prior $p_\pi := \E_\pi [ q ]$. We denote by $\Se$ the class of all experiments, by $\Ex := \Delta(\Delta(\Theta))$ the set of all random posteriors, and by $h_B : \Se \times \Delta(\Theta) \to \Ex$ the Bayesian map that takes experiments and priors to their induced random posteriors.\footnote{%As is standard in set theory, we call $\Se$ a ``class'' because the collection of all Polish spaces is not a well-defined set.
For any experiment $\sigma\in\Se$ and prior $p\in\Delta(\Theta)$, Bayes' rule specifies that the posterior $q^{\sigma,p}(\cdot \mid s) \in \Delta(\Theta)$ conditional on signal $s$ is given by $q^{\sigma,p}(\theta \mid s)= p(\theta) \frac{\d \sigma_{\theta}}{\d \rp{\sigma,p}}(s)$, where $\rp{\sigma,p} := \sum_{\theta \in \Theta} p(\theta) \sigma_\theta \in \Delta(S)$ is the unconditional signal distribution. The induced random posterior is then defined as $h_B(\sigma,p)(B):=\langle \sigma,p\rangle \left( \left\{s \in S \mid q^{\sigma,p}(\cdot \mid s) \in B\right\}\right)$ for all Borel $B\subseteq\Delta(\Theta)$.} We let $\Ex^{\varnothing}:= \bigcup_{p\in \Delta(\Theta)} \{\delta_p\}$ denote the set of \emph{trivial} random posteriors, which correspond to uninformative experiments (i.e., acquiring no information).

A \emph{cost function} is a map $C:\Ex\to \overbar{\R}_+$ that satisfies $C[\Ex^{\varnothing}]=\{ 0\}$, i.e., such that acquiring no information has zero cost.\footnote{We let $\overbar{\R}_+ := [0,+\infty]$ and adopt the usual conventions that $+\infty=+\infty$ and $a + \infty = +\infty$ for all $a \in \mathbb{R}$. \label{fn:infinity}} We make no other \emph{a priori} assumptions about the shape of $C$ or its \emph{effective domain} $\dom(C):= \{\pi \in \Ex  \mid C(\pi) < +\infty\}$, which represents the set of feasible random posteriors.\footnote{For any set $X$ and map $f : X \to (-\infty,+\infty]$, we let $\dom(f) := \{x \in X \mid f(x) <+\infty\}$.} This generality allows us to capture a wide range of settings, including those where $\dom(C)$ is highly restricted and those where $C$ is discontinuous.

Let $\C$ denote the set of all cost functions. We endow $\C$ with addition, multiplication by positive scalars, and the pointwise order $\succeq$.\footnote{That is, $C \succeq C'$ denotes that $C(\pi) \geq C'(\pi)$ for all $\pi \in \Ex$.} With this structure, $\C$ is a convex cone, a complete lattice, and closed under pointwise limits (\Cref{lem:structure:C} in \cref{sec:app:structure}).

Note that, by defining cost functions on random posteriors, we treat the underlying experiment and prior belief as implicit objects. While this ``belief-based approach'' is notationally convenient, it is %also 
often 
%can also be 
useful to make these objects explicit. To this end, note that each $C \in \C$ is equivalent to the corresponding function $C \circ h_B : \Se \times \Delta(\Theta) \to \overline{\R}_+$ over experiments and prior beliefs, where $C(h_B(\sigma,p))$ is the cost assigned to experiment $\sigma \in \Se$ when the DM's prior is $p \in \Delta(\Theta)$. We freely pivot between these conventions as needed. 

\begin{remark}\label{rmk:1}

    %For convenience, we adopt the ``belief-based approach'' by modeling the DM's information acquisition via the random posteriors that it induces and defining information costs as functions of these random posteriors. This approach 
    The ``belief-based approach'' involves two main implicit assumptions: (i) the cost of each experiment may---but need not---vary with the prior belief, and (ii) for each fixed prior, all experiments that generate the same random posterior are assigned the same cost. We revisit assumption (i) in \cref{section:applications}, where we study the special case of our framework with ``\hyperref[axiom:prior:invariant]{Prior Invariant}'' cost functions. We revisit assumption (ii) in \cref{ssec:beyond:belief}, where we show that it is inconsequential for our main analysis, clarify its potential limitations in settings where the prior belief has partial support, and extend our framework to address these limitations. 
\end{remark}

\subsection{Sequential Learning and Indirect Cost}\label{ssec:model-sequential}

Given any cost function $C \in \C$ and ``target'' random posterior $\pi\in \Ex$, the DM solves a cost-minimization problem: to find the cheapest information acquisition procedure that ``produces'' $\pi$. The DM can utilize general sequential learning strategies, in which the number of rounds may be arbitrary and the experiments chosen in later rounds may be contingent on the full history of previously acquired experiments and realized signals. 

We model such strategies recursively, using ``two-step strategies'' as the building blocks. Formally, a \emph{two-step strategy} is a distribution $\Pi \in \Delta(\Ex)$ over random posteriors.
%The building blocks of such procedures are \emph{two-step strategies}, which we model as distributions $\Pi \in \Delta(\Ex)$ over random posteriors. 
Every $\Pi$ specifies: (i) a ``first-round'' random posterior $\pi_1 \in \Ex$ defined as $\pi_1 (B) := \Pi(\{\pi\in \Ex \mid p_{\pi} \in B \})$ for all Borel $B \subseteq \Delta(\Theta)$, and (ii) a collection of ``second-round'' random posteriors $\pi_2 \in \Ex$ defined as the elements of $\supp(\Pi)$. In words, each $\Pi$ describes a two-step contingent plan in which the DM starts from the prior $p_{\pi_1}$, runs a first-round experiment that induces $\pi_1$, %observes the realized first-round signal, 
and then---contingent on the realized first-round signal---runs a second-round experiment that induces the corresponding $\pi_2 \in \supp(\Pi)$.\footnote{We use the formulation of strategies as contingent plans of experiments for our leading examples in \cref{ssec:leading-examples}.} By Bayes' rule, the ``interim'' posteriors $q_1$ drawn from $\pi_1$ and ``terminal'' posteriors $q_2$ drawn from the $\pi_2 \in \supp(\Pi)$ form a martingale process, where $q_1 = p_{\pi_2}$ serves as the ``prior'' for the second round.\footnote{Note that $\supp(\Pi)$ may contain distinct random posteriors $\pi_2 \neq \pi'_2$ corresponding to the same interim belief $q_1 = p_{\pi_2} = p_{\pi'_2}$. This occurs when the first-round experiment generates distinct signals $s_1 \neq s'_1$ that induce the same interim belief $q_1$, but which are used as a ``randomization device'' for determining whether to run $\pi_2$ or $\pi'_2$ in the second round.} The expected second-round random posterior, $\mathbb{E}_\Pi [\pi_2] \in \Ex$, describes the marginal distribution over terminal posteriors $q_2$, that is, the overall information acquired under $\Pi$.

For each target $\pi \in \Ex$, we are interested only in those two-step strategies that generate at least as much information as $\pi$. Formally, a two-step strategy $\Pi \in \Delta(\Ex)$ \emph{produces} the target $\pi \in \Ex$ if $\E_\Pi[\pi_2]$ is a mean-preserving spread (MPS) of $\pi$, which we denote as $\E_\Pi[\pi_2] \geq_\text{mps} \pi$.\footnote{Recall that for any $\pi, \pi' \in \Ex$, we have $\pi' \geq_\text{mps} \pi$ if and only if: (i) $\pi'$ and $\pi$ have the same prior (i.e., $p_{\pi'} = p_\pi$), and (ii) $\pi'$ is induced by an experiment that is Blackwell more informative than the one that induces $\pi$ \parencite{blackwell-experiment51}.} In other words, $\Pi$ produces $\pi$ if $\pi$ can be generated by first running $\Pi$ and then---potentially---``freely disposing'' some of the information acquired under $\Pi$. 

We now %formally 
define the DM's cost-minimization over two-step strategies. For technical convenience, we restrict attention to optimization over the subset of such strategies with finite non-trivial support, denoted as $\Delta^\dag (\Ex) := \{ \Pi \in \Delta(\Ex) \mid |\supp(\Pi) \backslash \Ex^\varnothing| < +\infty\}$.\footnote{This technical restriction ensures that various expectations (e.g., in \cref{defi:2slm}) are well-defined without requiring us to assume that cost functions are measurable; it is without loss of generality for weak$^*$-continuous $C \in \C$.}

\begin{definition} \label{defi:2slm}
The \dred{two-step learning map}\label{defi:slm} $\Psi:\C\to \C$ is defined, for every $C \in \C$ and $\pi \in \Ex$, as
	\begin{align*}
            \Psi(C)(\pi):=\inf_{\Pi \in \Delta^\dag(\Ex)} C(\pi_1)+\E_{\Pi}\left[C(\pi_2) \right] \quad \text{subject to} \quad \E_{\Pi}[\pi_2] \ge_\text{mps}\pi.%\footnotemark
	\end{align*}
\end{definition}
%\footnotetext{The map $\Psi : \C \to \C$ is well-defined because, by construction, we have $C(\pi) \geq \Psi(C)(\pi) \geq 0$ for all $C \in \C$ and $\pi \in \Ex$.}

In words, the cost function $\Psi(C)$ %$\Psi(C)(\pi)$ is 
represents the minimum total expected cost of producing any given target $\pi \in \Ex$ via two-step strategies under the primitive cost function $C$. Note that the minimization problem defining $\Psi$ involves two distinct margins of optimization: \emph{sequential decomposition} and \emph{free disposal} of information. 
%Moreover, a simple form of sequential decomposition is \emph{randomization}: when $\pi_1 \in \Ex^\varnothing$ is trivial, the strategy $\Pi \in \Delta^\dag(\Ex)$ merely randomizes over the experiment to be run in the second round. 
%\footnote{Such randomization is costless in our framework because every $C \in \C$ satisfies the normalization $C[\Ex^\varnothing]=\{0\}$.} 
We will clarify the separate roles played by each of these operations in \cref{ssec:leading-examples,ssec:indirect:cost} below.

\iffalse
An information cost is \emph{sequential learning-proof} if it is a fixed point of $\Psi$; namely, under direct cost $C$, acquiring information in two steps is never strictly better than doing so in one shot.

\begin{definition}[SLP]\label{axiom:slp}
	$C\in\C$ is \dred{Sequential Learning-Proof} (\nameref{axiom:slp}) if $\Psi(C)=C$. 
\end{definition}
\fi

Next, we extend the DM's optimization to sequential strategies of arbitrary length.

\begin{definition}\label{defi:slm}
\hspace{-0.45em}
The \dred{sequential learning map} $\Phi: \C\to \C$ is defined, for every $C \in \C$ and $\pi \in \Ex$, as 
\[
\Phi(C)(\pi):=\lim\limits_{n\to \infty} \Psi^n(C)(\pi).\footnotemark
\]
\end{definition}
\footnotetext{The map $\Phi : \C \to \C$ is well-defined because, by construction, $\Psi^n(C)(\pi) \geq \Psi^{n+1}(C)(\pi) \geq 0$ for all $C \in \C$, $\pi \in \Ex$, $n \in \mathbb{N}$.}

In words, the cost function $\Phi(C)$ %$\Psi(C)(\pi)$ is 
represents the minimum total expected cost of producing any given target $\pi \in \Ex$ via \emph{fully flexible} sequential learning under the primitive cost function $C$. We model this optimization by recursively applying the two-step learning map $\Psi$. 
%\footnote{We adopt this convention for notational and expositional simplicity. Our working paper \parencite{bloedel-zhong-2020} offers an equivalent but more cumbersome formulation of $\Phi$ that explicitly models the space of all sequential strategies.} 
Intuitively, each application of $\Psi$ doubles the number of rounds over which the DM can learn, so that $\Psi^n$ models optimization over ``$2^n$-round'' strategies and $\Phi$ represents the infinite-horizon limit. We note that the ``rounds'' of such strategies \emph{need not} correspond to fixed ``periods'' of calendar time. In particular, we will often interpret the $n \to \infty$ limit as approximating a continuous-time setting in which each unit of calendar time is subdivided into many small increments (e.g., see \cref{ssec:leading-examples} below).

We now use the two-step and sequential learning maps, $\Psi$ and $\Phi$, to state our two main definitions. Our first main definition distinguishes between the %inputs and outputs 
\emph{domain} and \emph{range} 
of $\Phi$.

\begin{definition}[Indirect Cost]\label{defi:IC}
    For any $C \in \C$, the cost function $\Phi(C) \in \C$ is the \dred{indirect cost} generated by the \dred{direct cost} $C$. The set of all indirect cost functions is %denoted by 
    $\C^* :=  \{\Phi(C) \mid C \in \C\}$.  
\end{definition}

We interpret the direct cost $C$ as the DM's ``primitive'' information acquisition technology and the indirect cost $\Phi(C)$ as his ``reduced-form'' cost of information. Under this interpretation, if the DM is endowed with the direct cost $C$ and can engage in sequential cost-minimization before facing a one-time decision problem, then his learning incentives and optimal behavior in that decision problem are determined by the indirect cost $\Phi(C)$.

These objects are analogous to classic concepts from producer theory, viz., the firm's cost-minimization problem. Under this analogy, the indirect cost $\Phi(C)$ corresponds to the firm's ``cost function'' for producing ``output bundles'' $\pi \in \Ex$ at fixed ``input prices'' given by the direct cost $C$, the sequential learning map $\Phi$ models optimization over a rich set of ``production plans,'' and $\C^*$ represents the set of all ``rationalizable'' cost functions.

Our second main definition adopts a distinct perspective: rather than distinguishing between primitive and reduced-form cost functions, it considers costs that are ``robust'' to the possibility of further optimization. We formalize this idea via the \emph{fixed points} of $\Psi$.

\begin{definition}[SLP]\label{axiom:slp}
	$C\in\C$ is \dred{Sequential Learning-Proof} (\nameref{axiom:slp}) if $\Psi(C)=C$. 
\end{definition}

In words, a cost function is \nameref{axiom:slp} if and only if it cannot be reduced through \emph{two-step} optimization. We interpret \nameref{axiom:slp} as a mild ``internal consistency'' desideratum for modeling reduced-form information costs. In particular, if the DM's cost function $C$ were \emph{not} \nameref{axiom:slp} then, at least for some target $\pi \in \Ex$, he would be able to pay the strictly lower cost $\Psi(C)(\pi) < C(\pi)$ by using a simple two-step strategy. Therefore, from the perspective of a modeler who may be either unwilling or unable to fully specify the DM's strategy space,\footnote{A modeler may be unwilling to do so for the sake of tractability, and unable to do so in settings where the strategy space is difficult to observe empirically (e.g., when studying the DM's cognitive costs of internal information processing).} non-\nameref{axiom:slp} cost functions necessarily have features that the DM might be able to optimize away---\emph{even if} the DM does not actually have access to the full set of sequential strategies. 

\begin{remark}\label{remark:sec-2-2}
    Our sequential learning framework imposes two main implicit assumptions: (i) it abstracts away from the DM's time-preference for decision-making (e.g., discounting), and (ii) it presumes that any restrictions on the DM's strategy space are ``stationary,'' i.e., can be represented as domain restrictions on the DM's direct cost, which is history-independent. We make assumption (i) deliberately and view it as important for the portability of our framework.\footnote{This assumption lets us focus on the DM's ``inner'' cost-minimization problem while remaining agnostic about his ``outer'' utility-maximization problem. Such separation between the cost and value of information is needed to ensure that the cost functions we study can be applied in any downstream decision problem that the DM might face, which is the standard interpretation of information cost functions in the literature. By contrast, nontrivial time preferences (e.g., discounting) would make it conceptually difficult to disentangle the gains from deferred learning effort and the losses from delayed action, which are inherently decision-problem-specific \parencite{moscarini-smith-ecma2001,zhong2022optimal}.} We revisit assumption (ii) in \cref{ssec:beyond:flexibility}, where we extend the framework to accommodate arbitrary restrictions on the DM's strategy space and general forms of history-dependent direct costs.
\end{remark}

\subsection{Illustrative Examples: Wald Sampling}\label{ssec:leading-examples}

We illustrate the framework via two simple examples, which we will revisit throughout the paper. In both, the state space $\Theta=\left\{ 0,1 \right\}$ is binary and the DM samples from a fixed parametric class of experiments, as in the classic setting of \textcite{wald-ams1945,wald-ecma1947}. The first example, which highlights the role of \emph{sequential decomposition}, resembles canonical models of diffusion learning \parencite{moscarini-smith-ecma2001,morris-strack-sampling,fudenberg2018speed}. The second example, which highlights the role of \emph{free disposal}, resembles canonical models of Poisson learning \parencite{che-mierendorff-aer2019}.\footnote{Diffusion and Poisson learning models also feature prominently in mathematical statistics \parencite[Ch. VI]{peskir2006optimal} and the literature on bandit experimentation \parencite{bolton-harris-ecma99,krc05}.}

\begin{eg}[Diffusion Sampling]\label{eg:Diffusion:0}
\begin{table}[t]
    \centering
    \renewcommand{\arraystretch}{1.2}
    \begin{tabular}{c|cc}
    \hline
         & $s_0$ & $s_1$\\
         \hline\hline
        $\theta=0$ & $e^\ell/(1+e^{\ell})$ & $1/(1+e^{\ell})$\\
        $\theta=1$ & $1/(1+e^{\ell})$ & $e^\ell/(1+e^{\ell})$\\
        \hline
    \end{tabular}
     \caption{\centering Bernoulli experiment with log-likelihood ratio $\ell$.}
     \label{tab:Bernoulli-diff}
\end{table}
    \textbf{Direct cost of Bernoulli signals:} The DM's primitive experiments generate the symmetric Bernoulli signals described in \cref{tab:Bernoulli-diff}. Let $\sigma^\ell \in \Se$ denote the Bernoulli experiment with log-likelihood ratio (LLR) parameter $\ell \in \R_+$. Conditional on each state $\theta$, the experiment $\sigma^\ell$ yields a signal $s_\theta$ in favor of the true state with probability $\sigma^\ell_\theta(s_\theta) = e^\ell / (1+e^\ell)$, and a signal $s_{1-\theta}$ in favor of the opposite state $1-\theta$ with the remaining probability. The information content of each signal $s \in \{s_0,s_1\}$ is summarized by the LLR
    \[
    \log\left( \frac{\sigma_1^{\ell}(s)}{\sigma_0^{\ell}(s)} \right)  \, = \, \begin{cases} 
    +\ell, & \text{if $s = s_1$} \\
    %%%
    -\ell, & \text{if $s = s_0$,}
    \end{cases}
    \]
    where more positive (resp., negative) values represent stronger evidence for (resp., against) state $\theta = 1$, and a value of zero corresponds to a completely uninformative signal.
    %\footnote{Given any prior belief, the Bayesian posterior that $\theta=1$ is strictly increasing in this LLR. We work with LLRs (rather than beliefs) in this example due to their convenient additivity properties, which we discuss further below.} 

    The DM's direct cost $C \in \C$ is defined, for such experiments, as
    \begin{equation}\label{eqn:bernoulli-DC}
    C(h_B(\sigma^\ell,p))  = f(\ell) \quad \text{ for all }  
\, \ell \in \R_+ \, \text{ and } \, p \in \Delta^\circ(\Theta),
    \end{equation}
    where $f : \R_+ \to \R_+$ is twice differentiable at $\ell =0$ with $f(0) = f'(0)  = 0$ and $f''(0)> 0$. All other non-trivial random posteriors are infeasible (i.e., excluded from $\dom(C)$). 
    
    %This technology can be interpreted as modeling the process of sampling from a population, where each Bernoulli experiment represents an individual draw and $f$ represents the cost of increasing each draw's precision. In what follows, we analyze the cost of the ``incremental learning'' strategy that sequentially acquires many low-precision draws.

    We can interpret each Bernoulli experiment as an independent draw from a large population, and $f$ as modeling the cost of each draw's precision. In what follows, we analyze the ``incremental learning'' strategy that sequentially acquires many low-precision draws.
    %%%%
    \\
    \textbf{A simple sequential learning strategy:} Suppose that the target experiment is $\sigma^\ell$ for some $\ell>0$. We begin by illustrating how to produce $\sigma^\ell$ using multiple copies of $\sigma^{\ell/2}$. 
    
    To this end, we recall that LLRs are additive under repeated experiments. For instance, if two i.i.d.~draws from $\sigma^{\ell/2}$ yield the signals $s$ and $s'$, then the LLR of the compound signal $(s,s')$ is the sum of the individual LLRs, which equals $+\ell$ if both signals are $s_1$, equals $-\ell$ if both signals are $s_0$, and equals $0$ if the signals disagree. Therefore, the target $\sigma^{\ell}$ can be produced via the following ``Bernoulli random walk'' strategy: (i) acquire two copies of $\sigma^{\ell/2}$, (ii) stop if their signals agree, and (iii) repeat the process if their signals disagree (see the dashed arrows in \cref{fig:random-walk}). Since the cost of acquiring two copies is $2 f(\ell/2)$ and their signals disagree with probability $ 2 e^{\ell/2} / \big(1+e^{\ell/2}\big)^2$, the expected cost of this strategy equals
    %%%
    \begin{equation}\label{eqn:bernoulli-twostep}
    2 f(\ell / 2) \cdot \sum_{k=0}^\infty \left(\frac{2 e^{\ell/2}}{ \big(1+e^{\ell/2}\big)^2 }\right)^{k} \, = \, 2 f(\ell / 2) \cdot \frac{\big( 1+e^{\ell/2} \big)^2}{1+e^\ell}. %\notag
    \end{equation}

    %%%%%
    %%%%%
    %%%%%
        \begin{figure}[t]
    \centering
    \include{Figure/random_walk}
    \vspace{-2.5em}
 \caption{\centering Producing a Bernoulli experiment via Bernoulli random walks.}\label{fig:random-walk}
\end{figure}
    %%%%%
    %%%%%
    %%%%%
    
    \noindent\textbf{The incremental learning limit:} Note that the target $\sigma^\ell$ can be further decomposed, as each copy of $\sigma^{\ell/2}$ in the above strategy can itself be  replicated by sampling from $\sigma^{\ell/4}$ via an analogous Bernoulli random walk. Therefore, by ``stitching together'' such replications, we can produce the target $\sigma^\ell$ via a finer Bernoulli random walk (with smaller step size $\pm \ell/4$) by sampling from $\sigma^{\ell/4}$ rather than $\sigma^{\ell/2}$ (see the solid arrows in \cref{fig:random-walk}). To compute the expected cost of this new strategy, we simply replace the $f(\ell/2)$ direct cost term in \eqref{eqn:bernoulli-twostep} with the expected cost of replicating $\sigma^{\ell/2}$, which equals $2 f(\ell/4) \cdot \big(1+e^{\ell/4}\big)^2/(1+e^{\ell/2})$.
    %
    %$2 f(\ell/4) \cdot \big(1+e^{\ell/4}\big)^2/(1+e^{\ell/2})$, the expected cost of replicating $\sigma^{\ell/2}$.

    Applying this logic recursively, we see that \emph{for any $n \in \mathbb{N}$}, the DM can produce the target $\sigma^\ell$ by sampling from $\sigma^{\ell/2^n}$ via a Bernoulli random walk with step size $\pm \ell / 2^n$, and the expected cost of this strategy equals $2^{n} f(\ell / 2^n) \cdot  \prod_{k=1}^n \big(1+e^{\ell/2^{k}}\big)^2 / \big( 1+e^{\ell/2^{k-1}}\big)$. As $n \to \infty$, each draw becomes vanishingly informative and the expected cost converges to\footnote{This limit can be evaluated by: (i) noting that $\lim_{n \to \infty} \left(2^n\right)^2 f(\ell/2^n) = \frac{1}{2}f''(0) \ell^2$ by Taylor's theorem (since $f(0) = f'(0) = 0$), and (ii) directly calculating that $\lim_{n \to \infty} \frac{1}{2^n} \cdot \prod_{k=1}^n \big(1+e^{\ell/2^{k}}\big)^2 /\big(1+e^{\ell/2^{k-1}}\big) = \frac{2}{\ell} \cdot \big(e^\ell-1\big) / \big(1+e^\ell\big)$.} 
     %%%%%%%
    %%%%%
    %%%%%
    \begin{equation}\label{eqn:bernoulli-incremental}
    \frac{1}{2} f''(0)  \cdot  \frac{2 \ell \big( e^\ell -1\big)}{1+ e^{\ell}} 
    %\, = \, f''(0)\cdot \bigg[\underbrace{\sigma^\ell_1(s_1)\log\left(\frac{\sigma^\ell_1(s_1)}{\sigma^\ell_0(s_1)}\right)+\sigma^\ell_1(s_0)\log\left(\frac{\sigma^\ell_1(s_0)}{\sigma^\ell_0(s_0)}\right)}_{D_{\text{KL}}(\sigma^\ell_1 \mid \sigma^\ell_0)}\bigg]
    %\, = \, f''(0)\cdot \big[\underbrace{\sigma^\ell_1(s_1) L^\ell(s_1)+\sigma^\ell_1(s_0)L^\ell(s_0)}_{D_{\text{KL}}(\sigma^\ell_1 \mid \sigma^\ell_0)}\big],
    \, = \, f''(0)\cdot \big[\sigma^\ell_1(s_1)\cdot (+\ell)+\sigma^\ell_1(s_0)\cdot (-\ell)\big] \, = \, f''(0)\cdot D_{\text{KL}}(\sigma^\ell_1 \mid \sigma^\ell_0),
    %\, = \, f''(0)\cdot D_{\text{KL}}(\sigma^\ell_1 \mid \sigma^\ell_0), \quad \text{ where } \quad D_{\text{KL}}(\sigma^\ell_1 \mid \sigma^\ell_0):= \sigma^\ell_1(s_1) L^\ell(s_1)+\sigma^\ell_1(s_0)L^\ell(s_0) 
    \end{equation}
    where $D_{\text{KL}}(\sigma^\ell_1 \mid \sigma^\ell_0)$ is the Kullback-Leibler (KL) divergence between the target experiment's state-contingent signal distributions, a well-known notion of statistical distance.%\footnote{The limiting cost in \eqref{eqn:bernoulli-incremental} can be derived by: (i) noting that $\lim_{n \to \infty} \left(2^n\right)^2 f(\ell/2^n) = \frac{1}{2}f''(0) \ell^2$ by Taylor's theorem (since $f(0) = f'(0) = 0$), and (ii) directly calculating that $\lim_{n \to \infty} \frac{1}{2^n} \cdot \prod_{k=1}^n \big(1+e^{\ell/2^{k}}\big)^2 /\big(1+e^{\ell/2^{k-1}}\big) = \frac{2}{\ell} \cdot \big(e^\ell-1\big) / \big(1+e^\ell\big)$.} 

    To interpret this expression, we note that, as $n \to \infty$, the Bernoulli random walk strategy converges to a continuous-time diffusion strategy under which: (i) the cumulative LLR process $(L_t)_{t\geq0}$ follows a standard Brownian motion $(W_t)_{t\geq 0}$ with state-dependent drift, %viz.,
     \begin{equation}\label{eqn:Brownian-motion}
    \d L_t = \left( \theta  - 1/2\right) \d t + \d W_t %\sim N(\theta\, t, t)
    \qquad \text{ for }  t \in \R_+; %\notag
    \end{equation}
    (ii) the DM pays a flow cost of $\frac{1}{2} f''(0) $ per instant; and (iii) the DM stops sampling at the first time $\tau \in \R_+$ such that $\left| L_\tau \right| \geq \ell$.\footnote{% In particular, by well-known approximation results, sampling the process \eqref{eqn:Brownian-motion} for $t >0$ units of continuous time is asymptotically equivalent to dividing the interval $[0,t]$ into discrete rounds of length $\delta^{(n)} := (\ell/2^n)^2$, acquiring $\big \lfloor t / \delta^{(n)} \big\rfloor\in \mathbb{N}$ draws from $\sigma^{\ell / 2^n}$, and taking $n \to \infty$. Since $f(0) = f'(0) = 0$, Taylor's theorem implies that, for any fixed $t>0$, the cost of this approximation converges to $\lim_{n \to \infty} f(\ell / 2^n) \cdot \left \lfloor \frac{t}{ \delta^{(n)} }\right\rfloor \, = \, \lim_{n \to \infty} \frac{1}{2} f''(0) \delta^{(n)} \cdot \left \lfloor \frac{t}{ \delta^{(n)} }\right\rfloor \, = \, \frac{1}{2} f''(0) \cdot t$. 
    Formally, for each $t \in \R_+$ and $n \in \mathbb{N}$, we divide the time interval $[0,t]$ into discrete rounds of length $\delta^{(n)} := (\ell/2^n)^2$, acquire $\big \lfloor t / \delta^{(n)} \big\rfloor\in \mathbb{N}$ draws from $\sigma^{\ell / 2^n}$ at the per-round cost $f(\ell/2^n) \approx \frac{1}{2} f''(0) \delta^{(n)}$, and define the random variable $L_t^{(n)}$ as the sum of the LLRs generated by the realized signals. Donsker's Theorem then yields $\lim_{n \to \infty}L_t^{(n)} \equiv L_t$ as defined in \eqref{eqn:Brownian-motion}.
    %[Formally, let $\{s^k\}$ be the realized signals of iid copies of $\sigma^{\ell/2^n}$. Let $S^n_t=\frac{\ell}{2^n}(\text{\# of $s^k=s_1$ -\# of $s^k=s_0$})_{k\le 2^{2n}\cdot t}$. Then, the Donsker's theorem implies that $S^n_t$ converges in distribution to $S_t-\frac{1}{2}t$, where the drift $\frac{1}{2}t$ is informationally inconsequential. Per $\d t =\frac{1}{2^{2n}}$ unit of time, the cost is $f(\ell/2^n)\approx\frac{1}{2}f''(0)\d t$.]
    } Thus, \eqref{eqn:bernoulli-incremental} represents the total expected cost of sampling the diffusion process \eqref{eqn:Brownian-motion}, where the expected stopping time is $\E[\tau] = 2 \cdot  D_\text{KL} (\sigma^\ell_1 \mid \sigma^\ell_0)$. \\
    %%%%
    %%%
     \textbf{An (\nameref{axiom:slp}) upper bound on the indirect cost:} The problem of sampling from the diffusion \eqref{eqn:Brownian-motion} at the constant flow cost $\frac{1}{2}f''(0) $ is studied in \textcite[Proposition 3]{morris-strack-sampling}. Their analysis implies that, by choosing a suitable stopping time, essentially \emph{any} target experiment can be produced at expected cost equal to $f''(0)$ times the \emph{\ref{eqn:MS}} cost function, which is defined as follows: for every $\sigma \in \Se$ and $p \in \Delta^\circ(\Theta)$ such that $\supp(h_B(\sigma,p))\subseteq \Delta^\circ(\Theta)$, 
     \begin{align}\label{eqn:MS}\tag{Wald}
     C_{\text{Wald}}(h_B(\sigma,p)) &:= p(0) \, D_{\text{KL}}(\sigma
     _0 \mid \sigma_1) + p(1) \, D_{\text{KL}}(\sigma_1 \mid \sigma_0),
    \end{align} 
    where the KL divergence $D_\text{KL}(\sigma_{\theta} \mid \sigma_{1-\theta}) :=\int_S \log \big(\frac{\d \sigma_{\theta}}{\d \sigma_{1-\theta}}(s) \big)\dd \sigma_{\theta}(s)$ represents the expected LLR conditional on state $\theta$. Note that, for any Bernoulli experiment $\sigma^\ell$, symmetry implies that $D_\text{KL}(\sigma^\ell_0\mid \sigma^\ell_1)  = D_\text{KL}(\sigma^\ell_1\mid \sigma^\ell_0)$, so \eqref{eqn:MS} reduces to the expression \eqref{eqn:bernoulli-incremental} derived above.

    Since the incremental learning strategy is not necessarily optimal, we conclude that the \ref{eqn:MS} cost yields an upper bound on the DM's indirect cost. That is, $\Phi(C) \preceq f''(0) \cdot C_\text{Wald}$.%\footnote{Formally, this upper bound will follow from (the proof of) our \cref{thm:qk}(i). See \cref{ssec:bounding-phi} for details.}
    %\footnote{See \cref{ssec:bounding-phi} for a self-contained proof of this bound via our \cref{thm:qk}(i), rather than \textcite{morris-strack-sampling}.}
    %\footnote{The formal proof of this bound is achieved (via the proof of the more general \cref{thm:qk} (i)) by approximating the continuous-time diffusion process in discrete-time.}
    \footnote{Formally, this upper bound follows from the proof of our \cref{thm:qk}(i), which generalizes the above random walk approximation of continuous-time diffusion strategies. See  \cref{ssec:bounding-phi} for discussion and \cref{sssec:app:thm:qk-pt1} for details.}

    In \cref{sec:C}, we characterize precisely when this bound is tight: $\Phi(C) = f''(0) \cdot C_\text{Wald}$. This characterization will hinge on the fact that, as we show there, the \ref{eqn:MS} cost is \nameref{axiom:slp}. %Informally, we will see that the \ref{eqn:MS} cost is \nameref{axiom:slp} because, being an expectation over LLRs, it is additive under repeated experiments and thus cannot be reduced via further sequential decomposition. 
    %This characterization will hinge on the \ref{eqn:MS} cost being \nameref{axiom:slp}, which follows from our analysis in next section. %Informally, we will see that the \ref{eqn:MS} cost is \nameref{axiom:slp} because, being an expectation over LLRs, it is additive under repeated experiments and thus cannot be reduced via further sequential decomposition. 

\end{eg}

\begin{eg}[Poisson Sampling]\label{eg:Poisson:0}
\textbf{Direct cost of Poisson signals:} The DM's primitive experiments are the ``Poisson dilutions'' of full information described in \cref{tab:Dilution-pois}. 
%\footnote{We call these experiments ``dilutions'' of full information in anticipation of the terminology in \cref{ssec:ups} below.} 
%%%
\begin{table}[t]
    \centering
   \centering
   \renewcommand{\arraystretch}{1.2}
    \begin{tabular}{c|ccc}
    \hline
         & $s_0$ & $s_1$ & $s_\varnothing$\\
         \hline\hline
        $\theta=0$ & $1-e^{-\lambda}$ & 0 & $e^{-\lambda}$ \\
        $\theta=1$ & 0 & $1-e^{-\lambda}$ & $e^{-\lambda}$ \\
        \hline
    \end{tabular}
    \caption{\centering Poisson dilution experiment with hazard rate $\lambda$.}
    \label{tab:Dilution-pois}
\end{table}
Let $\sigma^\lambda \in \Se$ denote the Poisson dilution with hazard rate $\lambda \in \overline{\R}_+$. Each experiment $\sigma^\lambda$ either generates a signal $s_\theta$ that fully reveals the state $\theta$, which occurs with probability $1-e^{-\lambda}$, or yields a completely uninformative ``null'' signal $s_\varnothing$, which occurs with probability $e^{-\lambda}$.  The DM's direct cost $C \in \C$ is defined, for such experiments, to be %linear in 
the probability of receiving a revealing signal:
 \begin{equation}\label{eqn:poisson-DC}
     C(h_B(\sigma^\lambda,p))  = 1-e^{-\lambda} \quad \text{ for all }  
 \, \lambda \in \overline{\R}_+ \, \text{ and } \, p \in \Delta^\circ(\Theta).
\end{equation}
     All other non-trivial random posteriors are infeasible (i.e., excluded from $\dom(C)$).

We interpret this technology as modeling the outcome of sampling from a continuous-time Poisson process that generates a single, fully revealing signal with unit arrival rate. Specifically, if the DM samples from such a process until the deterministic time $\lambda \in \overline{\R}_+$ and incurs a unit flow cost per instant (until the signal arrives), then both the probability of receiving the signal and the expected cost of sampling are given by $\int_0^\lambda e^{-t} \dd t = 1-e^{-\lambda}$.%\footnote{Versions of this direct cost and interpretation are common in the literature (e.g., \cite{dgs_aer2016}). To provide explicit foundations for this interpretation, one could adopt an approach analogous to \cref{eg:Diffusion:0} by: (i) specifying a ``more primitive'' direct cost function (e.g., a fixed cost of acquiring full information or a cost that is convex in the probability of doing so), and (ii) showing that the continuous-time Poisson sampling strategy (asymptotically) minimizes the cost of producing Poisson dilutions, at expected cost proportional to $C$. We bypass this step for brevity.}

\noindent\textbf{A simple Poisson-with-free-disposal strategy:} In contrast to the Bernoulli technology from \cref{eg:Diffusion:0}, the Poisson technology here generates discrete ``chunks'' of information that cannot be decomposed. Therefore, to produce any target experiment outside of the Poisson dilution class, the DM must acquire extra information and then use free disposal.

To illustrate how this can be done, fix any target experiment $\sigma = (S,\sigma_0,\sigma_1) \in \Se$ for which the signal space $S$ is finite (see \cref{tab:target-finite}). Consider the associated experiment $\widehat{\sigma} \in \Se$ described in \cref{tab:Poisson-covering}. First, by comparing  \cref{tab:target-finite,tab:Poisson-covering}, we see that $\widehat{\sigma}$ is Blackwell more informative than the target $\sigma$, as each pair of signals $\{s'_i,s''_i\}$ generated by the former can be ``pooled'' into the corresponding signal $s_i$ generated by the latter. 
%\footnote{Formally, this is witnessed by the garbling $\gamma : \widehat{S} \to \Delta(S)$ defined as $\gamma(s_i \mid s'_i) = \gamma(s_i \mid s''_i) = 1$ for all $i \in \{1,\dots, k\}$.} 
Second, note that, under $\widehat{\sigma}$, each signal $s'_i$ that arises with positive probability is fully informative and each signal $s''_i$ is uninformative. Therefore, by comparing \cref{tab:Dilution-pois,tab:Poisson-covering}, we see that $\widehat{\sigma}$ is Blackwell equivalent to the Poisson dilution experiment $\sigma^{\widehat{\lambda}}$, where $\widehat{\lambda}= - \log \big( \sum_{s \in S} \min\left\{\sigma_0(s), \sigma_1(s)\right\} \big)$.

We conclude that the DM can produce $\sigma$ via the two-step strategy that acquires $\sigma^{\widehat{\lambda}}$ in one round and then utilizes free disposal. For any prior $p \in \Delta^\circ(\Theta)$, this strategy costs
\begin{equation}\label{eqn:TV-1}
C(h_B(\sigma^{\widehat{\lambda}},p)) \, = \, 1 - \sum_{s \in S} \min\left\{\sigma_0(s), \sigma_1(s)\right\}  \, = \, \frac{1}{2} \, \sum_{s \in S} \left| \sigma_0(s) - \sigma_1(s) \right|. 
\end{equation}
Equivalently, under the continuous-time interpretation, the DM produces the target $\sigma$ by sampling from a fully revealing Poisson process for (up to) $\widehat{\lambda}$ units of continuous time.
\\
%%%
%%
\begin{table}[t]
    %\centering
\begin{minipage}{0.36\linewidth}\centering
    \label{tab:sigma}
    \renewcommand{\arraystretch}{1.1}
    \begin{tabular}{c|ccc}
    \hline
         &  $\cdots$ & $s_i$& $\cdots$\\
         \hline\hline
        $\theta=0$ &  $\cdots$& $\sigma_0(s_i)$& $\cdots$  \\
        $\theta=1$ &  $\cdots$& $\sigma_1(s_i)$& $\cdots$  \\
        \hline
    \end{tabular}   
    \caption{\centering Target experiment $\sigma$ \\ with signal space $S = \{s_1, \dots, s_k\}$.} \label{tab:target-finite}
\end{minipage}
\hspace{1.5em}
 \begin{minipage}{0.59\linewidth}\centering
        %%%
    \label{tab:sigma:1}
    \renewcommand{\arraystretch}{1.1}
    \begin{tabular}{c|cccc}
    \hline
         & $\cdots$ & $s'_i$ & $s''_i$ & $\cdots$\\
         \hline\hline
        $\theta=0$ & $\cdots$  & $\sigma_0(s_i)-\min_{\theta}\sigma_\theta(s_i)$ & $\min_{\theta}\sigma_\theta(s_i)$ & $\cdots$ \\
        $\theta=1$ & $\cdots$ &$\sigma_1(s_i)-\min_{\theta}\sigma_\theta(s_i)$ & $\min_{\theta}\sigma_\theta(s_i)$ & $\cdots$ \\
        \hline
    \end{tabular}    
    \caption{\centering More informative experiment $\widehat{\sigma}$  \\ with signal space $\widehat{S} = \{s_1',s_1'', \dots, s_k', s_k''\}$.}\label{tab:Poisson-covering}
 \end{minipage}
\end{table}
\textbf{An (\nameref{axiom:slp}) upper bound on the indirect cost:} More generally, an analogous Poisson-with-free-disposal strategy can be used to produce \emph{any} target experiment at cost equal to the total variation distance between the state-contingent signal distributions. Formally, this cost is described by the \emph{Total Variation} \eqref{eqn:TV} cost function: for every $\sigma \in \Se$ and $p \in \Delta^\circ(\Theta)$,
\begin{equation}
C_\text{TV}(h_B(\sigma,p)) \, := \, \| \sigma_0 - \sigma_1\|_\text{TV} \, = \, \sup_{\text{Borel} \, B\subseteq S} \left| \sigma_0(B) - \sigma_1(B) \right|. \tag{\text{TV}}\label{eqn:TV}
\end{equation}
Note that, for finite-support experiments, \eqref{eqn:TV} reduces to the expression \eqref{eqn:TV-1} derived above.\footnote{\textcite[Section V.A]{che-mierendorff-aer2019} and \textcite[Example 4]{zhong2022optimal} use special cases of \eqref{eqn:TV-1} to model the flow cost of Poisson signals in dynamic decision problems.}

Since the Poisson-with-free-disposal strategy is not necessarily optimal, we conclude that the \ref{eqn:TV} cost provides an upper bound on the DM's indirect cost. That is, $\Phi(C) \preceq C_\text{TV}$. 

We show in \cref{sec:C} that this bound is tight: $\Phi(C) = C_\text{TV}$. As in \cref{eg:Diffusion:0}, this will hinge on the fact that the \ref{eqn:TV} cost is \nameref{axiom:slp}, which follows from our general analysis below.

\end{eg}

\section{The Indirect Cost of Information}\label{sec:C}

In this section, we characterize the set $\C^*$ of indirect costs. \cref{ssec:indirect:cost} establishes that it equals the set of \nameref{axiom:slp} costs. \cref{ssec:ups} analyzes the subset of ``differentiable'' \nameref{axiom:slp} costs.

\subsection{Foundations for Sequential Learning-Proofness}\label{ssec:indirect:cost}

The notions of indirect cost and \nameref{axiom:slp} are distinct desiderata for modeling ``reduced-form'' information costs: indirect costs model the \emph{outcome} of fully flexible sequential learning, while \nameref{axiom:slp} costs model \emph{robustness} to the possibility of (two-step) sequential learning. Our first main result shows that these two notions are, in fact, equivalent and simplifies the task of determining whether a cost function is \nameref{axiom:slp}. We begin with two definitions.

\begin{axiom}\label{axiom:mono}
	$C\in \C$ is \hyperref[axiom:mono]{Monotone} if $C(\pi)\le C(\pi')$ for all $\pi, \pi'\in \Ex$ such that $\pi\le_\text{mps}\pi'$.
\end{axiom}

In words, a cost function is \hyperref[axiom:mono]{Monotone} if acquiring more information is always weakly more costly. \hyperref[axiom:mono]{Monotonicity} thus represents \emph{robustness to free disposal} of information.

\begin{axiom}\label{axiom:POSL}
	$C\in \C$ is \hyperref[axiom:POSL]{Subadditive} if $C(\E_{\Pi}[\pi_2])\le C(\pi_1)+\E_{\Pi}\left[C(\pi_2) \right]$ for all $\Pi\in \Delta^\dag(\Ex)$. 
\end{axiom}

In words, a cost function is \hyperref[axiom:POSL]{Subadditive} if acquiring information directly is always weakly cheaper than producing it via two-step strategies \emph{without free disposal}. \hyperref[axiom:POSL]{Subadditivity} thus represents \emph{robustness to sequential decomposition} of information. When restricted to trivial $\pi_1 \in \Ex^\varnothing$, \hyperref[axiom:POSL]{Subadditivity} reduces to \hyperref[lem:C^*:convex]{Convexity} with respect to mixtures of experiments (\cref{lem:C^*:convex} in \cref{ssec:app:thm:UPS}), which represents \emph{robustness to randomization}.\footnote{\hyperref[axiom:mono]{Monotonicity} and \hyperref[lem:C^*:convex]{Convexity} are often viewed as ``canonical'' properties of reduced-form cost functions because they cannot be falsified using standard data on the DM's choice behavior  \parencite{ddmo-te2017,caplin_dean_aer2015}.}

Given these definitions, we have two equivalent characterizations of indirect costs.

\begin{thm}\label{prop:1}
	For every $C\in\C$,  
 \[
 C\in\C^*  \ \ \iff   \ \ C  \ \ \text{is \nameref{axiom:slp}} \ \  \iff \ \ \text{$C$ is \hyperref[axiom:mono]{Monotone} and \hyperref[axiom:POSL]{Subadditive}}.
 \]
\end{thm}

\begin{proof}
    See \cref{ssec:app:prop:1}.
\end{proof}

First, \cref{prop:1} shows that $C$ is an indirect cost \emph{if and only if} $C$ is \nameref{axiom:slp}. In other words, \nameref{axiom:slp} fully characterizes the ``context-free'' implications of sequential optimization: $C$ is \nameref{axiom:slp} \emph{if and only if} there exists \emph{some} direct cost $C' \in \C$ such that $C = \Phi(C')$.\footnote{Moreover, it is easy to see from \cref{defi:slm} that $C \in \C$ is \nameref{axiom:slp} \emph{if and only if} $\Phi(C)=C$, i.e., $C$ is \emph{its own} indirect cost. 
} This characterization, which is analogous to the classic principle of dynamic programming, provides a foundation for using \nameref{axiom:slp} as a standalone definition for ``reduced-form'' cost functions.

Second, \cref{prop:1} shows that \nameref{axiom:slp} is \emph{equivalent} to the conjunction of \hyperref[axiom:mono]{Monotonicity} and \hyperref[axiom:POSL]{Subadditivity}. This decouples the operations of free disposal and sequential decomposition. It also reduces \nameref{axiom:slp}, a fixed-point property, to two ``simpler'' %axioms defined via 
systems of inequalities.

We highlight two useful implications of \cref{prop:1}. First, it delivers a variational characterization of $\Phi$: the indirect cost $\Phi(C)$ is the \emph{lower \nameref{axiom:slp} envelope} of the direct cost $C$.

\begin{cor}\label{cor:envelope}
    For any $C \in \C$, the indirect cost is $\Phi(C)=\max\left\{ C'\in \C \ | \ C' \preceq C \text{ and $C'$ is \nameref{axiom:slp}} \right\}$.
\end{cor}
\begin{proof} 
 Fix any $C \in \C$. We have $\Phi(C) \preceq C$ by definition, and \cref{prop:1} implies that $\Phi(C)$ is \nameref{axiom:slp}. Meanwhile, since $\Phi$ is isotone,\footnote{That is, $C \succeq C'$ implies $\Phi(C) \succeq \Phi(C')$. See \cref{sec:app:structure} for this and other structural facts about the $\Phi$ and $\Psi$ maps.} every \nameref{axiom:slp} $C' \preceq C$ satisfies $C' = \Phi(C') \preceq \Phi(C)$.
\end{proof}

We will use \cref{cor:envelope} to continue our analysis of \dred{Examples} \ref{eg:Diffusion:0} and \ref{eg:Poisson:0} in \cref{ssec:examples-revisit}. Second, \cref{prop:1} implies that the set $\C^*$ is closed under conical combinations and pointwise suprema (\cref{lem:structure:C*} in \cref{sec:app:structure}). This enables one to generate new \nameref{axiom:slp} costs from existing ones and to construct non-trivial variants of the $\Phi$ map (e.g., see \cref{ssec:bounding-phi}).

\subsection{Foundations for Uniform Posterior Separability}\label{ssec:ups}

The most widely applied class of cost functions in the flexible information acquisition literature is the class of \emph{uniformly posterior separable} costs \parencite{caplin2022rationally}. In this section, we show that this class characterizes the set of ``differentiable'' \nameref{axiom:slp} costs. 

%We begin with a definition: 
%For any $W \subseteq \Delta(\Theta)$, we let $\Delta(W) := \{\pi \in \Ex \mid \supp(\pi) \subseteq W\}$.

\begin{definition}[UPS] \label{defi:ups} 
$C \in \C$ is \dred{Uniformly Posterior Separable} (\nameref{defi:ups}) if there is a convex function $H:\Delta(\Theta)\to (-\infty, +\infty]$ such that $\dom(C) = \Delta(\dom(H)) \cup \Ex^\varnothing$ and, for every $\pi \in \Delta(\dom(H))$,  
\[
C (\pi) =\E_{\pi}[H(q)-H(p_{\pi})]. %\quad \text{ for all }\  \pi \in \Delta(\dom(H)).
\]
For any such convex function $H$, the associated \nameref{defi:ups} cost function is denoted as $C^H_\text{ups} \in \C$.\footnotemark
\iffalse
For any convex function $H:\Delta(\Theta)\to (-\infty, +\infty]$, the \dred{Uniformly Posterior Separable} (\nameref{defi:ups}) cost function $C_\text{ups}^H \in \C$ is defined as
    \begin{align*}
        C_\text{ups}^H(\pi):=\E_{\pi}[H(q)-H(p_{\pi})] \quad %\forall 
        \text{ for all }\  \pi \in \Delta(\dom(H)), %\footnotemark
    \end{align*} 
    with  $C_\text{ups}^H(\pi) := 0$ for all $\pi \in \Ex^\varnothing$ and $ C_\text{ups}^H(\pi):= +\infty$ otherwise.\footnotemark 
\fi
\end{definition}

\footnotetext{Most authors focus on \nameref{defi:ups} costs with full domain, i.e., $\dom(H) = \Delta(\Theta)$. \textcite{caplin2022rationally} also define a notion of ``weak UPS'' that generalizes \cref{defi:ups} by allowing the function $H$ to vary with the support of the prior; since our analysis of \nameref{defi:ups} costs (aside from \cref{prop:ups:additive}) focuses on full-support priors, this distinction is immaterial.}

Many well-known cost functions are \nameref{defi:ups}, including the classic \hyperref[MI]{Mutual Information} cost \parencite{sims_JME2003,matejka_mckay_AER2015} and the broader family of neighborhood-based costs \parencite{hebert2021neighborhood}. We will revisit these and other examples in \cref{section:applications}.

\nameref{defi:ups} costs are inherently related to sequential learning via the following property:

\begin{axiom}\label{axiom:additive}
    $C \in \C$ is \hyperref[axiom:additive]{Additive} if, for every $\Pi \in \Delta(\Ex)$ such that $\E_{\Pi}[\pi_2]\in\dom(C)$,
    \[
    C(\E_{\Pi}[\pi_2]) = C(\pi_1)+\E_{\Pi}\left[C(\pi_2) \right].
    \]
\end{axiom}

\hyperref[axiom:additive]{Additivity} represents \emph{indifference to sequential decomposition} of information: for each $\pi \in \dom(C)$, the cost of directly acquiring $\pi$ equals the expected cost of producing it via \emph{any} two-step strategy (without free disposal). By induction, \hyperref[axiom:additive]{Additivity} implies that \emph{all} sequential strategies that produce a given (feasible) target have the same expected cost. 

%It is easy to see that (i) every \nameref{defi:ups} cost is \hyperref[axiom:additive]{Additive} and (ii) every \hyperref[axiom:additive]{Additive} cost function is \nameref{axiom:slp}. 
It is easy to see that every \nameref{defi:ups} cost is \hyperref[axiom:additive]{Additive}
%, and hence that 
and that, consequently, every \nameref{defi:ups} cost \nameref{axiom:slp} (\cref{lem:ups:to:additive} in \cref{proof:ups:additive}). 
Conversely, for the special case of cost functions with full domain, \nameref{defi:ups} is known to be \emph{equivalent} to \hyperref[axiom:additive]{Additivity} \parencite[Theorem 3]{zhong2022optimal}.\footnote{Due to this equivalence, (full-domain) \nameref{defi:ups} costs are a workhorse tool for modeling flexible information acquisition in dynamic decision problems (e.g., \cite{zhong2022optimal,steiner-stewart-matejka-ecma2017,georgiadis2024preparing}). See also \textcite[Theorem 1]{frankel-kamenica-2018} for a special case of this equivalence in the context of the \emph{value} of information.} The following result extends this equivalence to \nameref{defi:ups} cost functions with general domains.

\begin{prop}\label{prop:ups:additive}
    For any open convex set $W \subseteq \Delta(\Theta)$ and $C \in \C$ with $\dom(C) = \Delta(W) \cup \Ex^{\varnothing}$,\footnote{Per \cref{defi:ups}, the only nontrivial hypothesis is that $W\subseteq \Delta(\Theta)$ is open. This hypothesis nests the special case of full-domain \nameref{defi:ups} costs because $W = \Delta(\Theta)$ is open in itself under the subspace topology on $\Delta(\Theta)$. 
    %(recall \cref{fn:subspace}).
    }
    \[
    C \text{ is \nameref{defi:ups}} \ \ %\xrightleftharpoons[\text{$W$ is open}]{} 
    \iff \ \ C \text{ is \hyperref[axiom:additive]{Additive}}.%\footnotemark
    \]
\end{prop}

\begin{proof}
See \cref{proof:ups:additive}.
\end{proof}

\cref{prop:ups:additive} suggests that \nameref{defi:ups} costs occupy a special place among \nameref{axiom:slp} costs. We now show that \hyperref[axiom:additive]{Additivity} is, in fact, implied by \nameref{axiom:slp} plus a mild form of ``local differentiability.'' 

We require a few definitions. First, for any $\pi \in \Ex$ and $\alpha \in [0,1]$, the \emph{$\alpha$-dilution of $\pi$} is defined as $\alpha \cdot \pi := \alpha \pi + (1-\alpha)\delta_{p_\pi} \in \Ex$. In words, $\alpha \cdot \pi$ is the random posterior produced by acquiring $\pi$ with probability $\alpha$ and learning nothing otherwise. Second, a \emph{divergence} is any map $D : \Delta(\Theta) \times \Delta(\Theta) \to \overline{\R}_+$ such that $D(p \mid p) = 0$ for all $p \in \Delta(\Theta)$, where $D(q\mid p)\geq0$ represents the ``distance'' of posterior $q$ from prior $p$. If $D(\cdot \mid p)$ is differentiable at $q$, its gradient is denoted as $\nabla_1 D(q \mid p) \in \R^{|\Theta|}$ and normalized so that $D(q\mid p) = q^\top \nabla_1 D(q\mid p)$.\footnote{
%We denote the inner product of (column) vectors $x,y \in \R^n$ as $x^\top y  \in \R$, where $x^\top$ is the transpose (row) vector of $x$. 
This normalization of gradients is obtained (without loss of generality) by extending functions from $\Delta(\Theta)$ to $\R^{|\Theta|}_+$ via homogeneity of degree $1$ and then defining derivatives in the usual way (cf. our normalization of Hessians in \cref{remark:kernels}).\label{fn:HD1-gradient}}

\begin{definition}[Regular]\label{defi:ll} $C\in\C$ is \nameref{defi:ll} if there is a divergence $D$ such that
	\begin{align}
		\lim_{\alpha\to 0}\frac{C(\alpha \cdot \pi)}{\alpha}=\E_{\pi}[D(q \mid p_{\pi})] \quad %\forall 
  \text{ for all } \ \pi \in \dom(C), \label{gateux}
	\end{align}  
and both $D$ and $\nabla_1 D$ are well-defined and jointly continuous on $\ri(\dom (D))$.\footnote{For any $X \subseteq \R^{n}$, we let $\ri(X) \subseteq \R^{n}$ denote its relative interior, i.e., its interior with respect to the subspace topology on the affine hull of $X$ \parencite{rock70}. Since gradients are only well-defined on subsets of $\Delta(\Theta)$ with nonempty interior (with respect to the subspace topology on $\Delta(\Theta)$), \cref{defi:ll} implicitly requires that $\ri(\dom (D))$ be open in $\Delta(\Theta)$. We note that, for any open $W \subseteq \Delta(\Theta)$, it holds that $\ri(W) = W \cap \Delta^\circ(\Theta)$. E.g., $\ri(\Delta(\Theta)) = \Delta^\circ(\Theta)$.} %We call any such divergence $D$ a \dred{derivative} of $C$ and denote $D_C := D$.
\end{definition}

In words, a cost function $C \in \C$ is \nameref{defi:ll} if it satisfies two conditions. First, \eqref{gateux} states that $C$ is \emph{Gateaux differentiable} at every trivial random posterior $\delta_p \in \Ex^\varnothing$ in the direction of any feasible random posterior $\pi \in \dom(C)$ with the same prior (i.e., $p_\pi = p$), where the divergence $D(\cdot \mid p)$ represents the ``derivative'' of $C$ evaluated at the prior $p$. Second, the divergence $D$ must itself be continuously differentiable with respect to the posterior. 

Economically, the limit in \eqref{gateux} represents the cost of producing $\pi$ via a ``Poisson sampling'' strategy (cf. \cref{eg:Poisson:0}) under which the dilution $\alpha \cdot \pi$ is acquired $1/\alpha$ times in expectation until a success is obtained, where taking $\alpha \to 0$ yields the continuous-time limit. From this perspective, $D(q \mid p)$ is the cost of a ``Poisson jump'' in beliefs from $p$ to $q$. 

We interpret \hyperref[defi:ll]{Regularity} as a mild ``tractability'' desideratum for applications. In particular, essentially all applications of flexible information acquisition restrict attention to \emph{\hyperref[eqn:PS]{Posterior Separable}} cost functions \parencite{caplin2022rationally}---that is, $C \in \C$ such that
\begin{equation}\label{eqn:PS}
C(\pi) = \E_\pi[D(q \mid p_\pi)] \quad \text{ for all } \, \pi \in \dom(C) = \Delta(W)\cup\Ex^\varnothing 
%\pi \in \Ex \, \text{ with $\supp(\pi)\times\{p_\pi\}\subseteq \dom(D)$}
\tag{PS}
\end{equation}
for some divergence $D$ and convex $W \subseteq \Delta(\Theta)$---and assume that the divergence in \eqref{eqn:PS} is smooth. These assumptions are common because they allow one to characterize the DM's optimal behavior via first-order conditions. \hyperref[defi:ll]{Regularity} is a much milder assumption.\footnote{The class of (smooth) \nameref{defi:ups} costs is a strict subset of the class of (smooth) \hyperref[eqn:PS]{Posterior Separable} costs. For first-order conditions arising from (smooth) \hyperref[eqn:PS]{Posterior Separable} costs, see \textcite{caplin2022rationally,bs2025scoring,bdp2025modeling}. \textcite{lipnowski2022predicting} independently propose a notion of ``iterative differentiability'' that is analogous to \hyperref[defi:ll]{Regularity} and show that it enables first-order characterizations of choice behavior.}

\begin{thm}\label{thm:UPS}
For any open convex set $W \subseteq \Delta^\circ(\Theta)$ and $C \in \C$ with $\dom(C) = \Delta(W) \cup \Ex^{\varnothing}$,
\[
C \text{ is \nameref{axiom:slp} and \nameref{defi:ll}} \ \  \iff \ \ C=C^H_\text{ups} \ \text{ for some convex } H \in \mathbf{C}^1(W).\footnotemark 
\]
\end{thm}
\begin{proof}
    See \cref{ssec:app:thm:UPS}.
\end{proof}
\footnotetext{For any $W \subseteq \Delta(\Theta)$ and $n \in \mathbb{N} \cup\{+\infty\}$, we let $\mathbf{C}^n(W) := \left\{ f : \Delta(\Theta) \to (-\infty,+\infty] \mid \dom(f) = W \text{ and $f$ is $\mathbf{C}^n$-smooth on $W$}\right\}$.}

\cref{thm:UPS} offers a novel optimality- and tractability-based foundation for  \nameref{defi:ups} cost functions: they are the \emph{only} cost functions that are both robust to sequential optimization (\nameref{axiom:slp}) and ``tractable'' (\nameref{defi:ll}). This provides a powerful practical rationale for using \nameref{defi:ups} cost functions as a modeling tool in applications, regardless of the specific economic context. In this respect, \cref{thm:UPS} is orthogonal to various behavioral characterizations of the \nameref{defi:ups} model in the literature (e.g., \cite{denti2022posterior,caplin2022rationally}). 

We note that a simple but central step in the proof of \cref{thm:UPS} shows that every \nameref{axiom:slp} cost is linear in the probability of running experiments (\cref{lem:convex-dl} in \cref{ssec:app:thm:UPS}). This property is formalized via the following axiom \parencite{pomatto2023cost}.

\begin{axiom}%[Dilution Linear]
\label{axiom:DL}
	$C \in \C$ is \hyperref[axiom:DL]{Dilution Linear} if $C(\alpha \cdot \pi ) =\alpha C(\pi)$ for every $\pi \in \dom(C)$ and $\alpha \in [0,1]$.
\end{axiom}

In the proof of \cref{thm:UPS}, \hyperref[axiom:DL]{Dilution Linearity} implies that---for \nameref{axiom:slp} costs---the ``local'' differentiability condition \eqref{gateux} is, in fact, equivalent to the ``global'' property \eqref{eqn:PS}. More broadly, \hyperref[axiom:DL]{Dilution Linearity} represents \emph{robustness to ``Poisson sampling''} (as defined above).

\begin{remark}\label{remark:full-support}
As in \cref{thm:UPS}, our subsequent analysis focuses mainly on cost functions for which only ``interior'' random posteriors are feasible, i.e., $C \in \C$ such that $\dom(C) \subseteq \Delta(W) \cup \Ex^\varnothing$ for some $W \subseteq \Delta^\circ(\Theta)$. This simplifies the exposition and, by ensuring that the DM's beliefs always have full support, lets us remain agnostic about how the cost of experiments varies with the support of the prior. To analyze cost functions $C \in \C$ with full domain (i.e., $\dom(C) = \Ex$), one can: (i) restrict $C$ to the ``rich domain'' $\Delta(\Delta^\circ(\Theta)) \cup \Ex^\varnothing \subsetneq \Ex$, (ii) apply our results to the rich-domain restriction of $C$, and (iii) then extend back to the full domain (e.g., by continuity). 
%%%%%
\end{remark}

\subsection{Examples Revisited }\label{ssec:examples-revisit}

With \cref{prop:1,thm:UPS} in hand, we can continue our analysis of \dred{Examples} \ref{eg:Diffusion:0} and \ref{eg:Poisson:0}. Recall from \cref{ssec:leading-examples} that these examples feature the binary state space $\Theta = \{0,1\}$.

\setcounter{eg}{0}
\begin{eg}[Diffusion Sampling---continued]\label{eg:Diffusion:1}
To begin, note that Bayes' rule implies the \ref{eqn:MS} cost equals the \nameref{defi:ups} cost $C^{H_\text{Wald}}_\text{ups} \in \C$, where 
\[
H_\text{Wald}(p) := p(1) \log\left( \frac{p(1)}{p(0)} \right) + p(0) \log\left( \frac{p(0)}{p(1)} \right) \quad \text{ for all $p \in \Delta^\circ(\Theta)$.}
\]
Since $H_\text{Wald} \in \mathbf{C}^\infty(\Delta^\circ(\Theta))$, \cref{thm:UPS} implies that the \ref{eqn:MS} cost is \nameref{axiom:slp} and \nameref{defi:ll}, where its derivative is given by the Bregman divergence associated with $H_\text{Wald}$.\footnote{The \emph{Bregman divergence} associated with a convex $H \in \mathbf{C}^1(\Delta^\circ(\Theta))$ is defined as $D(q \mid p) := H(q) - H(p) - (q-p)^\top\nabla H(p)$.
%Formally, this \emph{Bregman divergence} $D_\text{Wald}$ is defined as $D_\text{Wald}(q \mid p) := H_\text{Wald}(q) - H_\text{Wald}(p) - (q-p)^\top\nabla H_\text{Wald}(p)$.
} 
Intuitively, $H_\text{Wald}$ is smooth because it derives from the direct cost \eqref{eqn:bernoulli-DC} and incremental learning strategy \eqref{eqn:Brownian-motion}, under which the ``flow cost'' of producing small belief changes is second-order in the size of the belief change. %(see \cref{ssec:bounding-phi} for a more formal articulation of this point).
We will formalize and generalize this intuition in \cref{sec:phi}. 
%Intuitively, $H_\text{Wald}$ is smooth because it derives from the direct cost \eqref{eqn:bernoulli-DC} and incremental learning strategy \eqref{eqn:Brownian-motion}, under which the ``flow cost'' of producing small belief changes is ``second-order.''
%

We now characterize when the upper bound $\Phi(C) \preceq f''(0) \cdot C_\text{Wald}$ is tight for the direct cost $C$ in \eqref{eqn:bernoulli-DC}. %By \cref{cor:envelope}, $\Phi(C)$ is the largest \nameref{axiom:slp} cost that lies below $C$. 
Since $\Phi(C)$ is \nameref{axiom:slp} by \cref{prop:1} and $C_\text{Wald}$ is \nameref{axiom:slp} by the above, \cref{cor:envelope} implies that $\Phi(C) = f''(0) \cdot C_\text{Wald}$ if and only if $C \succeq f''(0) \cdot C_\text{Wald}$. Moreover, since only the Bernoulli experiments $\sigma^\ell$ are feasible under $C$, by \eqref{eqn:bernoulli-DC} and \eqref{eqn:MS} this inequality becomes
%In fact, we will show in \cref{sec:C} that this bound is tight (i.e., $\Phi(C) = f''(0) \cdot C_\text{Wald}$) if and only if  incremental learning is cheaper than direct learning for Bernoulli experiments:
    \begin{equation}\label{eqn:flies-eg1}
    f(\ell) \, \geq \, f''(0) \cdot D_\text{KL} (\sigma^\ell_1 \mid \sigma^\ell_0) \quad \text{ for all $\ell \in \R_+$.}
     \end{equation}
     %Thus, $\Phi(C) = f''(0) \cdot C_\text{Wald}$ \emph{if and only if} the direct cost exceeds the cost of incremental learning.
    We conclude that $\Phi(C) = f''(0) \cdot C_\text{Wald}$ \emph{if and only if} \eqref{eqn:flies-eg1} holds, i.e., the direct cost lies above the cost of incremental learning from \eqref{eqn:bernoulli-incremental}. We will revisit condition \eqref{eqn:flies-eg1} in \cref{sec:phi} below. %Henceforth, we assume that condition \eqref{eqn:flies-eg1} holds.
    %In \cref{sec:phi}, we will see that the natural generalization of \eqref{eqn:flies-eg1} characterizes the class of all direct costs that generate \nameref{defi:ll}/\nameref{defi:ups} indirect costs.
   \end{eg}

\begin{eg}[Poisson Sampling---continued]
    By Bayes' rule, the \ref{eqn:TV} cost can be represented as a \hyperref[eqn:PS]{Posterior Separable} cost with divergence
    \[
    D_\text{TV} (q \mid p) := 1 - \min\left\{  \frac{q(0)}{p(0)}, \frac{q(1)}{p(1)}\right\} \quad \text{ for all $q,p \in \Delta(\Theta)$.}
    \]
Thus, the \ref{eqn:TV} cost satisfies \eqref{gateux}, but it is not \nameref{defi:ll} because its derivative $D_\text{TV}$ is kinked at points where $q=p$. Intuitively, $D_\text{TV}$ is kinked because it derives from the Poisson-with-disposal strategy, under which the marginal cost of producing small belief changes equals the (strictly positive) marginal cost of increasing the probability of full information.
%Intuitively, $D_\text{TV}$ is kinked at these points because it derives from the Poisson-with-free-disposal strategy that acquires (and then garbles) discrete ``chunks'' of information, under which the cost of producing even small belief changes is ``first-order.'' 
%More broadly, the \ref{eqn:TV} cost is not \nameref{defi:ups} because it violates \hyperref[axiom:additive]{Additivity}.

We now show that the \ref{eqn:TV} cost is \nameref{axiom:slp}. First, note that Jensen's inequality implies that it is \hyperref[axiom:mono]{Monotone}, as $D_\text{TV}(\cdot \mid p)$ is convex for each prior $p$. Second, it can be verified that $D_\text{TV}$ is a \emph{quasi-metric}, and hence satisfies the triangle inequality; this implies that the \ref{eqn:TV} cost is \hyperref[axiom:POSL]{Subadditive}.\footnote{Recall that a divergence $D$ is a \emph{quasi-metric} if: (i) $D(q \mid p) = 0$ only if $q=p$, and (ii) $D(q \mid p) \leq D(r \mid p) + D(q \mid r) $ for all $p,q,r \in \Delta(\Theta)$. It is easy to see that any \hyperref[eqn:PS]{Posterior Separable} cost with a divergence satisfying the triangle inequality (condition (ii)) is \hyperref[axiom:POSL]{Subadditive}. We verify in \cref{sssec:proof-tri-pt3} that $D_\text{TV}$ is a quasi-metric. Moreover, since the triangle inequality is generically strict for $D_\text{TV}$, the \ref{eqn:TV} cost is not \hyperref[axiom:additive]{Additive} and hence not \nameref{defi:ups} (in addition to being non-\nameref{defi:ll}).} 
It then follows from \cref{prop:1} that the \ref{eqn:TV} cost is \nameref{axiom:slp}, as desired.

Finally, we show that $\Phi(C) = C_\text{TV}$ for the direct cost $C$ in \eqref{eqn:poisson-DC}. The logic is similar to that in \cref{eg:Diffusion:0}. In particular, since $C \succeq C_\text{TV}$ by construction, $C_\text{TV}$ is \nameref{axiom:slp} by the above, and $C_\text{TV} \succeq \Phi(C)$ by the upper bound from \cref{ssec:leading-examples}, \cref{cor:envelope} implies that $\Phi(C) = C_\text{TV}$. 

\end{eg}

\section{The Sequential Learning Map}\label{sec:phi}

In this section, we characterize the sequential learning map $\Phi$, building on \cref{eg:Diffusion:0}. %\cref{ssec:kern-def,ssec:bounding-phi} introduce a general notion of ``incremental learning'' and use it to bound the indirect cost given any direct cost.  \cref{ssec:flie-characteriation} shows that these bounds are tight precisely when incremental learning is optimal and the indirect cost is \nameref{defi:ll}/\nameref{defi:ups}.
\cref{ssec:kern-def,ssec:bounding-phi} introduce a general notion of ``incremental learning'' and use it bound the indirect cost given any direct cost. \cref{ssec:flie-characteriation} shows that these bounds are tight, and hence  the indirect cost is \nameref{defi:ll}/\nameref{defi:ups}, precisely when incremental learning is optimal.

\subsection{The Cost of Incremental Evidence}\label{ssec:kern-def}

In \cref{eg:Diffusion:0}, each incremental  diffusion signal generates only an infinitesimal change in the DM's belief. By analogy, we call a random posterior ``incremental evidence'' if its support is contained in an infinitesimal neighborhood of the prior. To model the ``flow cost'' of incremental evidence with general cost functions, we use the following definition.

\begin{definition}[Locally Quadratic] \label{defi:lq} 
For any $C\in \C$ and $W\subseteq \Delta(\Theta)$, a matrix-valued function $k: W\to \R^{|\Theta| \times |\Theta|}$ such that $k(p)$ is symmetric and $k(p)  p = \mathbf{0}$ for all $p \in W$ is called:
\begin{enumerate}
\item[(i)] An \dred{upper kernel} of $C$ on $W$ if, for every $p\in W$ and $\epsilon>0$, there exists a $\delta>0$ such that
\begin{align*}
    C(\pi)\le \int_{B_{\delta}(p)} (q-p_{\pi})^\top\left(\frac{1}{2}k(p)+\epsilon I\right)(q-p_{\pi}) \dd \pi( q) \quad \text{for all $\pi\in \Ex$ with $\supp(\pi) \subseteq B_{\delta}(p)$}.\footnotemark
\end{align*}\footnotetext{We denote by $B_{\delta}(p) := \{q\in \Delta(\Theta) \mid \|q-p\|<\delta \}$ the open ball in $\Delta(\Theta)$ of radius $\delta>0$ around $p \in \Delta(\Theta)$.} \vspace{-2em}
   \item[(ii)] A \dred{lower kernel} of $C$ on $W$ if, for every $p\in W$ and $\epsilon>0$, there exists a $ \delta>0$ such that %for all $ \pi\in \Ex$ with $p_{\pi}\in B_{\delta}(p)$,
    \begin{align*}
    C(\pi)\ge \int_{B_{\delta}(p)} (q-p_{\pi})^\top\left(\frac{1}{2}k(p)-\epsilon I\right)(q-p_{\pi}) \dd \pi(q) \quad \text{for all $ \pi\in \Ex$ with $p_{\pi}\in B_{\delta}(p)$}.\footnotemark
\end{align*}
\footnotetext{Note that this condition applies to all $\pi \in \Ex$ with $p_\pi \in B_\delta(p)$, including those with $\supp(\pi) \not\subseteq B_\delta(p)$. Nonetheless, when $\delta>0$ is small, it imposes almost no restrictions on the cost of such ``non-incremental'' evidence: (i) for any $\pi \in \Ex$ with $\supp(\pi) \cap B_\delta(p) = \emptyset$, the inequality holds trivially, and (ii) in general, the lower bound on $C(\pi)$ vanishes as $\delta \to 0$.}
\vspace{-2em}
\item[(iii)] The \dred{kernel} of $C$ on $W$ if it is both a lower kernel and an upper kernel on $W$.%\footnote{It is easy to verify that the kernel of $C$ is unique whenever it exists (cf. \cref{remark:kernels} below).} 
\end{enumerate}
If $C$ admits a kernel on $W$, we say that $C$ is \nameref{defi:lq} on $W$ and denote its kernel as $k_C$. In each case above, we omit the qualifier ``on $W$'' whenever $\Delta^\circ(\Theta)\subseteq W$.
\end{definition}

A cost function $C$ is \nameref{defi:lq} if it is ``locally twice continuously differentiable'' with respect to incremental evidence, where the kernel $k_C(p)$ is the ``Hessian matrix'' that quadratically approximates the cost of incremental evidence at the prior $p$. To relate this to the standard finite-dimensional notion of $\mathbf{C}^2$-smoothness, we note that a \nameref{defi:ups} cost $C^H_\text{ups}$ is \nameref{defi:lq} \emph{if and only if} $H$ is $\mathbf{C}^2$-smooth, in which case $k_{C^H_\text{ups}} = \H H$ and the quadratic approximation resembles a standard ``It\^{o} expansion'' for the flow cost of diffusion signals (cf. \cite{zhong2022optimal,hebert2023rational}). Observe that \hyperref[defi:lq]{Local Quadradicity} imposes \emph{no} smoothness conditions on the cost of ``non-incremental'' evidence.\footnote{For instance, a \hyperref[eqn:PS]{Posterior Separable} cost is \nameref{defi:lq} if its divergence $D(q\mid p)$ is $\mathbf{C}^2$-smooth in $q$ at points where $q=p$ \emph{even if it is non-smooth elsewhere}, in which case $k_C (p) \equiv \H_1 D(q\mid p)|_{q=p}$ (\cref{lem:ps-kernel-suff} in \cref{ssec:calc-kernel}).}

We define upper and lower kernels separately because, while not every cost function is \nameref{defi:lq}, these objects exist very generally (e.g., \emph{every} $C \in \C$ has lower kernels). 
%Namely, \emph{every} $C \in \C$ admits lower kernels (e.g., $k \equiv \mathbf{0}$) and any $C \in \C$ without ``fixed costs'' or ``kinks'' admits upper kernels.
The existence of upper and lower kernels will suffice for much of our subsequent analysis.

Finally, for any (upper/lower) kernel $k$ on $W \subseteq \Delta(\Theta)$, we call $k$ \emph{integrable} if $k = \H H$ for some $H \in \mathbf{C}^2(W)$. Integrable kernels will play an important role in our analysis. We note that \emph{every} (upper/lower) kernel is integrable when $|\Theta|=2$, but not when $|\Theta|\geq 3$.

\begin{remark}\label{remark:kernels}
\cref{defi:lq} requires (upper/lower) kernels $k$ to be symmetric and satisfy $k(p) p \equiv  \mathbf{0}$. These are merely normalizations to ensure that kernels are uniquely defined. For any $W \subseteq \Delta(\Theta)$ and $\beta : W \to  \R^{|\Theta| \times |\Theta|}$, we can normalize $\beta$ as $k(p) := \frac{1}{2}(I-\mathbf{1}p^\top)\left(\beta(p)+\beta(p)^\top \right)(I-p\mathbf{1}^\top)$. This normalization ensures that $k$ is symmetric and $k(p) p \equiv  \mathbf{0}$ without affecting the quadratic forms in \cref{defi:lq}; moreover, if $\beta$ is symmetric and $\beta(p) p \equiv \mathbf{0}$, then $k = \beta$. We apply the same normalization to all other (square) matrix-valued functions on the simplex (viz., Hessians).
\end{remark}

\subsection{Bounding the Sequential Learning Map}\label{ssec:bounding-phi}

Using these definitions, we now show that the cost of incremental evidence under the direct cost $C$ provides \emph{global upper} bounds and \emph{local lower} bounds on the indirect cost $\Phi(C)$.

Henceforth, we sometimes adopt a technical assumption to rule out ``degenerate'' cases:

\begin{definition}[Strongly Positive]\label{defi:sp}
 $C\in \C$ is \nameref{defi:sp} if there exists an $m>0$ such that $C(\pi)\ge m \cdot \text{Var}(\pi)$ for all $\pi \in \Ex$, where $\text{Var}(\pi) := \E_{\pi}\big[ \|q-p_{\pi}\|^2\big]$ is the ``variance'' of $\pi$.  %In this case, we say $C$ is $m$-\nameref{defi:sp}.
\end{definition}

Essentially all cost functions in the literature, including those in \dred{Examples} \ref{eg:Diffusion:0} and \ref{eg:Poisson:0}, are \hyperref[defi:sp]{Strongly Positive}. The \nameref{defi:ups} cost  $C^H_\text{ups}$ is \nameref{defi:sp} whenever $H$ is strongly convex.

\begin{thm}\label{thm:qk}
For any $C \in \C$ and $W \subseteq \Delta(\Theta)$, the following holds:
\begin{itemize}[noitemsep,leftmargin=2em]
    \item[(i)] If $W\subseteq \Delta^\circ(\Theta)$ is open and convex, $H \in \mathbf{C}^2(W)$, and $\H H$ is an upper kernel of $C$ on $W$, then $\Phi(C)(\pi)\le C^H_\text{ups}(\pi)$ for all $\pi\in \Delta(W)$.
    %\begin{align*}
    %\Phi(C)(\pi)\le C^H_\text{ups}(\pi) \quad \text{for all } \pi\in \Delta(W).
    %\end{align*}
    \item[(ii)] If $C$ is \nameref{defi:sp} and $k\gg_\text{psd} \mathbf{0}$ is a lower kernel of $C$ on $W$,\footnote{For any matrix $M \in \R^{|\Theta| \times |\Theta|}$, we let $M \gg_\text{psd} \mathbf{0}$ denote that $(q-q')^{\top} M (q-q')>0 $ for all $q,q' \in \Delta(\Theta)$ with $q \neq q'$. Analogously, for any $W\subseteq \Delta(\Theta)$ and function $k : W \to  \R^{|\Theta| \times |\Theta|}$, we let $k\gg_\text{psd} \mathbf{0}$ denote that $k(p) \gg_\text{psd} \mathbf{0}$ for every $p\in W$. It is easy to see that all \nameref{defi:sp} $C \in \C$ have (lower) kernels with this property (\cref{cor:ker-SP} in \cref{ssec:calc-kernel}).\label{fn:psd-strict}} then $k$ is also a lower kernel of $\Phi(C)$ on $W$.
\end{itemize}
\end{thm}
\begin{proof}
	See \cref{sssec:app:thm:qk}.
\end{proof}

First, \cref{thm:qk}(i) shows that the (integrable) upper kernels of any direct cost yield \emph{global} \nameref{defi:ups} \emph{upper} bounds for its indirect cost. These upper bounds are powerful because they apply even if non-incremental evidence is 
%very costly or 
infeasible under the direct cost. We prove this result by generalizing the incremental learning strategy construction in \cref{eg:Diffusion:0}.

Next, \cref{thm:qk}(ii) shows that the (positive) lower kernels of any \nameref{defi:sp} direct cost yield \emph{local} \emph{lower} bounds for its indirect cost. Formally, it establishes that these lower kernels are \emph{invariant} under $\Phi$. Consequently, for any \nameref{defi:sp} $C \in \C$, it holds that: (a) $C$ and $\Phi(C)$ have the \emph{same} sets of (positive) lower kernels, and (b) if $C$ is \nameref{defi:lq}, then $\Phi(C)$ is also \nameref{defi:lq} with the \emph{same} kernel $k_C = k_{\Phi(C)}$. 
%\footnote{These properties follow from \cref{thm:qk}(ii) and the elementary facts that, since $C \succeq \Phi(C)$ by construction: (a) every lower kernel of $\Phi(C)$ is also a lower kernel of $C$, and (b) every upper kernel of $C$ is also an upper kernel of $\Phi(C)$.} 
%This \emph{(lower) kernel invariance} result, which we have already seen illustrated in \cref{eg:Diffusion:2}, is a key methodological tool for the rest of our analysis. 

This (lower) kernel invariance result reflects the idea that the cost of incremental evidence cannot be reduced through optimization. Intuitively, the only way to decompose a piece of incremental evidence is into ``more incremental'' component pieces; since the direct cost of each component piece is bounded below by the direct cost's lower kernels, the indirect cost of the original piece must also be bounded below by the same lower kernels. 

Notably, the lower kernels of the direct cost are generally \emph{not} sufficient to yield \emph{global} lower bounds on its indirect cost: if non-incremental learning is ever optimal, then the full shape of $C$ is relevant for bounding $\Phi(C)$ from below. 
%\footnote{In contrast, the upper bounds in \cref{thm:qk}(i) are valid even if the strategies used to construct them are suboptimal.} 
%Thus, to extend the local lower bounds in \cref{thm:qk}(ii) to global ones, we must consider a more constrained setting in which the DM is exogenously restricted to learning via incremental evidence. 
%In what follows, we make a quick detour and show that the local lower bounds in \cref{thm:qk} exactly pin down the global lower bound in a more constrained setting in which only incremental evidence is feasible.
However, we now show that these lower kernels do, in fact, yield global lower bounds in an auxiliary setting where the DM can only learn via incremental evidence. %As we will see in \cref{ssec:flie-characteriation}, these global bounds are essential to fully characterize $\Phi$. To formalize them, we need some definitions.
To this end, we require a few definitions.

For every $C \in \C$, we denote by $\Delta_C := \{ p \in \Delta(\Theta) \mid \exists \pi \in \dom(C) \backslash \Ex^{\varnothing} \text{ s.t. } p_\pi = p\}$ the set of all priors at which nontrivial learning is feasible, and by $\Omega(C)$ the set of all open covers of $\Delta_C$.\footnote{That is, each $\mathbb{O} \in \Omega(C)$ comprises a collection of open sets $O \subseteq \Delta(\Theta)$ such that $\Delta_C\subseteq \cup_{O\in\mathbb{O}}O$.} 
Each open cover $\mathbb{O}\in\Omega(C)$ specifies a collection of neighborhoods that parameterize what it means for beliefs to shift ``locally'' away from the prior. For any direct cost $C \in \C$ and any such open cover, we define the restricted direct cost $C|_{\mathbb{O}}\in \C$ as
\begin{align*}
        C|_{\mathbb{O}}(\pi):=
    \begin{dcases}
        C(\pi), &\text{if $\exists O \in \mathbb{O}$ s.t. $\supp(\pi)\subseteq O$}\\
        0 , & \text{if $\pi \in \Ex^{\varnothing}$}\\
        +\infty, &\text{otherwise}.
    \end{dcases}
\end{align*}
That is, $C|_{\mathbb{O}}$ restricts the domain of $C$ so that only random posteriors inducing ``local'' belief shifts are feasible. We can then define the ``indirect cost'' generated by the direct cost $C$ when the DM is restricted to learning via incremental evidence as follows:

   \begin{definition}\label{defi:Phi:IE}  
   \hspace{-0.3em}
\mbox{The \dred{incremental learning map} $\Phi_\text{IE} : \C \to \C^*$ is defined, for all $C \in \C$ and $\pi \in \Ex$, as}
\begin{align*}
        \Phi_{\text{IE}}(C)(\pi) \defn \sup_{\mathbb{O} \in \Omega(C)} \Phi(C|_{\mathbb{O}})(\pi).\footnotemark \label{eqn:Phi-IE} \tag{IE}
    \end{align*}

    %where, for each open cover $\mathbb{O} \in \Omega(C)$, the cost function $C|_{\mathbb{O}} \in \C$ is given by

\end{definition}
\footnotetext{For every $C \in \C$, the cost function $\Phi_\text{IE}(C) \in \C^*$ is a well-defined indirect cost because $\Phi(C|_{\mathbb{O}}) \in \C^*$ for all $\mathbb{O}\in\Omega(C)$ (by construction) and the set $\C^*$ is closed under pointwise suprema (\cref{lem:structure:C*} in  \cref{sec:app:structure}).\label{fn:phi-ie-defn}}

%In words, $\Phi_\text{IE}(C)$ is the ``indirect cost'' generated by the direct cost $C$ when the DM is restricted to learning via incremental evidence. 
%\cref{defi:Phi:IE} models this restriction in three steps. First, each open cover $\mathbb{O}$ defines a collection of neighborhoods that parameterize what it means for beliefs to shift ``locally.'' Second, $\Phi(C|_{\mathbb{O}})$ defines the indirect cost generated by $C$ when the DM can only use strategies that shift beliefs ``locally'' in each round. Finally, the supremum in \eqref{eqn:Phi-IE} defines the limit in which all the neighborhoods in $\mathbb{O}$ become vanishingly small. 
In words, $\Phi(C|_{\mathbb{O}})$ represents the indirect cost generated by $C$ when the DM can only use strategies that shift beliefs locally in each round. The supremum in \eqref{eqn:Phi-IE} then defines the limit in which all the neighborhoods in $\mathbb{O}$ become vanishingly small, i.e., only incremental learning is feasible.  Intuitively, this limit approximates a continuous-time setting where the DM samples from a diffusion process as in \cref{eg:Diffusion:0} and \textcite{morris-strack-sampling}, but with full control over the drift and volatility (cf. \cite{hebert2023rational}).

 Importantly, the $\Phi_\text{IE}$ map is fully determined by ``integrating'' the direct cost's kernel:

\begin{prop}\label{lem:phi-ie} 
For any open convex set $W \subseteq \Delta^\circ(\Theta)$, strongly convex $H \in \mathbf{C}^2(W)$, and $C \in \C$ with $\dom(C)\subseteq\Delta(W)\cup \Ex^{\varnothing}$, the following properties hold:
%%%
\begin{itemize}[noitemsep,leftmargin=2em]
    \item[(i)] If $\H H$ is an upper kernel of $C$ on $W$, then $\Phi_\text{IE} (C) \preceq C^H_\text{ups}$.
    \vspace{.2em}
    \item[(ii)] If $\H H$ is a lower kernel of $C$ on $W$, then $\Phi_\text{IE} (C) \succeq C^H_\text{ups}$.
    \vspace{.2em}
    \item[(iii)] If $C$ is \nameref{defi:sp} and \nameref{defi:lq} on $W$, 
    \[
    k_C = \H H \ \  \iff \ \ \Phi_\text{IE}(C) = C^H_\text{ups}.
    \]
%%%
\end{itemize}

\end{prop}
\begin{proof}
    See \cref{ssec:app:phi-ie}.
\end{proof}

\cref{lem:phi-ie}(i) mirrors the upper bounds in \cref{thm:qk}(i). \cref{lem:phi-ie}(ii) provides the desired \emph{global} lower bounds on $\Phi_\text{IE}(C)$, extending the \emph{local} lower bounds on $\Phi(C)$ in \cref{thm:qk}(ii). \cref{lem:phi-ie}(iii) shows that $\Phi_\text{IE}$ defines a bijection between the integrable kernels of direct costs and \nameref{defi:ll}/\nameref{defi:ups} indirect costs; this can be viewed as the natural extension of \textcite[Theorem 1]{morris-strack-sampling} to general state spaces and direct costs.

We conclude this section by illustrating the definitions of upper/lower kernels and the bounds from \cref{thm:qk} and \cref{lem:phi-ie} in the context of \cref{eg:Diffusion:0} and \cref{eg:Poisson:0}.

\setcounter{eg}{0}
\begin{eg}[Diffusion Sampling---continued]
\noindent\textbf{Kernels of the direct cost:} Recall that the flow cost of sampling the diffusion signal process \eqref{eqn:Brownian-motion} for an instant ``$\d t$'' of continuous time is $\frac{1}{2} f''(0) \dd t$. Let $\rho_t \in [0,1]$ denote the DM's belief that $\theta =1$ after sampling until time $t \in \R_+$. As is well known, this belief evolves as $\d \rho_t = \rho_t (1-\rho_t) \dd Z_t$ for some standard Brownian motion $Z$, and its ``flow variance'' equals $\rho_t^2 (1-\rho_t)^2 \dd t$. Thus, the cost of sampling ``per unit of belief variance'' equals $\frac{1}{2} f''(0) / \rho_t^2 (1-\rho_t)^2$, which is the projection of $\frac{1}{2} f''(0) \cdot \H H_\text{Wald}$ onto the unit interval. 

This suggests that, as can be verified, $f''(0) \cdot \H H_\text{Wald}$ is a \emph{lower kernel} of $C$, the direct cost in \eqref{eqn:bernoulli-DC}.\footnote{To verify this directly, fix any $\epsilon>0$ and note that, for any prior $p \in \Delta^\circ(\Theta)$ and Bernoulli experiment $\sigma^\ell$ with $\ell \approx 0$, the inequality in \cref{defi:lq}(ii) with $k(p):= f''(0) \cdot \H H_\text{Wald}(p)$ requires that $f(\ell) \geq \frac{1}{2} f''(0)\ell^2 - \epsilon \text{Var}(h_B(\sigma^\ell,p)) + o(\ell^2)$, which holds by Taylor's theorem. Meanwhile, the lower bound in \cref{defi:lq}(ii) is trivial for non-Bernoulli experiments.} Since non-Bernoulli experiments are infeasible, $C$ does \emph{not} have \emph{upper kernels}.

\noindent\textbf{Bounds of the indirect cost:} We begin with lower bounds. Given the above, \cref{thm:qk}(ii) implies that $f''(0)\cdot\H  H_{\text{Wald}}$ is also a lower kernel of $\Phi(C)$. This yields a local lower bound on $\Phi(C)$ \emph{even if} condition \eqref{eqn:flies-eg1} fails, in which case our analysis in \cref{ssec:leading-examples,ssec:examples-revisit} does not pin down $\Phi(C)$. Meanwhile, \cref{lem:phi-ie}(ii) directly delivers $\Phi_\text{IE}(C)\succeq f''(0)\cdot C_\text{Wald}$.

As for upper bounds, the \emph{statements} of \cref{thm:qk}(i) and \cref{lem:phi-ie}(i) do not apply since $C$ does not admit upper kernels. Nonetheless, the \emph{proofs} of these results (which mirror our random walk construction in \cref{ssec:leading-examples}) imply $f''(0)\cdot C_\text{Wald} \succeq \Phi_\text{IE}(C) \succeq \Phi(C)$.\footnote{The proofs of \cref{thm:qk}(i) and \cref{lem:phi-ie}(i) only require the upper kernel inequality in \cref{defi:lq}(i) to hold for ``incremental Bernoulli'' random posteriors (see \cref{rmk:bernoulli:uqk} in \cref{sssec:app:thm:qk-pt1}). Thus, the definition of upper kernels is stronger than needed for these upper bounds, which can often be obtained by direct construction even when upper kernels do not exist. In contrast, we note that the more subtle lower bounds derived via lower kernels in \cref{thm:qk}(ii) and \cref{lem:phi-ie}(ii) cannot readily be obtained by other means.} 

Together, these upper and lower bounds uniquely determine the kernel of $\Phi(C)$ and the incremental learning map: we have $k_{\Phi(C)} = f''(0)\cdot\H  H_{\text{Wald}}$ and $\Phi_\text{IE}(C) = f''(0)\cdot C_\text{Wald}$.

\noindent\textbf{Tightness of the bounds:} As noted in \cref{ssec:examples-revisit}, we have $\Phi(C)= f''(0)\cdot C_\text{Wald}$ \emph{if and only if} $C \succeq f''(0)\cdot C_\text{Wald}$ (i.e., \eqref{eqn:flies-eg1} holds). We can now restate this as follows: the upper bound $f''(0)\cdot C_\text{Wald} \succeq \Phi(C)$ from (the proof of) \cref{thm:qk}(i) is tight \emph{if and only if} $C \succeq \Phi_\text{IE}(C)$.

\end{eg}

\begin{eg}[Poisson Sampling---continued]
Recall that the divergence $D_\text{TV}(q \mid p)$ is kinked at points where $q = p$. This implies that: (i) neither the direct cost in \eqref{eqn:poisson-DC} nor the \ref{eqn:TV} cost admit upper kernels, and (ii) \emph{every} $k$ is a lower kernel of these cost functions. Therefore, the ``flow cost'' of incremental evidence is infinite and, as a result, \cref{thm:qk} and \cref{lem:phi-ie} yield trivial bounds in this setting.
\end{eg}

\subsection{Determining the Sequential Learning Map}\label{ssec:flie-characteriation}

We now identify the precise condition under which the bounds in \cref{thm:qk} are tight:

\begin{definition}[FLIEs] \label{axiom:flie} $C\in\C$ \dred{favors learning via incremental evidence} (\nameref{axiom:flie}) if $C\succeq \Phi_\text{IE}(C)$.
\end{definition}
In words, a cost function \nameref{axiom:flie} if the expected cost of acquiring information via incremental learning is always weakly lower than the cost of acquiring it directly. Therefore, \nameref{axiom:flie} generalizes the inequality \eqref{eqn:flies-eg1} from \cref{eg:Diffusion:1} to arbitrary direct cost functions.

\begin{thm}\label{thm:flie} 
For any open convex set $W \subseteq \Delta^\circ(\Theta)$, strongly convex $H \in \mathbf{C}^2(W)$, and direct cost $C \in \C$ that is \nameref{defi:lq} on $W$ and satisfies $\dom(C)\subseteq\Delta(W)\cup \Ex^{\varnothing}$, 
\begin{align*}
	C \ \text{ \nameref{axiom:flie} \ and  } \ k_C=\H H \ \ \iff \ \  \Phi(C) = C^H_\text{ups}.
 \end{align*}
\end{thm}
\begin{proof}
    See \cref{proof:flie}.
\end{proof}

\cref{thm:flie} offers two characterizations. First, it fully determines the domain of (\nameref{defi:lq}) direct costs that give rise to \nameref{defi:ll}/\nameref{defi:ups} indirect costs: such direct costs \hyperref[axiom:flie]{FLIE} and have integrable kernels. Second, it fully determines the map $\Phi$ for the codomain of \nameref{defi:ll}/\nameref{defi:ups} indirect costs, which are pinned down by the kernels of their direct costs.\footnote{The hypothesis that $C$ is \nameref{defi:lq} is ``nearly'' without loss of generality for \cref{thm:flie} in two respects: (i) minor variants of both directions hold without it, and (ii) any $C \in \C$ for which $\Phi(C) = C^H_\text{ups}$ can be approximated arbitrarily well by a \nameref{defi:lq} $C' \in \C$ for which both directions hold exactly. See \cref{proof:flie-nonsmooth} for details. \label{fn:lq-thm4}} 

The first characterization provides a novel economic foundation for the \nameref{defi:ups} model. In particular, \cref{thm:flie} implies that $\Phi(C)$ is \nameref{defi:ups} \emph{if and only if} $ \Phi(C) = \Phi_\text{IE}(C)$ is \hyperref[axiom:additive]{Additive}; that is, incremental learning is globally optimal and, for each fixed target, all incremental learning strategies are equally costly.\footnote{In general, $C\in \C$ \nameref{axiom:flie} \emph{if and only if} $\Phi(C) = \Phi_\text{IE}(C)$. The ``if'' direction is immediate because $C \succeq \Phi(C)$. For the ``only if'' direction, note that \nameref{axiom:flie} implies $\Phi(C) \succeq \Phi(\Phi_\text{IE}(C)) = \Phi_\text{IE}(C) \succeq \Phi(C)$ because $\Phi$ is isotone and $\Phi_\text{IE}(C)$ is \nameref{axiom:slp}.} This helps delineate the range of applications in which \nameref{defi:ups} costs are economically reasonable. The following examples illustrate: 
\begin{itemize}[leftmargin=1em]
    %%%
    \item \emph{Cognitive costs of attention:} Following \textcite{sims_JME2003}, the rational inattention literature often interprets \nameref{defi:ups} costs as modeling the cognitive costs of processing freely available information (e.g., \cite{dean2023experimental,denti2022posterior}). In psychology and neuroscience, the \emph{drift-diffusion model (DDM)} models human cognition as a process of sequentially sampling diffusion signals (e.g., \cite{ratcliff2008diffusion}). \cref{thm:flie} suggests a bridge between these literatures: the indirect cost of attention is \nameref{defi:ups} \emph{if and only if} DDM-style sampling is the optimal cognitive process (cf. \cite{hebert2023rational}). 
    %%%
    %%%
    \item \emph{Statistical sampling:} Consider the problem of testing a hypothesis by sampling from a large population, where each draw is minimally informative on its own (e.g., political polling, market research, or clinical trials). These applications closely match the setting of \cref{eg:Diffusion:0}, in which the direct cost \nameref{axiom:flie} and the indirect cost is the \ref{eqn:MS} cost. 
    %%%
    \item \emph{Research \& development:} In industrial R\&D and scientific research, learning often occurs via discrete ``breakthroughs.'' Since under \nameref{axiom:flie} it is never (strictly) optimal to acquire such discrete ``chunks'' of information, \nameref{defi:ups} indirect costs are unnatural in these settings. Instead, these applications more closely match the setting of \cref{eg:Poisson:0}, where the optimal strategy samples from a Poisson process and the indirect cost is the \ref{eqn:TV} cost.
    %
    %Under \nameref{axiom:flie}, it is never (strictly) optimal to acquire ``discrete chunks'' of information that cause beliefs to jump. \cref{thm:flie} therefore suggests that \nameref{defi:ups} indirect costs are unnatural in settings where learning typically occurs via discrete ``breakthroughs'' (e.g., industrial R\&D or scientific research). Rather, such applications closely match the setting of \cref{eg:Poisson:0}, in which the optimal strategy involves sampling from a Poisson signal process and the indirect cost is the \ref{eqn:TV} cost.
    %this property is inappropriate for modeling settings where learning occurs via infrequent, discrete ``breakthroughs,'' which are common in industrial R\&D and the process of scientific research.%, where learning often occurs through infrequent and discrete ``breakthroughs'' that can be modeled as jumps of a Poisson signal process \parencite{che-mierendorff-aer2019,zhong2022optimal}. 
    %%%
\end{itemize}

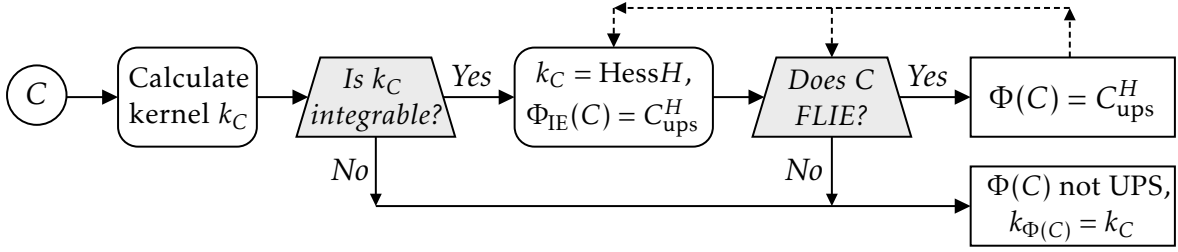
\begin{figure}[t]
    \centering
    \tikzset{every picture/.style={line width=0.75pt}} %set default line width to 0.75pt        

\begin{tikzpicture}[x=0.75pt,y=0.75pt,yscale=-1,xscale=1]
%uncomment if require: \path (0,201); %set diagram left start at 0, and has height of 201

%Rounded Rect [id:dp9805259259632102] 
\draw   (60,85) .. controls (60,79.48) and (64.48,75) .. (70,75) -- (120,75) .. controls (125.52,75) and (130,79.48) .. (130,85) -- (130,115) .. controls (130,120.52) and (125.52,125) .. (120,125) -- (70,125) .. controls (64.48,125) and (60,120.52) .. (60,115) -- cycle ;
%Shape: Trapezoid [id:dp3639184848454238] 
\draw  [fill={rgb, 255:red, 0; green, 0; blue, 0 }  ,fill opacity=0.08 ] (150,120) -- (162,80) -- (218,80) -- (230,120) -- cycle ;
%Rounded Rect [id:dp25669692811206724] 
\draw   (260,85) .. controls (260,79.48) and (264.48,75) .. (270,75) -- (350,75) .. controls (355.52,75) and (360,79.48) .. (360,85) -- (360,115) .. controls (360,120.52) and (355.52,125) .. (350,125) -- (270,125) .. controls (264.48,125) and (260,120.52) .. (260,115) -- cycle ;
%Shape: Trapezoid [id:dp33584300121378885] 
\draw  [fill={rgb, 255:red, 0; green, 0; blue, 0 }  ,fill opacity=0.08 ] (380,120) -- (392,80) -- (448,80) -- (460,120) -- cycle ;
%Shape: Circle [id:dp9301134321518603] 
\draw   (4,99) .. controls (4,90.72) and (10.72,84) .. (19,84) .. controls (27.28,84) and (34,90.72) .. (34,99) .. controls (34,107.28) and (27.28,114) .. (19,114) .. controls (10.72,114) and (4,107.28) .. (4,99) -- cycle ;
%Shape: Rectangle [id:dp6506234544453419] 
\draw   (490,80) -- (593,80) -- (593,120) -- (490,120) -- cycle ;
%Straight Lines [id:da4185735902749925] 
\draw    (190,155) -- (487,155) ;
\draw [shift={(490,155)}, rotate = 180] [fill={rgb, 255:red, 0; green, 0; blue, 0 }  ][line width=0.08]  [draw opacity=0] (8.04,-3.86) -- (0,0) -- (8.04,3.86) -- cycle    ;
%Straight Lines [id:da9368537519653215] 
\draw    (224,100) -- (257,100) ;
\draw [shift={(260,100)}, rotate = 180] [fill={rgb, 255:red, 0; green, 0; blue, 0 }  ][line width=0.08]  [draw opacity=0] (8.04,-3.86) -- (0,0) -- (8.04,3.86) -- cycle    ;
%Shape: Rectangle [id:dp08318048350052054] 
\draw   (490,135) -- (593,135) -- (593,175) -- (490,175) -- cycle ;
%Straight Lines [id:da7816024128715895] 
\draw    (360,100) -- (383,100) ;
\draw [shift={(386,100)}, rotate = 180] [fill={rgb, 255:red, 0; green, 0; blue, 0 }  ][line width=0.08]  [draw opacity=0] (8.04,-3.86) -- (0,0) -- (8.04,3.86) -- cycle    ;
%Straight Lines [id:da3315533726896439] 
\draw    (130,100) -- (153,100) ;
\draw [shift={(156,100)}, rotate = 180] [fill={rgb, 255:red, 0; green, 0; blue, 0 }  ][line width=0.08]  [draw opacity=0] (8.04,-3.86) -- (0,0) -- (8.04,3.86) -- cycle    ;
%Straight Lines [id:da48386168604071045] 
\draw    (34,100) -- (57,100) ;
\draw [shift={(60,100)}, rotate = 180] [fill={rgb, 255:red, 0; green, 0; blue, 0 }  ][line width=0.08]  [draw opacity=0] (8.04,-3.86) -- (0,0) -- (8.04,3.86) -- cycle    ;
%Straight Lines [id:da36868482098813615] 
\draw    (190,120) -- (190,152) ;
\draw [shift={(190,155)}, rotate = 270] [fill={rgb, 255:red, 0; green, 0; blue, 0 }  ][line width=0.08]  [draw opacity=0] (6.25,-3) -- (0,0) -- (6.25,3) -- cycle    ;
%Straight Lines [id:da15954648577343378] 
\draw    (420,120) -- (420,152) ;
\draw [shift={(420,155)}, rotate = 270] [fill={rgb, 255:red, 0; green, 0; blue, 0 }  ][line width=0.08]  [draw opacity=0] (6.25,-3) -- (0,0) -- (6.25,3) -- cycle    ;
%Straight Lines [id:da7473030552918111] 
\draw  [dash pattern={on 2.25pt off 1.5pt}]  (540,80) -- (540,58) ;
\draw [shift={(540,55)}, rotate = 90] [fill={rgb, 255:red, 0; green, 0; blue, 0 }  ][line width=0.08]  [draw opacity=0] (6.25,-3) -- (0,0) -- (6.25,3) -- cycle    ;
%Straight Lines [id:da07434258177442521] 
\draw  [dash pattern={on 2.25pt off 1.5pt}]  (540,55) -- (313,55) ;
\draw [shift={(310,55)}, rotate = 360] [fill={rgb, 255:red, 0; green, 0; blue, 0 }  ][line width=0.08]  [draw opacity=0] (6.25,-3) -- (0,0) -- (6.25,3) -- cycle    ;
%Straight Lines [id:da28456451676958106] 
\draw  [dash pattern={on 2.25pt off 1.5pt}]  (310,55) -- (310,72) ;
\draw [shift={(310,75)}, rotate = 270] [fill={rgb, 255:red, 0; green, 0; blue, 0 }  ][line width=0.08]  [draw opacity=0] (7.14,-3.43) -- (0,0) -- (7.14,3.43) -- cycle    ;
%Straight Lines [id:da4972584547710407] 
\draw  [dash pattern={on 2.25pt off 1.5pt}]  (420,55) -- (420,77) ;
\draw [shift={(420,80)}, rotate = 270] [fill={rgb, 255:red, 0; green, 0; blue, 0 }  ][line width=0.08]  [draw opacity=0] (7.14,-3.43) -- (0,0) -- (7.14,3.43) -- cycle    ;
%Straight Lines [id:da008056199845637657] 
\draw    (454,100) -- (487,100) ;
\draw [shift={(490,100)}, rotate = 180] [fill={rgb, 255:red, 0; green, 0; blue, 0 }  ][line width=0.08]  [draw opacity=0] (8.04,-3.86) -- (0,0) -- (8.04,3.86) -- cycle    ;

% Text Node
\draw (11.5,91) node [anchor=north west][inner sep=0.75pt]    {$C$};
% Text Node
\draw (96,100.5) node  [font=\small] [align=left] {Calculate\\ kernel $ k_{C}$};
% Text Node
\draw (190.5,101) node  [font=\small] [align=left] {$\ \ \ \,$ \textit{Is} $ k_{C}$\\\textit{integrable?}};
% Text Node
\draw (310,88.5) node  [font=\small] [align=left] {$k_{C} = \H H$,%\\  \\$\Phi_\text{IE}( C) =C_\text{ups}^{H}$
};
% Text Node
\draw (311.5,110.5) node  [font=\small] [align=left] {$\Phi_\text{IE}( C) =C_\text{ups}^{H}$
};
%% Text Node
%\draw (419.5,101) node  [font=\small] [align=left] {$\displaystyle C\succeq C_{ups}^{H}$?};
% Text Node
\draw (420,100) node  [font=\small] [align=center] {\textit{Does} $C$ \\  \textit{FLIE?}};
% Text Node
\draw (498,90) node [anchor=north west][inner sep=0.75pt]  [font=\normalsize] [align=left] {$\Phi ( C) =C_\text{ups}^{H}$};
% Text Node
\draw (497,137) node [anchor=north west][inner sep=0.75pt]  [font=\small] [align=left] {$\Phi ( C)$ not UPS, \\  $ \ \ \, \,  k_{\Phi ( C)} =k_{C}$};
% Text Node
\draw (226,82) node [anchor=north west][inner sep=0.75pt]  [font=\normalsize] [align=left] {\textit{Yes}};
% Text Node
\draw (456,82) node [anchor=north west][inner sep=0.75pt]  [font=\normalsize] [align=left] {\textit{Yes}};
% Text Node
\draw (166,129.5) node [anchor=north west][inner sep=0.75pt]  [font=\normalsize] [align=left] {\textit{No}};
% Text Node
\draw (395,129.5) node [anchor=north west][inner sep=0.75pt]  [font=\normalsize] [align=left] {\textit{No}};

\end{tikzpicture}
    \caption{%Flow diagram for calculating direct/indirect cost.
    An algorithm for calculating $\Phi$ (solid arrows) and $\Phi^{-1}$ (dashed arrows).}
    \label{fig:flow}
\end{figure}

The second characterization yields a methodological tool for analyzing the sequential learning map. In particular, the procedure depicted in \cref{fig:flow} can be used to tractably calculate both: (i) the indirect cost $\Phi(C)$ generated by any direct cost $C$ that satisfies the conditions of \cref{thm:flie}, and (ii) the full set of direct costs $\Phi^{-1}\big(C^H_\text{ups}\big)$ that rationalize any \nameref{defi:ll}/\nameref{defi:ups} indirect cost $C^H_\text{ups}$. We develop applications of this tool in \cref{section:applications} below.

%\begin{remark}\label{remark:lq-thm4}
%    The hypothesis that $C$ is \nameref{defi:lq} is ``nearly'' without loss of generality for \cref{thm:flie} in two respects: (i) minor variants of both directions hold without it, and (ii) any $C \in \C$ for which $\Phi(C) = C^H_\text{ups}$ can be approximated arbitrarily well by a \nameref{defi:lq} $C' \in \C$ for which both directions hold exactly. See \cref{proof:flie-nonsmooth} for additional details.
%\end{remark}

\section{Applications: Reduced-Form Information Costs}\label{section:applications}

In this section, we apply our framework by studying various reduced-form cost functions through the lens of optimization. \cref{ssec:illustrative} presents illustrative examples. \dred{Sections} \ref{ssec:axiomatic}--\ref{ssec:SPI} develop the information cost trilemma and its resolution (recall \cref{fig:trilemma,fig:wald-thm}).

For these applications, we focus mainly on cost functions with \emph{rich domain}, that is, $C \in \C$ such that $\dom(C) = \Delta(\Delta^\circ(\Theta))\cup \Ex^\varnothing$. This focus aligns with the emphasis placed on rich- and full-domain cost functions in the flexible information acquisition literature. %(Per \cref{remark:full-support}, we can analyze any full-domain cost function via its rich-domain restriction.) 

\subsection{Illustrative Examples}\label{ssec:illustrative}

We present two examples that illustrate how \cref{thm:flie} can be used in practice to calculate the sequential learning map (as depicted in \cref{fig:flow}). In the first example, we characterize the set of all direct costs that generate the classic \hyperref[MI]{Mutual Information} indirect cost \parencite{shannon-1948,sims_JME2003}. In the second example, we introduce a novel class of direct costs and show that their indirect costs include the families of neighborhood-based costs \parencite{hebert2021neighborhood} and pairwise separable costs \parencite{morris-yang-continuous}.

\setcounter{eg}{2}
\begin{eg}[Mutual Information]
Following \textcite{sims_JME2003}, much of the rational inattention literature focuses on the \emph{\hyperref[MI]{Mutual Information}} cost of \textcite{shannon-1948}, which is the full-domain \nameref{defi:ups} cost $C^{H_\text{MI}}_\text{ups} \in \C$ given by
    \begin{align}
       H_\text{MI} (p) := \sum_{\theta \in \Theta} p (\theta)  \log(p (\theta)) \quad \text{ and }  \quad \H H_\text{MI} (p) = \diag(p)^{-1} - \mathbf{1}\mathbf{1}^\top, \label{MI} \tag{MI}
    \end{align}
    where $H_\text{MI}$ is the negative of \emph{Shannon entropy} and $\H H_\text{MI}$ is the \emph{Fisher information matrix}. 
    
    To formally apply our results, we let $C^\circ_\text{MI} \in \C$ denote the rich-domain restriction of \hyperref[MI]{Mutual Information}. \cref{thm:flie} and \cref{lem:phi-ie} then directly imply the following:
    % that a \nameref{defi:lq} direct cost $C \in \C$ generates the indirect cost $\Phi(C) = C^\circ_\text{MI}$ if and only if $k_C = k_\text{MI}$ and $C$ \nameref{axiom:flie} (i.e., $C \succeq  C^\circ_\text{MI}$.
    \begin{cor}\label{cor:MI}
        For any \nameref{defi:lq} $C \in \C$,
        \[
        \Phi(C) = C^\circ_\text{MI} \ \ \iff \ \   C \ \text{ \nameref{axiom:flie} and } \ \Phi_\text{IE}(C) = C^\circ_\text{MI} \ \ \iff \ \ C \succeq C^\circ_\text{MI} \ \text{ and } \ k_C= \H H_\text{MI}.
        \]
    \end{cor}
    \begin{proof}
    See \cref{ssec:proofs-MI-combine}.
    \end{proof}
    
The classic information-theoretic foundation for \hyperref[MI]{Mutual Information} is that it \emph{approximates} the indirect cost generated by the direct cost under which: (i) all ``bits'' (i.e., binary partitions of the state space) are
equally costly and all other experiments are infeasible, and hence (ii) the optimal strategy is to sequentially ask deterministic yes-or-no questions about the state \parencite{shannon-1948,cover-thomas-2006}. However, since the approximation error vanishes only if the DM is able to ``block code'' many i.i.d.~draws of the state, this reasoning does not directly apply to economic settings where the DM faces a one-time decision problem. \cref{cor:MI} offers a novel and complementary foundation by characterizing all (\nameref{defi:lq}) direct costs that \emph{exactly} generate \hyperref[MI]{Mutual Information}. 
\end{eg}

\begin{eg}[Combining Technologies]\label{eg:combining}
It is common for DMs to learn via a combination of multiple ``basic technologies'' that each produce information about different aspects of the state. For instance, clinical trials often draw samples from multiple subpopulations and use a variety of measurement devices (e.g., distinct pieces of lab equipment). We propose a stylized model of this phenomenon.

Given a finite profile of cost functions $(C^i)_{i=1}^n$ and a map $g : \overline{\R}^n_+ \to \overline{\R}_+$, the direct cost $C \in \C$ is defined as $C(\pi):=g(C^1(\pi),\ldots,C^n(\pi))$. We interpret each $C^i$ as a ``basic technology'' and the map $g$ as a ``production function'' that combines them. We assume that each $C^i$ satisfies $\dom(C^i) \subseteq \Delta(\Delta^\circ(\Theta)) \cup \Ex^\varnothing$ and is \nameref{defi:lq} with kernel $k_{C^i}=\H H^i$ for some convex  $H^i\in \mathbf{C}^2(\Delta^\circ(\Theta))$. To allow for rich complementarities and substitutabilities among the technologies, we assume only that the map $g$ satisfies $g(\mathbf{0}) = 0$ and is both continuously differentiable and subdifferentiable at $\mathbf{0}$, with gradient $\nabla g(\mathbf{0}) = (\nabla_i g(\mathbf{0}))_{i=1}^n$.\footnote{Recall that $g$ is \emph{subdifferentiable at $\mathbf{0}$} if $g(x) \geq x^\top \nabla g(\mathbf{0}) $ for all $x \in \mathbb{R}^n_{+}$, which automatically holds if $g$ is convex.} For technical reasons, we also assume that $\underline{C} :=\sum_{i=1}^n \nabla_i g(\mathbf{0}) C^i \in \C$ is \nameref{defi:sp}. Under these assumptions, \cref{thm:qk,thm:flie} and \cref{lem:phi-ie} yield the following:

        \begin{cor}\label{prop:mult:tech}
            The direct cost $C \in \C$ is \nameref{defi:lq} and, for $H := \sum_{i=1}^n \nabla_i g(\mathbf{0}) H^i$, satisfies  
            \[
            \Phi(C) \preceq \Phi_\text{IE}(C) = \Phi_\text{IE}(\underline{C}) = C^H_\text{ups} \quad  \text{ and } \quad k_C = k_{\Phi(C)}= \H H = \sum_{i =1}^n \nabla_i g(\mathbf{0}) \, \H H^i.
            \]
            %Moreover, it holds that $\Phi(C)=C^{H}_\text{ups}$ if and only if $\underline{C}$ \nameref{axiom:flie} (i.e., $\underline{C}\succeq \Phi_\text{IE}(\underline{C}) = C^H_\text{ups}$).
            Moreover, if  $\underline{C}$ \nameref{axiom:flie} (i.e., $\underline{C}\succeq  C^H_\text{ups}$), then $C$ \nameref{axiom:flie} and  $\Phi(C)=C^{H}_\text{ups}$.
        \end{cor}
        \begin{proof}
            See \cref{ssec:proofs-MI-combine}.
        \end{proof}

        \cref{prop:mult:tech} offers a simple way to check whether the indirect cost is \nameref{defi:ll}/\nameref{defi:ups} and characterizes its form. Notably, optimization ``smooths away'' all non-linearities in the production function: the inequality $\underline{C}\succeq C^H_\text{ups}$ and the functional form $\Phi(C) = C^H_\text{ups}$ both depend on $g$ only through the gradient $\nabla g(\mathbf{0})$. We highlight two special cases of interest:

        \begin{itemize}[leftmargin=1em]
            \item Let each $i \in \{1,\dots,n\}$ index a nonempty ``neighborhood'' of states $N_i \subseteq \Theta$ such that $\{N_i\}_{i =1}^n$ covers $\Theta$, and define each $H^i$ as $H^i(p) := p(N_i) G^i (p(\cdot \mid N_i ))$ for some symmetric convex $G^i \in \mathbf{C}^2(\Delta^\circ(N_i))$. 
            %\footnote{For each $p \in \Delta^\circ(\Theta)$ and $N_i \subseteq \Theta$, we let $p(N_i) := \sum_{\theta \in N_i} p(\theta)$ and $p(\theta \mid N_i) := p(\theta) / p(N_i)$ for all $\theta \in N_i$.} 
            The resulting family of indirect costs $\Phi(C) = C^H_\text{ups}$ then equals the family of \emph{neighborhood-based costs} from \textcite{hebert2021neighborhood}.%, where $\nabla f_i(\mathbf{0})$ is the marginal cost of learning within neighborhood $N_i$.
            %%%
            \item Let each $i \in \{1, \dots, |\Theta|^2\}$ index a distinct ordered pair of states $(\theta, \theta') \in \Theta \times \Theta$. For each such pair, define $H^{(\theta,\theta')} \in \mathbf{C}^2(\Delta^\circ(\Theta))$ as $H^{(\theta,\theta')}(p) := p(\theta') \phi\left( \frac{p(\theta)}{p(\theta')} \right)$ for some convex $\phi \in \mathbf{C}^2(\R_{++})$ normalized such that $\phi(1) = 0$. Letting $\gamma_{\theta,\theta'} := \nabla g_{(\theta,\theta')}(\mathbf{0})$, we then have
        \[
        H(p) 
        = \sum_{\theta,\theta' \in \Theta} \gamma_{\theta,\theta'}  \, p(\theta') \phi\left( \frac{p(\theta)}{p(\theta')} \right).
        \]
        The resulting family of indirect costs $\Phi(C) = C^H_\text{ups}$ can be viewed as a natural finite-state analog to the family of \emph{pairwise-separable costs} from \textcite{morris-yang-continuous}.\footnote{Fixing the uniform prior $p^\star(\cdot) \equiv 1/|\Theta|$, Bayes' rule and $\phi(1) = 0$ yield $\Phi(C)(h_B(\sigma,p^\star)) \equiv \frac{1}{|\Theta|} \sum_{\theta,\theta'} \gamma_{\theta,\theta'} D_\phi(\sigma_\theta \mid \sigma_{\theta'})$, where $D_\phi(\sigma_\theta \mid \sigma_{\theta'}):= \int_S \phi\left( \frac{\d \sigma_{\theta}}{\d \sigma_{\theta'} }(s) \right) \dd \sigma_{\theta'} (s)$ is the \emph{$\phi$-divergence} between the signal distributions $\sigma_\theta, \sigma_{\theta'} \in \Delta(S)$ (e.g., \cite{csiszar1967information}). \textcite{morris-yang-continuous}, who also hold the prior fixed, study the continuous-state analog of this functional form (for a broader class of ``decreasing differences'' divergences) on the restricted  domain of binary-signal experiments.}  
        \end{itemize}

\end{eg}

\subsection{Context-Specific Properties}\label{ssec:axiomatic}

Next, we introduce two important context-specific axioms from the literature---\hyperref[axiom:prior:invariant]{Prior Invariance} and \hyperref[axiom:CMC:0]{Constant Marginal Cost}---and two novel \nameref{axiom:slp} costs that satisfy them.

\subsubsection{Prior Invariance}
\label{ssec:PI}

In many economic settings, information acquisition involves expending physical resources (e.g., for statistical sampling or R\&D) or money (e.g., in markets for news or data). The literature has proposed that, when modeling such applications, it is most natural to use cost functions that depend only on the ``objective'' experiment being acquired---\emph{not} on the DM's ``subjective'' prior beliefs. This property is formalized via the following axiom:

\begin{axiom}\label{axiom:prior:invariant}
    \hspace{-0.25em}
    \mbox{$C\in\C$ is \hyperref[axiom:prior:invariant]{Prior Invariant} if, for every $\sigma \in \Se$ and all $p,p'\in\Delta(\Theta)$ with common support,\footnotemark}
    \[
    C(h_B(\sigma,p))=C(h_B(\sigma,p')).
    \]
    %$C\in\C$ is (weakly) \hyperref[axiom:prior:invariant]{Prior Invariant} if $C(h_B(\sigma,p))=C(h_B(\sigma,p'))$ for every $\sigma \in \Se$ and all priors $p,p'\in\Delta(\Theta)$ with the same support.\footnotemark
\end{axiom}
%%%%
\footnotetext{\cref{axiom:prior:invariant} allows the cost of an experiment to vary across priors with \emph{different} supports. This is an artifact of the belief-based approach: since all experiments induce trivial random posteriors when the prior is concentrated on a single state, $C \in \C$ is \emph{fully} prior-independent \emph{if and only if} it is identically zero. However, this is \emph{inconsequential} for our analysis in two respects: (i) since we allow cost functions to be discontinuous, it is irrelevant whenever we restrict attention to full-support priors (recall \cref{remark:full-support}); and (ii) full prior-independence is easily incorporated in the richer ``experiment-based'' version of our framework, to which all of our results extend (see  \cref{ssec:beyond:belief,app:beyond:belief}).}

\hyperref[axiom:prior:invariant]{Prior Invariance} is a standard assumption on direct costs in models of statistical sampling, ranging from \textcite{wald-ams1945} to our \dred{Examples} \ref{eg:Diffusion:0} and \ref{eg:Poisson:0}. However, most reduced-form cost functions in the flexible information acquisition literature \emph{violate} \hyperref[axiom:prior:invariant]{Prior Invariance}. 
%For instance, it is well known that mutual information is not \hyperref[axiom:prior:invariant]{Prior Invariant} \parencite[Theorem 2.7.4]{cover-thomas-2006}. More broadly, 
For instance, it is immediate that any \emph{(weak$^*$) continuous} $C \in \C$---a class that includes most \emph{full-domain} \nameref{defi:ups} costs---is \hyperref[axiom:prior:invariant]{Prior Invariant} \emph{if and only if} it is identically zero.\footnote{The class of weak$^{*}$-continuous $C \in \C$ has the ``free at full information'' property highlighted in \textcite{denti2022experimental}  and includes all \nameref{defi:ups} costs $C^H_\text{ups}$ for which $H : \Delta(\Theta) \to \R$ is continuous, e.g., \hyperref[MI]{Mutual Information}.} 
%\footnote{For the ``only if'' direction, let $C \in \C$ be \hyperref[axiom:prior:invariant]{Prior Invariant} and weak$^*$-continuous. For any experiment $\sigma \in \Se$, state $\theta \in \Theta$, and sequence of priors $p^n \in \Delta(\Theta)$ such that $\lim_{n\to \infty} p^n = \delta_\theta$, continuity implies $\lim_{n\to\infty} C(h_B(\sigma,p^n)) = C(h_B(\sigma,\delta_\theta)) = C(\delta_{\delta_\theta}) =0$. By \hyperref[axiom:prior:invariant]{Prior Invariance}, it follows that $C[\Ex]=\{0\}$. This logic applies to any \nameref{defi:ups} cost $C^H_\text{ups}$ for which $H : \Delta(\Theta) \to \R$ is continuous, e.g., \hyperref[MI]{Mutual Information}. See also \textcite{denti2022experimental,mensch-cardinal-information}. }
%%LONGER VERSION OF FOOTNOTE BELOW
%\footnote{Let $C \in \C$ be \hyperref[axiom:prior:invariant]{Prior Invariant} and (weak$^*$) continuous. Fix any $\sigma \in \Se$ and $p \in \Delta(\Theta)$. Pick some $\theta\in \supp(p)$ and define the sequence $p^n := \frac{1}{n} p + \frac{n-1}{n} \delta_\theta$, which satisfies $\lim_{n\to\infty} p^n=\delta_\theta$ and $\supp(p^n) = \supp(p)$ for all $n \in \mathbb{N}$. Then $C(h_B(\sigma,p)) = \lim_{n\to\infty}C(h_B(\sigma,p^n)) = C(h_B(\sigma,\delta_\theta)) = C(\delta_{\delta_\theta}) = 0$, where the first equality is by \hyperref[axiom:prior:invariant]{Prior Invariance} and the second equality is by continuity. It follows that $C[\Ex]=\{0\}$. This argument---which also appears in \textcite[p. 3115]{denti2022experimental}---applies to any \nameref{defi:ups} cost $C^H_\text{ups}$ for which $H : \Delta(\Theta) \to \R$ is continuous.} 
The \ref{eqn:MS} cost from \cref{eg:Diffusion:0}, which is \nameref{defi:ups} but not full-domain, is also clearly prior-dependent.

Meanwhile, the \ref{eqn:TV} cost from \cref{eg:Poisson:0} implies that---once we look beyond the \nameref{defi:ll}/\nameref{defi:ups} class---there do exist nontrivial, full-domain \nameref{axiom:slp} cost functions that satisfy \hyperref[axiom:prior:invariant]{Prior Invariance}.  The following cost function extends the \ref{eqn:TV} cost to general state spaces:

\begin{definition}[MLR]\label{defi:MLR}
    %\hspace{-0.5em}
    The \dred{Minimal Likelihood Ratio} (\nameref{defi:MLR}) cost is defined, for all $\pi\in\Ex$, as
\[
    C_\text{MLR}(\pi) := \mathbb{E}_\pi \left[ D_\text{MLR} (q \mid p_\pi ) \right]  \quad \ \text{ where } \  \quad D_\text{MLR}(q \mid p) : = 1-\min_{\theta \in\supp(p)} \frac{q(\theta)}{p(\theta)}.
    \]
    %By Bayes' rule, for all $\sigma\in \Se$ with finite $S$ and %$p\in\Delta(\Theta)$,
    %\begin{align*}
    %           C_\text{MLR}(h_B(\sigma, p)) = 1- \sum_{s\in S}\min_{\theta\in \supp(p)} \sigma_\theta(s).\footnotemark
    %\end{align*}  
     Equivalently, for every $\sigma \in \Se$ and $p \in \Delta(\Theta)$, 
     %\begin{align*}
     %           C_\text{MLR}(h_B(\sigma, p)) = 1- \bigwedge_{\theta  \in \supp(p)} \sigma_\theta ( S),
     %\end{align*}    
     %where $\bigwedge_{\theta  \in \supp(p)} \sigma_\theta$ denotes the meet of the measures $\{\sigma_\theta\}_{\theta \in \supp(p)} \subseteq \Delta(S)$.\footnotemark
     \begin{align*}
                C_\text{MLR}(h_B(\sigma, p)) = 1- \int_S \, \min_{\theta \in \supp(p)} \left\{\frac{\d \sigma_\theta}{\d \overline{\sigma}}(s) \right\} \dd \overline{\sigma}(s), 
     \end{align*}    
     where $\overline{\sigma} := \sum_{\theta \in \Theta} \sigma_\theta $ denotes the sum of the state-contingent signal distributions.\footnotemark 
\end{definition}
\footnotetext{The second representation follows from the first one via Bayes' rule  and the fact that, for every  $\sigma \in \Se$, all of the state-contingent signal distributions $ \sigma_\theta \in \Delta(S)$ are absolutely continuous with respect to the (Borel) measure $\overline{\sigma}$.}
%\footnotetext{When $S$ is infinite, $C_\text{MLR}(h_B(\sigma, p)) = 1- \bigwedge_{\theta  \in \supp(p)} \sigma_\theta (S)$, where the meet (minimum) of two Radon measures $\mu\wedge\nu$ is $\mu-(\mu-\nu)^+$ per the Hahn decomposition theorem and the finite meet is defined accordingly. }
%\footnotetext{The meet (minimum) of two Radon measures $\mu\wedge\nu$ is $\mu-(\mu-\nu)^+$ per the Hahn decomposition theorem. The finite meet is defined accordingly. When $S$ is finite, $C_\text{MLR}(h_B(\sigma, p)) =1- \sum_{s\in S}\min_{\theta\in \supp(p)} \sigma_\theta(s)$.}
%The \nameref{defi:MLR} cost extends the \ref{eqn:TV} cost to general state spaces. 
The \nameref{defi:MLR} cost is \hyperref[axiom:prior:invariant]{Prior Invariant} by construction. Like the \ref{eqn:TV} cost, it is also \nameref{axiom:slp} because the divergence $D_\text{MLR}$ is convex with respect to the posterior and is a quasi-metric (\cref{lem:MLR-quasimetric} in \cref{sssec:proof-tri-pt3}).  
%, where the former property implies  \hyperref[axiom:mono]{Monotonicity} and the latter implies \hyperref[axiom:POSL]{Subadditivity}. 
Moreover, we note that the \nameref{defi:MLR} cost can be derived via the natural multi-state analog of the Poisson sampling technology from \cref{eg:Poisson:0}.

\subsubsection{Constant Marginal Cost}\label{ssec:CMC}

In statistical settings, the cost of drawing independent samples from a population is often linear in the sample size (e.g., the number of consumers being surveyed). \textcite{pomatto2023cost} propose formalizing this property via the following axiom. 

For any two experiments $\sigma = (S,(\sigma_\theta)_{\theta\in\Theta})$ and $\sigma' = (S',(\sigma'_\theta)_{\theta\in\Theta})$, define their \emph{product} as the experiment $\sigma \otimes \sigma' = ( S \times S', (\sigma_\theta \times \sigma'_\theta)_{\theta \in \Theta})$, where $\sigma_\theta \times \sigma'_\theta$ is the product of the measures $\sigma_\theta$ and $\sigma'_\theta$. In words, $\sigma \otimes \sigma'$ draws conditionally independent signals from both $\sigma$ and $\sigma'$.

\begin{axiom}\label{axiom:CMC:0}
    $C\in \C$ exhibits \dred{Constant Marginal Cost} (\hyperref[axiom:CMC:0]{CMC}) if, for every $\sigma, \sigma' \in \Se$ and $p \in \Delta(\Theta)$, 
    \[
    C(h_B(\sigma \otimes \sigma' , p)) = C(h_B(\sigma  , p)) + C(h_B(\sigma'  , p)). 
    \]
\end{axiom}

\hyperref[axiom:CMC:0]{CMC} posits that the cost of acquiring any two experiments together equals the total cost of acquiring them separately and \emph{simultaneously} (i.e., under the same prior). That is, \hyperref[axiom:CMC:0]{CMC} models indifference to \emph{simultaneous decomposition} of information: the cost of directly acquiring $\sigma \otimes \sigma'$ equals the cost of producing it via the non-contingent two-step strategy that acquires $\sigma$ and then, without observing the realized signal, acquires $\sigma'$. This is the natural ``static'' analog of the sequential \hyperref[axiom:additive]{Additivity} property that characterizes  \nameref{defi:ups} costs. 

We introduce a new family of \nameref{defi:ups} costs that satisfy this static additivity property:

\begin{definition}[Total Information]\label{defi:TI}
    $C_\text{TI} \in \C$ is a \nameref{defi:TI} cost function if it has rich domain and there exist non-negative coefficients $(\gamma_{\theta,\theta'})_{\theta, \theta' \in \Theta}$ such that
    \[
    C_\text{TI}%(\pi) 
    = C^{H_\text{TI}}_\text{ups} %(\pi) 
    \quad \text{ where }  \quad H_\text{TI} (p) := \sum_{\theta \in \Theta} p (\theta) \sum_{\theta' \in \Theta} \gamma_{\theta,\theta'}   \log \left(\frac{p (\theta)}{p (\theta')}\right).
    \]
    Equivalently, for all $\sigma \in \Se$ and $p \in \Delta^\circ(\Theta)$ such that $h_B(\sigma,p) \in \Delta(\Delta^\circ(\Theta))$,
    \[
    C_\text{TI}(h_B(\sigma, p)) = \sum_{\theta \in \Theta} p(\theta) \sum_{\theta' \in \Theta} \gamma_{\theta,\theta'}  D_\text{KL}(\sigma_\theta \mid \sigma_{\theta'}),
    \]
    where $D_\text{KL}$ denotes the Kullback-Leibler (KL) divergence.\footnote{The KL divergence between the signal distributions $\sigma_\theta, \sigma_{\theta'} \in \Delta(S)$ is defined as $D_\text{KL}(\sigma_{\theta} \mid \sigma_{\theta'}) := \int \log\left(\frac{\d \sigma_{\theta}}{\d \sigma_{\theta'}}(s)\right)\d \sigma_{\theta}(s)$.}
\end{definition}

\nameref{defi:TI} is of special interest for three reasons. First, it is both \nameref{defi:ups} and \hyperref[axiom:CMC:0]{CMC}, where the latter property holds because KL divergence is additive with respect to products of measures. In combination, these properties constitute a strong form of ``process invariance:'' the expected cost of each experiment is invariant under both sequential and simultaneous decomposition. In other words, \nameref{defi:TI} costs depend only on the \emph{total amount of information} that is acquired, not on the strategy that is used to acquire it.

Second, \nameref{defi:TI} encompasses two important \nameref{defi:ups} cost functions from the literature as limiting cases: the \ref{eqn:MS} cost of \textcite{morris-strack-sampling} and the \nameref{eg:FI} cost of \textcite{hebert2021neighborhood}.\footnote{\nameref{defi:TI} also intersects the \nameref{defi:ups} costs from \cref{eg:combining}: it is the pairwise-separable cost with $\phi(t) \equiv t \log(t)$ and, given symmetric coefficients (i.e., $\gamma_{\theta,\theta'} = \gamma_{\theta',\theta}$ for all $\theta,\theta' \in \Theta)$, it can be represented as a neighborhood-based cost.} The \ref{eqn:MS} cost, which we have already seen in \cref{eg:Diffusion:0}, is the special case of \nameref{defi:TI} when the state space is binary and the coefficients are symmetric (i.e., $\Theta =\{0,1\}$ and $\gamma_{0,1} = \gamma_{1,0}$). At the other extreme, the \nameref{eg:FI} cost emerges as a specific continuous-state limit of \nameref{defi:TI}.

\begin{eg}[Fisher Information]\label{eg:FI}
%%%
\textcite{hebert2021neighborhood} assume that the state space is an interval $[\underline{\theta},\overline{\theta}] \subseteq \R$, the prior belief admits a density $\rho$,  and each feasible experiment $\sigma$ has a finite signal space $S$ and is differentiable with respect to the state. The \emph{\nameref{eg:FI}} \eqref{eqn:FI} cost is then defined as  
\begin{equation}\label{eqn:FI}
\text{FI}(\sigma,\rho) := \int_{\underline{\theta}}^{\overline{\theta}} \mathcal{I} (\theta;\sigma) \, \rho(\theta) \dd\theta, \quad \ \text{ where } \quad \  \mathcal{I} (\theta; \sigma) := \sum_{s \in S} \sigma_\theta(s) \, \left[\frac{\partial}{\partial \theta} \log \left( \sigma_\theta(s)\right)\right]^2 \tag{FI}
\end{equation}
is the ``Fisher information'' of experiment $\sigma$ in state $\theta$, a standard notion in statistics. 
%\parencite[Ch. 11.10]{cover-thomas-2006}.
Since $\mathcal{I}(\theta; \cdot)$ is additive with respect to product experiments, the \ref{eqn:FI} cost is both \nameref{defi:ups} and \hyperref[axiom:CMC:0]{CMC}.

It is well known that Fisher information is the ``Hessian'' of KL divergence, namely, $ \mathcal{I}(\theta; \sigma) = \lim_{\theta' \to \theta} \frac{2}{(\theta - \theta')^2}\cdot D_\text{KL}(\sigma_\theta \mid \sigma_{\theta'}) $. Thus, we can approximate the \ref{eqn:FI} cost with \nameref{defi:TI} by (i) discretizing the state space to a finite grid and (ii) setting $\gamma_{\theta,\theta'} = 1/(\theta-\theta')^2$ for adjacent gridpoints and $\gamma_{\theta,\theta'} = 0$ for non-adjacent gridpoints. 
Conversely, this special case of \nameref{defi:TI} can be viewed as the finite-state analog of the \ref{eqn:FI} cost.
%\footnote{See \textcite[Section II.C]{hebert2021neighborhood} for the (continuous-state) \nameref{defi:ups} representation of \eqref{eqn:FI} and the details of an analogous finite-state approximation scheme (which, notably, approximates \eqref{eqn:FI} with \nameref{defi:ups} costs that \emph{violate} \hyperref[axiom:CMC:0]{CMC}).}
%%
%\footnote{More formally, for each $n \in \mathbb{N}$ we can define the: (i) state space $\{\theta_1, \dots, \theta_n\}$ where $\theta_i := \underline{\theta} + \frac{i-1}{n} \cdot (\overline{\theta} - \underline{\theta})$, (ii) prior belief $p(\theta_i):= \int_{\theta_i}^{\theta_{i+1}} \rho(\theta) \dd \theta$, and (iii) coefficients $\gamma_{\theta_i,\theta_j} := \mathbf{1}\left( |i-j| \leq 1\right) / (\theta_i-\theta_j)^2$. Under mild technical conditions, the resulting sequence of \nameref{defi:TI} costs converges, in a pointwise sense, to \eqref{eqn:FI} as $n \to \infty$. See \textcite[Section II.C]{hebert2021neighborhood} for discussion of the requisite technical conditions and the (continuous-state) \nameref{defi:ups} representation of \eqref{eqn:FI}. } 
%%
\end{eg}

Finally, \nameref{defi:TI} is %closely 
related to the \emph{\dred{Log-Likelihood Ratio} (\ref{eqn:LLR})} costs of \textcite{pomatto2023cost}. 
%, which are \hyperref[axiom:CMC:0]{CMC} and \hyperref[axiom:prior:invariant]{Prior Invariant}. 
A rich-domain cost function $C_\text{LLR} \in \C$ is an \ref{eqn:LLR} cost if there are coefficients $\beta_{\theta,\theta'}\geq0$ such that, for every $\sigma \in \Se$ and $p \in \Delta^\circ(\Theta)$ with $h_B(\sigma,p) \in \Delta(\Delta^\circ(\Theta))$,
\begin{align}
    C_\text{LLR}(h_B(\sigma, p)) = \sum_{\theta , \theta' \in \Theta} \beta_{\theta, \theta'}  \, D_\text{KL}(\sigma_{\theta} \mid \sigma_{\theta'}). \label{eqn:LLR} \tag{LLR}
\end{align}
By construction, \ref{eqn:LLR} costs are \hyperref[axiom:CMC:0]{CMC} and \hyperref[axiom:prior:invariant]{Prior Invariant}. Every \nameref{defi:TI} cost can be interpreted as the expectation, under the prior, of a
collection of \ref{eqn:LLR} costs---one for each possible state. Conversely, for every \emph{fixed} prior $p \in \Delta^\circ(\Theta)$, the \ref{eqn:LLR} cost with coefficients $\beta_{\theta,\theta'}$ coincides with the \nameref{defi:TI} cost with coefficients $\gamma_{\theta,\theta'} \equiv \beta_{\theta,\theta'}/p(\theta)$.\footnote{Concurrent to our working paper \parencite{bloedel-zhong-2020}, \textcite[Section 7]{pomatto2023cost} note that \ref{eqn:LLR} costs can be extended to a broader class of prior-dependent ``Bayesian LLR'' costs with $p$-dependent coefficients $\beta_{\theta,\theta'}(p)$. In this context, they independently observe that the \nameref{defi:TI} functional form is both \hyperref[axiom:CMC:0]{CMC} and \nameref{defi:ups}.}

%%%%%%%%%%%%%%%%%%%%%%
%%%%%%%%%%%%%%%%%%%%%%%
%%%%%%%%%%%%%%%%%%%%%%

\subsection{An Information Cost Trilemma}\label{ssec:trilemma}

We have seen by example that three key properties of information costs---\nameref{axiom:slp}, \hyperref[axiom:prior:invariant]{Prior Invariance}, and \hyperref[axiom:CMC:0]{CMC}---have nontrivial two-way intersections. %This raises two questions. First, do any \emph{other} cost functions satisfy two of these properties? Second, do \emph{any} cost functions satisfy \emph{all three} of them? In this section, we answer both questions by establishing a trilemma that characterizes the modeling tradeoffs among these properties. %This suggests modeling tradeoffs. 
In this section, we establish a trilemma that fully characterizes the modeling tradeoffs among these properties.

We begin with two definitions. First, we call $C \in \C$ \emph{nontrivial} if it is not identically zero on $\dom(C)$. Second, we follow \textcite[Axiom 4]{pomatto2023cost} by augmenting \hyperref[axiom:CMC:0]{CMC} with a mild but complex continuity condition, which we call \emph{uniform total variation-moment-continuity} (henceforth, \emph{\hyperref[defi:pst:cont]{uTVM-continuity}}). For brevity, we embed this condition in the following definition and relegate its formal description to \cref{app:utvm}.

\begin{definition}[CMC$^{\text{\textcopyright}}$]\label{axiom:CMC}
  $C\in \C$ is \nameref{axiom:CMC} if it is both \hyperref[axiom:CMC:0]{CMC} and \nameref{defi:pst:cont}.%\footnote{The distinction between \hyperref[axiom:CMC:0]{CMC} and \nameref{axiom:CMC} is purely technical. Notably, in \cref{thm:trilemma} below, the stronger \nameref{axiom:CMC} property is \emph{not} required for point (iii) or the final statement in \cref{thm:trilemma} below.} 
\end{definition}

We can now formally state the \emph{information cost trilemma} (recall \cref{fig:trilemma}). Point (ii) of the result is %due to 
a minor adaptation of Theorem 1 from \textcite{pomatto2023cost} and is stated here only for completeness; it uses the \hyperref[axiom:DL]{Dilution Linearity} axiom from \cref{ssec:ups}.

\begin{thm}\label{thm:trilemma}
    For any nontrivial $C\in \C$ with rich domain, the following hold:
    %%%
    \begin{enumerate}[noitemsep,label=(\roman*)]   
        %%%
        \item \label{item:TI} $C$ is \nameref{axiom:slp} and \nameref{axiom:CMC} if and only if $C$ is a \nameref{defi:TI} cost.
        %%%
         \item \label{item:pst} %\parencite{pomatto2023cost}\ 
        $C$ is \hyperref[axiom:prior:invariant]{Prior Invariant}, \nameref{axiom:CMC}, and \hyperref[axiom:DL]{Dilution Linear} if and only if $C$ is an \ref{eqn:LLR} cost.
         %%%
        \item \label{item:PI} If $C$ is (the rich-domain restriction of) an \nameref{defi:MLR} cost, then it is \nameref{axiom:slp} and \hyperref[axiom:prior:invariant]{Prior Invariant}. Conversely, if $C$ is \nameref{axiom:slp} and \hyperref[axiom:prior:invariant]{Prior Invariant}, then it is neither \hyperref[axiom:CMC:0]{CMC} nor \nameref{defi:ups}. 
    \end{enumerate}
\end{thm}
\begin{proof}
    See \cref{proof:trilemma}. 
\end{proof}

\cref{thm:trilemma} characterizes the modeling tradeoffs among \nameref{axiom:slp}, \hyperref[axiom:prior:invariant]{Prior Invariance}, and \hyperref[axiom:CMC:0]{CMC} by showing that: (a) their \emph{two-way} intersections are \emph{nearly uniquely} determined by the \nameref{defi:TI}, \ref{eqn:LLR}, and \nameref{defi:MLR} cost functions, and (b) their \emph{three-way} intersection is \emph{empty}. 
%\footnote{We clarify two aspects of this result. First, the \hyperref[axiom:DL]{Dilution Linearity} condition in \cref{thm:trilemma}(ii) is economically mild and implied by \nameref{axiom:slp} (see \cref{ssec:ups}). Second, \cref{thm:trilemma}(iii) does not use \hyperref[defi:pst:cont]{uTVM-continuity}: it invokes \hyperref[axiom:CMC:0]{CMC}, not \nameref{axiom:CMC}.} 
It also establishes, as a corollary, that the two-way intersection of \nameref{defi:ups} and \hyperref[axiom:prior:invariant]{Prior Invariance} is empty.\footnote{
We show that a \hyperref[axiom:prior:invariant]{Prior Invariant} cost is \nameref{defi:ups} \emph{only if} it is \nameref{axiom:slp} and \hyperref[axiom:CMC:0]{CMC}. Importantly, while it is clear that no \emph{continuous full-domain} \nameref{defi:ups} cost is \hyperref[axiom:prior:invariant]{Prior Invariant} (see \cref{ssec:axiomatic}), \cref{thm:trilemma}(iii) is needed to cover the broad class of rich-domain \nameref{defi:ups} costs $C^H_\text{ups}$ with unbounded $H$ (e.g., \nameref{defi:TI}; \cite{caplin2022rationally,hebert2021neighborhood}).
%The proof shows that a \hyperref[axiom:prior:invariant]{Prior Invariant} cost function is \nameref{defi:ups} \emph{only if} it is \nameref{axiom:slp} and \hyperref[axiom:CMC:0]{CMC}. We emphasize that, while the simple continuity argument noted in \cref{ssec:axiomatic} implies that no \emph{full-domain} \nameref{defi:ups} cost is \hyperref[axiom:prior:invariant]{Prior Invariant}, the more subtle proof of \cref{thm:trilemma}(iii) is needed to cover the broad class of rich-domain \nameref{defi:ups} costs $C^H_\text{ups}$ for which $H$ is unbounded, such as \nameref{defi:TI} and many specifications from  \textcite{caplin2022rationally,hebert2021neighborhood}.
} We interpret this trilemma as delivering three main lessons.

First, \cref{thm:trilemma}(i) shows that \nameref{defi:TI} \emph{uniquely} characterizes the two-way intersection of \nameref{axiom:slp} and \nameref{axiom:CMC}. This provides a foundation for using \nameref{defi:TI} to model reduced-form information costs in many applications (e.g., statistical sampling).

Second, a more subtle implication of \cref{thm:trilemma}(i) is that \hyperref[axiom:CMC:0]{CMC} admits two \emph{essentially distinct} interpretations: one as a \emph{primitive} property of direct costs, the other as an \emph{emergent} property of indirect costs. This dichotomy arises because \hyperref[axiom:CMC:0]{CMC} is typically \emph{not} preserved under optimization. Formally, we show that essentially any \nameref{axiom:CMC} and \hyperref[axiom:DL]{Dilution Linear} direct cost---aside from \nameref{defi:TI} itself---generates an indirect cost that is \emph{not} \nameref{axiom:CMC} (\cref{cor:preserve:CMC} in \cref{ssec:thm5-corollaries}).\footnote{Any such direct cost has a form similar to the \ref{eqn:LLR} cost, but with coefficients $\beta_{\theta,\theta'}(p)$ that may depend on the prior. Therefore, since these coefficients determine the direct cost's kernel, \dred{Theorems} \ref{thm:qk}(ii) and \ref{thm:trilemma}(i) imply that the indirect cost is \nameref{defi:TI} with coefficients $\gamma_{\theta,\theta'}$ if and only if $\beta_{\theta,\theta'}(p) \equiv p(\theta) \, \gamma_{\theta,\theta'}$, i.e., the direct and indirect costs coincide.} Thus, if we interpret \hyperref[axiom:CMC:0]{CMC} as a primitive property, \cref{thm:trilemma}(ii) (due to \cite{pomatto2023cost}) provides a foundation for using  \ref{eqn:LLR} costs to model the \hyperref[axiom:prior:invariant]{Prior Invariant} (e.g., physical or pecuniary) direct cost of information, but also implies that the associated indirect cost must violate \hyperref[axiom:CMC:0]{CMC}. Conversely, the fact that \nameref{defi:TI} satisfies \hyperref[axiom:CMC:0]{CMC} is most naturally viewed as an endogenous outcome of optimization, given that ``most'' of its underlying direct costs violate \hyperref[axiom:CMC:0]{CMC} (as in \cref{eg:Diffusion:0}). We therefore conclude that---despite the similarity of their functional forms---\ref{eqn:LLR} and \nameref{defi:TI} cost functions are economically distinct modeling tools. %More broadly, it highlights the conceptual clarity that comes from distinguishing between properties of direct and indirect cost functions---a theme that we return to in \cref{ssec:SPI} below.

Finally, and perhaps most importantly, \cref{thm:trilemma}(iii) highlights a strong tension between \nameref{axiom:slp} and \hyperref[axiom:prior:invariant]{Prior Invariance}. First, it directly shows that their two-way intersection does not contain any cost functions satisfying either \hyperref[axiom:CMC:0]{CMC} or \nameref{defi:ups}. Second, in conjunction with \cref{thm:UPS}, it implies that every \nameref{axiom:slp} and \hyperref[axiom:prior:invariant]{Prior Invariant} cost is non-\nameref{defi:ll}.

%First, a prominent subclass of \hyperref[axiom:prior:invariant]{Prior Invariant} costs---those that exhibit \hyperref[axiom:CMC:0]{CMC}---does not intersect the class of \nameref{axiom:slp} costs.  Second, the most prominent subclass of \nameref{axiom:slp} costs---the \nameref{defi:ups} costs---does not intersect the class of \hyperref[axiom:prior:invariant]{Prior Invariant} costs. By \cref{thm:UPS}, it follows that every \nameref{axiom:slp} and \hyperref[axiom:prior:invariant]{Prior Invariant} cost is non-\nameref{defi:ll}. 

This helps clarify how \hyperref[axiom:prior:invariant]{Prior Invariant} costs in the literature---most of which satisfy \hyperref[axiom:CMC:0]{CMC} or \hyperref[defi:ll]{Regularity}---should be interpreted. For instance, the \ref{eqn:LLR} cost satisfies both properties, the broader class of ``Renyi divergence'' costs from \textcite{mu2021blackwell} satisfy \hyperref[axiom:CMC:0]{CMC}, and the ``fixed-prior \nameref{defi:ups}'' costs from \textcite{gentzkow_kamenica_costly} and \textcite{denti2022experimental} are \nameref{defi:ll}. \cref{thm:trilemma}(iii) implies that it is natural to interpret these cost functions as ``primitive'' (direct) costs, but \emph{not} as ``reduced-form'' (\nameref{axiom:slp}) costs. 

This lesson also applies to many \hyperref[axiom:prior:invariant]{Prior Invariant} costs that are not \hyperref[axiom:CMC:0]{CMC} or \nameref{defi:ll}. Formally, we show that a \emph{full-domain} \hyperref[axiom:prior:invariant]{Prior Invariant} cost function is \nameref{axiom:slp} \emph{only if} it assigns equal cost to every experiment that reveals a nontrivial partition of the state space (\cref{cor:PI-part} in \cref{ssec:thm5-corollaries}). When $|\Theta| \geq 3$, this condition 
%For instance, it implies that such cost functions cannot be \emph{strictly} \hyperref[axiom:mono]{Monotone}, which 
rules out essentially all other \hyperref[axiom:prior:invariant]{Prior Invariant} costs in the literature, such as the ``channel capacity'' cost of \textcite{woodford-reference2012}.%\footnote{The \nameref{defi:MLR} cost satisfies this condition because $D_\text{MLR}(q \mid p) = 1$ for all $q,p \in \Delta(\Theta)$ such that $\supp(q) \subsetneq \supp(p)$,  which implies that the same cost is assigned to every experiment under which all signals rule out some ex-ante possible state.}

We conclude that the intersection of \nameref{axiom:slp} and \hyperref[axiom:prior:invariant]{Prior Invariance} is ``small.'' Indeed, it is perhaps surprising that the tension between these properties can be reconciled at all. We use the \nameref{defi:MLR} cost to illustrate this possibility because of its convenient functional form. While we conjecture that the \nameref{defi:MLR} cost does not \emph{uniquely} characterize the intersection of \nameref{axiom:slp} and \hyperref[axiom:prior:invariant]{Prior Invariance}: (a) it is currently the only known example in this class, and (b) its key properties are common to other cost functions that might exist in this class.\footnote{For instance, it can be shown that a \hyperref[eqn:PS]{Posterior Separable} cost function is \nameref{axiom:slp} and \hyperref[axiom:prior:invariant]{Prior Invariant} \emph{if and only if} its divergence $D$ satisfies two conditions: (i) the ``average-case triangle inequality'' in \cref{lem:SLP-divergence}, and (ii) there exists a sublinear function $G:\R^{|\Theta|}_{+} \to \overline{\R}_+$ such that $D(q \mid p) \equiv G\big(\frac{q}{p}\big)$. Moreover, if such a cost function has full domain, then the necessary condition noted above (\cref{cor:PI-part} in \cref{ssec:thm5-corollaries}) also implies that, for every prior $p$, $D(\cdot \mid p)$ must be affine on each face of the simplex $\Delta(\Theta)$. Note that the \nameref{defi:MLR} divergence satisfies all of these conditions by construction.
\label{fn:PI-SLP-gen}} 
%For these reasons, we view the \nameref{defi:MLR} cost as a canonical \nameref{axiom:slp} and \hyperref[axiom:prior:invariant]{Prior Invariant} cost function.

\subsection{A Resolution: Sequential Prior Invariance}\label{ssec:SPI} 

%A primary lesson of \cref{thm:trilemma} is that \hyperref[axiom:prior:invariant]{Prior Invariance}, while often an appealing property for ``primitive'' direct costs, is typically an overly restrictive assumption for modeling ``reduced-form'' indirect/\nameref{axiom:slp} costs. 
Our analysis indicates that \hyperref[axiom:prior:invariant]{Prior Invariance} is typically an overly restrictive assumption for modeling ``reduced-form'' indirect/\nameref{axiom:slp} costs. Intuitively, because indirect costs arise from sequential \emph{expected} cost-minimization, it is natural for them to \emph{endogenously} depend on prior beliefs---\emph{even if} their underlying direct costs are \hyperref[axiom:prior:invariant]{Prior Invariant} (as in \cref{eg:Diffusion:0}). This perspective motivates the following novel class of indirect costs:

\begin{definition}
[SPI]\label{defi:spi}
    $C\in\C$ is \dred{Sequentially Prior Invariant} (\nameref{defi:spi}) if $C=\Phi(C')$ for some \hyperref[axiom:prior:invariant]{Prior Invariant} $C'\in \C$.
\end{definition}

We view \nameref{defi:spi} cost functions as the natural modeling tool in most applications where the literature has advocated for using \hyperref[axiom:prior:invariant]{Prior Invariant} costs. Indeed, many real-world settings in which information costs are physical or pecuniary (e.g., clinical trials) feature at least some degree of flexible sequential learning (e.g., the design of multi-stage trials).

The class of \nameref{defi:spi} costs clearly includes all \hyperref[axiom:prior:invariant]{Prior Invariant} and  \nameref{axiom:slp} costs (viz., the \nameref{defi:MLR} cost). It also includes the \ref{eqn:MS} cost, which is \nameref{defi:ups} and \nameref{axiom:CMC} but not \hyperref[axiom:prior:invariant]{Prior Invariant} (\cref{eg:Diffusion:0}). Therefore, by relaxing \hyperref[axiom:prior:invariant]{Prior Invariance} to \nameref{defi:spi}, we obtain at least one resolution to the information cost trilemma. In fact, the \ref{eqn:MS} cost provides the \emph{unique} such resolution. Formally, we call $C \in \C$ ``a \ref{eqn:MS} cost'' if $C =\gamma \cdot C_\text{Wald}$ for some constant $\gamma \geq 0$.

\begin{thm}\label{thm:wald}
For any \nameref{defi:sp} $C \in \C$ with 
%$\dom(C) = \Delta(\Delta^\circ(\Theta)) \cup \Ex^\varnothing$
rich domain,
    \begin{align*}
           \hspace{-0.5em} &\text{$C$ is \nameref{defi:spi} and \nameref{axiom:CMC}}
            \iff \text{$C$ is \nameref{defi:spi}, \nameref{defi:ups}, and \nameref{defi:lq}}
            \iff \text{$|\Theta|=2$ and %$C^\circ\propto C_{\text{Wald}}$
             $C$ is a \ref{eqn:MS} cost.}
    \end{align*}
\end{thm}
\begin{proof}
   See \cref{proof:wald}.
\end{proof}

\cref{thm:wald} offers two characterizations (see \cref{fig:wald-thm}). First, it \emph{uniquely} resolves the information cost trilemma: the \ref{eqn:MS} cost is the \emph{only} \nameref{defi:spi} and \nameref{axiom:CMC} cost function. Second, it \emph{uniquely} resolves the tension between \hyperref[axiom:prior:invariant]{Prior Invariance} and \nameref{defi:ups}: the \ref{eqn:MS} cost is also the \emph{only} (smooth) \nameref{defi:spi} and \nameref{defi:ups}  cost function.\footnote{We emphasize that \cref{thm:wald} imposes \emph{no} smoothness assumptions on the underlying \hyperref[axiom:prior:invariant]{Prior Invariant} \emph{direct} cost.
%We emphasize that (the ``$\implies$'' direction of) this latter characterization assumes only that the \nameref{defi:spi} \emph{indirect} cost is \nameref{defi:lq}. In particular, it imposes \emph{no} smoothness assumptions on the underlying \hyperref[axiom:prior:invariant]{Prior Invariant} \emph{direct} cost.
} Both give the \ref{eqn:MS} cost strong foundations.

More broadly, \cref{thm:wald} highlights a new \emph{modeler's trilemma} (see \cref{fig:modeler-trilemma}): even after relaxing \hyperref[axiom:prior:invariant]{Prior Invariance} to \nameref{defi:spi}, there are inherent tradeoffs among the three key modeling desiderata of \emph{realism} (\nameref{defi:spi}), \emph{tractability} (\hyperref[defi:ll]{Regularity}/\nameref{defi:ups}), and \emph{scalability} (to general state spaces). While the \ref{eqn:MS} cost is both \nameref{defi:spi} and \hyperref[defi:ll]{Regular}/\nameref{defi:ups}, it is only well-defined for binary-state settings, which is restrictive.\footnote{This impossibility result does \emph{not} require the rich domain assumption in \cref{thm:wald}: when $|\Theta|\geq 3$, there do not exist \emph{any} (smooth) \nameref{defi:ups} cost functions $C^H_\text{ups}$ for which $\dom(H) \subseteq \Delta(\Theta)$ has nonempty interior (\cref{lem:ups-lpi-wald-local} in \cref{ssec:ups-lpi-wald-local}).
%%%%
%The rich-domain assumption in \cref{thm:wald} is \emph{not} needed to show that nontrivial \nameref{defi:ups} and \nameref{defi:spi} cost functions exist only when $|\Theta| =2$. Formally, we show that a (smooth) \nameref{defi:ups} cost function $C^H_\text{ups}$ is \nameref{defi:spi} \emph{only if} either $|\Theta| = 2$ or $\dom(H) \subseteq \Delta(\Theta)$ has empty interior (\cref{lem:ups-lpi-wald-local} in \cref{ssec:ups-lpi-wald-local}). This is consistent with the finding of \textcite{morris-strack-sampling} that, when $|\Theta| \geq 3$, there can exist \nameref{defi:spi} costs with the same functional form as \nameref{defi:TI}, but which are only defined on a restricted domain $\Ex^\star \subsetneq \Ex$ with ``empty interior'' (i.e., $\Ex^\star \not\supset\Delta(W)$ for \emph{any} nonempty open set $W \subseteq \Delta(\Theta)$).
%\label{fn:ms-multi-state}
} To scale to larger state spaces, one must sacrifice some of the tractability afforded by \hyperref[defi:ll]{Regularity}/\nameref{defi:ups} %(e.g., by using the \nameref{defi:MLR} cost) 
or some of the realism afforded by \nameref{defi:spi}. %(e.g., by using \hyperref[MI]{Mutual Information} or non-\nameref{defi:spi} versions of \nameref{defi:TI}). 
The optimal way to balance these tradeoffs depends on the application at hand.

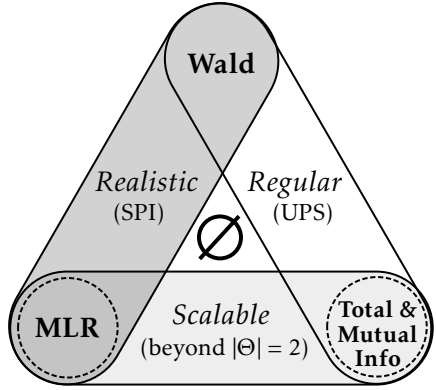
\begin{figure}[t]
\iffalse
\hspace{-3em}\centering
\begin{minipage}{.6\textwidth}
    \centering
    \include{Figure/SPI}
    \vspace{-2em}
    \end{minipage}
    \begin{minipage}{.4\textwidth}
    \centering
    \include{Figure/trilemma_2}
    \vspace{-2.5em}
    \end{minipage}
    \caption{{\small Resolving the information cost trilemma (left). The new modeler's trilemma (right).}}
    \label{fig:modeler-trilemma}
\fi
\vspace{-1em}
\centering
 \tikzset{every picture/.style={line width=0.75pt}} %set default line width to 0.75pt        

\begin{tikzpicture}[x=0.75pt,y=0.75pt,yscale=-1,xscale=1]

%Rounded Rect [id:dp7429499841978848] 
\draw  [fill={rgb, 255:red, 0; green, 0; blue, 0 }  ,fill opacity=0.18 ] (109.17,226.26) .. controls (95.15,218.18) and (90.34,200.26) .. (98.42,186.24) -- (175.59,52.41) .. controls (183.68,38.39) and (201.6,33.58) .. (215.62,41.66) -- (215.62,41.66) .. controls (229.64,49.75) and (234.45,67.67) .. (226.37,81.69) -- (149.19,215.51) .. controls (141.11,229.53) and (123.19,234.35) .. (109.17,226.26) -- cycle ;
%Rounded Rect [id:dp015062193874449958] 
\draw   (186.97,41.65) .. controls (200.27,33.97) and (217.28,38.53) .. (224.96,51.83) -- (303.7,188.21) .. controls (311.38,201.51) and (306.82,218.52) .. (293.52,226.19) -- (293.52,226.19) .. controls (280.22,233.87) and (263.21,229.32) .. (255.53,216.02) -- (176.79,79.64) .. controls (169.11,66.34) and (173.67,49.33) .. (186.97,41.65) -- cycle ;
%Rounded Rect [id:dp9404288890460293] 
\draw  [fill={rgb, 255:red, 0; green, 0; blue, 0 }  ,fill opacity=0.07 ] (95.06,201.78) .. controls (95.06,186.03) and (107.82,173.26) .. (123.57,173.26) -- (279.64,173.26) .. controls (295.38,173.26) and (308.15,186.03) .. (308.15,201.78) -- (308.15,201.78) .. controls (308.15,217.52) and (295.38,230.29) .. (279.64,230.29) -- (123.57,230.29) .. controls (107.82,230.29) and (95.06,217.52) .. (95.06,201.78) -- cycle ;
%Shape: Ellipse [id:dp6114719435244806] 
\draw  [dash pattern={on 2pt off 1pt}][line width=0.75]  (99.02,201.63) .. controls (99.02,187.76) and (110.26,176.52) .. (124.13,176.52) .. controls (137.99,176.52) and (149.23,187.76) .. (149.23,201.63) .. controls (149.23,215.49) and (137.99,226.73) .. (124.13,226.73) .. controls (110.26,226.73) and (99.02,215.49) .. (99.02,201.63) -- cycle ;
%Shape: Ellipse [id:dp07259130034810979] 
\draw  [line width=1.5]  (190.26,156.03) .. controls (190.26,150.18) and (195.01,145.43) .. (200.86,145.43) .. controls (206.71,145.43) and (211.45,150.18) .. (211.45,156.03) .. controls (211.45,161.88) and (206.71,166.62) .. (200.86,166.62) .. controls (195.01,166.62) and (190.26,161.88) .. (190.26,156.03) -- cycle ;
%Straight Lines [id:da5481822014483515] 
\draw [line width=1.5]    (210.45,145.31) -- (190.64,166.87) ;

%Shape: Ellipse [id:dp1668769084735755] 
\draw  [dash pattern={on 2pt off 1pt}][line width=0.75]  (254.91,201.14) .. controls (254.91,187.27) and (266.15,176.03) .. (280.02,176.03) .. controls (293.88,176.03) and (305.12,187.27) .. (305.12,201.14) .. controls (305.12,215) and (293.88,226.24) .. (280.02,226.24) .. controls (266.15,226.24) and (254.91,215) .. (254.91,201.14) -- cycle ;

% Text Node
\draw (183,61.19) node [anchor=north west][inner sep=0.75pt]  [font=\small] [align=left] {\textbf{Wald}};
% Text Node
\draw (106,195) node [anchor=north west][inner sep=0.75pt]  [font=\small] [align=left] {\textbf{MLR}};
% Text Node
\draw (203.17,194.58) node  [font=\small,color={rgb, 255:red, 0; green, 0; blue, 0 }  ,opacity=1 ] [align=left] {\textit{Scalable }};
% Text Node
\draw (203.17,212.58) node  [font=\footnotesize,color={rgb, 255:red, 0; green, 0; blue, 0 }  ,opacity=1 ] [align=left] {(beyond $|\Theta|=2$)};
% Text Node
\draw (238.74,129.22) node  [font=\small,color={rgb, 255:red, 0; green, 0; blue, 0 }  ,opacity=1 ] [align=left] {\textit{Regular}};
% Text Node
%\draw (150.01,125.22) node  [font=\small,color={rgb, 255:red, 0; green, 0; blue, 0 }  ,opacity=1 ,rotate=-299.3] [align=left] {\textit{Realistic}};
\draw (164.01,127.22) node  [font=\small,color={rgb, 255:red, 0; green, 0; blue, 0 }  ,opacity=1 ] [align=left] {\textit{Realistic}};
% Text Node
%\draw (167.98,135.42) node  [font=\small,color={rgb, 255:red, 0; green, 0; blue, 0 }  ,opacity=1 ,rotate=-299.3] [align=left] {(\nameref{defi:spi})};
\draw (159.98,144.42) node  [font=\footnotesize,color={rgb, 255:red, 0; green, 0; blue, 0 }  ,opacity=1 ] [align=left] {(SPI)};
% Text Node
\draw (242.69,144.42) node  [font=\footnotesize,color={rgb, 255:red, 0; green, 0; blue, 0 }  ,opacity=1 ] [align=left] {(UPS)};
\iffalse
% Text Node
\draw (256,186) node [anchor=north west][inner sep=0.75pt]  [font=\tiny] [align=left] {\begin{minipage}[lt]{35pt}\setlength\topsep{0pt}
\begin{center}
\textbf{Total Info,} \\ \vspace{.2em} \textbf{Mutual Info,}\\ \vspace{.15em} \textbf{etc.}
\end{center}
\end{minipage}};
\fi
% Text Node
\draw (280.5,204.5) node  [font=\footnotesize,color={rgb, 255:red, 0; green, 0; blue, 0 }  ,opacity=1 ] [align=center] 
{\begin{minipage}[lt]{35pt}\setlength\topsep{0pt}
\begin{center}
\textbf{Total \&} \\ \vspace{-.19em} \textbf{Mutual} \\ \vspace{-.19em} \textbf{Info}
\end{center}
\end{minipage}};

\end{tikzpicture}
    \vspace{-2.5em}
\caption{The modeler's trilemma implied by \cref{thm:wald}.}
\label{fig:modeler-trilemma}
\end{figure}

We sketch the proof of the most subtle part of \cref{thm:wald}---that the \ref{eqn:MS} cost is the \emph{only}  
 \nameref{defi:spi} and \nameref{defi:ups} cost---as it develops techniques that may be of broader interest. Our key methodological tool is a novel ``local'' characterization of \hyperref[defi:spi]{(Sequential)} \hyperref[axiom:prior:invariant]{Prior Invariance}. For any \nameref{defi:lq} $C \in \C$, we define the \emph{experimental kernel} of $C$ at $p \in \Delta(\Theta)$ as
 \begin{align*}
    \kappa_C(p):= \diag (p) \ k_C(p) \ \diag(p),
\end{align*}
which represents the image of the kernel $k_C(p)$ in the space of experiments after we ``change variables'' from posteriors to likelihood ratios. We show that prior-independence of the experimental kernel---a property we dub \hyperref[defi:lpi]{\emph{Local Prior Invariance}}---fully characterizes the local implications of both \hyperref[axiom:prior:invariant]{Prior Invariance} and \nameref{defi:spi}. This generalizes the fact that, in \cref{eg:Diffusion:0}, the flow cost of sampling the diffusion signal process \eqref{eqn:Brownian-motion} is prior-independent.

\begin{prop}\label{prop:LPI-main}
    For any $C \in \C$ that is \nameref{defi:sp} and \nameref{defi:lq},  
    \[
    \text{$C$ is \hyperref[axiom:prior:invariant]{Prior Invariant} or \nameref{defi:spi}} \ \ \implies \ \ \text{%$\kappa_C$ is constant, i.e., 
    $\kappa_C(p) = \kappa_C(p')$ for all $p,p' \in \Delta^\circ(\Theta)$.}
    \]
    %%%
    Conversely, for any symmetric $\kappa \in \R^{|\Theta|\times|\Theta|}$ with $\kappa\gg_\text{psd} \mathbf{0}$ and $\kappa \mathbf{1} = \mathbf{0}$,\footnote{We show in \cref{ssec:proof-LPI-main} that these assumptions on $\kappa$ are without loss of generality in a suitable sense.
    %These assumptions on $\kappa$ are without loss of generality, i.e., they are necessarily satisfied by the experimental kernel of any \nameref{defi:sp}, \nameref{defi:lq}, and \hyperref[axiom:prior:invariant]{Prior Invariant} cost function (see \cref{ssec:proof-LPI-main}).
    \label{fn:prop3-clarify}} 
    there exists a \nameref{defi:sp}, \nameref{defi:lq}, and \hyperref[axiom:prior:invariant]{Prior Invariant} $C \in \C$ such that
    %%%%
    \[
    \kappa_C(p) = \kappa_{\Phi(C)} (p) =  \kappa \quad \text{ for all $p \in \Delta^\circ(\Theta)$.}
    \]
    %%%
    \end{prop}
    \begin{proof}
        See \cref{ssec:proof-LPI-main}.
    \end{proof}

Basic calculus reveals that \hyperref[defi:lpi]{Local Prior Invariance}: (i) implies that the kernel cannot be integrable when $|\Theta|>2$, and (ii) uniquely pins down the \ref{eqn:MS} kernel among integrable kernels when $|\Theta|=2$. Thus, the \ref{eqn:MS} cost is the unique (smooth) \hyperref[defi:lpi]{Locally Prior Invariant} and \nameref{defi:ups} cost function. \cref{prop:LPI-main} then implies that it is uniquely \nameref{defi:spi} and \nameref{defi:ups}. 

Looking ahead, we expect that \hyperref[defi:lpi]{Local Prior Invariance} and \cref{prop:LPI-main} will be key tools for analyzing the full class of \nameref{defi:spi} costs, which is an important task for future work.\footnote{In \cref{proof:wald}, we establish two other results that may be useful for this task. First, we derive a ``non-smooth'' version of \cref{prop:LPI-main} that also applies to non-\nameref{defi:lq} cost functions (\cref{lem:pi-w-implies-lpi-w}). Second, we show that an indirect cost is \nameref{defi:spi} \emph{if and only if} it is generated by its \emph{\hyperref[axiom:prior:invariant]{Prior Invariant} upper envelope}, i.e., the \emph{smallest} \hyperref[axiom:prior:invariant]{Prior Invariant} cost that lies above it (\cref{lem:PIE}). For the \ref{eqn:MS} cost, this upper envelope is $\overline{C}_\text{Wald}(h_B(\sigma,p)) \equiv \max\{ D_\text{KL} (\sigma_1 \mid \sigma_0), D_\text{KL} (\sigma_0 \mid \sigma_1)\}$. 
%and does, in fact, generate the \ref{eqn:MS} cost (\cref{lem:wald-is-spi}).
}

\section{Extensions and Discussion}\label{section:discussion}

\subsection{Beyond the Belief-Based Framework}\label{ssec:beyond:belief}

For convenience, our baseline framework uses the \emph{belief-based approach} in which cost functions are defined directly on random posteriors. This approach involves two implicit assumptions: (i) all experiments that generate the same random posterior are assigned the same cost, and (ii) distinct experiments generate conditionally independent signals. In this section, we critically evaluate these assumptions and explain how to relax them.

\paragraph{Relaxing Assumption 1: Experiment-Based Framework.} The first assumption is irrelevant if the DM only cares about the random posterior produced by his information acquisition (e.g., if he faces a standard single-agent decision problem) because the optimization process implicitly selects the cheapest experiment to induce each random posterior. However, it can be consequential if the DM's prior belief has partial support \emph{and} he cares about the information acquired about zero-probability states. For example, when the prior $p = \delta_\theta$ is concentrated on a single state $\theta$, \emph{all} experiments induce the trivial random posterior $\delta_{p} \in \Ex^{\varnothing}$ and thus have \emph{zero cost}. This feature of the belief-based approach creates subtleties in applications to \emph{costly monitoring}, where the state represents another agent's action, the prior represents that agent's mixed strategy, and the DM needs to monitor for off-path deviations \parencite{ravid-aer2020,denti2022experimental}.
%\footnote{More broadly, in the spirit of classical statistics, it may be conceptually desirable to decouple the ``objective information'' generated by an experiment from the ``subjective uncertainty'' embodied by the DM's prior. %(e.g., \textcite{blackwell-experiment51} does not even presuppose the existence of prior beliefs).} 

In \cref{app:beyond:belief}, we address this limitation by developing a richer \emph{experiment-based framework} in which: (i) cost functions $\Ec : \Se \times \Delta(\Theta) \to \overline{\R}_+$ are defined directly on experiments and prior beliefs, and (ii) the MPS constraint in \cref{defi:2slm} is replaced by a Blackwell dominance constraint, which is more stringent (only) at partial support priors. This framework lets us distinguish between, and assign different costs to, experiments that are Blackwell non-equivalent but nevertheless induce the same random posterior. For instance, it accommodates \dred{\emph{Fully} Prior Invariant} cost functions, for which $\Ec(\sigma, p) = \Ec(\sigma, p')$ for every $\sigma \in \Se$ and \emph{all} priors $p ,p' \in \Delta(\Theta)$, even those with different supports (cf. \cref{axiom:prior:invariant}).
%\footnote{Per \textcite{blackwell-experiment51}, we say that $\sigma' \in \Se$ \emph{Blackwell dominates} $\sigma \in \Se$ if $h_B(\sigma',p) \geq_\text{mps} h_B(\sigma,p)$ for \emph{every} $p \in \Delta(\Theta)$, and that $\sigma, \sigma' \in \Se$ are \emph{Blackwell equivalent} if they Blackwell dominate each other. While the Blackwell order on $\Se$ and the MPS order on $\Ex$ are equivalent at full-support priors $p \in \Delta^\circ(\Theta)$, the former is more restrictive at partial-support priors.}

We develop a scheme for mapping between the belief- and experiment-based frameworks, which reveals that they are \emph{equivalent}, and hence our results directly extend, whenever the DM's (initial) prior belief has full support (\cref{prop:commute}). Moreover, although the experiment-based framework allows for strictly richer behavior of the sequential learning map at partial-support priors, \cref{prop:1} directly extends: experiment-based indirect costs are characterized by the natural experiment-based analog of \nameref{axiom:slp} (\cref{thm1-Se}).

We offer two examples of such experiment-based \nameref{axiom:slp} cost functions. First, for any collection of coefficients $\gamma_{\theta,\theta'}\geq 0$, we define \dred{\emph{experiment-based Total Information}} as
\begin{equation}
        \Ec_\text{TI}(\sigma, p) := \sum_{\theta \in \Theta} p (\theta)  \sum_{\theta' \in \Theta} \gamma_{\theta,\theta'} D_\text{KL} \left( \sigma_\theta  \mid \sigma_{\theta'} \right) \quad \ \ \text{for all $\sigma \in \Se_b$, $p \in \Delta(\Theta)$.}\footnotemark\label{E-TI} \tag{$\Se$-TI}
\end{equation}
\footnotetext{Here $\Se_b \subsetneq \Se$ is the subclass of \emph{bounded} experiments, where $\sigma \in \Se_b$ if and only if $h_B(\sigma,p) \in \Delta(\Delta^\circ(\Theta))$ for all $p \in \Delta^\circ(\Theta)$. }
Second, we define the \dred{\emph{experiment-based MLR cost}} as  
\begin{equation}
\Ec_\text{MLR}(\sigma,p) := 1 - \int_S \, \min_{\theta \in \Theta} \left\{\frac{\d \sigma_\theta}{\d \overline{\sigma}}(s) \right\} \dd \overline{\sigma}(s)
%:= 1-\bigwedge_{\theta  \in \Theta}  \sigma_\theta (S ) 
\quad \ \ \text{for all $\sigma \in \Se$, $p \in \Delta(\Theta)$}. 
\label{E-MLR} \tag{$\Se$-MLR}
\end{equation}
These expressions mirror those for the belief-based versions of these cost functions in \cref{section:applications}, except that here the inner summation in \eqref{E-TI} and the minimum in \eqref{E-MLR} quantify over \emph{all} states, rather than just those in $\supp(p)$.\footnote{To extend the definition of (belief-based) \nameref{defi:TI} to a partial-support prior $p$, we apply \cref{defi:TI} to the ``state space'' $\supp(p) \subsetneq \Theta$, which yields a \nameref{defi:ups} cost with domain $\Delta(\Delta^\circ(\supp(p))) \cup \Ex^\varnothing$ and coefficients $(\gamma_{\theta,\theta'})_{\theta,\theta' \in \supp(p)}$.\label{fn:TI-partial-support}} This difference ensures that it is costly to learn about all states, including those that have zero prior probability. It also implies that the experiment-based \hyperref[E-MLR]{MLR} cost is \dred{\emph{Fully} Prior Invariant} (as defined above).

Our analysis provides a foundation for using experiment-based \nameref{axiom:slp} cost functions in applications. For instance, \textcite{georgiadis2020optimal,wong2023dynamic} use experiment-based \hyperref[E-TI]{Total Information} to model costly monitoring in principal-agent settings.\footnote{
Specifically, \textcite{georgiadis2020optimal} derive a continuous-state analog of \eqref{E-TI} (cf. the \ref{eqn:FI} cost in \cref{eg:FI}).
%Specifically, \textcite{georgiadis2020optimal} derive a continuous-state analog of \eqref{E-TI}, which can be viewed as the experiment-based version of the \nameref{eg:FI} cost from \textcite{hebert2021neighborhood} (recall \cref{eg:FI}).
}

%Our analysis provides a foundation for using experiment-based \nameref{axiom:slp} costs in applications to costly monitoring. For instance, \textcite{georgiadis2020optimal,wong2023dynamic} use experiment-based \hyperref[E-TI]{Total Information} to model costly monitoring in principal-agent settings, where the principal's prior over the agent's actions is degenerate in equilibrium.\footnote{Specifically, \textcite{georgiadis2020optimal} derive a continuous-state analog of \eqref{E-TI}, which can be viewed as the experiment-based version of the \nameref{eg:FI} cost from \textcite{hebert2021neighborhood} (recall \cref{eg:FI}).}

\paragraph{Relaxing Assumption 2: Correlated Signals.} The second assumption is more substantive. In reality, the DM may learn from specific ``information sources'' (e.g., news outlets) that generate signals with ``latent'' state-contingent correlation (e.g., due to common sampling error). Such sources convey information about not only the state, but also the other sources' signals. For instance, two sources may be ``complements'' if their signals are informative only when combined, or ``substitutes'' if their signals are redundant. Our baseline model abstracts away from these possibilities (cf. \cite{brooks2024comparisons}).

We can address this limitation by expanding the state space. Formally, given the set $\Theta$ of payoff-relevant states and any set $Z$ of \emph{ancillary states}, we can define an \emph{expanded state space} as $\Omega := \Theta \times Z$. We can then define beliefs, experiments, random posteriors, and cost functions on $\Omega$ in the natural way. While the signals generated by distinct experiments on $\Omega$ must still be independent conditional on $(\theta, z) \in \Omega$, they may now be correlated conditional on $\theta \in \Theta$ alone. Therefore, since we are free to specify the set $Z$ and the joint prior on $\Omega = \Theta \times Z$, this scheme allows us to model arbitrary forms of latent correlation.\footnote{For instance, this scheme can be used to model:  (i) information sources with correlated ``biases'' \parencite{liang2022dynamically,mu-liang-qje2020}, and (ii) ``all remaining randomness'' conditional on $\theta \in \Theta$, including the noise in the signals generated by all available information sources \parencite{green1978two,green2022two,brooks2025representing,brooks2024comparisons}. See also \textcite{hebert2023information,denti2023robust,gentzkow2017bayesian}.
} 

Our belief- and experiment-based analyses both extend verbatim to cost functions defined on $\Omega$ (at least if $\Omega$ is a finite set). However, this extension comes with a caveat: to avoid trivialities, it is typically necessary to consider cost functions that price not only the ``first-order'' information that each source conveys about $\theta \in \Theta$, but also the ``higher-order'' information that it conveys about $z \in Z$ (and hence the other sources). For instance, if some information sources are complements, then assuming that cost functions only price first-order information (as in our baseline model) can force all \nameref{axiom:slp} cost functions to be trivial.\footnote{We illustrate this point with an example (suggested by Ian Jewitt), which can easily be generalized. Let $\Theta \subseteq Z = \R$. Suppose that $\theta$ and $z$ are independently distributed, where $z \sim N(0,v)$ and $v>0$ is very large. Consider two experiments on $\Omega = \Theta \times Z$, indexed by $i\in \{1,2\}$, that generate signals $s_1 = \theta + z$ and $s_2 = z$. These experiments are nearly perfect complements: each alone reveals (nearly) nothing about $\theta$, but together they fully reveal $\theta$. Therefore, any (continuous) cost function defined on $\Omega$ that only prices first-order information must assign (nearly) zero cost to each experiment $i \in \{1,2\}$. But since acquiring experiments $i\in \{1,2\}$ in sequence fully reveals $\theta$, if such a cost function is also \nameref{axiom:slp}, then it must assign (nearly) zero cost to \emph{all} experiments on $\Omega$. Note that this triviality can be avoided by also pricing higher-order information, viz., assigning a high cost to experiment $i=2$ based on its high informativeness about $z$.} This raises two subtle questions for future work. First, what are reasonable cost functions for pricing such higher-order information? Second, what do \nameref{axiom:slp} cost functions on $\Omega$ look like when ``projected'' back onto the space $\Theta$ of payoff-relevant states?

\subsection{Beyond Flexible Sequential Learning} \label{ssec:beyond:flexibility}

Our baseline model studies the \emph{full flexibility} benchmark in which the DM optimizes over all sequential learning strategies. Formally, it assumes that any restrictions on the DM’s strategy space can be modeled via domain restrictions on the direct cost function, which is history-independent. While this is a useful benchmark, real-world DMs may face richer ``non-stationary'' frictions that cannot be modeled in this way. In this section, we introduce a generalized framework that allows for \emph{arbitrary} optimization procedures.
%In this section, we introduce a generalized framework that permits \emph{arbitrary} optimization procedures, bringing our model closer to reality and helping to isolate the key forces driving our results.

%Our baseline framework endows the DM with \emph{full flexibility} to optimize over all sequential learning strategies. While this is a useful benchmark, real-world DMs often face additional constraints. In this section, we propose a generalized framework that can flexibly incorporate such constraints, which both brings our model closer to reality and helps to isolate the key forces driving our results. 

Central to the framework are generalized notions of indirect and \nameref{axiom:slp} costs:

\begin{definition}[GLM]\label{defi:gen:IC}
	A \dred{generalized learning map} (\nameref{defi:gen:IC}) is any isotone map $\widehat{\Phi}: \C \to \C$.\footnote{That is, we assume only that $C \succeq C'$ implies $\widehat{\Phi}(C) \succeq \widehat{\Phi}(C')$. This holds under any reasonable optimization procedure.} For any \nameref{defi:gen:IC} $\widehat{\Phi}$ and $C \in \C$, (i) $\widehat{\Phi}(C)$ is the \dred{$\widehat{\Phi}$-indirect cost} of $C$ and (ii) $C$ is \dred{$\widehat{\Phi}$-proof} if $\widehat{\Phi}(C)=C$.

 \end{definition}
    We interpret each \nameref{defi:gen:IC} as modeling ``some optimization procedure'' in which the DM may have less---or more---flexibility than in our baseline model. 
    %\footnote{Isotonicity is a mild consistency requirement on the strategy space and the way costs are aggregated across rounds.}  
    %
    This abstract approach lets us study a wide range of optimization procedures without explicitly modeling them.

    In \cref{app:beyond:flexible}, we identify mild sufficient conditions on the \nameref{defi:gen:IC} under which our main results generalize (see  \cref{tab:GSLM:both} for a summary). Each condition in \cref{tab:GSLM:both} holds for a broad class of optimization procedures. Informally, ADL holds whenever the procedure permits all one-shot strategies, AIE and GS hold whenever it permits a ``sufficiently rich'' set of sequential strategies, DUI holds whenever it is more constrained than the procedure in our baseline model, and EO holds whenever there is no benefit to running it multiple times.\footnote{For example: (i) the sequential learning map $\Phi$ satisfies all of these properties; (ii) the identity map $\text{Id}(C) \equiv C$ (which models ``no optimization'') satisfies ADL, DUI, and EO, but violates AIE and GS; (iii) the two-step learning map $\Psi$ satisfies ADL and DUI, but violates AIE, EO, and GS; and (iv) the incremental learning map $\Phi_\text{IE}$ satisfies AIE, DUI, and GS, but violates ADL because it disallows nontrivial one-shot learning (we do not know about EO). In \cref{app:beyond:flexible}, we also consider weaker versions of several properties in \cref{tab:GSLM:both}, e.g., a relaxation of ADL that accommodates $\Phi_\text{IE}$.
    } Since our generalized results rely only on these ``reduced-form'' properties of the \nameref{defi:gen:IC}, they are robust to many details of the underlying optimization procedure itself. Moreover, in applications, there is no need to re-derive our results for each procedure of interest: one can simply check if the associated \nameref{defi:gen:IC} satisfies the requisite properties.

\begin{table}[t]
    \hspace{1em}\begin{minipage}{.45\textwidth}
    \vspace{.5pt}
    \centering
    {\footnotesize \renewcommand{\arraystretch}{1.45}
    \begin{tabular}{c|c}
    \hline\hline 
        \emph{Properties of GLMs} & \emph{Definitions} \\
        \hline 
         Allows Direct Learning (ADL)  &  $\widehat{\Phi}(C) \, \preceq \, C$ \\
         Allows Incremental Evidence (AIE)  &  $\widehat{\Phi}(C) \, \preceq \, \Phi_\text{IE}(C) $ \\
         Disallows UPS Improvement (DUI) &   $\widehat{\Phi}(C^H_\text{ups}) \,  \succeq \, C^H_\text{ups}$ \\
         Exhausts Optimization (EO)  & $\widehat{\Phi} (C)\, =\, \widehat{\Phi} ( \widehat{\Phi}(C))$ \\
         Generates Subadditivity (GS) & $\widehat{\Phi}(C)$ is Subadditive 
         \\
         \hline\hline
    \end{tabular}
    }
    \end{minipage}%
    \begin{minipage}{.65\textwidth}
    \centering
    {\footnotesize \renewcommand{\arraystretch}{1.25}
    \begin{tabular}{c|c}
    \hline\hline
        \emph{Generalized Results} & \emph{Hold under} \\
        \hline
        Theorem 1(i) & EO \\
         Theorems 2($\Rightarrow$) \& 5($\Rightarrow$) & GS \\
         Theorem 3(i) & AIE \\
         Theorem 3(ii) & DUI \\ 
         Theorems 4($\Rightarrow$) \& 6(ii)($\Leftarrow$) & AIE \& DUI\\
         Theorems 4($\Leftarrow$) \& 6(ii)($\Rightarrow$) & ADL \& DUI\\
         \hline\hline
    \end{tabular}
    }
    \end{minipage}
\caption{\centering Properties of GLMs (left) and extensions of main results (right).\\ 
{\footnotesize Theorem 1(i) denotes the first equivalence in Theorem 1 (``$\widehat{\Phi}$-indirect cost$\iff$$\widehat{\Phi}$-proof'').}
\\
{\footnotesize Theorem 6(ii) denotes the second equivalence in Theorem 6 (``SPI \& UPS$\iff$Wald'').}
\\
{\footnotesize Theorems X($\Rightarrow$) and X($\Leftarrow$) denote, resp., the ``$\Rightarrow$'' and ``$\Leftarrow$'' directions of a generic ``Theorem X.''}
}
\label{tab:GSLM:both}
\vspace{-1em}
\end{table}

In \textcite{bz2025-GLM}, we apply the \nameref{defi:gen:IC} framework by relaxing two key assumptions of our baseline model. First, we study a more constrained setting \emph{without free disposal} and show that all of our main results extend (the only substantive difference being that indirect costs may be non-\hyperref[axiom:mono]{Monotone}). Second, we enrich our model to include \emph{history-dependent direct costs}, which may increase or decrease over time as the DM develops ``fatigue'' or ``expertise'' (cf. \cite{dillenberger2023subjective}), and show that our main results extend under mild assumptions on the form of history-dependence.
%We show that the \nameref{defi:gen:IC} framework is flexible enough encompass a broad range of such settings, and that the resulting \nameref{defi:gen:IC}: (i) always satisfies ADL, (ii) satisfies DUI under a permissive notion of ``fatigue,'' (iii) satisfies GS under a permissive notion of ``expertise,'' and (iv) satisfies AIE under a mild condition that is consistent with forms of both ``fatigue'' and ``expertise.'' Consequently, many of our main results are robust to rich forms of history-dependence. 

\subsection{Beyond Regularity} \label{ssec:beyond:regular}

\cref{thm:flie} characterizes the sequential learning map $\Phi$ for (i) the \emph{domain} of \nameref{defi:lq} direct costs and (ii) the \emph{co-domain} of \nameref{defi:ll}/\nameref{defi:ups} indirect costs. The domain restriction is mild and made for technical convenience (see \cref{proof:flie-nonsmooth}). Meanwhile, the co-domain restriction, which we have motivated via the tractability and ubiquity of \nameref{defi:ups} costs in applications (\cref{thm:UPS}), can be economically restrictive (\cref{thm:trilemma,thm:wald}). 

Therefore, perhaps the main question left open by our analysis is how to characterize $\Phi$ for the \emph{full co-domain} of indirect/\nameref{axiom:slp} costs. In particular, further progress on this question is needed to tackle the narrower but equally important task of characterizing the \emph{full class of \nameref{defi:spi} indirect costs}. While we have argued that \nameref{defi:spi} cost functions are natural in many economic applications, the \ref{eqn:MS} and \nameref{defi:MLR} costs are currently the only known examples. 

There are two obstacles to further progress beyond the \nameref{defi:ll}/\nameref{defi:ups} case.  First, although \cref{thm:qk}(ii) shows that (lower) kernels are always invariant under $\Phi$, when the direct cost $C$ has a kernel $k_C$ that is \emph{not integrable}, it is unclear how $k_C$ can be ``integrated'' to fully determine  $\Phi(C)$ or even $\Phi_\text{IE}(C)$ (cf. \cref{lem:phi-ie}). Second, when the direct cost does \emph{not \hyperref[axiom:flie]{FLIE}}, it is necessary to look beyond incremental learning strategies and also consider, e.g., variants of the Poisson strategy from \cref{eg:Poisson:0}. Further progress therefore requires new techniques, the development of which is an exciting task for future work.

\newpage

%\setstretch{}
\appendix

% \titlespacing\section{0pt}{6pt plus 2pt minus 2pt}{0pt plus 2pt minus 2pt}
% \titlespacing\subsection{0pt}{4pt plus 2pt minus 2pt}{0pt plus 2pt minus 2pt}
% \titlespacing\subsubsection{0pt}{2pt plus 2pt minus 2pt}{0pt plus 2pt minus 2pt}
% \setlength\abovedisplayskip{3pt}
% \setlength\belowdisplayskip{2pt}
% \titlespacing*{\paragraph}{0pt}{1.25ex plus 1ex minus .2ex}{0.5em}

\setlength\abovedisplayskip{4pt}
\setlength\belowdisplayskip{4pt}
\section{Appendix}\label{app:main}

\paragraph{Notation.} In this Appendix and \dred{Online Appendices} \ref{app:omitted-results}--\ref{app:additional-proofs}, we make frequent use of the following notation. Let $\mathcal{T} := \{y \in \R^{|\Theta|} \mid y^\top \mathbf{1} = 0\}$ denote the tangent space to the simplex. 

For any matrices $A, B \in \R^{|\Theta| \times |\Theta|}$, we let $A \geq_\text{psd} B$ denote that $y^\top A y \geq y^\top By$ for all $y \in \mathcal{T}$, and (consistent with \cref{fn:psd-strict}) let $A \gg_\text{psd} B$ denote that $y^\top A y > y^\top By$ for all $y \in \mathcal{T}\backslash\{\mathbf{0}\}$. For any matrix $A \in \mathbb{R}^{|\Theta|\times|\Theta|}$, note that the following properties are equivalent: (i) $A \gg_\text{psd} \mathbf{0}$, (ii) $\min\{y^\top A y \mid y \in \mathcal{T} \text{ s.t. } \|y\|=1\} >0$, and (iii) there exists $m>0$ such that $A\gg_\text{psd} m I$.\footnote{This equivalence holds because the map $y \mapsto y^\top A y$ is continuous and the set $\{y \in \mathcal{T} \mid \|y\|=1\}$ is compact.}

For any matrix $A \in \mathbb{R}^{|\Theta|\times|\Theta|}$, we denote $\|A\|:= \max\{|y^\top A y| \mid y \in \mathcal{T} \text{ s.t. } \|y \|=1\}$. The induced map $\|\cdot \| : \mathbb{R}^{|\Theta|\times|\Theta|} \to \R_+$ then defines a \emph{semi-norm} on this space of matrices.

For any $p \in \Delta(\Theta)$, we denote by $I(p):= (I-\mathbf{1}p^\top) (I-p\mathbf{1}^\top)$ the normalized identity matrix.

\subsection{Proof of \cref{prop:1} }\label{ssec:app:prop:1}
\begin{proof}
    We consider the two equivalences in turn. 
    
 \noindent   \textbf{Equivalence 1: \dred{Indirect Cost}$\iff$\nameref{axiom:slp}.} The ``$\impliedby$'' direction is trivial. For the ``$\implies$'' direction, let $C \in \C$ and the corresponding $\Phi(C)\in \C^*$ be given. We claim that $\Phi(C)$ is \nameref{axiom:slp}. Since $\Psi(\Phi(C)) \preceq \Phi(C)$ by definition, it suffices to show that $\Psi(\Phi(C)) \succeq \Phi(C)$. To this end, fix an arbitrary $\pi \in \Ex$. Let $\epsilon >0$ and $\Pi \in \Delta^\dag(\Ex)$ satisfying $\E_\Pi[\pi_2] \geq_\text{mps} \pi$ be given. Since $\supp(\Pi)\backslash\Ex^\varnothing$ is finite, by the definition of $\Phi$ there exists an $n \in \mathbb{N}$ such that
    \begin{align*}
        \Psi^n(C)(\pi') \leq \Phi(C)(\pi' ) +\epsilon \qquad  \forall \, \pi' \in \{\pi_1\}\cup \left[ \supp(\Pi)\backslash\Ex^\varnothing\right].\footnotemark
    \end{align*}
    \footnotetext{For any $\pi' \notin \dom(\Phi(C))$, we have $\Phi(C)(\pi') = \Psi^k(C)(\pi') = +\infty$ for all $k \in \mathbb{N}$, so the inequality $\Psi^k(C)(\pi') \leq \Phi(C)(\pi') + \epsilon $ automatically holds for all $k \in \mathbb{N}$ and $\epsilon>0$.}
    Moreover, the same inequality also trivially holds for all $\pi' \in \supp(\Pi) \cap \Ex^\varnothing$ because $\Psi^n(C), \Phi(C) \in \C$ implies that $\Psi^n(C)[\Ex^\varnothing] = \Phi(C)[\Ex^\varnothing] = \{0\}$. It follows that
    \begin{align*}
    2 \epsilon + \Phi(C)(\pi_1)+\E_{\Pi}[\Phi(C)(\pi_2)]
     \geq& \ \Psi^n(C)(\pi_1)+\E_{\Pi}[\Psi^n(C)(\pi_2)] \\  \geq& \ \Psi^{n+1}(C)(\pi) \\ \geq& \ \Phi(C)(\pi),
    \end{align*}
    where the first inequality is by the above choice of $n \in \mathbb{N}$, the second inequality is by the definitions of $\Pi$ and $\Psi$, and the final inequality is by the definition of $\Phi$. Since the given $\epsilon$ and $\Pi$ were arbitrary, we may then send $\epsilon \to 0$ and infimize over the $\Pi \in \Delta^\dag(\Ex)$ satisfying $\E_\Pi[\pi_2] \geq_\text{mps} \pi$, which delivers $\Psi(\Phi(C))(\pi) \geq \Phi(C)(\pi)$. Since the fixed $\pi \in \Ex$ was arbitrary, we conclude that $\Psi(\Phi(C)) \succeq \Phi(C)$ and thus that $\Phi(C)$ is \nameref{axiom:slp}, as claimed.

 \noindent   \textbf{Equivalence 2: \nameref{axiom:slp}$\iff$\hyperref[axiom:mono]{Monotone} and \hyperref[axiom:POSL]{Subadditive}.} For the ``$\implies$'' direction, let $C \in \C$ be \nameref{axiom:slp}. First, note that for any $\pi, \pi' \in \Ex$ satisfying $\pi' \geq_\text{mps} \pi$, the degenerate strategy $\Pi := \delta_{\pi'} \in \Delta^\dag(\Ex)$ (for which $\pi_1 = \delta_{p_\pi}$ and $\pi_2 = \pi'$ $\Pi$-a.s.) satisfies $\E_\Pi[\pi_2] = \pi' \geq_\text{mps} \pi$. Thus, \cref{defi:2slm} and \nameref{axiom:slp} imply that $C(\pi') \geq \Psi(C)(\pi) = C(\pi)$, i.e., $C$ is \hyperref[axiom:mono]{Monotone}. Next, note that each $\Pi \in \Delta^\dag(\Ex)$ trivially satisfies $\E_\Pi[\pi_2] \geq_\text{mps} \E_\Pi[\pi_2]$. Thus, \cref{defi:2slm} and \nameref{axiom:slp} imply that $C(\pi_1) + \E_\Pi[C(\pi_2)] \geq \Psi(C)( \E_\Pi[\pi_2]) = C(\E_\Pi[\pi_2])$, i.e., $C$ is \hyperref[axiom:POSL]{Subadditive}.

    For the ``$\impliedby$'' direction, let $C \in \C$ be \hyperref[axiom:mono]{Monotone} and \hyperref[axiom:POSL]{Subadditive}. Let $\pi \in \Ex$ be given. For any $\Pi \in \Delta^\dag(\Ex)$ satisfying $\E_\Pi[\pi_2] \geq_\text{mps} \pi$, we have $C(\pi_1) + \E_\Pi[C(\pi_2)] \geq C(\E_\Pi[\pi_2]) \geq C(\pi)$, where the first inequality is because $C$ is \hyperref[axiom:POSL]{Subadditive} and the second inequality is because $C$ is \hyperref[axiom:mono]{Monotone}. We conclude that $\Psi(C)(\pi) \geq C(\pi)$ and, since $\pi \in \Ex$ was arbitrary, that $\Psi(C) \succeq C$. Since $\Psi(C) \preceq C$ by definition, we obtain $\Psi(C) = C$, i.e., $C$ is \nameref{axiom:slp}.
    %%%
\end{proof}

\subsection{Proof of \cref{thm:UPS}}\label{ssec:app:thm:UPS}

The necessity (``$\impliedby$'') direction of \Cref{thm:UPS} is straightforward. We prove the sufficiency (``$\implies$'') direction of \Cref{thm:UPS} via a series of five lemmas. The first two lemmas establish two basic implications of \hyperref[axiom:POSL]{Subadditivity}, which every \nameref{axiom:slp} cost function satisfies.

\begin{lem}\label{lem:C^*:convex}
    If $C \in \C$ is \hyperref[axiom:POSL]{Subadditive}, then it is \dred{Convex}, i.e., 
    \[
    C(\alpha \pi + (1-\alpha) \pi')\leq \alpha C(\pi) + (1-\alpha) C(\pi')
    \]
    for all $\pi,\pi' \in \Ex$ such that $p_\pi = p_{\pi'}$ and every $\alpha \in [0,1]$.
\end{lem}
\begin{proof}
    Let any $\pi,\pi' \in \Ex$ with $p_\pi = p_{\pi'}$, $\alpha \in [0,1]$, and \hyperref[axiom:POSL]{Subadditive} $C \in \C$ be given. Define $\Pi \in \Delta^\dag(\Ex)$ as $\Pi(\{\pi\}) := \alpha$ and $\Pi(\{\pi'\}) := 1-\alpha$, so that $\pi_1 = \delta_{p_\pi} \in \Ex^\varnothing$ and $\mathbb{E}_\Pi[\pi_2] = \alpha \pi + (1-\alpha)\pi'$. Since $C(\pi_1) = 0$ and $C$ is \hyperref[axiom:POSL]{Subadditive}, $C(\alpha \pi + (1-\alpha) \pi') \leq  \alpha C(\pi) + (1-\alpha) C(\pi')$.
\end{proof}

\begin{lem}\label{lem:convex-dl}
    If $C \in \C$ is \hyperref[axiom:POSL]{Subadditive}, then it is %\nameref{axiom:RA}
    \hyperref[axiom:DL]{Dilution Linear}. 
\end{lem}
%\begin{proof}
%See \cref{app:thm2:extra}.
%\end{proof}
\begin{proof}%[Proof of \cref{lem:convex-dl}]
   Let $\pi \in \dom(C) \backslash\Ex^\varnothing$ and $\alpha \in [0,1]$ be given.\footnote{If $\pi \in \Ex^\varnothing$, then we trivially have $C(\alpha \cdot \pi) = \alpha C(\pi) = 0$ for all $C \in \C$ and $\alpha \in [0,1]$.} Since $C \in \C$ is %\nameref{axiom:RA} 
    \hyperref[lem:C^*:convex]{Convex} (\cref{lem:C^*:convex}), $C(\alpha \cdot \pi ) \leq \alpha C(\pi) + (1-\alpha) C(\delta_{p_\pi}) =\alpha C(\pi)$. Thus, it suffices to show that $\alpha C(\pi) \leq C(\alpha \cdot \pi)$. Define $\Pi \in \Delta^\dag(\Ex)$ as $\Pi (\{\pi\}) := 1- \alpha$ and $\Pi(\{\delta_q \mid q \in B\}) := \alpha \pi(B)$ for all Borel $B \subseteq \Delta(\Theta)$. By construction, $\mathbb{E}_\Pi [ \pi_2 ] = \pi$ and the induced $\pi_1 = \alpha \cdot \pi $. Thus, since $C \in \C$ is \hyperref[axiom:POSL]{Subadditive}, 
    \[
    C(\pi) \, \leq \,  C(\alpha \cdot \pi) + (1-\alpha) C(\pi) + \alpha \int_{\Delta(\Theta)} C(\delta_q) \d \pi(q) \, = \, C(\alpha \cdot \pi) + (1-\alpha) C(\pi).
    \]
    Since $\pi \in \dom(C)$, it follows that $\alpha C(\pi) \leq C(\alpha \cdot \pi)$. Therefore, $C$ is \hyperref[axiom:DL]{Dilution Linear}.
\end{proof}

Notably, for any convex $W\subseteq \Delta(\Theta)$ and $C \in \C$ with $\dom(C) = \Delta(W) \cup \Ex^\varnothing$, \cref{lem:convex-dl} implies that $C$ is \hyperref[axiom:POSL]{Subadditive} and satisfies the Gateaux differentiability condition \eqref{gateux} \emph{if and only if} $C$ is \hyperref[axiom:POSL]{Subadditive} and \hyperref[eqn:PS]{Posterior Separable}.\footnote{This implication holds because, for any convex $W\subseteq \Delta(\Theta)$ and $C \in \C$ with $\dom(C) = \Delta(W) \cup \Ex^\varnothing$, it follows directly from the definitions that $C$ is \hyperref[axiom:DL]{Dilution Linear} and satisfies \eqref{gateux} \emph{if and only if} $C$ is \hyperref[eqn:PS]{Posterior Separable}.} Thus, we can henceforth focus on \hyperref[axiom:POSL]{Subadditive} and \hyperref[eqn:PS]{Posterior Separable} cost functions. The third lemma shows that such cost functions are characterized by an ``average-case triangle inequality'' for the divergence.

\begin{lem}\label{lem:SLP-divergence}
    For any open convex $W \subseteq \Delta(\Theta)$ and \hyperref[eqn:PS]{Posterior Separable} $C \in \C$ with $\dom(C) = \Delta(W) \cup \Ex^{\varnothing}$ and divergence $D$, it holds that $C$ is \hyperref[axiom:POSL]{Subadditive} if and only if
    \begin{align}
    \mathbb{E}_\pi \left[ D(q \mid p) \right] \leq D(p_\pi \mid p) + \mathbb{E}_\pi \left[ D(q \mid p_\pi)\right] \quad  \forall \, \pi\in \Delta(W) \text{ and } p \in W \text{ s.t. } p_\pi \ll p.\footnotemark \label{triangle-avg}
     \end{align}
\end{lem}     \footnotetext{For any $p,q \in \Delta(\Theta)$, we let $q\ll p$ denote that $\supp (q)\subseteq \supp(p)$.}
\begin{proof}
See \cref{app:thm2:extra}.
\end{proof}

%\awb{Note that a PS $C$ is SLP iff $q \in \argmax_{p} \mathbb{E}_\pi \left[ D(\tilde{r} \mid p) \right] - D(q \mid p)$ for all $\pi \in \Pi(q)$. (It's UPS iff the maps $p \mapsto \mathbb{E}_\pi \left[ D(\tilde{r} \mid p) \right] - D(q \mid p)$ are constant.) The Banerjee et al characterization requires that  $\mathbb{E}_\pi \left[ \nabla_p D(\tilde{r} \mid p)|_{p=q} \right] = \mathbf{0}$, which is a special case of the argmax characterization for SLP when $\nabla_p D(q \mid p)|_{p=q}$ exists (in which case it must $= \mathbf{0}$).}

The final two lemmas show that, under certain smoothness conditions, any divergence $D$ satisfying \eqref{triangle-avg} is a \emph{Bregman divergence}, viz., there exists some convex $H \in \mathbf{C}^1(W)$ such that 
\begin{equation}\label{eqn:bregman}
D(q \mid p) = H(q) - H(p) - (q-p)^\top \nabla H(p) \quad \forall \, p,q\in W.
\end{equation}
We first establish this under a smoothness condition on $D$ that is stronger than  \hyperref[defi:ll]{Regularity}, and then invoke a mollification argument to establish the same result under \hyperref[defi:ll]{Regularity}.

To this end, for any divergence $D$, we denote by $\nabla_2 D(q \mid p) \in \R^{|\Theta|}$ its gradient with respect to the prior at every point $(q,p) \in \Delta(\Theta)\times\Delta(\Theta)$ where this gradient exists. Moreover, at such points, we normalize this gradient so that $p^\top \nabla_2 D(q \mid p) =0$. This normalization is obtained (without loss of generality) by extending the map $D(q \mid \cdot)$ from $\Delta(\Theta)$ to $\R^{|\Theta|}_+$ by homogeneity of degree $0$ (HD0) and then defining derivatives in the usual way.\footnote{For any Bregman divergence \eqref{eqn:bregman} with $H \in \mathbf{C}^2(W)$, this normalization is implied by our HD1 normalization for the Hessian of $H$ (\cref{remark:kernels}). Namely, $\nabla_2 D(q \mid p) 
\equiv -\H H(p) q$, which implies $p^\top \nabla_2 D(q \mid p) \equiv 0$ because $p^\top \H H(p) \equiv \mathbf{0}^\top$.} 

We first show that any divergence $D$ that satisfies \eqref{triangle-avg} and is $\mathbf{C}^1$ with respect to the prior takes the Bregman form \eqref{eqn:bregman} (and hence is also $\mathbf{C}^1$ with respect to the posterior). Formally, for any $W \subseteq \Delta(\Theta)$, we define $\mathcal{D}_W \subseteq \Delta(\Theta) \times \Delta(\Theta)$ as $\mathcal{D}_W:= [W\times W]\cup \left\{(p,p) \mid p \in \Delta(\Theta)\backslash W\right\}$. 

\begin{lem}\label{lem:thm2-prior-c1}
    Let $W \subseteq \Delta^\circ(\Theta)$ be open and convex. Let $C \in \C$ satisfy $\dom(C) = \Delta(W) \cup \Ex^{\varnothing}$, be \hyperref[axiom:POSL]{Subadditive}, and be \hyperref[eqn:PS]{Posterior Separable} with divergence $D$ such that $\dom(D) = \mathcal{D}_W$.\footnote{Since $\dom(C) = \Delta(W) \cup \Ex^\varnothing$, every divergence $D$ satisfying \eqref{eqn:PS} has $\dom(D) \supseteq \mathcal{D}_W$, but such divergences are not uniquely determined outside $\mathcal{D}_W$. Assuming that $\dom(D) = \mathcal{D}_W$ lets us (without loss of generality) abstract away from this form of indeterminacy in the divergence. This convention simplifies notation in (the proofs of) \cref{lem:thm2-prior-c1,lem:thm2-posterior-c1}.} If the prior-gradient $\nabla_2 D$ exists and is jointly continuous on $W \times W$, there exists convex $H \in \mathbf{C}^1(W)$ such that $D$ has the Bregman form \eqref{eqn:bregman}. Namely, for any $ p^*\in W$, it suffices to let $H = D(\cdot \mid p^*)$.
%Let $W \subseteq \Delta^\circ(\Theta)$ be open and convex. Let $C \in \C$ satisfy $\dom(C) = \Delta(W) \cup \Ex^{\varnothing}$, be \hyperref[axiom:POSL]{Subadditive}, and be \hyperref[eqn:PS]{Posterior Separable} with divergence $D$. If the prior-gradient $\nabla D_2$ exists and is jointly continuous on $W \times W$, then there exists $H \in \mathbf{C}^1(W)$ such that $D$ takes the Bregman form \eqref{eqn:bregman} on $W \times W$. In particular, given any $ p^*\in W$, it suffices to let $H = D(\cdot \mid p^*)|_{W}$.
%%%
\end{lem}

We prove \cref{lem:thm2-prior-c1}, which is the main technical step in the proof of \cref{thm:UPS}, at the end of this section. Next, we extend the conclusion of \cref{lem:thm2-prior-c1} to the broader class of divergences that are merely $\mathbf{C}^1$ with respect to the posterior, as implied by \hyperref[defi:ll]{Regularity}.%\footnote{We note that (the proofs of) \cref{lem:thm2-prior-c1,lem:thm2-posterior-c1} generalize the characterization of Bregman divergences in \textcite[Theorem 4]{bgw-bregman-ieee2005} by relaxing the smoothness assumptions therein. (\textcite{bgw-bregman-ieee2005} consider divergences $D$ that satisfy $p_\pi \in \arg\min_{p \in W} \E_\pi[D(q \mid p)]$ for all $\pi \in \Delta(W)$, rather than those that satisfy \eqref{triangle-avg}. However, both of these variational conditions yield the same necessary first-order condition \eqref{eqn:prior-grad-FOC} stated below in the proof of \cref{lem:thm2-prior-c1}, and therefore yield the same conclusions.) The key to this generalization is \cref{lem:thm2-prior-c1} and its proof; meanwhile, the more straightforward proof of \cref{lem:thm2-posterior-c1} builds on mollification arguments from \textcite{bgw-bregman-ieee2005}. } 

\begin{lem}\label{lem:thm2-posterior-c1}
    Let $W \subseteq \Delta^\circ(\Theta)$ be open and convex. Let $C \in \C$ satisfy $\dom(C) = \Delta(W) \cup \Ex^{\varnothing}$, be \hyperref[axiom:POSL]{Subadditive}, and be \hyperref[eqn:PS]{Posterior Separable} with divergence $D$ such that $\dom(D) = \mathcal{D}_W$. If the posterior-gradient $\nabla_1 D$ exists and is jointly continuous on $W\times W$, there exists convex $H \in \mathbf{C}^1(W)$ such that $D$ has the Bregman form \eqref{eqn:bregman}. Namely, for any $ p^*\in W$, it suffices to let $H = D(\cdot \mid p^*)$.
    %then $\forall p^*\in W$, $D(q|p) = D(q|p^*)+L(q,p)$, where $D(\cdot|p^*)\in \mathbf{C}^1(W)$ and $L(\cdot,p)$ is affine.
\end{lem}

\begin{proof}
See \cref{app:thm2:extra}.
\end{proof}

We remark that (the proofs of) \cref{lem:thm2-prior-c1,lem:thm2-posterior-c1} may be of independent interest, because they generalize the characterization of Bregman divergences in \textcite[Theorem 4]{bgw-bregman-ieee2005} by relaxing the smoothness assumptions imposed therein.\footnote{\textcite{bgw-bregman-ieee2005} consider divergences that are $\mathbf{C}^2$-smooth and satisfy the variational condition $p_\pi \in \arg\min_{p \in W} \E_\pi[D(q \mid p)]$ for all $\pi \in \Delta(W)$, while we consider divergences that are $\mathbf{C}^1$-smooth and satisfy the variational condition \eqref{triangle-avg}. Since both of these variational conditions yield the same necessary first-order condition \eqref{eqn:prior-grad-FOC} stated below (in the proof of \cref{lem:thm2-prior-c1}), our analysis also applies to the setting of \textcite{bgw-bregman-ieee2005}. Our proof of \cref{lem:thm2-posterior-c1} builds on the mollification Step 2 in the proof of \textcite[Theorem 3]{bgw-bregman-ieee2005}. 
%We note that the key to this generalization is \cref{lem:thm2-prior-c1}, which significantly relaxes the smoothness assumptions in Step 1 of the proof of \textcite[Theorem 4]{bgw-bregman-ieee2005}.
} 

We use \dred{Lemmas} \ref{lem:C^*:convex}--\ref{lem:thm2-posterior-c1} to prove \cref{thm:UPS}. We then present the proof of \cref{lem:thm2-prior-c1}.

\begin{proof}[Proof of \cref{thm:UPS}]
Let $W \subseteq \Delta^\circ(\Theta)$ be open and convex. Note that $\ri(\mathcal{D}_W) = W \times W$. 

\noindent\textbf{($\impliedby$ direction)} Let $C = C^H_\text{ups}$ for some $H \in \mathbf{C}^1(W)$. Then $C$ is \hyperref[eqn:PS]{Posterior Separable} with the Bregman divergence $D$ defined via $\dom(D)= \mathcal{D}_W$ and \eqref{eqn:bregman}, and the gradient $\nabla_1 D (q \mid p) = \nabla H(q) - \nabla H(p)$ is jointly continuous on $\ri(\dom(D)) = W \times W$. Thus, $C$ is \nameref{defi:ll}. 

\noindent\textbf{($\implies$ direction)} 
Let $C$ be \nameref{axiom:slp} and \nameref{defi:ll} with $\dom(C) = \Delta(W) \cup \Ex^\varnothing$ and divergence $\overline{D}$. Since $C$ is \nameref{axiom:slp}, it is \hyperref[axiom:POSL]{Subadditive} (\cref{prop:1}) and hence \hyperref[axiom:DL]{Dilution Linear} (\cref{lem:convex-dl}). Thus, since $C$ and $\overline{D}$ satisfy \eqref{gateux}, $C$ is \hyperref[eqn:PS]{Posterior Separable} with divergence $\overline{D}$. Since $\dom(C) = \Delta(W) \cup \Ex^\varnothing$ and $W \subseteq \Delta^\circ(\Theta)$ is open, this implies $\dom(\overline{D}) \supseteq \mathcal{D}_W$ and $\ri(\dom(\overline{D})) \supseteq W \times W$. Hence, $C$ is also \hyperref[eqn:PS]{Posterior Separable} with divergence $D := \overline{D}|_{\mathcal{D}_W}$, for which $\ri(\dom(D)) = W \times W$ and (by \hyperref[defi:ll]{Regularity}) $\nabla_1 D = \nabla_1 \overline{D}|_{W \times W}$ is jointly continuous on $W \times W$. Applying \cref{lem:thm2-posterior-c1} to $C$ and $D$, we obtain that $C = C^H_\text{ups}$ for some $H \in \mathbf{C}^1(W)$. 
\end{proof}

\begin{proof}[Proof of \cref{lem:thm2-prior-c1}]
    %Note that $\dom(D) = \ri(\dom(D)) = W \times W$ because $C$ and $D$ satisfy \eqref{eqn:PS}, $\dom(C) = \Delta(W) \cup \Ex^{\varnothing}$, and $W \subseteq \Delta^\circ(\Theta)$ is open.
    We prove the lemma in five steps: 

\noindent\textbf{Step 1: Linear prior-gradient.} \cref{lem:SLP-divergence} implies that, for every $\pi \in \Delta(W)$ and $p \in W$,
	\begin{align*}
	0\le f^\pi(p) := D(p_\pi \mid p)+ \mathbb{E}_{\pi} \left[ D(q \mid p_\pi) \right]- \E_\pi\left[ D(q \mid p) \right]. %\quad \forall \,  p \in W.
	\end{align*}
        The maps $f^\pi : W \to \mathbb{R}_+$ and $D(p_\pi \mid \cdot) : W \to \mathbb{R}_+$ are both minimized at $p = p_\pi$ (where they both equal $0$).
        Moreover, if $|\supp(\pi)|<+\infty$, then $f^\pi $ is differentiable and, for every $p \in W$,
        \begin{equation}\label{eqn:f-pi-grad}
        \nabla f^\pi(p) = \nabla_2 D(p_\pi \mid p) - \mathbb{E}_{\pi} \left[ \nabla_2 D(q \mid p) \right],
        \end{equation}
        where $\pi$ having finite support lets us interchange the order of differentiation and integration in the final term.  Thus, if $|\supp(\pi)|<+\infty$, the FOCs for minimization of $f^\pi$ and $D(p_\pi \mid \cdot)$ at $p=p_\pi$ yield $y^\top \nabla f^\pi (p_\pi) =y^\top \nabla_2 D(p_\pi \mid p_\pi) = 0$ for all $y \in \mathcal{T}$.  
        Hence, \eqref{eqn:f-pi-grad} implies
        \[
         \mathbb{E}_\pi \left[ y^\top \nabla_2 D(q \mid p_\pi) \right] = 0 \quad \forall \, y \in \mathcal{T} \, \text{ and finite-support } \pi \in \Delta(W).
        \]
        Moreover, our HD0 normalization for prior-gradients implies that
        \[
          \mathbb{E}_\pi \left[ p_\pi^\top\nabla_2 D(q \mid p_\pi) \right] = 0 \quad \forall  \text{ finite-support } \pi \in \Delta(W).
        \]
        Since $\text{span}(\{p\}\cup\mathcal{T}) = \R^{|\Theta|}$ for all $p \in W$, the preceding two displays together imply that
        \begin{equation}\label{eqn:prior-grad-FOC}
        \mathbb{E}_\pi \left[  \nabla_2 D(q \mid p_\pi) \right] =  \mathbf{0} \quad \forall \text{ finite-support } \pi \in \Delta(W).
        \end{equation}
This implies that, for each $p \in W$, %applying \cref{lem:matrix-rep} to 
the map $\nabla_2 D(\cdot \mid p) : W \to \mathbb{R}^{|\Theta|}$ can be represented as 
\begin{equation}
\nabla_2 D(q \mid p) = - A(p) q \quad \forall q \in W \label{prior-grad-linear}
\end{equation}
for some matrix $A(p) \in \R^{|\Theta|\times|\Theta|}$ satisfying $A(p) p = \mathbf{0}$ (see \cref{lem:matrix-rep} in \cref{app:thm2:extra}). In what follows, we denote by $A : W \to \R^{|\Theta|\times|\Theta|}$ the corresponding matrix-valued function. 

\noindent\textbf{Step 2: Directional posterior-derivatives.} For every $p,q \in W$, it holds that %the Gradient Theorem\footnote{\awb{[@SELF: should cite some book]}} and \eqref{prior-grad-linear} deliver
	\begin{equation}
	D(q\mid p) \, = \, \int_a^b \left(r'(x)\right)^\top\nabla_2 D(q \mid r(x))  \dd x \, = \, -  \int_a^b \left(r'(x)\right)^\top A \left( r(x) \right)  q   \dd x \label{eqn:reg-grad-thm}
	\end{equation}
	for all $a,b \in \mathbb{R}$ and $\mathbf{C}^1$-smooth curves $r : [a,b]  \to W$ such that $r(a) = q$ and $r(b) = p$, where the first equality is by the Gradient Theorem (which applies because $D(q \mid \cdot) \in \mathbf{C}^1(W)$) and the second one is by \eqref{prior-grad-linear}. %(For any such path, $r'(x) \in \mathcal{T}$ for all $x \in [a,b]$.) 
    We use \eqref{eqn:reg-grad-thm} to compute the directional derivatives of $D(\cdot \mid p)$. 
 
 To this end, fix any $q,p \in W$ and $y \in \mathcal{T}$. Fix any $\delta \in (0,1/2)$ sufficiently small that $q + \eta y \in W$ for all $\eta \in [-\delta,\delta]$, and consider any $\mathbf{C}^1$-smooth curve $r : [0,1] \to W$ for which (i) $r(x) = q + (x-\delta) y$ for all $x \in [0,2\delta]$ and (ii) $r(1) = p$.\footnote{Such $\delta>0$ and curves $r$ exist because $W \subseteq \Delta^\circ(\Theta)$ is open and convex.} Note that $r(\delta) = q$ and $r'(x) = y$ for all $x \in [0,2\delta]$. Thus, for any $\epsilon' \in (-\delta,\delta)$ and corresponding $\zeta:= \delta+ \epsilon'$, the (two-sided) directional derivative of $D(\cdot \mid p)$ at $r(\zeta ) = q + \epsilon' y$ in direction $y$ exists and is given by %$\frac{\d}{\d \epsilon} D(q + \epsilon y \mid p) \big|_{\epsilon = 0}$ as 
	\begin{align}
    \begin{split}\label{D-direct-deriv-thm2}
	\frac{\partial}{\partial \epsilon} D(q + \epsilon' y + \epsilon y \mid p) \big|_{\epsilon = 0} &= \frac{\d}{\d t} D(r(t) \mid p) \big|_{t=\zeta}  %\label{eqn:reg-9} 
 \\
	& = - \frac{\d}{\d t} \left[ \int_t^1 \left(r'(x) \right)^\top A \left( r(x) \right) r(t)   \dd x \right] \Big|_{t = \zeta}%\label{eqn:reg-10} 
 \\
	& = \left(r'(\zeta) \right)^\top A(r(\zeta))r(\zeta)  - \int_{\zeta}^1 \left(r'(x) \right)^\top A(r(x)) r'(\zeta)  \dd x %\label{eqn:reg-11} 
 \\
	& = - \int_{\delta + \epsilon'}^1 \left(r'(x) \right)^\top A(r(x)) y  \dd x, %\label{eqn:reg-12}
    \end{split}
	\end{align}
	where the first two lines hold by definition of the curve $r$ and the identity \eqref{eqn:reg-grad-thm}, the third line follows from the classic Leibniz rule,\footnote{Formally, define $f : [0,1]^2 \to \R$ as $f(t,x) := [r'(x)]^\top \nabla_2 D(r(t) \mid r(x)) = - [r'(x)]^\top A(r(x)) r(t)$. The map $f$ is continuous because the curve $r$ is $\mathbf{C}^1$-smooth and the prior-gradient $\nabla_2 D$ is  continuous on $W \times W$. Note that the partial derivative $f_1(t,x) :=\frac{\partial}{\partial t} f(t,x) = - [r'(x)]^\top A(r(x)) r'(t)$ for all $(t,x) \in [0,1]^2$. Since $\zeta  \in (0,2\delta)$, there exists $\chi >0$ such that $E:=[\zeta - \chi, \zeta + \chi] \subseteq (0,2\delta)$ and hence $r'(t) = y$ for all $t \in E$. Thus, $f_1(t,x) =  - [r'(x)]^\top A(r(x)) y$ for all $(t,x) \in E \times [0,1]$. Moreover, the map $x \mapsto - [r'(x)]^\top A(r(x)) y$ is continuous on $[0,1]$ because the curve $r$ is $\mathbf{C}^1$ smooth, the prior-gradient $\nabla_2 D$ is continuous on $W \times W$, and for any $\eta \in [-\delta, \delta]\backslash\{0\}$ it holds that $q+\eta y \in W$ and $- A(r(x)) y = \frac{1}{\eta} \left( \nabla_2 D(q +\eta y \mid r(x)) - \nabla_2 D(q  \mid r(x))\right)$. Thus, both $f$ and $f_1$ are continuous on $E \times [0,1]$, so the Leibniz rule implies that the map $t \mapsto  \int_t^1 f(t,x) \dd x$ is differentiable on $E$ and $\frac{\d }{\d t} \int_t^1 f(t,x) \dd x = - f(t,t) + \int_t^1 f_1(t,x) \dd x $ for all $t \in E$. For $t=\zeta \in E$, this yields the desired equality in \eqref{D-direct-deriv-thm2}.} and the final line holds because (by definition) $A(r(\zeta))r(\zeta) = \mathbf{0}$, $\zeta = \delta + \epsilon ' \in (0,2\delta)$,  and $r'(\zeta) = y$. Consequently, the (two-sided) second-order directional derivative of $D(\cdot \mid p)$ at $q$ in direction $y$ exists and is given by
	\begin{align}
	\frac{\partial^2}{\partial \epsilon' \partial \epsilon} D(q + \epsilon' y + \epsilon y \mid p) \big|_{\epsilon = \epsilon' = 0} \, = \,  -\frac{\d}{\d t}  \left[ \int_t^1 \left( r'(x)\right)^\top A(r(x)) y  \dd x \right]\Big|_{t= \delta} %\notag \\ %\label{eqn:reg-13} \\
	%%
	%& = \left( r'(\delta)\right)^\top A(r(\delta)) y  \notag \\ %\label{eqn:reg-14} \\
	%%
	\,  = \, y^\top A(q)  y, \label{prior-indep-Hess}
	\end{align}
	where the first equality is by the preceding display and the second equality follows from the Leibniz rule and the facts that $r(\delta) = q$ and $r'(\delta) = y$ (by definition of the curve $r$).

 \noindent\textbf{Step 3: Decomposition.} Let any $p^* \in W$ be given. We define the maps $H: W \to \mathbb{R}_+$ as $H(q) := D(q \mid p^*)$ and $L : W \times W \to \R$ as $L(q, p) := D(q \mid p) - D(q \mid p^*)$. By construction, 
\begin{equation}\label{eqn:breg-pre-pre}
 D(q \mid p) = H(q) + L(q, p) \quad \forall \, p,q \in W.
\end{equation}
 We claim that $H$ is convex and that, for each $p \in W$, the map $L(\cdot, p) : W \to \mathbb{R}$ is affine. %, i.e., $L(\alpha q_1 + (1-\alpha)q_0) = \alpha L(q_1 ,p) + (1-\alpha) L(q_0,p)$ for all $q_0,q_1 \in W$ and $\alpha \in [0,1]$. 
 
 First, to show affinity, let $p, q_0,q_1 \in W$ be given; the $q_0=q_1$ case is trivial, so let $q_0 \neq q_1$. Define $y := q_1 - q_0 \in \mathcal{T}$ and the map $f : [0,1] \to \mathbb{R}$ as $f(t):= L(q_0 + t y, p)$. The argument in Step 2 above implies that $f$ is twice differentiable and that, for every $t \in [0,1]$, 
\begin{align*}
f''(t) &= \frac{\partial^2}{\partial \epsilon' \partial \epsilon} L(q_0 +t y + \epsilon y + \epsilon' y,p)\big|_{\epsilon = \epsilon' = 0}\\
%%%
& = \frac{\partial^2}{\partial \epsilon' \partial \epsilon} D(q_0 + t y + \epsilon y + \epsilon' y \mid p) \big|_{\epsilon = \epsilon' = 0} - \frac{\partial^2}{\partial \epsilon' \partial \epsilon} D(q_0 + t y + \epsilon y + \epsilon' y \mid p^*) \big|_{\epsilon = \epsilon' = 0}
\end{align*}
Note that the second line is identically zero because \eqref{prior-indep-Hess} implies that, for every $t \in [0,1]$, 
	\[
	\frac{\partial^2}{\partial \epsilon' \partial \epsilon} D(q_0 + t y + \epsilon y + \epsilon' y \mid p) \big|_{\epsilon = \epsilon' = 0}  \, = \,   \frac{\partial^2}{\partial \epsilon' \partial \epsilon} D(q_0 + t y + \epsilon y + \epsilon' y \mid p^*) \big|_{\epsilon = \epsilon' = 0} \, = \, y^\top A(q_0 + t y) y.
	\]
Consequently, $f''(t) = 0$ for all $t \in [0,1]$. This implies $f(t) = t f(1) + (1-t) f(0)$ for all $t \in [0,1]$. Since $q_0,q_1 \in W$ were arbitrary, we conclude that $L(\cdot,p)$ is affine on $W$, as claimed. %Since $p, q_0,q_1 \in W$ were arbitrary, this establishes affinity of $L(\cdot, p)$ for all $p \in W$. %Since each $L(\cdot,p)$ is continuous, \cref{lem:matrix-rep} implies that $L(q,p) = a \cdot (q$

We now show that $H$ is convex. To this end, note that for every $\pi \in \Delta(W)$, 
\[
C(\pi) \, = \,  \mathbb{E}_\pi \left[ H(q) + L(q, p_\pi )\right]\,  = \, \mathbb{E}_\pi \left[ H(q) + L(p_\pi, p_\pi )\right] \, = \, \mathbb{E}_\pi \left[ H(q)- H(p_\pi)\right],
\]
where the first equality holds because $C$ and $D$ satisfy \eqref{eqn:PS} and \eqref{eqn:breg-pre-pre}, the second equality holds because $L(\cdot, p_\pi)$ is affine, and the final equality holds because $L(p,p) = - H(p)$ for all $p \in W$ (by construction). Since $C(\pi)\geq 0$ for all $\pi \in \Delta(W)$, it follows that $H$ is convex.

%It follows that $\mathbb{E}_\pi \left[ L(q , p_\pi)\right] = L(p_\pi, p_\pi)$ and hence $C(\pi) = \mathbb{E}_\pi \left[ H(q) + L(p_\pi, p_\pi )\right]$ for all $\pi \in \Delta(W)$. Since $C[\Ex^\varnothing] = \{0\}$, this implies $H(p) = -L(p,p)$ for all $p \in W$. Thus, $C(\pi) = \E_\pi[ H(q) - H(p_\pi)]$ for all $ \pi \in \Delta(W)$. Note that $H$ is convex, as $C(\pi)\geq 0$ for all $\pi \in \Delta(W)$.

\noindent\textbf{Step 4: Smooth Potential.} We now show that $H \in \mathbf{C}^1(W)$. Step 2 above establishes that $H  = D(\cdot \mid p^*)$ has two-sided directional derivatives at every point $q\in W$ and in every direction $y \in \mathcal{T}$. Therefore, since $H$ is convex (by Step 3) and $W\subseteq \Delta^\circ(\Theta)$ is open, \textcite[Theorem 25.2 and Corollary 2.5.5.1]{rock70} imply that $H \in \mathbf{C}^1(W)$, as desired.\footnote{Formally, since $\dom(H) = W\subseteq \Delta^\circ(\Theta)$ has empty interior with respect to the Euclidean topology on $\R^{|\Theta|}$, to apply \textcite{rock70} we consider the HD1 extension of $H$, viz., the map $G : \R^{|\Theta|}_+ \to \R \cup \{+\infty\}$ defined as $G(x) := (\mathbf{1}^\top x) H\left( \frac{x}{\mathbf{1}^\top x} \right)$. Since $H$ admits finite two-sided directional derivatives in all directions $y \in \mathcal{T}$ at every $q \in W$, it can be shown that $G$ admits finite two-sided directional derivatives in all directions $x \in \R^{|\Theta|}$ at every $q \in W$. Since all such $q$ are in the interior of $\dom(G) \subseteq \R^{|\Theta|}_{++}$ with respect to the Euclidean topology on $\R^{|\Theta|}$, Theorem 25.2 and Corollary 25.5.1 in \textcite{rock70} imply that the gradient map $q \in W \mapsto \nabla G(q) \in \R^{|\Theta|}$ is well-defined and continuous. Our HD1 normalization for posterior-gradients (\cref{fn:HD1-gradient}) then implies $\nabla H(q) = \nabla G(q)$ for all $q \in W$. Thus, $H \in \mathbf{C}^1\big( W\big)$.}

\noindent\textbf{Step 5: Bregman Representation.} Steps 3 and 4 imply that, for every $p \in W$,  
\begin{equation}\label{eqn:breg-pre}
D(q\mid p) = H(q) - H(p) + \left[ L(q,p) - L(p,p)\right] \quad  \forall \, q \in W,
\end{equation}
where $H = D(\cdot \mid p^*) \in \mathbf{C}^1(W)$ is convex and $L(\cdot,p) : W \to \R$ is affine. Since $D \geq 0$ on $W \times W$, this implies $L(q,p) - L(p,p) = -(q-p)^\top\nabla H(p)$ for all $p,q \in W$. Thus, \eqref{eqn:breg-pre} yields \eqref{eqn:bregman}.
%%%%
\end{proof}

%\subsection{Main Proofs for \cref{sec:phi}}

\subsection{Proof of \cref{thm:qk}}\label{sssec:app:thm:qk}

\subsubsection{Proof of \cref{thm:qk}(i) (\nameref{defi:ups} Upper Bound)}\label{sssec:app:thm:qk-pt1}

We begin by establishing a ``local'' version of the desired \nameref{defi:ups} upper bound. This local bound strengthens the definition of upper kernels via continuity-compactness arguments.

%We begin with a technical lemma, which establishes a ``local'' version of the desired \nameref{defi:ups} upper bound using the definition of upper kernels and continuity-compactness arguments. 

\begin{lem}\label{lem:loc-upper-bound}
    For any $C \in \C$, open convex $W \subseteq \Delta^\circ(\Theta)$, 
    %\footnote{\awb{[@SELF: The proof actually works for open convex $W \subseteq \Delta(\Theta)$ (i.e., allowing for partial support beliefs), provided that we define the matrix semi-norm appropriately at such beliefs. ]}} 
    and $H \in \mathbf{C}^2(W)$, if $\H H$ is an upper kernel of $C$ on $W$, then for every compact $V \subseteq W$ and $\epsilon >0$ there exists $\delta>0$ such that
    \begin{align}
C(\widehat{\pi}) \leq  C^H_\text{ups}(\widehat{\pi}) + 2 \epsilon \text{Var}(\widehat{\pi}) \qquad \text{$\forall\,\widehat{\pi} \in \Delta(V)$ with $\text{diam}(\supp(\widehat{\pi}))\leq \delta$}. \label{eqn:UK-p-pi}
\end{align}
\end{lem}
\begin{proof}
    See \cref{app:thm3-1:extra}.
\end{proof}

The key feature of \cref{lem:loc-upper-bound} is that the $\delta>0$ identified therein can depend on the subset $V \subseteq W$ and the error parameter $\epsilon>0$, but is uniform across all points $p \in V$.

We now turn to the main proof of \cref{thm:qk}(i), which: (a) constructs incremental learning strategies that approximate the target $\pi \in \Delta(W)$, and (b) iteratively applies \cref{lem:loc-upper-bound} to show that such strategies yield the desired ``global'' \nameref{defi:ups} upper bound.

\begin{proof}[Proof of \cref{thm:qk}(i)] Since the result holds trivially for $\pi\in\Ex^{\emptyset}$, we suppose throughout that the target random posterior satisfies $\pi\not\in \Ex^{\emptyset}$. We prove the result in three steps.
    
\noindent \textbf{Step 1: Let $\pi\in \Delta(W)$ have binary support.} Let $\supp(\pi)=\left\{ q_1,q_2 \right\}$.
%\footnote{We assume that $q_1 \neq q_2$, as the desired bound $\Phi(C)(\pi) \leq C^H_\text{ups}(\pi)$ trivially holds for $\pi \in \Ex^\varnothing$.} 
Since $W \subseteq \Delta^\circ(\Theta)$ is open, there exist $q'_1, q'_2 \in W$ such that $q_1, q_2 \in \ri(\text{conv}(\{q'_1,q'_2\}))$. In particular, let $q'_1 := q_1 - \eta (q_2 - q_1)$ and $q'_2 := q_2 + \eta (q_2 - q_1)$ for any sufficiently small $\eta>0$. Let $\pi' \in \Delta(W)$ be the unique random posterior with $p_{\pi'} = p_\pi$ and $\supp(\pi') = \{q'_1, q'_2\}$. Note that $\pi'\geq_\text{mps}\pi$.

Let $\epsilon>0$ be given. Since $H \in \mathbf{C}^2(W)$ and $\H H$ is an upper kernel of $C$ on $W$ (by hypothesis) and the compact set $V:= \text{conv}(\{q'_1,q'_2\})$ satisfies $V \subsetneq W$ (as $W$ is convex), \cref{lem:loc-upper-bound} delivers a corresponding $\delta>0$ such that \eqref{eqn:UK-p-pi} holds. Let $G := \{g_{i}\}_{i=1}^{N}$ be a finite grid on $\text{conv}(\{q'_1,q'_2\})$ that contains $\{p_\pi, q'_1, q'_2\}$ and has maximal step size of $\delta/2$; formally, let each $g_{i} := \alpha_{i} q'_1 + (1-\alpha_{i}) q'_2$ for weights $1 = \alpha_{1} > \cdots > \alpha_{N} = 0$ such that $g_{i^\star_\pi} = p_\pi$ for some $i^\star_\pi \notin\{1,N\}$ and $\| g_{i} - g_{i+1}\| \leq \delta/2$ for all $i \in \{1, \dots, N-1\}$. Let $\xi := \min_{i\in\{1, \dots, N-1\}} \|g_{i} - g_{i+1}\|>0$.

We now construct a sequence $(\pi^{(n)})_{n \in \mathbb{N}}$ in $\Delta(G)\subsetneq \Ex$ with the following properties: 
\begin{itemize}
    \item[(a)] $\pi^{(n)} \leq_\text{mps} \pi^{(n+1)} \leq_\text{mps}  \pi'$ and $\Phi(C)(\pi^{(n)}) \leq C^H_\text{ups}(\pi^{(n)}) + 2 \epsilon \text{Var}(\pi^{(n)})$ for all $n \in \mathbb{N}$; 
    \item[(b)] $\lim_{n \to \infty} \pi^{(n)} = \pi'$, and there exists an $\overline{n} \in \mathbb{N}$ such that $\pi^{(n)} \geq_\text{mps}\pi$ for all $n \geq \overline{n}$.
\end{itemize}
To this end, for each $i \notin\{1,N\}$, let $\widehat{\pi}_i \in \Ex$ be the unique random posterior with $p_{\widehat{\pi}_i} = g_i$ and $\supp(\widehat{\pi}_i) = \{g_{i-1},g_{i+1}\}$. For each $i \in \{1,N\}$, let $\widehat{\pi}_i := \delta_{g_i} \in \{\delta_{q'_1},\delta_{q'_2}\}$. Since $\text{diam}(\supp(\widehat{\pi}_i)) \leq \delta$ for all $i \in \{1,\dots, N\}$ (by definition of $G$), \eqref{eqn:UK-p-pi} and $\Phi(C)\preceq C$ together yield
\begin{align}
    \Phi(C)(\widehat{\pi}_i) \leq C(\widehat{\pi}_i) \leq C^H_\text{ups}(\widehat{\pi}_i) + 2 \epsilon \text{Var}(\widehat{\pi}_i) \qquad \forall i \in \{1,\dots, N\}. \label{eqn:grid-lub}
\end{align}
Now, let $\pi^{(1)} := \widehat{\pi}_{i^\star_\pi}$ and then inductively define $\pi^{(n)} := \sum_{i=1}^N \pi^{(n-1)}(\{g_i\}) \widehat{\pi}_i$ for all $n \geq 2$. In words, $\pi^{(n)}$ is the distribution at ``time $n$'' of a random walk on $G$ with initial condition $p_\pi$, transition probabilities $\{\widehat{\pi}_i\}_{i=1}^N$, and absorbing boundaries $\{g_1,g_N\} = \{q'_1, q'_2\}$. Since this process generalizes the Bernoulli random walk in \cref{eg:Diffusion:0} to asymmetric increments and (if $|\Theta|>2$) to arbitrary line segments in $\Delta(\Theta)$, we refer to the binary-support random posteriors $\{\widehat{\pi}_i\}_{i=2}^{N-1}$ as ``generalized Bernoulli'' random posteriors (see \cref{rmk:bernoulli:uqk} below). 
%\footnote{In words, $\pi^{(n)}$ is the marginal distribution at ``time $n$'' of a generalized Bernoulli random walk on $G$ with initial condition $p_\pi$, transition probabilities $\{\widehat{\pi}_i\}_{i=1}^N$, and absorbing boundaries $\{g_1,g_N\} = \{q'_1, q'_2\}$.} 

We verify that this sequence satisfies properties (a) and (b). For property (a), note first that $\pi^{(n)} \leq_\text{mps} \pi^{(n+1)} \leq_\text{mps}  \pi'$ for all $n \in \mathbb{N}$ by construction; we verify the other half by induction. For the base step, \eqref{eqn:grid-lub} implies that $\Phi(C)(\pi^{(1)}) \leq C^H_\text{ups}(\pi^{(1)}) + 2 \epsilon \text{Var}(\pi^{(1)})$. For the inductive step, let $n \geq 2$ and suppose that $\Phi(C)(\pi^{(n-1)}) \leq C^H_\text{ups}(\pi^{(n-1)}) + 2 \epsilon \text{Var}(\pi^{(n-1)})$. Define $\Pi^{(n)} \in \Delta^\dag(\Ex)$ as $\Pi^{(n)}(\{\widehat{\pi}_i\}) := \pi^{(n-1)}(\{g_i\})$, which induces first-round random posterior $\pi_1 = \pi^{(n-1)}$ and expected second-round random posterior $\E_{\Pi^{(n)}}[\pi_2] = \pi^{(n)}$. We then have
\begin{align*}
    \Phi(C)(\pi^{(n)}) &\leq \Phi(C)(\pi^{(n-1)}) + \sum_{i=1}^N \pi^{(n-1)}(\{g_i\}) \Phi(C)(\widehat{\pi}_i) \\
    %%%
    & \leq C^H_\text{ups}(\pi^{(n-1)}) + 2 \epsilon \text{Var}(\pi^{(n-1)}) + \sum_{i=1}^N \pi^{(n-1)}(\{g_i\}) \left[C^H_\text{ups}(\widehat{\pi}_i) + 2 \epsilon \text{Var}(\widehat{\pi}_i) \right] \\
    %%%
    & = C^H_\text{ups}(\pi^{(n)}) + 2 \epsilon \text{Var}(\pi^{(n)}),
\end{align*}
where the first line holds because $\Phi(C)$ is \hyperref[axiom:POSL]{Subadditive} (\cref{prop:1}), the second line is by the inductive hypothesis (first term) and \eqref{eqn:grid-lub} (second term), and the final line holds because $C^H_\text{ups} + 2 \epsilon \text{Var} \in \C$ is \nameref{defi:ups} and hence \hyperref[axiom:additive]{Additive} (\cref{prop:ups:additive}). This completes the induction. Next, to establish property (b), consider the sequence $(P_n, V_n)_{n \in \mathbb{N}}$ in $\mathbb{R}^2_+$ defined as $P_n := \pi^{(n)}(\{q'_1, q'_2\})$ and $V_n := \text{Var}(\pi^{(n)})$. By construction, $P_n\leq P_{n+1} \leq 1$ and $V_n \leq \text{Var}(\pi')$ (since $\text{Var}\in \C$ is \hyperref[axiom:mono]{Monotone} and $\pi' \geq_\text{mps} \pi^{(n)}$) for all $n \in \mathbb{N}$. We claim that $P_\infty := \lim_{n\to\infty} P_n = 1$. Suppose, towards a contradiction, that $P_\infty <1$. Then, we have 
\[
V_n\ \,  = \ \,V_{n-1} + \sum_{i=1}^N \pi^{(n-1)}(\{g_i\}) \text{Var}(\widehat{\pi_i}) \ \,   \geq \  \,  V_{n-1} + (1-P_{n-1}) \xi^2 \ \, \geq \ \, V_{n-1} + (1-P_\infty) \xi^2 , %\quad \text{ where $\xi := \min_{i\in\{1, \dots, N-1\}} \|g_{i} - g_{i+1}\|>0$.}
\]
where the first equality holds because $\text{Var}\in\C$ is \nameref{defi:ups} and hence \hyperref[axiom:additive]{Additive}, the second inequality is by definition of the $\widehat{\pi}_i$ and minimal grid-step size $\xi>0$, and the third inequality is by $P_n \nearrow P_\infty$. But this implies that $V_n \nearrow +\infty$, which contradicts $\sup_{n \in \mathbb{N}} V_n \leq \text{Var}(\pi') < +\infty$. Thus, $P_\infty = 1$ as claimed. Since $p_{\pi^{(n)}} = p_{\pi'}$ for all $n \in \mathbb{N}$, it then follows that $\lim_{n\to \infty}\pi^{(n)}(\{q'_i\}) = \pi'(\{q'_i\})$ for $i\in\{1,2\}$. 
%\footnote{Since $\Delta(G)$ is compact, there exists a convergent subsequence; for any such subsequence and corresponding limit point $\pi^\infty\in\Delta(G)$, the facts that $P_\infty = 1$ and $p_{\pi^{(n)}} = p_\pi$ for all $n\in\mathbb{N}$ imply that $p_\pi = \pi^\infty(\{q'_1\}) q'_1 + \pi^\infty(\{q'_2\}) q'_2$, and therefore $\pi^\infty(\{q'_i\}) = \pi'(\{q'_i\})$ for $i \in \{1,2\}$. Since this holds for all convergent subsequences, the result follows.} 
This has two implications. First, $\pi^{(n)} \to \pi'$ as desired. Second, $r^{(n)}_1 := \E_{\pi^{(n)}}[q \mid q \in\{g_1, \dots, g_{i^\star_\pi}\}] $ and $r^{(n)}_2 := \E_{\pi^{(n)}}[q \mid q \in\{g_{i^\star_\pi +1 }, \dots, g_{N}\}] $ satisfy $r^{(n)}_i \to q'_i$ for $i \in \{1,2\}$. Therefore, there exists an $\overline{n} \in\mathbb{N}$ such that $\text{conv}(\{q_1,q_2\}) \subseteq \text{conv}(\{r^{(n)}_1,r^{(n)}_2\})$ for all $n \geq \overline{n}$. Letting $\underline{\pi}^{(n)}\in \Ex$ denote the unique random posterior with $p_{\underline{\pi}^{(n)}} = p_\pi$ and $\supp(\underline{\pi}^{(n)}) = \{r^{(n)}_1 , r^{(n)}_2\}$, it follows that $\pi \leq_\text{mps} \underline{\pi}^{(n)} \leq_\text{mps} \pi^{(n)}$ for all $n \geq \overline{n}$. Since $\leq_\text{mps}$ is transitive, we conclude that $\pi \leq_\text{mps} \pi^{(n)}$ for all $n \geq \overline{n}$, as desired.

To conclude the proof of Step 1, observe that 
\[
\Phi(C)(\pi) \ \, \leq \ \, \Phi(C)(\pi^{(\overline{n})}) \ \, \leq \ \, C^H_\text{ups}(\pi^{(\overline{n})}) + 2 \epsilon \text{Var}(\pi^{(\overline{n})}) \ \, \leq \ \, C^H_\text{ups}(\pi') + 2 \epsilon \text{Var}(\pi'),
\]
where the first inequality holds because $\pi \leq_\text{mps} \pi^{(\overline{n})}$ (property (b)) and $\Phi(C)$ is \hyperref[axiom:mono]{Monotone} (\cref{prop:1}), the second inequality is by property (a), and the final inequality holds because $\pi^{(\overline{n})}\leq_\text{mps}  \pi'$ (property (a)) and $C^H_\text{ups} + 2\epsilon \text{Var} \in \C$ is \hyperref[axiom:mono]{Monotone}. Since the given $\epsilon>0$ was arbitrary, $\Phi(C)(\pi) \leq C^H_\text{ups}(\pi')$. Then, since $q'_1 = q_1 - \eta (q_2 - q_1)$ and $q'_2 = q_2 + \eta (q_2 - q_1)$ where $\eta>0$ is (sufficiently small but)  arbitrary, taking $\eta \to 0$ yields $\pi' \to \pi$ and $C^H_\text{ups}(\pi') \to C^H_\text{ups}(\pi)$ (as $H$ is continuous on $W$). Thus, $\Phi(C)(\pi) \leq C^H_\text{ups}(\pi)$, as desired.

\noindent \textbf{Step 2: Let $\pi \in \Delta(W)$ have finite support.} We proceed by induction on the size of $\supp(\pi)$. For the base step, Step 1 yields $\Phi(C)(\pi') \leq C^H_\text{ups}(\pi')$ for all $\pi' \in \Delta(W)$ with $|\supp(\pi')| \leq 2$. For the inductive step, let $N >2$ be given and suppose that $\Phi(C)(\pi') \leq C^H_\text{ups}(\pi')$ for all $\pi' \in \Delta(W)$ with $|\supp(\pi')| \leq N-1$. Let $\pi \in \Delta(W)$ satisfying $|\supp(\pi)| = N$ be given, and denote $\supp(\pi) = \{q_1, \dots, q_N\}$. Define the two-step strategy $\Pi\in \Delta^\dag (\Ex)$ as $\Pi(\{\delta_{q_i}\}) := \pi(\{q_i\}) $ for $i \in \{1, \dots, N-2\}$ and $\Pi(\{\widehat{\pi}\}) := \pi(\{q_{N-1},q_{N}\})$, where $\widehat{\pi}:= \pi(\cdot \mid \{q_{N-1}, q_N\}) \in \Ex$. Thus, $\Pi$ induces the expected second-round random posterior $\E_\Pi[\pi_2] = \pi$ and the first-round random posterior $\pi_1 = \sum_{i=1}^{N-2} \pi(\{q_i\}) \delta_{q_i} + \pi(\{q_{N-1},q_{N}\}) \delta_{\E_\pi[q \mid q \in \{q_{N-1}, q_N\}]}$. Note that $|\supp(\pi_1)| \leq N-1$ and $|\supp(\pi_2)| \leq 2$ for all $\pi_2 \in \supp(\Pi)$. Therefore, we obtain
\begin{align*}
    \Phi(C)(\pi) \ \, \leq \ \, \Phi(C)(\pi_1) + \E_\Pi \left[\Phi(C)(\pi_2)\right] \ \,  \leq \ \, C^H_\text{ups}(\pi_1) + \E_\Pi \left[ C^H_\text{ups}(\pi_2)\right] \ \, = \ \, C^H_\text{ups}(\pi),
\end{align*}
where the first inequality holds because $\Phi(C)$ is \hyperref[axiom:POSL]{Subadditive} (\cref{prop:1}), the second inequality is by the inductive hypothesis, and the final equality holds because $C^H_\text{ups}$ is \hyperref[axiom:additive]{Additive} (\cref{prop:ups:additive}). This completes the induction, as desired. %Thus, $\Phi(C)(\pi) \leq C^H_\text{ups}(\pi)$ for all finite-support $\pi \in \Delta(W)$.

\noindent \textbf{Step 3: Let $\pi \in \Delta(W)$ be arbitrary.} Let $\epsilon> 0$ be given. Since $\supp(\pi) \subseteq W$, for every $p \in \supp(\pi)$ there exists a linearly independent set $\{r_\theta(p)\}_{\theta \in \Theta}\subsetneq W$ such that: (a) $N(p):= \text{conv}\left(\{r_\theta(p)\}_{\theta \in\Theta}\right) \subseteq W$ and $N^\circ(p) := \ri(N(p))$ is an open neighborhood of $p$ (as $W\subseteq \Delta^\circ(\Theta)$ is open), and (b) $\max_{q,q'\in N(p)}|H(q) - H(q')|\leq \epsilon$ (as $H$ is continuous on $W$).\footnote{For instance, it suffices to let $r_\theta(p):= (1-\eta) p + \eta \delta_\theta$ for some sufficiently small $\eta>0$.} Since $\{N^\circ(p)\}_{p\in \supp(\pi)}$ is an open cover of the compact set $\supp(\pi)$, there exists a finite subcover $\{N^\circ(p_k)\}_{k=1}^K$. For every $q \in \supp(\pi)$, let $k(q):= \min\left\{k \mid q \in N^\circ(p_k) \right\}$ and note that, since $\{r_\theta(p_{k(q)})\}_{\theta\in\Theta}$ is  linearly independent, there exists a unique $\pi'(\cdot \mid q) \in \Ex$ such that $p_{\pi'(\cdot \mid q)} = q$ and $\supp(\pi'(\cdot \mid q)) = \{r_\theta(p_{k(q)})\}_{\theta\in\Theta}$. For all $q \notin\supp(\pi)$, let $\pi'(\cdot \mid q):= \delta_q$. Define the two-step strategy $\Pi \in \Delta(\Ex)$ as $\Pi(B) := \pi\left( \left\{q \in \Delta(\Theta) \mid \pi'(\cdot \mid q) \in B\right\} \right)$ for all Borel $B \subseteq \Ex$, 
%$\Pi\left(\left\{\pi'(\cdot \mid q)\in \Ex \mid q \in B\right\}\right)) := \pi(B)$ for all Borel $B \subseteq \Delta(\Theta)$, 
which induces the first-round random posterior $\pi_1 = \pi$ and the expected second-round random posterior $\pi_\epsilon:= \E_\Pi[\pi_2] = \E_\pi [\pi'(\cdot \mid q)]$.\footnote{In words, $\Pi\in \Delta(\Ex)$ is the pushforward of $\pi\in \Ex = \Delta(\Delta(\Theta))$ under the Borel measurable injection $q \in \Delta(\Theta) \mapsto \pi'(\cdot \mid q)\in \Ex$ (where injectivity is by construction and measurability is implied by piecewise continuity on the finite Borel partition $\{A_k\}_{k=0}^K$ of $\Delta(\Theta)$ defined as $A_0 := \Delta(\Theta) \backslash \supp(\pi)$ and $A_k := N_k^\circ \backslash\, \big[ \cup_{j=1}^{k-1} A_j\big]$ for $k \geq 1$). 
%However, note that $\Pi \in \Delta(\Ex) \backslash \Delta^\dag(\Ex)$ whenever $\supp(\pi)$ is infinite.
} By construction, $\supp\left( \pi_\epsilon\right) \subseteq \bigcup_{k=1}^K \{r_\theta (p_k)\}_{\theta \in \Theta} \subseteq W$ is finite. It follows that
\begin{align*}
    \Phi(C)(\pi) \ \, \leq \ \, \Phi(C)(\pi_\epsilon) \ \, \leq \ \, C^H_\text{ups}(\pi_\epsilon) \ \, = \ \, C^H_\text{ups}(\pi) + \E_\pi\left[C^H_\text{ups}(\pi'(\cdot\mid q)) \right] \ \, \leq \ \, C^H_\text{ups}(\pi) + \epsilon,
\end{align*}
where the first inequality is because $\Phi(C)$ is \hyperref[axiom:mono]{Monotone} (\cref{prop:1}), the second inequality is by Step 2, the third inequality is because $C^H_\text{ups}$ is \hyperref[axiom:additive]{Additive} (\cref{prop:ups:additive}), and the final inequality is because $C^H_\text{ups}(\pi'(\cdot \mid q)) \leq \max_{q' \in N(p_{k(q)})}|H(q') - H(q)| \leq \epsilon$ for all $q \in \supp(\pi)$ by construction. Since $\epsilon>0$ was arbitrary, we obtain $\Phi(C)(\pi) \leq C^H_\text{ups}(\pi)$, as desired. 
\end{proof}

\begin{remark}\label{rmk:bernoulli:uqk}
    In the proof of \Cref{thm:qk}(i), upper kernels are only used in Step 1 (via \cref{lem:loc-upper-bound}) to bound the direct cost of (incremental) generalized Bernoulli random posteriors.  Consequently, \Cref{thm:qk}(i) would continue to hold if we were to weaken \cref{defi:lq}(i) to require only that the upper kernel inequality holds for the restricted class of generalized Bernoulli random posteriors. Moreover, in the special case where $|\Theta|=2$, it is without loss of generality in Step 1 to restrict attention to Bernoulli random posteriors that are generated by the symmetric Bernoulli experiments from \cref{eg:Diffusion:0}. This construction therefore provides a formal proof of the upper bound $\Phi(C)\preceq f''(0)\cdot C_{\text{Wald}}$ derived in \cref{eg:Diffusion:0}, as claimed in \cref{ssec:bounding-phi}. 
    %In fact, we conjecture that the proof still goes through as long as the upper kernel bounds ``some'' local random posteriors. 
\end{remark}

\subsubsection{Proof of \cref{thm:qk}(ii) (Lower Kernel Invariance)}\label{app:thm3-2:proof}

The core of the proof is summarized in the following lemma. Informally, it establishes that, for any direct cost $C \in \C$, belief $p_0 \in \Delta(\Theta)$, and lower kernel $k(p_0)$ of $C$ at $p_0$, there exists a \nameref{defi:ups} cost  that: (i) provides a global lower bound on $C$, and (ii) is \nameref{defi:lq} and its kernel at $p_0$ provides an arbitrarily tight ``local'' lower bound on $k(p_0)$. Formally:

\begin{lem}\label{lem:lqk-invariance-lem}
    For any \nameref{defi:sp} $C \in \C$, $p_0 \in \Delta(\Theta)$, and $\xi >0$, if $k(p_0)$ is a lower kernel of $C$ at $p_0$ satisfying $k(p_0) - \xi I(p_0) \gg_\text{psd} \mathbf{0}$, then there exists a convex $H \in \mathbf{C}^2(\Delta(\Theta))$ such that (i) $C \succeq C^{H}_\text{ups}$ and (ii) $\H H(p_0) = k(p_0) - \xi I(p_0)$.
\end{lem}
\begin{proof}
    See \cref{sssec:lqk-invariance-lem-proof}.
\end{proof}

The formal proof of \cref{lem:lqk-invariance-lem} is technical and lengthy, but the basic idea is simple. In brief, we directly construct a convex function $H$ that has the desired Hessian at the point $p_0$, and which is ``approximately affine'' outside of an arbitrarily small neighborhood of $p_0$. The latter property, together with the \hyperref[defi:sp]{Strong Positivity} of $C$, ensures that $C\succeq C^H_\text{ups}$.

We now use \cref{lem:lqk-invariance-lem} to prove \cref{thm:qk}(ii):

\begin{proof}[Proof of \cref{thm:qk}(ii)]
Let $p_0 \in W$ be given. Since $k(p_0) \gg_\text{psd} \mathbf{0}$, there exists an $\overline{\epsilon}>0$ such that $k(p_0) - \epsilon I(p_0) \gg_\text{psd}\mathbf{0}$ for all $\epsilon \leq \overline{\epsilon}$. Let an $\epsilon  \in (0, \overline{\epsilon})$ be given. Setting $\xi := \epsilon$, \cref{lem:lqk-invariance-lem} then delivers the existence of an $H \in \mathbf{C}^2(\Delta(\Theta))$ such that (i) $C \succeq C^H_\text{ups}$ and (ii) $\H H(p_0) = k(p_0) - \epsilon I(p_0)$. Since $\Phi$ is isotone (\cref{lem:structure:Phi} in \cref{sec:app:structure}) and $C^H_\text{ups}$ is \nameref{axiom:slp} (\cref{lem:ups:to:additive} in \cref{proof:ups:additive}), it follows that $\Phi(C) \succeq \Phi(C^H_\text{ups}) = C^H_\text{ups}$. Since $\H H(p_0) = k(p_0) -\epsilon I(p_0)$ is a (lower) kernel of $C^H_\text{ups}$ at $p_0$ (by \cref{lem:ups-kernel-equiv} in \cref{ssec:calc-kernel}), it follows that $k(p_0) -\epsilon I(p_0)$ is also a lower kernel of $\Phi(C)$ at $p_0$. Thus, for $\epsilon' := \frac{\epsilon}{2}$, there exists a $\delta>0$ such that the lower kernel bound in \cref{defi:lq}(ii) holds for $\Phi(C)$ and $k(p_0)- \epsilon I(p_0)$ at $p_0$ with error parameters $\epsilon'$ and $\delta$. That is, for every $\pi \in \Ex$ with $p_\pi \in B_\delta(p_0)$,  
%%%%
 \begin{align*}
     \Phi(C)(\pi) & \ge \int_{B_{\delta}(p_0)}(q-p)^\top \left(\frac{1}{2}\left(k(p_0) -\epsilon I(p_0)\right) - \epsilon' I\right)(q-p) \dd \pi( q) \\
     %%%
     &= \int_{B_{\delta}(p_0)}(q-p)^\top \left(\frac{1}{2} k (p_0) - \epsilon I\right)(q-p) \dd \pi( q),
 \end{align*}
 where the equality uses the facts that (by definition) $\epsilon' = \frac{\epsilon}{2}$ and $I(p_0) \sim_\text{psd} I$. Since $\epsilon\in (0, \overline{\epsilon})$ was arbitrary, we conclude that $k(p_0)$ is a lower kernel of $\Phi(C)$ at $p_0$, as desired.  
\end{proof}

%To conclude, we now present the main proof of \cref{lem:lqk-invariance-lem}. (See \cref{app:thm3:extra} for a technical lemma invoked in the proof.)

\subsection{Proof of \cref{thm:flie}}\label{proof:flie}
\begin{proof}
Let $W \subseteq \Delta^\circ(\Theta)$ be open and convex. We prove each direction in turn. 

\noindent\textbf{($\implies$ direction)}
	First, we claim that $\Phi(C)\preceq C_\text{ups}^H$. Since $k_C = \H H$ is an upper kernel of $C$ on $W$, \cref{thm:qk}(i) implies that $\Phi(C)(\pi)\le C_\text{ups}^H(\pi)$ for all $\pi \in\Delta(W)$. Since $\Phi(C),  C^H_\text{ups} \in \C$ and $\dom(C^H_\text{ups}) = \Delta(W) \cup \Ex^\varnothing$, 
    we also have $\Phi(C)[\Ex^\varnothing] = C^H_\text{ups}[\Ex^\varnothing] = \{0\}$, $ C^H_\text{ups}[\Ex \backslash (\Delta(W) \cup \Ex^\varnothing)] = \{+\infty\}$, and $\sup\Phi(C)[\Ex \backslash (\Delta(W) \cup \Ex^\varnothing)] \leq +\infty$. It follows that %$\Phi(C)(\pi) \leq C^H_\text{ups}(\pi)$ for all $\pi \in \Ex$, i.e., 
    $\Phi(C) \preceq C^H_\text{ups}$.

    Next, we claim that $\Phi(C) \succeq C^H_\text{ups}$. Since $C$ \nameref{axiom:flie}, $C \succeq \Phi_\text{IE}(C)$. Since $\dom(C) \subseteq \Delta(W) \cup\Ex^\varnothing$, $H\in \mathbf{C}^2(W)$ is strongly convex, and $k_C = \H H$ is a lower kernel of $C$ on $W$, \cref{lem:phi-ie}(ii) implies $\Phi_\text{IE}(C) \succeq C^H_\text{ups}$. Thus, $C \succeq C^H_\text{ups}$. Therefore, since $\Phi$ is isotone (\cref{lem:structure:Phi} in \cref{sec:app:structure}) and $C^H_\text{ups}$ is \nameref{axiom:slp} (\cref{lem:ups:to:additive} in \cref{proof:ups:additive}), $\Phi(C) \succeq \Phi(C^H_\text{ups}) = C^H_\text{ups}$.
 
 Finally, combining these two inequalities, we conclude that $\Phi(C)=C_\text{ups}^H$.

 \noindent\textbf{($\impliedby$ direction)} To begin, note that $\Phi(C) = C^H_\text{ups}$ is \nameref{defi:sp} because $H$ is strongly convex. Since $C \succeq \Phi(C)$, it follows that $C$ is also \nameref{defi:sp}.

 First, we claim that $k_C = \H H$. Since $C$ is \nameref{defi:sp}, $k_C \gg_\text{psd} \mathbf{0}$ on $W$ (\cref{cor:ker-SP} in \cref{ssec:calc-kernel}). Hence, \cref{thm:qk}(ii) implies that $k_C$ is a lower kernel of $\Phi(C)$ on $W$. Since $C \succeq \Phi(C)$, $k_C$ is also an upper kernel of $\Phi(C)$ on $W$. Therefore, $\Phi(C)$ is \nameref{defi:lq} on $W$ with kernel $k_{\Phi(C)} = k_C$. Meanwhile, since $\Phi(C)=C_\text{ups}^H$ and $H \in \mathbf{C}^2(W)$, \cref{lem:ups-kernel-equiv} in \cref{ssec:calc-kernel} implies that $k_{\Phi(C)} = \H H$. It follows that $k_C = \H H$. 

 Next, we claim that $C$ \nameref{axiom:flie}. Since $C \succeq \Phi(C)$ and $\Phi(C) = C^H_\text{ups}$, we have $C \succeq C^H_\text{ups}$. Since $\dom(C)\subseteq \Delta(W)\cup\Ex^\varnothing$, $H \in \mathbf{C}^2(W)$ is strongly convex, and (as just shown) $k_C = \H H$ on $W$, \cref{lem:phi-ie}(iii) implies that $\Phi_\text{IE}(C) = C^H_\text{ups}$. Therefore, $C \succeq \Phi_\text{IE}(C)$, i.e., $C$ \nameref{axiom:flie}.
\end{proof}

\subsection{Definition of \hyperref[defi:pst:cont]{uTVM-Continuity}}\label{app:utvm}

We begin with some auxiliary definitions, which are adapted from \textcite{pomatto2023cost} (henceforth PST23). First, an experiment $\sigma \in \Se$ is \emph{bounded} if there exists an $m \in \R_+$ such that for every $\theta,\theta'\in \Theta$, the log-likelihood ratio $\log\left(\frac{\d \sigma_{\theta}}{\d \sigma_{\theta'}}\right)$ is $\sigma_\theta$-almost surely in $[-m,m]$. %\footnote{This definition presumes that, for every $\theta,\theta'\in \Theta$, the measures $\sigma_\theta, \sigma_{\theta'} \in \Delta(S)$ are mutually absolutely continuous.} 
We denote by $\Se_b \subsetneq \Se$ the class of all bounded experiments. It follows from Bayes' rule that $h_B[ \Se_b \times \Delta^\circ(\Theta)] = \Delta(\Delta^\circ(\Theta))$; we use this fact in what follows.

Next, for every $\sigma \in \Se_b$ and integral vector $\bm{\alpha}\in (\mathbb{N}\cup\{0\})^{|\Theta|}$, define $M^\sigma(\bm{\alpha}) \in \R_+^{|\Theta|}$ as 
\[
M_{\theta}^{\sigma}(\bm{\alpha}):=\int_S \Bigg|\prod_{\theta'\neq \theta}\log\left(\frac{\d \sigma_{\theta}}{\d \sigma_{\theta'}}(s)\right)^{\alpha_{\theta'}}\Bigg|\dd \sigma_{\theta}(s) \quad \text{for each $\theta \in \Theta$.}
\]
In words, $M_\theta^\sigma(\bm{\alpha}) \in \R_+$ is the $\bm{\alpha}$-moment of the vector of log-likelihood ratios $\log\left(\frac{\d \sigma_{\theta}}{\d \sigma_{\theta'}}\right)_{\theta'\in \Theta}$ conditional on state $\theta \in \Theta$ under experiment $\sigma \in \Se_b$. Moreover, for every $\sigma \in \Se_b$ and $\theta \in \Theta$, we denote by $\upsilon^{\sigma}_\theta \in \Delta ( \R^{|\Theta| \times |\Theta|})$ the distribution of the vector of all log-likelihood ratios $\log\left(\frac{\d \sigma_{\theta'}}{\d \sigma_{\theta''} }\right)_{\theta', \theta''\in \Theta}$ induced by the $\theta$-contingent signal distribution $\sigma_\theta \in \Delta(S)$.\footnote{That is, $\upsilon^{\sigma}_{\theta}(B):= \sigma_{\theta} \big( \big\{ s \in S \mid \log \left(\frac{\d \sigma_{\theta'}}{\d \sigma_{\theta''}}(s)\right)_{\theta',\theta'' \in \Theta}\in B  \big\} \big) $ for all Borel $B \subseteq \mathbb{R}^{|\Theta| \times |\Theta|}$.}

Finally, we adapt PST23's continuity Axiom 4 to our setting as follows:

\begin{definition}[uTVM-continuous]\label{defi:pst:cont}
    A cost function $C\in \C$ with rich domain is uniformly total variation-moment-continuous (\nameref{defi:pst:cont}) if the map $\Ec:\Se_b\times \Delta^\circ(\Theta)\to \overline{\R}_+$ defined as $\Ec(\sigma,p) := C (h_B(\sigma,p))$ satisfies the following condition:
    %\begin{itemize}[noitemsep]
       % \item[(i)] For every $\sigma\in \Se_b$ and $p \in \Delta^\circ(\Theta)$, it holds that $\Ec(\sigma,p) = C(h_B(\sigma,p))$.
        %%%
        \begin{quote}
        For every $p \in \Delta^\circ(\Theta)$, there exists an $N \in \mathbb{N}$ such that $\Ec(\cdot, p) : \Se_b \to \overline{\R}_+$ is uniformly continuous with respect to the pseudo-metric $d_N$ on $\Se_b$ defined as
        \begin{align*}
            d_N(\sigma,\sigma'):=\max_{\theta\in\Theta} d_\text{TV}(\upsilon^{\sigma}_{\theta} , \upsilon^{\sigma'}_{\theta}) +\max_{\theta\in\Theta} \max_{\bm{\alpha}\in\{0,\ldots, N\}^{|\Theta|}} |M_{\theta}^{\sigma}(\bm{\alpha})-M_{\theta}^{\sigma'}(\bm{\alpha})|,
        \end{align*}
        where $d_\text{TV}$ denotes the total variation metric on $\Delta(\mathbb{R}^{|\Theta|\times|\Theta|})$.\footnote{In particular, $d_\text{TV}(\upsilon^{\sigma}_{\theta}, \upsilon^{\sigma'}_\theta) := \sup_B |\upsilon^{\sigma}_{\theta}(B) - \upsilon^{\sigma'}_\theta(B)|$ where the supremum is taken over all Borel $B \subseteq \mathbb{R}^{|\Theta|\times|\Theta|}$.}
    %\end{itemize}
    \end{quote}
\end{definition}

In words, a cost function is \nameref{defi:pst:cont} if it satisfies PST23's Axiom 4 for each fixed full-support prior. Since PST23 implicitly hold the prior fixed and work directly with cost functions defined on experiments, in \cref{defi:pst:cont} we first define $K$ as the ``experiment-based'' version of $C$ and then impose PST23's Axiom 4 on $K$ prior-by-prior.\footnote{As we explain in the proof of \cref{lem:bayes-LLR} (see \cref{sssec:proof:thm5}), while PST23 define cost functions and their axioms on a larger domain of experiments that includes $\Se_b$ as a strict subclass, their main result (Theorem 1) applies verbatim if we restrict attention to the smaller domain $\Se_b$. See \cref{app:beyond:belief} for details on the relationship between ``belief-based'' cost functions defined on random posteriors and ``experiment-based'' cost functions defined on experiments.} As discussed in PST23, \hyperref[defi:pst:cont]{uTVM-continuity} is a mild continuity assumption because convergence under the $d_N$ pseudo-metric (for any $N \in \mathbb{N}$) is a demanding requirement.

\AtNextBibliography{\small}
\setlength\bibitemsep{1pt}
{\setstretch{1}
\printbibliography
}
\newpage
\setcounter{footnote}{0}
\pagenumbering{arabic}% resets `page` counter to 1
\section*{\center{\Large{Online Appendix to} \\ ``The Cost of Optimally Acquired Information'' \\ \vspace{4pt} \large{Alexander W. Bloedel \qquad Weijie Zhong}}}\label{sec:OA}

\setcounter{section}{1}
\setcounter{subsection}{0}
\setcounter{tocdepth}{2}

\vspace{20pt}
%\section*{Table of Contents}
\startcontents
\printcontents{ }{1}{}

\newpage

\section{Results Omitted from the Main Paper}\label{app:omitted-results}

\subsection{Facts about Cost Functions and Operators}\label{sec:app:structure}

This section presents three structural facts about the sets of direct and indirect cost functions  and the two-step and sequential learning maps. Recall that we endow the set of cost functions $\C$ with the operations of pointwise addition and multiplication by positive scalars, as well as the pointwise (partial) order $\succeq$. That is, for any $C,C' \in \C$ and $\alpha \geq 0$, we let: (i) $C+\C' \in \C$ be defined as $[C+C'](\pi) := C(\pi) + C'(\pi)$ for all $\pi \in \Ex$, (ii) $\alpha C \in \C$ be defined as $[\alpha C](\pi) := \alpha C(\pi)$ for all $\pi \in \Ex$, and (iii) $C\succeq C'$ denote that $C(\pi) \geq C'(\pi)$ for all $\pi \in \Ex$. The set of indirect costs $\C^* \subseteq \C$ is endowed with the same operations and order.

The first result records basic facts about $\C$, the space of all cost functions:

\begin{lem}\label{lem:structure:C}
    The set $\C$ is a convex cone, a complete lattice, and closed under pointwise limits.
\end{lem}
\begin{proof}
    First, $\C$ is clearly a convex cone: for any $C,C' \in\C$ and $\alpha,\beta \geq 0$, we have $\alpha C+\beta C' \in \C$.

    Second, for any nonempty $\mathcal{D}\subseteq \C$, note that $\sup_{C \in \mathcal{D}} C(\pi) \geq \inf_{C \in \mathcal{D}} C(\pi) \geq 0$ for all $\pi \in \Ex$ and $\sup_{C \in \mathcal{D}} C(\pi) = \inf_{C \in \mathcal{D}} C(\pi) = 0$ for all $\pi \in \Ex^\varnothing$. Therefore, given the poset $(\C,\succeq)$ and any nonempty $\mathcal{D}\subseteq \C$, the meet $\wedge \mathcal{D} \in \C$ and join $\vee \mathcal{D} \in \C$ are defined as $\wedge \mathcal{D}(\pi) := \inf_{C \in \mathcal{D}} C(\pi)$ and $\vee \mathcal{D}(\pi) := \sup_{C \in \mathcal{D}} C(\pi)$ for all $\pi \in \Ex$, respectively. Moreover, for the empty subset $\mathcal{D} = \emptyset$, we define  $\wedge \mathcal{D}\in \C$ as $\wedge \mathcal{D}[\Ex^\varnothing]=\{0\}$ and $\wedge \mathcal{D}[\Ex\backslash \Ex^\varnothing] = \{+\infty\}$, and we define $\vee \mathcal{D} \in \C$ as $\vee \mathcal{D}[\Ex]=\{0\}$. We conclude that $\C$ is a complete lattice.

    Third, take any sequence $(C_n)_{n \in \mathbb{N}}$ in $\C$ such that $\lim_{n \to\infty} C_n(\pi) \in \overline{\R}_+$ exists for all $\pi \in \Ex$. Define $C : \Ex \to \overline{\R}_+$ as $C(\pi):= \lim_{n \to\infty} C_n(\pi)$ for all $\pi \in \Ex$. Note that $C \in \C$, as $C_n[\Ex^\varnothing]=\{0\}$ for all $n \in \mathbb{N}$ implies $C[\Ex^\varnothing]=\{0\}$. We conclude that $\C$ is closed under pointwise limits. 
\end{proof}

The second result records basic facts about $\Psi$ and $\Phi$, the two-step and sequential learning maps. We say that a map $\widehat{\Phi} : \C \to \C$ is: (i) \emph{isotone} if $C \succeq \C'$ implies that $\widehat{\Phi}(C) \succeq \widehat{\Phi}(C')$, (ii) \emph{(positively) homogeneous of degree 1} (HD1) if $\widehat{\Phi}(\alpha C) = \alpha \widehat{\Phi}(C)$ for all $C \in \C$ and $\alpha \geq 0$, and (iii) \emph{concave} if $\widehat{\Phi}(\alpha C + (1-\alpha) C') \succeq \alpha\widehat{\Phi}(C) + (1-\alpha) \widehat{\Phi}(C')$ for all $C,C' \in \C$ and $\alpha \in [0,1]$.

\begin{lem}\label{lem:structure:Phi}
    The maps $\Psi$ and $\Phi$ are isotone, HD1, and concave.
\end{lem}
\begin{proof}
    First, for isotonicity, take any $C ,C' \in \C$ with $C\succeq C'$. By construction, for every $\Pi \in \Delta^\dag(\Ex)$, it holds that $C(\pi_1)+\E_{\Pi}[C(\pi_2)] \geq  C'(\pi_1)+\E_{\Pi}[C'(\pi_2)]$. This implies $\Psi(C)\succeq \Psi(C')$. By induction, we have $\Psi^n(C) \succeq \Psi^n(C')$ for all $n \in \mathbb{N}$. Taking $n \to \infty$ yields $\Phi(C)\succeq \Phi(C')$. 
    
    Second, for HD1, take any $C \in \C$ and $\alpha \geq 0$. For every $\Pi \in \Delta^\dag(\Ex)$, it holds that $\alpha C(\pi_1)+\E_{\Pi}[\alpha C(\pi_2)]=\alpha \left(C(\pi_1)+\E_{\Pi}[C(\pi_2)]\right)$. This implies $\Psi(\alpha C)=\alpha \Psi(C)$. By induction, we have $\Psi^n(\alpha C) = \alpha \Psi^n(C)$ for all $n \in \mathbb{N}$. Taking $n \to \infty$, we obtain $\Phi(\alpha C) = \alpha \Phi(C)$.

    Finally, for concavity, take any $C, C' \in \C$ and $\alpha \in [0,1]$. Define $C'' \in \C$ as $C'':= \alpha C+(1-\alpha)C'$. For every $\pi \in \Ex$ and $\Pi \in \Delta^\dag(\Ex)$ such that $\E_\Pi[\pi_2]\geq_\text{mps} \pi$, we have
    \begin{align*}
        C''(\pi_1)+\E_{\Pi}[C''(\pi_2)]  &=  \alpha \left( C(\pi_1)+\E_{\Pi}[C(\pi_2)] \right) +(1-\alpha) \left( C'(\pi_1)+\E_{\Pi}[C'(\pi_2)] \right) \\
        %%%
        & \geq \alpha \Psi(C)(\pi) + (1-\alpha) \Psi(C')(\pi).
    \end{align*}
    It follows that $\Psi(C'') \succeq \alpha \Psi(C) + (1-\alpha) \Psi(C')$. We conclude that $\Psi$ is concave. Since $\Psi$ is also isotone (as shown above), it then follows by induction that $\Psi^n(C'') \succeq \alpha \Psi^n(C) + (1-\alpha) \Psi^n(C')$ for all $n \in \mathbb{N}$. Taking $n \to \infty$, we obtain that $\Phi(C'') \succeq \alpha \Phi(C) + (1-\alpha) \Phi(C')$. 
\end{proof}

The final result records basic facts about $\C^*$, the set of indirect costs. Let $\vee$ denote the join (supremum) operation on $(\C,\succeq)$ defined in (the proof of) \cref{lem:structure:C}. We say that $\mathcal{D}\subseteq \mathcal{C}$ is \emph{closed under suprema} if, for every subset $\mathcal{D}'\subseteq \mathcal{D}$, the supremum satisfies $\vee \mathcal{D}' \in \mathcal{D}$. 

\begin{lem}\label{lem:structure:C*}
    The set $\C^*$ is a convex cone and closed under suprema.
\end{lem}
\begin{proof}
    First, take any $C,C' \in \C^*$ and $\alpha,\beta\geq 0$. Define $C'' \in \C$ as $C'' := \alpha C + \beta C'$. We have 
    \[
    \Phi(C'') \, \succeq \, \alpha \Phi(C)+\beta \Phi (C') \, = \, C'',
    \]
    where the inequality holds because $\Phi$ is HD1 and concave (\cref{lem:structure:Phi}) and the equality holds because $C,\C'\in\C^*$ are \nameref{axiom:slp} (\cref{prop:1}) and by definition of $C''$. Since $C'' \succeq \Phi(C'')$ by definition, it follows that $\Phi(C'') = C''$ and hence $C'' \in \C^*$. Therefore, $\C^*$ is a convex cone.

    Next, take any $\mathcal{D}\subseteq \C^*$. By definition, the supremum $\vee \mathcal{D} \in \C$ satisfies $\vee \mathcal{D} \succeq C$ for every $C \in \mathcal{D}$. Since $\Phi$ is isotone (\cref{lem:structure:Phi}) and each $C \in \mathcal{D}$ is \nameref{axiom:slp} (\cref{prop:1}), we have $\Phi\left( \vee \mathcal{D} \right) \succeq \Phi(C) = C$ for every $C \in \mathcal{D}$. Thus, $\Phi\left( \vee \mathcal{D} \right) \succeq \vee \mathcal{D}$. Since $\vee \mathcal{D} \succeq \Phi\left( \vee \mathcal{D} \right)$ by definition, it follows that $\vee \mathcal{D} = \Phi\left( \vee \mathcal{D} \right)$ and hence $\vee \mathcal{D} \in \mathcal{C}^*$. Thus, $\C^*$ is closed under suprema.
\end{proof}

\subsection{Facts about Kernels}\label{ssec:calc-kernel}

This section presents four technical results about kernels. The first two results provide practical tools to calculate the kernels of \hyperref[defi:ups]{(Uniformly)} \hyperref[eqn:PS]{Posterior Separable} cost functions. The latter two results are structural facts about kernels of general cost functions.  %We use all of these results in our other proofs.

\paragraph{Calculation Tools.} We first characterize the kernels of \hyperref[defi:ups]{(Uniformly)} \hyperref[eqn:PS]{Posterior Separable} cost functions. In addition to providing a practical method for calculating kernels in applications, this helps to clarify the connection between our notion of \nameref{defi:lq} cost functions and the standard (finite-dimensional) definition of twice differentiability. %These results serve two purposes. First, they provide a practical method to calculate kernels in applications. Second, they connect our notion of \nameref{defi:lq} cost functions to the standard (finite-dimensional) definition of twice differentiability. 

Our first result shows that every \hyperref[eqn:PS]{Posterior Separable} cost with a ``locally smooth'' divergence is \nameref{defi:lq}; moreover, its kernel equals the Hessian of the divergence with respect to the posterior, evaluated at the prior. Formally, we say that divergence $D$ is \emph{locally $\mathbf{C}^2$ at $p_0 \in \Delta(\Theta)$} if there exists $\delta>0$ such that (i) $\dom(D) \supseteq B_{\delta}(p_0) \times B_\delta(p_0)$ and (ii) the map $(q,p) \mapsto \H_1 D(q \mid p) \in \R^{|\Theta| \times |\Theta|}$ is well-defined and continuous on $B_{\delta}(p_0) \times B_\delta(p_0)$. 

\begin{lem}\label{lem:ps-kernel-suff}
For any $p_0 \in \Delta(\Theta)$ and \hyperref[eqn:PS]{Posterior Separable} $C \in \C$ with divergence $D$,
\[
\text{$D$ is locally $\mathbf{C}^2$ at $p_0$ }  \ \implies \   \text{ $C$ is \nameref{defi:lq} at $p_0$ and $k_C(p_0) = \H_1 D(p_0 \mid p_0)$.} 
\]
\end{lem}
\begin{proof}
    See \cref{ssec:proof-ps-kernel-suff}.
\end{proof}

Our second result refines \cref{lem:ps-kernel-suff} for the subclass of \nameref{defi:ups} costs by showing that twice continuous differentiability of the potential function $H$ is both sufficient \emph{and necessary} for $C^H_\text{ups}$ to be \nameref{defi:lq}. Formally: %, we have the following characterization: 

\begin{lem}\label{lem:ups-kernel-equiv}
For any open $W \subseteq \Delta(\Theta)$ and convex $H : \Delta(\Theta) \to \R \cup\{+\infty\}$ with $\dom(H) \supseteq W$, 
\[
\text{$C^H_\text{ups}$ is \nameref{defi:lq} on $W$} %\iff 
\quad \xrightleftharpoons[]{\text{$W \subseteq \Delta^\circ(\Theta)$}} \quad H|_W \in \mathbf{C}^2(W).\footnotemark
\]
Under either of these equivalent conditions, the kernel of $C^H_\text{ups}$ on $W$ is $k_{C^H_\text{ups}} = \H H$.
\end{lem}
\begin{proof}
    See \cref{ssec:proof-ups-kernel-equiv}.
\end{proof}
\footnotetext{That is, $H|_W \in \mathbf{C}^2(W)$ implies $C^H_\text{ups}$ is \nameref{defi:lq} on $W$, and the converse implication holds when $W \subseteq \Delta^\circ(\Theta)$.}

\paragraph{Structural Facts.} Our next result shows that, for any $C \in \C$ and any $p_0 \in \Delta(\Theta)$ at which it is \nameref{defi:lq}, its kernel is the ``largest'' lower kernel and the ``smallest'' upper kernel with respect to the $\geq_\text{psd}$ order. 
%is: (a) the ``largest'' lower kernel and the ``smallest'' upper kernel with respect to the $\geq_\text{psd}$ order, and (b) uniquely defined, modulo the normalization described in \cref{remark:kernels}. 
Formally, we let $\underline{K}_C(p_0)\subseteq \R^{|\Theta|\times|\Theta|}$ and $\overline{K}_C(p_0)\subseteq \R^{|\Theta|\times|\Theta|}$ denote, respectively, the set of all lower kernels and the set of all upper kernels of $C$ at $p_0$. We call $\underline{k}(p_0) \in \underline{K}_C(p_0)$ a \emph{largest lower kernel} of $C$ at $p_0$ if $\underline{k}(p_0) \geq_\text{psd} \underline{k}'(p_0)$ for all $\underline{k}'(p_0) \in \underline{K}_C(p_0)$. Symmetrically, we call $\overline{k}(p_0) \in \overline{K}_C(p_0)$ a \emph{smallest upper kernel} of $C$ at $p_0$ if $\overline{k}(p_0) \leq_\text{psd} \overline{k}'(p_0)$ for all $\overline{k}'(p_0) \in \overline{K}_C(p_0)$. Under the normalization noted in \cref{remark:kernels}, the largest lower kernel and smallest upper kernel are unique whenever they exist, in which case we denote them by $\max \underline{K}_C(p_0)$ and $
\min \overline{K}_C(p_0)$, respectively.

\begin{lem}\label{lem:kernel-rank}
    For any $C \in \C$ and $p_0 \in \Delta(\Theta)$, 
    \[
    \text{$C$ is \nameref{defi:lq} at $p_0$ with kernel $k_C(p_0)$} \ \ \implies \ \  \text{$k_C(p_0) = \max \underline{K}_C(p_0) = \min \overline{K}_C(p_0)$.}
    \]
    %Consequently, there exists at most one kernel $k_C(p_0) \in \underline{K}_C(p_0) \cap \overline{K}_C(p_0)$
    %Consequently, $k_C(p_0)$ is unique modulo normalization, i.e., for any other kernel $k'_C(p_0)$ of $C$ at $p_0$, it holds that $(I - \mathbf{1} p_0^\top)k_C(p_0) (I-p_0 \mathbf{1}^\top) = (I - \mathbf{1} p_0^\top)k'_C(p_0) (I-p_0 \mathbf{1}^\top)$.
\end{lem}
\begin{proof}
    See \cref{ssec:proof-kernel-rank}.
\end{proof}

Our final result states that the kernels of \nameref{defi:sp} cost functions are ``strictly positive definite.'' Formally, for any $C \in \C$ and $p_0 \in \Delta(\Theta)$, we define $\underline{K}^+_C(p_0) := \{\underline{k}(p_0) \in \underline{K}_C(p_0) \mid \underline{k}(p_0) \gg_\text{psd} \mathbf{0}\}$. Following the above notation, we also denote by $\max \underline{K}^+_C(p_0)$ the (unique) $\geq_\text{psd}$-largest element of $\underline{K}^+_C(p_0)$, if such an element exists. We then have:

\begin{lem}\label{cor:ker-SP}
    For any \nameref{defi:sp} $C \in \C$ and $p_0 \in \Delta(\Theta)$, we have $\underline{K}^+_C(p_0) \neq \emptyset$. Moreover,
    \[
    \text{$C$ is \nameref{defi:lq} at $p_0$ with kernel $k_C(p_0)$} \ \ \implies \ \  \text{$k_C(p_0) = \max \underline{K}^+_C(p_0) \gg_\text{psd}\mathbf{0}$.}
    \] 
    %Consequently, the kernel $k_C$ satisfies $k_C(p_0)\gg_\text{psd} \mathbf{0}$ at every $p_0 \in \Delta(\Theta)$ where it exists.
\end{lem}
\begin{proof}
    See \cref{ssec:proof-ker-SP}.
\end{proof}

\subsection{Beyond Locally Quadratic Direct Costs in Theorem \ref{thm:flie}
%Relaxing the \nameref{defi:lq} Hypothesis in \cref{thm:flie}
}\label{proof:flie-nonsmooth}

In \cref{ssec:flie-characteriation}, \cref{fn:lq-thm4} claims that the restriction to \nameref{defi:lq} direct costs in \cref{thm:flie} is ``nearly'' without loss of generality. To formalize this, we present a technical extension of \cref{thm:flie} that: (a) ``nearly'' characterizes the $\Phi$ map for the co-domain of \nameref{defi:ups}/\nameref{defi:ll} indirect costs without \emph{any} restrictions on the domain of direct costs, and (b) shows that every direct cost generating a \nameref{defi:ups}/\nameref{defi:ll} indirect cost can be ``approximated'' arbitrarily well by \nameref{defi:lq} direct costs with the same indirect cost.

Following the notation from \cref{ssec:calc-kernel}, for every $C \in \C$ and $W \subseteq \Delta(\Theta)$, let $\underline{K}_C(W)$ (resp., $\overline{K}_C(W)$) denote the set of all lower (resp., upper) kernels of $C$ on $W$. We say that $k \in \underline{K}_C(W)$ is a \emph{largest lower kernel of $C$ on $W$} if $k(p)\geq_\text{psd} k'(p)$ for all $k' \in \underline{K}_C(W)$ and $p \in W$ (i.e., $k(p) = \max\underline{K}_C(p)$ for all $p \in W$). Under the normalization noted in \cref{remark:kernels}, the largest lower kernel on $W$ is unique whenever it exists, in which we denote it by $\max \underline{K}_C(W)$.  %Just as the kernel $k_C$ on $W$ is unique modulo the normalization $k_C(p) p = \mathbf{0}$ for all $p \in W$ whenever it exists (\cref{remark:kernels}), the largest lower kernel on $W$ is also unique modulo this normalization whenever it exists. Therefore, we identify all largest lower kernels on $W$ with their unique normalized version, which we denote by $\max \underline{K}_C(W)$. 
With this notation in hand, we have the following result:

\begin{cor}\label{cor:flie-nonsmooth}
For any $C \in \C$, open convex $W \subseteq \Delta^\circ(\Theta)$, and strongly convex $H \in \mathbf{C}^2(W)$,
\begin{align*}
    \hspace{-1em}
    \text{$C\succeq C^H_\text{ups}$ and $\H H \in \overline{K}_C(W)$} \ \ \implies \  \ \Phi(C) = C^H_\text{ups} \ \ \implies \ \ \text{$C\succeq C^H_\text{ups}$ and $\max \underline{K}_C(W) = \H H$.}
\end{align*}
Furthermore, if $\Phi(C) = C^H_\text{ups}$, then the following holds: 
\begin{quote}
    For every open cover $\mathbb{O}$ of $W$, there exists a direct cost $\widehat{C} \in \Phi^{-1}(C^H_\text{ups})$ such that (i) $\widehat{C}$ is \nameref{defi:lq} on $W$ with kernel $k_{\widehat{C}} = \H H$, (ii) $C \succeq \widehat{C}$, and (iii) $C(\pi) \neq \widehat{C}(\pi)$ only if $\supp(\pi) \subseteq O$ for some $O \in \mathbb{O}$.
\end{quote}
\end{cor}
\begin{proof}
    See \cref{app:proof-cor-flie-nonsmooth}.
\end{proof}

The first implication in \cref{cor:flie-nonsmooth} extends the sufficiency (``$\implies$'') direction of \cref{thm:flie}, while the second implication extends the necessity (``$\impliedby$'') direction of \cref{thm:flie}. There are two technical differences from \cref{thm:flie}. First, when the direct cost $C$ is \emph{not} \nameref{defi:lq}, there is a ``gap'' between the set of upper kernels $\overline{K}_C(W)$ and the largest lower kernel $\max \underline{K}_C(W)$.\footnote{When $C$ is \nameref{defi:lq} on $W$, $k_C$ is both the smallest upper kernel and the largest lower kernel (\cref{lem:kernel-rank}).} Second, since this gap precludes a tight characterization of the $\Phi_\text{IE}$ map (cf. \cref{lem:phi-ie}(iii)), \cref{cor:flie-nonsmooth} replaces the \nameref{axiom:flie} inequality ``$C\succeq \Phi_\text{IE}(C)$'' with the alternative inequality ``$C \succeq C^H_\text{ups}$,'' which is weakly more (resp., less) restrictive when $\H H$ is an upper (resp., lower) kernel of $C$ (per points (i) and (ii) of \cref{lem:phi-ie}).  

With these technical caveats, \cref{cor:flie-nonsmooth} shows that the main lessons of \cref{thm:flie} are robust. Methodologically, it can be used to calculate $\Phi(C)$ and an ``outer bound'' for $\Phi^{-1}(C^H_\text{ups})$, extending the procedure from \cref{fig:flow}. Economically, it extends the lesson that \nameref{defi:ups}/\nameref{defi:ll} indirect costs can only be generated by direct costs for which ``incremental learning'' is optimal. First, since every $C$ for which $\Phi(C) 
= C^H_\text{ups}$ satisfies $\max \underline{K}_C(W) = \H H$, such $C$ cannot have ``fixed costs'' or ``kinks'' that would make non-incremental learning strictly optimal, as such features require the set of lower kernels $\underline{K}_C(W)$ to be ``unbounded above.''\footnote{If there is a fixed cost $\underline{c}>0$ such that $C(\pi) \geq \underline{c}$ for all $\pi \in \Ex\backslash \Ex^\varnothing$, then \emph{every} $k : W \to \R^{|\Theta| \times|\Theta|}$ (normalized as in \cref{remark:kernels}) is a lower kernel of $C$ on $W$. If $C$ is \hyperref[eqn:PS]{Posterior Separable} and its divergence $D(\cdot\mid p)$ is not differentiable at $q=p$ for any $p \in W$ (e.g., the \nameref{defi:MLR} cost), then a similar result holds because $D(q \mid p)$ and $\|q-p\|$ are of the same order.} Second, the final part of \cref{cor:flie-nonsmooth} implies (via \cref{thm:flie} and \cref{lem:phi-ie}) that every $C$ for which $\Phi(C) = C^H_\text{ups}$ can be approximated by \nameref{defi:lq} $\widehat{C}$ such that (a) $\widehat{C}$ \nameref{axiom:flie} and $\Phi_\text{IE}(\widehat{C}) = \Phi(\widehat{C}) = C^H_\text{ups}$ and (b) $\widehat{C}(\pi) = C(\pi)$ for all ``non-incremental'' $\pi \in \Ex$, suggesting that all such $C$ ``approximately \hyperref[axiom:flie]{FLIE}.''

\subsection{Supplementary Results for Theorem \ref{thm:trilemma}}\label{ssec:thm5-corollaries}

In this section, we present two auxiliary results discussed in \cref{ssec:trilemma}. The first result formalizes the observation that \hyperref[axiom:CMC:0]{CMC} is typically \emph{not} preserved under optimization: 

\begin{cor}\label{cor:preserve:CMC}
For any \nameref{defi:sp} and \nameref{defi:lq} $C\in C$ with rich domain, 
\[
\text{$C$ is \nameref{axiom:CMC} and \hyperref[axiom:DL]{Dilution Linear}} \ \ \implies \ \ \text{$\Phi(C)$ is \nameref{axiom:CMC} iff $C = \Phi(C)$ is a \nameref{defi:TI} cost.}
\]
\end{cor}
\begin{proof}
    See \cref{ssec:thm5-cor-proofs-1}.
\end{proof}

The technical assumptions in \cref{cor:preserve:CMC} are mild. %By \cref{lem:bayes-LLR}, a cost function with rich domain is \nameref{axiom:CMC} and \hyperref[axiom:DL]{Dilution Linear} if and only if it satisfies \eqref{F-beta}--\eqref{eqn:LLR-KLform}.
By \cref{lem:bayes-LLR} in \cref{sssec:proof:thm5}, a rich-domain cost function is \nameref{axiom:CMC} and \hyperref[axiom:DL]{Dilution Linear} if and only if it has the same form as an \ref{eqn:LLR} cost, except with potentially prior-dependent coefficients $\beta_{\theta,\theta'}(p)$. 
Hence, such a cost function is \nameref{defi:sp} and \nameref{defi:lq} if, for every $\theta \neq \theta'$, the map $p \mapsto \beta_{\theta,\theta'}(p)/p(\theta)$ is bounded away from zero and continuous on $\Delta^\circ(\Theta)$.\footnote{If these maps are bounded away from zero, then %it is straightforward to verify that 
there exists $m>0$ such that the divergence $D_{\bm{\beta}}$ defined in \cref{lem:bayes-LLR} satisfies $D_{\bm{\beta}}(q \mid p)\geq m \cdot \|q-p\|^2$ for all $q,p \in \Delta^\circ(\Theta)$, which implies that $C$ is \nameref{defi:sp}. If these maps are continuous, then $D_{\bm{\beta}}$ is ``locally $\mathbf{C}^2$'' on $\Delta^\circ(\Theta)$ (as defined in \cref{ssec:calc-kernel}) and hence \cref{lem:ps-kernel-suff} implies that $C$ is \nameref{defi:lq}.} This holds under any \nameref{defi:TI} (resp., \ref{eqn:LLR}) cost with $\gamma_{\theta,\theta'}>0$ (resp., $\beta_{\theta,\theta'}>0$) for all $\theta \neq \theta'$.

The second result supports the observation that---aside from the \nameref{defi:MLR} cost---no known (full- or rich-domain)  \hyperref[axiom:prior:invariant]{Prior Invariant} cost functions are \nameref{axiom:slp}. Essentially all full domain, \hyperref[axiom:prior:invariant]{Prior Invariant} costs in the literature are either: (a) \nameref{defi:ll}, or (b) \emph{strictly} \hyperref[axiom:mono]{Monotone}, i.e., satisfy $C(\pi) > C(\pi')$ whenever $\pi >_\text{mps} \pi'$. As discussed in \cref{ssec:trilemma}, \cref{thm:trilemma}(iii) precludes case (a). The next result, which is a corollary of \cref{prop:1}, precludes case (b).

Let $\mathcal{P}$ denote the set of all partitions of $\Theta$, i.e., the set of all $P = \{E_1, \dots, E_k\} \subseteq 2^\Theta \backslash\{\emptyset\}$ such that $E_i \cap E_j = \emptyset$ for all $i\neq j$ and $\cup_{i=1}^k E_i = \Theta$. We call $P_\varnothing := \{\Theta\} \in \mathcal{P}$ the \emph{trivial} partition, and we call $P \in \mathcal{P}$ \emph{nontrivial} if $P \neq P_\varnothing$. For every $P \in \mathcal{P}$, we define the experiment $\sigma^P = \left(P, (\sigma^P_\theta)_{\theta\in \Theta}\right) \in \Se$ as $\sigma_\theta(E_i) := \mathbf{1}(\theta \in E_i)$ for all $\theta \in \Theta$ and $E_i \in P$. In words, $\sigma^P$ reveals (only) which cell of $P$ contains the true state. We have the following necessary condition: 

\begin{cor}\label{cor:PI-part}
    For any \nameref{axiom:slp} and \hyperref[axiom:prior:invariant]{Prior Invariant} $C \in \C$ with full domain,
    \[
    C(h_B(\sigma^P,p)) = C(h_B(\sigma^{P'},p)) \quad \text{for all $P,P' \in \mathcal{P}\backslash\{P_\varnothing\}$ and $p \in \Delta^\circ (\Theta)$.}
    \]
\end{cor}
\begin{proof}
    See \cref{ssec:thm5-cor-proofs-2}.
\end{proof}

Plainly, this condition precludes \emph{strictly} \hyperref[axiom:mono]{Monotone} cost functions when $|\Theta|\geq 3$. However, the \nameref{defi:MLR} cost satisfies it because $D_\text{MLR}(q \mid p) = 1$ for all $q \in \Delta(\Theta) \backslash \Delta^\circ(\Theta)$ and $p \in \Delta^\circ(\Theta)$.

\subsection{Experiment-Based Framework}\label{app:beyond:belief}

%\subsection{The Experiment-Based Framework \awb{[NEW VERSION]}}\label{ssec:experiment}

%\cref{ssec:expermient-model} presents the framework. Our main results are in \cref{ssec:experiment-relate,ssec:e-slp}.

In this section, we formally develop the experiment-based framework described in \cref{ssec:beyond:belief}. \cref{ssec:expermient-model} introduces the model. \cref{ssec:experiment-relate} analyzes the relationship between the experiment- and belief-based frameworks. \cref{ssec:e-slp} presents the experiment-based analog of \cref{prop:1}.

\subsubsection{Model}\label{ssec:expermient-model}

The model closely mirrors the belief-based model from \cref{section:model}. We therefore proceed succinctly, with a focus on developing the requisite formal definitions and notation. 

\paragraph{Preliminaries.} Following \textcite{blackwell-experiment51}, we say that $\sigma' \in \Se$ \emph{Blackwell dominates} $\sigma \in \Se$ if $h_B(\sigma',p) \geq_\text{mps} h_B(\sigma,p)$ for \emph{every} $p \in \Delta(\Theta)$, and we call $\sigma, \sigma' \in \Se$ \emph{Blackwell equivalent} if they Blackwell dominate each other.  Let $\geq_\text{B}$ denote the \emph{Blackwell order} on $\Se$, whereby $\sigma' \geq_\text{B}\sigma$ denotes that $\sigma'$ Blackwell dominates  $\sigma$ and $\sigma' \sim_B \sigma$ denotes Blackwell equivalence. We call $\sigma \in \Se$  \emph{uninformative} if $\sigma' \geq_\text{B} \sigma$ for all $\sigma'\in \Se$; equivalently, if $h_B(\sigma,p) \in \Ex^\varnothing$ for all $p \in \Delta(\Theta)$. We denote by $\Se^\varnothing\subsetneq \Se$ the subclass of all uninformative experiments. 

\paragraph{Cost Functions.} An \emph{experiment-based cost function} is a map $\Ec : \Se \times \Delta(\Theta) \to \overline{\R}_+$ such that, for every prior belief $p \in \Delta(\Theta)$: (i) $\Ec(\sigma,p) =  0$ for all $\sigma \in \Se^\varnothing$, and (ii) $\Ec(\sigma,p) = \Ec(\sigma',p)$ for all $\sigma,\sigma'\in \Se$ such that $\sigma \sim_B \sigma'$. At full-support priors $p \in \Delta^\circ(\Theta)$, these conditions imply that the cost of an  experiment $\sigma\in \Se$ depends only on its induced random posterior $h_B(\sigma, p) \in \Ex$, and that $\sigma \in \Se$ has zero cost if it induces a trivial random posterior $h_B(\sigma,p) \in \Ex^\varnothing$.\footnote{Per \textcite{blackwell-experiment51}, we have $\sigma' \geq_\text{B} \sigma$ if and only $h_B(\sigma',p) \geq_\text{mps} h_B(\sigma,p)$ for \emph{some (full-support)} $p \in \Delta^\circ(\Theta)$.\label{fn:BW-equiv}} However, these implications \emph{need not} hold at partial-support priors $p \notin \Delta^\circ(\Theta)$. 

Let $\EC$ denote the space of all experiment-based cost functions. We endow $\EC$ with the pointwise order $\succeq_{\Se}$, whereby $\Ec \succeq_{\Se} \Ec'$ denotes  $\Ec(\sigma,p)\geq \Ec'(\sigma,p)$ for all $\sigma\in \Se , \, p \in \Delta(\Theta)$.

\paragraph{Sequential Learning.} A \emph{two-step sequential experiment} $\Sigma = \left(S_1 \times S_2, (\Sigma_\theta)_{\theta \in \Theta} \right)$ is an experiment for which the signal space is a product set $S_1 \times S_2$, where $S_1$ is a Polish space of ``first round'' signal realizations and $S_2$ is a Polish space of ``second round'' signal realizations. Equivalently, any such sequential experiment can be represented as a pair $\Sigma = (\sigma_1,\bm{\sigma}_2)$, where $\sigma_1 = \left(S_1, (\sigma_{1,\theta})_{\theta \in \Theta}\right)$ is the \emph{first-round experiment} defined as $\sigma_{1,\theta} (\widehat{S}_1) := \Sigma_\theta( \widehat{S}_1 \times S_2)$ for all Borel $\widehat{S}_1 \subseteq S_1$ and $\bm{\sigma}_2 : S_1 \to \Delta(S_2)^\Theta$ is a Borel measurable map from first-round signals $s_1 \in S_1$ to \emph{contingent second-round experiments} $\sigma_{2}^{s_1} \in \Delta(S_2)^\Theta$ defined as $\sigma_{2,\theta}^{s_1} (\widehat{S}_2):= \Sigma_\theta(\widehat{S}_2 \mid s_1)$ for all Borel $\widehat{S}_2 \subseteq S_2$.\footnote{We denote by $\Delta(S_2)^\Theta \subset \Se$ the subset of experiments defined on the common signal space $S_2$, and we denote by $\Sigma_\theta (\cdot \mid s_1) \in \Delta(S_2)$ an appropriate regular conditional probability of $\Sigma_\theta \in \Delta(S_1\times S_2)$ given $s_1 \in S_1$. These two formulations of sequential experiments are equivalent by standard distintegration arguments.} For technical convenience, we restrict attention to $\Sigma$ such that
\begin{equation}\label{eqn:finite-ND-support-exp}
\left| \left\{s_1 \in \cup_{\theta \in \Theta} \supp(\sigma_{1,\theta})  \mid \sigma^{s_1}_2 \in \Se \backslash\Se^\varnothing \right\}\right| < +\infty, 
\end{equation}
which is analogous to the ``finite non-degenerate support'' restriction that we impose on two-step strategies in the belief-based framework (see \cref{ssec:indirect:cost}). 

Let $\Se^2$ denote the class of all two-step sequential experiments that satisfy restriction \eqref{eqn:finite-ND-support-exp}. We extend the Blackwell order $\geq_\text{B}$ and the Bayesian map $h_B$ to $\Se \cup \Se^2$ in the natural way, by viewing each $\Sigma \in \Se^2$ as a ``one-shot'' experiment in $\Se$ with a product signal space.

%We model sequential optimization as follows. 
Given any ``target'' experiment $\sigma \in \Se$ and prior $p \in \Delta(\Theta)$, the DM constructs a sequential experiment of arbitrary length to ``produce'' $\sigma$ at minimal expected cost, using two-step sequential experiments as the building blocks. Formally, for any $\Sigma \in \Se^2$ and $p \in \Delta(\Theta)$, let $\langle \Sigma,p\rangle := \sum_{\theta \in \Theta} p(\theta) \sigma_{1,\theta} \in \Delta(S_1)$ denote the  marginal distribution over first-round signals, and let $q_{s_1}^{\sigma_1,p}\in\Delta(\Theta)$ denote the posterior belief conditional on observing (only) the first-round signal $s_1 \in S_1$. Our main definition is then as follows:

\begin{definition}\label{defi:E-main}
    The \dred{experiment-based two-step learning map} $\Psi_{\Se} : \EC \to \EC$ is defined as 
    \[
\Psi_{\Se} (\Ec) (\sigma, p) := \inf_{\Sigma \in \Se^2} \, \Ec(\sigma_1,p) + \E_{\langle \Sigma, p \rangle} \left[ \Ec\left(\sigma_{2 }^{s_1}, q_{s_1}^{\sigma_1,p} \right)\right] \quad \text{such that} \quad \Sigma\geq_\text{B} \sigma, 
\]
and the \dred{experiment-based sequential learning map} $\Phi_{\Se} : \EC \to \EC$ is defined as 
\[
\Phi_{\Se} (\Ec) := \lim_{n \to \infty} \Psi_{\Se}^n(\Ec).\footnotemark
\]
We call $\Ec \in \EC$ an \dred{indirect cost} if $\Ec \in \EC^* := \Phi_{\Se}[\EC]$, and we say that $\Ec$ is \dred{$\Se$-SLP} if $\Ec = \Psi_{\Se}(\Ec)$.\footnotetext{It is easy to verify that $\Psi_{\Se}$ is well-defined. It follows that $\Phi_{\Se}$ is well-defined, as $\Ec \succeq_{\Se} \Psi_{\Se}(\Ec)$ for all $\Ec \in \EC$.}
\end{definition}

Each object in \cref{defi:E-main} mirrors its belief-based counterpart from \cref{ssec:model-sequential}. We emphasize that the definition of $\Psi_{\Se}$ features a Blackwell dominance constraint, which is more restrictive than the MPS constraint in the definition of $\Psi$ at partial-support priors (but is equivalent to the MPS constraint at full-support priors, per \cref{fn:BW-equiv}).

\subsubsection{Relating the Belief- and Experiment-Based Frameworks}\label{ssec:experiment-relate}

%\paragraph{Relating the Frameworks.}

We connect the belief- and experiment-based frameworks using two morphisms between $\C$ and $\EC$, the respective spaces of cost functions.
%To connect these frameworks, we define two morphisms between $\EC$ and $\C$. 
We project $\EC$ onto $\C$ via the surjective map $\Lambda : \EC \to \C$ defined as $\Lambda(\Ec)(\pi) := \inf\left\{ \Ec(\sigma, p_\pi) \mid \sigma \in \Se \text{ s.t. } h_B(\sigma, p_\pi) = \pi\right\}$. In words, $\Lambda(\Ec) \in \C$ represents the cheapest way of generating random posteriors using $\Ec \in \EC$. Inversely, we embed $\C$ into $\EC$ via the injective map $\Upsilon : \C \to \EC$ defined as $\Upsilon(C)(\sigma,p) := C(h_B(\sigma,p))$. In words, $\Upsilon(C) \in \EC$ is the %experiment-based cost generated by composing 
composition of $C \in \C$ with the Bayesian map $h_B$.  

These morphisms are pseudo-inverses. Plainly, $\Lambda \circ \Upsilon : \C \to \C$ is the identity map. Meanwhile, $\Upsilon \circ \Lambda : \EC \to \EC$ need only coincide with the identity map \emph{at full-support priors}, where the Blackwell and MPS orders coincide; it may lie below the identity map at partial-support priors, where the Blackwell order is more restrictive. Formally, for every $\Ec \in \EC$, we have: (a) $\Ec \succeq_{\Se} [\Upsilon\circ\Lambda](\Ec)$ and (b) $\Ec(\cdot, p) \equiv [\Upsilon\circ\Lambda](\Ec)(\cdot, p)$ for all $p \in \Delta^\circ(\Theta)$.

Our first result uses these morphisms to characterize the relationship between the belief- and experiment-based sequential learning maps, $\Phi$ and $\Phi_{\Se}$. In particular, we characterize how these two maps commute with $\Lambda$ and $\Upsilon$ (see \cref{fig:commute} for an illustration).

\begin{prop}\label{prop:commute} 
The maps $\Phi$ and $\Phi_{\Se}$ satisfy the following commutative properties:\footnote{The same commutative properties hold when $\Phi$ and $\Phi_{\Se}$ are replaced with $\Psi$ and  $\Psi_{\Se}$, respectively.}
\begin{itemize}[noitemsep]
    \item[(i)] $\Upsilon\circ\Phi=\Phi_{\Se}\circ \Upsilon$.
    %%%
    \item[(ii)] $\Phi = \Lambda\circ \Phi_{\Se}\circ \Upsilon $.
    %%%
    \item[(iii)] $\Phi\circ \Lambda=\Lambda\circ \Phi_{\Se}$.
    %%%
    \item[(iv)] For every $\Ec \in \EC$ and $p \in \Delta^\circ(\Theta)$, it holds that $\Phi_{\Se}(\Ec)(\cdot,p) \equiv [\Upsilon\circ \Phi\circ \Lambda] (\Ec) (\cdot,p) $.
\end{itemize}
\end{prop}
\begin{proof}
    See \cref{ssec:proof-commute}.
\end{proof}

\begin{figure}[t]
        \centering
    \tikzset{every picture/.style={line width=0.75pt}} %set default line width to 0.75pt        

\begin{tikzpicture}[x=0.75pt,y=0.75pt,yscale=-1,xscale=1]

%Straight Lines [id:da9145232748510743] 
\draw    (222.6,104.82) -- (353,105) ;
\draw [shift={(356,105)}, rotate = 180.08] [fill={rgb, 255:red, 0; green, 0; blue, 0 }  ][line width=0.08]  [draw opacity=0] (5.36,-2.57) -- (0,0) -- (5.36,2.57) -- cycle    ;
%Straight Lines [id:da9050814666386089] 
\draw  [dash pattern={on 2.25pt off 0.75pt}]  (222.6,146) -- (353,146) ;
\draw [shift={(356,146)}, rotate = 180] [fill={rgb, 255:red, 0; green, 0; blue, 0 }  ][line width=0.08]  [draw opacity=0] (5.36,-2.57) -- (0,0) -- (5.36,2.57) -- cycle    ;
%Straight Lines [id:da2776988858031568] 
\draw    (207.27,138.14) -- (207.27,118.1) ;
\draw [shift={(207.27,115.1)}, rotate = 90] [fill={rgb, 255:red, 0; green, 0; blue, 0 }  ][line width=0.08]  [draw opacity=0] (5.36,-2.57) -- (0,0) -- (5.36,2.57) -- cycle    ;
%Straight Lines [id:da29528706386384784] 
\draw    (367.34,137.04) -- (367.34,117) ;
\draw [shift={(367.34,114)}, rotate = 90] [fill={rgb, 255:red, 0; green, 0; blue, 0 }  ][line width=0.08]  [draw opacity=0] (5.36,-2.57) -- (0,0) -- (5.36,2.57) -- cycle    ;
%Straight Lines [id:da21163906476298155] 
\draw    (373.23,114) -- (373.23,134.04) ;
\draw [shift={(373.23,137.04)}, rotate = 270] [fill={rgb, 255:red, 0; green, 0; blue, 0 }  ][line width=0.08]  [draw opacity=0] (5.36,-2.57) -- (0,0) -- (5.36,2.57) -- cycle    ;
%Straight Lines [id:da5416858019818891] 
\draw    (213.17,115.1) -- (213.17,135.14) ;
\draw [shift={(213.17,138.14)}, rotate = 270] [fill={rgb, 255:red, 0; green, 0; blue, 0 }  ][line width=0.08]  [draw opacity=0] (5.36,-2.57) -- (0,0) -- (5.36,2.57) -- cycle    ;

% Text Node
\draw (203.81,97.22) node [anchor=north west][inner sep=0.75pt]    {$C$};
% Text Node
\draw (364.41,93.75) node [anchor=north west][inner sep=0.75pt]    {$C^{*}$};
% Text Node
\draw (203.72,138.4) node [anchor=north west][inner sep=0.75pt]    {$\Ec$};
% Text Node
\draw (364.41,136.99) node [anchor=north west][inner sep=0.75pt]    {$\Ec^{*}$};
% Text Node
\draw (218.78,117.81) node [anchor=north west][inner sep=0.75pt]    {$\Upsilon$};
% Text Node
\draw (377.66,116.71) node [anchor=north west][inner sep=0.75pt]    {$\Upsilon$};
% Text Node
\draw (349.9,116.71) node [anchor=north west][inner sep=0.75pt]    {$\Lambda $};
% Text Node
\draw (190.45,117.5) node [anchor=north west][inner sep=0.75pt]    {$\Lambda $};
% Text Node
\draw (282,82.4) node [anchor=north west][inner sep=0.75pt]    {$\Phi $};
% Text Node
\draw (280,151.4) node [anchor=north west][inner sep=0.75pt]    {$\Phi _{\mathcal{E}}$};
% Text Node
\draw (242,130.4) node [anchor=north west][inner sep=0.75pt]  [font=\footnotesize]  {$\longrightarrow \ \text{if} \ p\in \Delta^{\circ } ( \Theta )$};

\end{tikzpicture}
    \caption{Commutative properties of the $\Phi$ and $\Phi_{\Se}$ maps (\cref{prop:commute}).}
    \label{fig:commute}
\end{figure}

\cref{prop:commute} delivers three lessons. First, \cref{prop:commute}(ii), which follows from point (i), shows that %the belief-based sequential learning map 
$\Phi$ is fully determined by 
%the experiment-based sequential learning map 
$\Phi_{\Se}$. Second, \cref{prop:commute}(iv), which follows from point (iii), shows that $\Phi_{\Se}$ is fully determined by $\Phi$ at \emph{full-support} prior beliefs $p \in \Delta^\circ(\Theta)$. In this sense, the belief- and experiment-based frameworks are \emph{equivalent} at such full-support priors. We emphasize that this equivalence only relies on the \emph{initial} prior belief having full support; that is, it permits the ``priors'' in later rounds of a sequential procedure, which are endogenously determined, to have partial support.

Third, \cref{prop:commute}(iii) further implies that, for many applications, we do \emph{not} need to separately analyze the $\Phi_{\Se}$ map, \emph{even if}: (a) the DM's primitive technology is modeled as an experiment-based direct cost $\Ec \in \EC$ and (b) the prior has partial support. To illustrate, suppose the DM acquires information to solve a canonical single-agent decision problem. %, in which case the value of information depends only on the induced random posterior. 
In such settings, since the value of information depends only on the induced random posterior, the DM's information acquisition incentives are determined by $[\Lambda \circ \Phi_{\Se}](\Ec)$. \cref{prop:commute}(iii) shows that, to characterize this object, it suffices to first compute the belief-based direct cost $\Lambda(\Ec)$ and then use our main analysis to calculate $[\Phi \circ \Lambda](\Ec)$.

\begin{remark}[Full Prior Invariance]\label{remark:full-PI}
    As noted in \cref{ssec:PI,ssec:beyond:belief}, our main belief-based definition of \hyperref[axiom:prior:invariant]{Prior Invariance} allows the cost of experiments to vary (only) with the support of the prior belief. This is an artifact of the belief-based approach: if $C \in \C$ is completely independent of prior beliefs, then it must be identically zero. Importantly, the experiment-based framework is not subject to this limitation. Formally, we say that $\Ec \in \EC$ is \hyperref[eqn:full-PI]{Fully Prior Invariant} if  
    \begin{equation}\label{eqn:full-PI}
    \Ec(\sigma, p) = \Ec(\sigma, p') \quad  \text{for all } \ \ \sigma \in \Se \, \text{ and } \, p, p' \in \Delta(\Theta). \tag{$\Se$-PI}
    \end{equation}
    We then say that $\Ec \in \EC^*$ is \dred{$\Se$-SPI} if $\Ec = \Phi_{\Se}(\Ec')$ for some \hyperref[eqn:full-PI]{Fully Prior Invariant} $\Ec' \in \EC$. As an example, the experiment-based \hyperref[E-MLR]{MLR} cost is both \hyperref[eqn:full-PI]{Fully Prior Invariant} and \dred{$\Se$-SPI}. 

    A key implication of \cref{prop:commute} is that all of our results for \hyperref[axiom:prior:invariant]{Prior Invariant} and \nameref{defi:spi} (belief-based) cost functions---viz., \cref{thm:trilemma,thm:wald} and \cref{cor:PI-part}---apply essentially verbatim to \hyperref[eqn:full-PI]{Fully Prior Invariant} and \dred{$\Se$-SPI} (experiment-based) cost functions. This implication follows from the above discussion and two simple observations: (i) these results restrict attention to full-support priors, and (ii) $\Ec \in \EC$ is \hyperref[eqn:full-PI]{Fully Prior Invariant} only if $\Lambda(\Ec) \in \C$ is \hyperref[axiom:prior:invariant]{Prior Invariant}.\footnote{A minor subtlety is that \cref{thm:trilemma,thm:wald} concern rich-domain belief-based cost functions, while it is easy to see that $\Lambda(\Ec) \in \C$ generally does not have rich domain when $\Ec \in \EC$ is \hyperref[eqn:full-PI]{Fully Prior Invariant}. Nevertheless, these results can be applied by viewing the rich-domain costs therein as restrictions of belief-based costs with larger domains, as described in \cref{remark:full-support}. See \dred{Theorems} \ref{thm:trilemma3:gen} and \ref{thm:wald:gen} and \cref{remark:wald:gen} in \cref{app:beyond:flexible} for versions of \cref{thm:trilemma,thm:wald} that (among other generalizations) explicitly distinguish between the domain of a cost function and its rich-domain restriction.}
\end{remark}

\subsubsection{Foundations for $\Se$-SLP Costs}\label{ssec:e-slp}

%\paragraph{Foundations for $\Se$-SLP Costs.} 
For other applications (e.g., to costly monitoring), it is important to directly consider experiment-based indirect costs at partial-support priors. Our second result shows that \hyperref[defi:E-main]{$\Se$-SLP} characterizes the reduced-form implications of such indirect costs, delivering an experiment-based analog of \cref{prop:1} and \cref{cor:envelope}. We thereby provide a foundation for using \hyperref[defi:E-main]{$\Se$-SLP} cost functions %---such as the experiment-based \hyperref[E-TI]{Total Information} and \hyperref[E-MLR]{MLR} costs from \cref{ssec:beyond:belief}---
in these applications.

We first define experiment-based analogues of the belief-based  \hyperref[axiom:mono]{Monotonicity} and \hyperref[axiom:POSL]{Subadditivity} conditions from \cref{ssec:indirect:cost}. Formally, an experiment-based cost $\Ec \in \EC$ is:
\begin{itemize}[noitemsep]
\item \label{E-mono} \dred{\emph{$\Se$-Monotone}} if, for every $p \in \Delta(\Theta)$, $\Ec(\sigma,p) \leq \Ec(\sigma',p)$ for all $\sigma, \sigma' \in \Se$ such that $\sigma \leq_\text{B} \sigma'$.
%%%
\item \label{E-POSL} \dred{\emph{$\Se$-Subadditive}} if, for every $p \in \Delta(\Theta)$, 
\[
\Ec(\Sigma, p) \leq \Ec(\sigma_1,p) + \E_{\langle \Sigma, p\rangle}\left[ \Ec(\sigma_2^{s_1}, p^{\sigma_1,p}_{s_1}) \right] \quad \text{for all } \ \ \Sigma \in \Se^2.\footnote{When writing $\Ec(\Sigma,p)$, we view $\Sigma \in \Se^2$ as a ``one-shot'' experiment in $\Se$ with the product signal space $S_1\times S_2$.}
\]
\end{itemize}

We then have the following characterization result:

\begin{customthm}{1--$\Se$}\label{thm1-Se}
    For all $\Ec \in \EC$, 
    \[
    \Ec \in \EC^* \ \ \iff \ \  \Ec \text{ is \hyperref[defi:E-main]{$\Se$-SLP}} \ \ \iff \ \ \Ec \text{ is \hyperref[E-mono]{$\Se$-Monotone} and \hyperref[E-POSL]{$\Se$-Subadditive}.}
    \]
    Moreover, the indirect cost  $\Phi_{\Se}(\Ec) = \max\{\Ec' \in \EC \mid  \Ec' \preceq_{\Se} \Ec \text{ and } \Ec' \text{ is $\Se$-\nameref{axiom:slp}}  \}$.
\end{customthm}
\begin{proof}
    The argument is identical to those from the proofs of \cref{prop:1} and \cref{cor:envelope}, modulo the obvious (minor) notational adjustments. We omit the details for brevity. 
\end{proof}

\begin{remark}\label{remark:e-slp-check}
    It can be verified that the experiment-based \hyperref[E-TI]{Total Information} and \hyperref[E-MLR]{MLR} costs are \hyperref[E-mono]{$\Se$-Monotone} and \hyperref[E-POSL]{$\Se$-Subadditive}. 
    %\footnote{This can be verified either directly, or indirectly via properties of these costs' belief-based analogues. To see the latter, let $C_{TI}\in \C$ be the partial-support version of \nameref{defi:TI} described in \cref{fn:TI-partial-support}. For every bounded experiment $\sigma \in \Se_b$: (i)  $\Ec_\text{TI}(\sigma,\cdot) : \Delta(\Theta) \to \R$ is the continuous extension of $L(C_\text{TI})(\sigma,\cdot)|_{\Delta^\circ(\Theta)}$ from $\Delta^\circ(\Theta)$ to $\Delta(\Theta)$, and (ii)  $\Ec_\text{TI}(\sigma, p) \geq L(C_\text{TI})(\sigma,p)$ for all $p \in \Delta(\Theta) \backslash\Delta^\circ(\Theta)$. These properties and the fact that $C_\text{TI}$ is \hyperref[axiom:mono]{Monotone} and \hyperref[axiom:POSL]{Subadditive} together imply that $\Ec_\text{TI}$ is \hyperref[E-mono]{$\Se$-Monotone} and \hyperref[E-POSL]{$\Se$-Subadditive}. Similar arguments apply to the experiment-based \hyperref[E-MLR]{MLR} cost.} 
    Hence, \cref{thm1-Se} implies that these costs are \hyperref[defi:E-main]{$\Se$-SLP}. In fact, experiment-based \hyperref[E-TI]{Total Information}, $\Ec_\text{TI}$, satisfies the additional property of \dred{$\Se$-Additivity}: %for every $p \in \Delta(\Theta)$, 
    \[
\Ec_\text{TI}(\Sigma, p) = \Ec_\text{TI}(\sigma_1,p) + \E_{\langle \Sigma, p\rangle}\left[ \Ec_\text{TI}(\sigma_2^{s_1}, p^{\sigma_1,p}_{s_1}) \right] %\quad \text{for all } \ \ \Sigma \in \Se^2 \text{ and } p \in \Delta(\Theta).
\]
for all sequential experiments $\Sigma$ and priors $p \in \Delta(\Theta)$ such that $(\Sigma,p) \in \dom(\Gamma_\text{TI})$.\footnote{Analogous to belief-based \hyperref[axiom:additive]{Additivity}, \dred{$\Se$-Additivity} allows for (feasible) sequential experiments that violate \eqref{eqn:finite-ND-support-exp}.} This property---which is an experiment-based analog of the belief-based \hyperref[axiom:additive]{Additivity} property from \cref{prop:ups:additive}---holds because (i) the KL divergences $\sigma \mapsto D_\text{KL}(\sigma_\theta \mid\sigma_{\theta'})$ are additive with respect to conditionally independent experiments and (ii) $\Ec_\text{TI}$ is linear with respect to the prior.
\end{remark}

\subsection{Generalized Learning Map Framework}\label{app:beyond:flexible}

In this section, we formally develop the generalized learning map (\hyperref[defi:gen:IC]{GLM}) framework described in \cref{ssec:beyond:flexibility}. \cref{ssec:GLM-properties} presents properties of \nameref{defi:gen:IC}s under which our main results extend. \cref{ssec:GLM-results} then presents the  extensions of our main results.

\subsubsection{Properties of Generalized Learning Maps}\label{ssec:GLM-properties}

Recall from \cref{defi:gen:IC} that a \emph{generalized learning map} (\hyperref[defi:gen:IC]{GLM}) is any map $\widehat{\Phi} : \C \to \C$ that is isotone, i.e., such that $C\succeq C'$ implies $\widehat{\Phi}(C) \succeq \widehat{\Phi}(C')$. The following definition presents additional properties of \nameref{defi:gen:IC}s---which any given \nameref{defi:gen:IC} may or may not satisfy---under which our main results can be extended (see   \cref{tab:GSLM:both} in \cref{ssec:beyond:flexibility} for a summary).

\begin{definition}\label{defi:GSLM:properties}
A \nameref{defi:gen:IC} $\widehat{\Phi}: \C \to \C$ satisfies:
    \begin{itemize}[noitemsep,topsep=0pt]
    \item[(i)] \label{ADL} \dred{Allows Direct Learning (ADL)} if $\widehat{\Phi}(C) \preceq C$ for all $C \in \C$.
    \item[(ii)] \label{AIE} \dred{Allows Incremental Evidence (AIE)} 
    %if $\widehat{\Phi}(C) \preceq \Phi_\text{IE}(C)$ for all $C \in \C$.
    if, for all $C \in \C$, open convex $W \subseteq \Delta^\circ(\Theta)$, and $H \in \mathbf{C}^2(W)$ such that $\H H$ is an upper kernel of $C$ on $W$, it holds that $\widehat{\Phi}(C)(\pi) \leq C^H_\text{ups}(\pi)$ for all $\pi \in \Delta(W)$.
    %%%
    %%
    \item[(iii)] \label{DUI} \dred{Disallows UPS Improvements (DUI)} if $\widehat{\Phi}(C^H_\text{ups}) \succeq C^H_\text{ups}$ for all lower semi-continuous convex functions $H : \Delta(\Theta) \to \R \cup\{+\infty\}$.
    %%%
    \item[(iv)] \label{EO} \dred{Exhausts Optimization (EO)} if $\widehat{\Phi}(C) = \widehat{\Phi} ( \widehat{\Phi}(C) )$ for all $C \in \C$.
    %%%
    \item[(v)] \label{GS} \dred{Generates Subadditivity (GS)} if $\widehat{\Phi}(C)$ is \hyperref[axiom:POSL]{Subadditive} for all $C \in \C$.
    %%%
    \end{itemize}
\end{definition}

\hyperref[ADL]{ADL}, \hyperref[EO]{EO}, and \hyperref[GS]{GS} are exactly as stated in \cref{tab:GSLM:both}. However, for technical convenience, the versions of \hyperref[AIE]{AIE} and \hyperref[DUI]{DUI} stated here are \emph{more permissive} than those from  \cref{tab:GSLM:both}. 

First, the version of \hyperref[AIE]{AIE} stated here is implied by the version from  \cref{tab:GSLM:both}, which states that $\widehat{\Phi}(C)\preceq \Phi_\text{IE}(C)$ for all $C \in \C$.\footnote{This implication follows from the proof of \cref{lem:phi-ie}(i) (see \cref{lem:phi-ie-upper} in  \cref{ssec:app:phi-ie}), which in turn follows from \cref{thm:qk}(i) and the definition of the $\Phi_\text{IE}$ map (\cref{defi:Phi:IE}).} The present version is simpler to verify in practice, e.g., by adapting the proof of \cref{thm:qk}(i) (from \cref{sssec:app:thm:qk-pt1}) to the \hyperref[defi:gen:IC]{GLM} $\widehat{\Phi}$.

Second, the version of 
\hyperref[DUI]{DUI} stated here posits that $\widehat{\Phi}(C^H_\text{ups}) \succeq C^H_\text{ups}$ for all  \nameref{defi:ups} costs such that \emph{$H$ is lower semi-continuous}, but imposes \emph{no restrictions} on $\widehat{\Phi}(C^H_\text{ups})$ when $H$ is \emph{not} lower semi-continuous. Since all \nameref{defi:ups} costs are \nameref{axiom:slp} (\cref{lem:ups:to:additive} in \cref{proof:ups:additive}),  every \nameref{defi:gen:IC} $\widehat{\Phi}$ that models a \emph{more constrained} procedure than our baseline $\Phi$ map (i.e., such that $\widehat{\Phi}(C) \succeq \Phi(C)$ for all $C \in \C$) satisfies the stronger version of \hyperref[DUI]{DUI} from \cref{tab:GSLM:both}, which states that $\widehat{\Phi}(C^H_\text{ups}) \succeq C^H_\text{ups}$ for \emph{every} \nameref{defi:ups} cost. The present version also permits some procedures with \emph{more flexibility} than  our baseline model, which is technically useful in certain applications (see \cite{bz2025-GLM}). 

These technical caveats aside, each property in \cref{defi:GSLM:properties} admits a simple economic interpretation (as discussed in \cref{ssec:beyond:flexibility}). We elaborate here on two of them. First, while \hyperref[ADL]{ADL} is a nearly innocuous assumption, it does rule out some \nameref{defi:gen:IC}s that preclude nontrivial one-shot strategies, such as the incremental learning map $\Phi_\text{IE}$. As we explain in the next subsection (see \cref{remark:local-ADL}), we can accommodate such procedures by instead imposing a ``local'' version of \hyperref[ADL]{ADL} that applies only to (upper) kernels. Second, we interpret \hyperref[GS]{GS}---which is perhaps the most ``reduced form'' property---as modeling procedures that feature a ``sufficiently rich'' space of sequential strategies. This interpretation is justified by the fact that \hyperref[GS]{GS} holds for our baseline $\Phi$ and $\Phi_\text{IE}$ maps, the ``no free disposal'' version of $\Phi$ studied in \textcite{bz2025-GLM}, and \nameref{defi:gen:IC}s that model flexible sequential learning with constraints on the ``rate'' of learning (cf. \cite{zhong2022optimal,hebert2023rational}).\footnote{Such ``rate constraints'' can be modeled with the \nameref{defi:gen:IC} $\Phi_\text{HWZ}$ defined as $\Phi_\text{HWZ}(C)(\pi):= \lim_{\delta \to 0} \Phi(C_\delta)$, where $C_\delta \in \C$ is defined as $C_\delta(\pi):= C(\pi) + \infty \cdot \mathbf{1}(C(\pi)>\delta)$. It can be shown that $\Phi_\text{HWZ}$ (like $\Phi_\text{IE}$) satisfies \hyperref[AIE]{AIE}, \hyperref[DUI]{DUI}, and \hyperref[GS]{GS}.}

%\paragraph{Extensions of Main Results.} 

\subsubsection{Extensions of Main Results to GLMs}\label{ssec:GLM-results}

Herein, we present  extensions of our main \dred{Theorems} \ref{prop:1}--\ref{thm:wald} to broad classes of \hyperref[defi:gen:IC]{GLM}s (see \cref{tab:GSLM:both} in \cref{ssec:beyond:flexibility} for a summary). To this end, we begin with two pieces of notation.

First, for several of our main equivalence results, each direction of implication relies on different properties of the \hyperref[defi:gen:IC]{GLM}. To streamline the statements of such results, we let
\[
``\text{Condition A} \ \   \xrightleftharpoons[\widehat{\Phi} \text{ satisfies Y}]{\widehat{\Phi} \text{ satisfies X}} \ \ \text{Condition B}''
\]
denote that: (i) ``Condition A'' implies ``Condition B'' if the \hyperref[defi:gen:IC]{GLM} $\widehat{\Phi}$ satisfies property ``X,'' and (ii) conversely,  ``Condition B'' implies ``Condition A'' if $\widehat{\Phi}$ satisfies property ``Y.'' 

Second, for any $C \in \C$ and $W \subseteq \Delta(\Theta)$, we denote by $C|_W \in \C$ the restriction of $C$ to the domain $\Delta(W)\cup\Ex^\varnothing \subseteq \Ex$,  that is, 
\[
C|_W(\pi) := \begin{cases}
    C(\pi), & \text{if $\pi \in \Delta(W) \cup \Ex^\varnothing$} \\
    %%%
    +\infty, & \text{otherwise.}
\end{cases}
\]
For any \hyperref[defi:gen:IC]{GLM} $\widehat{\Phi}$, we then let $C \in \widehat{\Phi}[\C]|_W$ denote that $C = \widehat{\Phi}(C')|_W$ for some $C' \in \C$. This notation serves two purposes: (i) it lets us formally demonstrate that our main results (viz., \dred{Theorems} \ref{thm:UPS}--\ref{thm:wald}) apply to the restrictions of full-domain indirect costs to smaller domains (e.g., as  in \cref{remark:full-support}), and (ii) it lets us treat abstract \hyperref[defi:gen:IC]{GLM}s $\widehat{\Phi}$, for which the relationship between $\dom(C)$ and $\dom(\widehat{\Phi}(C))$ may be complicated, in a simple unified manner.

With this notation in hand, we now state our extended results. Our first result extends the first equivalence in \cref{prop:1} and (the entirety of) \cref{cor:envelope}. The key observation is that the notions of \hyperref[defi:gen:IC]{$\widehat{\Phi}$-indirect} and  \hyperref[defi:gen:IC]{$\widehat{\Phi}$-proof} costs coincide whenever $\widehat{\Phi}$ satisfies \hyperref[EO]{EO}.

\begin{customthm}{$\widehat{1}$(i)}\label{thm1-hat}
For any \nameref{defi:gen:IC} $\widehat{\Phi}$ and $C \in \C$,
\begin{align*}
    C \in \widehat{\Phi}[\C] \ \   \xrightleftharpoons[]{\widehat{\Phi} \text{ satisfies \hyperref[EO]{EO}}} \ \ C \text{ is \hyperref[defi:gen:IC]{$\widehat{\Phi}$-proof}.}
\end{align*}
Consequently, if $\widehat{\Phi}$ satisfies both \hyperref[ADL]{ADL} and \hyperref[EO]{EO}, then for every $C\in \C$, 
\begin{align*}
    \widehat{\Phi}(C) = \max\{C' \in \C \mid \text{$C' \preceq C$ and $C'$ is \hyperref[defi:gen:IC]{$\widehat{\Phi}$-proof}}\,\}.
\end{align*}
\end{customthm}
\begin{proof}
    See \cref{ssec:proofs-GLM}.
\end{proof}

Our second result extends \cref{thm:UPS} by showing that \nameref{defi:ll} \hyperref[defi:gen:IC]{$\widehat{\Phi}$-indirect} costs are necessarily \nameref{defi:ups} whenever $\widehat{\Phi}$ satisfies \hyperref[GS]{GS} (while the converse holds under \hyperref[ADL]{ADL} and \hyperref[DUI]{DUI}). Moreover, this implication applies ``locally,'' i.e., to the restriction $\widehat{\Phi}(C)|_W$ for any open convex $W \subseteq \Delta^\circ(\Theta)$. Informally, \hyperref[defi:ll]{Regularity} and \nameref{defi:ups} are (locally) equivalent properties of \hyperref[defi:gen:IC]{$\widehat{\Phi}$-indirect} costs whenever the procedure has a  ``sufficiently rich'' strategy space. Formally:

\begin{customthm}{$\widehat{2}$}\label{thm2-hat}
For any \nameref{defi:gen:IC} $\widehat{\Phi}$, open convex $W \subseteq \Delta^\circ(\Theta)$, and $C \in \C$ with $\dom(C) = \Delta(W) \cup \Ex^{\varnothing}$,
\[
\text{$C \in \widehat{\Phi}[\C]|_W$ and is \nameref{defi:ll} %is \hyperref[defi:gen:IC]{generalized SLP} and \nameref{defi:ll} 
} \ \   \xrightleftharpoons[\widehat{\Phi} \text{ satisfies \hyperref[ADL]{ADL} and \hyperref[DUI]{DUI}}]{\widehat{\Phi} \text{ satisfies \hyperref[GS]{GS}}} \ \ C=C^H_\text{ups} \ \text{ for some convex } H \in \mathbf{C}^1(W). 
\]
\end{customthm}
\begin{proof}
    See \cref{ssec:proofs-GLM}.
\end{proof}

Our third result extends \cref{thm:qk}(ii) by showing that lower kernels are invariant under $\widehat{\Phi}$ whenever $\widehat{\Phi}$ satisfies \hyperref[DUI]{DUI}.\footnote{We do not present an extension of \cref{thm:qk}(i) because its conclusion is already built into the definition of \hyperref[AIE]{AIE}.} Informally, lower kernels of the direct cost yield local lower bounds for the \hyperref[defi:gen:IC]{$\widehat{\Phi}$-indirect} cost if $\widehat{\Phi}$ is weakly more restrictive than $\Phi$. Formally:

\begin{customthm}{$\widehat{3}$(ii)}\label{thm:qk:gen}
For any \nameref{defi:gen:IC} $\widehat{\Phi}$, $W \subseteq \Delta(\Theta)$, and \nameref{defi:sp} $C \in \C$, 
\begin{align*}
    \text{$k\gg_\text{psd}\mathbf{0}$ is a lower kernel of $C$ on $W$ } \ \   \xRightarrow{\widehat{\Phi}  \text{ satisfies \hyperref[DUI]{DUI}}} \ \ \text{ $k$ is a lower kernel of $\widehat{\Phi}(C)$ on $W$}.
\end{align*}
%If $C$ is \nameref{defi:sp} and $k\gg_\text{psd}\mathbf{0}$ is a lower kernel of $C$ on $W$, then $k$ is also a lower kernel of $\widehat{\Phi}(C)$ on $W$.
\end{customthm}
\begin{proof}
    See \cref{ssec:proofs-GLM}.
\end{proof}

Our fourth result extends \cref{thm:flie} by showing that the \hyperref[defi:gen:IC]{$\widehat{\Phi}$-indirect} cost is \nameref{defi:ups} if and only if the direct cost \nameref{axiom:flie} whenever $\widehat{\Phi}$ satisfies \hyperref[ADL]{ADL}, \hyperref[AIE]{AIE}, and \hyperref[DUI]{DUI}. Moreover, the  \emph{necessity} of \nameref{axiom:flie} does not require \hyperref[AIE]{AIE}, while the \emph{sufficiency} of \nameref{axiom:flie} does not require \hyperref[ADL]{ADL}. We thereby extend both the economic (recall \cref{ssec:flie-characteriation}) and methodological (recall \cref{fig:flow}) implications of \cref{thm:flie} to broad classes of optimization procedures. Formally:

\begin{customthm}{$\widehat{4}$}\label{thm:flie:gen}
For any \nameref{defi:gen:IC} $\widehat{\Phi}$, open convex set $W \subseteq \Delta^\circ(\Theta)$, strongly convex $H \in \mathbf{C}^2(W)$, and $C \in \C$ that is \nameref{defi:lq} on $W$ and satisfies $\dom(C)\subseteq\Delta(W)\cup \Ex^{\varnothing}$, 
\begin{align*}
	C \text{ \nameref{axiom:flie} and }k_C=\H H \ \ \xrightleftharpoons[\widehat{\Phi}  \text{ satisfies \hyperref[ADL]{ADL} and \hyperref[DUI]{DUI}}]{\widehat{\Phi} \text{ satisfies \hyperref[AIE]{AIE} and \hyperref[DUI]{DUI}}} \ \  \widehat{\Phi}(C)|_W = C^H_\text{ups}.	%\label{eq:flie}
 \end{align*}
\end{customthm}
\begin{proof}
    See \cref{ssec:proofs-GLM}.
\end{proof}

Our fifth result extends the characterization of \nameref{defi:TI} in \cref{thm:trilemma}(i). Informally, we show that \nameref{defi:TI} is the unique \hyperref[defi:gen:IC]{$\widehat{\Phi}$-indirect} cost exhibiting \nameref{axiom:CMC} whenever $\widehat{\Phi}$ satisfies \hyperref[GS]{GS}, i.e., optimizes over a ``sufficiently rich'' strategy space. Formally:

\begin{customthm}{$\widehat{5}$(i)}\label{thm:trilemma1:gen}
For any \nameref{defi:gen:IC} $\widehat{\Phi}$ and nontrivial $C \in \C$ with rich domain, 
\begin{align*}
\text{$C \in \widehat{\Phi}[\C]|_{\Delta^\circ(\Theta)}$ and is \nameref{axiom:CMC}}   \ \ \xrightleftharpoons[\widehat{\Phi}  \text{ satisfies \hyperref[ADL]{ADL} and \hyperref[DUI]{DUI}}]{\widehat{\Phi} \text{ satisfies \hyperref[GS]{GS}}} \ \ \text{$C$ is a \nameref{defi:TI} cost.}
\end{align*}
\end{customthm}
\begin{proof}
    See \cref{ssec:proofs-GLM}.
\end{proof}

Our sixth result extends the main ``converse'' direction of \cref{thm:trilemma}(iii). Informally, we show that the ``negative'' portion of the information cost trilemma---the mutually inconsistency among \hyperref[defi:gen:IC]{$\widehat{\Phi}$-proofness}, \hyperref[axiom:prior:invariant]{Prior Invariance}, and \hyperref[axiom:CMC:0]{CMC}---holds whenever $\widehat{\Phi}$ satisfies \hyperref[GS]{GS}, i.e., optimizes over a ``sufficiently rich'' strategy space. Formally:

\begin{customthm}{$\widehat{5}$(iii)}\label{thm:trilemma3:gen}
For any \nameref{defi:gen:IC} $\widehat{\Phi}$ and 
nontrivial %and \hyperref[axiom:prior:invariant]{Prior Invariant} 
$C \in \C$ with rich domain, 
\begin{align*}
\text{$C \in \widehat{\Phi}[\C]|_{\Delta^\circ(\Theta)}$ and is \hyperref[axiom:prior:invariant]{Prior Invariant} and \hyperref[axiom:mono]{Monotone}} \ \ 
\xRightarrow{\text{$\widehat{\Phi}$ satisfies \hyperref[GS]{GS}} \, } \ \  \text{$C$ is not \hyperref[axiom:CMC:0]{CMC}.}
\end{align*}
\end{customthm}
\begin{proof}
    See \cref{ssec:proofs-GLM}.
\end{proof}

Our final result extends the second equivalence in \cref{thm:wald} by showing that the \ref{eqn:MS} cost is the unique (smooth) \nameref{defi:ups} \hyperref[defi:gen:IC]{$\widehat{\Phi}$-indirect} cost generated by a \hyperref[axiom:prior:invariant]{Prior Invariant} direct cost whenever $\widehat{\Phi}$ satisfies \hyperref[ADL]{ADL}, \hyperref[AIE]{AIE}, and \hyperref[DUI]{DUI}.\footnote{As for the first equivalence in \cref{thm:wald}: the ``$\impliedby$'' direction also extends under the same conditions on $\widehat{\Phi}$ (as the \ref{eqn:MS} cost is \nameref{axiom:CMC} by construction), while the ``$\implies$'' direction extends whenever $\widehat{\Phi}$ satisfies \hyperref[GS]{GS} (per \cref{thm:trilemma1:gen}).} First, under \hyperref[AIE]{AIE} and \hyperref[DUI]{DUI}, the \ref{eqn:MS} cost can indeed be generated in this manner; this extends our ``positive'' finding that relaxing \hyperref[axiom:prior:invariant]{Prior Invariance} to \nameref{defi:spi} resolves the information cost trilemma (recall \cref{fig:wald-thm}). Second, under \hyperref[ADL]{ADL} and \hyperref[DUI]{DUI}, the \ref{eqn:MS} cost is the unique \emph{candidate} for such a cost function; this extends the ``negative'' conclusion of the modeler's trilemma (recall \cref{fig:wald-thm}). Formally, we say that $C \in \C$ is \emph{\dred{$\widehat{\Phi}$-PI}} if $C = \widehat{\Phi}(C')$ for some \hyperref[axiom:prior:invariant]{Prior Invariant} $C' \in \C$. We then have:

\begin{customthm}{$\widehat{6}$(ii)}\label{thm:wald:gen}
For any \nameref{defi:gen:IC} $\widehat{\Phi}$ and \nameref{defi:sp} $C \in \C$ with rich domain,
    \begin{align*}
            %\text{$C$ is \nameref{defi:spi} and \nameref{axiom:CMC}} \iff 
           \text{$C$ is \dred{$\widehat{\Phi}$-PI}, \nameref{defi:ups}, and \nameref{defi:lq}}
            \ \ \xrightleftharpoons[\widehat{\Phi}  \text{ satisfies \hyperref[AIE]{AIE} and \hyperref[DUI]{DUI}}]{\widehat{\Phi} \text{ satisfies \hyperref[ADL]{ADL} and \hyperref[DUI]{DUI}}} \ \ \text{$|\Theta|=2$ and %$C^\circ\propto C_{\text{Wald}}$
             $C$ is a \ref{eqn:MS} cost.}
    \end{align*}
%%%
\end{customthm}
\begin{proof}
    See \cref{ssec:proofs-GLM}.
\end{proof}

We conclude this section with two technical remarks:

\begin{remark}\label{remark:wald:gen}
    The ``$\rightharpoonup$'' direction of \cref{thm:wald:gen} holds under the weaker domain assumption that $\dom(C) \supseteq \Delta(\Delta^\circ(\Theta)) \cup \Ex^\varnothing$, provided we also weaken the conclusion to ``$|\Theta| = 2$ and $C|_{\Delta^\circ(\Theta)}$ is a \ref{eqn:MS} cost'' (see \cref{ssec:proofs-GLM} for details). This formalizes the claim in \cref{remark:full-PI} that \cref{thm:wald} applies to the rich-domain restrictions $C|_{\Delta^\circ(\Theta)}$ of full-domain \nameref{defi:spi} cost functions $C$. 
\end{remark}

%We conclude this section with a technical remark about relaxing the \hyperref[ADL]{ADL} property:

\begin{remark}[Local ADL]\label{remark:local-ADL}
    As noted above in \cref{ssec:GLM-properties}, \hyperref[ADL]{ADL} is violated by some \nameref{defi:gen:IC}s that preclude nontrivial one-shot learning (e.g., the $\Phi_\text{IE}$ map). To accommodate them, we say that a \nameref{defi:gen:IC} $\widehat{\Phi}$ satisfies \nameref{remark:local-ADL} if, for every $C \in \C$ and $p \in \Delta^\circ(\Theta)$, every upper kernel of $C$ at $p$ is also an upper kernel of $\widehat{\Phi}(C)$ at $p$. \nameref{remark:local-ADL} is implied by \hyperref[ADL]{ADL}, but is much weaker; it holds under $\Phi_\text{IE}$ (by \cref{lem:phi-ie-kernel}(i)) and all other optimization procedures that we know of. If we relax \hyperref[ADL]{ADL} to \nameref{remark:local-ADL}, then: (a) the ``$\leftharpoondown$'' direction of \cref{thm:flie:gen} partially extends, viz., if $C$ is \nameref{defi:sp}, then $\widehat{\Phi}(C)|_W = C^H_\text{ups}$ implies that $k_C = \H H$, but not necessarily that $C$ \nameref{axiom:flie} (cf. \cref{lem:phi-ie}(iii)); and (b) the ``$\rightharpoonup$'' direction of \cref{thm:wald:gen} fully extends, provided that we slightly strengthen the hypotheses that $C$ is \dred{$\widehat{\Phi}$-PI} and \nameref{defi:lq} by requiring that $C = \widehat{\Phi}(C')$ for some  \hyperref[axiom:prior:invariant]{Prior Invariant} $C' \in \C$ that is itself \nameref{defi:lq} and \nameref{defi:sp}. We describe the requisite adjustments to the proofs in \cref{ssec:proofs-GLM}. 
\end{remark}

\section{Remaining Proofs of Theorems
%Proofs of Theorems \ref{thm:trilemma} and \ref{thm:wald}
}\label{app:proofs-theorems-5-6}

\subsection{Proofs of Lemmas for Theorem \ref{thm:UPS}}\label{app:thm2:extra}

\subsubsection{Proof of \cref{lem:SLP-divergence}}

\begin{proof}
        %Since every \awb{[Posterior Separable]} cost is \hyperref[axiom:mono]{Monotone}, \cref{prop:1} implies that suffices to show that $C$ is \hyperref[axiom:POSL]{Subadditive} iff $D$ satisfies \eqref{triangle-avg}. 
    We prove each direction in turn.

    \noindent \textbf{($\impliedby$ direction)} Let the divergence $D$ satisfy \eqref{triangle-avg}. Take any $\Pi \in \Delta^\dag(\Ex)$. Let $\pi_1$ be the induced first-round random posterior and $p := p_{\pi_1}$. There are two cases to consider:

    \emph{Case 1: Suppose $\left[ \{\pi_1\} \cup \supp(\Pi) \right] \not\subseteq \dom(C)$.} This implies that $C(\pi_1)=+\infty$ or that there exists $\pi_2\in \supp(\Pi)$ with $C(\pi_2)=+\infty$. Therefore, $C(\E_\Pi[\pi_2])  \leq +\infty =  C(\pi_1) + \E_\Pi[C(\pi_2)]$. 

    \emph{Case 2: Suppose $\left[ \{\pi_1\} \cup \supp(\Pi) \right] \subseteq \dom(C) = \Delta(W)\cup\Ex^\varnothing$.} There are two sub-cases.

    First, let $p \notin W$. This implies $\pi_1 \notin \Delta(W)$ (as $W$ is convex), and thus the supposition implies $\pi_1 = \delta_p \in \Ex^\varnothing$. It follows that $p_{\pi_2} = p \notin W$ for all $\pi_2 \in \supp(\Pi)$, and hence that $\supp(\Pi)\cap \Delta(W) = \emptyset$ (as $W$ is convex). The supposition then implies $\supp(\Pi) = \{\delta_p\}$. It follows that $\E_\Pi[\pi_2] = \delta_p \in\Ex^\varnothing$. We therefore obtain $C(\E_\Pi[\pi_2]) = C(\pi_1) + \E_\Pi[C(\pi_2)] = 0$.

    Next, let $p \in W$. The supposition implies $\pi_1 \in \Delta(W) \cup\{\delta_p\} = \Delta(W)$. Define the Borel measure $\mu_1$ on $\Delta(\Theta)$ as $\mu_1(B) := \Pi(\{\pi_2 \in \Ex \mid p_{\pi_2} \in B\} \cap \Ex^\varnothing)$ for all Borel $B \subseteq \Delta(\Theta)$. By the definition of $\pi_1$ and the finiteness of $\supp(\Pi) \backslash \Ex^\varnothing$, it follows that
    \begin{equation}\label{eqn:mu1-decomp-1-PS}
    \pi_1 \, = \,  \mu_1 \,  +  \, \sum_{\pi_2 \in \supp(\Pi)\backslash \Ex^\varnothing} \Pi(\{\pi_2\}) \, \delta_{p_{\pi_2}}. %\quad \forall \, \text{ Borel $B \subseteq \Delta(\Theta)$.}
    \end{equation}
     By construction, $\supp(\mu_1) \subseteq \supp(\pi_1)$. Moreover, since $\supp(\Pi) \backslash \Ex^\varnothing$ is finite, it holds that 
    \begin{equation}\label{eqn:mu1-decomp-2-PS}
    \E_\Pi[\pi_2] \, =  \,  \int_{\Ex^\varnothing} \pi_2 \dd \Pi(\pi_2) \, + \, \sum_{\pi_2 \in \supp(\Pi) \backslash \Ex^\varnothing} \Pi (\{\pi_2\}) \cdot \pi_2  \, =  \,  \mu_1 \, + \, \sum_{\pi_2 \in \supp(\Pi) \backslash \Ex^\varnothing} \Pi (\{\pi_2\}) \cdot \pi_2 ,
    \end{equation}
    where the second equality is by a change of variables. Thus, since the supposition implies $\supp(\mu_1)\cup \big[ \bigcup_{\pi_2 \in \supp(\Pi)\backslash \Ex^\varnothing} \supp(\pi_2)\big] \subseteq W$ and the union is finite, $\supp(\E_\Pi[\pi_2]) \subseteq W$. Hence, $\E_\Pi[\pi_2] \in \Delta(W) \subseteq \dom(C)$. Therefore, since $C$ is \hyperref[eqn:PS]{Posterior Separable}, it follows that
    \begin{align*}
        C( \E_\Pi [\pi_2]  ) %&= \E_{\E_{\Pi}[\pi_2]} \left[ D(q\mid p)\right] \\
        &= \E_{\mu_1} [D(q \mid p)] \, + \,  \sum_{\pi_2 \in \supp(\Pi)\backslash \Ex^\varnothing} \Pi(\{\pi_2\}) \cdot  \mathbb{E}_{\pi_2} [ D(q \mid p ) ] \\
        %%%
        & \leq \E_{\mu_1} [D(q \mid p)] \,  + \, \sum_{\pi_2 \in \supp(\Pi)\backslash \Ex^\varnothing} \Pi(\{\pi_2\}) \cdot \left[ D(p_{\pi_2} \mid p) + \mathbb{E}_{\pi_2} [ D(q \mid p_{\pi_2})]  \right] \\
        & = \E_{\pi_1} [ D(q \mid p)] \, + \,  \sum_{\pi_2 \in \supp(\Pi)\backslash \Ex^\varnothing} \Pi(\{\pi_2\}) \cdot \mathbb{E}_{\pi_2} [ D(q \mid p_{\pi_2})]   \\
        & = C(\pi_1) \, + \,  \mathbb{E}_\Pi \left[ C(\pi_2)\right],
    \end{align*}
    where the first line follows from \eqref{eqn:PS}, the fact that $\E_\Pi[\pi_2] \in \Delta(W) \subseteq \dom(C)$, and \eqref{eqn:mu1-decomp-2-PS}; the second line holds because $D$ satisfies \eqref{triangle-avg} (by hypothesis),  $\supp(\Pi)\backslash \Ex^\varnothing \subseteq \Delta(W)$ (by supposition), and $p_{\pi_2} \ll p$ for all $\pi_2 \in \supp(\Pi)\backslash\Ex^\varnothing$ (by the definition of $\pi_1$ and Bayes' rule); the third line follows from  \eqref{eqn:mu1-decomp-1-PS}; and the final line follows from \eqref{eqn:PS} and the supposition.

    Since the given $\Pi \in \Delta^\dag(\Ex)$ was arbitrary, we conclude that $C$ is \hyperref[axiom:POSL]{Subadditive}.

    \noindent \textbf{($\implies$ direction)} Let $C$ be \hyperref[axiom:POSL]{Subadditive}. Let $p \in W$ and $\pi \in \Delta(W)$ with $p_\pi \ll p$ be given. The case in which $p = p_\pi$ is trivial, so suppose $p \neq p_\pi$. Note that $p_\pi \in W$ because: (a) $\supp(\pi) \subseteq W$ and $p_\pi \in \mathrm{conv}(\supp(\pi))$ by construction, and (b) $W$ is convex. Since $p_\pi \ll p$ and $W$ is open, it follows that there exist $r \in W \backslash \{p\}$ and $\alpha \in (0,1)$ such that $p = \alpha p_\pi + (1-\alpha)r$.\footnote{Let $\Theta' := \supp(p)$. Since $p,p_\pi \in W$ and $p_\pi \ll p$, we have $p, p_\pi \in W \cap \Delta(\Theta') $. Define $y := p_\pi - p \in \mathcal{T}\backslash\{\mathbf{0}\}$. Since $W$ is open, there exists an $\epsilon>0$ such that $r:= p - \epsilon y \in W \cap \Delta(\Theta')$. Then, for $\alpha := \epsilon / (1+\epsilon) \in (0,1)$, we have $p = \alpha p_\pi + (1-\alpha) r$.} Define $\Pi \in \Delta^\dag(\Ex)$ as $\Pi(\{\pi\}) := \alpha$ and $\Pi(\{\delta_{r}\}) := 1-\alpha$, which induces $\pi_1 := \alpha \delta_{p_\pi} + (1-\alpha) \delta_{r}$ and $\mathbb{E}_\Pi \left[ \pi_2 \right] = \alpha \pi + (1-\alpha) \delta_r$, where $p_{\pi_1} = p$. Since $C$ is \hyperref[eqn:PS]{Posterior Separable}, it follows that
    \begin{align*}
        C\left( \mathbb{E}_\Pi \left[ \pi_2 \right]\right) &= \alpha \mathbb{E}_{\pi} \left[ D(q \mid p )  \right] + (1-\alpha) D(r \mid p),  \\
        C(\pi_1) & = \alpha D(p_\pi \mid p) + (1-\alpha) D(r \mid p), \\
        C(\pi) &= \mathbb{E}_{\pi} \left[ D(q \mid p_\pi )  \right].
    \end{align*}
    Since $C$ is \hyperref[axiom:POSL]{Subadditive}, $C\left(\mathbb{E}_\Pi \left[ \pi_2 \right] \right) \leq C(\pi_1) + \alpha C(\pi) + (1-\alpha) C(\delta_r)$. Plugging the above display and $C(\delta_r) = 0$ into this inequality and simplifying, we obtain \eqref{triangle-avg}, as desired.
    %The $p = p_\pi$ case is trivial, so suppose that $p \neq p_\pi$. Because $W$ is open and convex, $p_\pi \in W$ and $p =  \sum_{i=1}^{|\Theta|} \alpha_i p_i$ for some $\{p_i\}_{i=1}^{|\Theta|} \subseteq W$ with $p_1 := p_\pi$ and $\alpha \in \Delta(\{1,\dots,|\Theta|\})$ with $\alpha_1  \in (0,1)$. Define the finite-support $\Pi \in \Delta(\Delta(W))$ as $\Pi(\pi) := \alpha_1$ and $\Pi(\delta_{p_i}) := \alpha_i$ for all $2 \leq i \leq |\Theta|$, which induces $\pi_1 = \sum_{i=1}^{|\Theta|} \alpha_i \delta_{p_i}$ and $\pi' := \mathbb{E}_\Pi \left[ \pi_2 \right] = \alpha_1 \pi + \sum_{i=2}^{|\Theta|} \alpha_i \delta_{p_i}$. Then
    %\begin{align*}
     %   C(\pi') = \mathbb{E}_{\pi'} \left[ D(q \mid p) \right] = \alpha_1 \mathbb{E}_\pi \left[ D(q \mid p) \right] + \sum_{i=2}^{|\Theta|} \alpha_i D(p_i \mid p)
   % \end{align*}
\end{proof}

\subsubsection{Auxiliary Lemma for the Proof of  \cref{lem:thm2-posterior-c1}}

The following lemma is invoked in the proof of \cref{lem:thm2-posterior-c1} (in \cref{ssec:app:thm:UPS}). 

\begin{lem}\label{lem:matrix-rep}
    Let $W \subseteq \Delta^\circ(\Theta)$ be open and convex. For any $p \in W$, if $f : W \to \mathbb{R}^{|\Theta|}$ satisfies 
    \begin{equation}\label{eqn:f-0-p}
    \mathbb{E}_\pi \left[ f(q) \right] = \mathbf{0} \quad \forall \,  \text{ finite-support $\pi \in \Delta(W)$ with $p_\pi = p$,}
    \end{equation}
    then there exists $A(p) \in \mathbb{R}^{|\Theta|\times|\Theta|}$ such that  $f(q) = - A(p) q$ for all $q \in W$, and hence $A(p) p = \mathbf{0}$.
\end{lem}

Within the proof of \cref{lem:thm2-posterior-c1}, for each fixed $p \in W$, we define $f := \nabla_2 D(\cdot \mid p)$ and use \cref{lem:matrix-rep} to deduce \eqref{prior-grad-linear} from \eqref{eqn:prior-grad-FOC}. In what follows, we prove \cref{lem:matrix-rep} itself. 

\begin{proof}
    Let $p \in W$ and such an $f : W \to \R^{|\Theta|}$ be given. We first show that $f$ is affine. 

    To this end, suppose towards a contradiction that there exist $q_1,q_0 \in W$ and $\alpha \in (0,1)$ such that $f( q_\alpha) \neq \alpha f(q_1) + (1-\alpha) f(q_0)$, where we denote $q_\alpha:= \alpha q_1 + (1-\alpha) q_0$. Define the finite-support $\pi \in \Delta(W)$ as $\pi := \alpha \delta_{q_1} + (1-\alpha)\delta_{q_0}$. Note that $p_\pi = q_\alpha$. There are two cases. First, if $q_\alpha =p$, then the supposition implies $\E_\pi[f(q)] \neq f(q_\alpha) =\E_{\delta_p}[f(q)]$, which contradicts \eqref{eqn:f-0-p}. Second, suppose $q_\alpha \neq p$. Since $W\subseteq \Delta^\circ(\Theta)$ is open, there exists $\epsilon>0$ such that $r := p + \epsilon (p-q_\alpha) \in W$. Define the finite-support $\pi',\pi'' \in \Delta(W)$ as $\pi' := \frac{1}{1+\epsilon} r + \frac{\epsilon}{1+\epsilon} q_\alpha$ and $\pi'' := \frac{1}{1+\epsilon} r + \frac{\epsilon}{1+\epsilon} \pi$. Note that $p_{\pi'} = p_{\pi'} = p$. These definitions and the supposition imply 
    \[
    \E_{\pi''}[f(q)] - \E_{\pi'}[f(q)] \,  = \, \frac{\epsilon}{1+\epsilon} \left(\E_\pi[f(q)] - f(q_\alpha) \right) \,  \neq \, 0,
    \]
    which again contradicts \eqref{eqn:f-0-p}. We conclude that $f$ must be affine, as desired. 

    We now show that $f$ has the desired matrix representation. Let $V:= \cup_{\alpha >0} \alpha W \subseteq \mathbb{R}^{|\Theta|}_{++}$ be the (open) convex cone generated by $W$. Define $g : V \to \R^{|\Theta|}$ as $g(x) := (\mathbf{1}^\top x) f\left( \frac{x}{\mathbf{1}^\top x} \right)$, viz., $g$ is the HD1 extension of $f$ to $V$. By construction, $g$ is affine and HD1. Hence, it is continuous (as $\dom(g) = V$ is open in $\R^{|\Theta|}$) and additive, i.e., $g(x+y) = g(x) + g(y)$ for all $x,y \in V$. For each $x \in V$, we denote by $g_i (x)$ the $i$th component of the vector $g(x) \in \R^{|\Theta|}$. Then, for each $i \in \{1,\dots,|\Theta|\}$, the map $g_i : V \to \mathbb{R}$ is a continuous solution to the restricted Cauchy equation on domain $V \times V$ \parencite[Ch. 13.6]{kuczma2009introduction}. By Corollary 13.6.2 and Theorem 5.5.2 in \textcite{kuczma2009introduction}, there exists $a_i \in \R^{|\Theta|}$ such that $g_i(x) = -  a_i^\top x$ for all $x \in V$. Let $A(p) := [a_i]_{i=1}^{|\Theta|} \in \R^{|\Theta|\times |\Theta|}$ be the $|\Theta| \times |\Theta|$ matrix with rows $a_i$. Then, by construction, $g(x) = - A(p) x$ for all $x \in V$. This implies that $f(q) = - A(p) q$ for all $q \in W$, as desired. Moreover, since \eqref{eqn:f-0-p} implies that $f(p) = \E_{\delta_p}[f(q)] = \mathbf{0}$, we conclude that $A(p) p = \mathbf{0}$.
\end{proof}

\subsubsection{Proof of \cref{lem:thm2-posterior-c1}}

The proof supplements \cref{lem:thm2-prior-c1} with a mollification argument adapted from Step 2 in the proof of \textcite[Theorem 3]{bgw-bregman-ieee2005}. We begin by recalling some standard definitions and facts about mollification, following \textcite[Chapter 7.2]{gilbarg2001elliptic}. For each $\epsilon >0$, let $\mathcal{F}(\epsilon) := \{y \in \mathcal{T} \mid \|y \| \leq \epsilon \}$ denote the closed ball in $\mathcal{T}$ of radius $\epsilon$ centered at $\mathbf{0}$. A map $\xi : \mathcal{T} \to \mathbb{R}_+$ is called a \emph{mollifier} if: (i) $\xi \in \mathbf{C}^\infty(\mathcal{T})$, (ii) $\supp(\xi) \subseteq \mathcal{F}(1)$, and (iii) $\int_{\mathcal{T}} \xi(y) \dd y =1$. It is a standard fact that mollifiers exist. For any mollifier $\xi$ and $\epsilon>0$, the map $\xi_\epsilon : \mathcal{T} \to \R_+$ is defined as $\xi_\epsilon (y) := \epsilon^{-(|\Theta|-1)} \xi (y /\epsilon)$. By construction, we have $\supp(\xi_\epsilon) \subseteq \mathcal{F}(\epsilon)$ for every $\epsilon>0$ and $\lim_{\epsilon \to 0} \xi_\epsilon(y) = \delta(y)$ for every $y \in \mathcal{T}$, where $\delta(y)$ is the Dirac delta function at $y$. We now proceed to prove \cref{lem:thm2-posterior-c1}.

\begin{proof}
    Let $W\subseteq \Delta^\circ(\Theta)$ be open and convex. \cref{lem:SLP-divergence} implies that $D$ satisfies \eqref{triangle-avg}. Since $\nabla_1 D$ exists and is jointly continuous on $W \times W$ (by hypothesis), our HD1 normalization $D(q\mid p) \equiv q^\top \nabla_1 D(q\mid p)$ (\cref{fn:HD1-gradient}) implies that $D$ is also jointly continuous on $W \times W$.

    For every $\epsilon>0$, we define $W_\epsilon := \{p \in W \mid \overline{B}_\epsilon(p) \subseteq W\}$.\footnote{For any $X \subseteq \Delta(\Theta)$, we denote its closure as $\overline{X}$. Thus,  $\overline{B}_\epsilon(p) := \{q \in \Delta(\Theta) \mid \|p-q\|\leq \epsilon\}$ is the closed $\epsilon$-ball around $p$.} Note that, because $W \subseteq \Delta^\circ(\Theta)$, we can equivalently write $W_\epsilon = \{p \in W \mid \{p\} +\mathcal{F}(\epsilon)\subseteq W\}$. It holds that: (i) $W_{\epsilon'} \subseteq W_{\epsilon}$ for all $\epsilon' \geq \epsilon>0$ (by definition); (ii) $W = \cup_{\epsilon>0} W_\epsilon$, and hence there exists $\overline{\epsilon}>0$ such that $W_\epsilon \neq \emptyset$ for all $\epsilon \in (0,\overline{\epsilon})$ (as $W$ is open); and (iii) $W_\epsilon$ is convex for all $\epsilon>0$ (as $W$ is convex).

    Let any mollifier $\xi$ be given. Fix any $\epsilon \in (0,\overline{\epsilon}/2)$. We define the divergence $D_\epsilon$ as %$\dom(D_\epsilon) = \mathcal{D}_{W_{2\epsilon}}$ and 
    \begin{equation} \label{div-moll-0}
\dom(D_\epsilon) = \mathcal{D}_{W_{2\epsilon}} \quad \text{ and } \quad D_\epsilon( q \mid p) := \int_{\supp(\xi_\epsilon)} D( q + y \mid p + y) \xi_\epsilon (y) \dd y \quad \forall \, p,q \in W_{2\epsilon},
    \end{equation}
    where the integral is well-defined and finite because 
    \begin{equation}\label{eqn:W2e-nest}
W_{2\epsilon }\,  \subseteq \, \left\{p \in W \mid \{p\} + \mathcal{F}(\epsilon)\subseteq W_\epsilon \right\} \, \subseteq \, \left\{p \in W \mid \{p\} + \supp(\xi_\epsilon) \subseteq W_\epsilon \right\}
    \end{equation} 
and $D$ is uniformly continuous on the compact set $\overline{W}_\epsilon \times \overline{W}_\epsilon$ (being that $\overline{W}_\epsilon \subseteq W$ and $D$ is continuous on $W \times W$). Moreover, by \eqref{div-moll-0}--\eqref{eqn:W2e-nest} and the uniform continuity of $D$ on $\overline{W}_\epsilon \times \overline{W}_\epsilon$, the Dominated Convergence Theorem implies that $D_\epsilon$ is jointly continuous on $W_{2\epsilon} \times W_{2\epsilon}$.

First, we claim that $D_\epsilon$ satisfies the inequality \eqref{triangle-avg} (from \cref{lem:SLP-divergence}) for all $\pi \in \Delta(W_{2\epsilon})$ and $p \in W_{2\epsilon}$.\footnote{Since $W_{2\epsilon} \subseteq  \Delta^\circ(\Theta)$, every such $\pi$ and $p$ satisfy $\supp(p_\pi) = \supp(p) = \Theta$, and hence $p_\pi \ll p$.} To this end, let $\pi \in \Delta(W_{2\epsilon})$ and $p \in W_{2\epsilon}$ be given. For every $y \in \supp(\xi_\epsilon )$, we define $\pi_y \in \Delta(W_\epsilon)$  as $\pi_y(E) := \pi\left(\{q \in W_{2\epsilon} \mid q + y \in E\} \right)$ for all Borel $E \subseteq W_\epsilon$. Note that $\pi_y$ is well-defined (viz., $\pi_y(W_\epsilon) = 1$) by \eqref{eqn:W2e-nest}, and that $p_{\pi_y} = p_\pi +y$ by construction. We have
\begin{align*}
    \mathbb{E}_\pi \left[ D_\epsilon (q \mid p) \right] &= \int_{\supp(\xi_\epsilon)} \mathbb{E}_\pi \left[ D( q + y \mid p +y) \right] \xi_\epsilon(y) \dd y \\
    %%%
    & = \int_{\supp(\xi_\epsilon)} \mathbb{E}_{\pi_y} \left[ D( q  \mid p +y ) \right] \xi_\epsilon(y) \dd y \\
    %%%
    & \leq \int_{\supp(\xi_\epsilon)} \left[D (p_{\pi_y} \mid p+y ) + \mathbb{E}_{\pi_{y}} [ D(q \mid p_{\pi_y} )] \right] \xi_\epsilon(y) \dd y \\
    %%%
    & = \int_{\supp(\xi_\epsilon)} \big[D (p_\pi + y \mid p+y ) + \mathbb{E}_{\pi} \left[ D(q +y \mid p_\pi + y) \right] \big] \xi_\epsilon(y) \dd y \\
    %%%
    & = D_\epsilon (p_\pi \mid p) + \mathbb{E}_\pi \left[ D_\epsilon (q \mid p_\pi )\right],
\end{align*}
where the first line is by definition of $D_\epsilon$ and Fubini's Theorem, the second line is by definition of $\pi_y$, the third line holds because $D$ satisfies \eqref{triangle-avg} (on $W\supseteq W_\epsilon$) and $\xi_\epsilon  \geq0$, the fourth line is by definition of $\pi_y$ and  $p_{\pi_y} = p_\pi + y$, and the final line is again by definition of $D_\epsilon$ and Fubini's Theorem. We conclude that $D_\epsilon$ satisfies \eqref{triangle-avg} on $W_{2\epsilon}$, as claimed.

Next, we claim that $\nabla_2 D_\epsilon$ exists and is jointly continuous on $W_{2\epsilon} \times W_{2\epsilon}$. To this end, note that by changing variables from $y\in \mathcal{T}$ to $r := p +y \in \Delta(\Theta)$, we have
\begin{align}
D_\epsilon( q \mid p) = \int_{\{p\}+\supp(\xi_\epsilon)} D(q - p + r \mid r) \xi_\epsilon (r  - p ) \dd r \quad \forall \, p,q \in W_{2\epsilon}. \label{div-moll}  
\end{align}
It is useful to define the sets $F_\epsilon \subseteq G_\epsilon := W_{2\epsilon} \times W_{2 \epsilon} \times \Delta(\Theta)$ and the map $f_\epsilon :  G_\epsilon \to \R_+$ as 
\begin{align*}
%\hspace{-0.5em}
F_\epsilon  &:=  \left\{(q,p,r) \in G_\epsilon \mid  r \in W_\epsilon \ \ \text{and} \ \ q-p+r \in W_{\epsilon/2}\right\}, \\ %\quad \text{ and } \quad f(q,p,r)  :=  D(q - p + r \mid r) \xi_\epsilon (r  - p ).
%%%
f_\epsilon(q,p,r)  &:= \begin{cases}
    D(q - p + r \mid r) \xi_\epsilon (r  - p ) , & \text{if $(q,p,r) \in F_\epsilon$} \\
    0, & \text{if $(q,p,r) \in G_\epsilon \backslash F_\epsilon$.}
\end{cases}
\end{align*}
Note three properties: (i) $F_\epsilon = \ri(F_\epsilon)$ is open (by construction); (ii) $F_\epsilon$ satisfies 
%\begin{equation}\label{eqn:domain-nest-moll-feps}
\[
\left\{ (q,p, r) \in W_{2\epsilon} \times W_{2\epsilon} \times \Delta(\Theta) \mid r-p \in \supp(\xi_\epsilon) \right\} \, \subseteq \, F_\epsilon
\]
%\end{equation}
(by $\supp(\xi_\epsilon)\subseteq \mathcal{F}(\epsilon)$ and the first inclusion in \eqref{eqn:W2e-nest}); and (iii) it can be verified that
\[
\text{$\forall \, q,p \in W_{2\epsilon}$, \ $\exists \, \eta >0$ } \quad \text{ s.t. } \quad R_{\epsilon,\eta}(q,p):=\overline{B}_\eta(q)\times \overline{B}_\eta(p) \times \left[\overline{B}_\eta(p) + \supp(\xi_\epsilon) \right] \subseteq F_\epsilon.
\]

Property (ii) implies that \eqref{div-moll} can be equivalently rewritten as 
\[
%\begin{align}
D_\epsilon( q \mid p) = \int_{\Delta(\Theta)} f_\epsilon(q,p,r) \dd r \quad \forall \, p,q \in W_{2\epsilon}.   
%\end{align}
\]
Note that $f_\epsilon$ is uniformly continuous on $F_\epsilon$ (being that $\xi_\epsilon \in \mathbf{C}^\infty(\mathcal{T})$ and $D$ is uniformly continuous on $\overline{W}_\epsilon \times \overline{W}_\epsilon$). Also note that $\nabla_1 D$ exists and is jointly uniformly continuous on the compact set $\overline{W}_{\epsilon}\times \overline{W}_\epsilon$ (being that $\overline{W}_\epsilon\subseteq W$ and $\nabla_1 D$ exists and is continuous on $W\times W$). This observation, the fact that $\xi_\epsilon \in \mathbf{C}^\infty(\mathcal{T})$, and property (i) together imply that $\nabla_2 f_\epsilon$ (the $p$-gradient of $f_\epsilon$) exists and is jointly uniformly continuous on $F_\epsilon$.\footnote{In particular, $\nabla_2 f_\epsilon(q,p,r) = - \nabla_1 D (q-p+r \mid r) \cdot \xi_\epsilon(r-p) -  D (q-p+r \mid r) \cdot \nabla\xi_\epsilon(r-p)$ for all $(q,p,r)\in F_\epsilon$.} Given property (iii), the classic Leibniz rule then implies that $\nabla_2 D_\epsilon$ exists on $W_{2\epsilon}\times W_{2\epsilon}$ and is given by\footnote{Property (iii) ensures that, for any given $(q,p) \in W_{2\epsilon} \times W_{2\epsilon}$, there exists $\eta>0$ and $U:=\overline{B}_\eta(p) + \supp(\xi_\epsilon)$ such that
\[
D_\epsilon(q'\mid p') \, = \, \int_{\Delta(\Theta)} f_\epsilon(q', p',r) \dd r \, = \,  \int_{U} D(q' - p' + r \mid r) \xi_\epsilon (r  - p' ) \dd r \quad \forall \, (q',p') \in \overline{B}_\eta(q)\times \overline{B}_\eta(p).
\]
The Dominated Convergence Theorem then yields the Leibniz rule for $\nabla_2 D_\epsilon(q \mid p)$ and the continuity of $\nabla_2 D_\epsilon$ at $(q,p)$.}
%\footnote{\awb{[HD0 (re)normalization]}} 
\[
\nabla_2 D_\epsilon(q \mid p) \, = \,  \int_{\Delta(\Theta)} \nabla_2 f_\epsilon(q,p,r) \dd r \, \quad \forall \, p,q \in W_{2\epsilon},
\]
and the Dominated Convergence Theorem implies $\nabla_2 D_\epsilon$ is jointly continuous on $W_{2\epsilon}\times W_{2\epsilon}$.

Now, let $C_\epsilon \in \C$ be the \hyperref[eqn:PS]{Posterior Separable} cost with divergence $D_\epsilon$. By construction, $\dom(C_\epsilon) = \Delta(W_{2\epsilon}) \cup \Ex^\varnothing$. Since $D_\epsilon$ satisfies \eqref{triangle-avg} on $W_{2\epsilon}$ (as shown above), \cref{lem:SLP-divergence} implies $C_\epsilon$ is \hyperref[axiom:POSL]{Subadditive}. Therefore, since $W_{2\epsilon} \subseteq \Delta^\circ(\Theta)$ is open and convex (as noted above) and $\nabla_2 D_\epsilon$ exists and is continuous on $W_{2\epsilon} \times W_{2 \epsilon}$ (as shown above), \cref{lem:thm2-prior-c1} implies that there exists convex $H_\epsilon \in \mathbf{C}^1(W_{2\epsilon})$ such that $D_\epsilon$ has the Bregman form \eqref{eqn:bregman} on $W_{2\epsilon}\times W_{2\epsilon}$. In particular, for any given $p^* \in W_{2\epsilon}$, \cref{lem:thm2-prior-c1} implies that it suffices to let $H_\epsilon:= D_\epsilon(\cdot \mid p^*)$.

For every $\delta \in (0,\epsilon)$, we can define a divergence $D_\delta$ with $\dom(D_\delta) = \mathcal{D}_{2\delta}$ as in \eqref{div-moll-0} (with $\delta$ replacing $\epsilon$ everywhere the latter appears). By the same arguments as above, it follows that $D_\delta$ has the Bregman form \eqref{eqn:bregman} on $W_{2\delta}\times W_{2\delta}$ for the convex function $H_\delta := D_\delta(\cdot \mid p^*) \in \mathbf{C}^1(W_{2\delta})$ (where the same $p^* \in W_{2\delta}$ because $W_{2\delta} \supseteq W_{2\epsilon}$). Define $H : \Delta(\Theta) \to \overline{\R}_+$ as $H := D(\cdot \mid p^*)$. Since $D$ is jointly continuous on $W \times W$ (as noted above), a standard result on mollification \parencite[Lemma 7.1]{gilbarg2001elliptic} implies that $\lim_{\delta\to 0} D_\delta(q\mid p) = D(q \mid p)$ for all $q,p\in W$.\footnote{\textcite[Lemma 7.1]{gilbarg2001elliptic} directly implies that $\lim_{\delta \to 0} \sup_{q,p \in W_{2\zeta}} |D_\delta (q\mid p) - D(q\mid p)| = 0$ for every $\zeta >0$. Since $W = \cup_{\zeta>0} W_{2\zeta}$, it follows that $\lim_{\delta\to 0} D_\delta(q\mid p) = D(q \mid p)$ for all $q,p\in W$, as desired.} This directly implies that $\lim_{\delta \to 0} H_\delta(q) = H(q)$ for all $q \in W$.

We show that $D$ is the Bregman divergence generated by $H$, i.e., $D$ and $H$ satisfy \eqref{eqn:bregman}. First, note that $H \in \mathbf{C}^1(W)$ because $D(\cdot \mid p^*) \in \mathbf{C}^1(W)$ (by hypothesis). Second, note that $H$ is convex, being the pointwise limit of the convex functions $H_\delta$ as $\delta \to0$ \parencite[Theorem 10.8]{rock70}.\footnote{Formally, \textcite[Theorem 10.8]{rock70} requires all of the approximating functions to be finite-valued on $\dom(H) =W$, which does not hold here. We can accommodate this as follows. First, for every $\zeta>0$, directly apply the result on $W_{2\zeta}$ to show that the restriction $H|_{W_{2\zeta}} = \lim_{\delta \to 0} H_{\delta}|_{W_{2\zeta}}$ is convex on $W_{2\zeta}$. Next, take any $q_1,q_0 \in W$ and $\alpha \in (0,1)$. Let $q_\alpha := \alpha q_1 + (1-\alpha) q_0$. There exists $\zeta>0$ such that $q_1,q_0 \in W_{2\zeta}$ (as $W$ is open) and hence $q_\alpha \in W_{2\zeta}$ (as $W_{2\zeta}$ is convex). Since $H \equiv H|_{W_{2\zeta}}$ on $W_{2\zeta}$, it follows that $\alpha H(q_1) + (1-\alpha) H(q_0) \geq H(q_\alpha)$. We conclude that $H$ is convex.} Third, we claim that $\nabla H_\delta$ converges pointwise to $\nabla H$ as $\delta \to 0$.

To establish the claim, let $p \in W$ be given. Since $W \subseteq \Delta^\circ(\Theta)$ is open, there exist $\eta,\zeta >0$ such that $p + y \in W_{2\zeta}$ for all $y \in \mathcal{F}(\eta)$. Thus, since each $D_\delta$ and $H_\delta$ with $\delta \in (0,\zeta)$ satisfy  \eqref{eqn:bregman} on $W_{2\delta}\times W_{2\delta}\supseteq W_{2\zeta}\times W_{2\zeta}$, it holds that
\begin{align}
\begin{split}\label{eqn:lim-grad-moll-1}
\lim_{\delta\to0} y^\top \nabla H_\delta(p) &= \lim_{\delta\to0} \big[H_\delta(p+y)  - H_\delta(p) - D_\delta(p+y \mid p) \big] \\
&= H(p+y)  - H(p) - D(p+y \mid p)
\end{split}
\qquad \forall \, y \in \mathcal{F}(\eta).
\end{align}
Meanwhile, our HD1 normalization for gradients (\cref{fn:HD1-gradient}) implies that
\begin{equation}\label{eqn:lim-grad-moll-2}
\lim_{\delta\to0} p^\top \nabla H_\delta(p) \,  = \, \lim_{\delta \to 0} H_\delta(p) \, = \,  H(p).
\end{equation}
Since $\text{span}\left(\{p\} \cup\mathcal{F}(\eta) \right) = \mathbb{R}^{|\Theta|}$, \eqref{eqn:lim-grad-moll-1} and \eqref{eqn:lim-grad-moll-2} imply that there exists $v_p \in \R^{|\Theta|}$ such that $\lim_{\delta \to 0} \nabla H_\delta(p) = v_p$. Moreover, \eqref{eqn:lim-grad-moll-2} implies that $p^\top v_p = H(p)$ and \eqref{eqn:lim-grad-moll-1} implies that
\[
D(q \mid p) = H(q) - H(p) - (q-p)^\top v_p \quad \forall \, q \in B_\eta(p),
\]
where we use the fact that $B_\eta(p)  \subseteq \{p+y \mid y \in \mathcal{F}(\eta)\}$. Since $D(\cdot \mid p)\geq 0$, it follows that $v_p$ is a subgradient of (the HD1 extension of) the convex function $H|_{B_\eta(p)} \in \mathbf{C}^1\big(B_\eta(p) \big)$, i.e., the restriction of $H$ to the (relatively) open ball $B_\eta(p)$. This implies $v_p = \nabla H(p)$. Therefore, $\lim_{\delta \to 0} \nabla H_\delta(p) = \nabla H(p)$. Since the given $p \in W$ was arbitrary, this establishes the claim. 

Now, to complete the proof that $D$ and $H$ satisfy \eqref{eqn:bregman}, let any $p,q \in W$ be given. By construction, there exists some $\zeta>0$ such that $p,q \in W_{2\zeta}$. Since each $D_\delta$ and $H_\delta$ with $\delta \in (0,\zeta)$ satisfy  \eqref{eqn:bregman} on $W_{2\delta}\times W_{2\delta}\supseteq W_{2\zeta}\times W_{2\zeta}$, it follows from the above work that
\[
D(q\mid p) \, = \,  \lim_{\delta \to 0} D_\delta(q \mid p) \, = \, \lim_{\delta \to 0} \big[ H_\delta(q) - H_\delta(p) - (q-p)^\top \nabla H_\delta(p) \big] \, = \, H(q) - H(p) - (q-p)^\top \nabla H(p).
\]
We conclude that $D$ and the convex function $H = D(\cdot \mid p^*) \in \mathbf{C}^1(W)$ satisfy \eqref{eqn:bregman}.  
%%%
\end{proof}

\subsection{Proofs of Lemmas for Theorem \ref{thm:qk}}\label{app:thm3:extra}

\subsubsection{Proof of \cref{lem:loc-upper-bound}}\label{app:thm3-1:extra}

\begin{proof}
    Let compact $V \subseteq W$ and $\epsilon >0$ be given. Since $\H H : W \to \mathbb{R}^{|\Theta|\times|\Theta|}$ is an upper kernel of $C$ on $W$ and is continuous, for every $p \in V$ there exists $\delta(p)>0$ such that: (a) the upper kernel bound in \cref{defi:lq}(i) holds for $C$ and $k(p):= \H H(p)$ at $p$ with error parameters $\epsilon ' := \epsilon/2$ and $\delta':= \delta(p)$, and (b) $\| \H H(q) - \H H(p)\| \leq \epsilon$ for all $q \in B_{\delta(p)}(p) \subseteq W$. Moreover, since $\{B_{\delta(p)}(p)\}_{p \in V}$ is an open cover of the compact set $V \subseteq \Delta(\Theta)$, by the Lebesgue Number Lemma \parencite[Lemma 27.5]{munkres-2000} there exists $\delta >0$ such that, for every $V' \subseteq V$ with $\text{diam}(V')\leq \delta$, there exists some $p \in V$ such that $V' \subseteq B_{\delta(p)}(p)$.

    Now, let $\widehat{\pi}\in \Delta(V)$ with $\text{diam}(\supp(\widehat{\pi}))\leq \delta$ be given. First, observe that 
\begin{align*}
\hspace{-1em} C(\widehat{\pi}) &\leq \E_{\widehat{\pi}} \left[ (q-p_{\widehat{\pi}})^\top \left( \frac{1}{2}\H H(p) + \frac{\epsilon}{2} I\right) (q-p_{\widehat{\pi}})\right] \qquad \text{for some $ p \in V$}\\
& = \E_{\widehat{\pi}} \left[ (q-p_{\widehat{\pi}})^\top \left( \frac{1}{2}\H H(p_{\widehat{\pi}}) + \frac{\epsilon}{2} I\right) (q-p_{\widehat{\pi}})\right] + \frac{1}{2} \E_{\widehat{\pi}} \left[ (q-p_{\widehat{\pi}})^\top \left( \H H(p) - \H H(p_{\widehat{\pi}})\right) (q-p_{\widehat{\pi}})\right] \\
& \leq \E_{\widehat{\pi}} \left[ (q-p_{\widehat{\pi}})^\top \left( \frac{1}{2}\H H(p_{\widehat{\pi}}) + \frac{\epsilon}{2} I\right) (q-p_{\widehat{\pi}})\right] + \frac{\epsilon}{2} \E_{\widehat{\pi}} \left[ \|q-p_{\widehat{\pi}}\|^2\right]  \\
& = \frac{1}{2} \E_{\widehat{\pi}} \left[ (q-p_{\widehat{\pi}})^\top \H H(p_{\widehat{\pi}})  (q-p_{\widehat{\pi}})\right] + \epsilon \text{Var}(\widehat{\pi}),
\end{align*}
where the first line holds because $\supp(\widehat{\pi})\subseteq B_{\delta(p)}(p)$ for some $p \in V$ (by definition of $\delta$) and by property (a) in the definition of $\delta(p)$, the second line rearranges terms, the third line is by property (b) in the definition of $\delta(p)$ (since $p_{\widehat{\pi}} \in B_{\delta(p)}(p)$ by convexity of the ball), and the final line rearranges terms (recall that $\text{Var}(\widehat{\pi}) = \E_{\widehat{\pi}} [ \|q-p_{\widehat{\pi}}\|^2] $). Next, observe that 
\begin{align*}
    \hspace{-2em} C^H_\text{ups}(\widehat{\pi}) & = \E_{\widehat{\pi}} \left[ H(q) - H(p_{\widehat{\pi}}) - \nabla H(p_{\widehat{\pi}}) \cdot (q-p_{\widehat{\pi}}) \right] \\
    & = \E_{\widehat{\pi}} \left[ \int_0^1 (1-t) (q-p_{\widehat{\pi}})^\top \H H(r_{q}(t) ) (q-p_{\widehat{\pi}}) \dd t \right] \qquad \text{where \, $r_{q}(t) := p_{\widehat{\pi}} + t (q-p_{\widehat{\pi}})$} \\
    & = \frac{1}{2} \E_{\widehat{\pi}} \left[  (q-p_{\widehat{\pi}})^\top \H H(p_{\widehat{\pi}} ) (q-p_{\widehat{\pi}}) \right] + \E_{\widehat{\pi}} \left[ \int_0^1 (1-t) (q-p_{\widehat{\pi}})^\top \left( \H H(r_{q}(t)) - \H H(p_{\widehat{\pi}}) \right) (q-p_{\widehat{\pi}}) \dd t \right] \\
    %%%
    & \geq \frac{1}{2} \E_{\widehat{\pi}} \left[  (q-p_{\widehat{\pi}})^\top \H H(p_{\widehat{\pi}} ) (q-p_{\widehat{\pi}}) \right] - \frac{1}{2}  \E_{\widehat{\pi}}\left[ \sup_{t \in[0,1]}\|\H H(r_{q}(t)) - \H H(p_{\widehat{\pi}}) \| \cdot \|q-p_{\widehat{\pi}}\|^2\right] \\
    %- \sup_{t\in[0,1]} \|\H(r(t)) - \H H(p_{\widehat{\pi}}) \| \cdot \frac{1}{2}\text{Var}(\widehat{\pi}) \\
    %%%
    & \geq  \frac{1}{2} \E_{\widehat{\pi}} \left[  (q-p_{\widehat{\pi}})^\top \H H(p_{\widehat{\pi}} ) (q-p_{\widehat{\pi}}) \right] - \epsilon  \text{Var}(\widehat{\pi}) ,
    \end{align*}
    where the first line is by definition of $C^H_\text{ups}$ and $p_{\widehat{\pi}} = \E_{\widehat{\pi}}[q]$, the second line is by the Fundamental Theorem of Calculus,\footnote{Take any $q \in \supp(\widehat{\pi})$. Define $D_H(r_q(t) \mid p_{\widehat{\pi}}) := H(r_q(t)) - H(p_{\widehat{\pi}}) - \nabla H(p_{\widehat{\pi}}) \cdot (r_q(t)-p_{\widehat{\pi}})$ and $f : [0,1] \to \mathbb{R}$ as $f(t):= D_H(r_q(t) \mid p_{\widehat{\pi}})$. Note that $f \in \mathbf{C}^2([0,1])$ because $H\in\mathbf{C}^2(W)$ and $r_q(t) \in B_{\delta(p)}(p)\subseteq W$ for all $t \in[0,1]$ (by convexity of the ball). Namely, $f(0) = f'(0) = 0$ and $f''(t) = (q-p_{\widehat{\pi}})^\top \H H(r_q(t)) (q-p_{\widehat{\pi}})$ for all $t \in [0,1]$. Meanwhile, the Fundamental Theorem of Calculus applied to $f \in \mathbf{C}^2([0,1])$ and $f' \in \mathbf{C}^1([0,1])$ yields $f(1) = f(0) + \int_0^1 f'(s) \dd s$ and $f'(s) = f'(0) + \int_0^s f''(t) \dd t$, respectively.  Therefore, we obtain $D_H(q \mid p_{\widehat{\pi}}) = f(1) = \int_0^1 \big[ \int_0^s f''(t) \dd t\big] \dd s = \int_0^1 \big[ \int_t^1 f''(t) \dd s\big] \dd t = \int_0^1 (1-t) f''(t) \dd t = \int_0^1 (1-t) (q-p_{\widehat{\pi}})^\top \H H(r_q(t)) (q-p_{\widehat{\pi}}) \dd t$, where the third equality changes the order of integration. \label{fn:FTC-hessian}} 
    the third line rearranges terms and uses $\int_0^1 (1-t) \dd t = \frac{1}{2}$, the fourth line uses the definition of the matrix semi-norm and $\int_0^1 (1-t) \dd t = \frac{1}{2}$, and the final line holds because $\|\H H(r_{q}(t)) - \H H(p_{\widehat{\pi}})\| \leq \|\H H(r_{q}(t)) - \H H(p)\| + \|\H H(p_{\widehat{\pi}}) - \H(p)\| \leq 2\epsilon$ for all $q \in \supp(\widehat{\pi})$ and $t \in [0,1]$ by the triangle inequality and property (b) in the definition of $\delta(p)$ (where $p \in V$ is the same as in the preceding display, and $r_{q}(t) \in B_{\delta(p)}(p)$ for all $q \in \supp(\widehat{\pi})$ and $t \in [0,1]$ by $\supp(\widehat{\pi}) \subseteq B_{\delta(p)}(p)$ and convexity of the ball). Combining the two displays above yields $C(\widehat{\pi}) \leq C^H_\text{ups}(C)(\widehat{\pi}) + 2\epsilon \text{Var}(\widehat{\pi})$. Since the given $\widehat{\pi}$ was arbitrary, we conclude that \eqref{eqn:UK-p-pi} holds, as desired. 
\end{proof}

\subsubsection{Proof of \cref{lem:lqk-invariance-lem}}\label{sssec:lqk-invariance-lem-proof}

%We invoke the following lemma during the proof of \cref{lem:lqk-invariance-lem} in \cref{app:thm3-2:proof}:

To prove \cref{lem:lqk-invariance-lem}, we require the following technical lemma:

\begin{lem}\label{lem:convex}
    For any $p_0 \in \Delta(\Theta)$, symmetric matrix $M \in \mathbb{R}^{|\Theta| \times |\Theta|}$ such that $M p_0 = \mathbf{0}$ and $M \gg_\text{psd} \mathbf{0}$, and scalar $\chi >0$, there exists an $H_\chi \in \mathbf{C}^2(\Delta(\Theta))$ such that (a) $\mathbf{0} \leq_\text{psd} \H H_\chi (p) \leq_\text{psd} M$ for all $p \in \Delta(\Theta)$, (b) $\H H_\chi (p_0) = M$, and (c) $\| \H H_\chi (p) \| \leq \chi$ for all $p \notin B_\chi (p_0)$. 
\end{lem}
\begin{proof}
See \cref{ssec:proof-lem-convex} below. The proof is by construction. 
\end{proof}

In words, \cref{lem:convex} states that the function $H_\chi$ is convex and $\mathbf{C}^2$-smooth on the entire simplex; its Hessian is maximized at the given belief $p_0$, where it equals the given matrix $M$; and $H_\chi$ is approximately linear outside the ball of radius $\chi$ around $p_0$.

\begin{proof}[Proof of \cref{lem:lqk-invariance-lem}]
    Let a \nameref{defi:sp} $C \in \C$, $p_0 \in \Delta(\Theta)$, $\xi >0$, and lower kernel $k(p_0)$ of $C$ at $p_0$ with $k(p_0) - \xi I(p_0) \gg_\text{psd} \mathbf{0}$ be given. For every $\chi >0$, \cref{lem:convex} implies that there exists an $H_\chi \in \mathbf{C}^2(\Delta(\Theta))$ such that (a) $\mathbf{0} \leq_\text{psd} \H H_\chi (p) \leq_\text{psd} k(p_0) - \xi I(p_0)$ for all $p \in \Delta(\Theta)$, (b) $\H H_\chi (p_0) = k(p_0) - \xi I(p_0)$, and (c) $\| \H H_\chi (p) \| \leq \chi$ for all $p \notin B_\chi (p_0)$. 
    %\footnote{In words,  $H_\chi$ is convex and $\mathbf{C}^2$-smooth on the entire simplex; its Hessian is maximized at $p_0$, where it equals $k(p_0) - \xi I(p_0)$; and $H_\chi$ is approximately linear outside the ball of radius $\chi$ around $p_0$.} 
    By property (b), every such $H_\chi$ satisfies the desired condition (ii). Thus, it suffices to show that we can choose $\chi >0$ small enough that $C \succeq C^{H_\chi}_\text{ups}$, i.e., condition (i) %in the present lemma 
    also holds.

To this end, first observe that, since $\xi >0$, there exists an $\epsilon>0$ such that\footnote{In particular, \eqref{eqn:lqk} holds if and only if $\epsilon>0$ satisfies $\xi \geq \|k(p_0)\| \epsilon + 2 (1-\epsilon) \epsilon$.}
    \begin{align}
		(1-\epsilon)\left(k(p_0) -2\epsilon I(p_0)\right) \ge_\text{psd} k(p_0) -\xi I(p_0). \label{eqn:lqk}
	\end{align}
    Since $k(p_0)$ is a lower kernel of $C$ at $p_0$, there exists a $\delta>0$ such that the lower kernel bound in \cref{defi:lq}(ii) holds %for $C$ and $k(p_0)$ 
    at $p_0$ with error parameters $\epsilon$ and $\delta$. Fix some $\delta' \in (0, \delta)$. We show that $\chi>0$ can be chosen small enough (relative to $\delta$ and $\delta'$) in three steps.

    \noindent \textbf{Step 1: Useful Bounds.} Given any $\chi >0$ and $p,q\in \Delta(\Theta)$, define $D_\chi(q \mid p) := H_\chi(q) - H_\chi (p) - \nabla H_\chi(p) (q-p)$. By the Fundamental Theorem of Calculus,\footnote{Take any  $p,q\in\Delta(\Theta)$, define $f : [0,1] \to \mathbb{R}$ as $f(t):= D_\chi(r(t) \mid p)$, and reason as in \cref{fn:FTC-hessian} (see \cref{app:thm3-1:extra}). 
    %Note that $f \in \mathbf{C}^2([0,1])$ because $H\in\mathbf{C}^2(\Delta(\Theta))$. In particular, $f(0) = f'(0) = 0$ and $f''(t) = (q-p)^\top \H H_\chi(r(t)) (q-p)$ for all $t \in [0,1]$. Meanwhile, the fundamental theorem of calculus applied to $f \in \mathbf{C}^2([0,1])$ and $f' \in \mathbf{C}^1([0,1])$ yields $f(1) = f(0) + \int_0^1 f'(s) \dd s$ and $f'(s) = f'(0) + \int_0^s f''(t) \dd t$, respectively.  Therefore, we obtain $D_\chi(q \mid p) = f(1) = \int_0^1 \big[ \int_0^s f''(t) \dd t\big] \dd s = \int_0^1 \big[ \int_t^1 f''(t) \dd s\big] \dd t = \int_0^1 (1-t) f''(t) \dd t = \int_0^1 (1-t) (q-p)^\top \H H_\chi(r(t)) (q-p) \dd t$ as desired, where the third equality changes the order of integration. \label{fn:FTC-hessian}
    }
    \begin{align}
    D_\chi(q \mid p) = \int_0^1 (1-t) (q-p)^\top \H H_\chi \left( r(t) \right) (q-p) \dd t , \quad \text{where } \ r(t) := p + t (q-p). \label{ftc-D-chi}
    \end{align}
    Together, \eqref{ftc-D-chi} and properties (a)--(c) of $H_\chi$ above yield three upper bounds on $D_\chi(q\mid p)$. 

    First, plugging property (a) into \eqref{ftc-D-chi} and noting that $\int_0^1 (1-t) \dd t = \frac{1}{2}$ delivers
    \begin{align}
        D_\chi (q \mid p) \leq \frac{1}{2} (q-p)^\top \left( k(p_0) - \xi I(p_0) \right) (q-p) \quad \forall \, \chi >0 \text{ and } p,q \in \Delta(\Theta). \label{D-chi-bound1}
    \end{align}

    Second, plugging properties (a) and (c) into \eqref{ftc-D-chi} delivers
    \[
    \hspace{-1em} D_\chi (q \mid p) \leq \|k(p_0) - \xi I(p_0)\| \, \cdot \, \|q-p\|^2   \int_0^1 (1-t)  \mathbf{1}\left( r(t) \in B_\chi (p_0) \right) \dd t \ + \ \chi \cdot \|q-p\|^2   \int_0^1 (1-t)  \mathbf{1}\left( r(t) \not \in B_\chi (p_0) \right) \dd t.
    \]
    Consider the nontrivial case in which $q \neq p$. (For $q=p$, we trivially have $D_\chi(q \mid p) = 0$.) In the first term, the integral is bounded above by $\int_0^1 \mathbf{1}\left( r(t) \in B_\chi (p_0) \right) \dd t \leq \frac{2 \chi}{\|q-p\|}$, where the inequality holds by the definition of the path $r(\cdot)$ and the fact that $\text{diam}(B_\chi(p_0)) = 2\chi$.\footnote{Let $\underline{t} := \inf\{t \in [0,1] \mid r(t) \in B_\chi(p_0)\}$ and $\overline{t} := \sup\{t \in [0,1] \mid r(t) \in B_\chi(p_0)\}$. Since $B_\chi(p_0)$ is convex, $r(t) \in B_\chi(p_0)$ for all $t \in (\underline{t},\overline{t})$. Since $\text{diam}(B_\chi(p_0)) = 2\chi$, $\|r(\overline{t}) - r(\underline{t})\| = ( \overline{t} - \underline{t})  \| q-p\| \leq 2 \chi$. Therefore, $ \int_0^1 \mathbf{1}\left( r(t) \in B_\chi (p_0)\right) \dd t =  \int_{\underline{t}}^{\overline{t}} \dd t = \overline{t} - \underline{t} \leq \frac{2\chi}{\|q-p\|}$.} In the second term, the integral is clearly bounded above by $\int_0^1 (1-t) \dd t = \frac{1}{2}$ and, since $\text{diam}(\Delta(\Theta)) = \sqrt{2}$, we have $\|q-p\|^2 \leq \sqrt{2} \, \|q-p\|$. It follows that
    \begin{align}
    \begin{split}\label{D-chi-bound2}
        \hspace{-2em} D_\chi (q \mid p) \leq \chi \cdot A \cdot \|q-p\| \quad &\forall \, \chi >0 \text{ and } p,q \in \Delta(\Theta), 
        \\
        &\text{ where }  A  := 2\,  \|k(p_0) - \xi I(p_0) \| + \frac{1}{\sqrt{2}} >0. 
        \end{split}
    \end{align}

    Third, consider any $p \notin B_{\delta'}(p_0)$, $\chi \in (0,\delta')$, and $q \in B_{\delta'-\chi}(p)$. Since $B_{\delta'-\chi}(p) \cap B_\chi (p_0) = \emptyset$, we have $r(t) \notin B_\chi(p_0)$ for all $t \in [0,1]$. Thus, the display below \eqref{D-chi-bound1} delivers 
    \begin{align}
    D_\chi(q \mid p) \leq \frac{\chi}{2} \cdot \|q - p\|^2 \quad \forall\, \chi \in (0, \delta') \text{ and } p \notin B_{\delta'}(p_0), \ q \in B_{\delta'-\chi}(p). \label{D-chi-bound3}
    \end{align}
    We use the upper bounds \eqref{D-chi-bound1}, \eqref{D-chi-bound2}, and \eqref{D-chi-bound3} in Steps 2 and 3 below. 
    
    \noindent \textbf{Step 2: Let $\pi \in \Ex$ satisfy $p_\pi \in B_{\delta'}(p_0)$.} For every $\chi >0$, we have
        \begin{align*}
	\hspace{-1em}	C^{H_\chi}_\text{ups} (\pi) %&=\E_{\pi}  \left[ D_\chi(q \mid p_\pi) \right] \\
		&=\int_{q\in B_{\delta}(p_0)}  D_\chi(q \mid p_\pi)  \dd \pi(q)+\int_{q\not\in B_{\delta}(p_0)} D_\chi(q \mid p_\pi) \dd \pi(q)\\
		&\le \int_{q\in B_{\delta}(p_0)}\frac{1}{2}(q-p_\pi)^\top \left( k(p_0) -\xi I(p_0) \right)(q-p_\pi ) \dd \pi(q) + \int_{q\not\in B_{\delta}(p_0)}\chi \cdot A \cdot \| q - p_\pi \| \dd \pi(q)\\
		%& \le \int_{q\in B_{\delta}(p_0)} \frac{1}{2} (q-p_\pi)^\top \left( k(p_0) -\xi I \right)(q-p_\pi )\dd \pi(q) + \chi \cdot \sqrt{2} \, A \cdot \pi(\Delta(\Theta)\setminus B_{\delta}(p_0))\\
		& \le (1-\epsilon) \int_{q\in B_{\delta}(p_0)}(q-p_\pi )^\top \left(\frac{1}{2}k(p_0) -\epsilon I \right)(q-p_\pi) \dd \pi(q) + \frac{\chi \cdot  A}{\delta-\delta'} \int_{q\not\in B_{\delta}(p_0)} \|q-p_\pi\|^2 \dd \pi(q)\\
		&\le  (1-\epsilon) C(\pi)+ \frac{\chi \cdot  A}{m \cdot (\delta-\delta')} C(\pi) \qquad \text{for some $m >0$,}
	\end{align*}
        where the first line holds because $C^{H_\chi}_\text{ups}(\pi) = \E_{\pi}[ D_\chi (q \mid p_{\pi})]$, the second line is by \eqref{D-chi-bound1} (first term) and \eqref{D-chi-bound2} (second term), the third line is by \eqref{eqn:lqk} and $I(p_0) \sim_\text{psd} I$ (first term) and the fact that $\|q - p_\pi\| \geq \delta - \delta' >0$ for all $p_\pi \in B_{\delta'} (p_0) \subsetneq B_\delta (p_0)$ and $q \not\in B_\delta (p_0)$ (second term), and the final line holds because, by definition of $\delta$, the lower kernel bound in \cref{defi:lq}(ii) holds for $C$ and $k(p_0)$ at $p_0$ with error parameters $\epsilon$ and $\delta$ (first term) and because $\int_{q\not\in B_{\delta}(p_0)}\|q-p_\pi\|^2\dd \pi(q) \leq \text{Var}(\pi)$ and $C$ is \nameref{defi:sp} (second term). Thus, for any $\chi \in (0,\chi_1]$ where $\chi_1 :=  \frac{m \cdot (\delta -\delta')}{A} \epsilon>0$, $C^{H_{\chi}}_\text{ups}(\pi) \leq C(\pi)$ for all $\pi \in \Ex$ with $p_\pi \in B_{\delta'}(p_0)$. 

     \noindent \textbf{Step 3: Let $\pi \in \Ex$ satisfy $p_\pi \not\in B_{\delta'}(p_0)$.} For every $\chi \in (0, \delta')$, we have 
	\begin{align*}
		D_\chi (q \mid p_\pi) & \leq \mathbf{1}\left( q\in B_{\delta'-\chi}(p_\pi)\right) \frac{\chi}{2} \cdot \| q-p_\pi\|^2 + \mathbf{1}\left(q\not\in B_{\delta'-\chi}(p_\pi)\right) \chi \cdot A \cdot \|q-p_\pi\|\\
		%\le&\mathbf{1}\left(q\in B_{\delta'-\chi}(p_\pi)\right) \chi\cdot \|q-p_\pi \|^2 + \mathbf{1}\left( q\not\in B_{\delta'-\chi}(p_\pi)\right) \chi (||k-\xi||+1)\cdot \|q-p_\pi\| \\
		& \leq \mathbf{1}\left(q\in B_{\delta'-\chi}(p_\pi)\right) \frac{\chi}{2} \cdot \|q-p_\pi \|^2 + \mathbf{1}\left(q\not\in B_{\delta'-\chi}(p_\pi)\right)\chi \cdot A \cdot\frac{\|q-p_\pi\|^2}{\delta'-\chi}\\
        & \leq  \max\left\{ \frac{\chi}{2}, \frac{ \chi \cdot A}{\delta' - \chi} \right\} \|q-p_\pi\|^2,
		%\implies &\E_{\pi}[H_{\chi}(q)-H_{\chi}(p)]\le  \max\left\{\chi,\frac{\chi (||k-\xi||+1) }{\delta'-\chi}\right\}\cdot \frac{C(\pi)}{m} 
	\end{align*}
        where the first line is by \eqref{D-chi-bound3} (first term) and \eqref{D-chi-bound2} (second term), the second line holds because $\|q-p_\pi\| \geq \delta'-\chi>0$ for all $q \notin B_{\delta'-\chi}(p_\pi)$, and the final line consolidates terms. Since $C$ is \nameref{defi:sp}, it follows that, for the same $m>0$ as in Step 2 above, 
        \begin{align*}
            C^{H_\chi}_\text{ups}(\pi) = \mathbb{E}_\pi \left[ D_\chi(q\mid p)\right] \leq \max\left\{ \frac{\chi}{2}, \frac{ \chi \cdot A}{\delta' - \chi} \right\} \text{Var}(\pi) \leq \max\left\{ \frac{\chi}{2 m}, \frac{ \chi \cdot A}{m\cdot (\delta' - \chi)} \right\} C(\pi).
        \end{align*}
        Thus, for any $\chi \in (0,\chi_2]$ where $\chi_2 := \min\left\{2 m, \frac{m \delta'}{m+A}\right\} \in (0,\delta')$, we have $C^{H_{\chi}}_\text{ups}(\pi) \leq C(\pi)$ for all $\pi \in \Ex$ with $p_\pi \not\in B_{\delta'}(p_0)$.
	%Fixing $\epsilon,\delta$ and $\delta'$ and picking $\chi$ s.t. $\max\left\{\frac{\chi}{m},\frac{\chi (||k-\xi||+1) }{m(\delta'-\chi)}\right\}\le1$, the last line is lower than $C(\pi)$.
 
    \noindent \textbf{Wrapping up.} Combining Steps 2 and 3 above, we conclude that $C^{H_{\chi}}_\text{ups} \preceq C$ for any $\chi \in (0, \min\{\chi_1, \chi_2\}]$. Thus, for any such $\chi$, setting $H:= H_\chi$ completes the proof.  
\end{proof}

\subsubsection{Proof of \cref{lem:convex}}\label{ssec:proof-lem-convex}

To prove \cref{lem:convex}, it is technically useful to first establish a slightly different result in which we construct a smooth function on the entirety of $\R^{|\Theta|}$ and use a more demanding version of the $\geq_\text{psd}$ order. For symmetric matrices $A,B \in  \R^{|\Theta|\times|\Theta|}$, we let $A \geq^\star_\text{psd} B$ denote that $x^\top A x \geq x^\top B x$ for all $x \in \R^{|\Theta|}$ and let $A >^\star_\text{psd} B$ denote that $x^\top A x > x^\top B x$ for all $x \in \R^{|\Theta|}\backslash\{\mathbf{0}\}$. Observe that $A \geq^\star_\text{psd} B$ (resp. $A >^\star_\text{psd} B$) implies that $A \geq_\text{psd} B$ (resp. $A >_\text{psd} B$), but not necessarily conversely, because $\mathcal{T} =\{x \in \R^{|\Theta|} \mid \mathbf{1}^\top x = 0\} \subsetneq \R^{|\Theta|}$. We then have: 

\begin{lem}\label{lem:convex-fullspace}
    For every $x_0 \in \R^{|\Theta|}$, symmetric matrix $M \in \R^{|\Theta|\times|\Theta|}$ such that $M >^\star_\text{psd} \mathbf{0}$, and scalar $\chi >0$, there exists an $F_\chi \in \mathbf{C}^2(\R^{|\Theta|})$ such that (a) $\mathbf{0} \le^\star_\text{psd} \H F_\chi(x) \le^\star_\text{psd} M$ for all $x \in \R^{|\Theta|}$, (b) $\H F_\chi(x_0) = M$, and (c) $\H F_\chi(x) \le^\star_\text{psd} \chi I$ for all $x \in \R^{|\Theta|}$ such that $\|x - x_0\|\geq \chi$.
\end{lem}

In what follows, we first prove \cref{lem:convex-fullspace} and then use it to prove \cref{lem:convex}.

\begin{proof}[Proof of \cref{lem:convex-fullspace}]
    Let $\chi>0$ be given. We construct the desired function in two steps.

    \noindent \textbf{Step 1: Let $x_0 = \mathbf{0}$ and $M = I$.} For every $\epsilon>0$, define the univariate functions $f_\epsilon  \in \mathbf{C}(\R_+)$, $g_\epsilon \in \mathbf{C}^1(\R_+)$, and $h_\epsilon \in \mathbf{C}^2(\R_+)$ as
     \begin{align*}
		f_\epsilon(t) := 
		\begin{dcases}
			1,  &\text{if } t\in[0, \epsilon/2]\\
			2 - 2t/\epsilon, &\text{if } t\in(\epsilon/2,\epsilon)\\
			0,  &\text{if } t \in [\epsilon, \infty)
		\end{dcases}
        \, ,
        \qquad g_\epsilon(t) := \frac{1}{2 \sqrt{t}} \int_0^t \frac{f_\epsilon(u)}{\sqrt{u}} \dd u, \qquad h_\epsilon(t) := \frac{1}{2} \int_0^t g_\epsilon(u) \dd u.
	\end{align*}
        It can be verified that: (i) $h'_\epsilon(t) = \frac{1}{2}g_\epsilon(t) \in [0,1/2]$ for all $t \geq0$, with $h'_\epsilon(t) = \frac{1}{2}$ for $t \in[0,\epsilon/2)$; (ii) $h_\epsilon''(t) \leq 0$ for all $t \geq0$, with $h_\epsilon''(t) = 0$ for $t \in[0,\epsilon/2)$;  (iii) $2h_\epsilon'(t) + 4 t \cdot h_\epsilon''(t)  = f_\epsilon(t)$ for all $t \geq0$; and (iv) $h_\epsilon(t) = c_0(\epsilon) + c_1 \sqrt{\epsilon} \cdot \sqrt{t}$ for all $t \in[\epsilon, \infty)$, where $c_0(\epsilon) \in \mathbb{R}$ is an $\epsilon$-dependent constant and $c_1 \in\R_{++}$ is an $\epsilon$-independent constant.\footnote{Facts (i) and (iii) hold because, by construction, $h'_\epsilon(t) = \frac{1}{2} g_\epsilon(t)\geq0$ and $4 t \cdot h''_\epsilon(t) = f_\epsilon (t) - g_\epsilon(t)$ for all $t \geq0$. Direct calculation yields $g_\epsilon(t) = 1$ for $t \in [0,\epsilon/2)$, $g_\epsilon(t) = 2 - \frac{2 t}{3 \epsilon} - \frac{2}{3}\sqrt{\frac{\epsilon}{2 t}}$ for $t \in (\epsilon/2, \epsilon]$, and $g_\epsilon(t) = g_\epsilon(\epsilon) \sqrt{\frac{\epsilon}{t}} $ for $t \in (\epsilon, \infty)$. Therefore: (a) $f_\epsilon (t) \leq g_\epsilon(t)$ for all $t \geq 0$, (b) $h_\epsilon(t) = t/2$ for all $t \in [0,\epsilon/2)$, and (c) $g_\epsilon(\epsilon) =  c_1 := \frac{2}{3}\left( 2 - 1/\sqrt{2}\right)>0$ for all $\epsilon>0$. Fact (ii) follows from (a) and (b). Fact (iv) follows from (c) and direct calculation, where $c_0(\epsilon) := h_\epsilon(\epsilon) - \epsilon c_1 $.} We use these facts (i)--(iv) below. 

        For every $\epsilon>0$, define the multivariate function $F^0_\epsilon \in \mathbf{C}^2(\R^{|\Theta|})$ as $F^0_\epsilon(x) := h_\epsilon (\|x\|^2)$. We claim that, for $\epsilon>0$  sufficiently small, $F^0_\epsilon$ satisfies the desired properties (a)--(c).
        
        To this end, let $\epsilon>0$ be a parameter to be chosen later. For all $x \in \R^{|\Theta|}$, we have 
        \begin{align}
        \H F^0_\epsilon(x) =  2 h_\epsilon'(\|x\|^2) I + 4 h_\epsilon'' (\|x\|^2) x x^\top \quad \text{ and } \quad x x^\top \geq^\star_\text{psd}\mathbf{0}. \label{eqn:hess-bump}
        \end{align}
        Facts (i) and (ii) then imply that $
        \H F^0_\epsilon (x) \le^\star_\text{psd} \H F^0_\epsilon(\mathbf{0}) = I$ for all $x \in \R^{|\Theta|}$. Meanwhile, for all $x \in \R^{|\Theta|}\backslash\{\mathbf{0}\}$, facts (i) and (iii) imply that $\mathbf{0} \le^\star_\text{psd} \H F^0_\epsilon (x)$ because, for all $z \in \R^{|\Theta|}$, 
        \begin{align*}
            z^\top \H F^0_\epsilon(x) z  &= 2 h_\epsilon'(\|x\|^2) \cdot \|z\|^2 + 4 h_\epsilon'' (\|x\|^2) \cdot (z^\top x)^2 \\
            & = 2 h_\epsilon'(\|x\|^2) \cdot \|z\|^2 + 4 \|x\|^2 \cdot h_\epsilon'' (\|x\|^2)   \cdot \frac{(z^\top x)^2}{\|x\|^2 }, \\
            %%%     
            & = 2 h_\epsilon'(\|x\|^2) \cdot \left\{ \|z\|^2 - \frac{(z^\top x)^2}{\|x\|^2 }\right\} + f_\epsilon(\|x\|^2) \cdot \frac{(z^\top x)^2}{\|x\|^2} \\
            %%%
            & \geq 0,
        \end{align*}
        where the first line is by definition, the second line rearranges the second term, the third  line uses fact (iii) to substitute out for $4  \|x\|^2 \cdot h''_\epsilon(\|x\|^2)$, and the final line holds by fact (i) and the Cauchy-Schwarz inequality (first term) and because $f_\epsilon(\cdot)\geq 0$ by construction (second term). We conclude that, for any $\epsilon>0$, $F^0_\epsilon$ satisfies properties (a) and (b). As for property (c), observe that \eqref{eqn:hess-bump}, fact (ii), and fact (iv) imply that
        \begin{align}
            \H F^0_\epsilon(x) \le^\star_\text{psd}  2 h_\epsilon'(\|x\|^2) I =   \frac{c_1 \cdot \sqrt{\epsilon}}{\|x\|} \, I \qquad \forall \, x\in \R^{|\Theta|} \ \text{ s.t. } \ \|x\| \geq \sqrt{\epsilon}. \label{eqn:hess-bump-bound}
        \end{align}
        Let $\overline{\epsilon}(\chi):= \min\{\chi^2, \chi^4/c_1^2\}>0$. Then property (c) also holds for any $\epsilon \in (0, \overline{\epsilon}(\chi)]$.%, as desired. 

    \noindent \textbf{Step 2: Let $x_0 \in \R^{|\Theta|}$ and $M >^\star_\text{psd} \mathbf{0}$ be arbitrary.} By the Spectral Theorem, there exists a diagonal matrix $\Lambda \in \R^{|\Theta|\times|\Theta|}$ with $\Lambda >^\star_\text{psd} \mathbf{0}$ %(i.e., its diagonal entries are all strictly positive) 
    and an orthonormal matrix $U \in \R^{|\Theta|\times|\Theta|}$ such that $M = U \Lambda^2 U^\top$. For every $\epsilon>0$, define $F_\epsilon \in \mathbf{C}^2(\R^{|\Theta|})$ as $F_\epsilon(x) := F^0_\epsilon (\Lambda U^\top (x-x_0))$. %, so that
   % \[
    %\H F_\epsilon(x) = U^\top \Lambda \H %F^0_\epsilon (x) \Lambda U.
   % \]
    We claim that, for $\epsilon>0$ sufficiently small, $F_\epsilon$ satisfies the desired properties (a)--(c).

    To this end, let $\epsilon>0$ be a parameter to be chosen later. For all $x \in \R^{|\Theta|}$,  we have
    \begin{align}
    \H F_\epsilon(x) = U \Lambda \H F^0_\epsilon (\Lambda U^\top (x - x_0)) \Lambda U^\top. \label{eqn:hess-bump-trans}
    \end{align}
    Since $\mathbf{0}\le^\star_\text{psd} \H F_\epsilon^0(z) \le^\star_\text{psd} I$ for all $z \in \R^{|\Theta|}$ by Step 1, it follows that $\mathbf{0}\le^\star_\text{psd} \H F_\epsilon (x) \le^\star_\text{psd} U \Lambda^2 U^\top = M$ for all $x \in \R^{|\Theta|}$. Likewise, since $\H F_\epsilon^0(\mathbf{0}) = I$ by Step 1, it follows that $ \H F_\epsilon (x_0) = U \Lambda^2 U^\top = M$. We conclude that, for all $\epsilon>0$, $F_\epsilon$ satisfies properties (a) and (b). As for property (c), define $\xi := \min\{z^\top M z \mid z \in \R^{|\Theta|} \text{ s.t. } \|z\|^2 =1\}>0$ and 
    %let $\lambda>0$ denote the smallest diagonal entry of $\Lambda$ and define
    $\delta(\epsilon) := \epsilon / \xi > 0$, so that (i) $\|\Lambda U^\top (x-x_0)\|^2 \geq \xi \|x-x_0\|^2$ and (ii) $\|x-x_0\|^2\geq \delta(\epsilon)$ implies that $\|\Lambda U^\top (x-x_0)\|^2 \geq \epsilon$ for all $x \in \mathbb{R}^{|\Theta|}$.\footnote{We have $\xi>0$ because $z \mapsto z^\top M z$ is continuous and strictly positive (as $M >^\star_\text{psd} \mathbf{0}$) on the compact set $\{z \in \R^{|\Theta|} \mid \|z\|^2 = 1\}$. Fact (i) holds because $\|\Lambda U^\top z\|^2 = z^\top M z\geq \xi \|z\|^2$ for all $z\in\R^{|\Theta|}$. Fact (ii) then follows directly from fact (i).} Then, \eqref{eqn:hess-bump-bound}, \eqref{eqn:hess-bump-trans}, and these facts (i) and (ii) imply %that
    \[
    \H F_\epsilon (x) \le^\star_\text{psd} \frac{c_1 \cdot\sqrt{\epsilon}}{\|\Lambda U^\top (x-x_0)\|} \ M \le^\star_\text{psd}  \frac{c_1 \cdot\sqrt{\delta(\epsilon)}}{\|x-x_0\|} \ M \qquad \forall \, x\in \R^{|\Theta|} \ \text{ s.t. } \ \|x-x_0\| \geq \sqrt{\delta(\epsilon)}.
    \]
    Let $\|M\|^\star := \max\{z^\top M z \mid z \in \mathbb{R}^{|\Theta|} \text{ s.t. } \|z\|^2=1\}  >0$. Since $M \leq^\star_\text{psd} \|M\|^\star \cdot I$, it follows that
    \[
    \H F_\epsilon (x)  \le^\star_\text{psd}  \frac{c_1 \cdot\sqrt{\delta(\epsilon)}\cdot \|M\|^\star}{\|x-x_0\|} \ I \qquad \forall \, x\in \R^{|\Theta|} \ \text{ s.t. } \ \|x-x_0\| \geq \sqrt{\delta(\epsilon)}.
    \]
    Letting $\widehat{\epsilon}(\chi):= \min\left\{1, \left(1/\|M\|^\star\right)^2\right\} \cdot \overline{\epsilon}(\chi)$ (where $\overline{\epsilon}(\chi)>0$ is from Step 1), it follows from the above display that $F_\epsilon$ also satisfies property (c) for any $\epsilon \in (0, \xi \cdot \widehat{\epsilon}(\chi)  ]$.
    %Thus, property (c) also holds for any $\epsilon \in (0, \xi \cdot \overline{\epsilon}(\chi)  ]$ (where $\overline{\epsilon}(\chi)>0$ is from Step 1). 
\end{proof}

\begin{proof}[Proof of \cref{lem:convex}]
    Let such $p_0$, $M$, and $\chi$ be given. Define $\overline{M} \in \R^{|\Theta|\times|\Theta|}$ as $\overline{M} := %(I - \mathbf{1} p_0^\top) M (I - p_0 \mathbf{1}^\top) 
    M + \mathbf{1} \mathbf{1}^\top$. By construction, $\overline{M}$ is symmetric and $\overline{M}\sim_\text{psd} M$. We claim that $\overline{M} >^\star_\text{psd} \mathbf{0}$. To this end, let $x \in \R^{|\Theta|}\backslash\{\mathbf{0}\}$ be given. First, if $x \in \mathcal{T}$, then $x^\top \overline{M} x = x^\top M x >0$ because $\overline{M} \sim_\text{psd} M$ and $M \gg_\text{psd} \mathbf{0}$. Second, if $x \notin \mathcal{T}$, then 
    \[
    x^\top \overline{M}x = (x \cdot \mathbf{1})^2 \left( \frac{x^\top}{x \cdot \mathbf{1}} - p_0^\top \right) M \left(\frac{x}{x \cdot \mathbf{1}} - p_0 \right) + (x \cdot \mathbf{1})^2 \geq (x \cdot \mathbf{1})^2 >0,
    \]
    where the equality is by $Mp_0 = \mathbf{0}$ (and symmetry of $M$), the weak inequality holds because $\frac{x}{x \cdot \mathbf{1}} - p_0 \in \mathcal{T}$ and $M \gg_\text{psd} \mathbf{0}$, and the strict inequality is by $x \notin \mathcal{T}$. This proves the claim.

    \cref{lem:convex-fullspace} then implies that there exists an $F_\chi \in \mathbf{C}^2 (\R^{|\Theta|})$ such that: (a') $\mathbf{0} \le^\star_\text{psd} \H F_\chi(x) \le^\star_\text{psd} \overline{M}$ for all $x \in \R^{|\Theta|}$, (b') $\H F_\chi(p_0) = \overline{M}$, and (c') $\H F_\chi(x) \le^\star_\text{psd} \chi I$ for all $x \in \R^{|\Theta|}$ such that $\|x - p_0\|\geq \chi$. Let $H_\chi := F_\chi |_{\Delta(\Theta)} \in \mathbf{C}^2(\Delta(\Theta))$ be the restriction of $F_\chi$ to $\Delta(\Theta) \subsetneq \R^{|\Theta|}$, and normalize $\H H_\chi(p) := A(p)^\top \, \H F_\chi(p) \, A(p)$ for all $p \in \Delta(\Theta)$, where we define $A(p) := I - p \mathbf{1}^\top \in \R^{|\Theta| \times |\Theta|}$ (per \cref{remark:kernels}). Then property (a') of $F_\chi$ implies that $H_\chi$ satisfies the desired property (a) because, for every $p \in \Delta(\Theta)$, it holds that:
    \begin{itemize}
        \item[(i)]  $\H F_\chi (p) \le^\star_\text{psd} \overline{M}$ $\implies$ $\H H_\chi (p) \le^\star_\text{psd} A(p)^\top \, \overline{M} \, A(p) $ $\implies$ $\H H_\chi (p) \leq_\text{psd} A(p)^\top \, \overline{M} \, A(p) $,  
        \item[(ii)] $A(p)^\top\,  \overline{M} \, A(p)   = A(p)^\top \, M A(p) \, \sim_\text{psd} M$,
    \end{itemize}
    where point (i) follows from the definitions of $\H H_\chi$ and the $\le^\star_\text{psd}$ and $\leq_\text{psd}$ orders, and point (ii) holds because $A(p)^\top \mathbf{1} \mathbf{1}^\top A(p) = \mathbf{0} \in \R^{|\Theta| \times |\Theta|}$ (first equivalence) and $A(p) y = y$ for all $y \in \mathcal{T}$ (second equivalence). Next, property (b') of $F_\chi$ implies that $H_\chi$ satisfies the desired property (b) because $M = A(p_0)^\top \, \overline{M} \, A(p_0)$ (by the first equality in point (ii) above and the facts that $M$ is symmetric and $Mp_0 = \mathbf{0}$). Finally, property (c') of $F_\chi$ implies that $\H H_\chi(p) \leq_\text{psd} \chi A(p)^\top A(p) = \chi I(p)$ (cf. point (i) above) and therefore that $\|\H H_\chi (p) \| \leq \chi \|I(p)\| = \chi $ for all $p \in \Delta(\Theta)$, i.e., $H_\chi$ satisfies the desired property (c).
    %Thus, properties (a')--(c') of $F_\chi$ imply that $H_\chi$ has the desired properties (a)--(c).\footnote{Property (b) follows directly from pre- and post-multiplying }
\end{proof}

\subsection{Proof of Theorem \ref{thm:trilemma} 
%and Auxiliary Results for \cref{ssec:trilemma}
}\label{proof:trilemma}

Throughout this section, it is convenient to change variables between random posteriors and experiments. To this end, recall that for any experiment $\sigma\in\Se$ and prior $p\in\Delta(\Theta)$, Bayes' rule specifies that the posterior $q^{\sigma,p}(\cdot \mid s) \in \Delta(\Theta)$ conditional on signal $s$ is given by $q^{\sigma,p}(\theta \mid s)= p(\theta) \frac{\d \sigma_{\theta}}{\d \rp{\sigma,p}}(s)$, where $\rp{\sigma,p} := \sum_{\theta \in \Theta} p(\theta) \sigma_\theta \in \Delta(S)$ is the unconditional signal distribution. To streamline notation, we denote $q^{\sigma,p}_s := q^{\sigma,p}(\cdot \mid s)$. The induced random posterior is then defined as $h_B(\sigma,p)(B):=\langle \sigma,p\rangle \left( \left\{s \in S \mid q^{\sigma,p}_s \in B\right\}\right)$ for all Borel $B\subseteq\Delta(\Theta)$.

\subsubsection{Preliminaries}\label{sssec:proof:thm5}

Recall that $C \in \C$ is \nameref{axiom:CMC} if it is both \hyperref[axiom:CMC:0]{CMC} and \nameref{defi:pst:cont}. Also recall from \cref{app:utvm} that $\Se_b\subsetneq \Se$ denotes the subclass of bounded experiments and that Bayes' rule implies $h_B[ \Se_b \times \Delta^\circ(\Theta)] = \Delta(\Delta^\circ(\Theta))$. Consequently, $C \in \C$ has rich domain if and only if $\dom(C)\backslash\Ex^\varnothing = h_B[ \Se_b \times \Delta^\circ(\Theta)]$. We will use this fact freely throughout the proof.

We build on the following lemma, which adapts Theorems 1 and 5 of PST23 to our setting. It states that: (a) all \nameref{axiom:CMC} and \hyperref[axiom:DL]{Dilution Linear} cost functions with rich domain are ``prior-dependent \ref{eqn:LLR} costs,'' and (b) \hyperref[defi:pst:cont]{uTVM-continuity} is automatic when $|\Theta|=2$.

\begin{lem}\label{lem:bayes-LLR}
    %For every $\sigma \in \Se_\circ$ and $p \in \Delta^\circ(\Theta)$, there exists a sequence $(\sigma^n)_{n \in \mathbb{N}} \subseteq \Se_B$ such that (a) $h_B(\sigma^{n+1},p) \geq_\text{mps} h_B(\sigma^n,p)$ for all $n \in \mathbb{N}$ and (b) $h_B (\sigma^n, p) \xrightarrow{w^*} h_B(\sigma, p)$.
    Let $C \in \C$ have rich domain. % (i.e., $\dom(C) = \Delta(\Delta^\circ(\Theta)) \cup \Ex^\varnothing$). 
    Suppose that either: (a) $C$ is \nameref{axiom:CMC} and \hyperref[axiom:DL]{Dilution Linear}, or (b) $|\Theta|=2$ and $C$ is \hyperref[axiom:mono]{Monotone}, \hyperref[axiom:CMC:0]{CMC}, and \hyperref[axiom:DL]{Dilution Linear}. Then there exist functions $\bm{\beta} : \Delta^\circ(\Theta) \to \mathbb{R}_+^{|\Theta| \times |\Theta|}$ and $F_{\bm{\beta}} : \Delta^\circ(\Theta) \times \Delta^\circ(\Theta) \to \mathbb{R}$, where 
     \begin{align}
    F_{\bm{\beta}} (q \mid p) := \sum_{\theta, \theta' \in \Theta} \frac{\beta_{\theta, \theta'}(p)}{p(\theta)} \, q(\theta) \log \left( \frac{q(\theta)}{q(\theta')} \right), \label{F-beta}
     \end{align}
    such that $C$ is \hyperref[eqn:PS]{Posterior Separable} with the divergence $D_{\bm{\beta}}$ defined as
    \begin{align}
    D_{\bm{\beta}}(q \mid p) := F_{\bm{\beta}} (q \mid p) - F_{\bm{\beta}} (p \mid p)  - (q-p)^\top \nabla_1 F_{\bm{\beta}} (p \mid p)  \quad \forall \, p,q \in \Delta^\circ(\Theta). \label{D-beta}
    \end{align}
    Equivalently, for every $\sigma \in \Se_b$ and $p \in \Delta^\circ(\Theta)$,
    \begin{align}
    C\left( h_B(\sigma,p)\right) = \sum_{\theta,\theta' \in \Theta} \beta_{\theta,\theta'}(p) D_\text{KL}(\sigma_\theta \mid \sigma_{\theta'}). \label{eqn:LLR-KLform}
    \end{align}
    Moreover, the coefficients $\beta_{\theta,\theta'} : \Delta^{\circ}(\Theta) \to \R_+$ are unique for all $\theta,\theta' \in\Theta$ such that $\theta\neq\theta'$.
\end{lem}
\begin{proof}
    Let $C \in \C$ with rich domain be given. It suffices to show that \eqref{eqn:LLR-KLform} holds and that the coefficients are unique for $\theta \neq \theta$'; the \hyperref[eqn:PS]{Posterior Separable} representation in \eqref{F-beta}--\eqref{D-beta} then follows from a standard calculation via Bayes' rule. To this end, fix any $p \in \Delta^\circ(\Theta)$ and define $\Ec_p : \Se_b \to \R_+$ as $\Ec_p(\sigma):= C(h_B(\sigma,p))$. (Recall that $h_B[ \Se_b \times \Delta^\circ(\Theta)] = \Delta(\Delta^\circ(\Theta))$.)

    \noindent \textbf{Case (a).} Let $C$ be \nameref{axiom:CMC} and \hyperref[axiom:DL]{Dilution Linear}. By construction, $\Ec_p$ satisfies Axioms 1--4 of PST23 on the domain $\Se_b$. While the statement of PST23's Theorem 1 assumes these axioms hold on the larger domain of finite-moment experiments (i.e., all $\sigma \in \Se$ for which $\max_{\theta\in \Theta}M^\sigma_\theta(\bm{\alpha})<+\infty$ for every $\bm{\alpha} \in (\mathbb{N}\cup\{0\})^{|\Theta|}$), it can be verified that PST23's proof of Theorem 1 applies nearly verbatim when restricted to the smaller domain $\Se_b$.\footnote{We summarize the requisite adjustments here. First, letting $\mathcal{M} \subseteq \R^d$ and $\mathcal{K}\subseteq \R^d$ (for suitable $d \in \mathbb{N}$) be the sets of admissible moments and cumulants defined in PST23's Appendix B.4, define $\widehat{\mathcal{M}} \subseteq \mathcal{M}$ and $\widehat{\mathcal{K}} \subseteq \mathcal{K}$ as the subsets of moments and cumulants that are inducible by bounded experiments. Since the proof of PST23's Lemma 6 only utilizes finite-moment and -support experiments, which are necessarily bounded, it follows that $\widehat{\mathcal{M}} \subseteq \R^d$ has nonempty interior. Hence, by the proof of PST23's Lemma 7, $\widehat{\mathcal{K}} \subseteq \R^d$ also has nonempty interior. Second, observe that $\Se_b$ is closed under finite products and dilutions of experiments. This implies, among other things, that $\widehat{\mathcal{K}}\subseteq\R^d$ is a subsemigroup (as defined in PST23's Appendix C). Third, since the proof of PST23's Lemma 2 only utilizes bounded experiments, we have $\R^{|\Theta| \, (|\Theta|-1)}_{++} \subseteq \{(D_\text{KL}(\sigma_\theta \mid \sigma_\theta'))_{\theta \neq \theta'} \mid \sigma \in \Se_b \}$. Given these facts, it is straightforward to verify that all other results and arguments in PST23's Appendices B--D apply verbatim under the restriction to the domain $\Se_b$ and the corresponding sets $\widehat{\mathcal{M}}\subseteq \mathcal{M}$ and $\widehat{\mathcal{K}}\subseteq\mathcal{K}$ of moments and cumulants. This yields the desired version of PST23's Theorem 1.} Therefore, by applying this minor variant of PST23's Theorem 1 (i.e., under the restriction to domain $\Se_b$) to $\Ec_p$, we obtain the existence and uniqueness of coefficients $\beta_{\theta,\theta'}(p) \in \R_+$ for all $\theta\neq\theta'$ such that $\Ec_p(\sigma) = \sum_{\theta,\theta'\in \Theta} \beta_{\theta,\theta'}(p) D_\text{KL}(\sigma_\theta \mid \sigma_{\theta'})$ for all $\sigma \in \Se_b$.\footnote{Since $D_\text{KL}(\sigma_\theta \mid \sigma_\theta) =0$ for all $\sigma \in \Se_b$ and $\theta \in \Theta$, we can define $\beta_{\theta,\theta}(p) \in\R_+$ arbitrarily for each $\theta \in \Theta$.} %Since the fixed $p \in \Delta^\circ(\Theta)$ was arbitrary, we conclude that \eqref{eqn:LLR-KLform} holds and that the implied maps $\beta_{\theta,\theta'}: \Delta^\circ(\Theta) \to \R_+$ are unique for all $\theta\neq \theta'$, as desired. 

    \noindent \textbf{Case (b).} Let $|\Theta| = 2$ and $C$ be \hyperref[axiom:mono]{Monotone}, \hyperref[axiom:CMC:0]{CMC}, and \hyperref[axiom:DL]{Dilution Linear}. By construction, $\Ec_p$ is Blackwell monotone and satisfies Axioms 2--3 of PST23 on the domain $\Se_b$. Hence, by applying PST23's Theorem 5 to $\Ec_p$, we again obtain the existence and uniqueness of $\beta_{\theta,\theta'}(p) \in \R_+$ for all $\theta\neq\theta'$ such that $\Ec_p(\sigma) = \sum_{\theta,\theta'\in \Theta} \beta_{\theta,\theta'}(p) D_\text{KL}(\sigma_\theta \mid \sigma_{\theta'})$ for all $\sigma \in \Se_b$. 

    \noindent \textbf{Wrapping Up.} In both cases, since the fixed $p \in \Delta^\circ(\Theta)$ was arbitrary, we conclude that \eqref{eqn:LLR-KLform} holds and the implied maps $\beta_{\theta,\theta'}: \Delta^\circ(\Theta) \to \R_+$ are unique for all $\theta\neq \theta'$, as desired.
\end{proof}

With \cref{lem:bayes-LLR} in hand, we now turn to the main proof of \cref{thm:trilemma}. To prove points (i) and (ii), we build on case (a) of \cref{lem:bayes-LLR}. To prove point (iii), which does not assume \hyperref[defi:pst:cont]{uTVM-continuity}, we instead build on case (b) of \cref{lem:bayes-LLR}.

%We note that only points (i) and (ii) require \hyperref[defi:pst:cont]{uTVM-continuity}; the proof of point (iii) does not.

\subsubsection{Proof of Theorem \ref{thm:trilemma}(i) (\nameref{axiom:slp} \& \nameref{axiom:CMC})}\label{sssec:proof-tri-pt1}

\begin{proof}%[Proof of \cref{thm:trilemma}(i) (\nameref{axiom:slp} \& \nameref{axiom:CMC})]

The ``if'' direction is immediate because \nameref{defi:TI} is \nameref{defi:ups} and \nameref{axiom:CMC} (by definition) and every \nameref{defi:ups} cost is \nameref{axiom:slp} (\cref{lem:ups:to:additive}). Here we prove the ``only if'' direction.

Let $C \in \C$ have rich domain and be \nameref{axiom:CMC} and \nameref{axiom:slp}. Since $C$ is \nameref{axiom:slp}, it is \hyperref[axiom:POSL]{Subadditive} (\cref{prop:1}) and thus \hyperref[axiom:DL]{Dilution Linear} (\cref{lem:convex-dl}). By case (a) of \cref{lem:bayes-LLR}, there exists $\bm{\beta} : \Delta^\circ (\Theta) \to \mathbb{R}_+^{|\Theta| \times |\Theta|}$ such that $C$ is \hyperref[eqn:PS]{Posterior Separable} with divergence $D_{\bm{\beta}}$ given by \eqref{F-beta} and \eqref{D-beta}; moreover, the coefficients $\beta_{\theta,\theta'} : \Delta^{\circ}(\Theta) \to \R_+$ are unique for all $\theta,\theta' \in\Theta$ such that $\theta\neq\theta'$. For each $\theta \in \Theta$, we normalize $\beta_{\theta,\theta}: \Delta^{\circ}(\Theta) \to \R_+$ by setting $\beta_{\theta,\theta}(p) := p(\theta)$ for all $p \in \Delta^\circ(\Theta)$; this is without loss of generality, as these terms do not affect \eqref{F-beta} or \eqref{D-beta}. To show that $C$ is a \nameref{defi:TI} cost, it suffices to show that the function $\bm{\gamma} : \Delta^\circ(\Theta) \to \mathbb{R}_+^{|\Theta| \times |\Theta|}$, defined componentwise as $\gamma_{\theta,\theta'} (p) := \beta_{\theta,\theta'}(p) / p(\theta)$, is constant. Note that, by our normalization, we automatically have $\gamma_{\theta,\theta}(\cdot) = 1$ for every $\theta \in \Theta$.

To this end, let $p, p' \in \Delta^\circ(\Theta)$ be given. Fix an arbitrary $\sigma \in \Se_b$ and let $\pi := h_B(\sigma, p')$. By construction, we have $p_\pi = p'$. Since $C$ is \hyperref[axiom:POSL]{Subadditive} and \hyperref[eqn:PS]{Posterior Separable}, \cref{lem:SLP-divergence} (for $W := \Delta^\circ(\Theta)$) implies that $\E_\pi [ D_{\bm{\beta}}(q \mid p)] \leq D_{\bm{\beta}}(p' \mid p) + \E_\pi[ D_{\bm{\beta}}(q \mid p')]$. Letting $\ell_{\theta,\theta'}(q):= q(\theta) \log\left( \frac{q(\theta)}{q(\theta')}\right)$, this inequality is equivalent to
\begin{align}
\sum_{\theta, \theta' \in \Theta} \left( \gamma_{\theta,\theta'}(p) - \gamma_{\theta,\theta'}(p')\right) \cdot \left( \E_\pi \left[ \ell_{\theta,\theta'}(q) \right] - \ell_{\theta,\theta'}(p') \right) \leq 0. \label{LLR-ineq}
\end{align}
We can then compute
\begin{align*}
\E_\pi \left[ \ell_{\theta,\theta'}(q) \right] & = \int_S q_{s}^{\sigma,p'
}(\theta) \log \left(\frac{q_{s}^{\sigma,p'} (\theta)}{q_{s}^{\sigma,p'} (\theta')} \right) \dd \langle \sigma, p'\rangle (s) \\
%%%
& = \int_S p'(\theta) \left[ \log\left( \frac{p'(\theta)}{p'(\theta')}\right) + \log \left( \frac{\dd \sigma_\theta}{\dd \sigma_{\theta'} } (s) \right) \right] \, \dd \sigma_\theta (s) \\
%%%%
& = \ell_{\theta,\theta'}(p') + p'(\theta) 
\, D_\text{KL}(\sigma_\theta \mid \sigma_{\theta'}),
\end{align*}
where the first line is a change of variables, the second line follows from Bayes' rule and the chain rule for Radon-Nikodym derivatives, and the third line follows from definitions. Plugging this into \eqref{LLR-ineq} and recalling that $\sigma \in \Se_b$ was arbitrary, we obtain:
\begin{align}
    \sum_{\theta, \theta' \in \Theta} \left( \gamma_{\theta,\theta'}(p) - \gamma_{\theta,\theta'}(p')\right) \cdot p'(\theta) \cdot  D_\text{KL}(\sigma_\theta \mid \sigma_{\theta'}) \leq 0 \qquad \forall \sigma \in \Se_b. \label{LLR-ineq2}
\end{align}

Suppose, towards a contradiction, that $\gamma_{\tau,\tau'}(p) > \gamma_{\tau,\tau'}(p')$ for some $\tau, \tau' \in \Theta$ with $\tau \neq \tau'$. Since $p' \in \Delta^\circ(\Theta)$, we have $\left( \gamma_{\tau,\tau'}(p) - \gamma_{\tau,\tau'}(p')\right) \cdot p'(\tau)>0$. By (the proof of) Lemma 2 in PST23, for every $\epsilon>0$ and $M>0$ there exists some $\sigma^{\epsilon, M} \in \Se_b$ such that $D_\text{KL}(\sigma^{\epsilon, M}_\tau \mid \sigma^{\epsilon, M}_{\tau'}) = M$ and $D_\text{KL}(\sigma^{\epsilon, M}_\theta \mid \sigma^{\epsilon, M}_{\theta'}) = \epsilon$ for all ordered pairs of (distinct) states $(\theta,\theta') \neq (\tau,\tau')$. Thus, for fixed $\epsilon>0$ and sufficiently large $M>0$, $\sigma^{\epsilon,M}$ yields the desired contradiction to \eqref{LLR-ineq2}. 

We conclude that $\gamma_{\theta,\theta'}(p) \leq \gamma_{\theta,\theta'}(p')$ for all $\theta,\theta' \in \Theta$. By interchanging the roles of $p$ and $p'$ in the above argument, we also obtain $\gamma_{\theta,\theta'}(p) \geq \gamma_{\theta,\theta'}(p')$ for all $\theta,\theta' \in \Theta$. Since $p,p'\in\Delta^\circ(\Theta)$ were arbitrary, it follows that $\bm{\gamma}(\cdot)$ is a constant function, as desired. 
\end{proof}

\subsubsection{Proof of Theorem \ref{thm:trilemma}(ii) (\nameref{axiom:CMC} \& \hyperref[axiom:prior:invariant]{Prior Invariant})}

\begin{proof}%[Proof of \cref{thm:trilemma}(ii) (\nameref{axiom:CMC} \& \hyperref[axiom:prior:invariant]{Prior Invariant})]
The ``if'' direction is immediate. Here, we prove the ``only if'' direction.

Let $C \in \C$ be \hyperref[axiom:prior:invariant]{Prior Invariant}, \nameref{axiom:CMC}, and \hyperref[axiom:DL]{Dilution Linear}. By case (a) of \cref{lem:bayes-LLR}, there exists $\bm{\beta} : \Delta^\circ(\Theta) \to \R^{|\Theta| \times|\Theta|}_+$ such that $C$ has the representation \eqref{eqn:LLR-KLform};  moreover, the coefficients $\beta_{\theta,\theta'} : \Delta^{\circ}(\Theta) \to \R_+$ are unique for all $\theta,\theta' \in\Theta$ such that $\theta\neq\theta'$. To show that $C$ is an \ref{eqn:LLR} cost, it suffices to show that, for every pair $\theta,\theta' \in \Theta$ with $\theta \neq \theta'$, the function $\beta_{\theta,\theta'} : \Delta^\circ(\Theta) \to \R_+$ is constant. To this end, let $\tau, \tau' \in \Theta$ with $\tau \neq \tau'$ and $p, p ' \in \Delta^\circ(\Theta)$ be given. Suppose, towards a contradiction, that $\beta_{\tau,\tau'}(p) \neq \beta_{\tau,\tau'}(p')$. Since $C$ is \hyperref[axiom:prior:invariant]{Prior Invariant}, it follows from \eqref{eqn:LLR-KLform} that
\begin{equation}
\sum_{\theta,\theta' \in \Theta} \left( \beta_{\theta,\theta'}(p) - \beta_{\theta,\theta'}(p') \right) D_\text{KL}(\sigma_\theta \mid \sigma_{\theta'}) = 0 \qquad \forall \, \sigma \in \Se_b. \label{eqn:LLR-PI-eq}
\end{equation}
By (the proof of) Lemma 2 in PST23, for every $\epsilon>0$ and $M>0$ there exists some $\sigma^{\epsilon, M} \in \Se_b$ such that $D_\text{KL}(\sigma^{\epsilon, M}_\tau \mid \sigma^{\epsilon, M}_{\tau'}) = M$ and $D_\text{KL}(\sigma^{\epsilon, M}_\theta \mid \sigma^{\epsilon, M}_{\theta'}) = \epsilon$ for all ordered pairs of (distinct) states $(\theta,\theta') \neq (\tau,\tau')$. For fixed $\epsilon>0$ and sufficiently large $M>0$, $\sigma^{\epsilon,M}$ yields the desired contradiction to \eqref{eqn:LLR-PI-eq}. We conclude that $\beta_{\tau,\tau'}(p) = \beta_{\tau,\tau'}(p')$. Thus, $C$ is an \ref{eqn:LLR} cost. 
\end{proof}

\subsubsection{Proof of Theorem \ref{thm:trilemma}(iii) (\nameref{axiom:slp} \& \hyperref[axiom:prior:invariant]{Prior Invariant})}\label{sssec:proof-tri-pt3}

We begin with two lemmas that are used in the proof and may be of separate interest. The first lemma shows that the \nameref{defi:MLR} divergence $D_\text{MLR}$ in \cref{defi:MLR} is a quasi-metric. 

\begin{lem}\label{lem:MLR-quasimetric}
    The \nameref{defi:MLR} divergence $D_\text{MLR}$ is a quasi-metric. 
\end{lem}

Note that, since the \ref{eqn:TV} divergence $D_\text{TV}$ in \cref{eg:Poisson:0} is a special case of \nameref{defi:MLR} divergence, \cref{lem:MLR-quasimetric} also implies that $D_\text{TV}$ is a quasi-metric, as claimed in \cref{ssec:examples-revisit}.

%We begin with a lemma that 
The second lemma lets us ``bootstrap'' case (b) of \cref{lem:bayes-LLR} from binary state spaces to general state spaces. For each $\theta \in \Theta$, we denote by $\Se(\theta)$ the subclass of experiments $\sigma\in \Se$ such that $\sigma_{\theta'} = \sigma_{\theta''}$ for all $\theta',\theta'' \in \Theta\backslash\{\theta\}$ (i.e., experiments that may distinguish between the events $\{\theta\}$ and $\Theta \backslash\{\theta\}$, but are uninformative about the state within $\Theta \backslash\{\theta\}$). For each $\theta \in \Theta$, we also denote by $\Se_b(\theta) := \Se(\theta) \cap \Se_b$ the subclass of such experiments that are bounded. We call $C\in \C$ \emph{statewise trivial} if, for every $\theta \in \Theta$, it holds that $C(h_B(\sigma, p)) = 0$ for all $\sigma \in \Se(\theta)$ and $p \in \Delta(\Theta)$ such that $h_B(\sigma,p) \in \dom(C)$.

%Since $C$ has rich domain and is \nameref{axiom:slp} and nontrivial, \cref{lem:pairwise-trivial} in \cref{ssec:lem:trilemma} implies that there exist some $\tau \in \Theta$ and $p^* \in \Delta^\circ(\Theta)$ such that $C(h_B(\cdot, p^*))$ is not identically zero on $\Se_b(\tau) := \Se(\tau) \cap \Se_b$. 

%We invoke the following lemma in (Case 1, Step 2 of) the proof of \cref{thm:trilemma}(iii). Recall that, for every $\theta \in \Theta$: (a) $\Se(\theta)$ denotes the subclass of experiments $\sigma \in \Se$ such that $\sigma_{\theta'} = \sigma_{\theta''}$ for all $\theta', \theta'' \in \Theta \backslash\{\theta\}$, and (b) $\Se_b(\theta)= \Se(\theta) \cap \Se_b$ denotes the subclass of such experiments that are also bounded. We call $C\in \C$ \emph{statewise trivial} if, for every $\theta \in \Theta$, $C(h_B(\sigma, p)) = 0$ for all $\sigma \in \Se(\theta)$ and $p \in \Delta(\Theta)$ such that $h_B(\sigma,p) \in \dom(C)$. 

\begin{lem}\label{lem:pairwise-trivial}
    For any \nameref{axiom:slp} $C \in \C$ with rich domain, %then it is trivial if and only if it is statewise trivial. 
    \[
    \text{$C$ is trivial} \quad \iff \quad \text{$C$ is statewise trivial.}
    \]
    Consequently, such $C$ is nontrivial iff $C(h_B(\sigma,p^*)) \neq 0$ for some $p^* \in \Delta^\circ(\Theta)$ and $\sigma \in \cup_{\theta \in \Theta} \Se_b(\theta)$.
\end{lem}

We first use \cref{lem:MLR-quasimetric,lem:pairwise-trivial} to prove \cref{thm:trilemma}(iii), and then prove the lemmas.

\begin{proof}[Proof of \cref{thm:trilemma}(iii)]
We first show that the rich-domain restriction of any \nameref{defi:MLR} cost is  \hyperref[axiom:prior:invariant]{Prior Invariant} and \nameref{axiom:slp}. Every (full-domain) \nameref{defi:MLR} cost is \hyperref[axiom:prior:invariant]{Prior Invariant} by construction, \hyperref[axiom:mono]{Monotone} by Jensen's inequality because $D_\text{MLR}(\cdot \mid p)$ is convex for each $p \in \Delta(\Theta)$, and \hyperref[axiom:POSL]{Subadditive} because $D_\text{MLR}$ is a quasi-metric and hence satisfies the triangle inequality (\cref{lem:MLR-quasimetric}). Thus, \cref{prop:1} implies that every (full-domain) \nameref{defi:MLR} cost is \hyperref[axiom:prior:invariant]{Prior Invariant} and \nameref{axiom:slp}. It is easy to see that the rich-domain restriction of any full-domain \hyperref[axiom:prior:invariant]{Prior Invariant} cost is also \hyperref[axiom:prior:invariant]{Prior Invariant}.\footnote{Here are the details: Let $C' \in \C$ be \hyperref[axiom:prior:invariant]{Prior Invariant} and have full domain. Define $C \in \C$ as $\dom(C) := \Delta(\Delta^\circ(\Theta))\cup\Ex^\varnothing$ and $C(\pi) := C'(\pi)$ for all $\pi \in \dom(C)$. Since Bayes' rule implies that $h_B[\Se_b \times \Delta^\circ(\Theta)] = \Delta(\Delta^\circ(\Theta))$ and $h_B[\Se\backslash\Se_b \times \Delta^\circ(\Theta)] \cap \Delta(\Delta^\circ(\Theta)) =\emptyset$, we have $C(h_B(\sigma,p)) = C(h_B(\sigma,p'))$ for every $\sigma \in \Se$ and $p,p' \in \Delta^\circ(\Theta)$. Meanwhile, for every $\sigma \in \Se$ and $p \in \Delta(\Theta)$, we have $h_B(\sigma,p) \in \Ex^\varnothing$ if and only if $\sigma_\theta = \sigma_{\theta'}$ for all $\theta,\theta' \in \supp(p)$. Thus, for every $\sigma \in \Se$ and $p,p' \in \Delta(\Theta)\backslash \Delta^\circ(\Theta)$ with $\widehat{\Theta} := \supp(p) = \supp(p')$, we have: (i) $C(h_B(\sigma,p)) = C(h_B(\sigma,p')) = 0$ if $\sigma_\theta = \sigma_{\theta'}$ for all $\theta,\theta' \in \widehat{\Theta}$, and (ii) $C(h_B(\sigma,p)) = C(h_B(\sigma,p')) = +\infty$ if there exists $\theta,\theta' \in \widehat{\Theta}$ such that $\sigma_\theta \neq \sigma_{\theta'}$. We conclude that  $C$ is \hyperref[axiom:prior:invariant]{Prior Invariant}, as desired. \label{fn:PI-rich-dom}} Moreover, \cref{lem:cost-domain-restrict}(iii) in \cref{ssec:proofs-GLM} implies that the rich-domain restriction of any \nameref{axiom:slp} cost is also \nameref{axiom:slp}. The result follows. 

%The sufficiency of the \nameref{defi:MLR} cost is straightforward.\footnote{Per \cref{defi:MLR}, the \nameref{defi:MLR} cost $C_\text{MLR}$ is \hyperref[axiom:prior:invariant]{Prior Invariant} by construction. Let $C \in \C$ be defined as $\dom(C) := \Delta(\Delta^\circ(\Theta))\cup\Ex^\varnothing$ and $C(\pi) := C_\text{MLR}(\pi)$ for all $\pi \in \dom(C)$. Since Bayes' rule implies that $h_B[\Se_b \times \Delta^\circ(\Theta)] = \Delta(\Delta^\circ(\Theta))$ and $h_B[\Se\backslash\Se_b \times \Delta^\circ(\Theta)] \cap \Delta(\Delta^\circ(\Theta)) =\emptyset$, we have $C(h_B(\sigma,p)) = C(h_B(\sigma,p'))$ for every $\sigma \in \Se$ and $p,p' \in \Delta^\circ(\Theta)$. Meanwhile, for every $\sigma \in \Se$ and $p \in \Delta(\Theta)$, we have $h_B(\sigma,p) \in \Ex^\varnothing$ if and only if $\sigma_\theta = \sigma_{\theta'}$ for all $\theta,\theta' \in \supp(p)$. Thus, for every $\sigma \in \Se$ and $p,p' \in \Delta(\Theta)\backslash \Delta^\circ(\Theta)$ with $\widehat{\Theta} := \supp(p) = \supp(p')$, we have: (i) $C(h_B(\sigma,p)) = C(h_B(\sigma,p')) = 0$ if $\sigma_\theta = \sigma_{\theta'}$ for all $\theta,\theta' \in \widehat{\Theta}$, and (ii) $C(h_B(\sigma,p)) = C(h_B(\sigma,p')) = +\infty$ if there exists $\theta,\theta' \in \widehat{\Theta}$ such that $\sigma_\theta \neq \sigma_{\theta'}$. Hence,  $C$ is \hyperref[axiom:prior:invariant]{Prior Invariant}. \label{fn:PI-rich-dom}} 

We now prove the ``converse'' direction. Let $C \in \C$ have rich domain and be \nameref{axiom:slp},  \hyperref[axiom:prior:invariant]{Prior Invariant}, and nontrivial. We proceed by contradiction; there are two cases to consider.

\noindent \textbf{Case 1: Suppose, towards contradiction, that $C$ is \hyperref[axiom:CMC:0]{CMC}.} We first prove the result for the special case of $|\Theta|=2$. We then use this special case to prove the result for general $|\Theta|\geq 2$.

\textbf{\emph{Step 1: Let $|\Theta| =2$.}} Since $C$ is \nameref{axiom:slp}, it is \hyperref[axiom:mono]{Monotone} and \hyperref[axiom:POSL]{Subadditive} (\cref{prop:1}) and thus \hyperref[axiom:DL]{Dilution Linear} (\cref{lem:convex-dl}). Since $C$ is also \hyperref[axiom:CMC:0]{CMC} and \hyperref[axiom:prior:invariant]{Prior Invariant}, case (b) of \cref{lem:bayes-LLR} and the argument from the above proof of \cref{thm:trilemma}(ii) imply that $C$ is an \ref{eqn:LLR} cost, i.e., has the representation in \cref{lem:bayes-LLR} with $\bm{\beta}(\cdot) \equiv \bm{b}$ for some $\bm{b} \in \mathbb{R}^{|\Theta| \times |\Theta|}_+$. Since $C$ is nontrivial, there exist $\tau,\tau' \in \Theta$ with $\tau \neq \tau'$ such that $b_{\tau, \tau'} >0$. Thus, $\gamma_{\tau, \tau'}(p) := b_{\tau,\tau'} / p(\tau)$ is not constant on $\Delta^\circ(\Theta)$. The argument from the above proof of \cref{thm:trilemma}(i) then implies that $C$ is not \nameref{axiom:slp}, yielding the desired contradiction. Thus, $C$ is not \hyperref[axiom:CMC:0]{CMC}.

\textbf{\emph{Step 2: Let $\Theta$ be any finite set.}} We proceed via a reduction to the binary-state case from Step 1. 
%For each $\theta \in \Theta$, denote by $\Se(\theta)$ the subclass of experiments $\sigma\in \Se$ such that $\sigma_{\theta'} = \sigma_{\theta''}$ for all $\theta',\theta'' \in \Theta\backslash\{\theta\}$ (i.e., experiments that may distinguish between the events $\{\theta\}$ and $\Theta \backslash\{\theta\}$, but are uninformative about the state within $\Theta \backslash\{\theta\}$). 
To begin, observe that, since $C$ has rich domain and is \nameref{axiom:slp} and nontrivial, \cref{lem:pairwise-trivial} implies that there exist some $\tau \in \Theta$ and $p^* \in \Delta^\circ(\Theta)$ such that $C(h_B(\cdot, p^*))$ is not identically zero on $\Se_b(\tau) = \Se(\tau) \cap \Se_b$. Fix any such $\tau \in \Theta$ and $p^* \in \Delta^\circ(\Theta)$.

Construct an auxiliary binary state space $\widehat{\Theta} := \{0,1\}$ via the projection $f : \Theta \to \widehat{\Theta}$ defined as $f(\theta) := \mathbf{1}(\theta = \tau)$. Let $\widehat{\Se}_b$ (resp., $\widehat{\Se}$) denote the class of all  %experiments on $\widehat{\Theta}$, and by $\widehat{\Se}_b$ the subclass of 
bounded (resp., all) experiments on $\widehat{\Theta}$. Then the map $F : \Se_b(\tau) \to \widehat{\Se}_b$ given by $F(\sigma)_{f(\theta)} := \sigma_\theta$ is a well-defined bijection. Let  $\Delta^\circ_\tau(\Theta) := \left\{p \in \Delta^\circ(\Theta) \mid \frac{p(\theta)}{p(\theta')} = \frac{p^*(\theta)}{p^*(\theta')} \ \forall \theta,\theta' \in \Theta\backslash\{\tau\}\right\}$. Then the map $G : \Delta^\circ_\tau(\Theta) \to \Delta^\circ(\widehat{\Theta})$ given by $G(p)(1) := p(\tau)$ and $G(p)(0) := 1-p(\tau)$ is a well-defined bijection.\footnote{To see this, note that for every $p \in \Delta_\tau^\circ(\Theta)$ and fixed $\theta' \in \Theta\backslash\{\tau\}$, we have $p(\theta) = \frac{p(\theta')}{p^*(\theta')} p^*(\theta)$ for all $\theta \in \Theta \backslash\{\tau\}$ by construction; summing over $\theta \in \Theta \backslash \{\tau\}$ yields $1-p(\tau) = \frac{p(\theta')}{p^*(\theta')} (1-p^*(\tau))$. The bijectivity of $G$ then directly follows.} Next, define $\widehat{\Ex}:= \Delta(\Delta(\widehat{\Theta}))$ and $\widehat{\Ex}^\varnothing := \bigcup_{\widehat{p} \in \Delta(\widehat{\Theta})} \{\delta_{\widehat{p}}\}$, and let $\widehat{h}_B : \widehat{\Se}\times\Delta(\widehat{\Theta}) \to \widehat{\Ex}$ be the associated Bayesian map. Note that $\widehat{h}_B[\widehat{\Se}_b \times \Delta^\circ(\widehat{\Theta})] = \Delta(\Delta^\circ(\widehat{\Theta}))$. Then the map $\widehat{C} : \widehat{\Ex} \to \overline{\R}_+$ defined as
\[
\widehat{C}\big(\widehat{h}_B(\widehat{\sigma},\widehat{p})\big) := \begin{cases}
    C\left(h_B\left(F^{-1}(\widehat{\sigma}), G^{-1}(\widehat{p}) \right)\right), & \text{if $\widehat{\sigma} \in \widehat{\Se}_b$ and $\widehat{p} \in \Delta^\circ(\widehat{\Theta})$}\\
    0, & \text{if $\widehat{h}_B(\widehat{\sigma},\widehat{p}) \in \widehat{\Ex}^\varnothing$} \\
    +\infty, & \text{otherwise}
\end{cases}
\]
is a well-defined cost function on the auxiliary state space $\widehat{\Theta}$.\footnote{In words, $\widehat{C}$ is the projection onto this auxiliary space of the restriction of $C$ to $h_B[\Se_b(\tau) \times \Delta_\tau^\circ(\Theta)] \cup \Ex^\varnothing \subseteq \Ex$.} By construction, $\widehat{C}$ is nontrivial (viz., $\{0\} \subsetneq \widehat{C}[\Delta(\Delta^\circ(\widehat{\Theta}))] \subseteq \R_+$) and has rich domain (i.e., $\dom(\widehat{C}) = \Delta(\Delta^\circ(\widehat{\Theta})) \cup \widehat{\Ex}^\varnothing$). Since $C$ is \hyperref[axiom:prior:invariant]{Prior Invariant}, $\widehat{C}$ is also \hyperref[axiom:prior:invariant]{Prior Invariant}. Since $C$ is \nameref{axiom:slp} and $\supp(h_B(\sigma,p)) \subseteq \Delta^\circ_\tau(\Theta)$ for all $\sigma \in \Se_b(\tau)$ and $p \in \Delta^\circ_\tau(\Theta)$, it follows that $\widehat{C}$ is also \nameref{axiom:slp}. Finally, since $C$ is \hyperref[axiom:CMC:0]{CMC} and $F^{-1}(\widehat{\sigma}\otimes\widehat{\sigma}') = F^{-1}(\widehat{\sigma}) \otimes F^{-1}(\widehat{\sigma}')$ for all $\widehat{\sigma},\widehat{\sigma}' \in \widehat{\Se}_b$, it follows that $\widehat{C}$ is also \hyperref[axiom:CMC:0]{CMC}. Thus, applying the binary-state argument from Step 1 to $\widehat{C}$ yields the desired contradiction.

\noindent \textbf{Case 2: Suppose, towards contradiction, that $C$ is \nameref{defi:ups}.} We show that this implies that $C$ is \hyperref[axiom:CMC:0]{CMC}; the argument from Case 1 above then yields the desired contradiction. 

To this end, let $p \in \Delta^\circ(\Theta)$ and $\sigma, \sigma' \in \Se_b$ be given. Let $\Pi_{(\sigma,\sigma',p)} \in \Delta(\Ex)$ denote the non-contingent two-step strategy induced by running $\sigma$ at prior $p$ and then running $\sigma'$ regardless of the first-round signal realization; formally, let $\Pi_{(\sigma,\sigma',p)}(B) := \int_{\Delta(\Theta)} \mathbf{1} \big(h_B(\sigma', q) \in B\big) \, \dd h_B(\sigma,p) (q)$ for all Borel $B \subseteq\Ex$%, so that $\pi_2 \in \supp(\Pi_{(\sigma,\sigma',p)})$ iff $\pi_2 = h_B(\sigma',q)$ for some $q \in \supp(\pi_1)$
. Since this strategy is non-contingent, Bayes' rule implies $\E_{\Pi_{(\sigma,\sigma',p)}}[\pi_2] = h_B(\sigma \otimes \sigma', p)$, i.e., the random posterior induced by the two-step strategy equals that induced by running $\sigma$ and $\sigma'$ simultaneously.\footnote{Formally, letting $S$ and $S'$ denote the respective signal spaces of $\sigma$ and $\sigma'$, for all Borel $B \subseteq \Delta(\Theta)$ we have 
\[
\E_{\Pi_{(\sigma,\sigma',p)}}[\pi_2](B) \, = \, \int_{\Delta(\Theta)} h_B(\sigma', q)(B) \dd h_B(\sigma,p)(q) \, = \, \int_{S} \left( \int_{S'} \mathbf{1}\big( q^{\sigma', q^{\sigma,p}_s}_{s'} \in B\big) \dd \langle \sigma', q^{\sigma,p}_s\rangle(s') \right) \dd \langle \sigma,p \rangle(s),
\]
where the first equality holds by definition and the second equality is by a change of variable. Moreover, Bayes' rule implies that: (a) $q^{\sigma', q^{\sigma,p}_s}_{s'} = q^{\sigma\otimes \sigma', p}_{(s,s')}$ for all $(s,s') \in S \times S'$, and (b) $\d \langle \sigma', q^{\sigma,p}_s\rangle(s') \, \dd\langle \sigma,p \rangle (s) = \dd\langle \sigma \otimes \sigma', p\rangle(s,s')$ for all $(s,s') \in S \times S'$. Plugging these identities into the above display then delivers: for all Borel $B \subseteq \Delta(\Theta)$, 
\[
\E_{\Pi_{(\sigma,\sigma',p)}}[\pi_2](B)  \, = \, \int_{S \times S'} \mathbf{1}\big(q^{\sigma\otimes \sigma', p}_{(s,s')} \in B \big) \dd \langle \sigma \otimes \sigma', p\rangle(s,s') \, = \, h_B(\sigma\otimes \sigma',p) (B).
\]
} 
Note that, because $\sigma,\sigma' \in \Se_b$ implies $\sigma \otimes \sigma' \in \Se_b$, it holds that $h_B(\sigma\otimes \sigma',p) \in \Delta^\circ(\Theta)\subseteq \dom(C)$. We then have
\begin{align*}
    C(h_B(\sigma \otimes \sigma', p) )  \, = \, C(\mathbb{E}_{\Pi_{(\sigma,\sigma',p)}}[\pi_2]) \, &= \, C(h_B(\sigma, p)) + \mathbb{E}_{\Pi_{(\sigma,\sigma',p)}}[ C(\pi_2)] \\
    %%%
    &= \, C(h_B(\sigma, p)) + \int_{\Delta(\Theta)} C(h_B(\sigma',q)) \, \dd h_B(\sigma,p)(q) \\
    & = \, C(h_B(\sigma, p)) + C(h_B(\sigma', p)), 
\end{align*}
where the second equality holds by definition of $\Pi_{(\sigma,\sigma',p)}$ and because $C$ is \nameref{defi:ups} and thus \hyperref[axiom:additive]{Additive} (\cref{lem:ups:to:additive}), the third equality is by definition of $\Pi_{(\sigma,\sigma',p)}$, and the final equality holds because $\supp(h_B(\sigma,p)) \subseteq \Delta^\circ(\Theta)$ (since $\sigma \in \Se_b$) and $C$ is \hyperref[axiom:prior:invariant]{Prior Invariant}.\footnote{The expectation in the first line and the integral in the second line are well-defined by \cref{lem:ups:to:additive} (or, alternatively, because $C$ being \hyperref[axiom:prior:invariant]{Prior Invariant} implies that $C(h_B (\sigma',\cdot))$ is constant on $\supp(h_B(\sigma,p)) \subseteq \Delta^\circ(\Theta)$).} Since $p \in \Delta^\circ(\Theta)$ and $\sigma, \sigma' \in \Se_b$ were arbitrary, $\dom(C) = \Delta(\Delta^\circ(\Theta)) \cup\Ex^\varnothing$, and $\Delta(\Delta^\circ(\Theta)) = h_B[ \Se_b \times \Delta^\circ(\Theta)]$, we conclude that $C$ is \hyperref[axiom:CMC:0]{CMC}. This completes the proof. 
\end{proof}

\begin{proof}[Proof of \cref{lem:MLR-quasimetric}]
    It is easy to see that $D_\text{MLR}(q \mid p) = 0$ if and only if $q =p$. We show that $D_\text{MLR}$ satisfies the triangle inequality. That is, for any given $p,q, r \in \Delta(\Theta)$ we claim that
    \begin{equation}\label{eqn:MLR-triangle}
    D_\text{MLR}(q \mid p) \leq D_\text{MLR}(r \mid p) + D_\text{MLR}(q \mid r).
    \end{equation}
    Plugging in the definition of $D_\text{MLR}$, we see that \eqref{eqn:MLR-triangle} holds if and only if
     \begin{equation}\label{eqn:MLR-triangle-2}
    \min_{\theta \in \supp(p)} \frac{r(\theta)}{p(\theta)} + \min_{\theta \in \supp(r)} \frac{q(\theta)}{r(\theta)} \leq  \min_{\theta \in \supp(p)} \frac{q(\theta)}{p(\theta)} + 1.
    \end{equation}
    Thus, it suffices to show that \eqref{eqn:MLR-triangle-2} holds. To this end, we note that
    \begin{align*}
    \min_{\theta \in \supp(p)} \frac{q(\theta)}{p(\theta)} & = \min \left\{ \inf_{\theta \in \supp(p) \cap \supp(r) } \left[\frac{q(\theta)}{r(\theta)} \cdot \frac{r(\theta)}{p(\theta)}\right], \inf_{\theta \in \supp(p) \backslash \supp(r)} \frac{q(\theta)}{p(\theta)}\right\} \\
    %%%
    &\geq \min \left\{ \inf_{\theta \in \supp(p) \cap \supp(r) } \left[\frac{q(\theta)}{r(\theta)} \right] \, \cdot \,  \inf_{\theta \in \supp(p) \cap \supp(r) } \left[\frac{r(\theta)}{p(\theta)}\right], \inf_{\theta \in \supp(p) \backslash \supp(r)} \frac{q(\theta)}{p(\theta)}\right\} \\
    %%%
    & \geq \min \left\{ \min_{\theta \in \supp(r) } \left[\frac{q(\theta)}{r(\theta)} \right] \, \cdot \,  \min_{\theta \in \supp(p)  } \left[\frac{r(\theta)}{p(\theta)}\right], \inf_{\theta \in \supp(p) \backslash \supp(r)} \frac{q(\theta)}{p(\theta)}\right\} \\
    %%%
    & = \min_{\theta \in \supp(r) } \left[\frac{q(\theta)}{r(\theta)} \right] \, \cdot \,  \min_{\theta \in \supp(p)  } \left[\frac{r(\theta)}{p(\theta)}\right],
    \end{align*}
    where the final line holds because $\supp(p)\backslash \supp(r) \neq \emptyset$ only if $\min_{\theta \in \supp(p)  }\frac{r(\theta)}{p(\theta)} = 0$.\footnote{The infima in the first three lines reflect the fact that either $\supp(p)\cap\supp(r)$ or $\supp(p)\backslash \supp(r)$ may be empty.} Consequently, a sufficient condition for \eqref{eqn:MLR-triangle-2} to hold is that
    %\begin{equation}\label{eqn:MLR-triangle-3}
    \[
    \min_{\theta \in \supp(p)} \frac{r(\theta)}{p(\theta)} + \min_{\theta \in \supp(r)} \frac{q(\theta)}{r(\theta)} \, \leq \, \min_{\theta \in \supp(r) } \left[\frac{q(\theta)}{r(\theta)} \right] \, \cdot \,  \min_{\theta \in \supp(p)  } \left[\frac{r(\theta)}{p(\theta)}\right] \,  + \,  1.
    \]
    Moreover, this inequality is equivalent to  
    \[
    0 \, \leq \, \left( 1 - \min_{\theta \in \supp(p)} \frac{r(\theta)}{p(\theta)}\right) \cdot \left( 1 - \min_{\theta \in \supp(r)} \frac{q(\theta)}{r(\theta)}\right),
    \]
    which holds because $p,q,r \in \Delta(\Theta)$ implies that $\max\left\{\min_{\theta \in \supp(p)} \frac{r(\theta)}{p(\theta)}, \min_{\theta \in \supp(r)} \frac{q(\theta)}{r(\theta)}\right\} \leq 1$. We conclude that \eqref{eqn:MLR-triangle-2} holds, and therefore that \eqref{eqn:MLR-triangle} holds. This proves the claim.
\end{proof}

\begin{proof}[Proof of \cref{lem:pairwise-trivial}]
    The ``$\implies$'' direction is immediate (for any $\dom(C)\supseteq\Ex^\varnothing$). For the ``$\impliedby$'' direction, let $C \in \C$ have rich domain, be \nameref{axiom:slp}, and be statewise trivial. If $|\Theta|=2$, the result is immediate. So, suppose that $n:= |\Theta| \geq 3$. For each $\theta \in \Theta$, we define  $\Ex^\circ_b(\theta):= h_B[\Se_b(\theta)\times \Delta^\circ(\Theta)] \subseteq \Delta(\Delta^\circ(\Theta))$. By Bayes' rule, for every $\theta \in \Theta$,
    \[
    \Ex^\circ_b(\theta) = \big\{\pi \in \Delta(\Delta^\circ(\Theta)) \mid \frac{q(\theta')}{q(\theta'')} = \frac{p_\pi(\theta')}{p_\pi(\theta'')} \ \ \ \forall \, \theta', \theta'' \in \Theta \backslash\{\theta\} \ \, \text{ and} \ \, q \in \supp(\pi)\big\}. 
    \]
    Since $C$ has rich domain and is statewise trivial, % (i.e., $\dom(C) = \Delta(\Delta^\circ(\Theta))$), 
    we have $C[\Ex_b(\theta)] = \{0\}$ for all $\theta \in \Theta$. 

    Now, let $\pi \in \dom(C) = \Delta(\Delta^\circ(\Theta))$ be given. We must show that $C(\pi) = 0$. 

    To this end, note that since $\supp(\pi) \subseteq \Delta^\circ(\Theta)$, there exists  $\epsilon>0$ such that $q(\theta) \in [\epsilon, 1-\epsilon]$ for all $q \in \supp(\pi)$ and $\theta \in \Theta$. %It follows that $q(\theta) \in [\epsilon, 1-\epsilon]$ for all $q \in \text{conv}(\supp(\pi))$ and $\theta \in \Theta$. 
    Hence, there exists $\delta \in (0,\epsilon)$ such that, for every $\{p_\theta\}_{\theta \in \Theta} \subseteq \Delta^\circ(\Theta)$ with $p_\theta(\theta) \geq 1-\delta$ for all $\theta \in \Theta$, it holds that: (i) $\text{conv}\left( \{p_\theta\}_{\theta \in \Theta}\right) \supseteq \supp(\pi)$, and (ii) $\{p_\theta\}_{\theta \in \Theta}$ is a linearly independent set.\footnote{Property (ii) holds for sufficiently small $\delta>0$ because $\{\delta_\theta\}_{\theta\in\Theta} \subseteq \Delta(\Theta)$ is a linearly independent set.} For every such $\{p_\theta\}_{\theta \in \Theta} \subseteq \Delta^\circ(\Theta)$, there exists a unique $\pi' \in \Ex$ such that $\supp(\pi') = \{p_\theta\}_{\theta\in\Theta}$ and $p_{\pi'} = p_\pi$. %; moreover, it satisfies $\pi' \geq_\text{mps}\pi$.
    We denote by %the set of such random posteriors as
    \[
    \Ex_\delta := \left\{ \pi' \in \Ex \mid p_{\pi'} = p_\pi \ \ \text{and} \ \ \supp(\pi') = \{p_\theta\}_{\theta\in\Theta} \subseteq \Delta^\circ(\Theta), \, \, \text{where} \, \, p_\theta(\theta) \geq 1-\delta  \ \forall \, \theta\in \Theta\right\}.
    \]
    the set of all such random posteriors. By construction, every $\pi' \in \Ex_\delta$ satisfies $\pi' \geq_\text{mps}\pi$.\footnote{By properties (i) and (ii) above, for each $q \in \supp(\pi)$ there exists a unique $\pi''(\cdot \mid q) \in \Delta(\supp(\pi'))$ with $p_{\pi''(\cdot \mid q)} = q$.}
    
    In what follows, we construct a sequential strategy that implements some $\pi' \in \Ex_\delta$ at zero cost. Enumerate the state space as $\Theta = \{\theta_1, \dots, \theta_n\}$. First, pick any binary-support $\widehat{\pi}_1 \in \Ex^\circ_b(\theta_1)$ such that $p_{\widehat{\pi}_1} = q_0 := p_\pi$ and $\supp(\widehat{\pi}_1) = \{p_1, q_1\}$, where $p_1(\theta_1) \geq 1-\delta$ and $\max_{i \neq 1} \frac{q_1(\theta_1)}{q_1(\theta_i)} \leq \delta/(n-1)$. Next, for every $k \in \{2, \dots, n-1\}$, inductively pick any binary-support $\widehat{\pi}_k \in \Ex^\circ_b(\theta_k)$ such that $p_{\widehat{\pi}_k} = q_{k-1}$ and $\supp(\widehat{\pi}_k) = \{p_k, q_k\}$, where $p_k (\theta_k) \geq 1-\delta$ and $\max_{\ell = k+1, \dots, n} \frac{q_k(\theta_k)}{q_k(\theta_\ell)} \leq \delta / (n-1)$.\footnote{Explicitly, for each $\ell \in \{1,\dots, n\}$, we can construct $\widehat{\pi}_\ell \in \Ex$ satisfying the desired properties as follows: (i) let $p_\ell := \alpha \delta_{\theta_\ell} + (1-\alpha)q_{\ell-1}$ with $\alpha \in (0,1)$ sufficiently close to $1$, (ii) let $q_{\ell} := q_{\ell-1} - \eta (\delta_{\theta_\ell} - q_{\ell-1})$ with $\eta \in \left( 0, \frac{q_{\ell-1}(\theta_\ell)}{1-q_{\ell-1}(\theta_{\ell})} \right)$ sufficiently close to the upper bound, and (iii) then picking $\widehat{\pi}_\ell(\{p_\ell\}) \in (0,1)$  to uniquely solve $p_{\widehat{\pi}_\ell} = q_{\ell-1}$.} We claim that $q_{n-1}(\theta_n) \geq 1-\delta$. To show this, first note that since $\widehat{\pi}_k \in \Ex^\circ_b(\theta_k)$ and $p_{\widehat{\pi}_k} = q_{k-1}$ for every $k \in \{1,\dots, n-1\}$, we have $\frac{q_{n-1}(\theta_\ell)}{q_{n-1}(\theta_n)} = \frac{q_{\ell}(\theta_\ell)}{q_{\ell}(\theta_n)}$ for all $\ell \in \{1,\dots, n-2\}$. It follows that $\max_{k \neq n}\frac{q_{n-1}(\theta_k)}{q_{n-1}(\theta_n)} \leq \delta/(n-1)$, which then implies that $1-q_{n-1}(\theta_n) = \sum_{k=1}^{n-1} q_{n-1}(\theta_k) \leq \delta \cdot q_{n-1}(\theta_n)$. Hence, $q_{n-1}(\theta_n) \geq 1/(1+\delta) \geq 1-\delta$ as desired. 

    Next, inductively define $\{\pi^{(k)}\}_{k=1}^{n-1} \subseteq \Delta(\Delta^\circ(\Theta))$ as follows: let $\pi^{(n-1)} := \widehat{\pi}_{n-1}$ and, for each $k \in \{1, \dots, n-2\}$, let $\pi^{(k)} := \widehat{\pi}_k (p_k) \delta_{p_k} + \left(1-\widehat{\pi}_k (p_k)\right) \pi^{(k+1)}$. By construction, we have $p_{\pi^{(\ell)}}= q_{\ell-1}$ for every $\ell \in \{1,\dots, n-1\}$. For every $k \in \{1,\dots, n-2\}$, also define $\Pi^{(k)} \in \Delta^\dag(\Ex)$ as $\Pi^{(k)}\big(\{\delta_{p_k}\}\big):= \widehat{\pi}_k (p_k)$ and $\Pi^{(k)}\big(\{\pi^{(k+1)}\}\big):= 1-\widehat{\pi}_k (p_k)$, which induces the first-round random posterior $\pi_1 = \widehat{\pi}_k$ and the expected second-round random posterior $\E_{\Pi^{(k)}}[\pi_2 ] = \pi^{(k)}$. Therefore, since $C \in \C$ is \nameref{axiom:slp} and hence \hyperref[axiom:POSL]{Subadditive} (\cref{prop:1}), we obtain:
    \[
    C(\pi^{(k)}) \leq C(\widehat{\pi}_k) %+ \widehat{\pi}_k (p_k) \cdot C(\delta_{p_k}) 
    +  \left(1-\widehat{\pi}_k (p_k)\right)\cdot C(\pi^{(k+1)}) \quad \forall \, k\in \{1, \dots, n-2\}.
    \]
    Since $\widehat{\pi}_\ell \in \Ex_b^\circ(\theta_\ell)$ for all $\ell \in \{1,\dots, n-1\}$ and $C$ is statewise trivial, it follows by induction that $C(\pi^{(\ell)}) = 0$ for all $\ell \in \{1,\dots, n-1\}$. Moreover, we have $\pi^{(1)} \in \Ex_\delta$ because $p_{\pi^{(1)}}= q_0 =  p_\pi$ and $\supp(\pi^{(1)}) = \{p_1, \dots, p_{n-1}, q_{n-1}\}$ by construction. Therefore, $\pi^{(1)} \geq_\text{mps} \pi$. Since $C$ is \nameref{axiom:slp} and hence \hyperref[axiom:mono]{Monotone} (\cref{prop:1}), it follows that $C(\pi) =0$, as desired.
    
    Since the given $\pi \in \dom(C) = \Delta(\Delta^\circ(\Theta))$ was arbitrary, we conclude that $C$ is trivial.
    \end{proof}

\subsection{Proof of Theorem \ref{thm:wald}}\label{proof:wald}

The proof consists of four main steps. First, in \cref{ssec:wald-is-spi}, we show that the \ref{eqn:MS} cost is \nameref{defi:spi}. Second, in \cref{ssec:lpi}, we introduce our main notion of \hyperref[defi:lpi]{Local Prior Invariance} and show that it is satisfied by all \hyperref[axiom:prior:invariant]{Prior Invariant} and \nameref{defi:spi} costs. Third, in \cref{ssec:ups-lpi-wald-local}, we show that the \ref{eqn:MS} cost is the only cost function that is both \nameref{defi:ups} and \hyperref[defi:lpi]{Local Prior Invariant}. Finally, in \cref{ssec:thm6-wrapping-up}, we consolidate these steps into a proof of \cref{thm:wald}. Auxiliary technical facts and proofs are in \dred{Appendices} \ref{ssec:thm6-technical-facts}--\ref{ssec:proof-thm6-technical-facts}.

Following the notation in \cref{proof:trilemma}, for any experiment $\sigma\in\Se$ and prior $p\in\Delta(\Theta)$, we denote by $q^{\sigma,p}_s \in \Delta(\Theta)$ the Bayesian posterior conditional on signal $s$, so that induced random posterior is given by $h_B(\sigma,p)(B)=\langle \sigma,p\rangle \left( \left\{s \in S \mid q^{\sigma,p}_s \in B\right\}\right)$ for all Borel $B\subseteq\Delta(\Theta)$.

\subsubsection{Step 1: \ref{eqn:MS} Costs are \nameref{defi:spi}}\label{ssec:wald-is-spi}

%In this section, we show that the \ref{eqn:MS} cost is \nameref{defi:spi}. Moreover, we characterize the full set of (\nameref{defi:lq}) \hyperref[axiom:prior:invariant]{Prior Invariant} direct costs that generate the \ref{eqn:MS} indirect cost.

%We proceed in two steps. First, we highlight a general approach for checking whether a given cost function is \nameref{defi:spi}. Second, we apply this approach to the \ref{eqn:MS} cost.

We begin with a general approach for checking whether a given cost function is \nameref{defi:spi}. For any $C \in \C$, its \emph{\hyperref[axiom:prior:invariant]{Prior Invariant} upper envelope} (\ref{eqn:PIE}) is the cost function $\overline{C} \in \C$ defined as 
\begin{equation}\label{eqn:PIE}
\overline{C}(h_B(\sigma,p)) := \sup_{p'\in \Delta(\Theta)} C(h_B(\sigma,p')) \quad \text{s.t.} \quad \supp(p') = \supp(p).\footnotemark \tag{PIE}
\end{equation}
\footnotetext{To see that $\overline{C} \in \C$ is a well-defined cost function, it suffices to note that for any $\sigma,\sigma'\in \Se$ and $p \in \Delta(\Theta)$, we have $h_B(\sigma,p) = h_B(\sigma',p)$ \emph{if and only if} $h_B(\sigma,p') = h_B(\sigma',p')$ for every $p' \in \Delta(\Theta)$ with $\supp(p') = \supp(p)$ (cf. \cite{blackwell-experiment51}).}
\ref{eqn:PIE}s satisfy two key properties. First, the \ref{eqn:PIE} of $C \in \C$ is the smallest \hyperref[axiom:prior:invariant]{Prior Invariant} cost that lies above $C$. Second, to check whether $C \in \C$ is \nameref{defi:spi}, it suffices to check whether $C$ is the indirect cost generated by its \ref{eqn:PIE}; no other direct costs need be considered. Formally:

\begin{lem}\label{lem:PIE}
    For any $C \in \C$, the following hold:
    \begin{itemize}[noitemsep]
        \item[(i)] Its \ref{eqn:PIE} satisfies $\overline{C} = \min\{C' \in \C \mid C' \succeq C \text{ and $C'$ is \hyperref[axiom:prior:invariant]{Prior Invariant}}\}$.
        %%%
        \item[(ii)] $C$ is \nameref{defi:spi} if and only if $C = \Phi(\overline{C})$.
    \end{itemize}
    %\[
    %\text{$C$ is \nameref{defi:spi}} \quad \iff \quad C = \Phi(\overline{C}).
   % \]
\end{lem}
\begin{proof}
    We prove each point in turn:

    \noindent \textbf{Point (i).} By construction, $\overline{C}$ is \hyperref[axiom:prior:invariant]{Prior Invariant} and $\overline{C} \succeq C$. Hence, it suffices to show that $ C' \succeq \overline{C}$ for every \hyperref[axiom:prior:invariant]{Prior Invariant} $C' \in \C$ satisfying $C' \succeq C$. To this end, fix any such $C' \in \C$. Let $\pi \in \Ex$, and corresponding $\sigma \in \Se$ such that $h_B(\sigma,p_\pi) = \pi$, be given. We then have
    \[
    C'(h_B(\sigma,p_\pi)) \, = \, C'(h_B(\sigma,p)) \, \geq \, C(h_B(\sigma,p)) \qquad \forall\, p \in \Delta(\Theta) \ \text{ s.t. } \supp(p) = \supp(p_\pi), 
    \]
    where the equality holds because $C'$ is \hyperref[axiom:prior:invariant]{Prior Invariant} and the inequality is by $C'\succeq C$. Taking the supremum over such $p \in \Delta(\Theta)$, we obtain $ C'(h_B(\sigma,p_\pi)) \geq \overline{C}(h_B(\sigma,p_\pi))$. That is, $C'(\pi) \geq \overline{C}(\pi)$. Since the given $\pi \in \Ex$ was arbitrary, we conclude that $C' \succeq \overline{C}$, as desired.

    \noindent \textbf{Point (ii).} The ``if'' direction is trivial. For the ``only if'' direction, suppose that $C$ is \nameref{defi:spi}. Then, by definition, there exists some \hyperref[axiom:prior:invariant]{Prior Invariant} $C' \in \C$ such that $C = \Phi(C')$. Since $C' \succeq \Phi(C')$ by construction, we have $C' \succeq C$. Therefore, point (i) (proved above) implies that $C' \succeq \overline{C}\succeq C$. Since $\Phi$ is isotone (\cref{lem:structure:Phi}) and $C$ is \nameref{axiom:slp} (\cref{prop:1}), it follows that
    \[
    C \, = \, \Phi(C') \, \succeq \Phi(\overline{C}) \, \succeq \, \Phi(C) \, =\, C,
    \]
    where the first equality is by hypothesis. We conclude that $C = \Phi(\overline{C})$, as desired.
\end{proof}

Following the approach suggested by \cref{lem:PIE}, we now prove that the \ref{eqn:MS} cost is \nameref{defi:spi} by showing that it is generated by its \ref{eqn:PIE}. Incidentally, we also characterize the full set of (\nameref{defi:lq}) \hyperref[axiom:prior:invariant]{Prior Invariant} direct costs that generate the \ref{eqn:MS} indirect cost.

%We prove \cref{lem:PIE} at the end of this section, after applying it to the \ref{eqn:MS} cost.

To this end, fix the binary state space $\Theta =\{0,1\}$. Recall from \cref{eg:Diffusion:0} (\cref{ssec:examples-revisit}) that the \ref{eqn:MS} cost $C_\text{Wald} = C^{H_\text{Wald}}_\text{ups}$ is \nameref{defi:ups}, where $H_\text{Wald} \in \mathbf{C}^2(\Delta^\circ(\Theta))$ is defined as
\[
H_\text{Wald} (p) = p(0) \, \log\left(\frac{p(0)}{p(1)} \right) + p(1) \, \log\left(\frac{p(1)}{p(0)} \right) \qquad \text{ for all $p \in \Delta^\circ(\Theta)$.}
\]
Therefore, by \cref{lem:ups-kernel-equiv}, $C_\text{Wald}$ is \nameref{defi:lq} and its kernel,  $k_\text{Wald}$, is given by 
\[
k_\text{Wald}(p) = \H H_\text{Wald}(p) = \diag(p)^{-1}  \begin{bmatrix}
    1 & -1 \\ 
    -1 & 1
\end{bmatrix}
\diag(p)^{-1} \qquad \text{ for all $p \in \Delta^\circ(\Theta)$,}
\]
where the final equality is by direct calculation of the Hessian.

Per \eqref{eqn:MS} in \cref{ssec:leading-examples}, the \ref{eqn:MS} cost $C_\text{Wald}$ can be equivalently represented as
\[
C_\text{Wald}(h_B(\sigma,p)) = p(0) \, D_\text{KL} (\sigma_0 \mid \sigma_1) + p(1) \, D_\text{KL} (\sigma_1 \mid \sigma_0) \qquad \forall \sigma \in \Se_b \, \text{ and } \, p \in \Delta^\circ(\Theta),
\]
where $\Se_b \subsetneq \Se$ is the class of bounded experiments and $h_B[\Se_b \times \Delta^\circ(\Theta)] =  \Delta(\Delta^\circ(\Theta))$ (recall \cref{app:utvm,sssec:proof:thm5}).\footnote{Since $C_\text{Wald}$ has the rich domain $\dom(C_\text{Wald}) = \Delta(\Delta^\circ(\Theta)) \cup \Ex^\varnothing$, this is a full description of the \ref{eqn:MS} cost.} Therefore, by inspection, its \ref{eqn:PIE}  $\overline{C}_\text{Wald}\in\C$ is given by 
\[
    \overline{C}_\text{Wald}(h_B(\sigma, p)) := \max\big\{ D_\text{KL}(\sigma_1 \mid \sigma_0) ,  D_\text{KL}(\sigma_0 \mid \sigma_1) \big\} \qquad \forall \sigma \in \Se_b \, \text{ and } \, p \in \Delta^\circ(\Theta)
    \]
on the rich domain $\dom(\overline{C}_\text{Wald}) = \Delta(\Delta^\circ(\Theta)) \cup \Ex^\varnothing = h_B[\Se_b \times \Delta^\circ(\Theta)] \cup\Ex^\varnothing$. 

By \cref{thm:flie}, to show that $\Phi(\overline{C}_\text{Wald}) = C_\text{Wald}$, it suffices to show that $\overline{C}_\text{Wald}$ \nameref{axiom:flie} and is \nameref{defi:lq} with the same kernel as $C_\text{Wald}$. We verify this in the next lemma.

\begin{lem}\label{lem:wald-is-spi}
The \ref{eqn:MS} cost, $C_\text{Wald} \in \C$, is \nameref{defi:spi}. In particular:
\begin{itemize}[noitemsep]
    \item[(i)] Its \ref{eqn:PIE}, $\overline{C}_\text{Wald}\in\C$, is \nameref{defi:lq} and satisfies
    \[
      \Phi(\overline{C}_\text{Wald}) = C_\text{Wald} \qquad \text{ and } \qquad k_{\overline{C}_\text{Wald}} = k_\text{Wald}.
    \]
    \item[(ii)] For any \hyperref[axiom:prior:invariant]{Prior Invariant} and \nameref{defi:lq} $C \in \C$, 
    \[
    \text{$\Phi(C) =  C_\text{Wald}$} \quad \iff \quad \text{$C \succeq \overline{C}_\text{Wald}$ \ and \ $k_C = k_\text{Wald}$.}
    \]
\end{itemize}
\end{lem}
\begin{remark}
    In \cref{lem:wald-is-spi}, we normalize the coefficient $\gamma \geq 0$ on the \ref{eqn:MS} cost $\gamma \, C_\text{Wald}$ to $\gamma =1$. Since $\Phi$ is positively HD1 (\cref{lem:structure:Phi}), this normalization is without loss of generality.
\end{remark}
\begin{proof}
Since point (i) implies that $C_\text{Wald}$ is \nameref{defi:spi}, it suffices to prove points (i) and (ii). 

\noindent \textbf{Point (i).} We first show that $\overline{C}_\text{Wald}$ is \nameref{defi:lq} with kernel $k_{\overline{C}_\text{Wald}} = k_\text{Wald}$. 

To this end, define the cost functions $C_0 \in \C$ and $C_1 \in \C$ with rich domain as %follows: let $\dom(C_1)  = \dom(C_2)  = \Delta(\Delta^\circ(\Theta)) \cup \Ex^\varnothing$ and
\[
C_0(h_B(\sigma,p)) := D_\text{KL} (\sigma_0 \mid \sigma_1) \quad \text{ and } \quad C_1(h_B(\sigma,p)) := D_\text{KL} (\sigma_1 \mid \sigma_0) \qquad \forall \, \sigma \in \Se_b \ \text{ and } \ p \in \Delta^\circ(\Theta).
\]
We claim that $C_0$ and $C_1$ are both \nameref{defi:lq} with kernel $k_{C_0} = k_{C_1} = k_\text{Wald}$. Given this claim, since we have $\overline{C}_\text{Wald}(\pi) = \max\{C_0(\pi),C_1(\pi)\}$ for all $\pi \in \Ex$ by construction, it follows directly from the definition of kernels (\cref{defi:lq}) that $\overline{C}_\text{Wald}$ is also \nameref{defi:lq} with kernel $k_{\overline{C}_\text{Wald}} = k_\text{Wald}$.\footnote{Since $\overline{C}_\text{Wald} \succeq C_0$ and $\overline{C}_\text{Wald}\succeq C_1$, the claim trivially implies that $k_\text{Wald}$ is a lower kernel of $\overline{C}_\text{Wald}$. To see that $k_\text{Wald}$ is also an upper kernel of $\overline{C}_\text{Wald}$, fix any $p \in \Delta^\circ(\Theta)$ and $\epsilon >0$. Given the claim, for each $i \in\{0,1\}$, there exists $\delta_i >0$ such that the upper kernel inequality in \cref{defi:lq}(i) holds for $C_i$ and $k_\text{Wald}(p)$ at $p$ with error parameters $\epsilon$ and $\delta_i$. Consequently, the upper kernel inequality in \cref{defi:lq}(i) holds for $\overline{C}_\text{Wald}$ and $k_\text{Wald}(p)$ at $p$ with error parameters $\epsilon$ and $\delta:= \min\{\delta_0,\delta_1\}>0$. Thus, since the fixed $p \in \Delta^\circ(\Theta)$ and $\epsilon >0$ were arbitrary,  $k_\text{Wald}$ is an upper kernel of $\overline{C}_\text{Wald}$.} Therefore, it suffices to prove the claim.

To this end, note that $C_0$ is the \ref{eqn:LLR} cost with coefficients $\beta_{01} = 1$ and $\beta_{10} =0$; symmetrically, $C_1$ is the \ref{eqn:LLR} cost with $\beta_{10} = 1$ and $\beta_{01} =0$. Therefore, \cref{lem:bayes-LLR} implies that, for each $i \in \{0,1\}$, the cost function $C_i$ is \hyperref[eqn:PS]{Posterior Separable} with divergence $D_i$ generated (as in \eqref{D-beta}) by the map $F_i : \Delta^\circ(\Theta) \times \Delta^\circ(\Theta) \to \R$ defined (as in \eqref{F-beta}) by
\[
F_0(q \mid p) : = \frac{q(0)}{p(0)} \log \left( \frac{q(0)}{q(1)}\right) \qquad \text{ and } \qquad F_1(q \mid p) : = \frac{q(1)}{p(1)} \log \left( \frac{q(1)}{q(0)}\right).
\]
Note that, for each $i \in \{0,1\}$, $(q,p) \mapsto  \H_1 D_i(q \mid p) = \H_1 F_i(q \mid p)$ is well-defined and continuous on $\Delta^\circ(\Theta) \times \Delta^\circ(\Theta)$. Thus, \cref{lem:ps-kernel-suff} implies that, for each $i \in \{0,1\}$, $k_{C_i}(p) = \H_1 F_i (p \mid p)$ for all $p \in \Delta^\circ(\Theta)$. By direct calculation of the Hessians, we then obtain $k_{C_i}(p) = \H_1 F_i (p \mid p) = k_\text{Wald}(p)$ for all $i \in \{0,1\}$ and $p \in\Delta^\circ{\Theta}$. This proves the claim.

%Now, by construction, $\overline{C}_\text{Wald}(\pi) = \max\{C_0(\pi),C_1(\pi)\}$ for all $\pi \in \Ex$. Therefore, since $C_0$ and $C_1$ are both \nameref{defi:lq} with kernel $k_{C_0} = k_{C_1} = k_\text{Wald}$, it follows directly from \cref{defi:lq} that $\overline{C}_\text{Wald}$ is also \nameref{defi:lq} with kernel $k_{\overline{C}_\text{Wald}} = k_\text{Wald}$, as desired.\footnote{\awb{explain}}

Next, we show that $\Phi(\overline{C}_\text{Wald}) = C_\text{Wald}$. Note that $H_\text{Wald} \in   \mathbf{C}^2(\Delta^\circ(\Theta))$ is strongly convex. Therefore, since $k_\text{Wald} = \H H_\text{Wald}$ (by \cref{lem:ups-kernel-equiv}) and $k_{\overline{C}_\text{Wald}} = k_\text{Wald}$ (as shown above), points (i) and (ii) of \cref{lem:phi-ie} (with $W:= \Delta^\circ(\Theta)$) imply that $\Phi_\text{IE}(\overline{C}_\text{Wald}) = C_\text{Wald}$. Since $\overline{C}_\text{Wald} \succeq C_\text{Wald}$ by construction, it follows that $\overline{C}_\text{Wald}$ \nameref{axiom:flie}. Therefore, the ``$\implies$'' direction of \cref{thm:flie} (again with $W:= \Delta^\circ(\Theta)$) implies that $\Phi(\overline{C}_\text{Wald}) = C_\text{Wald}$, as desired.

\noindent \textbf{Point (ii).} Let $C \in C$ be \hyperref[axiom:prior:invariant]{Prior Invariant} and \nameref{defi:lq}.

\emph{\textbf{($\implies$ direction)}} Suppose that $\Phi(C) = C_\text{Wald}$. Since $C$ is \hyperref[axiom:prior:invariant]{Prior Invariant} and $C \succeq \Phi(C)$, \cref{lem:PIE} implies that $C\succeq \overline{C}_\text{Wald}$, as desired. It follows that $\dom(C) \subseteq \dom(\overline{C}_\text{Wald}) = \Delta(\Delta^\circ(\Theta))\cup\Ex^\varnothing$. Thus, since $C$ is \nameref{defi:lq} and $H_\text{Wald} \in \mathbf{C}^2(\Delta^\circ(\Theta))$ is strongly convex, the ``$\impliedby$'' direction of \cref{thm:flie} %(for $W := \Delta^\circ(\Theta)$) 
yields $k_C = k_\text{Wald}$, as desired. 

\emph{\textbf{($\impliedby$ direction)}} Suppose that $C \succeq \overline{C}_\text{Wald}$ and $k_C = k_\text{Wald}$. The former hypothesis implies that $C \succeq C_\text{Wald}$ and $\dom(C) \subseteq \Delta(\Delta^\circ(\Theta)) \cup\Ex^\varnothing$. Since $H_\text{Wald} \in \mathbf{C}^2(\Delta^\circ(\Theta))$ is strongly convex and the latter hypothesis implies that $k_C = k_\text{Wald} = \H H_\text{Wald}$ (by \cref{lem:ups-kernel-equiv}), points (i) and (ii) of \cref{lem:phi-ie} deliver $\Phi_\text{IE}(C) = C_\text{Wald}$. Hence, $C$ \nameref{axiom:flie} and $k_C = \H H_\text{Wald}$. The ``$\implies$'' direction of \cref{thm:flie} then yields $\Phi(C) = C_\text{Wald}$, as desired. 
\end{proof}

\subsubsection{Step 2: Local Characterization of \hyperref[defi:spi]{(Sequential)} \hyperref[axiom:prior:invariant]{Prior Invariance}
}\label{ssec:lpi}

For any $C \in \C$ and $W \subseteq \Delta^\circ(\Theta)$, we call a matrix-valued function $\kappa:W\to \R^{|\Theta|\times|\Theta|}$ an \dred{\emph{experimental upper (resp., lower) kernel}} of $C$ on $W$ if the map $p \in W \mapsto \diag(p)^{-1} \, \kappa(p) \, \diag(p)^{-1}$ is an upper (resp., lower) kernel of $C$ on $W$. If $\kappa$ is both an upper and lower experimental kernel of $C$ on $W$, then it is the \dred{\emph{experimental kernel}} of $C$ on $W$ and denoted as $\kappa_C := \kappa$. Note that $C$ admits an experiment kernel on $W$ if and only if $C$ is \nameref{defi:lq} on $W$, in which case $\kappa_C(p) = \diag (p) \, k_C(p) \, \diag(p)$ for all $p \in W$ (as described in \cref{ssec:SPI}).\footnote{To illustrate these definitions, let $C \in\C$ be \nameref{defi:lq}. For every $\sigma \in \Se$ and $p \in \Delta^\circ(\Theta)$, Bayes' rule yields
\[
\hspace{-0.5em} \E_{h_B(\sigma,p)}\left[(q -p)^\top k_C(p) (q-p)\right] \, = \, \int_{S} \left(\bm{\ell}^{\sigma,p}(s) - \mathbf{1} \right)^\top \, \kappa_C(p) \, \left(\bm{\ell}^{\sigma,p}(s) - \mathbf{1} \right) \dd \langle \sigma,p \rangle(s), 
\]
where $\bm{\ell}^{\sigma,p}(s) \in \overline{\R}^{|\Theta|}_+$ is the vector of likelihood ratios $\ell_\theta^{\sigma,p}(s) := \frac{\d \sigma_\theta}{\d\langle \sigma,p\rangle}(s)$ between the $\theta$-contingent and unconditional signal distributions at realization $s \in S$. Hence, whereas the kernel $k_C$ provides a local quadratic approximation of $C$ in the space of beliefs, the experimental kernel $\kappa_C$ provides an analogous approximation in the space of such likelihood ratios (wherein ``incremental evidence'' corresponds to experiments for which $\sup_{s \in \supp(\langle \sigma,p\rangle)} \| \bm{\ell}^{\sigma,p}(s) - \mathbf{1}\|\approx 0$).}

\begin{remark}\label{remark:exp-kernel}
    Each experimental upper (resp., lower) kernel $\kappa$ of $C$ on $W$ inherits properties from its corresponding upper (resp., lower) kernel $k$, i.e., $k(p) := \diag(p)^{-1} \, \kappa(p) \, \diag(p)^{-1}$. In particular, under the normalization noted in \cref{remark:kernels}, for every $p \in W$ it holds that: (i) $\kappa(p)$ is symmetric, (ii) $\kappa(p) \mathbf{1} = \mathbf{0}$, and (iii) $k(p) \geq_\text{psd} \mathbf{0}$ only if $x^\top \kappa(p) x\geq0$ for all $x \in \R^{|\Theta|}$.\footnote{Properties (i) and (ii) are immediate. Property (iii) is easy to verify (see, e.g., \cref{lem:psd-star} in \cref{ssec:thm6-technical-facts} below).}
\end{remark}

For any $C \in \C$ and $p \in \Delta^\circ(\Theta)$, we denote by $\underline{\K}_C^+(p) \subseteq \R^{|\Theta|\times|\Theta|}$ the set of all experiment lower kernels $\kappa(p)$ of $C$ at $p$ satisfying $\diag(p)^{-1}\kappa(p) \diag(p)^{-1}\gg_\text{psd}\mathbf{0}$. Note that, if $C \in \C$ is \nameref{defi:sp}, then $\underline{\K}_C^+(p) \neq \emptyset$ for all $p \in \Delta^\circ(\Theta)$.\footnote{This follows from the definition of experimental lower kernels and the fact that, if $C \in \C$ is \nameref{defi:sp}, then there exists a lower kernel $\underline{k}$ of $C$ on $\Delta(\Theta)$ such that $\underline{k}(p)\gg_\text{psd}\mathbf{0}$ for all $p \in \Delta(\Theta)$ (\cref{cor:ker-SP} in \cref{ssec:calc-kernel}).\label{fn:sp-exp-kern}} Our main definition is then:

\begin{definition}[LPI]\label{defi:lpi}
    For any $W \subseteq \Delta^\circ(\Theta)$, $C \in \C$ is \dred{Locally Prior Invariant (LPI)} on $W$ if $\underline{\K}_C^+(p) = \underline{\K}_C^+(p')$ for all $p,p' \in W$.
\end{definition}

This setwise definition of \nameref{defi:lpi} applies to all cost functions, including those that are non-smooth. This generality is  essential for the purpose of proving \cref{thm:wald}, which does not impose any smoothness assumptions on the underlying direct cost. %For the special case of \nameref{defi:lq} costs, being \nameref{defi:lpi} is equivalent to the more intuitive property of having a constant experimental kernel (as in \cref{ssec:SPI}): 
However, for \nameref{defi:lq} cost functions, we note that this definition reduces to the more intuitive requirement that the experimental kernel is constant (as described in \cref{ssec:SPI}):

\begin{lem}\label{lem:lpi-kernel}
    \hspace{-.35em}For any $W \subseteq \Delta^\circ(\Theta)$ and \nameref{defi:sp} $C \in \C$ that is \nameref{defi:lq} on $W$, 
    \[
    \hspace{-1em}\text{$C$ is \nameref{defi:lpi} on $W$%and \nameref{defi:lq} on $W$
    } \ \ \iff \ \ \text{$\kappa_C(p) = \kappa_C(p')$ for all $p, p' \in W$.}
    %\text{$C$ is \hyperref[LPI-W]{LPI on $W$}} \ \ \implies \ \ \text{$\exists \,\kappa \in \R^{|\Theta| \times |\Theta|}$ such that $\kappa = \diag(p) k_C(p) \diag(p)$ for all $p \in W$.}
    \]
    %Moreover, the matrix $\kappa$ is symmetric and satisfies $\kappa \mathbf{1} = \mathbf{0}$ and $x^\top \kappa x \geq 0$ for all $x\in \R^{|\Theta|}$.
    %%%%%
    %$C$ is \hyperref[LPI-W]{LPI on $W$} if and only if there exists a symmetric matrix $\kappa \in \R^{|\Theta| \times |\Theta|}$ such that: 
    %\[
    %\text{(i) $\kappa = \diag(p) k_C(p) \diag(p)$ for all $p \in W$, \ \ (ii) $x^\top \kappa x \geq 0$ for all $x\in \R^{|\Theta|}$, \ \ and (iii) $\kappa \mathbf{1} = \mathbf{0}$.}
    %\]
\end{lem}
\begin{proof}
See \cref{ssec:proof-lpi-kernel} below.
\end{proof}

Our key methodological result is that, in general, \nameref{defi:lpi} is a necessary condition for both \hyperref[axiom:prior:invariant]{Prior Invariance} and \nameref{defi:spi}:

\begin{lem}\label{lem:pi-w-implies-lpi-w}
    For any $W \subseteq \Delta^\circ(\Theta)$ and \nameref{defi:sp} $C \in \C$,
    \[
    \text{$C$ is \hyperref[axiom:prior:invariant]{Prior Invariant}} \ \ \implies \ \ \text{$C$ and $\Phi(C)$ are both \nameref{defi:lpi} on $W$.}
    \]
\end{lem}
\begin{proof}
See \cref{ssec:proof-pi-w-implies-lpi-w} below.
\end{proof}

We defer the proofs of \cref{lem:lpi-kernel,lem:pi-w-implies-lpi-w} until after the main proof of \cref{thm:wald} because they are technical and lengthy. Here, we note two aspects of these results. First, the proof of \cref{lem:pi-w-implies-lpi-w} consists of two main steps: (i) we show directly that \hyperref[axiom:prior:invariant]{Prior Invariance} implies \nameref{defi:lpi} on every $W\subseteq \Delta^\circ(\Theta)$, and then (ii) we use lower kernel invariance (\cref{thm:qk}(ii)) to show that $C$ is \nameref{defi:lpi} on $W$ only if $\Phi(C)$ is also \nameref{defi:lpi} on $W$. Second, we remark that \cref{lem:lpi-kernel,lem:pi-w-implies-lpi-w} together imply the ``$\implies$'' direction of \cref{prop:LPI-main}.

\subsubsection{Step 3: \ref{eqn:MS} Costs are Uniquely \nameref{defi:ups} and \hyperref[defi:lpi]{Locally Prior Invariant}}\label{ssec:ups-lpi-wald-local}

We now show that: (i) if $|\Theta|>2$, there do not exist any (smooth, rich domain) \nameref{defi:ups} and \nameref{defi:lpi} cost functions, and (ii) if $|\Theta|=2$, the \ref{eqn:MS} cost is the unique such cost function. In fact, we establish much stronger ``local'' versions of these facts that apply to any (smooth) \nameref{defi:ups} cost $C^H_\text{ups}$ for which $\dom(H) \subseteq \Delta(\Theta)$ has nonempty interior. %This extra generality facilitates comparison to \textcite{morris-strack-sampling} (see \cref{fn:ms-multi-state} in \cref{ssec:SPI}).

%We emphasize that these conclusions are \emph{not} specific to \nameref{defi:ups} cost functions with rich domain. Rather, they apply to any \nameref{defi:ups} cost function $C^H_\text{ups}$ for which $\dom(H) \cap \Delta^\circ(\Theta)$ has nonempty interior. 

\begin{lem}\label{lem:ups-lpi-wald-local}
    For any open convex $W \subseteq \Delta^\circ(\Theta)$ and strongly convex $H \in \mathbf{C}^2(W)$, 
    \[
    \text{$C^H_\text{ups}$ is \nameref{defi:lpi} on $W$} \ \  \implies \ \ \text{$|\Theta| = 2$ and $\exists \, \gamma>0$ such that $C^H_\text{ups}(\pi) = \gamma \,C_\text{Wald}(\pi)$ for all $\pi \in \Delta(W)$.}%$H = \gamma \, H_\text{Wald}|_W$ for some $\gamma >0$.}
    \]
\end{lem}
\begin{proof}
Since $H \in \mathbf{C}^2(W)$, \cref{lem:ups-kernel-equiv} in \cref{ssec:calc-kernel} implies that $C^H_\text{ups}$ is \nameref{defi:lq} on $W$ with kernel $k_C = \H H$. Since $C^H_\text{ups}$ is \nameref{defi:sp} (as $H$ is strongly convex) and \nameref{defi:lpi} on $W$, \cref{lem:lpi-kernel} then implies that 
\begin{align}
    \H H(p) = \diag(p)^{-1} \, \kappa \, \diag(p)^{-1} \qquad  \forall \, p \in W \label{eqn:lpi-hess}
\end{align}
for some matrix $\kappa \in \R^{|\Theta|\times|\Theta|}$ that (per \cref{remark:exp-kernel}) is symmetric with $\kappa \mathbf{1} = \mathbf{0}$ and $x^\top \kappa x \geq 0$ for all $x \in \R^{|\Theta|}$. By \eqref{eqn:lpi-hess}, we have $H \in \mathbf{C}^\infty(W)$ (as each component of $\H H$ is itself $\mathbf{C}^\infty(W)$).

We now use \eqref{eqn:lpi-hess} to prove the lemma in two steps. To simplify notation, we let $n := |\Theta|\geq 2$ denote the number of states, enumerate the state space as $\Theta := \{1,\dots, n\}$, and denote beliefs $p \in \Delta(\Theta) \subsetneq \R^n$ as vectors $p = (p_1, \dots, p_n)$ for the remainder of this proof.

\noindent \textbf{Step 1: Necessity of $\bm{n = 2}$.} Let $n \geq 2$ be given. %Suppose, towards a contradiction, that $n >2$. 
For every $p = (p_1, \dots, p_n) \in \Delta(\Theta)$, we denote by $p_{-n} := (p_1, \dots, p_{n-1})\in \R^{n-1}$ the vector consisting of the first $(n-1)$ components of $p$. Define $V \subseteq \R^{n-1}_{++}$ as $V:= \{ p_{-n}\in \R^{n-1} \mid p  \in W\}$, $\zeta : V \to W$ as $\zeta(p_{-n}) := (p_1, \dots, p_{n-1}, 1- \sum_{\ell=1}^{n-1}p_\ell)$, and $G : V \to \R$ as $G(p_{-n}) := H(\zeta(p_{-n}))$. Note that $V$ is open (in the Euclidean topology on $\R^{n-1}$) because $W$ is open (in the subspace topology on $\Delta(\Theta)\subsetneq \R^n$), while $G \in \mathbf{C}^\infty(V)$ because $H \in \mathbf{C}^\infty(W)$ (as implied by \eqref{eqn:lpi-hess}) and $\zeta$ is a (linear) $\mathbf{C}^\infty$-diffeomorphism. 
%%%
%and $G \in \mathbf{C}^\infty(V)$, because $W\subseteq \Delta^\circ(\Theta)$ is open (in the subspace topology on $\Delta(\Theta)\subset \R^n$) and $H \in \mathbf{C}^\infty(W)$ (as implied by \eqref{eqn:lpi-hess}).
%%%%
%(being the projection of the open set $W \subseteq \Delta^\circ(\Theta)$) and $G \in \mathbf{C}^\infty(V)$.
%because the embedding $\zeta : V \to W$ defined as $\zeta(p_{-n}):= (p_1, \dots, p_{n-1}, 1- \sum_{\ell=1}^{n-1}p_\ell)$ is a linear $\mathbf{C}^\infty$-diffeomorphism. 
For every $p \in V$ and $i,j \in \{1,\dots, n-1\}$, we have
\begin{align}
\hspace{-2em}
\begin{split}
 \frac{\partial^2}{\partial p_i \partial p_j}G(p_{-n})  &=  [\H H(\zeta(p))]_{ij} \, - \, [\H H(\zeta(p))]_{in} \, - \, [\H H(\zeta(p))]_{jn} \, + \, [\H H(\zeta(p))]_{nn}  \\
%%%
 & =  \frac{\kappa_{ij}}{p_i \, p_j}-\frac{\kappa_{in}}{p_i\, (1-\sum_{\ell=1}^{n-1}p_{\ell})}-\frac{\kappa_{jn}}{p_j\, (1-\sum_{\ell=1}^{n-1}p_{\ell})}+\frac{\kappa_{nn}}{(1-\sum_{\ell=1}^{n-1}p_{\ell})^2}, 
\end{split}
\label{eqn:LPI-cross-1}
\end{align}
where the first line is by the chain rule and the second line is by \eqref{eqn:lpi-hess}.\footnote{For every $k,\ell \in \{1, \dots, n\}$, $[\H H(\zeta(p))]_{k\ell} := \frac{\partial^2}{\partial x_k  \partial x_\ell} H(x)\big|_{x=\zeta(p)}$ denotes the $(k,\ell)^\text{th}$ entry of the matrix $\H H(\zeta(p))$.} 

Now, suppose towards a contradiction that $n >2$. Then, for every $p_{-n} \in V$ and $i,j \in \{1,\dots, n-1\}$ such that $i \neq j$ (which exist because $n>2$), it holds that
\begin{align}
  \hspace{-2em}  \frac{\partial}{\partial p_i} \frac{\partial^2}{\partial p_i \partial p_j}G(p_{-n})  &= - \frac{\kappa_{ij}}{p_i^2 p_j} + \frac{\kappa_{in} \cdot (1-p_i  - \sum_{\ell=1}^{n-1} p_\ell)}{p_i^2 (1-\sum_{\ell=1}^{n-1} p_\ell)^2} - \frac{\kappa_{jn}}{p_j (1-\sum_{\ell=1}^{n-1} p_\ell)^2} + \frac{2 \kappa_{nn}}{(1-\sum_{\ell=1}^{n-1} p_\ell)^3},  \label{eqn:LPI-cross-2-1}\\
    %%%
   \hspace{-2em}  \frac{\partial}{\partial p_j} \frac{\partial^2}{\partial p_i \partial p_i}G(p_{-n}) &= - \frac{2\kappa_{in}}{p_i (1-\sum_{\ell=1}^{n-1} p_\ell)^2} + \frac{2 \kappa_{nn}}{(1-\sum_{\ell=1}^{n-1} p_\ell)^3}. \label{eqn:LPI-cross-2-2}
    \end{align}
Since $G \in \mathbf{C}^\infty(W)$ implies that $G$ has symmetric cross-partial derivatives of all orders, the third-order cross-partials in \eqref{eqn:LPI-cross-2-1} and \eqref{eqn:LPI-cross-2-2} must be equal. Equating these expressions and using the definition of $V$ and the identity $p_n = 1- \sum_{\ell = 1}^{n-1} p_\ell$, we obtain the condition:
\begin{align}
    0 = \kappa_{ij} \, p_n^2 + \kappa_{jn} \, p_i^2 - \kappa_{in} \, p_j  (p_i+ p_n) \quad \ \ \forall\, p\in W\ \text{ and } \ i,j \in \{1\dots, n-1\} \ \text{ s.t. } \ i \neq j. \label{eqn:LPI-cross-3}
\end{align}
Since $W \subseteq \Delta^\circ(\Theta)$ is open and $n >2$, by varying $p \in W$ we see that \eqref{eqn:LPI-cross-3} holds if and only if $\kappa_{ij} = \kappa_{in} = \kappa_{jn} = 0$ for all $i,j\in \{1,\dots, n-1\}$ with $i\neq j$.\footnote{In particular, fix any $\widehat{p} \in W$ and $i,j \in \{1,\dots, n-1\}$ such that $i \neq j$. Since $W \subseteq \Delta^\circ(\Theta)$ is open, there exists an  $\epsilon>0$ such that $\widehat{p} + t (\delta_i -\delta_n) \in W$ for all $t \in (-\epsilon, \epsilon)$ (where $\delta_i,\delta_n \in \Delta(\Theta)$ are the Dirac measures on states $i,n \in \Theta$). Therefore, \eqref{eqn:LPI-cross-3} implies that $0 = \kappa_{ij} (\widehat{p}_n -t)^2 + \kappa_{jn} (\widehat{p}_i + t)^2 - \kappa_{in} \widehat{p}_j (\widehat{p}_i + \widehat{p}_n)$ for all $t \in (-\epsilon, \epsilon)$. We claim that this condition holds if and only if $\kappa_{ij} = \kappa_{jn} = \kappa_{in} = 0$. The ``if'' direction is immediate; we show the ``only if'' direction in two steps. First, note that the condition requires $0 = \frac{d}{dt} \left[\kappa_{ij} (\widehat{p}_n -t)^2 + \kappa_{jn} (\widehat{p}_i + t)^2 \right]$ for all $t \in (-\epsilon,\epsilon)$. By a short calculation, this holds (if and) only if $\kappa_{ij} = \kappa_{jn} = 0$. Second, plugging $\kappa_{ij} = \kappa_{jn} = 0$ back into the original condition implies $\kappa_{in}=0$. This proves the claim.} Since $\kappa \in \R^{n\times n}$ is symmetric (as noted above), it follows that $\kappa$ is a diagonal matrix. But since $\kappa \mathbf{1} = \mathbf{0} \in \R^n$ (as also noted above), this implies that $\kappa_{ii} =0$ for all $i \in \{1,\dots,n\}$, as well. Hence, $\kappa = \mathbf{0} \in \R^{n \times n}$ is the zero matrix. But then \eqref{eqn:lpi-hess} implies that $\H H(p) = \mathbf{0} \in \R^{n \times n}$ for all $p \in W$, which contradicts the hypothesis that $H$ is strongly convex. We conclude that $ n =  2$, as desired.

\noindent \textbf{Step 2: Necessity of Wald.} Let $n = 2$. As in Step 1, define $V \subseteq (0,1)$ as $V := \{p_1 \in \R \mid (p_1,1-p_1) \in W\}$ and $G : V \to \R$ as $G(p_1) := H(p_1, 1-p_1)$. Note that $V$ is convex and open (in the Euclidean topology on $\R$) because $W$ is convex and open (in the subspace topology on $\Delta(\Theta) \subsetneq \R^2$), while $G \in \mathbf{C}^2(V)$ because  $H\in \mathbf{C}^2(W)$. Also define $G_\text{Wald} \in \mathbf{C}^2\left( (0,1) \right)$ as 
\[
G_\text{Wald} (p_1):= H_\text{Wald}(p_1,1-p_1) = p_1 \log\left( \frac{p_1}{1-p_1} \right) + (1-p_1) \log\left( \frac{1-p_1}{p_1} \right),
\]
where $H_\text{Wald} \in \mathbf{C}^2(\Delta^\circ(\Theta))$ is as defined in \cref{eg:Diffusion:1} (\cref{ssec:examples-revisit}).

Now, since the $\kappa \in \R^{2\times 2}$ in \eqref{eqn:lpi-hess} is symmetric and satisfies $x^\top \kappa x\geq 0$ for all $x \in \R^2$ and $\kappa \mathbf{1} = \mathbf{0}$, there exists a $\gamma \geq 0$ such that $\kappa_{11} = \kappa_{22} = - \kappa_{12} = - \kappa_{21} =\gamma$. Since $H$ is strongly convex, \eqref{eqn:lpi-hess} also implies that $\gamma>0$. Hence, we obtain %\eqref{eqn:LPI-cross-1} (with $n=2$, $p_{-n} := p_1$, and $i=j=1$) reduces to\footnote{Note that \eqref{eqn:LPI-cross-1} is valid when $n=2$ as well as when $n>2$ (unlike \eqref{eqn:LPI-cross-2-1}--\eqref{eqn:LPI-cross-3}, which apply only when $n>2$).}
\[
G''(p_1) = \frac{\gamma}{p_1^2 (1-p_1)^2} = \gamma \, G''_\text{Wald}(p_1) \qquad \forall \, p_1 \in V,
\]
where the first equality follows from  \eqref{eqn:LPI-cross-1} (with $n=2$, $p_{-n} := p_1$, and $i=j=1$) and the second equality is by direct calculation.\footnote{Note that \eqref{eqn:LPI-cross-1} is valid when $n=2$ as well as when $n>2$ (unlike \eqref{eqn:LPI-cross-2-1}--\eqref{eqn:LPI-cross-3}, which apply only when $n>2$).} Therefore, for every $\pi \in \Delta(V)$,  we have
\begin{align*}
    C^H_\text{ups}(\pi)  & = \E_\pi \left[ G(q_1) - G(p_{\pi,1}) - G'(p_{\pi,1}) (q_1-p_{\pi,1}) \right] \\
    & = \E_\pi \left[ \int_{p_{\pi,1}}^{q_1} \left( \int_{p_{\pi,1}}^s G''(t) \dd t \right) \dd s \right] \\
    & = \E_\pi \left[ \int_{p_{\pi,1}}^{q_1} \left( \int_{p_{\pi,1}}^s \gamma \, G_\text{Wald}''(t) \dd t \right) \dd s \right] \\
    %%%
    & = \gamma \, \E_\pi \left[ G_\text{Wald}(q_1) - G_\text{Wald}(p_{\pi,1}) - G_\text{Wald}'(p_{\pi,1}) (q_1-p_{\pi,1}) \right] \, =  \, \gamma \, C_\text{Wald}(\pi),
\end{align*}
where the first line is by the definition of $G$ and $p_{\pi,1} := \E_\pi[q_1]$, the second line is by the Fundamental Theorem of Calculus (as $G\in \mathbf{C}^2(V)$ and $V\subseteq (0,1)$ is convex), the third line is by the preceding display, and the final line is again by the Fundamental Theorem of Calculus and the definitions of $G_\text{Wald}$ and $p_{\pi,1} = \E_\pi[q_1]$. This completes the proof.
%%%
\end{proof}

\subsubsection{Step 4: Wrapping Up}\label{ssec:thm6-wrapping-up}

\begin{proof}[Proof of \cref{thm:wald}]
Let $C\in \C$ be \nameref{defi:sp} and have rich domain. (Note that a \ref{eqn:MS} cost satisfies these conditions if and only if it is nontrivial, i.e., has coefficient $\gamma>0$.) 
%To begin, we note that every nontrivial \ref{eqn:MS} cost (i.e., with $\gamma>0$) is \nameref{defi:sp} and has rich domain. Now, let a $C\in \C$ 
%
We prove the desired three-way equivalence by establishing a cycle of implications:

\noindent \textbf{(1) $C$ is \nameref{defi:spi} and \nameref{axiom:CMC}$\implies$$C$ is \nameref{defi:spi}, \nameref{defi:ups}, and \nameref{defi:lq}.} Let $C$ be \nameref{defi:spi} and \nameref{axiom:CMC}. Since every \nameref{defi:spi} cost is \nameref{axiom:slp} (\cref{prop:1}), the ``only if'' direction of \cref{thm:trilemma}(i) implies that $C$ is a \nameref{defi:TI} cost. Hence, $C$ is \nameref{defi:ups} by definition. Since $H_\text{TI} \in \mathbf{C}^2(\Delta^\circ(\Theta))$, \cref{lem:ups-kernel-equiv} in \cref{ssec:calc-kernel} implies that $C$ is also \nameref{defi:lq}.

\noindent \textbf{(2) $C$ is \nameref{defi:spi}, \nameref{defi:ups}, and \nameref{defi:lq}$\implies$$|\Theta|=2$ and $C$ is a \ref{eqn:MS} cost.} Let $C$ be \nameref{defi:spi}, \nameref{defi:ups} (with $C = C^H_\text{ups}$ for some $H : \Delta^\circ(\Theta) \to \R$), and \nameref{defi:lq}. Since $C$ is also \nameref{defi:sp} (by hypothesis), it follows that: (a) $C$ is \nameref{defi:lpi} on $W := \Delta^\circ(\Theta)$ (by \cref{lem:pi-w-implies-lpi-w}), (b) $H \in \mathbf{C}^2(\Delta^\circ(\Theta))$ (by \cref{lem:ups-kernel-equiv} in \cref{ssec:calc-kernel}), and (c) $H$ is strongly convex. The desired conclusion then follows from \cref{lem:ups-lpi-wald-local} (with $W := \Delta^\circ(\Theta)$).

\noindent \textbf{(3) $|\Theta|=2$ and $C$ is a \ref{eqn:MS} cost$\implies$$C$ is \nameref{defi:spi} and \nameref{axiom:CMC}.} Let $C$ be a \ref{eqn:MS} cost. Then $C$ is \nameref{defi:spi} by \cref{lem:wald-is-spi}(i) and \nameref{axiom:CMC} by construction (per \eqref{eqn:MS} and \cref{thm:trilemma}(i)).
\end{proof}

\subsubsection{Technical Facts for \cref{lem:lpi-kernel,lem:pi-w-implies-lpi-w}}\label{ssec:thm6-technical-facts}

In this section, we state several definitions and technical facts that are used below in \cref{ssec:proof-lpi-kernel,ssec:proof-pi-w-implies-lpi-w} during the proofs of \cref{lem:lpi-kernel,lem:pi-w-implies-lpi-w}. Proofs of these facts are deferred until \cref{ssec:proof-thm6-technical-facts}, after the main proofs of \cref{lem:lpi-kernel,lem:pi-w-implies-lpi-w}.

\paragraph{Notation.} As in \cref{ssec:calc-kernel}, for each $p \in W$, we let $\underline{K}_C(p)\subseteq \R^{|\Theta| \times|\Theta|}$ denote the set of all lower kernels of $C$ at $p$, and let $\underline{K}^+_C(p) := \{\underline{k}(p) \in \underline{K}_C(p) \mid \underline{k}(p)\gg_\text{psd} \mathbf{0}\}$. By construction, 
    \begin{equation}\label{eqn:kappa-plus-equiv}
    \hspace{-0.25em} \underline{\K}^+_C(p) = \diag(p) \, \underline{K}^+_C(p) \, \diag(p) \quad \text{ and } \quad \underline{K}^+_C(p)  = \diag(p)^{-1}  \underline{\K}^+_C(p) \, \diag(p)^{-1} \quad \forall \, p \in W,
    \end{equation}
    where the multiplication of these sets by $\diag(p), \diag(p)^{-1} \in \R^{|\Theta|\times|\Theta|}_{++}$ is elementwise.\footnote{Viz.,  $\underline{\K}^+_C(p) = \{\diag(p)\underline{k}(p)\diag(p) \mid \underline{k}(p) \in \underline{K}^+_C(p)\}$ and $\underline{K}^+_C(p) = \{\diag(p)^{-1} \underline{\kappa}(p) \diag(p)^{-1} \mid \underline{\kappa}(p) \in \underline{\K}^+_C(p)\}$ for all $p \in W$.}

\paragraph{Facts about Matrices.} Next we record several miscellaneous facts about matrices.  

For any symmetric matrix $M \in \R^{|\Theta| \times |\Theta|}$, let $M \geq^\star_\text{psd} \mathbf{0}$ denote that $x^\top M x\geq 0$ for all $x \in \R^{|\Theta|}$. Note that $M \geq^\star_\text{psd} \mathbf{0}$ implies that $M \geq_\text{psd} \mathbf{0}$, but not necessarily conversely since $\mathcal{T} =\{x \in \R^{|\Theta|} \mid \mathbf{1}^\top x = 0\} \subsetneq \R^{|\Theta|}$. However, the converse holds for ``normalized'' matrices:

\begin{lem}\label{lem:psd-star}
    For any symmetric matrix $M \in \R^{|\Theta| \times |\Theta|}$ and $p_0 \in \Delta(\Theta)$,
    %\begin{equation}\label{eqn:psd-psd_full}
    \[
    M \geq_\text{psd} \mathbf{0} \ \ \text{ and } \ \  M p_0 = \mathbf{0} \quad \implies \quad M \geq^\star_\text{psd} \mathbf{0}.
    \]
\end{lem}
\begin{proof}
    See \cref{ssec:proof-thm6-technical-facts}.
\end{proof}

The following lemma provides a way to verify the ``strict positive definiteness'' of (experimental) kernels when pivoting across different prior beliefs:

\begin{lem}\label{lem:exp-ker-prior-pivot}
    For any $p, p' \in \Delta^\circ(\Theta)$ and any symmetric matrix $M\in \R^{|\Theta|\times|\Theta|}$ that satisfies $M p = \mathbf{0}$ and $M\gg_\text{psd} \mathbf{0}$, the matrix $\widehat{M} \in \R^{|\Theta|\times|\Theta|}$ defined as
    \[
    \widehat{M} := \diag(p')^{-1}\, \diag(p) \, M \,  \diag(p) \, \diag(p')^{-1}
    \]
    is symmetric and satisfies $\widehat{M} p' = \mathbf{0}$ and $\widehat{M} \gg_\text{psd} \mathbf{0}$.
\end{lem}
\begin{proof}
    See \cref{ssec:proof-thm6-technical-facts}.
\end{proof}

The following lemma shows that pre- and post-multiplying a given positive semi-definite matrix by an ``approximately identity'' diagonal matrix generates ``approximately'' the same quadratic forms as the original matrix:

\begin{lem}\label{lem:diag-matrix-approx}
    For any symmetric matrix $M \in \R^{|\Theta|\times|\Theta|}$ such that $M\gg_\text{psd} \mathbf{0}$, there exists $\chi \in \R_{++}$ such that:
    \begin{equation}\label{eqn:diag-matrix-approx}
    \inf_{v \in V(\epsilon)} y^\top  \diag(v) \, M \, \diag(v) \, y \, \geq \,  \left( 1 - \epsilon \cdot \chi \right) \, y^\top M y \quad \text{ for all $\epsilon \in (0,1)$ and $y \in \mathcal{T}$,}
    \end{equation}
    where $V(\epsilon) := \left\{v \in \R^{|\Theta|}_{+} \ \big|\ \sqrt{1-\epsilon} \leq v(\theta) \leq \sqrt{1+\epsilon} \ \ \forall \, \theta \in \Theta \right\}$.
\end{lem}
\begin{proof}
    See \cref{ssec:proof-thm6-technical-facts}.
\end{proof}

\paragraph{Incremental Evidence Bounds.} The following lemma establishes several facts that are useful for approximating the cost of incremental evidence across different prior beliefs: 

\begin{lem}\label{lem:cross-prior-small}
    \hspace{-0.275em}For every $p_0, p_1 \in \Delta^\circ(\Theta)$ and $\delta_0>0$, there exist constants $\overline{\delta}_1, \beta>0$ and a function $g: (0,\overline{\delta}_1) \to (0,1)$ with $\lim_{\delta_1 \to 0} g(\delta_1) = 0$ such that, for every $\delta_1 \in (0,\overline{\delta}_1)$, the following hold: 
    \begin{itemize}
        \item[(i)] For every $p \in B_{\delta_1}(p_1)$, experiment $\sigma = (S,(\sigma_\theta)_{\theta\in\Theta})\in \Se$, and signal $s \in \bigcup_{\theta \in \Theta}\supp(\sigma_\theta)$, 
        \begin{equation}\label{eqn:cross-prior-small-1}
        q_s^{\sigma,p} \in B_{\delta_1}(p_1)  \quad \implies \quad  q_{s}^{\sigma,p_0} \in B_{\delta_0}(p_0) \ \ \text{ and } \ \ \max_{\theta \in \Theta} \left| \frac{\d \sigma_\theta }{\d \langle \sigma ,p\rangle}(s) - 1   \right| \, \leq \, \beta \, \delta_1.
        \end{equation}
        \item[(ii)] For every $p \in B_{\delta_1}(p_1)$ and $\theta \in \Theta$, 
        \begin{equation}\label{eqn:cross-prior-small-2}
        \sqrt{ 1- g(\delta_1)} \, \leq \, \frac{p_1(\theta)}{p(\theta)} \, \leq \, \sqrt{ 1 +  g(\delta_1)}.
        \end{equation}
    \end{itemize}
\end{lem}
\begin{proof}
    See \cref{ssec:proof-thm6-technical-facts}.
\end{proof}

\subsubsection{Proof of \cref{lem:lpi-kernel}}\label{ssec:proof-lpi-kernel}

\begin{proof}%[Proof of \cref{lem:lpi-kernel}]
    Let $C \in \C$ be \nameref{defi:sp} and \nameref{defi:lq} on $W \subseteq \Delta^\circ(\Theta)$. 
    
    \noindent \textbf{($\implies$ direction)} Let $C$ be \nameref{defi:lpi} on $W$. Fix any $p_0 \in W$. It suffices to show that $\kappa_C(p) = \kappa_C(p_0)$ for all $p \in W$. So, let $p \in W \backslash\{p_0\}$ be given. Since $C$ is \nameref{defi:sp}, \cref{cor:ker-SP} implies that $k_C(p_0) \in \underline{K}_C^+(p_0)$ and $k_C(p_0) \geq_\text{psd} \underline{k}(p_0)$ for all $\underline{k}(p_0) \in \underline{K}_C^+(p_0)$. 
    %(i.e., $k_C(p_0) = \max \underline{K}_C^+(p_0)$). 
    Since $k_C(p_0) p_0 = \underline{k}(p_0) p_0 = \mathbf{0}$ for all $\underline{k}(p_0) \in \underline{K}_C^+(p_0)$, \cref{lem:psd-star} (with $M := k_C(p_0) - \underline{k}(p_0)\geq_\text{psd}\mathbf{0}$) then implies that $k_C(p_0) \geq^\star_\text{psd} \underline{k}(p_0)$ for all $\underline{k}(p_0) \in \underline{K}_C^+(p_0)$. These facts and \eqref{eqn:kappa-plus-equiv} together imply:
    \begin{equation}\label{eqn:lpi-kern-pf-0}
    \kappa_C(p_0) \in \underline{\K}^+_C(p_0) \qquad \text{and} \qquad \kappa_C(p_0) \geq^\star_\text{psd} \underline{\kappa}(p_0) \quad \forall\, \underline{\kappa}(p_0) \in \underline{\K}^+_C(p_0).
    \end{equation}
    Since $C$ is \nameref{defi:lpi} on $W$, and hence $\underline{\K}^+_C(p_0) = \underline{\K}^+_C(p)$, it follows that
    \begin{equation}\label{eqn:lpi-kern-pf-1}
    %\[
    \kappa_C(p_0) \in \underline{\K}^+_C(p) \qquad \text{and} \qquad \kappa_C(p_0) \geq^\star_\text{psd} \underline{\kappa}(p)  \quad \forall\, \underline{\kappa}(p) \in \underline{\K}^+_C(p). 
    %\]
     \end{equation}
     By definition of the experimental kernel $\kappa_C$, it holds that
     \begin{align}\label{eqn:lpi-kern-pf-2}
     \kappa_C(p) = \kappa_C(p_0) \quad \iff \quad k_C(p) = \widehat{k}_C(p) := \diag(p)^{-1} \, \kappa_C(p_0) \, \diag(p)^{-1}.
     \end{align}
     Moreover, \eqref{eqn:kappa-plus-equiv} implies that condition \eqref{eqn:lpi-kern-pf-1} is equivalent to the condition:
    \begin{align}\label{eqn:lpi-kern-pf-3}
    \widehat{k}_C(p_0) \in \underline{K}^+_C(p) \qquad \text{and} \qquad  \widehat{k}_C(p_0) \geq^\star_\text{psd} \underline{k}(p) \quad \forall\, \underline{k}(p) \in \underline{K}_C^+(p). 
    \end{align}
    Meanwhile, %since $C$ is \nameref{defi:sp} and \nameref{defi:lq} at $p \in W$, 
    the same argument that led to \eqref{eqn:lpi-kern-pf-0} (with $p$ appearing in place of $p_0$) implies that $k_C(p) \in \underline{K}^+_C(p)$ and $k_C(p) \geq^\star_\text{psd} \underline{k}(p)$ for all $\underline{k}(p) \in \underline{K}_C^+(p)$. 
    %\cref{cor:ker-SP} implies that $k_C(p) \in \underline{K}^+_C(p)$ and $k_C(p)\geq_\text{psd} \underline{k}(p)$ for all $\underline{k}(p) \in \underline{K}^+_C(p)$. Since $k_C(p) p = \underline{k}(p) p = \mathbf{0}$ for all $\underline{k}(p) \in \underline{K}_C^+(p)$, condition \eqref{eqn:psd-psd_full} (with $A := k_C(p) - \underline{k}(p)\geq_\text{psd}\mathbf{0}$) then implies that $k_C(p) \geq^\star_\text{psd} \underline{k}(p)$ for all $\underline{k}(p) \in \underline{K}_C^+(p)$. 
    Taken together, this fact and \eqref{eqn:lpi-kern-pf-3} imply that $k_C(p)\geq^\star_\text{psd} \widehat{k}_C(p) \geq^\star_\text{psd} k_C(p)$. It follows $k_C(p) = \widehat{k}_C(p)$ (since these matrices are symmetric). We therefore obtain from \eqref{eqn:lpi-kern-pf-2} that $\kappa_C(p) = \kappa_C(p_0)$. Since the given $p \in W$ was arbitrary, we conclude that $\kappa_C$ is constant on $W$, as desired. %this completes the proof.

    \noindent \textbf{($\impliedby$ direction)} Let $\kappa_C(p) = \kappa_C(p')$ for all $p,p' \in W$. Equivalently, fix any $p_0 \in W$ and suppose that $\kappa_C(p) = \kappa_C(p_0)$ for all $p \in W$. To show that $C$ is \nameref{defi:lpi} on $W$, it suffices to show that $\underline{\K}^+_C(p) = \underline{\K}^+_C(p_0)$ for all $p \in W$. So, let $p \in W \backslash\{p_0\}$ be given. 
    
    We first show that $\underline{\K}^+_C(p_0)\subseteq \underline{\K}^+_C(p)$. Let $\underline{\kappa}(p_0) \in \underline{\K}^+_C(p_0)$ be given. Since \eqref{eqn:lpi-kern-pf-0} holds (by the same argument as above) and $\kappa_C(p) = \kappa_C(p_0)$ (by hypothesis), we have %$\kappa_C(p) \in \underline{\K}^+_C(p_0)$ and 
    $\kappa_C(p) \geq^\star_\text{psd} \underline{\kappa}(p_0)$. By definition of $\kappa_C(p)$, this implies that $k_C(p) \geq^\star_\text{psd} \widehat{k}(p):= \diag(p)^{-1} \underline{\kappa}(p_0) \diag(p)^{-1}$. Since $\widehat{k}(p)$ is symmetric and satisfies $\widehat{k}(p) p = \mathbf{0}$ by construction, it follows that $\widehat{k}(p)$ is a well-defined lower kernel of $C$ at $p$. Moreover, since $\underline{\kappa}(p_0) \in \underline{\K}^+_C(p_0)$, \eqref{eqn:kappa-plus-equiv} implies that %there exists $\underline{k}(p_0) \in \underline{K}^+_C(p_0)$ such that $\underline{\kappa}(p_0) = \diag(p_0) \underline{k}(p_0) \diag(p_0)$, we have
    \[
    \widehat{k}(p) =  \diag(p)^{-1} \diag(p_0) \underline{k}(p_0) \diag(p_0) \diag(p)^{-1} \quad \text{ for some } \quad \underline{k}(p_0) \in \underline{K}^+_C(p_0).
    \]
    \cref{lem:exp-ker-prior-pivot} then yields $\widehat{k}(p) \gg_\text{psd} \mathbf{0}$. Hence, $\widehat{k}(p) \in K^+_C(p)$. It then follows from \eqref{eqn:kappa-plus-equiv} that $\underline{\kappa}(p_0) = \diag(p) \widehat{k}(p) \diag(p) \in \underline{\K}^+_C(p)$. Since $\underline{\kappa}(p_0) \in \underline{\K}^+_C(p_0)$ was arbitrary, $\underline{\K}^+_C(p_0)\subseteq \underline{\K}^+_C(p)$.
    
    Now, by interchanging the roles of $p_0$ and $p$ in the above argument, we also obtain $\underline{\K}^+_C(p)\subseteq \underline{\K}^+_C(p_0)$. It follows that $\underline{\K}^+_C(p) = \underline{\K}^+_C(p_0)$. Since the given $p \in W\backslash\{p_0\}$ was arbitrary, we therefore conclude that $C$ is \nameref{defi:lpi} on $W$, as desired.
\end{proof}

\subsubsection{Proof of \cref{lem:pi-w-implies-lpi-w}}\label{ssec:proof-pi-w-implies-lpi-w}

\begin{proof}%[Proof of \cref{lem:pi-w-implies-lpi-w}]
Let $W \subseteq \Delta^\circ(\Theta)$ and let $C \in \C$ be \nameref{defi:sp}. We proceed in two steps. First, we show that if $C$ is \hyperref[axiom:prior:invariant]{Prior Invariant}, then $C$ is \nameref{defi:lpi} on $W$. Second, we show that if $C$ is \nameref{defi:lpi} on $W$, then $\Phi(C)$ is also \nameref{defi:lpi} on $W$. Taken together, these two steps yield the lemma.

    \noindent \textbf{Step 1: Let $C$ be \hyperref[axiom:prior:invariant]{Prior Invariant}.} Let $p_0, p_1 \in W$ be given. In what follows, we prove that $\underline{\K}^+_C(p_0) \subseteq  \underline{\K}^+_C(p_1)$. By interchanging the roles of $p_0$ and $p_1$, the same argument also yields the opposite inclusion $\underline{\K}^+_C(p_0) \supseteq  \underline{\K}^+_C(p_1)$, and hence the equality $\underline{\K}^+_C(p_0) =  \underline{\K}^+_C(p_1)$. Since the given $p_0,p_1 \in W$ are arbitrary, this suffices to prove that $C$ is \nameref{defi:lpi} on $W$.
    
    To this end, let $\underline{\kappa}(p_0) \in \underline{\K}^+_C(p_0)$ be given. Let  $k(p_0) := \diag(p_0)^{-1} \underline{\kappa}(p_0) \diag(p_0)^{-1} \in \underline{K}^+_C(p_0)$ denote the corresponding lower kernel of $C$ at $p_0$. We define $k(p_1) \in \R^{|\Theta|\times|\Theta|}$ as 
    \begin{equation}\label{eqn:lpi-lq-pivot-0}
    k(p_1) \, := \, \diag(p_1)^{-1} \underline{\kappa}(p_0) \diag(p_1)^{-1} \, = \, \diag(p_1)^{-1}  \diag(p_0) k(p_0) \diag(p_0)  \diag(p_1)^{-1}.
    \end{equation}
    Since $k(p_0)$ is symmetric and satisfies $k(p_0)p_0 = \mathbf{0}$ and $k(p_0)\gg_\text{psd} \mathbf{0}$ (by definition), it follows from \cref{lem:exp-ker-prior-pivot} that $k(p_1)$ is symmetric, $k(p_1) p_1 =\mathbf{0}$, and $k(p_1)\gg_\text{psd} \mathbf{0}$. 
    
    We claim that $k(p_1)$ is a lower kernel of $C$ at $p_1$. Given this claim, it follows that $k(p_1) \in \underline{K}_C^+(p_1)$, and hence (by \eqref{eqn:kappa-plus-equiv}) that $\underline{\kappa}(p_0) = \diag(p_1) k(p_1) \diag(p_1) \in \underline{\K}^+_C(p_1)$. Since the given $\underline{\kappa}(p_0) \in \underline{\K}^+_C(p_0)$ is arbitrary, this implies that $\underline{\K}^+_C(p_0) \subseteq  \underline{\K}^+_C(p_1)$, as desired.
    
    Hence, it suffices to prove the claim. To this end, we proceed in four (sub)steps.

    \emph{\textbf{Step 1(a): Preliminaries:}} Since $k(p_0)\gg_\text{psd} \mathbf{0}$, there exists $\overline{\epsilon}>0$ such that $k(p_0) - 2 \epsilon I(p_0) \geq_\text{psd} \mathbf{0} $ for all $\epsilon \leq \overline{\epsilon}$. Let $\epsilon \in (0, \overline{\epsilon})$ be given. Since $C$ is \hyperref[axiom:prior:invariant]{Prior Invariant} and $k(p_0)$ is a lower kernel of $C$ at $p_0$, there exists $\delta_0>0$ such that, for every $\sigma \in \Se$ and $p \in \Delta^\circ(\Theta)$, 
    \begin{align*}
    \hspace{-1em}
        C(h_B (\sigma, p)) \, =\,  C(h_B (\sigma, p_0))  \, &\geq \int_{B_{\delta_0}(p_0)} (q - p_0)^\top \left( \frac{1}{2} k(p_0) - \epsilon I \right) (q-p_0) \dd h_B (\sigma, p_0) (q) \\
        %%%
        & = \int_{S} \mathbf{1}\left( q_{s}^{\sigma,p_0} \in B_{\delta_0} (p_0) \right) \cdot (q_{s}^{\sigma,p_0} - p_0)^\top \left( \frac{1}{2} k(p_0) - \epsilon I \right)(q_{s}^{\sigma,p_0} - p_0) \dd \langle \sigma, p_0 \rangle (s),
    \end{align*}
    where the first equality is by \hyperref[axiom:prior:invariant]{Prior Invariance} (as $p_0 \in W \subseteq \Delta^\circ(\Theta)$), the inequality is by \cref{defi:lq}(ii) and the inclusion $h_B[\Se \times \{p_0\}] \subseteq \big\{\pi \in \Ex \mid p_\pi \in B_{\delta_0}(p_0)\big\}$, and the final equality is a change of variables. Given this $\delta_0>0$, \cref{lem:cross-prior-small} yields the existence of constants $\overline{\delta}_1, 
    \, \beta>0$ and a function $g : (0,\overline{\delta}_1) \to (0,1)$ with $\lim_{\delta_1 \to 0} g(\delta_1) = 0$ such that conditions \eqref{eqn:cross-prior-small-1}--\eqref{eqn:cross-prior-small-2} hold for all $\delta_1 \in (0,\overline{\delta}_1)$. Without loss of generality (by making $\overline{\delta}_1>0$ smaller if needed), we  assume that $B_{\overline{\delta}_1} (p_1) \subseteq \Delta^\circ(\Theta)$ (as $p_1 \in W\subseteq \Delta^\circ(\Theta)$) and that $\sqrt{|\Theta|} \cdot \beta \, \overline{\delta}_1 <1$.\footnote{We will use the former assumption throughout, and the latter assumption to obtain \eqref{eqn:step1b-3} in Step 1(c) below.} Henceforth, we let $\delta_1 \in (0, \overline{\delta}_1)$ denote a parameter to be chosen at the end.

    By condition \eqref{eqn:cross-prior-small-1}, we have $\mathbf{1}\left( q_{s}^{\sigma,p} \in B_{\delta_1} (p_1) \right) \leq \mathbf{1}\left( q_{s}^{\sigma,p_0} \in B_{\delta_0} (p_0) \right)$ for all $p \in B_{\delta_1}(p_1)$, $\sigma \in \Se$, and $s \in \cup_{\theta \in \Theta} \supp(\sigma_\theta)$. Plugging this into the above display and noting that the integrand is non-negative (because $\frac{1}{2}k(p_0)- \epsilon I \geq_\text{psd} \mathbf{0}$ by definition of $\epsilon \in (0,\overline{\epsilon})$), we obtain 
    \begin{equation}\label{eqn:lpi-lq-pivot-1}
    \begin{split}
        C(h_B(\sigma,p)) &\geq \int_{S} \mathbf{1}\left( q_{s}^{\sigma,p} \in B_{\delta_1} (p_1) \right) \cdot  (q_{s}^{\sigma,p_0} - p_0)^\top \left( \frac{1}{2} k(p_0) - \epsilon I \right)(q_{s}^{\sigma,p_0} - p_0) \dd \langle \sigma, p_0 \rangle (s) \\
        %%%
        & = \frac{1}{2} \, A(\sigma,p; \delta_1) - \epsilon \cdot B(\sigma,p; \delta_1),
    \end{split}
    \end{equation}
    for every $\sigma \in \Se$ and $p \in B_{\delta_1}(p_1)$, where we define
    \begin{align}
        A(\sigma,p; \delta_1) &:= \int_{S} \mathbf{1}\left( q_{s}^{\sigma,p} \in B_{\delta_1} (p_1) \right) \cdot (q_{s}^{\sigma,p_0} - p_0)^\top \,  k(p_0) \, (q_{s}^{\sigma,p_0} - p_0) \dd \langle \sigma, p_0 \rangle (s), \label{eqn:A-term-lpi} \\
        %%%
        B(\sigma,p; \delta_1) &:= \int_{S} \mathbf{1}\left( q_{s}^{\sigma,p} \in B_{\delta_1} (p_1) \right) \cdot \|  q_{s}^{\sigma,p_0} - p_0\|^2 \dd \langle \sigma, p_0 \rangle (s) . \label{eqn:B-term-lpi}
    \end{align}

In the remaining three (sub)steps of the proof, we bound the error terms that arise when we ``change priors from $p_0$ to $p_1$'' in the integrals defining $A(\sigma, p; \delta_1)$ and $B(\sigma,p; \delta_1)$. In Steps 1(b) and 1(c), we obtain separate bounds on each of the terms $A(\sigma,p; \delta_1)$ and $B(\sigma,p; \delta_1)$, respectively. In Step 1(d), we then combine these bounds with \eqref{eqn:lpi-lq-pivot-1} to show that the matrix $k(p_1) \in \R^{|\Theta|\times|\Theta|}$ defined in \eqref{eqn:lpi-lq-pivot-0} is, in fact, a lower kernel of $C$ at $p_1$. 

In Steps 1(b) and 1(c), to simplify notation, we let $\sigma \in \Se$ and $p \in B_{\delta_1}(p_1)$ be given. The bounds we obtain will be uniform across these objects, i.e., depend only on  $|\Theta|\in\mathbb{N}$, the given $p_0,p_1 \in W$, the constants $\overline{\delta}_1, \beta>0$ and function $g$, and the parameter $\delta_1 \in (0,\overline{\delta}_1)$.  

\textbf{\emph{Step 1(b): Lower bound on $A(\sigma,p;\delta_1)$.}} First, solving \eqref{eqn:lpi-lq-pivot-0} for $k(p_0)$ delivers
\[
k(p_0) = \diag(p_0)^{-1} \diag(p_1) k(p_1) \diag(p_1) \diag(p_0)^{-1}.
\]
Then, by plugging this expression into \eqref{eqn:A-term-lpi} and rearranging, we obtain
\begin{align*}
    \hspace{-2em} A(\sigma,p; \delta_1) &= \int_{S} \mathbf{1}\left( q_{s}^{\sigma,p} \in B_{\delta_1} (p_1) \right) \cdot \left( \diag\left( \frac{p_1}{p_0}\right) q_{s}^{\sigma,p_0} - p_1\right)^\top \,  k(p_1) \, \left( \diag\left( \frac{p_1}{p_0}\right) q_{s}^{\sigma,p_0} - p_1\right) \dd \langle \sigma, p_0 \rangle (s) \\
    %%%
    & = \int_{S} \mathbf{1}\left( q_{s}^{\sigma,p} \in B_{\delta_1} (p_1) \right) \cdot \left( \diag\left( \frac{p}{p_0}\right) q_{s}^{\sigma,p_0} \right)^\top \,  \diag\left(\frac{p_1}{p} \right)k(p_1) \diag\left(\frac{p_1}{p} \right)\, \left( \diag\left( \frac{p}{p_0}\right) q_{s}^{\sigma,p_0} \right) \dd \langle \sigma, p_0 \rangle (s), 
\end{align*}
where the first line is by direct substitution and the second line holds because $k(p_1) p_1 = \mathbf{0}$ (by construction) and $\diag\left( \frac{p_1}{p_0}\right) = \diag\left(\frac{p_1}{p} \right)\diag\left( \frac{p}{p_0}\right)$. By Bayes' rule and the chain rule for Radon-Nikodym derivatives, for every $s \in \cup_{\theta \in \Theta} \supp(\sigma_\theta)$ and $\theta \in \Theta$, it holds that
\begin{equation}\label{eqn:bayes-changeofmeasure}
\frac{p(\theta)}{p_0(\theta)} q^{\sigma,p_0}_s(\theta) \, = \, p(\theta) \frac{\d \sigma_\theta}{\d \langle \sigma,p_0\rangle}(s)\, = \, p(\theta) \frac{\d \sigma_\theta}{\d \langle \sigma,p\rangle}(s) \cdot \frac{\d \langle \sigma,p\rangle}{\d \langle \sigma,p_0\rangle}(s) \, = \, q^{\sigma,p}_s (\theta)\cdot \frac{\d \langle \sigma,p\rangle}{\d \langle \sigma,p_0\rangle}(s).
\end{equation}
That is, $\diag\left( \frac{p}{p_0}\right) q_{s}^{\sigma,p_0} = q_{s}^{\sigma,p} \, \frac{\d \langle \sigma,p\rangle}{\d \langle \sigma,p_0\rangle}(s)$ for all $s \in \cup_{\theta \in \Theta} \supp(\sigma_\theta)$. Plugging this identity into the preceding display, we then obtain
\begin{align}
    \hspace{-2em} A(\sigma,p; \delta_1) &= \int_{S} \mathbf{1}\left( q_{s}^{\sigma,p} \in B_{\delta_1} (p_1) \right) \cdot \left( q_{s}^{\sigma,p} \right)^\top \,  \diag\left(\frac{p_1}{p} \right)k(p_1) \diag\left(\frac{p_1}{p} \right)\, \left(  q_{s}^{\sigma,p} \right) \cdot \left(\frac{\d \langle \sigma,p\rangle}{\d \langle \sigma,p_0\rangle}(s) \right)^2 \dd \langle \sigma, p_0 \rangle (s) \notag \\
    %%%
    & = \int_{S} \mathbf{1}\left( q_{s}^{\sigma,p} \in B_{\delta_1} (p_1) \right) \cdot \left( q_{s}^{\sigma,p} \right)^\top \,  \diag\left(\frac{p_1}{p} \right)k(p_1) \diag\left(\frac{p_1}{p} \right)\, \left(  q_{s}^{\sigma,p} \right) \cdot \left(\frac{\d \langle \sigma,p\rangle}{\d \langle \sigma,p_0\rangle}(s) \right)\, \dd \langle \sigma, p \rangle (s) \notag\\
    %%%
    & \geq \frac{1}{1 + \beta\, \delta_1} \cdot \int_{S} \mathbf{1}\left( q_{s}^{\sigma,p} \in B_{\delta_1} (p_1) \right) \cdot \left( q_{s}^{\sigma,p} \right)^\top \,  \diag\left(\frac{p_1}{p} \right)k(p_1) \diag\left(\frac{p_1}{p} \right)\, \left(  q_{s}^{\sigma,p} \right) \, \dd \langle \sigma, p \rangle (s) \notag \\
    %%%%
    & = \frac{1}{1 + \beta\, \delta_1} \cdot \int_{B_{\delta_1} (p_1)} (q-p)^\top \diag\left(\frac{p_1}{p} \right) k(p_1) \diag\left(\frac{p_1}{p} \right) (q-p) \dd h_B(\sigma,p)(q), \label{eqn:lpi-lq-pivot-2}
\end{align}
where the first line is by direct substitution, the second line uses the change of measure $\d\langle \sigma,p\rangle = \frac{\d\langle \sigma,p\rangle}{\d\langle \sigma,p_0\rangle}\d\langle \sigma,p_0\rangle$, the third line holds because \cref{lem:psd-star} (for $M := \diag\left(\frac{p_1}{p} \right)k(p_1) \diag\left(\frac{p_1}{p} \right)$) implies that the integrand is non-negative and (the second implication in) condition \eqref{eqn:cross-prior-small-1} implies that $\frac{\d \langle \sigma,p\rangle}{\d \langle \sigma,p_0\rangle}(s) \geq \frac{1}{1+\beta\, \delta_1}$ on the event $\{s \in S \mid q_{s}^{\sigma,p} \in B_{\delta_1} (p_1) \}$,\footnote{To elaborate: First, the properties of $k(p_1)$ noted below \eqref{eqn:lpi-lq-pivot-0} and \cref{lem:exp-ker-prior-pivot} imply that $M := \diag\left(\frac{p_1}{p} \right)k(p_1) \diag\left(\frac{p_1}{p} \right)$ is symmetric, $M p = \mathbf{0}$, and $M \gg_\text{psd}\mathbf{0}$. Therefore, \cref{lem:psd-star} implies that $M\geq^\star_\text{psd} \mathbf{0}$. Second, condition \eqref{eqn:cross-prior-small-1} implies that, on the stated event, we have $\frac{\d \langle \sigma,p_0\rangle}{\d\langle \sigma,p\rangle}(s) \leq 1 + \beta\, \delta_1$, which is equivalent (by a short calculation) to $\frac{\d\langle \sigma,p\rangle}{\d\langle \sigma,p_0\rangle}(s) \geq \frac{1}{1 + \beta\, \delta_1}$.} and the final line follows from a change of variables and the fact that $k(p_1)p_1 = \mathbf{0}$ (by construction). 

Next, we bound the contribution from the $\diag\left(\frac{p_1}{p} \right)$ terms in \eqref{eqn:lpi-lq-pivot-2}. Since $k(p_1)\gg_\text{psd} \mathbf{0}$ (as noted above) and $g(\delta_1) \in (0,1)$ for all $\delta_1 \in (0,\overline{\delta}_1)$ (by construction), condition \eqref{eqn:cross-prior-small-2} and \cref{lem:diag-matrix-approx} imply that there exists a constant $\chi>0$ such that\footnote{Per the construction in \cref{lem:diag-matrix-approx}, the constant $\chi>0$ depends only on the matrix $k(p_1) \in \R^{|\Theta|\times|\Theta|}$.}
\[
y^\top \diag\left(\frac{p_1}{p} \right) k(p_1) \diag\left(\frac{p_1}{p} \right) y \, \geq \, (1- g(\delta_1) \cdot \chi) \ y^\top k(p_1) y \qquad \forall y \in \mathcal{T}. %\text{ and } \delta_1\in (0,\overline{\delta}_1).
\]
Plugging this bound into \eqref{eqn:lpi-lq-pivot-2} then delivers 
\begin{align*}%\label{eqn:lpi-lq-pivot-3}
    %\hspace{-1em} 
    A(\sigma,p; \delta_1) \, \geq \, \frac{1- g(\delta_1) \cdot \chi}{1+ \beta \, \delta_1} \cdot \int_{B_{\delta_1} (p_1)} (q-p)^\top  k(p_1)  (q-p) \dd h_B(\sigma,p)(q). %\quad \ \ \forall \delta_1 \in (0,\overline{\delta}_1).
\end{align*}

Finally, define the map $\xi : (0,\overline{\delta}_1) \to \R_{++}$ as $\xi(\delta_1) := 1 -  \frac{1- g(\delta_1) \cdot \chi}{1+ \beta \, \delta_1}$. By plugging this definition into the above display and using the inequality $y^\top k(p_1) y \leq \|k(p_1)\| \cdot \|y\|^2$ for all $y \in \mathcal{T}$ (by definition of the matrix semi-norm), we then obtain the desired lower bound: 
\begin{align}\label{eqn:step1a-lower-bound}
\begin{split}
    A(\sigma,p; \delta_1) \, \geq \, &\int_{B_{\delta_1} (p_1)} (q-p)^\top  k(p_1)  (q-p) \dd h_B(\sigma,p)(q) \\    & - \xi( \delta_1) \cdot \| k(p_1)\| \cdot \int_{B_{\delta_1} (p_1)} \|q-p\|^2 \dd h_B(\sigma,p)(q). 
    \end{split}
    %\qquad \ \  \forall \delta_1 \in (0,\overline{\delta}_1).
\end{align}
Moreover, note that $\xi$ satisfies $\lim_{\delta_1 \to 0} \xi(\delta_1) = 0$ by construction. 

\textbf{\emph{Step 1(c): Upper Bound on $B(\sigma,p;\delta_1)$.}} We begin by rewriting \eqref{eqn:B-term-lpi} as
\begin{align}\label{eqn:step1b-1}
\begin{split}
\hspace{-0.5em}
B(\sigma,p; \delta_1) %&=  \int_{S} \mathbf{1}\left( q_{s}^{\sigma,p} \in B_{\delta_1} (p_1) \backslash\{p\}\right) \cdot \|  q_{s}^{\sigma,p_0} - p_0\|^2 \dd \langle \sigma, p_0 \rangle (s) \\
%%%
& = \int_{S} \mathbf{1}\left( q_{s}^{\sigma,p} \in B_{\delta_1} (p_1) \backslash\{p\}\right) \cdot \|  q_{s}^{\sigma,p} - p\|^2 \cdot \frac{\|  q_{s}^{\sigma,p_0} - p_0\|^2}{\|q_{s}^{\sigma,p} - p\|^2} \dd \langle \sigma, p_0 \rangle (s) \\
%%%
& = \int_{S} \mathbf{1}\left( q_{s}^{\sigma,p} \in B_{\delta_1} (p_1) \backslash\{p\}\right) \cdot \|  q_{s}^{\sigma,p} - p\|^2 \cdot \frac{\|  q_{s}^{\sigma,p_0} - p_0\|^2}{\|q_{s}^{\sigma,p} - p\|^2} \cdot \frac{\d \langle \sigma, p_0 \rangle (s)}{\d \langle \sigma, p \rangle (s)} \, \dd \langle \sigma, p \rangle (s), 
\end{split}
\end{align}
where the first line holds because $q^{\sigma,p}_s = p$ if and only if $q^{\sigma,p_0}_s = p_0$ by Bayes' rule (e.g., see \eqref{eqn:bayes-changeofmeasure}) and the second line uses the change of measure $\d \langle \sigma, p_0 \rangle = \frac{\d \langle \sigma, p_0 \rangle (s)}{\d \langle \sigma, p \rangle}  \dd \langle \sigma, p \rangle$. 

Next, fix any $s \in \cup_{\theta \in \Theta} \supp(\sigma_\theta)$ such that $q_{s}^{\sigma,p} \in B_{\delta_1} (p_1) \backslash\{p\}$. For any $r \in \Delta^\circ(\Theta)$, we denote $\frac{\d \bm{\sigma}}{\d \langle \sigma, r\rangle}(s)  := \big( \frac{\d \sigma_\theta}{\d \langle \sigma, r\rangle} (s)\big)_{\theta \in \Theta} \in \R^{|\Theta|}_+$. We also define the constant $\overline{p_0}:=\max_{\theta\in\Theta} p_0(\theta) \in (0,1)$ and the map $\underline{m} : (0,\overline{\delta}_1) \to (0,1)$ as $\underline{m}(\delta_1):= \min_{\theta \in \Theta} \inf_{p \in B_{\delta_1}(p_1)} p(\theta)$. It holds that  
\begin{align}\label{eqn:step1b-2}
\begin{split}
    \frac{\|  q_{s}^{\sigma,p_0} - p_0\|}{\|q_{s}^{\sigma,p} - p\|} \ &\leq \ \left(\frac{\overline{p_0}}{\underline{m}(\delta_1)}\right) \cdot \frac{\big\|\frac{q_{s}^{\sigma,p_0}}{p_0} - \mathbf{1}\big\|}{\big\|\frac{q_{s}^{\sigma,p}}{p} - \mathbf{1}\big\|} \\
    %%%
    & =  \left(\frac{\overline{p_0}}{\underline{m}(\delta_1)}\right) \cdot \frac{\big\|\frac{\d \bm{\sigma}}{\d \langle \sigma, p_0 \rangle}(s) - \mathbf{1}\big\|}{\big\|\frac{\d \bm{\sigma}}{\d \langle \sigma, p \rangle}(s) - \mathbf{1}\big\|} \ \leq \ \left(\frac{\overline{p_0}}{\underline{m}(\delta_1)}\right) \cdot \left( 1 + \frac{\big\|\frac{\d \bm{\sigma}}{\d \langle \sigma, p_0 \rangle}(s) - \frac{\d \bm{\sigma}}{\d \langle \sigma, p \rangle}(s)\big\|}{\big\|\frac{\d \bm{\sigma}}{\d \langle \sigma, p \rangle}(s) - \mathbf{1}\big\|}\right),
\end{split}
\end{align}
where the first inequality follows from the above definitions and the identities $q_{s}^{\sigma,p_0} - p_0 = \diag(p_0) \big( \frac{q_{s}^{\sigma,p_0}}{p_0} - \mathbf{1}\big)$ and $q_{s}^{\sigma,p} - p = \diag(p) \big( \frac{q_{s}^{\sigma,p}}{p} - \mathbf{1}\big)$, the equality is by Bayes' rule, and the final inequality follows from applying the triangle inequality to the numerator. Towards bounding the final term in \eqref{eqn:step1b-2}, we define $z(s):= \big\|\frac{\d \bm{\sigma}}{\d \langle \sigma, p \rangle}(s) - \mathbf{1}\big\|$ and note that
\begin{equation}\label{eqn:step1b-3}
0 \, < \, \max_{\theta\in\Theta} \left| \frac{\d \sigma_\theta}{\d \langle \sigma,p\rangle}(s) - 1 \right| \, \leq \,  z(s) \, \leq \, \sqrt{|\Theta|} \cdot \max_{\theta\in\Theta} \left| \frac{\d \sigma_\theta}{\d \langle \sigma,p\rangle}(s) - 1 \right| \, \leq \, \sqrt{|\Theta|} \cdot \beta \, \delta_1 \, < \, 1
\end{equation}
where the first inequality is by Bayes' rule and the hypothesis that $q^{\sigma,p}_s \neq p$, the second and third inequalities follow from the definition of the Euclidean norm, the fourth inequality is by condition \eqref{eqn:cross-prior-small-1}, and the final inequality holds because $\sqrt{|\Theta|} \cdot \beta \,\overline{\delta}_1<1$ (by assumption) and $\delta_1 \in (0,\overline{\delta}_1)$. The definition of $z(s)$ and the inequalities in \eqref{eqn:step1b-3} then imply that 
\begin{equation}\label{eqn:step1b-4}
\left\| \frac{\d \bm{\sigma}}{\d \langle \sigma, p \rangle}(s)\right\| \, \leq \, \sqrt{|\Theta|} \cdot (1+z(s)) \qquad \text{ and } \qquad \left| \frac{\d \langle \sigma,p\rangle}{\d \langle \sigma,p_0\rangle}(s)-1\right| \, \leq \, \frac{z(s)}{1-z(s)}.\footnote{The first bound in \eqref{eqn:step1b-4} holds because $\left\| \frac{\d \bm{\sigma}}{\d \langle \sigma, p \rangle}(s)\right\| \leq \| \mathbf{1} \| + z(s) \leq \sqrt{|\Theta|} \cdot (1 + z(s))$ by the triangle inequality, the definition of $z(s)$, and the fact that $|\Theta|\geq 2$. The second bound in \eqref{eqn:step1b-4} holds because the second and final inequalities in \eqref{eqn:step1b-3} imply that $\left| \frac{\d \langle \sigma, p_0\rangle}{\d \langle \sigma,p\rangle}(s) -1 \right|\leq z(s)<1$, which upon rearrangement implies that $\frac{-z(s)}{1-z(s)} \leq \frac{-z(s)}{1+z(s)} \leq \frac{\d \langle \sigma,p\rangle}{\d \langle \sigma,p_0\rangle}(s) -1\leq \frac{z(s)}{1-z(s)}$.}
\end{equation}
Moreover, by the chain rule for Radon-Nikodym derivatives, we have
\begin{equation}\label{eqn:step1b-5}
\left\|\frac{\d \bm{\sigma}}{\d \langle \sigma, p_0 \rangle}(s) - \frac{\d \bm{\sigma}}{\d \langle \sigma, p \rangle}(s)\right\| \, = \, \left\| \frac{\d \bm{\sigma}}{\d \langle \sigma, p \rangle}(s)\right\| \cdot \left|\frac{\d\langle \sigma, p\rangle}{\d\langle \sigma, p_0\rangle}(s) - 1 \right|.
\end{equation}
Therefore, it follows that
\[
\hspace{-1em}
\frac{\big\|\frac{\d \bm{\sigma}}{\d \langle \sigma, p_0 \rangle}(s) - \frac{\d \bm{\sigma}}{\d \langle \sigma, p \rangle}(s)\big\|}{\big\|\frac{\d \bm{\sigma}}{\d \langle \sigma, p \rangle}(s) - \mathbf{1}\big\|} \ = \ \frac{\left\| \frac{\d \bm{\sigma}}{\d \langle \sigma, p \rangle}(s)\right\| \cdot \left|\frac{\d\langle \sigma, p\rangle}{\d\langle \sigma, p_0\rangle}(s) - 1 \right|}{z(s)} \ \leq \ \sqrt{|\Theta|} \cdot \frac{1+z(s)}{1-z(s)} \ \leq \ \sqrt{|\Theta|} \cdot \frac{1+\sqrt{|\Theta|} \cdot \beta \, \delta_1}{1-\sqrt{|\Theta|} \cdot \beta \, \delta_1},
\]
where the equality is by \eqref{eqn:step1b-5} and the definition of $z(s)$, the next inequality is by \eqref{eqn:step1b-4}, and the final inequality holds because the function $z \in (0,1) \mapsto \frac{1+z}{1-z}$ is increasing on $(0,1)$ and \eqref{eqn:step1b-3} implies that $ 0 < z(s) \leq \sqrt{|\Theta|} \cdot \beta \, \delta_1 <1$. Plugging the above display into \eqref{eqn:step1b-2}, we obtain
\[
\frac{\|  q_{s}^{\sigma,p_0} - p_0\|}{\|q_{s}^{\sigma,p} - p\|} \, \leq \, \eta(\delta_1):= \left(\frac{\overline{p_0}}{\underline{m}(\delta_1)}\right) \cdot\left( 1 + \ \sqrt{|\Theta|} \cdot \frac{1+\sqrt{|\Theta|} \cdot \beta \, \delta_1}{1-\sqrt{|\Theta|} \cdot \beta \, \delta_1}\right).
\]
Since \eqref{eqn:step1b-3} also implies that $\frac{\d \langle \sigma, p_0 \rangle }{\d \langle \sigma, p \rangle }(s) \leq 1 + \sqrt{|\Theta|} \cdot \beta \, \delta_1$, we conclude that 
\begin{equation}\label{eqn:step1b-6}
\frac{\|  q_{s}^{\sigma,p_0} - p_0\|^2}{\|q_{s}^{\sigma,p} - p\|^2} \cdot \frac{\d \langle \sigma, p_0 \rangle }{\d \langle \sigma, p \rangle }(s) \, \leq \, \zeta(\delta_1) \, := \, \left[\eta(\delta_1)\right]^2 \cdot \big( 1 + \sqrt{|\Theta|} \cdot \beta \, \delta_1\big).
\end{equation}
Note that this bound is uniform across all $s \in \cup_{\theta\in\Theta} \supp(\sigma_\theta)$ such that $q^{\sigma,p}_s \in B_{\delta_1}(p_1)\backslash\{p\}$. %Moreover, by construction, the map $\zeta :(0,\overline{\delta}_1) \to \R_{++}$ satisfies $\lim_{\delta_1 \to 0} \zeta(\delta_1) = \zeta$

Therefore, plugging \eqref{eqn:step1b-6} into \eqref{eqn:step1b-1} delivers the desired upper bound: 
\begin{align}\label{eqn:step1b-upper-bound}
\begin{split}
B(\sigma, p; \delta_1) \, & \leq \, \zeta(\delta_1) \cdot \int_{S} \mathbf{1}\left( q_{s}^{\sigma,p} \in B_{\delta_1} (p_1) \backslash\{p\}\right) \cdot \|  q_{s}^{\sigma,p} - p\|^2 \dd \langle \sigma, p \rangle (s) \\
%%%
& = \, \zeta(\delta_1) \cdot \int_{B_{\delta_1}(p_1)} \|q - p\|^2 \dd h_B(\sigma,p)(q), %\qquad \forall\, \delta_1 \in (0,\overline{\delta}_1),
\end{split}
\end{align}
where the second line is by a change of variables. Moreover, note that $\zeta$ satisfies
\begin{equation}\label{eqn:step1b-7}
\lim_{\delta_1\to0} \zeta(\delta_1) \ = \  \underline{\zeta} \, := \, \left(  \frac{\overline{p_0}}{\underline{p_1}}\right)^2 \cdot \left( 1+ \sqrt{|\Theta|} \right)^2 \in \R_{++}, \quad \text{where } \ \  \underline{p_1} := \min_{\theta \in \Theta} p_1(\theta)>0, 
\end{equation}
because $\lim_{\delta_1 \to 0 } \underline{m}(\delta_1) = \underline{p_1}$ (by construction) and $p_1 \in W \subseteq \Delta^\circ(\Theta)$ (by hypothesis).

\textbf{\emph{Step 1(d): Wrapping Up.}} By plugging the lower bound on $A(\sigma, p;\delta_1)$ from \eqref{eqn:step1b-upper-bound} and the upper bound on $B(\sigma,p;\delta_1)$ from \eqref{eqn:step1b-upper-bound} (which yields a lower bound on $-\epsilon \cdot B(\sigma,p;\delta_1)$) into \eqref{eqn:lpi-lq-pivot-1}, we obtain an overall lower bound: for all $\sigma \in \Se$, $\delta_1 \in (0,\overline{\delta}_1)$, and $p \in B_{\delta_1}(p_1)$, we have
\begin{align*}
\begin{split}
    C(h_B(\sigma,p)) \, %& \geq \, \frac{1}{2} \, A(\sigma,p; \delta_1) - \epsilon \cdot B(\sigma,p; \delta_1) \\
    %%%
     \geq \, \frac{1}{2} \cdot &\int_{B_{\delta_1} (p_1)} (q-p)^\top  k(p_1)  (q-p) \dd h_B(\sigma,p)(q) \\    &   - \left( \frac{\xi( \delta_1) \cdot \| k(p_1)\|}{2} + \epsilon \cdot \zeta(\delta_1) \right) \cdot \int_{B_{\delta_1} (p_1)} \|q-p\|^2 \dd h_B(\sigma,p)(q).
\end{split}
%\qquad \forall \, \sigma \in \Se, \, \delta_1 \in (0,\overline{\delta}_1), \, p \in B_{\delta_1}(p_1).
\end{align*}
Moreover, since $\lim_{\delta_1\to0} \xi(\delta_1)=0$ and $\lim_{\delta_1 \to0} \zeta(\delta_1) = \underline{\zeta} \in \R_{++}$ (for the constant $\underline{\zeta}$ defined in \eqref{eqn:step1b-7}), there exists $\widehat{\delta}_1 \in (0,\overline{\delta}_1)$ such that $\frac{\xi( \widehat{\delta}_1) \cdot \| k(p_1)\|}{2} + \epsilon \cdot \zeta(\widehat{\delta}_1) \, \leq \, 2 \, \underline{\zeta} \, \epsilon$. Plugging this into the above display and using the identity $h_B[\Se \times B_{\widehat{\delta_1}}(p_1)] = \big\{\pi \in \Ex \mid p_\pi \in B_{\widehat{\delta_1}}(p_1)\big\}$, we obtain:
\begin{align*}
C(\pi) \, \geq \,  \int_{B_{\widehat{\delta}_1} (p_1)} (q-p)^\top  \left( \frac{1}{2} k(p_1)  - 2 \underline{\zeta}  \epsilon \,  I\right) (q-p) \dd \pi(q)  \qquad \forall \, \pi \in \Ex \ \, \text{ s.t. } \ \, p_\pi \in  B_{\widehat{\delta}_1} (p_1).
\end{align*}
Since the given $\epsilon \in (0,\overline{\epsilon})$ was arbitrary and the constant $\underline{\zeta}>0$ depends only on $|\Theta|\in \mathbb{N}$ and the given $p_0,p_1 \in W$, we conclude that $k(p_1)$ is a lower kernel of $C$ at $p_1$, as claimed.

This completes the proof of the claim, and thereby the proof of Step 1.

\noindent\textbf{Step 2: Let $C$ be \nameref{defi:lpi} on $W$.} Let $p, p' \in W$ be given. Since $C \succeq \Phi(C)$, we have $\underline{K}^+_{\Phi(C)}(p) \subseteq \underline{K}^+_C(p)$. Since $C \in C$ is \nameref{defi:sp}, \cref{thm:qk}(ii) yields $\underline{K}^+_{\Phi(C)}(p) \supseteq \underline{K}^+_C(p)$. Therefore, $\underline{K}^+_{\Phi(C)}(p) = \underline{K}^+_C(p)$. By the same argument,  $\underline{K}^+_{\Phi(C)}(p') = \underline{K}^+_C(p')$. Hence, we obtain
    \begin{align*}
 \hspace{-2em} \underline{\K}_{\Phi(C)}^+(p) \, = \,  \diag(p) \underline{K}^+_{\Phi(C)} (p) \diag(p) \, & =  \, \diag(p) \underline{K}^+_C (p) \diag(p) \\ 
     & = \, \underline{\K}_C^+(p) \\
    & = \, \underline{\K}_C^+(p') \\ 
    & = \, \diag(p') \underline{K}^+_C (p') \diag(p') \,  = \, \diag(p') \underline{K}^+_{\Phi(C)} (p') \diag(p') \, = \, \underline{\K}_{\Phi(C)}^+(p'), 
    \end{align*}
    where the first equality is by \eqref{eqn:kappa-plus-equiv} (applied to $\Phi(C)$), the second equality is by the above, the third equality is by \eqref{eqn:kappa-plus-equiv} (applied to $C$), the fourth equality holds because $C$ is \nameref{defi:lpi} on $W$ (by hypothesis), and remaining equalities follow from the same reasoning applied in reverse. Since the given $p,p' \in W$ were arbitrary, we conclude that $\Phi(C)$ is \nameref{defi:lpi} on $W$.
\end{proof}

    \subsubsection{Proofs of Technical Facts from \cref{ssec:thm6-technical-facts} (\dred{Lemmas} \ref{lem:psd-star}--\ref{lem:cross-prior-small})}\label{ssec:proof-thm6-technical-facts}
   %\subsection{Supporting Details for Theorem \ref{thm:wald}}\label{ssec:lem:wald}

\begin{proof}[Proof of \cref{lem:psd-star}]
    Fix any $p_0 \in W$, symmetric $M \in \R^{|\Theta| \times |\Theta|}$ with $M \geq_\text{psd} \mathbf{0}$ and $M p_0 = \mathbf{0}$, and $x \in \R^{|\Theta|}$. If $x \in \mathcal{T}$, then $x^\top M x \geq0$ since $M \geq_\text{psd} \mathbf{0}$. If $x \notin \mathcal{T}$ (i.e., $\mathbf{1}^\top x \neq 0$), then 
     \begin{align*}
    x^\top M x \ = \ (\mathbf{1}^\top x)^2  \cdot \left(\frac{x^\top}{\mathbf{1}^\top x}\right)\,  M \, \left(\frac{x}{\mathbf{1}^\top x}\right) \ = \ (\mathbf{1}^\top x)^2  \cdot \left(\frac{x^\top}{\mathbf{1}^\top x} - p_0^\top \right) \,  M \,  \left(\frac{x}{\mathbf{1}^\top x} - p_0\right) \ \geq \ 0, 
    \end{align*}
    where the second equality is by $M p_0 = \mathbf{0}$ (and symmetry of $M$) and the final inequality holds because $\frac{x}{\mathbf{1}^\top x} - p_0 \in \mathcal{T}$ and $M \geq_\text{psd} \mathbf{0}$. Since $x \in \R^{|\Theta|}$ was arbitrary, $M \geq^\star_\text{psd} \mathbf{0}$.
\end{proof}

%\subsubsection{Proof of \cref{lem:exp-ker-prior-pivot}}\label{ssec:proof-lef-exp-ker-prior-pivot}

\begin{proof}[Proof of \cref{lem:exp-ker-prior-pivot}]
    Plainly, $\widehat{M}$ is symmetric and $\widehat{M} p' = \mathbf{0}$. Here, we show that $\widehat{M} \gg_\text{psd} \mathbf{0}$. 
    
    We begin with some preliminaries. Define $\xi := \min\{y^\top M y \mid y \in \mathcal{T} \text{ s.t. } \|y\| =1\}$. Note that $\xi>0$ because the map $y \mapsto y^\top M y$ is continuous and strictly positive (as $M\gg_\text{psd} \mathbf{0}$) on the compact set $Y:=\{y \in \mathcal{T} \mid \|y\|=1\}\subseteq \mathcal{T}$. Thus, for every $x \in \R^{|\Theta|}$, it holds that
    \begin{equation}\label{eqn:psd-exp-pivot}
    x^\top M x \,\,  = \,\, \left(x - (\mathbf{1}^\top x) \, p \right)^\top \, M  \,  \left(x - (\mathbf{1}^\top x) \, p \right) \, \, \geq \, \, \xi \cdot \left\| x - (\mathbf{1}^\top x) \, p\right\|^2 ,
    \end{equation}
    where the equality holds because $M p = \mathbf{0}$ and $M$ is symmetric, and the inequality follows from the fact that $x - (\mathbf{1}^\top x) \, p\in \mathcal{T}$ (by construction) and the definition of $\xi>0$.

    Now, define $\eta := \min\{y^\top \widehat{M} y \mid y \in Y\}$. Note that $\widehat{M} \gg_\text{psd}\mathbf{0}$ if and only if $\eta >0$. Moreover, since the map $y \mapsto y^\top \widehat{M} y $ is continuous on the compact set $Y$, we have $\eta>0$ if and only if $y^\top \widehat{M} y>0$ for all $y \in Y$. We claim that the latter property holds. To this end, let $y \in Y$ be given and define $z := \diag(p) \diag(p')^{-1}  y \in \R^{|\Theta|} \backslash\{\mathbf{0}\}$.\footnote{We have $z \neq \mathbf{0}$ because $y \neq \mathbf{0}$ (by definition of $Y$) and $\diag(p) \diag(p')^{-1}  \in \R^{|\Theta|\times|\Theta|}$ is nonsingular (as $p,p'\in \Delta^\circ(\Theta)$).} There are two cases: 

    \textbf{\emph{Case 1: Let $z \in \mathcal{T}$.}} Then, by definition of $\xi>0$, we have $y^\top \widehat{M} y = z^\top M z \geq \xi \cdot \|z\|^2  >0$.

    \textbf{\emph{Case 2: Let $z \notin \mathcal{T}$.}} Then, by \eqref{eqn:psd-exp-pivot}, we have $y^\top \widehat{M} y = z^\top M z \geq \xi \cdot \|z-(\mathbf{1}^\top z)p\|^2 $. Since $\xi>0$, the lower bound is strictly positive if and only if $z \neq (\mathbf{1}^\top z)p$. Suppose, towards a contradiction, that $z =(\mathbf{1}^\top z)p$. By definition of $z$, this is equivalent to $y = (\mathbf{1}^\top z) p'$. Since $y \in Y\subseteq\mathcal{T}$ and $p' \in \Delta^\circ(\Theta)$, it follows that $0 = \mathbf{1}^\top y = (\mathbf{1}^\top z) \mathbf{1}^\top p' = (\mathbf{1}^\top z)$, i.e., $z \in \mathcal{T}$. This contradicts the hypothesis that $z \notin \mathcal{T}$, as desired. We conclude that $y^\top \widehat{M} y>0$.

    Therefore, since the given $y \in Y$ was arbitrary, we conclude that $\eta>0$ as desired. 
\end{proof}

%\subsubsection{Proof of \cref{lem:diag-matrix-approx}}\label{ssec:proof-diag-matrix-approx}

\begin{proof}[Proof of \cref{lem:diag-matrix-approx}]
    Let $M \in \R^{|\Theta|\times|\Theta|}$ such that $M \gg_\text{psd} \mathbf{0}$ be given. Define $\chi \in \overline{\R}_{++}$ as
    \[
    \chi := \sup_{y \in \mathcal{T}\backslash\{\mathbf{0}\}} \, \frac{\sum_{\theta,\theta'\in\Theta} \left|M_{\theta,\theta'} \,   y(\theta) y(\theta')\right|}{y^\top M y},
    \]
     where $M_{\theta,\theta'}\in\R$ denotes the $(\theta,\theta')^\text{th}$ entry of the matrix $M$. We proceed in two steps. 
     
     First, we claim that $\chi < +\infty$ (i.e., $\chi \in \R_{++}$). Define $Y:= \{ y \in \mathcal{T} \mid \|y \| =1\}$. Note the following facts: (a) $\xi := \min\{y^\top M y \mid y \in Y\}>0$ because $M\gg_\text{psd}\mathbf{0}$, (b) $\mathcal{T}\backslash\{\mathbf{0}\} = \{\alpha y \mid y \in Y, \, \alpha \in \R_{++}\}$, and (c) the maps $y \in \mathcal{T} \mapsto \|y\|^2$ and $y \in \mathcal{T} \mapsto \sum_{\theta,\theta'\in\Theta} \left|M_{\theta,\theta'} \,   y(\theta) y(\theta')\right|$ are both positively homogeneous of degree $2$. Therefore, it holds that
     \[
     \chi \, \leq \, \frac{1}{\xi} \cdot \sup_{y \in \mathcal{T}\backslash\{\mathbf{0}\}} \frac{\sum_{\theta,\theta'\in\Theta} \left|M_{\theta,\theta'} \,   y(\theta) y(\theta')\right|}{\|y\|^2} \, = \, \frac{1}{\xi} \cdot \sup_{y \in Y} \, \sum_{\theta,\theta'\in\Theta} \left|M_{\theta,\theta'} \,   y(\theta) y(\theta')\right| <+\infty,
     \]
    where the first inequality is by fact (a), the second equality is by facts (b) and (c) and the definition of $Y \subsetneq \mathcal{T}$, and the final inequality follows from fact (a) and the fact that $y \mapsto \sum_{\theta,\theta'\in\Theta} \left|M_{\theta,\theta'} \,   y(\theta) y(\theta')\right|$ is continuous on the compact set $Y$. This proves the claim. 

    Next, we claim that $\chi \in \R_{++}$ so-defined yields the desired bound \eqref{eqn:diag-matrix-approx}. To this end, let $\epsilon \in (0,1)$ and $y \in  \mathcal{T}\backslash\{\mathbf{0}\}$ be given.\footnote{Note that the inequality in \eqref{eqn:diag-matrix-approx} trivially holds if $y = \mathbf{0}$, so there is nothing to prove in that case.} By definition, every $v \in V(\epsilon)$ satisfies $1-\epsilon \, \leq \, v(\theta)v(\theta') \, \leq 1+\epsilon$ for all $\theta,\theta' \in \Theta$. Therefore, we have
    \[
    \zeta \cdot v(\theta) v(\theta') \geq \zeta - \epsilon \cdot|\zeta| \qquad \text{for all $\zeta \in \R$, $v \in V(\epsilon)$, and $\theta,\theta' \in \Theta$.}
    \]
    Consequently, for every $v \in V(\epsilon)$, it holds that
    \begin{align*}
    y^\top  \diag(v) \, M \, \diag(v) \, y \, & = \, \sum_{\theta,\theta' \in \Theta}  M_{\theta,\theta'} y(\theta) y(\theta') \cdot v(\theta) v(\theta') \\
    %%%
    & \geq \, \sum_{\theta,\theta' \in \Theta}  M_{\theta,\theta'} y(\theta) y(\theta')  - \epsilon \cdot \sum_{\theta,\theta' \in \Theta} \left| M_{\theta,\theta'} y(\theta) y(\theta')\right| \\
    %%%
    %& = \, y^\top Ay - \epsilon \cdot \sum_{\theta,\theta' \in \Theta} \left| A_{\theta,\theta' } y(\theta) y(\theta')\right| \\
    %%%
    & \geq \, y^\top M y - (\epsilon \cdot \chi) \ y^\top M y
    \end{align*}
    where the first line is by definition of the quadratic form, the second line follows from the inequality in the preceding display applied to each term in the sum, and the final line follows from the definitions of the quadratic form and $\chi>0$. This proves the claim. 

    Since the given $\epsilon \in (0,1)$ and $y \in  \mathcal{T}\backslash\{\mathbf{0}\}$ were arbitrary, we conclude that \eqref{eqn:diag-matrix-approx} holds. 
    \end{proof}

   % \subsubsection{Proof of \cref{lem:cross-prior-small}}\label{ssec:proof-cross-prior-small}

\begin{proof}[Proof of \cref{lem:cross-prior-small}]
Let $p_0, p_1 \in \Delta^\circ(\Theta)$ and $\delta_0 >0$ be given. %Since $p_1 \in \Delta^\circ(\Theta)$, there 
There exists $\overline{\eta}>0$ such that $B_{\overline{\eta}}(p_1)\subseteq \Delta^\circ(\Theta)$. Define the constant $\underline{p_1}\in (0,1)$ and the map $\underline{m} : (0,\overline{\eta}) \to (0,1)$ as
\[
%\hspace{-1em}
\underline{p_1} := \min_{\theta \in \Theta} p_1(\theta) \qquad \text{and} \qquad \underline{m}(\eta) := \min_{\theta \in \Theta} \inf_{p\in B_{\eta}(p_1)} p(\theta).
\]
By construction, we have $\lim_{\eta\to 0} \underline{m}(\eta) = \underline{p_1}$. Let $\eta \in (0,\overline{\eta})$ be a parameter to be chosen at the end. We proceed in three steps.

\noindent \textbf{Step 1: Ensuring condition \eqref{eqn:cross-prior-small-1}.} Let $q,p \in B_\eta(p_1)$ be given. It holds that $\|q-p\| \leq \text{diam}(B_\eta(p_1)) = 2\eta$. Moreover, since $q - p =\diag(p) \left(\frac{q}{p} -\mathbf{1}\right)$ and $\min_{\theta\in\Theta} p(\theta)\geq \underline{m}(\eta)$, we also have $\underline{m}(\eta) \cdot \|\frac{q}{p} -\mathbf{1}\| \leq \|q-p\|$. Combining these inequalities, we obtain
    %\begin{equation}\label{eqn:cross-prior-small-3}
    \[
    \max_{\theta \in \Theta} \left| \frac{q(\theta)}{p(\theta)} - 1 \right| \, \leq \, \left\| \frac{q}{p} -\mathbf{1} \right\|  \, \leq \,  \frac{2\eta}{\underline{m}(\eta)}. 
    %\end{equation}
    \]
Moreover, since $\lim_{\eta\to 0} \underline{m}(\eta) = \underline{p_1}$,  there exists $\widehat{\eta} \in (0,\overline{\eta})$ such that $\underline{m}(\eta) \geq \underline{p_1}/2 >0$ for all $ \eta \in (0,\widehat{\eta})$. Without loss of generality (by making $\widehat{\eta}>0$ smaller if necessary), we may further assume that $4 \widehat{\eta} / \underline{p_1} < 1$. Plugging this into the above display, it follows that
\begin{equation}\label{eqn:cross-prior-small-3}
\max_{\theta \in \Theta} \Big| \frac{q(\theta)}{p(\theta)} - 1 \Big|   \, \leq \,  \beta \, \eta \, < \, 1 \quad \ \ \forall \eta \in \big(0,\widehat{\eta}\big), \qquad \text{ where } \beta := 4 / \underline{p_1} \in \R_{++}.
\end{equation}

    Now, let $\eta \in \big(0,\widehat{\eta}\big)$, $p \in B_\eta (p_1)$, and $\sigma \in \Se$ be given. Fix any $s \in \cup_{\theta\in\Theta} \supp(\sigma_\theta)$ such that $q^{\sigma,p}_s \in B_\eta(p_1)$. By Bayes' rule, $\frac{q_s^{\sigma,p}(\theta)}{p(\theta)} = \frac{\d \sigma_\theta }{\d \langle \sigma ,p\rangle}(s)$ for all $\theta \in \Theta$. Then \eqref{eqn:cross-prior-small-3} (for $q := q^{\sigma,p}_s$) yields
    \begin{equation}\label{eqn:cross-prior-small-3-2}
    \max_{\theta \in \Theta} \Big| \frac{\d \sigma_\theta }{\d \langle \sigma ,p\rangle}(s) - 1   \Big| \, \leq \, \beta \, \eta .
    \end{equation}
    This inequality implies that $\Big| \frac{\d \langle \sigma, p_0\rangle }{\d \langle \sigma ,p\rangle}(s) - 1   \Big| \, \leq \, \beta \, \eta $, which (since $0<\beta \, \eta <1$) is equivalent to 
    \[
    \frac{1}{1+ \beta \, \eta} \, \leq \, \frac{\d \langle \sigma, p\rangle }{\d \langle \sigma ,p_0\rangle}(s) \, \leq \, \frac{1}{1- \beta\, \eta}.
    \]
    By the chain rule for Radon-Nikodym derivatives, we have $ \frac{\d \sigma_\theta }{\d \langle \sigma ,p\rangle}  = \frac{\d \sigma_\theta }{\d \langle \sigma ,p_0\rangle} \cdot \frac{\d \langle \sigma, p_0\rangle }{\d \langle \sigma ,p\rangle}$. Plugging this into the two preceding displays and simplifying, we obtain
   \[
   \max_{\theta \in \Theta} \Big| \frac{\d \sigma_\theta }{\d \langle \sigma ,p_0\rangle}(s) - 1   \Big| \, \leq \, \frac{2 \beta \, \eta}{1- \beta \, \eta}. 
   \]
   Therefore, letting $\overline{p_0} := \max_{\theta \in \Theta} p_0(\Theta)$, it follows that
  \begin{equation}\label{eqn:cross-prior-small-3-3}
   \|q^{\sigma,p_0}_s - p_0\| \, \leq \, \overline{p_0} \cdot \Big\| \frac{q^{\sigma,p_0}_s}{p_0}  - \mathbf{1}\Big\| \, \leq \, \overline{p_0} \cdot\sqrt{|\Theta|} \cdot \max_{\theta \in \Theta} \Big| \frac{q^{\sigma,p_0}_s}{p_0}  - \mathbf{1}\Big| \,  \leq \,  f(\eta) := \frac{2 \overline{p_0}  \sqrt{|\Theta|} \cdot \beta\,\eta}{1- \beta\,\eta},
   \end{equation}
   where the first inequality is by $q^{\sigma,p_0}_s - p_0 = \diag(p_0) \left( \frac{q^{\sigma,p_0}_s}{p_0} -\mathbf{1}\right)$, the second inequality is by definition of the Euclidean norm, and the third inequality is by the preceding display and Bayes' rule, viz., the identity $  \frac{q^{\sigma,p_0}_s(\theta)}{p_0(\theta)} = \frac{\d \sigma_\theta }{\d \langle \sigma ,p_0\rangle}(s)$ for all $\theta \in \Theta$. Hence, $q^{\sigma,p_0}_s \in \overline{B}_{f(\eta)}(p_0)$.

   Since the data given above were arbitrary, %given $\eta \in \big(0,\widehat{\eta}\big)$, $\sigma \in \Se$, and $p, q^{\sigma,p}_s \in B_\eta(p_1)$ were arbitrary,
   \eqref{eqn:cross-prior-small-3-2} and \eqref{eqn:cross-prior-small-3-3} imply that
   \[
   q^{\sigma,p_0}_s \in \overline{B}_{f(\eta)}(p_0) \quad \text{ and } \quad \max_{\theta \in \Theta} \Big| \frac{\d \sigma_\theta }{\d \langle \sigma ,p\rangle}(s) - 1   \Big| \, \leq \, \beta \, \eta %\quad \forall \, \eta \in \big(0,\widehat{\eta}\big), \ \sigma \in \Se, \, \text{ and } \, p, q^{\sigma,p}_s \in B_\eta(p_1).
   \]
   for every $\eta \in \big(0,\widehat{\eta}\big)$, $p \in B_\eta(p_1)$, $\sigma \in \Se$, and $s \in \cup_{\theta \in \Theta} \supp(\sigma_\theta)$ such that $q^{\sigma,p}_s \in B_{\eta}(p_1)$. 
    Moreover, since $\lim_{\eta \to 0} f(\eta) =0$ (by construction), there exists $\eta_1 \in \big(0, \widehat{\eta}\big)$ such that $f(\eta) \in (0, \delta_0)$ for all $\eta \in (0,\eta_1)$. We conclude that condition \eqref{eqn:cross-prior-small-1} holds for any $\delta_1 := \eta \in (0,\eta_1)$.

\noindent \textbf{Step 2: Ensuring condition \eqref{eqn:cross-prior-small-2}.} By construction, for every $\eta \in (0,\overline{\eta})$, we have 
\[
 \left| \frac{p_1(\theta)}{p(\theta)} -1 \right| \, \leq \,    \frac{\eta}{p(\theta)}  \, \leq \,  \frac{\eta}{\underline{m}(\eta)}\qquad \forall \, p \in B_{\eta}(p_1) \text{ and } \theta \in \Theta.
\]
Thus, for the $\widehat{\eta} \in (0,\overline{\eta})$ and $\beta \in \R_{++}$ defined above in Step 1, it follows (cf. \eqref{eqn:cross-prior-small-3}) that
\[
0 \, < \, \frac{\beta\,\eta}{2} \, < \,  \frac{1}{2} \quad \text{ and } \quad 1 - \frac{\beta\,\eta}{2} \, \leq \, \frac{p_1(\theta)}{p(\theta)}   \, \leq \, 1 + \frac{\beta\,\eta}{2}  \qquad \forall \, \eta \in \big(0,\widehat{\eta}\big), \ p \in B_{\eta}(p_1), \, \text{ and } \theta \in \Theta.
\]
Define the maps $\overline{g} : \big(0,\widehat{\eta}\big) \to \big(0,\frac{5}{4}\big)$ and $\underline{g} : \big(0,\widehat{\eta}\big) \to \big(0,\frac{3}{4}\big)$ as
\[
\overline{g}(\eta) := \left( 1 + \frac{\beta\,\eta}{2}\right)^2 -1 \qquad \text{ and } \qquad \underline{g}(\eta) := 1 -  \left( 1 - \frac{\beta\,\eta}{2}\right)^2.
\]
Plugging these definitions into the preceding display, we have
\[
\sqrt{1-\underline{g}(\eta)} \, \leq \, \frac{p_1(\theta)}{p(\theta)}   \, \leq \, \sqrt{1+ \overline{g}(\eta)} \qquad \forall \, \eta \in \big(0,\widehat{\eta}\big), \ p \in B_{\eta}(p_1), \, \text{ and } \theta \in \Theta.
\]
By construction, $\lim_{\eta \to 0} \max\{\overline{g}(\eta), \underline{g}(\eta)\} = 0$. Hence, there exists $\eta_2 \in \big(0,\widehat{\eta}\big)$ such that $\sup_{\eta \in (0,\eta_2)} \max\{\overline{g}(\eta), \underline{g}(\eta)\} <1$. Therefore, the map $g : (0,\eta_2) \to (0,1)$ given by $g(\eta):= \max\{\overline{g}(\eta), \underline{g}(\eta)\}$ is well-defined and $\lim_{\eta\to0} g(\eta)=0$. Moreover, the above display implies
\[
\sqrt{1-g(\eta)} \, \leq \, \frac{p_1(\theta)}{p(\theta)}   \, \leq \, \sqrt{1+ g(\eta)} \qquad \forall \, \eta \in (0, \eta_2), \ p \in B_{\eta}(p_1), \, \text{ and } \theta \in \Theta.
\]
We conclude that condition \eqref{eqn:cross-prior-small-2} holds for this map $g$ and all $ \delta_1 := \eta \in (0,\eta_2)$.

\noindent \textbf{Step 3: Wrapping Up.} Define $\overline{\delta}_1 := \min\{\eta_1, \eta_2\}>0$. Then conditions \eqref{eqn:cross-prior-small-1} and \eqref{eqn:cross-prior-small-2} both hold for the constant $\beta\in\R_{++}$ defined in Step 1, the (restriction to $(0,\overline{\delta}_1)\subseteq (0,\eta_2)$ of the) map $g : (0,\overline{\delta}_1) \to (0,1)$ with $\lim_{\delta_1 \to 0} g(\delta_1) = 0$ defined in Step 2, and every $\delta_1 \in (0,\overline{\delta}_1)$. 
\end{proof}

\section{Proofs of Additional Results}\label{app:additional-proofs}

\subsection{Proof of Proposition \ref{prop:ups:additive} }\label{proof:ups:additive}

We prove \cref{prop:ups:additive} via a series of lemmas. The first lemma implies the ``$\implies$'' direction of \cref{prop:ups:additive}. It also formalizes the important fact that all \nameref{defi:ups} costs are \nameref{axiom:slp}. 

\begin{lem}\label{lem:ups:to:additive}
    For any $C \in \C$, 
    \[
    \hspace{-0.5em}
    \text{$C$ is \nameref{defi:ups}} \ \ \implies \ \ \text{$C$ is \hyperref[axiom:additive]{Additive} and $\dom(C) = \Delta(W)\cup\Ex^\varnothing$ for convex $W \subseteq \Delta(\Theta)$} \ \ \implies \ \ \text{$C$ is \nameref{axiom:slp}.}
    \]
\end{lem}
\begin{proof}
    We prove each of the two implications in turn.

    \noindent \textbf{Implication 1: \nameref{defi:ups}$\implies$\hyperref[axiom:additive]{Additive}.} Let $C = C^H_\text{ups}$ for some convex $H : \Delta(\Theta) \to (-\infty,+\infty]$. By definition, $W:=\dom(H)\subseteq \Delta(\Theta)$ is convex and $\dom(C^H_\text{ups}) = \Delta(W) \cup \Ex^\varnothing$. If $W=\emptyset$, then \hyperref[axiom:additive]{Additivity} holds trivially. Thus, in what follows, we focus on the generic case $W\neq\emptyset$. 

    Take any $\Pi \in \Delta(\Ex)$ such that $\E_\Pi[\pi_2] \in \dom(C^H_\text{ups})= \Delta(W) \cup \Ex^\varnothing$. First, consider the trivial case where $\E_\Pi[\pi_2] \in \Ex^\varnothing$. This implies that $\pi_1 \in \Ex^\varnothing$ and $\Pi(\Ex^\varnothing) = 1$, and hence that $
    C^H_\text{ups}(\E_\Pi[\pi_2])  = C^H_\text{ups}(\pi_1)  =\E_\Pi[C^H_\text{ups}(\pi_2)] =0$. Thus, $C^H_\text{ups}(\E_\Pi[\pi_2])  = C^H_\text{ups}(\pi_1)  +\E_\Pi[C^H_\text{ups}(\pi_2)]$.

    Next, consider the nontrivial case where $\E_\Pi[\pi_2] \in \Delta(W) \backslash \Ex^\varnothing$. In this case, $\Pi$ satisfies: (i) $\pi_1 \in \Delta(W)$, and (ii) $\pi_2 \in \Delta(W)$ $\Pi$-almost surely. Property (i) holds because $\supp(\pi_1) \subseteq \text{conv} \left(\supp(\E_\Pi[\pi_2])\right) \subseteq W$, where the first inclusion holds because $ \pi_1 \leq_\text{mps}\E_\Pi[\pi_2]$ (by definition) and the second inclusion holds because $\supp(\E_\Pi[\pi_2])\subseteq W$ and $W$ is convex. Property (ii) holds since $\supp\left(\E_\Pi[\pi_2]\right) \subseteq W$ and because the definition of $\E_\Pi[\pi_2]$ implies
     \begin{align*}
         \int_{\Ex} \pi_2\left(\supp (\E_\Pi[\pi_2])\right)\d \Pi(\pi_2) \, = \, \E_\Pi[\pi_2]\left( \supp(\E_\Pi[\pi_2]) \right) \, = \, 1,
     \end{align*}
     which in turn implies that $\supp(\pi_2) \subseteq \supp (\E_\Pi[\pi_2])$ for $\Pi$-almost every $\pi_2 \in \Ex$ (as $\supp (\E_\Pi[\pi_2])$ is closed and the integrand on the left-hand side must $\Pi$-a.s. equal $1$). 
     
     By \cref{defi:ups}, the supposition that $\E_\Pi[\pi_2] \in \Delta(W)$ and properties (i)--(ii) imply that 
    \begin{align}
        C^H_\text{ups}\left( \E_\Pi[\pi_2]\right) \, &= \, \E_{\E_\Pi[\pi_2]} [ H(q) - H(p_{\pi_1})] \,  \, = \, \E_\Pi\left[ \E_{\pi_2}[H(q)] - H(p_{\pi_1}) \right], \label{eqn:ups-add-exp1} \\
        %%%
        C^H_\text{ups}(\pi_1) \, &= \, \E_{\pi_1} [ H(q) - H(p_{\pi_1})] \, \ \ \ \ \ \ \, = \, \E_\Pi[H(p_{\pi_2})  - H(p_{\pi_1})], \label{eqn:ups-add-exp2} \\
        %%%
        \E_\Pi[C^H_\text{ups}(\pi_2)] \, &= \, %\E_\Pi[\mathbf{1}(\pi_2 \in \Delta(W)) \cdot C^H_\text{ups}(\pi_2)] \, = \, 
        \E_\Pi\left[ \E_{\pi_2} [ H(q)] - H(p_{\pi_2}) \right], \label{eqn:ups-add-exp3}
    \end{align}
    where in \eqref{eqn:ups-add-exp1} we use the identity $p_{\E_\Pi[\pi_2]} = p_{\pi_1}$ and the second equality is by the Law of Iterated Expectations, and in \eqref{eqn:ups-add-exp2} the second equality is by definition of $\pi_1$. Note that $p_{\pi_1} \in W$ (as $\pi_1 \in \Delta(W)$ and $W$ is convex) and that the convex function $H$ is bounded below (as $\dom(H) \neq \emptyset$ and $\Delta(\Theta)$ is bounded).\footnote{Note that $H$ is a proper convex function because $-\infty \notin H[\Delta(\Theta)]$ (by definition) and $W = \dom(H) \neq \emptyset$ (by hypothesis). Hence, there exists $p^* \in \ri(\dom(H))$ such that $\partial H(p^*) \neq \emptyset$ \parencite[Theorem 23.4]{rock70}. For any $v \in \partial H(p^*)$, we have $H(p) \geq H(p^*) + (p-p^*)^\top v$ for all $p \in \Delta(\Theta)$. Since $\Delta(\Theta)$ is a bounded set, it follows that $H$ is bounded below.} Thus, since the expressions in \eqref{eqn:ups-add-exp1} and \eqref{eqn:ups-add-exp2} are finite (by the supposition and property (i)), it follows that the maps $\pi_2 \mapsto \E_{\pi_2}[H(q)]$ and $\pi_2 \mapsto H(p_{\pi_2})$ are $\Pi$-integrable. Therefore, combining \eqref{eqn:ups-add-exp1}, \eqref{eqn:ups-add-exp2} and \eqref{eqn:ups-add-exp3} yields 
    \[
    C^H_\text{ups}(\E_\Pi[\pi_2]) = C^H_\text{ups}(\pi_1) + \E_\Pi[C^H_\text{ups}(\pi_2)].
    \]
    %and \hyperref[axiom:additive]{Additivity} is satisfied.

    Since the given $\Pi \in \Delta(\Ex)$ with $\E_\Pi[\pi_2]\in\dom(C^H_\text{ups})$ was arbitrary, $C^H_\text{ups}$ is \hyperref[axiom:additive]{Additive}.

    \noindent \textbf{Implication 2: \hyperref[axiom:additive]{Additive}$\implies$\nameref{axiom:slp}.} Let $C \in \C$ be \hyperref[axiom:additive]{Additive} and satisfy $\dom(C) = \Delta(W) \cup \Ex^\varnothing$ for some convex $W \subseteq \Delta(\Theta)$. In what follows, we show that this implies that $C$ is \hyperref[axiom:mono]{Monotone} and \hyperref[axiom:POSL]{Subadditive}. 
    \cref{prop:1} then delivers the desired conclusion that $C$ is \nameref{axiom:slp}.

    We first show that $C$ is \hyperref[axiom:mono]{Monotone}. Take any $\pi,\pi' \in \Ex$ such that $\pi'\geq_\text{mps}\pi$. There are two cases. First, if $\pi' \notin \dom(C)$, then $C(\pi') = +\infty \geq C(\pi)$. Second, suppose $\pi' \in \dom(C)$. By definition of the MPS order, there exists some two-step strategy $\Pi \in \Delta(\Ex)$ such that $\pi' = \E_\Pi[\pi_2]$ and $\pi = \pi_1$.\footnote{Formally, by definition $\pi'\geq_\text{mps}\pi$ if and only if there exists a Borel map $q \in \supp(\pi) \mapsto r(\cdot \mid q) \in \Delta(\supp(\pi'))\subseteq \Ex$ such that: (i) $p_{r(\cdot\mid q)}  = q$ for all $q \in \supp(\pi)$, and (ii) $\pi'(B) = \int r(B \mid q) \dd \pi(q)$ for all Borel $B \subseteq \Delta(\Theta)$. We can then define $\Pi \in \Delta(\Ex)$ as $\Pi(B) := \pi\left(\{q \in \supp(\pi) \mid r(\cdot \mid q) \in B\} \right)$ for all Borel $B\subseteq \Ex$. By construction, we have $\E_\Pi[\pi_2] = \pi'$ and $\pi_1 = \pi$.} Therefore, since $\pi' \in \dom(C)$ and $C \in \C$ is \hyperref[axiom:additive]{Additive}, we have
    \[
    C(\pi') \, = \, C(\E_\Pi[\pi_2]) \, = \, C(\pi_1) + \E_\Pi[C(\pi_2)] \, \geq \, C(\pi_1) \, = \, C(\pi). 
    \]
    Since the given $\pi' \geq_\text{mps} \pi$ were arbitrary, we conclude that $C$ is \hyperref[axiom:mono]{Monotone}.

    We now show that $C$ is \hyperref[axiom:POSL]{Subadditive}. Take any $\Pi \in \Delta^\dag(\Ex)$. There are two cases:

    \emph{Case 1: Let $\E_{\Pi}[\pi_2]\in \dom (C)$.} \hyperref[axiom:additive]{Additivity} directly implies $C(\E_{\Pi}[\pi_2]) \leq C(\pi_1) + \E_\Pi[C(\pi_2)]$. 

    \emph{Case 2: Let $\E_{\Pi}[\pi_2]\notin \dom (C)$.} In this case, \hyperref[axiom:additive]{Additivity} has no bite. Instead, we claim that $\left[\{\pi_1\}\cup \supp(\Pi)\right] \not\subseteq \dom(C)$. Note that the claim implies that $C(\pi_1)=+\infty$ or that there exists some $\pi_2\in \supp(\Pi)$ with $C(\pi_2)=+\infty$, either of which in turn implies the inequality $C(\E_\Pi[\pi_2])  \leq +\infty =  C(\pi_1) + \E_\Pi[C(\pi_2)]$. Therefore, it suffices to prove the claim.%; we do so by contradiction. Let $p:=p_{\pi_1}$.

    Suppose, towards a contradiction, that $\left[\{\pi_1\}\cup \supp(\Pi)\right] \subseteq \dom(C)=\Delta(W)\cup\Ex^\varnothing$. There are two sub-cases to consider, depending on whether $p:= p_{\pi_1}$ is contained in $W$.

    First, consider the case $p \notin W$. This implies $\pi_1 \notin \Delta(W)$ (as $W$ is convex), and thus the supposition implies $\pi_1 = \delta_p \in \Ex^\varnothing$. It follows that $p_{\pi_2} = p \notin W$ for all $\pi_2 \in \supp(\Pi)$, and hence that $\supp(\Pi)\cap \Delta(W) = \emptyset$ (as $W$ is convex). The supposition then implies $\supp(\Pi) = \{\delta_p\}$. We thus obtain  $\E_\Pi[\pi_2] = \delta_p \in\Ex^\varnothing \subseteq\dom (C)$, which yields the desired contradiction.

    Second, consider the case $p \in W$. By the supposition, we have $\pi_1 \in \Delta(W) \cup\{\delta_p\} = \Delta(W)$. Define the Borel measure $\mu_1$ on $\Delta(\Theta)$ as $\mu_1(B) := \Pi(\{\pi_2 \in \Ex \mid p_{\pi_2} \in B\} \cap \Ex^\varnothing)$ for all Borel $B \subseteq \Delta(\Theta)$. By construction, we have $\mu_1(B) \leq \pi_1(B)$ for all Borel $B \subseteq \Delta(\Theta)$, which implies  $\supp(\mu_1) \subseteq \supp(\pi_1)\subseteq W$. Moreover, since $\supp(\Pi) \backslash \Ex^\varnothing$ is finite, it holds that 
    \[
    \E_\Pi[\pi_2] \, =  \, \sum_{\pi_2 \in \supp(\Pi) \backslash \Ex^\varnothing} \Pi (\{\pi_2\}) \cdot \pi_2 \, + \, \int_{\Ex^\varnothing} \pi_2 \dd \Pi(\pi_2) \, =  \, \sum_{\pi_2 \in \supp(\Pi) \backslash \Ex^\varnothing} \Pi (\{\pi_2\}) \cdot \pi_2 \, + \, \mu_1,
    \] 
    where the second equality is by a change of variables. Since the supposition implies $\supp(\mu_1)\cup \big[ \bigcup_{\pi_2 \in \supp(\Pi)\backslash \Ex^\varnothing} \supp(\pi_2)\big] \subseteq W$ and the union is finite, it follows that $\supp(\E_\Pi[\pi_2]) \subseteq W$. We thus obtain $\E_{\Pi}[\pi_2]\in \Delta(W)\subseteq \dom (C)$, which yields the desired contradiction.

    Since the given $\Pi \in \Delta^\dag(\Ex)$ was arbitrary, we conclude that $C$ is \hyperref[axiom:POSL]{Subadditive}. 
\end{proof}

The remaining four lemmas imply the more subtle ``$\impliedby$'' direction of \cref{prop:ups:additive}. For any $X \subseteq \Delta(\Theta)$, we denote by $\intr (X) \subseteq X$ the interior of $X$ with respect to the subspace topology on $\Delta(\Theta)$. (Recall that $\ri(X) \subseteq X$ denotes the \emph{relative} interior of $X$, i.e., with respect to the subspace topology on the affine hull of $X$.) We begin with a technical fact:

\begin{lem}\label{lem:span}
    Let $X\subseteq\Delta(\Theta)$ be open and convex. For every $p\in X$, there is a set $\{q_i(p)\}_{i=1}^{|\Theta|} \subseteq X$ of $|\Theta|$ (distinct) linearly independent points such that $p\in \intr \left( \mathrm{conv}\left(\{q_i(p)\}_{i=1}^{|\Theta|} \right) \right)\subseteq X$. 
\end{lem}
\begin{proof}
    Let $p \in X$ be given. Enumerate the state space as $\Theta = \{\theta_i\}_{i=1}^{|\Theta|}$. Since $X$ is open, there exists sufficiently small $\eta \in (0,1)$ such that, letting $q_{i}(p) := \eta \delta_{\theta_i} + (1-\eta) p$, we have $Q:= \{q_{i}(p)\}_{i=1}^{|\Theta|} \subseteq X$. By construction, the set $Q$ comprises $|\Theta|$ distinct, linearly independent points. Since $X$ is convex, we also have $\intr \left( \text{conv}(Q) \right) \subseteq \mathrm{conv}\left(Q \right) \subseteq X$. Thus, it suffices to show that $ p \in \intr \left( \text{conv}(Q) \right)$. To this end, note that $\text{conv}(Q) = \left\{\eta q + (1-\eta)p \mid q \in \Delta(\Theta)\right\}$. 

    Let $m := \min\{p(\theta) \mid \theta \in \supp(p)\}$ and $\epsilon:= m \cdot \eta/2$. Note that $m > \epsilon>0$. We claim that $B_\epsilon(p)\subseteq \text{conv}(Q)$. Take any $r \in B_\epsilon(p)$, and define $q := p + (r-p)/\eta$. Observe that $r \in \text{conv}(Q)$ if and only if $q \in \Delta(\Theta)$. Thus, in what follows, we show that $q \in \Delta(\Theta)$. First, note that $\mathbf{1}^\top q = \mathbf{1}^\top p = 1$. Next, we show that $\min_{\theta \in \Theta}q(\theta)\geq 0$. For every $\theta \notin \supp(p)$, we have $q(\theta) = r(\theta)/\eta\geq 0$ (because $r \in B_\epsilon(p) \subseteq\Delta(\Theta)$). For every $\theta \in \supp(p)$, we have
    \[
    q(\theta) \, = \, p(\theta) + \frac{r(\theta)-p(\theta)}{\eta}\, \geq \,  m - \frac{\|r(\theta) - p(\theta)\|}{\eta}  \, \geq \,  m - \frac{\epsilon}{\eta} \, = \, \frac{m}{2} \, > \, 0,
    \]
    where the first (in)equality is by definition of $q$, the second inequality is by definition of $m$ (first term) and the fact that $r(\theta)-p(\theta) \geq - \max_{\tau\in \Theta} |r(\tau)-p(\tau)| \geq - \|r-p\|$ (second term), the third inequality holds because $r \in B_\epsilon(p)$, and the last two inequalities are by definition of $m$ and $\epsilon$. Consequently, we have $q \in \Delta(\Theta)$, as desired. Since the given $r \in B_\epsilon(p)$ was arbitrary, it follows that $B_\epsilon(p) \subseteq \text{conv}(Q)$, as claimed. This implies $ p \in \intr \left( \text{conv}(Q) \right)$. 
\end{proof}

The next lemma uses \cref{lem:span} to show that any open convex set $W \subseteq \Delta(\Theta)$ can be covered by (the interiors of) a nested sequence of polytopes. As is standard, we call $K \subseteq \Delta(\Theta)$ a \emph{polytope} if $K = \text{conv}(\{p_1, \dots,p_n\})$ for some finite set $\{p_1, \dots,p_n\} \subseteq \Delta(\Theta)$. 

\begin{lem}\label{lem:polygon:approx}
    Let $W \subseteq \Delta(\Theta)$ be open and convex. There is a sequence $(W_n)_{n=1}^{\infty}$ of polytopes with nonempty interiors such that: (i) $W_n \subseteq W_{n+1}\subseteq W$ for all $n \in \mathbb{N}$, and (ii) $W = \cup_{n\in\mathbb{N}} \intr (W_n).$
\end{lem}
\begin{proof}
\Cref{lem:span} (with $X:= W$) implies that, for every $p \in W$, there exists a polytope $K_p \subseteq W$ such that $p \in \intr (K_p)$. Thus, $\{\intr(K_p)\}_{p \in W}$ is an open cover of $W$. Since $W$ is an open subset of the separable metric space $\Delta(\Theta)$, it is Lindel\"{o}f (i.e., every open cover of $W$ admits a countable subcover). Thus, there exists a sequence $(p_n)_{n=1}^\infty$ in $W$ such that $\{\intr(K_{p_n})\}_{n=1}^\infty$ is an open cover of $W$. We recursively define $(W_n)_{n=1}^\infty$ as $W_1 := K_{p_1}$ and $W_n := \text{conv}(W_{n-1} \cup K_{p_n})$ for all $n \geq 2$. By induction, each $W_n$ is a polytope (as each $K_{p_n}$ is a polytope). Property (i) holds because, for every $n \in \mathbb{N}$, $W_n \subseteq W_{n+1}$ (by construction) and $W_n \subseteq W$ (as $\cup_{m=1}^\infty K_{p_m}\subseteq W$ and $W$ is convex). Moreover, property (ii) holds because $\cup_{n \in \mathbb{N}}\intr(W_n) \subseteq \cup_{n \in \mathbb{N}}W_n \subseteq W$ (by property (i)) and $W = \cup_{n\in\mathbb{N}} \intr(K_{p_n}) \subseteq \cup_{n\in\mathbb{N}} \intr(W_n)$ (as $\intr(K_{p_n}) \subseteq \intr(W_n)$ for all $n \in \mathbb{N}$). 
\end{proof}

The next lemma, which is the main step in the proof, establishes that any (finite-valued) \hyperref[axiom:additive]{Additive} cost function defined on a polytope with nonempty interior is \nameref{defi:ups}.

\begin{lem}\label{lem:ups:polygon}
    Let $W \subseteq \Delta(\Theta)$ be open and convex, and $C \in \C$ be \hyperref[axiom:additive]{Additive} with $\dom(C) = \Delta(W) \cup \Ex^\varnothing$. For any polytope $W_0 \subseteq W$ such that $\intr(W_0) \neq \emptyset$, there exists a convex function $H : W_0 \to \R$ such that $C(\pi) = C^H_\text{ups}(\pi)$ for all $\pi \in \Delta(W_0)$. 
\end{lem}
\begin{proof}
The proof consists of three steps:

\noindent \textbf{Step 1: Project $C$ onto Auxiliary State Space.} Denote the vertices of \(W_0\) by \(\{q_1,\ldots,q_k\}\). Since $\intr(W_0) \neq \emptyset$, we have \(k\ge |\Theta|\). First, define the auxiliary state space \(\widehat{\Theta}:=\{1,\ldots,k\}\) (where each $i \in \widehat{\Theta}$ indexes the associated vertex $q_i \in W_0$). Next, define the affine, many-to-one map \(\gamma:\Delta(\widehat{\Theta})\to W_0\) as $\gamma(\widehat{q}) :=\sum_{i=1}^k \widehat{q}(i) q_i$ for all \(\widehat{q}\in \Delta(\widehat{\Theta})\). Let $\widehat{\Ex}:=\Delta(\Delta(\widehat{\Theta}))$. For every $\widehat{\pi} \in\widehat{\Ex} $, define the pushforward measure $\widehat{\pi}^\gamma \in \Delta(W_0)$ as $\widehat{\pi}^\gamma(B) := \widehat{\pi}(\gamma^{-1}(B))$ for all Borel $B \subseteq W_0$. Finally, define the cost function $\widehat{C}: \widehat{\Ex} \to \R_+$ as $\widehat{C}(\widehat{\pi}) := C(\widehat{\pi}^\gamma)$ for all $\widehat{\pi} \in \widehat{\Ex}$.\footnote{$\widehat{C}$ is a well-defined cost function on $\widehat{\Theta}$ because, for every $\widehat{q}\in \Delta(\widehat{\Theta})$ and $\widehat{\pi} = \delta_{\widehat{q}}$, we have $\widehat{\pi}^\gamma = \delta_{\gamma(\widehat{q})}$ and hence $C(\widehat{\pi}) = 0$.}

\noindent \textbf{Step 2: The Projection of $C$ is UPS.} First, we show that \(\widehat{C}\) is \hyperref[axiom:additive]{Additive}. For any $\widehat{\Pi} \in \Delta(\widehat{\Ex})$, 
\begin{align*}
    \widehat{C}(\E_{\widehat{\Pi}}[\widehat{\pi}_2]) \, = \, C(\E_{\widehat{\Pi}}[\widehat{\pi}^\gamma_2]) \, = \, C\left(\widehat{\pi}^\gamma_1\right)+\E_{\widehat{\Pi}}\left[C(\widehat{\pi}^\gamma_2)\right] \, = \, \widehat{C}\left(\widehat{\pi}_1\right)+\E_{\widehat{\Pi}}\left[\widehat{C}(\widehat{\pi})\right],
\end{align*}
where the first and third equalities are by definition of $\widehat{C}$, and the second equality holds because $C$ is \hyperref[axiom:additive]{Additive} and $\gamma$ is affine. This establishes that $\widehat{C}$ is \hyperref[axiom:additive]{Additive}, as desired. 

Next, we show that $\widehat{C}$ is \nameref{defi:ups}.\footnote{For this part of Step 2, our argument mirrors that in the proof of \textcite[Theorem 3]{zhong2022optimal}.} For any prior $\widehat{p} \in \Delta(\widehat{\Theta})$, we denote by $\widehat{\pi}^\text{full}_{\widehat{p}} := \sum_{i=1}^k \widehat{p}(i) \delta_i$ the associated fully revealing random posterior. Define $\widehat{H} : \Delta(\widehat{\Theta}) \to \R$ as $\widehat{H}(\widehat{p}) := - \widehat{C}(\widehat{\pi}^\text{full}_{\widehat{p}})$. Since $\widehat{C}$ is \hyperref[axiom:additive]{Additive}, for every $\widehat{\pi} \in \widehat{\Ex}$ it holds that $\widehat{C}(\widehat{\pi}^\text{full}_{\widehat{p}_{\widehat{\pi}}}) \, = \, \widehat{C}( \widehat{\pi}) + \E_{\widehat{\pi}} \big[\widehat{C}(\widehat{\pi}^\text{full}_{\widehat{q}}) \big]$, and hence
\[
 \widehat{C}( \widehat{\pi}) \, = \,  \widehat{C}(\widehat{\pi}^\text{full}_{\widehat{p}_{\widehat{\pi}}})  - \E_{\widehat{\pi}} \big[\widehat{C}(\widehat{\pi}^\text{full}_{\widehat{q}}) \big] \, = \, \E_{\widehat{\pi}} \big[\widehat{H}(\widehat{q}) - \widehat{H}(\widehat{p}_{\widehat{\pi}}) \big].
\]
Since $\widehat{\C}\geq0$, this implies that $\widehat{H}$ is convex. Consequently, we have $\widehat{C} = \widehat{C}^{\widehat{H}}_\text{ups}$ as desired.

\noindent \textbf{Step 3: $C$ is UPS.} Note that, for every $q \in W_0$ and $\widehat{\pi} \in \Delta(\gamma^{-1}(q))$, the pushforward measure $\widehat{\pi}^\gamma = \delta_q \in \Ex^\varnothing$ is trivial. This implies that $\widehat{H}$ is affine on each (convex) subset $\gamma^{-1}(q) \subseteq \Delta(\widehat{\Theta})$. Thus, for each $q \in W_0$, there exists $\beta(q) \in \R^{k}$ such that $\widehat{H}(\widehat{q}) = \beta(q) ^\top \widehat{q}$ for all $\widehat{q} \in \gamma^{-1}(q)$.
%we denote the slope of this affine function as $\beta(q) \in \R^{k}$. 

Take any \(\widehat{q}_0\in\ri(\Delta(\widehat{\Theta}))\) and let \(q_0:=\gamma(\widehat{q}_0)\). Define the map $\widetilde{H}: \Delta(\widehat{\Theta}) \to \R$ as $\widetilde{H}(\widehat{q}):=\widehat{H}(\widehat{q})-\beta(q_0)^\top \widehat{q}$. Note that, by construction, $\widetilde{H}$ is constant on $\gamma^{-1}(q_0)$. We claim that $\widetilde{H}$ is constant on $\gamma^{-1}(q)$, for every $q \in W_0$. 

Suppose, towards a contradiction, that there exist $q \in W_0$ and \(\widehat{q}_1,\widehat{q}_2\in\gamma^{-1}(q)\) with \(\widetilde{H}(\widehat{q}_1)\neq \widetilde{H}(\widehat{q}_2)\). For every $\epsilon \in (0,1)$, define $\widehat{p}_{\epsilon} \in \Delta(\widehat{\Theta})$ as $\widehat{p}_{\epsilon}:=\epsilon \widehat{q}_0+(1-\epsilon) \widehat{q}_1$, and let \(\widehat{q}_{0\epsilon} \in \R^k\) satisfy \(\widehat{p}_{\epsilon}=\epsilon \widehat{q}_{0\epsilon}+(1-\epsilon) \widehat{q}_2\). Since $\widehat{q}_0 \in \ri \big(\Delta(\widehat{\Theta})\big)$, there exists sufficiently large $\epsilon \in (0,1)$ such that \(\widehat{q}_{0\epsilon}\in \Delta(\widehat{\Theta})\). Fix this value of $\epsilon \in (0,1)$ henceforth. Since \(\gamma\) is affine, \(\gamma (\widehat{q}_{0\epsilon})=\frac{1}{\epsilon}(\gamma(\widehat{p}_{\epsilon})-(1-\epsilon)\gamma(\widehat{q}_2))=q_0\). Thus, \(\widehat{q}_{0\epsilon}\in \gamma^{-1}(q_0)\). Now, define $\widehat{\pi}_1, \widehat{\pi}_2 \in \widehat{\Ex}$ as \(\widehat{\pi}_1 :=\epsilon \delta_{\widehat{q}_0}+(1-\epsilon)\delta_{\widehat{q}_1}\) and \(\widehat{\pi}_2 :=\epsilon \delta_{\widehat{q}_{0\epsilon}}+(1-\epsilon)\delta_{\widehat{q}_2}\). By construction, $\widehat{p}_{\widehat{\pi}_1} = \widehat{p}_{\widehat{\pi}_2} = \widehat{p}_\epsilon$. Evidently, \(\widehat{\pi}_1^\gamma=\widehat{\pi}_2^\gamma=\epsilon \delta_{q_0}+(1-\epsilon)\delta_{q}\), which implies \(\widehat{C}(\widehat{\pi}_1)=\widehat{C}(\widehat{\pi}_2)\). However, the supposition implies 
\begin{align*}
    \widehat{C}(\widehat{\pi}_1) =& \epsilon\widehat{H}(\widehat{q}_0)+(1-\epsilon)\widehat{H}(\widehat{q}_1)-\widehat{H}(\widehat{p}_{\epsilon})\\
    =&\epsilon\widetilde{H}(\widehat{q}_0)+(1-\epsilon)\widetilde{H}(\widehat{q}_1)-\widetilde{H}(\widehat{p}_{\epsilon})\\
    \neq&\epsilon\widetilde{H}(\widehat{q}_{0\epsilon})+(1-\epsilon)\widetilde{H}(\widehat{q}_2)-\widetilde{H}(\widehat{p}_{\epsilon})\\
    =&\epsilon\widehat{H}(\widehat{q}_{0\epsilon})+(1-\epsilon)\widehat{H}(\widehat{q}_2)-\widehat{H}(\widehat{p}_{\epsilon}) = \widehat{C}(\widehat{\pi}_2),
\end{align*}
which yields the desired contradiction. Thus, $\widetilde{H}$ is constant on each \(\gamma^{-1}(q)\), as claimed.  

 We now use $\widetilde{H}$ to construct a function $H : W_0 \to \mathbb{R}$ such that $C (\pi) = C^H_\text{ups} (\pi)$ for all $\pi \in \Delta(W_0)$. Since $\gamma$ is affine and surjective, it is an open mapping (\cite[Theorem 2.11]{Rudin1973}). Thus, the inverse correspondence $\gamma^{-1} : W_0 \rightrightarrows \Delta(\widehat{\Theta})$ is lower hemi-continuous and nonempty-, convex-, and compact-valued (\cite{aliprantis-border99}, Theorem 17.7). 
 %\footnote{\awb{[@SELF: These theorems, as stated in Aliprantis-Border, require that the codomain be a vector space. They will also hold if the codomain is a convex subset of a vector space (as is the case here), but I need to find the right references.]}} 
 The Michael Selection Theorem (\cite{aliprantis-border99}, Theorem 17.66) then yields the existence of a continuous map $f : W_0 \to \Delta(\widehat{\Theta})$ such that $f(q) \in \gamma^{-1}(q)$ for all $q \in W_0$. 
 
 Define $H: W_0 \to \mathbb{R}$ as $H(q) := \widetilde{H}(f(q))$. Note that $H$ is integrable on $W_0$ (as $\widetilde{H}$ is convex and bounded on $\Delta(\widehat{\Theta})$ and $f$ is continuous). Take any $\pi \in \Delta(W_0)$. Denote by $f_*(\pi) \in \Delta(\Delta(\widehat{\Theta}))$ the pushforward measure defined as $f_*(\pi)(\widehat{B}) := \pi (f^{-1}(\widehat{B}))$ for all Borel $\widehat{B} \subseteq \Delta(\widehat{\Theta})$. By construction, we have $[f_* (\pi)]^\gamma = \pi$ and $\widehat{p}_{f_*(\pi)} = \E_\pi [ f(q)]$.\footnote{For any Borel $B \subseteq W_0$, we have $[f_* (\pi)]^\gamma (B) = \pi\left( f^{-1}( \gamma^{-1}(B)) \right) = \pi(B)$, where the second equality holds because $\gamma \circ f : W_0 \to W_0$ is the identity map, which implies $f^{-1}( \gamma^{-1}(B)) = \left\{q \in W_0 : \gamma( f(q) ) \in B\right\} = B$.} Therefore, 
  \begin{align*}
 C(\pi) = \widehat{C}(f_* (\pi)) &= \E_{f_* (\pi)} \left[ \widetilde{H}(\widehat{q}) - \widetilde{H}(\widehat{p}_{f_*(\pi)}) \right] \\
 &= \E_\pi \left[ \widetilde{H}(f(q))  - \widetilde{H}(\widehat{p}_{f_*(\pi)}) \right]\\
 & = \E_\pi \left[ \widetilde{H}(f(q))  - \widetilde{H}( f(p_\pi) ) \right]\\
 & = \E_\pi \left[ H(q) - H(p_\pi) \right] ,
 \end{align*}
where the first two lines hold by construction; the third line holds because $\gamma$ is affine and $\gamma\circ f$ is the identity map, which implies $\gamma(\widehat{p}_{f_*(\pi)}) = \E_\pi [ \gamma ( f(q) )] = \E_\pi [ q] = p_\pi$ and $\gamma ( f(p_\pi)) = p_\pi$, and because $\widetilde{H}$ is constant on $\gamma^{-1}(p_\pi)$; and the fourth line holds by construction. Since the given $\pi \in \Delta(W_0)$ was arbitrary and $C\succeq 0$, this implies that $H$ is convex. Consequently, the constructed $H$ satisfies $C (\pi) = C^H_\text{ups} (\pi)$ for all $\pi \in \Delta(W_0)$, as desired. 
\end{proof}

The final lemma lets us extend the domain of a \nameref{defi:ups} cost function in a consistent way. 

\begin{lem}\label{lem:ups:two:sets}
    Let $W \subseteq \Delta(\Theta)$ be open and convex, and $C \in \C$ be \nameref{defi:ups} with $\dom(C) = \Delta(W) \cup \Ex^\varnothing$. For any convex subsets $W_1, W_2 \subseteq W$ with $\intr(W_1) \neq \emptyset$ and $W_1 \subseteq W_2$, and any convex functions $H_1 , H_2 : \Delta(\Theta) \to (-\infty,+\infty]$ with $\dom(H_1) = W_1$ and $\dom(H_2) = W_2$, the following holds:

    \begin{quote}
    If $C(\pi) = C^{H_1}_\text{ups}(\pi)$ for all $\pi \in \Delta(W_1)$ and $C(\pi) = C^{H_2}_\text{ups}(\pi)$ for all $\pi \in \Delta(W_2)$, then there exists a convex function $H : \Delta(\Theta) \to (-\infty,+\infty]$ with $\dom(H) \supseteq W_2$ such that: (i) $C(\pi) = C^H_\text{ups}(\pi)$ for all $\pi \in \Delta(W_2)$, and (ii) $H(p) = H_1(p)$ for all $p \in W_1$. 
    \end{quote}
\end{lem}
\begin{proof}
    Fix any $p \in \intr(W_1)$. \cref{lem:span} (with $X:= \intr(W_1)$) implies that there exists a set $Q:=\{q_i\}_{i=1}^{|\Theta|} \subseteq \intr (W_1)$ of $|\Theta|$ linearly independent points with $p \in \intr\left( \text{conv}(Q) \right)\subseteq \intr(W_1)$. %Let $K:= \text{conv}(Q )\subseteq \intr(W_1)$. 
    
    Define $L : \text{conv}(Q) \to \R$ as $L(q):= H_2(q) - H_1(q)$. For every $\pi \in \Delta(\text{conv}(Q))$, we have
    \[
    0  \, = \, C(\pi) - C(\pi)\, = \, C^{H_2}_\text{ups}(\pi) - C^{H_1}_\text{ups}(\pi) \, = \, \E_\pi[L(q) - L(p_\pi)].
    \]
    This implies that $L$ is affine. Since $Q$ comprises $|\Theta|$ linearly independent points, for every $q \in \text{conv}(Q)$ there exists a unique $\alpha_q \in \Delta(\{1,\dots,|\Theta|\})$ such that $q = \sum_{i=1}^{|\Theta|} \alpha_q(i) q_i$. Thus, since $L$ is affine, we can write $L(q) = \sum_{i=1}^{|\Theta|} \alpha_q(i) L(q_i)$ for all $q \in \text{conv}(Q)$. Moreover, since $Q$ comprises $|\Theta|$ linearly independent points, for every $q \in W_2$ there exists a unique $\beta_q \in \R^{|\Theta|}$ (with $\mathbf{1}^\top \beta_q = 1$) such that $q = \sum_{i=1}^{|\Theta|} \beta_q(i) q_i$, where $\beta_q = \alpha_q$ for all $q \in  \text{conv}(Q)$. Thus, we can extend $L$ to the affine function $\overline{L}: W_2 \to \R$ defined as $\overline{L}(q) := \sum_{i=1}^{|\Theta|} \beta_q(i) L(q_i)$ for all $q \in W_2$. 
    
    We claim that $H_2(q) = H_1(q) + \overline{L}(q)$ for all $q \in W_1$. Given the claim, we can define $H: W_2 \to \R$ as $H(q):= H_2(q) - \overline{L}(q)$ for all $q \in W_2$, which satisfies the desired property (i) (as $\overline{L}$ is affine) and property (ii) (by construction).\footnote{We can define $H(q) \in (-\infty,+\infty]$ arbitrarily for $q \in \Delta(\Theta)\backslash W_2$.} Thus, it suffices to prove the claim. 

    To this end, first note that $H_2(q) = H_1(q) + \overline{L}(q)$ for all $q \in \text{conv}(Q)$ because $\overline{L}|_{\text{conv}(Q)} = L$. Next, take any $q \in W_1 \backslash \text{conv}(Q)$. By construction, we have $\ri (\text{conv}(Q)) \neq \emptyset$. Pick any $q'' \in \ri (\text{conv}(Q))$. Then there exists $\lambda \in (0,1)$ such that $q':= \lambda q + (1-\lambda)q'' \in \ri (\text{conv}(Q))$. Define $\widehat{\pi} \in \Delta(W_1)$ as $\widehat{\pi}:= \lambda \delta_q + (1-\lambda) \delta_{q''}$. By construction, we have $p_{\widehat{\pi}} = q'$. We now calculate the cost of $\widehat{\pi}$ in two different ways. First, we have
    \[
    C(\widehat{\pi}) \, = \, C^{H_1}_\text{ups}(\widehat{\pi}) \, = \, C^{H_1+\overline{L}}_\text{ups}(\widehat{\pi}) \, = \, \lambda \cdot (H_1+\overline{L})(q) + (1-\lambda) \cdot(H_1+\overline{L})(q'') - (H_1 + \overline{L})(q'),
    \]
    where the second equality holds because $\overline{L}$ is affine. Second, we also have
    \[
     C(\widehat{\pi}) \, = \, C^{H_2}_\text{ups}(\widehat{\pi}) \, = \, \lambda H_2(q) + (1-\lambda) H_2(q'') - H_2(q').
    \]
    Now, because $H_2|_{\text{conv}(Q)} = H_1|_{\text{conv}(Q)} + L = (H_1 + \overline{L})|_{\text{conv}(Q)} $ and $q', q'' \in\text{conv}(Q)$ by construction, combining the two displays above implies that
    \[
    0 \, = \, C^{H_2}_\text{ups}(\widehat{\pi}) -  C^{H_1+\overline{L}}_\text{ups}(\widehat{\pi}) \, = \,  \lambda H_2(q) - \lambda \cdot (H_1+\overline{L})(q).
    \]
    Since $\lambda \neq 0$, this implies that $H_2(q) = (H_1+\overline{L})(q)$, as desired. Thus, since the given $q \in W_1 \backslash \text{conv}(Q)$ was arbitrary, this completes the proof of the claim, and of the lemma.
\end{proof}

With \dred{Lemmas} \ref{lem:ups:to:additive}--\ref{lem:ups:two:sets} in hand, we now use them to prove \cref{prop:ups:additive}. 

\begin{proof}[Proof of \cref{prop:ups:additive}]
        The ``$\implies$'' direction follows directly from \cref{lem:ups:to:additive}. 

        For the ``$\impliedby$'' direction, we proceed as follows. Let $(W_n)_{n=1}^\infty$ be the nested sequence of polytopes given by \cref{lem:polygon:approx}. Since $C$ is \hyperref[axiom:additive]{Additive}, \cref{lem:ups:polygon} implies that $C$ \nameref{defi:ups} on each $W_n$. Namely, for every $n \in \mathbb{N}$, there exists convex \(H_n: W_n\to \R\) such that \(C(\pi)=C_\text{ups}^{H_n}(\pi)\) for all \(\pi\in\Delta (W_n)\). \Cref{lem:ups:two:sets} then implies that we can (without loss of generality) select the functions $\{H_n\}_{n=1}^\infty$ such that, for every $n \in \mathbb{N}$, \(H_{n+1}|_{W_n} =H_n\). Thus, we can define the convex $H : W \to \R$ as \(H(q)=H_{N(q)}(q)\) for \(N(q):=\min\{n \in \mathbb{N} \mid q\in W_n\}\). Finally, we verify that \(C (\pi)=C_\text{ups}^H(\pi)\) for all \( \pi \in \Delta(W)\). Take any $\pi \in \Delta(W)$. Since \(\supp (\pi) \subseteq W\) is compact and \(\{ \intr(W_n)\}_{n=1}^\infty\) is an open cover of \(\supp (\pi)\), there exists a finite subcover. Consequently, there exists \(n \in \mathbb{N}\) such that \(\supp (\pi)\subseteq \intr(W_n)\). This implies that  $C(\pi)=C_\text{ups}^{H_n}(\pi)=C_\text{ups}^H(\pi)$. Since the given $\pi \in \Delta(W)$ was arbitrary, this completes the proof.
\end{proof}

\subsection{Proof of Proposition \ref{lem:phi-ie}}\label{ssec:app:phi-ie}
%\begin{proof}

We begin by proving three lemmas that, taken together, imply \cref{lem:phi-ie}. The first lemma shows that each integrable upper kernel of $C$ implies a global \nameref{defi:ups} upper bound on $\Phi_\text{IE}(C)$, extending \cref{thm:qk}(i) from the $\Phi$ map to the more restrictive $\Phi_\text{IE}$ map. 

\begin{lem}\label{lem:phi-ie-upper}
    For any $C \in \C$, open convex $W \subseteq \Delta^\circ(\Theta)$, and $H \in \mathbf{C}^2(W)$, 
    \[
    \text{$\H H$ is an upper kernel of $C$ on $W$ } \ \ \implies \ \text{ $\Phi_\text{IE}(C) (\pi) \leq C^H_\text{ups} (\pi)$ for all $\pi \in \Delta(W)$.}
    \]
\end{lem}
\begin{proof}
    We begin with two preliminary facts. Since $\H H$ is an upper kernel of $C$ on $W$: (a) $\H H(p)\geq_\text{psd} \mathbf{0}$ for all $p \in W$, and (b) for every $p \in W$, there exists a $\delta(p)>0$ such that $\Delta(B_{\delta(p)} (p)) \subseteq \dom(C)$. Fact (a) implies that $H$ is convex, so $C^H_\text{ups}\in \C$ is a well-defined \nameref{defi:ups} cost. Fact (b) implies that $W \subseteq \Delta_C$, so every $\mathbb{O} \in \Omega(C)$ is also an open cover of $W$. 
    
    Now, fix an arbitrary $\mathbb{O} \in \Omega(C)$. Since $\mathbb{O}$ is an open cover of $W$, for every $p \in W$ there exists an $O \in \mathbb{O}$ and a $\overline{\delta}(p) >0$ such that $B_{\overline{\delta}(p)}(p) \subseteq O$. Therefore, for every $p \in W$, we have $C|_{\mathbb{O}}(\pi) = C(\pi)$ for all  $\pi \in \Delta(B_{\overline{\delta}(p)}(p))$. Since $\H H$ is an upper kernel of $C$ on $W$ and, for each $p \in W$, we are free to choose the ($p$-dependent) $\delta>0$ in \cref{defi:lq}(i) small enough that $\delta \leq \overline{\delta}(p)$, it follows that $\H H$ is also an upper kernel of $C|_{\mathbb{O}}$ on $W$. Since $W \subseteq \Delta^\circ(\Theta)$ is open and convex, \Cref{thm:qk}(i) then implies that $\Phi(C|_{\mathbb{O}})(\pi) \leq C_\text{ups}^H(\pi)$ for all $\pi \in \Delta(W)$. 
    
    Finally, since the fixed $\mathbb{O} \in \Omega(C)$ was arbitrary,  $\Phi_\text{IE}(C)(\pi) = \sup_{\mathbb{O}' \in \Omega(C)} \Phi( C|_{\mathbb{O}'})(\pi) \leq  C_\text{ups}^H(\pi)$ for all $\pi \in \Delta(W)$, as desired.
\end{proof}

Next, the second lemma shows that each integrable lower kernel of $C$ implies a \emph{global} \nameref{defi:ups} lower bound on $\Phi_\text{IE}(C)$, strengthening the \emph{local} lower bound on $\Phi(C)$ in \cref{thm:qk}(ii).\footnote{Note that, in contrast to \cref{lem:phi-ie-upper}, in \cref{lem:phi-ie-lower} we: (a) do not require that $W \subseteq \Delta^\circ(\Theta)$, but (b) impose additional strong convexity and domain assumptions on $H$ and $C$, respectively.}

\begin{lem}\label{lem:phi-ie-lower}
    For any $C \in \C$, open convex $W \subseteq \Delta(\Theta)$, and strongly convex $H \in \mathbf{C}^2(W)$, %and $C \in \C$ with $\dom(C) \subseteq \Delta(W) \cup\Ex^\varnothing$,
     \[
    \text{$\H H$ is lower kernel of $C$ on $W$ and $\dom(C) \subseteq \Delta(W) \cup\Ex^\varnothing$ } \ \ \implies \ \text{ $\Phi_\text{IE}\succeq C^H_\text{ups} $.}
    \]
\end{lem}
\begin{proof}
    %Since $W$ is convex and $\dom(C) \subseteq \Delta(W) \cup \Ex^\varnothing$, we have $\Delta_C \subseteq W$.\footnote{In particular, $\dom(C) \subseteq \Delta(W) \cup \Ex^\varnothing$ implies that $\dom(C) \backslash \Ex^\varnothing \subseteq \Delta(W)$, and $W$ being convex implies that $p_\pi \in W$ for every $\pi \in \Delta(W)$. Therefore, we have $p_\pi \in W$ for every $\pi \in \dom(C)\backslash \Ex^\varnothing$, i.e., $\Delta_C \subseteq W$.} 
    Since $H$ is strongly convex, there exists an $\overline{m}>0$ such that $\H H (p) -  2 m I(p_0) \ge_\text{psd} \mathbf{0}$ for all $p \in  W$ and $m \leq \overline{m}$. Let an $m \in (0, \overline{m})$ be given. 
    
    First, define $H_m \in \mathbf{C}^2(W)$ as $H_m(p) := H(p) - m \, \|p\|^2$ for all $p \in \dom(H_m) =\dom(H) = W$. Note that: (i) $\H H_{m}(p) =\H H(p) - 2m  I(p) \geq_\text{psd} \mathbf{0}$ (by definition of $m$), (ii) $H_m$ is convex and thus $C^{H_m}_\text{ups} \in \C$ is a well-defined \nameref{defi:ups} cost (by property (i) and since $W$ is convex), and (iii) $C^{H_m}_\text{ups}(\pi) = C^H_\text{ups}(\pi) - m \text{Var}(\pi)$ for all $\pi \in \dom(C^{H_m}_\text{ups}) = \dom(C^H_\text{ups}) = \Delta(W) \cup \Ex^\varnothing$.
    
    Next, note that, for every $p \in W$, there exists a $\delta(p)>0$ such that: (a) the lower kernel bound in \cref{defi:lq}(ii) holds for $C$ and $k(p) := \H H(p)$ with error parameters $\epsilon := m/2$ and $\delta := \delta(p)$ (by the lower kernel hypothesis), (b) $B_{\delta(p)}(p) \subseteq W$ (as $W$ is open), and (c) $\|\H H(p') - \H H(p) \| \leq m$ for all $p' \in B_{\delta(p)} (p)$ (as $H \in \mathbf{C}^2(W)$). Let $\mathbb{O} := \{B_{\delta(p)}(p)\}_{p \in W}$ denote the corresponding open cover of $W$. Since $W$ is convex and $\dom(C) \subseteq \Delta(W) \cup \Ex^\varnothing$, we have $\Delta_C \subseteq W$.\footnote{In particular, $\dom(C) \subseteq \Delta(W) \cup \Ex^\varnothing$ implies that $\dom(C) \backslash \Ex^\varnothing \subseteq \Delta(W)$, while the convexity of $W$  implies that $p_\pi \in W$ for every $\pi \in \Delta(W)$. Therefore, we have $p_\pi \in W$ for every $\pi \in \dom(C)\backslash \Ex^\varnothing$, i.e., $\Delta_C \subseteq W$.} Therefore, $\mathbb{O}$ is also an open cover of $\Delta_C$, i.e., $\mathbb{O} \in \Omega(C)$. 
    
    We claim that $C|_{\mathbb{O}} \succeq C^{H_m}_\text{ups}$. Since $C|_{\mathbb{O}}[\Ex^\varnothing] = C^{H_m}_\text{ups}[\Ex^\varnothing]=\{0\}$ and $+\infty \geq \sup C^{H_m}_\text{ups}[\Ex]$, to prove the claim it suffices to show that $C|_{\mathbb{O}}(\pi) \geq C^{H_m}_\text{ups}(\pi)$ for all $\pi \in \dom(C|_{\mathbb{O}}) \backslash \Ex^\varnothing$. In turn, because  $\dom(C|_{\mathbb{O}})\backslash \Ex^\varnothing \subseteq \bigcup_{p \in W} \Delta(B_{\delta(p)}(p))$ and $C(\pi) = C|_{\mathbb{O}}(\pi)$ for all $\pi \in  \bigcup_{p \in W} \Delta(B_{\delta(p)}(p))$ (by definition of $\mathbb{O}$ and $C|_{\mathbb{O}}$), it suffices to show that
    \begin{align}
    C(\pi) \geq C^{H_m}_\text{ups}(\pi) \qquad \forall \, \pi \in \bigcup_{p \in W} \Delta(B_{\delta(p)}(p)).
    \label{eqn:ie-approx-lb}
      \end{align}
    To this end, let $p \in W$ and $\pi \in \Delta(B_{\delta(p)}(p))$ be given. By property (a) of $\delta(p)>0$, we have
    \begin{align*}
        C(\pi) \geq \frac{1}{2} \, \E_\pi \big[ (q-p_\pi)^\top \left( \H H(p) -  m I \right) (q-p_\pi) \big].
    \end{align*}
    Now, observe that
    \begin{align*}
       \hspace{-2em} C^H_\text{ups}(\pi) & = \E_\pi \left[ H(q) - H(p_\pi) - (q-p_\pi)^\top \nabla H(p_\pi)\right] \\
        %%%
        & = \E_\pi \left[ \int_0^1 (1-t) (q-p_\pi)^\top \H H(r_q(t)) (q-p_\pi) \dd t\right] \quad \ \ \text{where $r_q(t) := p_\pi + t(q-p_\pi)$} \\
        %%%
        & = \frac{1}{2} \E_\pi \left[ (q-p_\pi)^\top \H H(p) (q-p_\pi) \right] + \E_\pi \left[ \int_0^1 (1-t) (q-p_\pi)^\top \left( \H H(r_q(t))- \H H(p) \right) (q-p_\pi) \dd t\right] \\
        %%%
        & \leq \frac{1}{2} \E_\pi \left[ (q-p_\pi)^\top \H H(p) (q-p_\pi) \right] + \frac{1}{2} \E_\pi \left[\sup_{t\in[0,1]} \|\H H(r_q(t)) - \H H(p)\| \cdot \|q-p_\pi\|^2 \right] \\
        & \leq \frac{1}{2} \E_\pi \left[ (q-p_\pi)^\top \H H(p) (q-p_\pi) \right] + \frac{m}{2} \text{Var}(\pi), 
    \end{align*}
    where the first line is by definition of $C^H_\text{ups}$ and $p_{\pi} = \E_{\pi}[q]$, the second line is by the Fundamental Theorem of Calculus,\footnote{In particular, the argument is a minor modification of that from \cref{fn:FTC-hessian} in \cref{app:thm3-1:extra}, where we now use that $H\in \mathbf{C}^2(W)$ and $r_q(t) \in B_{\delta(p)}(p) \subseteq W$ for all $t \in [0,1]$ (by property (b) of $\delta(p)$ and convexity of the ball).} 
    the third line rearranges terms and uses $\int_0^1 (1-t) \dd t = \frac{1}{2}$, the fourth line uses the definition of the matrix semi-norm and $\int_0^1 (1-t) \dd t = \frac{1}{2}$, and the final line follows from property (c) in the definition of $\delta(p)$ (where $r_q(t) \in B_{\delta(p)}(p)$ for all $t \in[0,1]$ by convexity of the ball). Combining the two displays above, we obtain
    \[
    C(\pi) \geq C^H_\text{ups}(\pi) - m \text{Var}(\pi) = C^{H_m}_\text{ups}(\pi),
    \]
    where the final equality is by property (iii) of $H_m$ (which applies because $\Delta(B_{\delta(p)}(p)) \subseteq \Delta(W)$ by property (b) in the definition of $\delta(p)$). Since the given $p \in W$ and $\pi \in \Delta(B_{\delta(p)}(p))$ were arbitrary, we conclude that \eqref{eqn:ie-approx-lb} holds. Thus,  $C|_{\mathbb{O}} \succeq C^{H_m}_\text{ups}$ as claimed. 

    We now complete the proof of the lemma. Since $C|_{\mathbb{O}} \succeq C^{H_m}_\text{ups}$ (as just shown) and $\Phi$ is isotone (\cref{lem:structure:Phi}), we have $\Phi(C|_{\mathbb{O}}) \succeq \Phi(C^{H_m}_\text{ups})$. Since $C^{H_m}_\text{ups}$ is \nameref{axiom:slp} (\cref{lem:ups:to:additive}), we also have $\Phi(C^{H_m}_\text{ups}) = C^{H_m}_\text{ups}$. Hence, $\Phi(C|_{\mathbb{O}}) \succeq  C^{H_m}_\text{ups}$. Then, since $\mathbb{O}\in \Omega(C)$ (as noted above),  
    \[
    \Phi_\text{IE}(C) = \sup_{\mathbb{O}' \in \Omega(C)} \Phi(C|_{\mathbb{O}'}) \succeq \Phi(C|_{\mathbb{O}}) \succeq C^{H_m}_\text{ups}.
    \]
    Finally, since the given $m \in (0,\overline{m})$ was arbitrary and $C^{H_m}_\text{ups} (\pi) = C^H_\text{ups}(\pi) - m \text{Var}(\pi)$ for all $\pi \in  \dom(H_m) = \dom(H)$ (by property (iii) of $H_m$), taking $m \to0$ yields $\Phi_\text{IE}(C) \succeq C^H_\text{ups}$.
\end{proof}

Finally, the third lemma shows that (upper) kernels are invariant under the $\Phi_\text{IE}$ map. While \cref{thm:qk}(ii) readily implies that lower kernels are invariant under $\Phi_\text{IE}$, the invariance of upper kernels is nontrivial because $\Phi_\text{IE}(C)$ need not satisfy $\Phi_\text{IE}(C) \preceq C$.

    \begin{lem}\label{lem:phi-ie-kernel}
        For any $C \in \C$ and $p_0 \in \Delta^\circ(\Theta)$, the following hold:
        \begin{itemize}
        \item[(i)] If $\overline{k}_C(p_0)$ is an upper kernel of $C$ at $p_0$, then $\overline{k}_C(p_0)$ is an upper kernel of $\Phi_\text{IE}(C)$ at $p_0$.
        %%%
        \item[(ii)] If $C$ is \nameref{defi:sp} and \nameref{defi:lq} at $p_0$ with kernel $k_C(p_0)$, then $\Phi_\text{IE}(C)$ is \nameref{defi:lq} at $p_0$ with kernel $k_{\Phi_\text{IE}(C)}(p_0) = k_C(p_0)$.
        \end{itemize}
        %\[
        %\text{$\overline{k}_C(p_0)$ is an upper kernel of $C$ at $p_0$ } \ \ \implies \ \ \text{ $\overline{k}_C(p_0)$ is an upper kernel of $\Phi_\text{IE}(C)$ at $p_0$.}
        %\]
    \end{lem}
    \begin{proof}
        We prove each point in turn.

        \noindent \textbf{Point (i) (Upper Kernel Invariance).} Since $\overline{k}_C(p_0)$ is an upper kernel of $C$ at $p_0 \in \Delta^\circ(\Theta)$, for every $\epsilon>0$ there exists a $\overline{\delta}(\epsilon)>0$ such that $B_{\overline{\delta}(\epsilon)}(p_0) \subseteq \Delta^\circ(\Theta)$ and 
        \[
        C(\pi) \leq  \E_\pi\left[(q-p_\pi)^\top \left( \frac{1}{2} \overline{k}_C (p_0) +  \epsilon I \right) (q-p_\pi) \right] \quad \ \ \forall \, \pi \in \Delta(B_{\overline{\delta}(\epsilon)} (p_0)).
        \]
        For every $\epsilon>0$, define $H_\epsilon \in \mathbf{C}^2 (B_{\overline{\delta}(\epsilon)} (p_0))$ as $H_\epsilon (p) :=  p^\top \left( \frac{1}{2} \overline{k}_C (p_0) +  \epsilon I \right) p$ for all $p \in \dom(H_\epsilon) =  B_{\overline{\delta}(\epsilon)} (p_0)$. Each $H_\epsilon$ is convex since $B_{\overline{\delta}(\epsilon)}(p_0)$ is convex and $\H H_\epsilon(\cdot) \sim_\text{psd}  \overline{k}_C(p_0) + 2\epsilon I(p_0) \geq_\text{psd} \mathbf{0}$ (as $\overline{k}_C(p_0) \geq_\text{psd}\mathbf{0}$ by \cref{defi:lq}).\footnote{In particular, per the normalization in \cref{remark:kernels}, $\H H_\epsilon  (p) = (I- \mathbf{1}p^\top) (\overline{k}_C  (p_0) + 2\epsilon I) (I- p \mathbf{1}^\top)$ for all $p\in B_{\overline{\delta}(\epsilon)} (p_0)$.} Thus, for every $\epsilon>0$, $C^{H_\epsilon}_\text{ups} \in \C$ is a well-defined \nameref{defi:ups} cost; moreover, since $p_\pi = \E_\pi[q]$ for all $\pi \in \Ex$, direct calculation yields
        %for all $\pi \in \dom(H_\epsilon) = \Delta(B_{\overline{\delta}(\epsilon)} (p_0))\cup\Ex^\varnothing$, we have
        \begin{align}
            C^{H_\epsilon}_\text{ups}(\pi) %&= \E_\pi \left[ \frac{1}{2} q^\top  \left( \overline{k}_C (p_0) + 2 \epsilon I \right) q - \frac{1}{2} p_\pi^\top \left( \overline{k}_C (p_0) + 2 \epsilon I \right)  p_\pi \right] \\
            &% = \E_\pi \left[ \frac{1}{2} q^\top  \left( \overline{k}_C (p_0) + 2 \epsilon I \right) q + \frac{1}{2} p_\pi^\top \left( \overline{k}_C (p_0) + 2 \epsilon I \right)  p_\pi - q^\top \left( \overline{k}_C (p_0) + 2 \epsilon I \right)  p_\pi \right] \\
            %%%
            %&= 
            =  \E_\pi\left[(q-p_\pi)^\top \left( \frac{1}{2} \overline{k}_C (p_0) + \epsilon I \right) (q-p_\pi) \right] \quad \ \ \forall \, \pi \in \Delta(B_{\overline{\delta}(\epsilon)} (p_0)). \label{eqn:ie-upper-ups-loc}
        \end{align}
        %where the first line is by definition, the second line holds because $p_\pi = \E_\pi[q]$, and the final line rearranges terms. 
        Combining the two displays above, it follows that, for every $\epsilon>0$,
        \begin{align}
        C(\pi) \leq C^{H_\epsilon}_\text{ups}(\pi) \quad \ \ \forall \, \pi \in \Delta(B_{\overline{\delta}(\epsilon)} (p_0)). \label{eqn:ie-upper-loc}
        \end{align}
        Now, for every $\epsilon>0$, since $\H H_\epsilon$ is an upper kernel of $C^{H_\epsilon}_\text{ups}$ on the open convex set $B_{\overline{\delta}(\epsilon)} (p_0) \subseteq \Delta^\circ(\Theta)$ (\cref{lem:ups-kernel-equiv}), \eqref{eqn:ie-upper-loc} implies that $\H H_\epsilon$ is also an upper kernel of $C$ on $B_{\overline{\delta}(\epsilon)} (p_0)$. Therefore, \cref{lem:phi-ie-upper} and \eqref{eqn:ie-upper-ups-loc} then imply that, for every $\epsilon >0$,
        \[
        \Phi_\text{IE}(C)(\pi) \leq C^{H_\epsilon}_\text{ups}(\pi) =  \E_\pi\left[(q-p_\pi)^\top \left( \frac{1}{2} \overline{k}_C (p_0) + \epsilon I \right) (q-p_\pi) \right] \quad \ \ \forall \, \pi \in \Delta(B_{\overline{\delta}(\epsilon)} (p_0)).
        \]
        %$\Phi_\text{IE}(C)(\pi) \leq C^{H_\epsilon}_\text{ups}(\pi)$ for all $\pi \in \Delta(B_{\overline{\delta}(\epsilon)} (p_0))$. It follows that $\H H_\epsilon(p_0) = \overline{k}_C(p_0) + 2\epsilon I$ is an upper kernel of $\Phi_\text{IE}(C)$ at $p_0$; in particular, for every $\eta >0$ there exists a $\delta(\eta,\epsilon) \in (0, \overline{\delta}(\epsilon))$ such that
        %\begin{align}
        %\Phi_\text{IE}(C)(\pi) \leq \E_\pi \left[ (q-p_\pi)^\top \left(\frac{1}{2} \left( \overline{k}_C(p_0) + 2\epsilon I \right) + \eta I \right) (q-p_\pi) \right] \quad \ \ \forall \, \pi \in \Delta(B_{\delta(\eta, \epsilon)} (p_0)). \label{eqn:ie-upper-loc-2}
        %\end{align}
%
        %Since the fixed $\epsilon>0$ was arbitrary, %\eqref{eqn:ie-upper-loc-2} implies the following: for every $\xi >0$, there exists a $\delta >0$ such that\footnote{In particular, given any $\xi >0$, this follows from  \eqref{eqn:ie-upper-loc-2} with any $\epsilon, \eta>0$ such that $\epsilon + \eta \leq \xi$ and  $\delta := \delta(\eta, \epsilon)$.}
        %\[
        %\Phi_\text{IE}(C)(\pi) \leq \E_\pi \left[ (q-p_\pi)^\top \left(\frac{1}{2}  \overline{k}_C(p_0) + \xi I \right) (q-p_\pi) \right] \quad \ \ \forall \, \pi \in \Delta(B_{\delta} (p_0)).
        %\]
        We conclude that $\overline{k}_C(p_0)$ is an upper kernel of $\Phi_\text{IE}(C)$ at $p_0$, as desired.

        \noindent \textbf{Point (ii) (Kernel Invariance).} Since $k_C(p_0)$ is an upper kernel of $C$ at $p_0$, point (i) (proved above) implies that $k_C(p_0)$ is an upper kernel of $\Phi_\text{IE}(C)$ at $p_0$. Since $C$ is \nameref{defi:sp} and thus $k_C(p_0) \gg_\text{psd} \mathbf{0}$ (by \cref{cor:ker-SP}), \cref{thm:qk}(ii) implies that $k_C(p_0)$ is a lower kernel of $\Phi(C)$ at $p_0$; since $\Phi(C) \preceq \Phi_\text{IE}(C)$ (by construction), it follows $k_C(p_0)$ is also a lower kernel of $\Phi_\text{IE}(C)$ at $p_0$. We conclude that $k_C(p_0)$ is the kernel of $\Phi_\text{IE}(C)$ at $p_0$.
        %, as desired. 
    \end{proof}

We now use \cref{lem:phi-ie-upper,lem:phi-ie-lower,lem:phi-ie-kernel} to prove \cref{lem:phi-ie}.

\begin{proof}[Proof of \cref{lem:phi-ie}]
    We prove each point in turn:

    \noindent \textbf{Point (i).} By \cref{lem:phi-ie-upper}, $\Phi_\text{IE}(C)(\pi) \leq C^H_\text{ups}(\pi)$ for all $\pi \in \Delta(W)$. Since %$\Phi_\text{IE}(C)[\Ex^\varnothing] = C^H_\text{ups}[\Ex^\varnothing] = \{0\}$, 
    $C^H_\text{ups}[\Ex \backslash(\Delta(W) \cup \Ex^\varnothing)] = \{+\infty\}$ and $\sup \Phi_\text{IE}(C)[\Ex \backslash(\Delta(W) \cup \Ex^\varnothing)] \leq +\infty$, it follows that $\Phi_\text{IE}(C) \preceq C^H_\text{ups}$.
    %, as desired.

    \noindent \textbf{Point (ii).} Immediate from \cref{lem:phi-ie-lower}.

    \noindent \textbf{Point (iii).} The ``$\implies$'' direction follows from points (i) and (ii). For the ``$\impliedby$'' direction, \cref{lem:phi-ie-kernel}(ii) and \cref{lem:ups-kernel-equiv} together imply that $k_C = k_{\Phi_\text{IE}(C)} = \H H$ on $W$.
\end{proof}

\subsection{Proofs of Corollaries \ref{cor:MI} and \ref{prop:mult:tech}}\label{ssec:proofs-MI-combine}

\begin{proof}[Proof of \cref{cor:MI}]
    Note that $H_\text{MI}$ is strongly convex and hence $C^\circ_\text{MI}$ is \nameref{defi:sp}. Thus, $C \in \C$ satisfies $C \succeq C^\circ_\text{MI}$ only if $C$ is \nameref{defi:sp} and $\dom(C) \subseteq \Delta(\Delta^\circ(\Theta))\cup\Ex^\varnothing$. The result then follows directly from applying \cref{thm:flie} and \cref{lem:phi-ie}(iii).
\end{proof}

\begin{proof}[Proof of \cref{prop:mult:tech}]
    We first verify that $C = g\circ (C^i)_{i =1}^n$ is \nameref{defi:lq} with $k_C = k_{\Phi(C)} = \H H$. Note that, by construction, $\underline{C}=\sum_{i =1}^n \nabla_i g(\mathbf{0}) C^i$ is \nameref{defi:lq} with kernel $k_{\underline{C}} = \H H$. Since $g$ is subdifferentiable at $\mathbf{0}$ and satisfies $g(\mathbf{0}) = 0$, we have $C \succeq \underline{C}$. This directly implies $\H H$ is a lower kernel of $C$. Now, take any $p \in \Delta^\circ(\Theta)$. For every $\epsilon>0$, there exists $\delta>0$ such that, for all $i \in \{1,\dots,n\}$ and $\pi \in \Ex$ with $\supp(\pi)\subseteq B_\delta(p)$, 
    \[
     C^i(\pi) \leq \int_{B_{\delta}(p)} (q-p_{\pi})^\top\left(\frac{1}{2}\H H^i(p)+\epsilon I\right)(q-p_{\pi}) \dd \pi( q) \quad \text{ and } \quad C(\pi) - \underline{C}(\pi) \leq \epsilon \cdot \left \| (C^j(\pi))_{j =1}^n \right\|,
    \]
    where the first inequality holds because each $C^i$ is \nameref{defi:lq} with $k_{C^i} = \H H^i$, and the second inequality follows from the first and the fact that $g$ is continuously differentiable at $\mathbf{0}$. Let $M := \max_{i =1,\dots,n} \|\H H^i(p)\|$. Then, for all $\pi \in \Ex$ with $\supp(\pi)\subseteq B_\delta(p)$,
    \begin{align*}
    \hspace{-2em}
        C(\pi) &= C(\pi) - \underline{C}(\pi) + \underline{C}(\pi) \\
        %%%
        & \leq \epsilon \cdot \left \| (C^i(\pi))_{i =1}^n \right\| + \int_{B_{\delta}(p)} (q-p_{\pi})^\top\left(\frac{1}{2}\H H(p)+\epsilon I\right)(q-p_{\pi}) \dd \pi( q) \\
        %%%
        & \leq \eta(\epsilon)\cdot \text{Var}(\pi) + \int_{B_{\delta}(p)} (q-p_{\pi})^\top\left(\frac{1}{2}\H H(p)+\epsilon I\right)(q-p_{\pi}) \dd \pi( q), \ \ \text{ where } \ \ \eta(\epsilon):= \epsilon \cdot \sqrt{n} \cdot \left( \frac{1}{2} M + \epsilon\right) \\
        %%%
        & = \int_{B_{\delta}(p)} (q-p_{\pi})^\top\left(\frac{1}{2}\H H(p)+ \left( \epsilon + \eta (\epsilon) \right) I\right)(q-p_{\pi}) \dd \pi( q),
    \end{align*}
    where the second and third lines follow from the preceding display and the fact that $ \left \| (C^i(\pi))_{i =1}^n \right\| \leq \sqrt{n} \cdot \max_{i =1,\dots, n} C^i(\pi)$, and the final line rearranges terms. Since $\lim_{\epsilon \to 0} \left[\epsilon  + \eta(\epsilon)\right] =0$, it follows that $\H H$ is also an upper kernel of $C$. Thus, $C$ is \nameref{defi:lq} with $k_C = \H H$. Now, since $\underline{C}$ is \nameref{defi:sp}, $C$ is \nameref{defi:sp} (as $C \succeq \underline{C}$) and hence $k_C = \H H \gg_\text{psd} \mathbf{0}$ (\cref{cor:ker-SP}). \cref{thm:qk}(ii) then implies that $k_{\Phi(C)} =  \H H$.

    Next, we verify that $\Phi(C) \preceq \Phi_\text{IE}(C)  =  C^H_\text{ups}$. The inequality holds by definition. The equality follows from \cref{lem:phi-ie} because $\dom(C) \subseteq \Delta(\Delta^\circ(\Theta)) \cup \Ex^\varnothing$ by construction, $k_C = \H H$ (shown above), and $\H H$ is strongly positive since  $\underline{C}$ being \nameref{defi:sp} implies that there exists $m >0$ such that $k_{\underline{C}} = \H H \geq_\text{psd} mI$ (\cref{cor:ker-SP} and its proof). 
    
    Note that an analogous argument also establishes that $\Phi_\text{IE}(\underline{C}) = C^H_\text{ups}$.
    
    Finally, we verify that $\Phi(C) = C^H_\text{ups}$ if $\underline{C}$ \nameref{axiom:flie}. To this end, suppose $\underline{C}$ \nameref{axiom:flie}. Then, by the above, $\underline{C} \succeq \Phi_\text{IE}(\underline{C})  = C^H_\text{ups}$. This implies $C \succeq C^H_\text{ups} = \Phi_\text{IE}(C)$ because $C \succeq \underline{C}$ (as noted above). Therefore, $C$ \nameref{axiom:flie} and hence \cref{thm:flie} implies that $\Phi(C) = C^H_\text{ups}$, as desired.
\end{proof}

\subsection{Proof of Proposition \ref{prop:LPI-main}}\label{ssec:proof-LPI-main}

We begin by formalizing the claim from \cref{fn:prop3-clarify} in \cref{ssec:SPI}:

\begin{lem}
    If $C \in \C$ is \nameref{defi:sp}, \nameref{defi:lq}, and \hyperref[axiom:prior:invariant]{Prior Invariant}, then there exists a symmetric $\kappa \in \mathbb{R}^{|\Theta| \times |\Theta|}$ with $\kappa \gg_\text{psd} \mathbf{0}$ and $\kappa \mathbf{1} = \mathbf{0}$ such that $\kappa_C(p) = \kappa$ for all $p \in \Delta^\circ(\Theta)$.
\end{lem}
\begin{proof}
    Let such $C \in \C$ be given. \cref{lem:lpi-kernel,lem:pi-w-implies-lpi-w} (with $W := \Delta^\circ(\Theta))$ and \cref{remark:exp-kernel} imply that there exists a symmetric $\kappa \in \mathbb{R}^{|\Theta| \times |\Theta|}$ with $\kappa \mathbf{1} = \mathbf{0}$ such that $\kappa_C(p) = \kappa$ for all $p \in \Delta^\circ(\Theta)$. It remains to show that $\kappa \gg_\text{psd} \mathbf{0}$. To this end, fix any $p \in \Delta^\circ(\Theta)$ and let $p^\star := \frac{1}{|\Theta|} \mathbf{1} \in \Delta^\circ(\Theta)$ denote the uniform prior. We have $\kappa = \kappa_C(p) = \diag(p) k_C(p) \diag(p)$ by the above and $k_C(p)  \gg_\text{psd}\mathbf{0}$ by \cref{cor:ker-SP}. Thus, \cref{lem:exp-ker-prior-pivot} (with $M:= k_C(p)$ and $p' := p^\star$) implies that $|\Theta|^2 \cdot \kappa = \diag(p^\star)^{-1} \kappa \, \diag(p^\star)^{-1} \gg_\text{psd}\mathbf{0}$. It follows that $\kappa \gg_\text{psd} \mathbf{0}$.  
\end{proof}

We now proceed to prove the proposition. Following the notation in \cref{proof:trilemma}, for any experiment $\sigma\in\Se$ and prior $p\in\Delta(\Theta)$, we denote by $q^{\sigma,p}_s \in \Delta(\Theta)$ the Bayesian posterior conditional on signal $s$, so that induced random posterior is given by $h_B(\sigma,p)(B)=\langle \sigma,p\rangle \left( \left\{s \in S \mid q^{\sigma,p}_s \in B\right\}\right)$ for all Borel $B\subseteq\Delta(\Theta)$.

\begin{proof}[Proof of \cref{prop:LPI-main}]
    We prove the necessity and sufficiency directions in turn.

    \noindent \textbf{Necessity.} Let $C \in \C$ be \nameref{defi:sp} and \nameref{defi:lq}. There are two cases:
    
    \textbf{\emph{Case 1:}} If $C$ is \hyperref[axiom:prior:invariant]{Prior Invariant}, then \cref{lem:lpi-kernel,lem:pi-w-implies-lpi-w} (with $W := \Delta^\circ(\Theta)$) directly imply that $\kappa_C(p) = \kappa_C(p')$ for all $p, p' \in \Delta^\circ(\Theta)$, as desired.
    
    \textbf{\emph{Case 2:}} If $C$ is \nameref{defi:spi}, then pick any \hyperref[axiom:prior:invariant]{Prior Invariant} $C' \in \Phi^{-1}(C)$. Note that $C'$ is \nameref{defi:sp} because $C$ is \nameref{defi:sp} and $C' \succeq \Phi(C') = C$. Hence, \cref{lem:pi-w-implies-lpi-w} (applied to $C'$ and $W := \Delta^\circ(\Theta)$) implies that $C = \Phi(C')$ is \nameref{defi:lpi} on $\Delta^\circ(\Theta)$. In turn, \cref{lem:lpi-kernel} (applied to $C$ and $W := \Delta^\circ(\Theta)$) implies that $\kappa_C(p) = \kappa_C(p')$ for all $p, p' \in \Delta^\circ(\Theta)$, as desired. 

    \noindent{\textbf{Sufficiency.}} Let $\kappa \in \R^{|\Theta|\times|\Theta|}$ be symmetric and satisfy $\kappa \gg_\text{psd} \mathbf{0}$ and $\kappa \mathbf{1}= \mathbf{0}$. We establish the existence of a cost function with the desired properties by construction. 
    
    To this end, we begin by defining the map $G \in \mathbf{C}^2(\R^{|\Theta|}_{++})$ as
    \[
    G(x) := \frac{|\Theta|}{2} \cdot \frac{x^\top \kappa \, x}{\mathbf{1}^\top x}.
    \]
    Note that $G$ satisfies three properties: (i) $G$ is non-negative, since $\kappa \mathbf{1} = \mathbf{0}$ and \cref{lem:psd-star} (with $p_0:= \frac{1}{|\Theta|}\mathbf{1}$) imply that $x^\top \kappa x \geq 0$ for all $x \in \mathbb{R}^{|\Theta|}$; (ii) $\H G(\mathbf{1}) = \kappa$ by a routine calculation; and (iii) $G$ is positively homogeneous of degree 1 (HD1) by construction. 

    By property (i) and $\kappa \mathbf{1} = \mathbf{0}$, the map $D : \Delta^\circ(\Theta) \times \Delta^\circ(\Theta) \to \R_+$ given by $D(q \mid p) := G\big(\frac{q}{p}\big)$ is a well-defined divergence. Define the \hyperref[eqn:PS]{Posterior Separable} cost function $C \in \C$ as
    \[
    C(\pi) = \begin{cases}
        \E_\pi \left[ D(q \mid p_\pi) \right], & \text{if $\pi \in \Delta(\Delta^\circ(\Theta))$} \\
        0, & \text{if $\pi \in \Ex^\varnothing$} \\
        %%%
        +\infty, & \text{otherwise.}
    \end{cases}
    \]
    We now verify that $C$ satisfies each of the desired properties. We proceed in four steps.  

    \emph{\textbf{Step 1:} $C$ is \hyperref[axiom:prior:invariant]{Prior Invariant}.} Let $p^\star := \frac{1}{|\Theta|} \mathbf{1} \in \Delta^\circ(\Theta)$ denote the uniform prior. Recall from \cref{app:utvm,sssec:proof:thm5} that $\Se_b \subsetneq \Se$ denotes the class of bounded experiments and that $h_B[\Se_b \times \Delta^\circ(\Theta)] = \Delta (\Delta^\circ(\Theta))$. Therefore, since $\dom(C) = \Delta(\Delta^\circ(\Theta))\cup\Ex^\varnothing$, it suffices to show that $C(h_B(\sigma,p)) = C(h_B(\sigma,p^\star))$ for all $\sigma \in \Se_b$ and $p \in \Delta^\circ(\Theta)$ (cf. \cref{fn:PI-rich-dom} in \cref{proof:trilemma}). To this end, let $\sigma \in \Se_b$ and $p \in \Delta^\circ(\Theta)$ be given. For each $s \in \cup_{\theta\in\Theta} \supp(\sigma_\theta)$ and $r \in \Delta^\circ(\Theta)$, we denote $\frac{\d \bm{\sigma}}{\d\langle \sigma,r \rangle}(s) := \big( \frac{\d \sigma_\theta}{\d\langle \sigma,r \rangle}(s) \big)_{\theta\in\Theta} \in \R^{|\Theta|}_+$. We then have
    \begin{align*}
        C(h_B(\sigma,p)) %\, = \, \int_{\Delta(\Theta)} G\left( \frac{q}{p} \right) \dd h_B(\sigma,p)(q) 
        \, = \, \int_S G\left( \frac{q^{\sigma,p}_s}{p}\right) \dd \langle \sigma,p \rangle(s) \, & = \, \int_S G\left( \frac{\d \bm{\sigma}}{\d\langle \sigma,p\rangle} (s)\right) \cdot \frac{\d \langle \sigma,p \rangle}{\d \langle \sigma,p^\star \rangle}(s) \cdot \dd \langle \sigma,p^\star \rangle(s) \\
        %%%%
        & = \, \int_S G\left( \frac{\d \bm{\sigma}}{\d\langle \sigma,p\rangle} (s) \cdot \frac{\d \langle \sigma,p \rangle}{\d \langle \sigma,p^\star \rangle}(s) \right)  \dd \langle \sigma,p^\star \rangle(s) \\
        %%%
        & = \, \int_S G\left( \frac{q^{\sigma,p^\star}_s}{p^\star}\right) \dd \langle \sigma,p^\star \rangle(s) \, = \, C(h_B(\sigma,p^\star)),
    \end{align*}
    where the first equality is by definition of $C$ and a change of variables, the second equality is by Bayes' rule and the change-of-measure $\d \langle \sigma, p\rangle = \frac{\d\langle \sigma,p\rangle}{\d\langle \sigma,p^\star \rangle}  \dd \langle \sigma,p^\star \rangle$, the third equality is by property (iii) above (i.e., $G$ is HD1), the fourth equality is by the chain rule for Radon-Nikodym derivatives and Bayes' rule, and the final equality is again by  definition of $C$. Since the given $\sigma \in \Se_b$ and $p \in \Delta^\circ(\Theta)$ were arbitrary, we conclude that $C$ is \hyperref[axiom:prior:invariant]{Prior Invariant}.
    
    \emph{\textbf{Step 2:} $C$ is \nameref{defi:lq} with $\kappa_C(p) = \kappa$ for all $p \in \Delta^\circ(\Theta)$.} By construction, $(q,p) \mapsto \H_1 D(q \mid p)$ is well-defined and continuous on $\Delta^\circ(\Theta) \times \Delta^\circ(\Theta)$. Hence, \cref{lem:ps-kernel-suff} implies that $C$ is \nameref{defi:lq} and $k_C(p) = \H_1 D(p\mid p)$ for all $p \in \Delta^\circ(\Theta)$. Thus, by the chain rule and property (ii) above (i.e., $\H G(\mathbf{1})=\kappa$), we have $\H_1 D(p\mid p) = \diag(p)^{-1} \H G(\mathbf{1}) \diag(p)^{-1} = \diag(p)^{-1} \kappa  \diag(p)^{-1}$ for all $p \in \Delta^\circ(\Theta)$. Therefore, we obtain 
    \[
    k_C(p) \, = \,  \diag(p)^{-1} \kappa \, \diag(p)^{-1} \qquad \forall\, p \in \Delta^\circ(\Theta). 
    \] 
    We conclude that $\kappa_C(p) = \diag(p) k_C(p) \diag(p) = \kappa$ for all $p \in \Delta^\circ(\Theta)$, as desired. 

    \emph{\textbf{Step 3:} $C$ is \nameref{defi:sp}.} Recall that $\text{Var} \in \C$ is defined as $\text{Var}(\pi):= \E_\pi[\|q-p_\pi\|^2]$ for all $\pi \in \Ex$. Since $\dom(C)  = \Delta(\Delta^\circ(\Theta)) \cup   \Ex^\varnothing \subsetneq \Ex = \dom(\text{Var})$, it suffices to show that there exists an $m>0$ such that $C(\pi) \geq m \text{Var}(\pi)$ for all $\pi \in \Delta(\Delta^\circ(\Theta))\backslash\Ex^\varnothing$. 
    
    To this end, let $\Se^\varnothing \subsetneq \Se_b$ denote the class of experiments $\sigma$ such that $\sigma_\theta = \sigma_{\theta'}$ for all $\theta,\theta'\in\Theta$. %$h_B(\sigma,p^\star) = \delta_{p^\star}$. 
    By Bayes' rule, it holds that $h_B[\Se_b \backslash\Se^\varnothing \times\Delta^\circ(\Theta)] = \Delta(\Delta^\circ(\Theta))\backslash\Ex^\varnothing$. %and $\text{Var}(\pi)>0$ for all $\pi \in \Delta(\Delta^\circ(\Theta))\backslash\Ex^\varnothing$

    Define $Y := \{y \in \mathcal{T} \mid \|y \|= 1\}$ and $\xi:= \min\{y^\top \kappa y \mid y \in Y\}$. Note that $\xi>0$ because $\kappa \gg_\text{psd} \mathbf{0}$ (by hypothesis), which implies that the continuous map $y \mapsto y^\top \kappa y$ is strictly positive on the compact set $Y$. Moreover, for every $\sigma \in \Se_b$ and $p \in \Delta^\circ(\Theta)$, it holds that
    \begin{align*}
    C(h_B(\sigma,p)) \, = \, C(h_B(\sigma,p^\star)) \, &= \, \frac{|\Theta|}{2} \cdot \E_{h_B(\sigma,p^\star)} \left[ \frac{q^\top \diag(p^\star)^{-1} \kappa \, \diag(p^\star)^{-1} q }{\mathbf{1}^\top \diag(p^\star)^{-1} q}\right]  \\
    %%%
    & = \, \frac{|\Theta|^2}{2} \cdot \E_{h_B(\sigma,p^\star)} \left[ \frac{q^\top \kappa\,  q }{\mathbf{1}^\top q}\right] \\
    %%%
    & = \, \frac{|\Theta|^2}{2} \cdot \E_{h_B(\sigma,p^\star)} \left[ (q-p^\star)^\top \kappa ( q  - p^\star)\right] \, \geq \, \frac{\xi \, |\Theta|^2}{2} \cdot \text{Var}(h_B(\sigma,p^\star)),
    \end{align*}
    where the first equality holds because $C$ is \hyperref[axiom:prior:invariant]{Prior Invariant} (Step 1), the second equality is by definition of $C$, the third equality is by $\diag(p^\star)^{-1} = |\Theta| \cdot I$, the fourth equality is by $\mathbf{1}^\top q = 1$ (as $q \in \Delta(\Theta)$) and $\kappa p^\star = \kappa \mathbf{1} = \mathbf{0}$ (by hypothesis), and the final inequality is by definition of $\xi >0$ and $\text{Var}\in \C$. Since $\text{Var}(\pi)>0$ for all $\pi \in \Ex \backslash \Ex^\varnothing$, it follows that
    \[
    \hspace{-1.5em}
    C(h_B(\sigma,p)) \, \geq \,\frac{\xi \, |\Theta|^2}{2} \cdot \text{Var}(h_B(\sigma,p^\star)) \, = \, \frac{\xi \, |\Theta|^2}{2} \cdot \left[ \frac{\text{Var}(h_B(\sigma,p^\star))}{\text{Var}(h_B(\sigma,p))} \right] \cdot \text{Var}(h_B(\sigma,p)) \qquad \forall \sigma \in \Se_b\backslash\Se^\varnothing, \, p \in \Delta^\circ(\Theta).
    \]
    We conclude that the following condition is sufficient for $C$ to be \nameref{defi:sp}:
    \begin{equation}\label{eqn:sp-verify-1}
    R := \inf_{\sigma \in \Se_b\backslash\Se^\varnothing} \inf_{p \in \Delta^\circ(\Theta)}\frac{\text{Var}(h_B(\sigma,p^\star))}{\text{Var}(h_B(\sigma,p))} >0.
    \end{equation}
Therefore, in what follows, we verify that condition \eqref{eqn:sp-verify-1} holds. 

For each $\sigma \in \Se_b$, let $S_\sigma^{\neg\varnothing}:= \{s \in \cup_{\theta\in\Theta}\supp(\sigma_\theta) \mid q^{\sigma,p^\star}_s \neq p^\star\}$ denote the set of ``nontrivial signals'' generated by $\sigma$. For every $\sigma \in \Se_b$, Bayes' rule implies that $s \in S_\sigma^{\neg\varnothing}$ if and only if $q^{\sigma,p}_s \neq p$ for all $p \in \Delta^\circ(\Theta)$. Thus, for every $\sigma \in \Se_b\backslash\Se^\varnothing$ and $p \in \Delta^\circ(\Theta)$, we have $\text{Var}(h_B(\sigma,p)) = \int_{S_\sigma^{\neg\varnothing}} \|q^{\sigma,p}_s - p\|^2 \dd\langle \sigma,p\rangle(s)>0$. Now, let $\sigma \in \Se_b\backslash\Se^\varnothing$ and $p \in \Delta^\circ(\Theta)$ be given. It holds that
\begin{align*}
\hspace{-1em}
    \frac{\text{Var}(h_B(\sigma,p^\star))}{\text{Var}(h_B(\sigma,p))} \, = \, \frac{\int_{S_\sigma^{\neg\varnothing}} \|q^{\sigma,p^\star}_s - p^\star \|^2 \dd \langle \sigma,p^\star\rangle(s)}{\int_{S_\sigma^{\neg\varnothing}} \|q^{\sigma,p}_s - p \|^2 \dd \langle \sigma,p\rangle(s)} \, &= \, \frac{\int_{S_\sigma^{\neg\varnothing}} \left\{\frac{\|q^{\sigma,p^\star}_s - p^\star \|^2}{\|q^{\sigma,p}_s - p \|^2} \cdot \frac{\d \langle \sigma, p^\star\rangle}{\d \langle \sigma, p\rangle}(s) \right\} \cdot  \|q^{\sigma,p}_s - p \|^2\dd \langle \sigma,p\rangle(s)}{\int_{S_\sigma^{\neg\varnothing}} \|q^{\sigma,p}_s - p \|^2 \dd \langle \sigma,p\rangle(s)} \\
    %%%
    & \geq \, \inf_{s \in S_\sigma^{\neg\varnothing}} \frac{\|q^{\sigma,p^\star}_s - p^\star \|^2}{\|q^{\sigma,p}_s - p \|^2} \cdot \frac{\d \langle \sigma, p^\star\rangle}{\d \langle \sigma, p\rangle}(s),
\end{align*}
where the first equality is by the above, the second equality rearranges terms and uses the change of measure $\d \langle \sigma, p^\star\rangle = \frac{\d \langle \sigma, p^\star\rangle}{\d \langle \sigma, p\rangle} \dd \langle \sigma, p \rangle$, and the final inequality follows from taking the infimum of the bracketed term in the numerator of the penultimate expression. Moreover, by the additivity of Radon-Nikodym derivatives, it holds that
    \[
    \frac{\d \langle \sigma, p^\star\rangle}{\d \langle \sigma, p\rangle} \, = \, \sum_{\theta \in \Theta} p^\star(\theta) \frac{\d \sigma_\theta}{\d \langle \sigma, p\rangle} \, = \, \sum_{\theta \in \Theta} \frac{p^\star(\theta)}{p(\theta)}  \cdot p(\theta)\frac{\d \sigma_\theta}{\d \langle \sigma, p\rangle} \, \geq \, \min_{\theta \in \Theta} \frac{p^\star(\theta)}{p(\theta)} \cdot \frac{\d \langle \sigma,p\rangle}{\d \langle \sigma, p\rangle}\, \geq \, \frac{1}{|\Theta|},
    \]
    where the final inequality uses that $p^\star(\theta) = 1/|\Theta|$ and $p(\theta) \leq 1$ for all $\theta \in \Theta$.
Since the given $\sigma \in \Se_b\backslash\Se^\varnothing$ and $p \in \Delta^\circ(\Theta)$ were arbitrary, combining the two displays above yields
\begin{align*}
R \, \geq \, \frac{1}{|\Theta|} \cdot \rho,  \quad \text{where} \quad \rho := \inf \left\{\frac{\|q^{\sigma,p^\star}_s - p^\star \|^2}{\|q^{\sigma,p}_s - p \|^2} \ \Bigg| \ \sigma \in \Se_b \backslash\Se^\varnothing, \ s \in S_\sigma^{\neg\varnothing}, \ p \in \Delta^\circ(\Theta)  \right\} .
\end{align*}
We claim that $\rho >0$. Since this directly implies \eqref{eqn:sp-verify-1}, it suffices to prove the claim. 

Suppose, towards a contradiction, that $\rho= 0$. Fix any $\epsilon >0$ small enough that 
\begin{equation}\label{eqn:sp-verify-2}
1 > \sqrt{ 2 \epsilon} \cdot  |\Theta|  \qquad \text{ and } \qquad    \frac{1-\sqrt{2 \epsilon} \cdot |\Theta|}{ 1-\sqrt{2 \epsilon} \cdot |\Theta| + |\Theta| } > \sqrt{\epsilon} \cdot |\Theta|.
\end{equation}
By the supposition, there exist $\sigma \in \Se_b\backslash\Se^\varnothing$, $s \in S_\sigma^{\neg\varnothing}$, and $p \in \Delta^\circ(\Theta)$ such that  
\begin{equation}\label{eqn:sp-verify-3}
\epsilon \, > \, \frac{\|q^{\sigma,p^\star}_s - p^\star \|^2}{\|q^{\sigma,p}_s - p \|^2} \, = \, \frac{1}{|\Theta|^2} \cdot \frac{z^2}{\|q^{\sigma,p}_s - p \|^2}, \quad \text{ where } \quad z := \left\| \frac{\d \bm{\sigma}}{\d \langle \sigma, p^\star \rangle}(s) - \mathbf{1}  \right\|>0,
\end{equation}
where the equality follows from the identity $q^{\sigma,p^\star}_s - p^\star = \frac{1}{|\Theta|}\big( \frac{q^{\sigma,p^\star}_s}{p^\star}- \mathbf{1}\big)$ and Bayes' rule. Since $\text{diam}(\Delta(\Theta)) = \sqrt{2}$, the first inequality in \eqref{eqn:sp-verify-2} and condition \eqref{eqn:sp-verify-3} together imply
\begin{equation}\label{eqn:sp-verify-4}
1 \, > \, \sqrt{2 \epsilon} \cdot |\Theta| \, > \, z \, \geq \, \max_{\theta \in \Theta} \left| \frac{\d\sigma_\theta}{\d\langle \sigma, p^\star\rangle}(s) -1 \right|,    
\end{equation}
where the final inequality is by definition of $z$. This implies, via a short calculation, that 
\[
\left| \frac{\d \langle \sigma, p^\star\rangle}{\d\langle \sigma, p\rangle} - 1 \right| \, \leq \, \frac{z}{1-z}. %\, \leq \, \frac{\sqrt{2 \epsilon} \cdot |\Theta|}{1- \sqrt{2 \epsilon} \cdot |\Theta|}.
\]
Therefore, we obtain the following bound:
\begin{align*}
\hspace{-1.5em}
\|q^{\sigma,p}_s - p \| \, \leq \, \left\| \frac{\d \bm{\sigma}}{\d \langle \sigma, p \rangle}(s) - \mathbf{1}  \right\| \, &= \, \left\| \frac{\d \bm{\sigma}}{\d \langle \sigma, p^\star \rangle}(s) \cdot \frac{\d \langle \sigma, p^\star \rangle}{\d \langle \sigma, p \rangle}(s)- \mathbf{1}  \right\| \\
& \leq \, z + \left| \frac{\d \langle \sigma, p^\star\rangle}{\d\langle \sigma, p\rangle} - 1 \right| \cdot\left\| \frac{\d \bm{\sigma}}{\d \langle \sigma, p^\star \rangle}(s) \right\| \,  \leq \, z + \frac{z}{1-z} \cdot \left\| \frac{q^{\sigma,p^\star}_s}{p^\star} \right\| \, \leq \, z + \frac{z}{1-z} \cdot |\Theta|,
\end{align*}
where the first inequality uses $q^{\sigma,p}_s - p = \diag(p) \big( \frac{q^{\sigma,p}_s}{p} - \mathbf{1}\big)$, the fact that $\max_{\theta\in\Theta}p(\theta)\leq 1$, and Bayes' rule; the second equality uses the chain rule for Radon-Nikodym derivatives; the third inequality uses the triangle inequality and the definition of $z$; the fourth inequality uses the preceding display and Bayes' rule; and the final inequality uses $\frac{q^{\sigma,p^\star}_s}{p^\star} = |\Theta|\cdot q^{\sigma,p^\star}_s$ and $\max_{q \in \Delta(\Theta)}\|q\| =1$. Plugging this bound into \eqref{eqn:sp-verify-3}, we then obtain 
\[
\sqrt{\epsilon}\cdot |\Theta| \, > \, \frac{z}{\|q^{\sigma,p}_s - p \|} \, \geq \, \frac{z}{z + \frac{z }{1-z}\cdot |\Theta|} \, = \, \frac{1-z}{1 -z  + |\Theta|} \, \geq \, \frac{1-\sqrt{2 \epsilon} \cdot |\Theta|}{ 1-\sqrt{2 \epsilon} \cdot |\Theta| + |\Theta| } > \sqrt{\epsilon} \cdot |\Theta|,
\]
where the first (strict) inequality is equivalent to \eqref{eqn:sp-verify-3}, the second inequality is by the preceding display, the third equality follows from rearranging terms, the fourth inequality follows from \eqref{eqn:sp-verify-4} and the fact that the map $x \in(0,1)\mapsto \frac{1-x}{1-x + |\Theta|}$ is decreasing, and the final (strict) inequality is by the second inequality in \eqref{eqn:sp-verify-2}. This delivers the desired contradiction. We conclude that $\rho>0$, and hence that $C$ is \nameref{defi:sp}, as desired.

\emph{\textbf{Step 4:} $\Phi(C)$ is \nameref{defi:spi}, \nameref{defi:sp}, and \nameref{defi:lq} with $\kappa_{\Phi(C)}(p) = \kappa$ for all $p \in \Delta^\circ(\Theta)$.} First, $\Phi(C)$ is \nameref{defi:spi} because $C$ is \hyperref[axiom:prior:invariant]{Prior Invariant} (Step 1). Second, $\Phi(C)$ is \nameref{defi:sp} because $C$ is \nameref{defi:sp} (Step 3), $\Phi$ is isotone and HD1 (\cref{lem:structure:Phi}), and $\text{Var} \in \C$ is \nameref{axiom:slp} (\cref{lem:ups:to:additive}). Finally, to show that $\Phi(C)$ is \nameref{defi:lq}, note that since $C$ is \nameref{defi:lq} and \nameref{defi:sp} (Steps 2--3) and $k_C(p)\gg_\text{psd} \mathbf{0}$ for all $p \in \Delta^\circ(\Theta)$ (by \cref{cor:ker-SP}), \cref{thm:qk}(ii) implies that $k_C$ is a lower kernel of $\Phi(C)$ on $\Delta^\circ(\Theta)$. Meanwhile, since $C \succeq \Phi(C)$, $k_C$ is also an upper kernel of $C$ on $\Delta^\circ(\Theta)$. We conclude that $\Phi(C)$ is \nameref{defi:lq} with kernel $k_{\Phi(C)} = k_C$. Therefore, since $\kappa_C(p) = \kappa$ for all $p \in \Delta^\circ(\Theta)$ (Step 2), it follows that $\kappa_{\Phi(C)}(p) = \kappa_C(p) = \kappa$ for all $p \in \Delta^\circ(\Theta)$, as desired.
\end{proof}

\subsection{Proofs of Lemmas \ref{lem:ps-kernel-suff}--\ref{cor:ker-SP}}\label{ssec:calc-kernel-proofs}

\subsubsection{Proof of \cref{lem:ps-kernel-suff}}\label{ssec:proof-ps-kernel-suff}

\begin{proof}%[Proof of \cref{lem:ps-kernel-suff}]
    First, we claim that, for every $\epsilon>0$, there exists a $\delta>0$ such that
    \begin{align}
        \left| D(q \mid p) - \frac{1}{2} (q-p)^\top \H_1 D(p_0 \mid p_0) (q-p) \right| \leq \epsilon \|q-p\|^2 \qquad \forall \, p , q \in B_{\delta}(p_0). \label{eqn:kernel-ps-pointwise}
    \end{align}
    To this end, let $\epsilon>0$ be given. Let $\overline{\delta}>0$ be the radius of some ball around $p_0$ witnessing that $D$ is locally $\mathbf{C}^2$ at $p_0$. Then, for all $p,q \in B_{\overline{\delta}}(p_0)$ we have
    %By the fundamental theorem of calculus (cf. \cref{fn:FTC-hessian} in \cref{app:thm3-2:proof}), for all $p,q \in B_{\overline{\delta}}(p_0)$ it holds that
    \begin{align*}
        D(q \mid p) &= \int_0^1 (1-t) (q-p)^\top \H D_1 (r(t) \mid p) (q-p) \dd t  \qquad \text{ where $r(t) := p + t (q-p)$} \\
        %%%%
        & = \frac{1}{2} (q-p)^\top \H_1 D(p_0 \mid p_0) (q-p) \\
        & \qquad \qquad \ \ \   + \int_0^1 (1-t) (q-p)^\top \big[ \H_1 D(r(t) \mid p)  - \H_1 D (p_0 \mid p_0) \big] (q-p) \dd t,
    \end{align*}
    where the first equality is by the Fundamental Theorem of Calculus\footnote{In particular, the argument is a minor modification of that from \cref{fn:FTC-hessian} in \cref{app:thm3-1:extra}, where we now define $f(t) := D(r(t) \mid p)$ and use the facts that $r(t) \in B_{\overline{\delta}}(p_0)$ and $f''(t) = (q-p)^\top \H_1 D (r(t) \mid p)(q-p)$ for all $t \in [0,1]$.} and the second equality rearranges terms and uses the fact that $\int_0^1 (1-t) \dd t = \frac{1}{2}$. Since $(q,p) \mapsto \H_1 D(q\mid p) \in \R^{|\Theta| \times|\Theta|}$ is continuous on $B_{\overline{\delta}}(p_0) \times B_{\overline{\delta}}(p_0)$, there exists a $\delta \in (0, \overline{\delta}]$ such that $\| \H_1 D(q \mid p) - \H_1 D(p_0 \mid p_0)\| \leq 2 \epsilon$ for all $p,q \in B_\delta(p_0)$. Thus, for all $p,q \in B_\delta (p_0)$ we have
    %Plugging this bound into the above display, we obtain that, for all $p,q \in B_\delta(p_0)$, it holds that 
    \begin{align*}
        &\left| D(q \mid p)  - \frac{1}{2} (q-p)^\top \H_1 D(p_0 \mid p_0) (q-p) \right| \\
        %%%
       & \qquad \qquad  = 
       \, \left| \int_0^1 (1-t) (q-p)^\top \big[ \H_1 D(r(t) \mid p)  - \H_1 D (p_0 \mid p_0) \big] (q-p) \dd t \right| \\
       %%%
       & \qquad \qquad \leq   \int_0^1 (1-t) 2 \epsilon  \|q-p\|^2 \dd t  \  = \  \epsilon \, \|q-p\|^2,
    \end{align*}
    where the first equality is by the preceding display, the inequality is by the definition of $\delta>0$ (where $r(t) \in B_\delta(p)$ for all $t \in [0,1]$ by convexity of the ball), and the final equality uses $\int_0^1 (1-t) \dd t = \frac{1}{2}$. Since the given $\epsilon>0$ was arbitrary, this establishes the claim. 

    We now use \eqref{eqn:kernel-ps-pointwise} to prove the lemma. Let $\epsilon >0$ be given and let $\delta>0$ be such that \eqref{eqn:kernel-ps-pointwise} holds. Then, for every $\pi \in \Delta(B_\delta(p_0))$, we have
    \begin{align}
       \hspace{-1em} C(\pi) \ = \  \E_\pi \big[ D(q \mid p_\pi) \big] \  &= \int_{B_\delta(p_0)} D(q \mid p_\pi) \dd \pi (q) \notag \\
       & \leq \int_{B_\delta(p_0)} (q-p_\pi)^\top \left( \frac{1}{2} \H_1 D(p_0 \mid p_0) + \epsilon I \right) (q-p_\pi) \dd \pi (q), \label{eqn:ps-upper-kernel} %\quad \forall \, \pi \in \Delta(B_\delta(p_0)),
    \end{align}
    where the first equality is by definition of $C$, the second equality is by $\supp(\pi) \subseteq B_\delta(p_0)$, and the final inequality follows from applying \eqref{eqn:kernel-ps-pointwise} to each $p_\pi, q \in \text{conv}(\supp(\pi)) \subseteq B_\delta(p)$ (where we use convexity of the ball). Meanwhile, for every $\pi \in \Ex$ with $p_\pi \in B_\delta(p_0)$, %we have
    \begin{align}
        C(\pi)  & = \int_{B_\delta(p_0)} D(q \mid p_\pi) \dd \pi(q) + \int_{ \Delta(\Theta) \backslash B_\delta(p_0)} D(q \mid p_\pi) \dd \pi(q) \notag \\
        %%%
        & \geq \int_{B_\delta(p_0)} D(q \mid p_\pi) \dd \pi(q) \notag \\
        %%%
        & \geq  \int_{B_\delta(p_0)} (q-p_\pi)^\top \left( \frac{1}{2} \H_1 D(p_0 \mid p_0) - \epsilon I \right) (q-p_\pi)  \dd \pi(q), \label{eqn:ps-lower-kernel}
    \end{align}
    where the first line is by definition of $C$, the second line holds because $D(q \mid p) \geq 0$ for all $p,q\in \Delta(\Theta)$, and the final line follows from applying \eqref{eqn:kernel-ps-pointwise} pointwise to each $p_\pi , q \in \text{conv}(\supp(\pi)) \cap B_\delta(p_0)$. Since the given $\epsilon >0$ was arbitrary, \eqref{eqn:ps-upper-kernel} and \eqref{eqn:ps-lower-kernel} imply that $\H_1 D(p_0 \mid p_0)$ is both an upper and lower kernel of $C$ at $p_0$. In other words, $C$ is \nameref{defi:lq} at $p_0$ and its kernel is $k_C(p_0) = \H_1 D(p_0 \mid p_0)$, as desired.  
\end{proof}

\subsubsection{Proof of \cref{lem:ups-kernel-equiv}}\label{ssec:proof-ups-kernel-equiv}

 \begin{proof}%[Proof of \cref{lem:ups-kernel-equiv}]
Let $W \subseteq \Delta(\Theta)$ be open, and let $H$ be convex with $\dom(H) \supseteq W$.% We prove each direction in turn.

\noindent \textbf{($\impliedby$ direction)} Let $H|_W \in \mathbf{C}^2(W)$. Define the divergence $D$ as $D (q \mid p) := H(q) - H(p) - (q-p)^\top \nabla H(p)$ for all $(p,q) \in W \times W$ and $D(q\mid p) := 0$ for all $(p,q) \notin W \times W$.\footnote{Note that $D$ is a well-defined divergence because the convexity of $H$ ensures that $D(q \mid p) \geq 0$ for all $p,q \in W$.} Since $W$ is open, at every $p_0 \in W$, $D$ is locally $\mathbf{C}^2$ and satisfies $\H_1 D(p_0 \mid p_0) =  \H H(p_0)$. %Since $W \subseteq \Delta^\circ(\Theta)$, 
Hence, \cref{lem:ps-kernel-suff} implies that the \hyperref[eqn:PS]{Posterior Separable} cost function $C \in \C$,  defined as $C(\pi) := \E_\pi [ D(q\mid p_\pi)]$ for all $\pi \in \Ex$, is \nameref{defi:lq} on $W$ with kernel $k_{C} = \H H$. By construction: (i) $C^H_\text{ups}\succeq C$ and (ii) $C^H_\text{ups}(\pi) = C(\pi)$ for all $\pi \in \Delta(W)$.\footnote{Formally, for any $\pi \in \Ex$ with $p_\pi \notin W$, we have $C^H_\text{ups}(\pi) \geq 0 = C(\pi)$ because $C^H_\text{ups} \in \C$ and $D(\cdot \mid p_\pi) \equiv 0$. Meanwhile, for any $\pi \in \Ex$ with $p_\pi \in W$, we have 
\[
C^H_\text{ups}(\pi) = \E_\pi \left[ H(q) - H(p_\pi) - (q-p_\pi)^\top \nabla H(p_\pi) \right] = C(\pi) + \int_{\Delta(\Theta) \backslash W} \left( H(q) - H(p_\pi) - (q-p_\pi)^\top \nabla H(p_\pi)\right) \dd \pi(q) \geq C(\pi), 
\]
where the inequality is by the convexity of $H$ and becomes an equality if $\supp(\pi) \subseteq W$. Properties (i) and (ii) follow.} Property (i) implies that every lower kernel of $C$ on $W$ is also a lower kernel of $C^H_\text{ups}$ on $W$. Since $W$ is open, property (ii) implies that every upper kernel of $C$ on $W$ is also an upper kernel of $C^H_\text{ups}$ on $W$.\footnote{Openness ensures that, at every $p_0 \in W$, we can choose the $\delta>0$ in \cref{defi:lq}(i) small enough that $B_\delta(p_0) \subseteq W$.} We conclude that $C^H_\text{ups}$ is \nameref{defi:lq} with kernel $k_{C^H_\text{ups}} = \H H$ on $W$.

\noindent \textbf{($\implies$ direction)} Let $C^H_\text{ups}$ be \nameref{defi:lq} on $W \subseteq \Delta^\circ(\Theta)$ with kernel $k := k_{C^H_\text{ups}}$. Let $p_0 \in W$ be given. For every $\epsilon >0$, there exists a $\delta(\epsilon)>0$ such that $B_{\delta(\epsilon)}(p_0) \subseteq W$ and 
\begin{align}
\left| C^H_\text{ups}(\pi) - \frac{1}{2} \E_\pi\left[ (q-p_\pi)^\top k(p_0)(q-p_\pi)\right]   \right| \leq \epsilon \text{Var}(\pi) \qquad \forall \, \pi \in \Delta(B_{\delta(\epsilon)}(p_0)), \label{eqn:kernel-ups}
\end{align}
where the set inclusion holds because $W$ is open and \eqref{eqn:kernel-ups} is implied by \cref{defi:lq}. 

For every $\epsilon>0$ and $p \in B_{\delta(\epsilon)}(p_0)$, define $\delta'(p,\epsilon):= \delta(\epsilon) - \|p - p_0\|$ and let $\mathcal{F}(p,\epsilon) := \{y \in \mathcal{T}\mid \, \|y\| < \delta'(p,\epsilon)\}$ denote the ball in $\mathcal{T}$ of radius $\delta'(p,\epsilon)$, so that $p + z \in B_{\delta'(p,\epsilon)}(p) \subseteq B_{\delta(\epsilon)}(p_0)$ for all $z \in \mathcal{F}(p,\epsilon)$.\footnote{Namely, $p + z \in B_{\delta'(p,\epsilon)}(p)$ for all $z \in \mathcal{F}(p,\epsilon)$ because  $B_{\delta'(p,\epsilon)}(p) \subseteq W$ by construction and $W \subseteq \Delta^\circ(\Theta)$ by hypothesis.} Then, for every $\epsilon>0$, $p \in B_{\delta(\epsilon)}(p_0)$, $z \in \mathcal{F}(p,\epsilon)$, and $t\in(0,1]$, define $\pi_{p,z,t} \in \Delta ( B_{\delta(\epsilon)}(p_0))$ as $\pi_{p,z,t} := \frac{t}{1+t} \delta_{p+z} + \frac{1}{1+t} \delta_{p - t z }$ and note that $p_{\pi_{p,z,t}} = p$. Plugging these $\pi_{p,z,t}$ into \eqref{eqn:kernel-ups}, multiplying through by $(1+t)/t >0$, and simplifying yields
\[
\left| H(p + z ) - H(p) + \frac{H(p- t z ) - H(p)}{t} - \frac{1+t}{2} \, z^\top k (p_0) z  \right| \leq \epsilon (1+t) \,  \|z\|^2 %\qquad \forall \, p \in B_\delta(p_0), \, z \in \mathcal{F}(p), \, t \in (0,1],
\]
for all $\epsilon>0$, $p \in B_{\delta(\epsilon)}(p_0)$, $z \in \mathcal{F}(p,\epsilon)$, and $t \in (0,1]$. 
%, where we have multiplied through by $\frac{1+t}{t}>0$ and simplified. 
Taking  $t \searrow 0$, we obtain
\begin{align}
\hspace{-2.5em}
\left| H(p + z) - H(p)  + H' (p; -z ) - \frac{1}{2} z^\top k(p_0) z  \right| \leq \epsilon  \,  \|z\|^2 \quad \ \ \forall \, \epsilon>0, \ p \in B_{\delta(\epsilon)}(p_0), \  z \in \mathcal{F}(p,\epsilon), \label{eqn:kernel-ups-binary}
\end{align}
where $H' (p; -z ) : = \lim_{t\searrow 0}\frac{H(p- t z ) - H(p)}{t} \in \mathbb{R}$ is the one-sided directional derivative of $H$ at $p$ in direction $-z$, which exists because  $H$ is convex \parencite[Theorem 23.1]{rock70} and is finite because all other terms in \eqref{eqn:kernel-ups-binary} are finite (recall that $W\subseteq \dom(H)$ by hypothesis).%\footnote{\awb{[and (ii) $D'_{-y} (p \mid p) = \lim_{t\searrow 0}\frac{D(p- t y \mid p)}{t} \in \R_+$ because $D(\cdot \mid p)$ has minimum $D(p\mid p) = 0$ (by definition),]}}

First, we claim that $H$ is continuously differentiable on $B:= \bigcup_{\epsilon>0} B_{\delta(\epsilon)}(p_0)$. To this end, let $\epsilon>0$, $p \in B_{\delta(\epsilon)}(p_0)$ and $y \in \mathcal{T}$ be given. Since $\tau y \in \mathcal{F}(p,\epsilon)$ and $H' (p; - \tau y) = \tau H' (p; -y)$ for all $\tau \in \left(0, \delta'(p,\epsilon)/\|y\|\right)$, the triangle inequality and \eqref{eqn:kernel-ups-binary} (with $z = \tau y$) imply that
\[
\left| H(p + \tau y ) - H(p) + \tau \, H' (p; -y ) \right| \leq \left( \epsilon + \frac{1}{2}\|k(p_0)\| \right) \cdot \tau^2 \cdot \|y\|^2  \qquad \forall \, \tau \in \left(0, \delta'(p,\epsilon)/\|y\|\right).
\]
Dividing through by $\tau>0$ and then taking $\tau \searrow 0$ yields $H'(p; y ) = - H'(p; -y)$. Thus, the corresponding two-sided directional derivative exists \parencite[p. 213]{rock70}. Since the given $p \in B_{\delta(\epsilon)}(p_0)$ and $y \in \mathcal{T}$ were arbitrary and $H$ is convex, it follows that $H$ is continuously differentiable on $B_{\delta(\epsilon)}(p_0)$ \parencite[Theorem 25.2 and Corollary 25.5.1]{rock70}.\footnote{Formally, since $\dom(H) \subseteq \Delta^\circ(\Theta)$ has empty interior with respect to the Euclidean topology on $\R^{|\Theta|}$, to apply \textcite{rock70} we consider the HD1 extension of $H$, viz., the map $G : \R^{|\Theta|}_+ \to \R \cup \{+\infty\}$ defined as $G(x) := (\mathbf{1}^\top x) H\left( \frac{x}{\mathbf{1}^\top x} \right)$. Since $H$ admits finite two-sided directional derivatives in all directions $y \in \mathcal{T}$ at every $p \in B_{\delta(\epsilon)}(p_0)$, it can be shown that $G$ admits finite two-sided directional derivatives in all directions $x \in \R^{|\Theta|}$ at every $p \in B_{\delta(\epsilon)}(p_0)$. Since all such $p$ are in the interior of $\dom(G) \subseteq \R^{|\Theta|}_{++}$ with respect to the Euclidean topology on $\R^{|\Theta|}$, Theorem 25.2 and Corollary 25.5.1 in \textcite{rock70} imply that the gradient map $p \in B_{\delta(\epsilon)}(p_0) \mapsto \nabla G(p) \in \R^{|\Theta|}$ is well-defined and continuous. For every $p \in B_{\delta(\epsilon)}(p_0)$, $\nabla H(p) := \nabla G(p)$ is then the gradient of $H$ at $p$, so $H \in \mathbf{C}^1\big( B_{\delta(\epsilon)}(p_0)\big)$ as claimed.} Since $\epsilon>0$ was arbitrary, $H$ is continuously differentiable on $B$, as desired. 

It follows that the gradient map $p \in B \mapsto \nabla H(p) \in \R^{|\Theta|}$ is well-defined and continuous. Therefore, we can equivalently rewrite \eqref{eqn:kernel-ups-binary} as
\begin{align}
\hspace{-2em}
\left| H(p + z ) - H(p) - z^\top \nabla H(p) - \frac{1}{2} z^\top k (p_0) z  \right| \leq \epsilon \|z\|^2 \quad \ \ \forall \, \epsilon>0, \ p \in B_{\delta(\epsilon)}(p_0), \  z \in \mathcal{F}(p,\epsilon). \label{eqn:kernel-ups-binary-2}
\end{align}
We will use the expansion \eqref{eqn:kernel-ups-binary-2} repeatedly below. 

Next, we claim that $H$ is twice differentiable at $p_0$ and $\H H(p_0) = k(p_0)$. To this end, note that because $p_0 \in \bigcap_{\epsilon>0}B_{\delta(\epsilon)}(p_0)$, \eqref{eqn:kernel-ups-binary-2} implies that
\begin{align}
\left| H(p_0 + z ) - H(p_0) - z^\top \nabla H(p_0) - \frac{1}{2} z^\top k (p_0) z  \right| \leq \epsilon \|z\|^2 \quad \ \ \forall \, \epsilon>0, \ z \in \mathcal{F}(p_0,\epsilon). \label{eqn:kernel-ps-binary-3}
\end{align}
Since $\nabla H(p) \in \R^{|\Theta|}$ exists for all $p \in  B$, \eqref{eqn:kernel-ps-binary-3} and \textcite[Theorem 2.8]{rockafellar1999second} deliver
\[
\lim_{y \in \mathcal{T}, \,  y \to \mathbf{0}} \frac{\| \nabla H(p_0 + y) - \nabla H(p_0) - k(p_0) y \|}{\|y\|} = 0 ,
\]
meaning that $\nabla H$ is differentiable at $p_0$ and its derivative is $k(p_0)$.\footnote{Formally, to apply \textcite[Theorem 2.8]{rockafellar1999second}, we again consider the HD1 extension of $H$ defined as $x \in \R^{|\Theta|}_+ \mapsto G(x):= (\mathbf{1}^\top x) H\left( \frac{x}{\mathbf{1}^\top x}\right)$. Since our convention for normalizing gradients and Hessians of functions on $\Delta(\Theta)$ ensures that $\nabla H (q) = \nabla G(q)$ and $\H H(q) = \H G(q)$ at all $q \in \Delta(\Theta)$ for which $\nabla H (q)$ and $\H H(q)$ are well-defined, it can be shown that \eqref{eqn:kernel-ps-binary-3} implies that, for every $\epsilon>0$, there exists a $\widehat{\delta}(\epsilon)>0$ such that $\left| H(p_0 + x ) - H(p_0) - x^\top \nabla H(p_0) - \frac{1}{2} x^\top k (p_0) x\right| \leq \epsilon \, \|x\|^2$ for all $x \in \R^{|\Theta|}$ such that $\|x\| < \widehat{\delta}(\epsilon)$. Then, since $p_0 \in W$ is in the interior of $\dom(G) \subseteq \R^{|\Theta|}_{++}$ with respect to the Euclidean topology on $\R^{|\Theta|}$, \textcite[Theorem 2.8]{rockafellar1999second} implies that $G$ is twice differentiable at $p_0$ and $\H G(p_0) = k(p_0)$. Thus, by the aforementioned normalization, $\H H(p_0) = k(p_0)$.}   Equivalently, $H$ is twice differentiable at $p_0$ and $\H H(p_0) = k(p_0)$, as desired.

Now, since the given $p_0 \in W$ was arbitrary, we conclude that $H$ is twice differentiable with $\H H = k$ on $W$. It remains to show that $\H H : W \to  \R^{|\Theta| \times |\Theta|}$ is continuous. To this end, let $\epsilon >0$, $p_0 \in W$, and $\widehat{p}_0 \in B_{\delta(\epsilon)}(p_0) \subseteq W$ be given.\footnote{Note that the $p_0 \in W$ and corresponding $\delta(\epsilon)>0$ given here may differ from those in the preceding paragraphs; we recycle the same symbols here with a minor abuse of notation.} For all $z \in \mathcal{F}(\widehat{p}_0,\epsilon)$, we have
\begin{align*}
    \frac{1}{2} \left| z^\top 
    \left( \H H(\widehat{p}_0)  - \H H(p_0) \right) z \right| & \leq \left|  H(\widehat{p}_0 + z )  - H(\widehat{p}_0) - z^\top \nabla H(\widehat{p}_0) - \frac{1}{2} z^\top \H H (p_0) z \right| \\
    %%%
    & \ \ \ \ \ \ \ \ \ \ \ \ \  + \left| H(\widehat{p}_0 + z ) - H(\widehat{p}_0) - z^\top \nabla H(\widehat{p}_0) - \frac{1}{2} z^\top \H H (\widehat{p}_0) z\right| \\
    %%%
    & \leq \epsilon \|z\|^2 + \left| H(\widehat{p}_0 + z ) - H(\widehat{p}_0) - z^\top \nabla H(\widehat{p}_0) - \frac{1}{2} z^\top \H H (\widehat{p}_0) z\right|,
\end{align*}
where the first inequality is by the triangle inequality and the second inequality is by \eqref{eqn:kernel-ups-binary-2} and $k(p_0) = \H H(p_0)$. Now, by replicating the derivation of \eqref{eqn:kernel-ps-binary-3} with $\widehat{p}_0$ in place of $p_0$ and using the fact that $k(\widehat{p}_0) = \H H (\widehat{p}_0)$, we conclude that there exists a $\widehat{\delta}>0$ such that 
\[
\left| H(\widehat{p}_0 + z ) - H(\widehat{p}_0) - z^\top \nabla H(\widehat{p}_0) - \frac{1}{2} z^\top \H H (\widehat{p}_0) z\right| \leq \epsilon \|z\|^2 \quad \ \ \forall \, z \in \mathcal{T} \  \text{ s.t. } \ \|z\|< \widehat{\delta}.
\]
By combining the two displays above, we conclude that there exists a $\overline{\delta}>0$ such that
%Recalling that $\mathcal{F}(p'_0,\epsilon) = \{y \in \mathcal{T} \mid \, \|y\| < \delta'(p'_0,\epsilon)\}$ for an appropriate parameter $\delta'(p'_0,\epsilon)>0$, we combine the two displays above to obtain 
\[
\left| z^\top \left( \H H(\widehat{p}_0) - \H H(p_0) \right) z  \right| \leq 4 \epsilon \|z\|^2 \quad \ \ \forall \, z \in \mathcal{T} \ \text{ s.t. } \  \|z \| < \overline{\delta}.
\]
It follows that $\|\H H(\widehat{p}_0) - \H H(p_0)\| \leq 4 \epsilon$. Since the given $\epsilon >0$ and $\widehat{p}_0 \in B_{\delta(\epsilon)}(p_0)$ were arbitrary, we conclude that $\H H$ is continuous at $p_0$. Since the given $p_0 \in W$ was arbitrary, it follows that $\H H$ is continuous on $W$, as desired. 
\end{proof}

\subsubsection{Proof of \cref{lem:kernel-rank}}\label{ssec:proof-kernel-rank}

\begin{proof}%[Proof of \cref{lem:kernel-rank}]
First, we show that $k_C(p_0) =  \max\underline{K}_C(p_0)$. Suppose, towards a contradiction, that this is not true, i.e., there exist $\underline{k}(p_0) \in \underline{K}_C(p_0)$ and $y \in \mathcal{T}$ such that $y^\top k_C(p_0) y < y^\top \underline{k}(p_0) y$. Thus, there exists an $\eta >0$ such that $y^\top \left( k_C(p_0) + 2 \eta I \right) y \leq y^\top \underline{k}(p_0) y$. Fix an arbitrary $\epsilon \in \left(0, \eta /2\right)$. Since $k_C(p_0)$ is an upper kernel of $C$ at $p_0$ and $\underline{k}(p_0)$ is a lower kernel of $C$ at $p_0$, there exists a $\delta>0$ such that, for all $\pi \in \Delta(B_\delta(p_0))$,
\[
\E_\pi \left[ (q-p_\pi)^\top \left( \frac{1}{2} k_C(p_0) + \epsilon I \right) (q-p_\pi)] \right] \ \geq \ C(\pi) \  \geq \ \E_\pi \left[ (q-p_\pi)^\top \left( \frac{1}{2} \underline{k}(p_0) - \epsilon I \right) (q-p_\pi)] \right].
\]
Fix any $p \in B_\delta(p_0) \cap \Delta^\circ(\Theta)$. Since $B_\delta(p_0) \cap \Delta^\circ(\Theta)\neq \emptyset$ is open, there exists $t>0$ such that $p \pm ty \in B_\delta(p_0)\cap\Delta^\circ(\Theta)$. Then, defining $\widehat{\pi} \in \Delta(B_\delta(p_0))$ as $\widehat{\pi} :=\frac{1}{2} \delta_{p + t y} + \frac{1}{2} \delta_{p - t y}$, we obtain:
%arbitrary $t \in \left(0, \delta / \|y\|\right)$ and define $\widehat{\pi} \in \Delta(B_\delta(p_0))$ as $\widehat{\pi} :=\frac{1}{2} \delta_{p_0 + t y} + \frac{1}{2} \delta_{p_0 - t y}$. Then
\[
t^2 y^\top \left( \frac{1}{2} k_C(p_0) + \epsilon I \right) y \ \geq \ C(\widehat{\pi}) \ \geq \ t^2 y^\top \left( \frac{1}{2} \underline{k}(p_0) - \epsilon I \right) y \ \geq \ t^2 y^\top \left( \frac{1}{2} k_C(p_0) +(\eta- \epsilon) I \right) y,
\]
where the first two inequalities follow from the preceding display and the final inequality is by the definition of $y$ and $\eta$. But this implies that $ \epsilon \cdot t^2 \|y\|^2 \geq (\eta- \epsilon) \cdot t^2 \|y\|^2$, and hence that $2 \epsilon \geq \eta$, contradicting that $\epsilon < \eta / 2$, as desired. We conclude that $k_C(p_0) = \max\underline{K}_C(p_0)$.

Next, it can be shown that $k_C(p_0) = \min\overline{K}_C(p_0)$ using a symmetric argument (as $k_C(p_0)$ is also a lower kernel of $C$ at $p_0$). We omit the straightforward details. %Finally, the uniqueness claim is immediate. 
\end{proof}

\subsubsection{Proof of \cref{cor:ker-SP}}\label{ssec:proof-ker-SP}

\begin{proof}
    Since $C$ is \nameref{defi:sp}, there exists $m >0$ such that $C(\pi) \succeq m \text{Var}(\pi)$ for all $\pi \in \Ex$. Hence, for every $p \in \Delta(\Theta)$ and $\delta>0$, the matrix $\widehat{k}(p) := 2 m I \in \R^{|\Theta| \times |\Theta|}$ satisfies
    \begin{align*}
        C(\pi) \geq m \text{Var}(\pi) \,  = \, \int_{\Delta(\Theta)} (q-p_\pi)^\top \, \frac{1}{2} \widehat{k}(p) \, (q-p_\pi) \dd \pi(q) \, \geq \, \int_{B_\delta(p)} (q-p_\pi)^\top \, \frac{1}{2} \widehat{k}(p) \, (q-p_\pi) \dd \pi(q) 
    \end{align*}
    for all $\pi \in \Ex$, where the final inequality holds because $\widehat{k}(p)\geq_\text{psd} \mathbf{0}$ and $B_\delta(p)\subseteq \Delta(\Theta)$. It is then easy to verify from \cref{defi:lq}(ii) that the normalized (as in \cref{remark:kernels}) matrix-valued function $p \mapsto k(p) := (I - \mathbf{1}p^\top) \widehat{k}(p) (I-p \mathbf{1}^\top)$ is a lower kernel of $C$ on $\Delta(\Theta)$. Moreover, $k \gg_\text{psd} \mathbf{0}$ on $\Delta(\Theta)$ by construction. We conclude that $\underline{K}^+_C(p) \neq \emptyset$ for all $p \in \Delta(\Theta)$.

    Now, let $C$ be \nameref{defi:lq} at $p_0 \in \Delta(\Theta)$. Since $\underline{K}^+_C(p_0) \subseteq \underline{K}_C(p_0)$, \cref{lem:kernel-rank} implies that $k_C(p_0) \geq_\text{psd} \underline{k}(p_0)$ for all $\underline{k}(p_0) \in \underline{K}^+_C(p_0)$. Since $\underline{K}^+_C(p_0) \neq \emptyset$ (as shown above), it follows that $k_C(p_0) \gg_\text{psd} \mathbf{0}$, and hence $k_C(p_0) \in \underline{K}^+_C(p_0)$. Thus, $k_C(p_0) = \max \underline{K}^+_C(p_0)$.
\end{proof}

\subsection{Proof of Corollary \ref{cor:flie-nonsmooth}}\label{app:proof-cor-flie-nonsmooth}

\begin{proof}
    We prove each part of the result in turn.

    \noindent\textbf{Sufficiency.} Let $H \in \mathbf{C}^2(W)$ and $\H H $ be an upper kernel of $C$ on the open convex set $W\subseteq \Delta^\circ(\Theta)$. By \cref{thm:qk}(i), $\Phi (C)(\pi) \leq C^H_\text{ups}(\pi)$ for all $\pi \in \Delta(W)$. Since $\dom(C^H_\text{ups})= \Delta(W) \cup\Ex^\varnothing$, it follows that $\Phi(C) \preceq C^H_\text{ups}$. Since $C \succeq C^H_\text{ups}$, $\Phi$ is isotone (\cref{lem:structure:Phi}), and $C^H_\text{ups}$ is \nameref{axiom:slp} (\cref{lem:ups:to:additive}), we also have $\Phi(C) \succeq \Phi(C^H_\text{ups}) = C^H_\text{ups}$. We conclude that $\Phi(C) = C^H_\text{ups}$.

    \noindent\textbf{Necessity.} Let $\Phi(C) = C^H_\text{ups}$. This immediately implies that $C \succeq C^H_\text{ups}$ (as $C \succeq \Phi(C)$ by construction). Moreover, since $H$ is strongly convex, $C$ and $\Phi(C)$ are \nameref{defi:sp}.

    Next, we claim that $\max \underline{K}_C(W) = \H H$. To this end, let $\underline{K}^+_C(W) \subseteq \underline{K}_C(W)$ (resp., $\underline{K}^+_{\Phi(C)}(W) \subseteq \underline{K}_{\Phi(C)}(W)$) denote the set of all lower kernels $k$ of $C$ (resp., of $\Phi(C)$) on $W$ such that $k(p) \gg_\text{psd} \mathbf{0}$ for all $p \in W$. First, observe that $\underline{K}^+_{\Phi(C)}(W) = \underline{K}^+_C(W)$, 
    %\[
    %\underline{K}^+_{\Phi(C)}(W) = \underline{K}^+_C(W)
    %\]
    because $C \succeq \Phi(C)$ implies that $\underline{K}^+_{\Phi(C)}(W) \subseteq \underline{K}^+_C(W)$ and (since $C$ is \nameref{defi:sp}) \cref{thm:qk}(ii) implies that $\underline{K}^+_{\Phi(C)}(W) \supseteq \underline{K}^+_C(W)$. Second, observe that 
    %\[
    $\H H = k_{\Phi(C)} %= \max \underline{K}_{\Phi(C)}(W) 
    = \max \underline{K}^+_{\Phi(C)}(W)$, 
    %\]
    where the first equality is by \cref{lem:ups-kernel-equiv} (as $\Phi(C) = C^H_\text{ups}$ and $H \in \mathbf{C}^2(W)$) and the second equality is by \cref{cor:ker-SP} (as $\Phi(C)$ is \nameref{defi:sp}). 
    %, the second equality is by \cref{lem:kernel-rank}, and the final equality holds because $\H H \in \underline{K}^+_{\Phi(C)}(W)$ (as $H$ is strongly convex) and $\underline{K}^+_{\Phi(C)}(W) \subseteq \underline{K}_{\Phi(C)}(W)$ (by definition). 
    Together, these two observations yield $\H H = \max \underline{K}^+_C(W)$. Finally, suppose towards a contradiction that $\H H \neq \max \underline{K}_C(W)$, i.e., there exist $p \in W$, $\underline{k}(p) \in \underline{K}_C(p)\backslash \underline{K}^+_C(p)$, and $y\in \mathcal{T}\backslash\{\mathbf{0}\}$ such that $y^\top \H (p) y < y^\top \underline{k}(p) y$. Since $\H H (p) \in \underline{K}^+_C(p)$, this implies that there exists $\alpha \in (0,1)$ sufficiently close to $1$ such that $\widehat{k}(p) := \alpha \H H(p) + (1-\alpha) \underline{k}(p) \in \underline{K}^+_C(p)$ and $y^\top \H H(p) y < y^\top \widehat{k}(p) y$.\footnote{It follows from \cref{defi:lq}(ii) that $\underline{K}_C(p)\subseteq \R^{|\Theta|\times|\Theta|}$ is convex, which implies that $\alpha \H H(p) + (1-\alpha) \underline{k}(p) \in \underline{K}_C(p)$ for all $\alpha \in [0,1]$. Define $\zeta, \eta \in \R$ as $\zeta := \min\{z^\top \H H(p) z \mid z \in \mathcal{T} \text{ s.t. } \|z\|^2=1\}$ and $\eta := \min\{z^\top \underline{k}(p) z \mid z \in \mathcal{T} \text{ s.t. } \|z\|^2=1\}$. Note that $\zeta$ and $\eta$ are well-defined and $\min\{\zeta,\eta\}>0$ because $\H H(p)\gg_\text{psd} \mathbf{0}$ (as $H$ is strongly convex). Hence, for $\alpha \in (0,1)$ sufficiently close to $1$, we have $\alpha \zeta +(1-\alpha) \eta >0$ and therefore $z^\top \left( \alpha \H H(p) + (1-\alpha) \underline{k}(p) \right) z \geq \left(\alpha \zeta +(1-\alpha) \eta\right) \cdot \|z\|^2>0$ for all $z \in \mathcal{T}$, i.e., $\alpha \H H(p) + (1-\alpha) \underline{k}(p) \in \underline{K}^+_C(p)$.} This contradicts $\H H = \max \underline{K}^+_C(W)$, as desired. We conclude that $\H H = \max \underline{K}_C(W)$.

\noindent \textbf{Approximation.} Given any open cover $\mathbb{O}$ of $W$, define $\widehat{C}\in \C$ as
\[
\widehat{C}(\pi) := \begin{cases}
    C^H_\text{ups}(\pi), & \text{if $\exists O \in \mathbb{O}$ s.t. $\supp(\pi) \subseteq O $} \\
    %%%
    C(\pi), & \text{otherwise.}
\end{cases}
\]
First, note that $\widehat{C}$ satisfies the desired property (iii) by construction. Second, note that $C \succeq \widehat{C} \succeq C^H_\text{ups}$ because $C \succeq \Phi(C) = C^H_\text{ups}$. Hence, $\widehat{C}$ satisfies the desired property (ii). 

Third, we claim that $\widehat{C}$ satisfies the desired property (iii), i.e., it is \nameref{defi:lq} on $W$ with kernel $k_{\widehat{C}} = \H H$. To this end, note that  $C^H_\text{ups}$ is \nameref{defi:lq} on $W$ with kernel $k_{C^H_\text{ups}} = \H H$ (by \cref{lem:ups-kernel-equiv} in \cref{ssec:calc-kernel}). Since $\widehat{C} \succeq C^H_\text{ups}$, it follows that $\H H$ is a lower kernel of $\widehat{C}$ on $W$. Moreover, because $\mathbb{O}$ is an open cover of $W$, for every $p \in W$ there exists an $O \in \mathbb{O}$ and a $\overline{\delta}(p)>0$ such that $B_{\overline{\delta}(p)}(p) \subseteq O$; hence, $\widehat{C}(\pi) = C^H_\text{ups}(\pi)$ for all $\pi \in \bigcup_{p\in W}\Delta(B_{\overline{\delta}(p)}(p))$. Since $\H H$ is an upper kernel of $C^H_\text{ups}$ on $W$ and, for every $p \in W$, we are free to choose the ($p$-dependent) $\delta>0$ in \cref{defi:lq}(i) small enough that $\delta \leq \overline{\delta}(p)$, it follows that $\H H$ is also an upper kernel of $\widehat{C}$ on $W$. This proves the claim. %We conclude that $k_{\widehat{C}} = \H H$ on $W$, as claimed. 

Finally, we claim that $\widehat{C} \in \Phi^{-1} (C^H_\text{ups})$, i.e., $\Phi(\widehat{C}) = C^H_\text{ups}$. To this end, observe that, since $\widehat{C} \succeq C^H_\text{ups}$ (as noted above), it holds that $\widehat{C}$ is \nameref{defi:sp} and $\dom(\widehat{C}) \subseteq \dom(C^H_\text{ups})  = \Delta(W)\cup\Ex^\varnothing$. Hence, \cref{lem:phi-ie}(iii) implies that $\Phi_\text{IE}(\widehat{C}) = C^H_\text{ups}$ (by the preceding observation and the facts that $k_{\widehat{C}}  = \H H$ on $W$, $H$ is strongly convex, and $W \subseteq \Delta^\circ(\Theta)$ is open and convex). Therefore, $\widehat{C}$ \nameref{axiom:flie} and \cref{thm:flie} yields $\Phi(\widehat{C}) = C^H_\text{ups}$, as claimed.
\end{proof}

\subsection{Proof of Corollary \ref{cor:preserve:CMC}}\label{ssec:thm5-cor-proofs-1}

    \begin{proof}%[Proof of \cref{cor:preserve:CMC}]
    Let $C \in \C$ have rich domain and be \nameref{defi:sp}, \nameref{defi:lq}, \nameref{axiom:CMC}, and \hyperref[axiom:DL]{Dilution Linear}. The ``if'' direction is immediate: if $C$ is a \nameref{defi:TI} cost, then it is \nameref{axiom:slp} and therefore $C = \Phi(C)$ is \nameref{axiom:CMC}. For the ``only if'' direction, suppose that $\Phi(C)$ is \nameref{axiom:CMC}. We claim that $C$ is a \nameref{defi:TI} cost and $C = \Phi(C)$.

    To this end, note that  $\Phi(C)$ has rich domain because $C$ has rich domain. Thus, since $\Phi(C)$ is \nameref{axiom:slp} (\cref{prop:1}), \cref{thm:trilemma}(i) implies that $\Phi(C)$ is a \nameref{defi:TI} cost. We denote by $(\gamma_{\theta,\theta'})_{\theta,\theta'\in\Theta} \in \R^{|\Theta|\times|\Theta|}_+$ its coefficients, and by $H_\text{TI} \in \mathbf{C}^2(\Delta^\circ(\Theta))$ the function from \cref{defi:TI} for which $\Phi(C) = C^{H_\text{TI}}_\text{ups}$. By direct calculation, for all $p \in \Delta^\circ(\Theta)$ we have:
    \begin{align}
    [\H H_\text{TI}(p)]_{\theta,\theta'} = - \frac{1}{p(\theta) p(\theta')}\cdot \left( p(\theta) \gamma_{\theta,\theta'} + p(\theta') \gamma_{\theta',\theta} \right)  \quad  \forall\, \theta \neq \theta' \label{ti-hess}
    \end{align}
    and $\H H_\text{TI}(p) p = \mathbf{0}$. By \cref{lem:ups-kernel-equiv}, $\Phi(C)$ is \nameref{defi:lq} with $k_{\Phi(C)} = \H H_\text{TI}$.
    
    It remains to show that $C = \Phi(C)$. Since $C$ satisfies the hypotheses of case (a) of \cref{lem:bayes-LLR}, there exists $\bm{\beta} : \Delta^\circ (\Theta) \to \mathbb{R}_+^{|\Theta| \times |\Theta|}$ such that $C$ has the representations in \eqref{F-beta}--\eqref{eqn:LLR-KLform}; moreover, the $\beta_{\theta,\theta'} : \Delta^{\circ}(\Theta) \to \R_+$ are unique for all $\theta\neq\theta'$. Since $C \succeq \Phi(C)$ by definition, it follows from \eqref{eqn:LLR-KLform} and \cref{defi:TI} that, for all $\sigma \in \Se_b$ and $p \in \Delta^\circ(\Theta)$,
    \[
    C(h_B(\sigma,p)) - \Phi(C)(h_B(\sigma,p)) = \sum_{\theta,\theta'\in \Theta} \left( \beta_{\theta,\theta'}(p) - p(\theta) \gamma_{\theta,\theta'} \right) D_\text{KL}(\sigma_\theta \mid \sigma_{\theta'}) \geq 0.
    \]
    By the same argument as in the proof of \cref{thm:trilemma}(i) (see \cref{sssec:proof:thm5}), we obtain:
    \begin{align}
        \beta_{\theta,\theta'}(p) \geq p(\theta) \gamma_{\theta,\theta'} \qquad \forall\, p \in \Delta^\circ(\Theta) \, \text{ and } \, \theta \neq \theta'. \label{flies-ti}
    \end{align} 
    Therefore, to show that $C = \Phi(C)$, it suffices to establish that the inequalities in \eqref{flies-ti} all hold as equalities. We do this in two steps:
    
    \emph{Step 1: Calculate the kernel $k_C$.} By inspection, the divergence $D_{\bm{\beta}}$ in \eqref{F-beta}--\eqref{D-beta} satisfies $D_{\bm{\beta}}(\cdot \mid p) \in \mathbf{C}^2(\Delta^\circ(\Theta))$ for all $p \in \Delta^\circ(\Theta)$. By direct calculation, for all $p \in \Delta^\circ(\Theta)$ we have:
    \begin{align}
    \hspace{-1em}
      [\H_1 D_{\bm{\beta}}(p \mid p)]_{\theta,\theta'} = - \frac{1}{p(\theta) p(\theta')} \cdot \left(\beta_{\theta,\theta'}(p) + \beta_{\theta' , \theta}(p)\right) \quad  \forall\, \theta \neq \theta' \label{llr-hess}
    \end{align}
    and $\H_1 D(p \mid p) p = \mathbf{0}$. We assert that $k_C(p) = \H_1 D(p \mid p) $ for all $p \in \Delta^\circ(\Theta)$. To this end, let $p \in \Delta^\circ(\Theta)$ be given. Fix any $y \in \mathcal{T}$ and $\epsilon>0$. Since $D(\cdot \mid p) \in \mathbf{C}^2(\Delta^\circ(\Theta))$ and $p \in \arg\min_{q \in \Delta^\circ(\Theta)} D_{\bm{\beta}} (q \mid p)$, there exists $\delta>0$ such that
    \[
    \big| D(p \pm t y\mid p) - \frac{1}{2} \cdot t^2 \cdot y^\top \H_1D(p\mid p)y \big|^2  \leq \epsilon \cdot t^2 \cdot \|y\|^2 \qquad \forall \, t \in \left[0, \delta/\|y\|\right).
    \]
    Meanwhile, since $k_C$ is the kernel of $C$, by \cref{defi:lq} there exists $\delta'>0$ such that
    \[
    \hspace{-1em}
    \big| \E_{\pi_t}\left[D(q \mid p)\right] - \frac{1}{2} \cdot t^2 \cdot y^\top k_C(p) y \big|^2  \leq \epsilon \cdot t^2 \cdot \|y\|^2 \qquad \forall \, \pi_t := \frac{1}{2}\delta_{p+ ty} + \frac{1}{2}\delta_{p-ty} \text{ with } t \in \left[0,  \delta' / \|y\| \right).
    \]
    Combining these two inequalities via the triangle inequality and simplifying, we obtain
    \[
    \big| y^\top \left( \H_1 D(p\mid p) - k_C(p) \right) y \big| \leq 4 \epsilon \, \|y\|^2.
    \]
    Since the fixed $y \in \mathcal{T}$ and $\epsilon>0$ were arbitrary, we conclude that $y^\top \H_1 D(p\mid p) y = y^\top k_C(p) y$ for all $y \in \mathcal{T}$. Since $\H_1 D(p \mid p) p = k_C(p) p = \mathbf{0}$, it follows that $x^\top \H_1 D(p \mid p)x = x^\top k_C(p) x$ for all $x \in \R^\Theta$. Hence, being symmetric matrices, $\H_1 D(p \mid p) = k_C(p)$. 

    \emph{Step 2: Kernel Invariance.} Since $C$ is \nameref{defi:sp}, we have $k_C \gg_\text{psd}\mathbf{0}$ on $\Delta^\circ(\Theta)$ (\cref{cor:ker-SP}). Thus, \cref{thm:qk}(ii) yields $k_C = k_{\Phi(C)}$. Then \eqref{ti-hess}, \eqref{llr-hess}, and Step 1 yield:
    \[
     \beta_{\theta,\theta'}(p) +   \beta_{\theta',\theta}(p) = p(\theta) \gamma_{\theta,\theta'} + p(\theta')  \gamma_{\theta',\theta} \quad \forall \, p \in \Delta^\circ(\theta) \, \text{ and } \, \theta \neq \theta'. 
    \]
    Plugging in \eqref{flies-ti} then yields $\beta_{\theta,\theta'}(p) = p(\theta) \gamma_{\theta,\theta'}$ for all $p \in \Delta^\circ(\Theta)$ and $\theta \neq \theta'$, as desired. 
\end{proof}

\subsection{Proof of Corollary \ref{cor:PI-part}}\label{ssec:thm5-cor-proofs-2}

\begin{proof}%[Proof of \cref{cor:PI-part}]
Let $\overline{P} := \{\{\theta\}\}_{\theta\in \Theta} \in \mathcal{P}$ be the fully revealing partition. If $|\Theta|=2$, the result is trivial since $\mathcal{P} = \{P_\varnothing, \overline{P}\}$. So let $|\Theta|>2$. It suffices to show that, for all $P \in \mathcal{P}\backslash\{P_\varnothing, \overline{P}\}$, 
\begin{align}
C\big(h_B(\sigma^{P},p)\big) = C\big(h_B(\sigma^{\overline{P}},p)\big) \quad \forall \, p \in \Delta^\circ(\Theta). \label{eqn:part-cont}
\end{align}

To this end, let $P = \{E_1, \dots, E_k\}\in \mathcal{P}\backslash\{P_\varnothing, \overline{P}\}$ be given. By definition, $2 \leq k< |\Theta|$ and there exists $\ell \in \{1,\dots,k\}$ with $|E_\ell|\geq 2$. Define $P' := \{E_i\}_{1\leq i\leq k, i \neq \ell} \cup \{\{\theta\}\}_{\theta \in E_\ell} \in \mathcal{P}$ (i.e., $P'$ refines $P$ by revealing the state within $E_\ell$). For each $p \in \Delta^\circ(\Theta)$, define $\Pi_p \in \Delta^\dag(\Ex)$ as
\[
\Pi_p (\{\delta_{p(\cdot \mid E_i)}\}) := p(E_i) \ \ \forall\, i \in \{1,\dots,k\}\backslash\{\ell\} \quad \text{and} \quad \Pi_p\Big(\Big\{ \sum_{\theta \in E_\ell} p(\theta \mid E_\ell) \delta_{\delta_\theta}\Big\}\Big) := p(E_\ell).
%:=\,  \sum_{i \in \{1,\dots,k\}\backslash\{\ell\}} p(E_i) \delta_{\delta_{p(\cdot\mid E_i)}} + p(E_\ell) \delta_{\sum_{\theta \in E_\ell} p(\theta \mid E_\ell) \delta_{\delta_\theta}}.
\]
By construction, the two-step strategy $\Pi_p$ induces $\pi_1 = h_B(\sigma^P,p)$ and $\E_{\Pi_p}[\pi_2] = h_B(\sigma^{P'} ,p)$. Moreover, note that $h_B(\sigma^{\overline{P}}, p(\cdot \mid E_\ell)) = \sum_{\theta \in E_\ell} p(\theta \mid E_\ell) \delta_{\delta_\theta}$, and therefore $\left\{h_B(\sigma^{\overline{P}}, p(\cdot \mid E_\ell))\right\} = \supp(\Pi_p)\backslash\Ex^\varnothing$. Since $C \in \C$ is \hyperref[axiom:POSL]{Subadditive} (\cref{prop:1}), it follows that
\begin{align*}
    C\big(h_B(\sigma^{P'},p)\big) \leq C\big(h_B(\sigma^P ,p)\big) + p(E_\ell) \cdot C\big(h_B(\sigma^{\overline{P}},p(\cdot\mid E_\ell))\big) \quad \forall \, p \in \Delta^\circ(\Theta).
\end{align*}
Since $C$ is \hyperref[axiom:prior:invariant]{Prior Invariant} and $\supp\left( p(\cdot \mid E_\ell)\right) = E_\ell$ for all $p \in \Delta^\circ(\Theta)$, there exist $x_0, x_1, x_2 \in \overline{\R}_+$ such that $x_0 = C\big(h_B(\sigma^{P'},p)\big)$, $x_1 = C\big(h_B(\sigma^P ,p)\big)$, and $x_2 = C\big(h_B(\sigma^{\overline{P}},p(\cdot\mid E_\ell)\big)$ for all $p \in \Delta^\circ(\Theta)$. Since $C$ has full domain, we have $x_0, x_1,x_2 < +\infty$. Therefore, for every $p \in \Delta^\circ(\Theta)$, the above display implies that $x_0 \leq  \inf_{p \in \Delta^\circ(\Theta)}\left(x_1 + p(E_\ell) \cdot x_2\right)  = x_1$. Meanwhile, since $C$ is \hyperref[axiom:mono]{Monotone} (\cref{prop:1}) and $h_B(\sigma^{P'},p)\geq_\text{mps} h_B(\sigma^{P},p)$ for all $p \in \Delta^\circ(\Theta)$, we also have $x_0 \geq x_1$. We conclude that $C\big(h_B(\sigma^{P'},p)\big) = x_0 = x_1 = C\big(h_B(\sigma^{P},p)\big)$ for every $p \in \Delta^\circ(\Theta)$.

Now, if $P' = \overline{P}$, we immediately obtain \eqref{eqn:part-cont}. Meanwhile, if $P' \neq \overline{P}$, then there exists $m \in \{1,\dots, k\}\backslash\{\ell\}$ such that $|E_m| \geq 2$. We can then mimic the preceding argument with $P'$ taking the place of $P$ and $P'' \in \mathcal{P}$ taking the place of $P'$, where $P'':= \{E_i\}_{1\leq i\leq k, i \notin\{ \ell, m\}} \cup \{\{\theta\}\}_{\theta \in E_\ell \cup E_m}$ (i.e., $P''$ refines $P'$ by revealing the state within $E_m$). This argument then yields $C\big(h_B(\sigma^{P''},p)\big) = C\big(h_B(\sigma^{P'},p)\big) = C\big(h_B(\sigma^{P},p)\big)$ for all $p \in \Delta^\circ(\Theta)$. If $P'' = \overline{P}$, then we obtain \eqref{eqn:part-cont}. If $P'' \neq \overline{P}$, then we can further refine some cell of $P''$ and repeat the same argument; proceeding iteratively in this way, we eventually obtain \eqref{eqn:part-cont} since $|\Theta|<+ \infty$.

Since $P \in \mathcal{P}\backslash\{P_\varnothing, \overline{P}\}$ was arbitrary, we conclude that \eqref{eqn:part-cont} holds for all $P \in \mathcal{P}\backslash\{P_\varnothing, \overline{P}\}$. 
\end{proof}

\subsection{Proof of Proposition \ref{prop:commute}}\label{ssec:proof-commute}

%\begin{proof}[Proof of \cref{prop:commute}]
\begin{proof}
    Note that point (ii) follows directly from point (i) and the fact that $\Lambda \circ \Upsilon : \C \to \C$ is the identity map. Similarly, point (iv) follows directly from point (iii) and the fact that, for every $\Ec \in \EC$, $\Ec(\cdot,p) \equiv [\Upsilon\circ \Lambda](\Ec)(\cdot,p)$ for all $p \in \Delta^\circ(\Theta)$. We now prove points (i) and (iii).

    To begin, note that for any $\Sigma \in \Se^2$ and $p \in \Delta(\Theta)$, observing $s_1$ induces the random (interim) posterior $q^{\sigma_1,p}_{s_1} \sim h_B(\sigma_1,p) \in \Ex$, and observing $(s_1,s_2)$ induces the random (terminal) posterior $q^{\Sigma,p}_{(s_1,s_2)}\sim h_B(\Sigma,p) \in \Ex$. The joint distribution of these posteriors induces a two-step (belief-based) strategy, which we denote by  $h_B^2(\Sigma,p) \in \Delta^\dag(\Ex)$. By standard arguments, the implied \emph{two-step Bayesian map} $h^2_B : \Se^2 \times \Delta(\Theta) \to \Delta^\dag(\Ex)$ is well-defined and surjective. 

    \noindent \textbf{Point (i).} We proceed in two steps:

    \emph{Step 1: We assert that $\Upsilon \circ \Psi = \Psi_{\Se} \circ \Upsilon$.} To this end, let $C \in \C$, $\sigma \in \Se$, and $p \in \Delta(\Theta)$ be given. For every $\Sigma \in \Se^2$ with $\Sigma \geq_\text{B} \sigma$, letting $\Pi:=h_B^2(\Sigma,p) \in \Delta^\dag (\Ex)$, we have
    \begin{align*}
    [\Upsilon\circ C](\sigma_1,p) + \E_{\langle \Sigma, p \rangle}\left[ [\Upsilon\circ C](\sigma_2^{s_1}, q^{\sigma_1,p}_{s_1})\right]  & = C\left( h_B(\sigma_1, p) \right) + \E_{\langle \Sigma, p \rangle}\left[ C\left( h_B(\sigma_2^{s_1}, q^{\sigma_1,p}_{s_1} \right) \right]\\ 
    %%%
    & = C(\pi_1) + \E_\Pi\left[C(\pi_2) \right]
    \end{align*}
    by the definitions of $\Upsilon$ and $\Pi$, respectively. Letting $\Se(\sigma) := \{\Sigma \in \Se^2 \mid \Sigma \geq_\text{B} \sigma\}$, we obtain
    \begin{align*}
        [\Psi_{\Se} \circ \Upsilon](C)(\sigma,p) \, & = \, \inf_{\Sigma \in \Se^2 (\sigma)} 
    \ [\Upsilon\circ C](\sigma_1,p) + \E_{\langle \Sigma, p \rangle}\left[ [\Upsilon\circ C]\left(\sigma_2^{s_1}, q^{\sigma_1,p}_{s_1}\right)\right] \\
        %%%
        & = \, \inf_{\Pi \in h_B^2[\Se^2(\sigma) \times \{p\}]} \ C(\pi_1) + \E_\Pi\left[C(\pi_2) \right] \\
        %%%%
        & = \, \inf_{\Pi \in \Delta^\dag(\Ex)}  \ C(\pi_1) + \E_\Pi\left[C(\pi_2) \right] \quad \text{s.t.} \quad \E_\Pi[\pi_2]\geq_\text{mps} h_B(\sigma,p) \\
        %%%
        & = \, [\Upsilon \circ \Psi](C)(\sigma,p), 
    \end{align*}
    where the first line is by definition of $\Psi_{\Se}$, the second line is by the preceding display, the third line holds because (by standard arguments) $\Pi\in \Delta^\dag(\Ex)$ satisfies $\E_\Pi[\pi_2] \geq_\text{mps} h_B(\sigma,p)$ if and only if there exists $\Sigma \in \Se^2(\sigma)$ such that $\Pi = h^2_B(\Sigma,p)$, and the final line is by definition of $\Upsilon\circ\Psi$. Since $\sigma \in \Se$ and $p \in \Delta(\Theta)$ were arbitrary, this establishes Step 1. 

    \emph{Step 2: We assert that $\Upsilon \circ \Phi = \Phi_{\Se} \circ \Upsilon$.} To this end, let $C \in \C$ be given. 

    First, we claim that $[\Upsilon \circ \Psi^n](C) = [\Psi^n_{\Se}\circ \Upsilon](C)$ for all $n \in \mathbb{N}$. We proceed by induction. Step 1 establishes the base ($n=1$) step. For the inductive step, let $n \geq 2$ be given and suppose that $[\Upsilon \circ \Psi^{n-1}](C) = [\Psi^{n-1}_{\Se}\circ \Upsilon](C)$. Define $\widehat{C} \in \C$ as $\widehat{C} := \Psi^{n-1}(C)$. Then, we have
    \[
    [\Upsilon \circ \Psi^n](C) \, = \, [\Upsilon \circ \Psi](\widehat{C}) \, = \, [\Psi_{\Se} \circ \Upsilon](\widehat{C}) \, = \, [\Psi_{\Se} \circ \Upsilon \circ \Psi^{n-1}](C) \, = \, [\Psi_{\Se}^n \circ \Upsilon](C),
    \]
    where the first equality is by definition of $\widehat{C}$, the second equality is by Step 1, the third equality is again by definition of $\widehat{C}$, and the final equality is by the inductive hypothesis. This completes the induction and thus proves the claim. 

    Next, let $\sigma \in \Se$ and $p \in \Delta(\Theta)$ be given. Observe that 
    \begin{align*}
    \hspace{-2em}
    [\Phi_{\Se} \circ \Upsilon](C)(\sigma,p) \, &= \, \lim_{n \to \infty} [\Psi^n_{\Se} \circ \Upsilon](C) (\sigma, p) \\
    %%%
    & = \, \lim_{n \to\infty} [\Upsilon \circ \Psi^n](C)(\sigma,p) \, = \, \lim_{n \to \infty} \Psi^n(C)\left( h_B(\sigma,p)\right) \, = \, \Phi(C)\left( h_B(\sigma,p)\right) \, = \, [\Upsilon\circ \Phi](C)(\sigma,p),
    \end{align*}
    where the first equality is by definition of $\Phi_{\Se}$, the second equality is by the above claim, the remaining equalities are by the definitions of $\Upsilon$ and $\Phi$. Since $\sigma \in \Se$ and $p \in \Delta(\Theta)$ were arbitrary, we conclude that $[\Phi_{\Se} \circ \Upsilon](C) = [\Upsilon\circ \Phi](C)$. This completes the proof of Step 2.

    \noindent \textbf{Point (iii).} We proceed in two steps:

    \emph{Step 1: We assert that $\Psi \circ \Lambda = \Lambda \circ \Psi_{\Se}$.} To this end, let $\Ec \in \EC$ be given. Since $\succeq$ is a partial order, it suffices to show that $[\Lambda \circ \Psi_{\Se}](\Ec) \succeq [\Psi\circ \Lambda](\Ec)$ and $[\Lambda \circ \Psi_{\Se}](\Ec) \preceq [\Psi\circ \Lambda](\Ec)$.

    First, we claim that $[\Lambda \circ \Psi_{\Se}](\Ec) \succeq [\Psi\circ \Lambda](\Ec)$. To this end, note that 
    \[
    \Psi_{\Se}(\Ec) \, \succeq_{\Se} \, \Psi_{\Se}([\Upsilon \circ \Lambda](\Ec)) \, = \, [\Psi_{\Se} \circ \Upsilon]( \Lambda(\Ec)) \, = \, [\Upsilon \circ \Psi](\Lambda(\Ec)),
    \]
    where the first inequality holds because $\Ec \succeq_{\Se} [\Upsilon\circ \Lambda](\Ec)$ and $\Psi_{\Se}$ is isotone, and the final equality follows from Step 1 in the above proof of point (i). Since $\Lambda$ is isotone and $\Lambda \circ \Upsilon : \C \to \C$ is the identity map, the claim then follows by applying $\Lambda$ to the above display.

    Next, we claim that $[\Lambda \circ \Psi_{\Se}](\Ec) \preceq [\Psi\circ \Lambda](\Ec)$. To this end, let $\pi \in \Ex$ and $\epsilon>0$ be given. Let $p := p_\pi$. By the definition of $\Psi$, there exists a $\Pi \in \Delta^\dag(\Ex)$ such that $\E_\Pi[\pi_2] \geq_\text{mps}\pi$ and
    \[
    [\Psi\circ \Lambda](\Ec)(\pi) + \epsilon \geq \Lambda(\Ec)(\pi_1) + \E_\Pi\left[ \Lambda(\Ec)(\pi_2) \right].
    \]
    By the definition of $\Lambda$: (a) there exists $\sigma^{\pi_1}\in \Se$ such that $h_B(\sigma^{\pi_1},p) = \pi_1$ and $\Lambda(\Ec)(\pi_1) + \epsilon \geq \Ec(\sigma^{\pi_1},p)$, and (b) for every $\pi_2 \in \supp(\Pi) \backslash\Ex^\varnothing$, there exists $\sigma^{\pi_2} \in \Se$ such that $h_B(\sigma^{\pi_2}, p_{\pi_2}) = \pi_2$ and $\Lambda(\Ec)(\pi_2) + \epsilon \geq \Ec(\sigma^{\pi_2}, p_{\pi_2})$. Therefore, we can then use these experiments to construct a $\Sigma = (\sigma_1, \bm{\sigma}_2) \in \Se^2$ such that $h^2_B(\Sigma,p) = \Pi$ (hence, $h_B(\Sigma,p) = \E_\Pi[\pi_2]$) and 
    %By \awb{[LemmaXX]},\footnote{\awb{[simply sketch the argument in-line here]}} there exists $\Sigma \in \Se^2$ such that $h^2_B(\Sigma,p) = \Pi$ (hence, $h_B(\Sigma,p) = \E_\Pi[\pi_2]$) and
    \[
    \Lambda(\Ec)(\pi_1) + \E_\Pi\left[ \Lambda(\Ec)(\pi_2) \right] + 2 \epsilon \, \geq \, \Ec(\sigma_1,p) + \E_{\langle \Sigma,p\rangle} \left[ \Ec(\sigma_2^{s_1}, p_{s_1}^{\sigma_1,p} )\right].\footnote{Formally, to construct such $\Sigma = (\sigma_1, \bm{\sigma}_2) \in \Se^2$ we must suitably ``stitch together'' the $\sigma^{\pi_1}$ and $\sigma^{\pi_2}$ experiments. This can be done in three steps; we sketch the argument here. First, we may assume (modulo Blackwell equivalence) that $\sigma^{\pi_1}$ and each $\sigma^{\pi_2}$ are defined on the same signal space, $\Omega := \Delta(\Theta)$, where signals are identified with their induced posterior beliefs (given the respective priors, $p_{\pi_1} = p$ and $p_{\pi_2}$). Second, construct $\sigma_1 = (S_1,(\sigma_{1,\theta})_{\theta\in\Theta})\in \Se$ such at $\sigma_1 \sim_B \sigma^{\pi_1}$ and $S_1 := \Omega \times \{1, \dots, N\}$, where: (a) $N \in \mathbb{N}$ is chosen to satisfy $N \geq \sup_{\omega \in \supp(\pi_1)} |\zeta(\omega)|$ for $\zeta(\omega):=\{\pi_2 \in \supp(\Pi)\backslash\Ex^\varnothing \mid p_{\pi_2} = \omega\}$ (which is possible because $\Pi \in \Delta^\dag(\Ex)$); (b) each $s_1 = (\omega,i) \in \supp(\langle\sigma_1,p\rangle)$ determines the posterior belief $q^{\sigma_1,p}_{s_1} = \omega$ and a corresponding second-round random posterior $\pi^\omega_{2,i} \in \zeta(\omega) = \{\pi^\omega_{2,j}\}_{j=1}^{|\zeta(\omega)|}$; and (c) for each of the finitely-many $\omega \in \Omega$ for which $\zeta(\omega)\neq \emptyset$, we have $\sigma_{1,\theta}(\{(\omega,i)\}) = \sigma_{\theta}^{\pi_1}(\{\omega\}) \cdot \Pi(\{\pi^\omega_{2,i}\} \mid \{p_{\pi_2} = \omega\})$ for all $\theta \in \Theta$. (In words, construct $\sigma_1$ to be Blackwell equivalent to $\sigma^{\pi_1}$ and such that it also encodes any additional randomness in the random variable $\pi_2 \sim \Pi$ conditional on the realized value of $p_{\pi_2} \in \supp(\pi_1)$.) Third, let $S_2 := \Omega$ and construct $\bm{\sigma}_2 : S_1 \to \Delta(S_2)$ as follows: (a) for each of the finitely-many $s_1 = (\omega,i) \in S_1$ such that $\zeta(\omega) \neq \emptyset$ and $i \leq |\zeta(\omega)|$, let $\sigma_2^{(\omega,i)} = \sigma^{\pi^\omega_{2,i}}$; and (b) for all other $s_1\in S_1$, let $\sigma_{2}^{s_1} := \underline{\sigma}_2$ for some fixed $\underline{\sigma}_2\in \Delta(S_2)^\Theta \cap \Se^\varnothing$. By construction, the resulting $\Sigma = (\sigma_1,\bm{\sigma}_2)$ is a well-defined element of $\Se^2$, induces the given two-step strategy $h_B^2(\Sigma, p) = \Pi$, and satisfies: (a) $\Ec(\sigma_1,p) = \Ec(\sigma^{\pi_1},p)$, (b) $\Ec(\sigma_2^{(\omega,i)},p_{(\omega,i)}^{\sigma_1,p}) = \Ec(\sigma^{\pi^\omega_{2,i}},p_{\pi^\omega_{2,i}})$ for all $(\omega,i) \in S_1$ such that $\zeta(\omega)\neq \emptyset$ and $i\leq|\zeta(\omega)|$, and (c) $\Ec(\sigma_2^{(\omega,i)},p_{(\omega,i)}^{\sigma_1,p})=0$ otherwise.}
    \]
    Therefore, we obtain
    \begin{align*}
    \hspace{-1em}
    [\Psi\circ \Lambda](\Ec)(\pi) + 3\epsilon & \geq \inf\left\{ \Ec(\hat{\sigma}_1,p) + \E_{\langle \widehat{\Sigma},p\rangle} \left[ \Ec(\hat{\sigma}_2^{\hat{s}_1}, p^{\hat{\sigma}_1, p}_{\hat{s}_1} ) \right]  \ \Big| \ \widehat{\Sigma} \in \Se^2 \ \text{s.t.} \ \widehat{\Sigma} \geq_\text{B} \Sigma \right\} \\
    %%%
    & \geq \inf\left\{ \Ec(\hat{\sigma}_1,p) + \E_{\langle \widehat{\Sigma},p\rangle} \left[ \Ec(\hat{\sigma}_2^{\hat{s}_1}, p^{\hat{\sigma}_1, p}_{\hat{s}_1} ) \right]  \ \Big| \ \sigma \in \Se, \, \widehat{\Sigma} \in \Se^2 \ \text{s.t.} \ \widehat{\Sigma} \geq_\text{B} \sigma ,\, h_B(\sigma,p) = \E_\Pi[\pi_2] \right\} \\
    %%%
    & = [\Lambda \circ \Psi_{\Se}](\Ec)(\E_\Pi[\pi_2]) \\
    %%%
    & \geq [\Lambda \circ \Psi_{\Se}](\Ec)(\pi),
    \end{align*}
   %%%%
   where the first line is by the two preceding displays, the second line is by $h_B(\Sigma,p) = \E_\Pi[\pi_2]$, the third line is by definition of $\Lambda \circ \Psi_{\Se}$, and the final line holds because $\Psi_{\Se}(\Ec)$ is \hyperref[E-mono]{$\Se$-Monotone} (by construction) and therefore $[\Lambda\circ\Psi_{\Se}](\Ec)$ is \hyperref[axiom:mono]{Monotone}.\footnote{In particular, $\Lambda(\widehat{\Ec})$ is \hyperref[axiom:mono]{Monotone} for any $\Se$-Monotone $\widehat{\Ec} \in \EC$. To see this, let $\pi,\pi' \in \Ex$ such that $\pi' \geq_\text{mps} \pi$ be given; define $p \in \Delta(\Theta)$ as $p := p_\pi = p_{\pi'}$. Fix any $\sigma' \in \Se$ such that $h_B(\sigma',p) = \pi'$. By standard arguments, there exists $\sigma \in \Se$ such that: (a) $h_B(\sigma,p)=\pi$ and (b) $\sigma' \geq_\text{B} \sigma$. Since $\widehat{\Ec}$ is $\Se$-Monotone, we have $\widehat{\Ec}(\sigma',p) \geq \widehat{\Ec}(\sigma,p) \geq \Lambda(\widehat{\Ec})(\pi)$. Then, infimizing over all $\sigma' \in \Se$ such that $h_B(\sigma',p) = \pi'$, we obtain $\Lambda(\widehat{\Ec})(\pi') \geq  \Lambda(\widehat{\Ec})(\pi)$. We conclude that $\Lambda(\widehat{\Ec})$ is \hyperref[axiom:mono]{Monotone}, as desired.} Since $\pi \in \Ex$ and $\epsilon>0$ were arbitrary, this proves the claim and thus completes the proof of Step 1.

   \emph{Step 2: We assert that $\Phi \circ \Lambda = \Lambda \circ \Phi_{\Se}$.} To this end, let $\Ec \in \EC$ be given.

   First, we claim that $[\Psi^n \circ \Lambda](\Ec) = [\Lambda \circ \Psi_{\Se}^n](\Ec)$ for all $n \in \mathbb{N}$. We proceed by induction. Step 1 establishes the base ($n=1$) step. For the inductive step, let $n \geq 2$ be given and suppose that $[\Psi^{n-1} \circ \Lambda](\Ec) = [\Lambda \circ \Psi_{\Se}^{n-1}](\Ec)$. Define $\widehat{\Ec} \in \EC$ as  $\widehat{\Ec}:= \Psi^{n-1}_{\Se}(\Ec)$. Then, we have 
   \[
   [\Psi^n \circ \Lambda](\Ec) \, = \, %[\Psi\circ \Psi^{n-1} \circ \Lambda](\Ec)  =  
   [\Psi \circ \Lambda \circ \Psi^{n-1}_{\Se}] (\Ec) \, = \,  [\Psi \circ \Lambda](\widehat{\Ec}) \,  = \, [\Lambda \circ \Psi_{\Se}](\widehat{\Ec}) \, = \, [\Lambda \circ \Psi^n_{\Se}](\Ec),
   \]
    where the first equality is by the inductive hypothesis, the second equality is by definition of $\widehat{\Ec}$, the third equality is by Step 1, and the final equality is again by definition of $\widehat{\Ec}$. This completes the induction and thus proves the claim.

    Next, observe that the definition of $\Phi$ and the above claim imply that
    \[
    [\Phi \circ \Lambda](\Ec) \, = \, \lim_{n\to\infty} [\Psi^n \circ \Lambda](\Ec) \, = \, \lim_{n \to \infty} [\Lambda \circ \Psi^n_{\Se}](\Ec) \, \succeq \, [\Lambda \circ \Phi_{\Se}](\Ec),
    \]
    where the final inequality holds because $\Lambda$ is isotone and $\Psi^n_{\Se}(\Ec)\succeq_{\Se} \Psi^{n+1}_{\Se}(\Ec) \succeq_{\Se} \Phi_{\Se}(\Ec)$ for all $n \in \mathbb{N}$. We claim that, in fact, $\lim_{n \to \infty} [\Lambda \circ \Psi^n_{\Se}](\Ec) = [\Lambda \circ \Phi_{\Se}](\Ec)$. Suppose, towards a contradiction, that there exists $\pi \in \Ex$ and $\epsilon >0$ such that $[\Lambda \circ \Psi^n_{\Se}](\Ec)(\pi)\geq [\Lambda \circ \Phi_{\Se}](\Ec)(\pi) + \epsilon$ for all $n \in \mathbb{N}$. By definition of $\Lambda$, there exists $\sigma \in \Se$ such that: (a) $h_B(\sigma,p_\pi) = \pi$, (b) $[\Lambda \circ \Phi_{\Se}](\Ec)(\pi) + \epsilon/2 \geq \Phi_{\Se}(\Ec)(\sigma,p_\pi)$, and (c) $\Psi^n_{\Se}(\Ec)(\sigma,p_\pi) \geq [\Lambda \circ \Psi^n_{\Se}](\Ec)(\pi)$ for all $n \in \mathbb{N}$. Therefore, it follows that $\Psi^n_{\Se}(\Ec)(\sigma,p_\pi) \geq \Phi_{\Se}(\Ec)(\sigma,p_\pi) + \epsilon/2$ for all $n \in \mathbb{N}$, which contradicts the definition of $\Phi_{\Se}$, as desired. This completes the proof of Step 2. 
\end{proof}

\subsection{Proofs of Theorems \ref{thm1-hat}--\ref{thm:wald:gen}}\label{ssec:proofs-GLM}

%\cref{sssec:GLM-prelim} collects useful preliminary facts. The main proofs are in \cref{sssec:GLM-mainproofs}.

%\subsubsection{Preliminaries}\label{sssec:GLM-prelim}

To prove these results, we require the following technical lemma:  

\begin{lem}\label{lem:cost-domain-restrict}
    For every $C \in \C$ and $W \subseteq \Delta(\Theta)$, it holds that: (i) $C|_W \succeq C$; (ii) if $C' \in \C$ satisfies $C \succeq C'$, then $C'|_W$ satisfies $C|_W \succeq C'|_W$; and (iii) if $C$ is \hyperref[axiom:POSL]{Subadditive} (resp., \hyperref[axiom:mono]{Monotone}) and the set $W$ is convex, then $C|_W$ is also \hyperref[axiom:POSL]{Subadditive} (resp., \hyperref[axiom:mono]{Monotone}).
\end{lem}

%We first use \cref{lem:cost-domain-restrict} to prove \dred{Theorems} \ref{thm1-hat}--\ref{thm:wald:gen}, and then prove \cref{lem:cost-domain-restrict} itself. 

We also require a few standard definitions and facts from convex analysis. For any convex function $H : \Delta(\Theta) \to \R \cup\{+\infty\}$, the \emph{closure of $H$} is the function $\overline{H} : \Delta(\Theta) \to \R \cup\{+\infty\}$ defined as $\overline{H}(p) := \liminf_{q \to p} H(q)$. By construction, $\overline{H}$ is the pointwise largest lower semi-continuous convex function that is majorized by $H$, and it satisfies $\overline{H}(p) = H(p)$ for all $p \in \ri (\dom(H))$ \parencite[Theorem 7.4]{rock70}. Thus, we always have $\dom(\overline{H}) \supseteq \dom(H)$ and $C^H_\text{ups}|_W = C^{\overline{H}}_\text{ups}|_W \succeq C^{\overline{H}}_\text{ups}$ for $W = \ri (\dom(H))$. For the leading special case in which $\dom(H)  \subseteq \Delta^\circ(\Theta)$ is open, and therefore $\dom(H) = \ri(\dom(H))$, it follows that: (i) $\overline{H}(p) = H(p)$ for all $p \in \dom(H)$, and (ii) $C^H_\text{ups} = C^{\overline{H}}_\text{ups}|_W \succeq C^{\overline{H}}_\text{ups}$ for $W = \dom(H)$.

In what follows, we first use \cref{lem:cost-domain-restrict} and the above facts to prove each of \dred{Theorems} \ref{thm1-hat}--\ref{thm:wald:gen} in turn. We then conclude by proving \cref{lem:cost-domain-restrict} itself.

%\subsubsection{Main Proofs}\label{sssec:GLM-mainproofs}

\begin{proof}[Proof of \cref{thm1-hat}]
    We begin with the first statement. The ``$\leftharpoondown$'' direction follows directly from the definition of \hyperref[defi:gen:IC]{$\widehat{\Phi}$-proofness} (i.e., $ C = \widehat{\Phi}(C)$). For the ``$\rightharpoonup$'' direction, suppose that $C \in \widehat{\Phi}[\C]$ and $\widehat{\Phi}$ satisfies \hyperref[EO]{EO}. Then there exists some $C' \in \C$ such that $C = \widehat{\Phi}(C')$ (by definition) and $\widehat{\Phi}(C') = \widehat{\Phi}(\widehat{\Phi}(C')) $ (by \hyperref[EO]{EO}). It follows that $C = \widehat{\Phi}(C)$, i.e., $C$ is \hyperref[defi:gen:IC]{$\widehat{\Phi}$-proof}.

    For the second statement, suppose that $\widehat{\Phi}$ satisfies \hyperref[ADL]{ADL} and \hyperref[EO]{EO}. Let $C \in \C$ be given. \hyperref[ADL]{ADL} implies that $\widehat{\Phi}(C) \preceq C$, and \hyperref[EO]{EO} implies that $\widehat{\Phi}(C)$ is \hyperref[defi:gen:IC]{$\widehat{\Phi}$-proof} (via the first statement). Moreover, since $\widehat{\Phi}$ is isotone, for any \hyperref[defi:gen:IC]{$\widehat{\Phi}$-proof} $C' \preceq C$ it holds that $C' = \widehat{\Phi}(C') \preceq \widehat{\Phi}(C)$.
\end{proof}

\begin{proof}[Proof of \cref{thm2-hat}]
    For the ``$\rightharpoonup$'' direction, suppose that $\widehat{\Phi}$ satisfies \hyperref[GS]{GS}. Let $C \in \widehat{\Phi}[\C]|_W$ be \nameref{defi:ll}. Since $W \subseteq \Delta^\circ(\Theta)$ is convex, \hyperref[GS]{GS} and \cref{lem:cost-domain-restrict}(iii) imply that $C$ is \hyperref[axiom:POSL]{Subadditive}. Since $W$ is also open, the proof of the  ``$\implies$'' direction of \cref{thm:UPS} (see \cref{ssec:app:thm:UPS}) then applies verbatim and delivers the desired conclusion. 
    
    For the ``$\leftharpoondown$'' direction, suppose that $\widehat{\Phi}$ satisfies \hyperref[ADL]{ADL} and \hyperref[DUI]{DUI}. Let $C = C^H_\text{ups}$ for some $H \in \mathbf{C}^1(W)$. We have $\widehat{\Phi}(C^H_\text{ups}) \succeq \widehat{\Phi}(C^{\overline{H}}_\text{ups}) \succeq C^{\overline{H}}_\text{ups}$ because (i) $C^H_\text{ups} \succeq C^{\overline{H}}_\text{ups}$ (as $W = \dom(H) \subseteq\Delta^\circ(\Theta)$ is open) and $\widehat{\Phi}$ is isotone and (ii) $\overline{H}$ is lower semi-continuous and $\widehat{\Phi}$ satisfies \hyperref[DUI]{DUI}. Since $C^{\overline{H}}_\text{ups}|_W = C^{H}_\text{ups}$ (as $W = \dom(H) \subseteq\Delta^\circ(\Theta)$ is open), \cref{lem:cost-domain-restrict}(ii) then implies that $\widehat{\Phi}(C^H_\text{ups})|_W \succeq C^{H}_\text{ups}$. Meanwhile, \hyperref[ADL]{ADL} implies that $ C^H_\text{ups}\succeq \widehat{\Phi}(C^H_\text{ups})$. Since $C^{H}_\text{ups}|_W = C^{H}_\text{ups}$ (as $\dom(H) = W$), \cref{lem:cost-domain-restrict}(ii) then implies that $C^H_\text{ups} \succeq \widehat{\Phi}(C^H_\text{ups})|_W$. We conclude that $C^H_\text{ups} = \widehat{\Phi}(C^H_\text{ups})|_W$, and therefore that $C^H_\text{ups} \in \widehat{\Phi}[\C]|_W$. Finally, the fact that $C^H_\text{ups}$ is \nameref{defi:ll} follows directly from \cref{defi:ll}, as in the proof of the ``$\impliedby$'' direction of \cref{thm:UPS}.
\end{proof}

\begin{proof}[Proof of \cref{thm:qk:gen}]
    Suppose that $\widehat{\Phi}$ satisfies \hyperref[DUI]{DUI}. \cref{lem:lqk-invariance-lem} applies verbatim, as its statement and proof rely only the definitions of \nameref{defi:ups} costs and lower kernels (\cref{defi:ups,defi:lq}). The main proof of \cref{thm:qk}(ii) (see \cref{app:thm3-2:proof}) then applies verbatim (with $\widehat{\Phi}$ used in place of $\Phi$) and delivers the desired conclusion after one minor adjustment. Specifically, given any convex $H \in \mathbf{C}^2(\Delta(\Theta))$ such that $C \succeq C^H_\text{ups}$, we now show that $\widehat{\Phi}(C) \succeq C^H_\text{ups}$ as follows: we observe that $\widehat{\Phi}(C) \succeq \widehat{\Phi}(C^H_\text{ups}) \succeq C^H_\text{ups}$ because (i) $\widehat{\Phi}$ is isotone and (ii) $H$ is lower semi-continuous (as $H \in \mathbf{C}^2(\Delta(\Theta))$) and $\widehat{\Phi}$ satisfies \hyperref[DUI]{DUI}.
\end{proof}

\begin{proof}[Proof of \cref{thm:flie:gen}]
    For the ``$\rightharpoonup$'' direction, suppose that $\widehat{\Phi}$ satisfies \hyperref[AIE]{AIE} and \hyperref[DUI]{DUI}. Since $C$ \nameref{axiom:flie} and $\widehat{\Phi}$ is isotone, we have $\widehat{\Phi}(C) \succeq \widehat{\Phi}(\Phi_\text{IE}(C))$. Since $\dom(C) \subseteq \Delta(W) \cup \Ex^\varnothing$ and $k_C = \H H$ on $W$, points (i) and (ii) of \cref{lem:phi-ie} imply that $\Phi_\text{IE}(C) = C^H_\text{ups}$. Thus, $\widehat{\Phi}(C) \succeq \widehat{\Phi}(C^H_\text{ups}) \succeq \widehat{\Phi}(C^{\overline{H}}_\text{ups}) \succeq C^{\overline{H}}_\text{ups}$, where the latter two inequalities hold because (i) $C^H_\text{ups} \succeq C^{\overline{H}}_\text{ups}$ (as $W = \dom(H) \subseteq\Delta^\circ(\Theta)$ is open) and $\widehat{\Phi}$ is isotone and (ii) $\overline{H}$ is lower semi-continuous and $\widehat{\Phi}$ satisfies \hyperref[DUI]{DUI}. Since $C^{\overline{H}}_\text{ups}|_W = C^{H}_\text{ups}$ (as $W = \dom(H) \subseteq\Delta^\circ(\Theta)$ is open), \cref{lem:cost-domain-restrict}(ii) then implies that $\widehat{\Phi}(C)|_W \succeq C^{H}_\text{ups}$. Meanwhile, we have $\widehat{\Phi}(C) \preceq C^H_\text{ups}$ because $k_C = \H H$ on $W$, $\widehat{\Phi}$ satisfies \hyperref[AIE]{AIE}, and $\dom(C^H_\text{ups}) = \Delta(W) \cup \Ex^\varnothing$. Since $C^{H}_\text{ups}|_W = C^{H}_\text{ups}$ (as $W = \dom(H)$), \cref{lem:cost-domain-restrict}(ii) then implies that $\widehat{\Phi}(C)|_W \preceq C^H_\text{ups}$. We conclude that $\widehat{\Phi}(C)|_W = C^H_\text{ups}$.

    For the ``$\leftharpoondown$'' direction, suppose that $\widehat{\Phi}$ satisfies \hyperref[ADL]{ADL} and \hyperref[DUI]{DUI}. Let $\widehat{\Phi}(C)|_W = C^H_\text{ups}$. The proof of the ``$\impliedby$'' direction of \cref{thm:flie} (see \cref{proof:flie}) applies verbatim (with $\widehat{\Phi}$ used in place of $\Phi$) and delivers the desired conclusion after two minor adjustments. First, we now use \hyperref[ADL]{ADL} to obtain $C\succeq \widehat{\Phi}(C)$. Since the assumption that $\dom(C) \subseteq \Delta(W)\cup\Ex^\varnothing$ implies that $C = C|_W$, \cref{lem:cost-domain-restrict}(ii) then implies that $C  \succeq \widehat{\Phi}(C)|_W = C^H_\text{ups}$. It follows that $C$ is \nameref{defi:sp} (as $H$ is strongly convex) and that $k_C$ is an upper kernel of $\widehat{\Phi}(C)|_W = C^H_\text{ups}$ on $W$. Second, since $\widehat{\Phi}$ satisfies \hyperref[DUI]{DUI}, we now use \cref{thm:qk:gen} to show that $k_C$ is a lower kernel of $\widehat{\Phi}(C)$ on $W$, which implies that $k_C$ is also a lower kernel of $\widehat{\Phi}(C)|_W = C^H_\text{ups}$ on $W$ (as $\widehat{\Phi}(C)|_W \succeq \widehat{\Phi}(C)$ by \cref{lem:cost-domain-restrict}(i)). We conclude that $k_C = k_{\widehat{\Phi}(C)|_W} = \H H$ on $W$, as desired.\footnote{\cref{remark:local-ADL} describes a variant of the ``$\leftharpoondown$'' direction of \cref{thm:flie:gen} that applies when $\widehat{\Phi}$ satisfies \nameref{remark:local-ADL}, rather than \hyperref[ADL]{ADL}. To obtain this result, we further modify the above adjustments as follows. Suppose that $\widehat{\Phi}$ satisfies \nameref{remark:local-ADL} and \hyperref[DUI]{DUI}. Let $C$ be \nameref{defi:sp} and satisfy $\widehat{\Phi}(C)|_W = C^H_\text{ups}$. Using \cref{thm:qk:gen} exactly as above, we find that $k_C$ is a lower kernel of $\widehat{\Phi}(C)|_W$. To show that $k_C$ is also an upper kernel of $\widehat{\Phi}(C)|_W$, we now proceed in two steps. First, \nameref{remark:local-ADL} implies that $k_C$ is an upper kernel of $\widehat{\Phi}(C)$. Second, since $W\subseteq\Delta^\circ(\Theta)$ is open,  it then follows that $k_C$ is also an upper kernel of $\widehat{\Phi}(C)|_W$ (since, for every $p \in W$, the $\delta>0$ in \cref{defi:lq}(i) for $\widehat{\Phi}(C)$ can be chosen such that $B_\delta(p) \subseteq W$). We conclude that $k_C = k_{\widehat{\Phi}(C)|_W} = \H H$, as desired. (The direct cost $C$ need not \hyperref[axiom:flie]{FLIE} when \hyperref[ADL]{ADL} is relaxed to \nameref{remark:local-ADL}.)} \cref{lem:phi-ie}(iii) then implies that $\Phi_\text{IE}(C) = C^H_\text{ups}$, and therefore $C \succeq \Phi_\text{IE}(C)$ (since, as noted above, \hyperref[ADL]{ADL} implies $C \succeq C^H_\text{ups}$). We conclude that $C$ \nameref{axiom:flie}, as desired.
\end{proof}

\begin{proof}[Proof of \cref{thm:trilemma1:gen}]
    For the ``$\rightharpoonup$'' direction, suppose that $\widehat{\Phi}$ satisfies \hyperref[GS]{GS}. \cref{lem:cost-domain-restrict}(iii) then implies that $C \in \widehat{\Phi}[\C]|_{\Delta^\circ(\Theta)}$ is \hyperref[axiom:POSL]{Subadditive}. The proof of the ``only if'' direction of \cref{thm:trilemma}(i) (see \cref{sssec:proof-tri-pt1}) then applies verbatim, as it only requires $C$ to be \hyperref[axiom:POSL]{Subadditive} and \nameref{axiom:CMC} and have rich domain (the latter two properties hold by assumption). 
    
    For the ``$\leftharpoondown$'' direction, suppose that $\widehat{\Phi}$ satisfies \hyperref[ADL]{ADL} and \hyperref[DUI]{DUI}. Let any \nameref{defi:TI} cost $C_\text{TI} = C^{H_\text{TI}}_\text{ups}$ be given. It is \nameref{axiom:CMC} by construction. Thus, it suffices to show that $C_\text{TI} = \widehat{\Phi}(C_\text{TI})|_{\Delta^\circ(\Theta)}$, which then implies that $C_\text{TI} \in \widehat{\Phi}[\C]|_{\Delta^\circ(\Theta)}$. To this end, note that $C_\text{TI} \succeq \widehat{\Phi}(C_\text{TI}) \succeq  \widehat{\Phi}(C^{\overline{H}_\text{TI}}_\text{ups}) \succeq C^{\overline{H}_\text{TI}}_\text{ups}$ because (i) $\widehat{\Phi}$ satisfies \hyperref[ADL]{ADL}, (ii) $C_\text{TI} \succeq C^{\overline{H}_\text{TI}}_\text{ups}$ (as $\dom(H_\text{TI}) = \Delta^\circ(\Theta)$ is open) and $\widehat{\Phi}$ is isotone, and (iii) $\overline{H}_\text{TI}$ is lower semi-continuous and $\widehat{\Phi}$ satisfies \hyperref[DUI]{DUI}. Since $C^{\overline{H}_\text{TI}}_\text{ups}|_{\Delta^\circ(\Theta)} = C_\text{TI}|_{\Delta^\circ(\Theta)} = C_\text{TI}$ (as $\dom(H_\text{TI}) = \Delta^\circ(\Theta)$), \cref{lem:cost-domain-restrict}(ii) then implies that $C_\text{TI} \succeq \widehat{\Phi}(C_\text{TI})|_{\Delta^\circ(\Theta)} \succeq C_\text{TI}$. We conclude that $C_\text{TI} = \widehat{\Phi}(C_\text{TI})|_{\Delta^\circ(\Theta)}$, as desired.
\end{proof}

\begin{proof}[Proof of \cref{thm:trilemma3:gen}]
    Suppose that $\widehat{\Phi}$ satisfies \hyperref[GS]{GS}. \cref{lem:cost-domain-restrict}(iii) then implies that $C \in \widehat{\Phi}[\C]|_{\Delta^\circ(\Theta)}$ is \hyperref[axiom:POSL]{Subadditive}. Since $C$ is assumed \hyperref[axiom:mono]{Monotone}, it follows that $C$ is \nameref{axiom:slp} (\cref{prop:1}). The proof of the ``converse'' direction of \cref{thm:trilemma}(iii) (see \cref{sssec:proof-tri-pt3}) then applies verbatim and yields the desired conclusion: \cref{lem:pairwise-trivial} applies since $C$ is \nameref{axiom:slp} and assumed to have rich domain, and Case 1 applies since $C$ is assumed \hyperref[axiom:prior:invariant]{Prior Invariant}.
\end{proof}

\begin{proof}[Proof of \cref{thm:wald:gen}]
    For the ``$\rightharpoonup$'' direction, suppose that $\widehat{\Phi}$ satisfies \hyperref[ADL]{ADL} and \hyperref[DUI]{DUI}. The proof of the ``$\implies$'' direction of \cref{thm:wald} (see Steps 2--3 and item (2) of Step 4 in \cref{proof:wald}) applies verbatim (with $\widehat{\Phi}$ used in place of $\Phi$) and delivers the desired conclusion after two minor adjustments. Specifically, it suffices to adjust Step 2 in the proof of \cref{lem:pi-w-implies-lpi-w} (see \cref{ssec:proof-pi-w-implies-lpi-w}) in two ways.\footnote{All other results used to prove the ``$\implies$'' direction of \cref{thm:wald}---viz., \cref{lem:lpi-kernel}, Step 1 in the proof of \cref{lem:lpi-kernel},  \cref{lem:ups-lpi-wald-local}, and the technical facts in \cref{ssec:thm6-technical-facts}---continue to apply verbatim here.} To this end, fix any \hyperref[axiom:prior:invariant]{Prior Invariant} $C' \in \widehat{\Phi}^{-1}(C)$, and note that Step 1 in the proof of \cref{lem:pi-w-implies-lpi-w} implies that $C'$ is \nameref{defi:lpi} on $\Delta^\circ(\Theta)$. First, we now use \hyperref[ADL]{ADL} to obtain $C' \succeq \widehat{\Phi}(C') = C$. This implies that $\underline{K}^+_{C}(p) \subseteq \underline{K}^+_{C'}(p) $ for all $p \in \Delta^\circ(\Theta)$ and that $C'$ is \nameref{defi:sp} (as $C$ is \nameref{defi:sp}). Second, since $\widehat{\Phi}$ satisfies \hyperref[DUI]{DUI}, we now use \cref{thm:qk:gen} to show that $\underline{K}^+_{C}(p) \supseteq \underline{K}^+_{C'}(p) $ for all $p \in \Delta^\circ(\Theta)$.\footnote{\cref{remark:wald:gen} describes a variant of the ``$\rightharpoonup$'' direction of \cref{thm:wald:gen} that applies when $\dom(C)\supseteq \Delta(\Delta^\circ(\Theta))\cup\Ex^\varnothing$. To obtain this result, we proceed as follows. The above work implies that $C$ is \nameref{defi:lpi} on $\Delta^\circ(\Theta)$. Since $C$ is \nameref{defi:ups} and \nameref{defi:lq}, applying \cref{lem:ups-lpi-wald-local} to $C|_{\Delta^\circ(\Theta)}$ (rather than $C$) then implies $|\Theta|=2$ and $C|_{\Delta^\circ(\Theta)}$ is a \ref{eqn:MS} cost, as desired. 
    
    Moreover, \cref{remark:local-ADL} describes another variant of the ``$\rightharpoonup$'' direction of \cref{thm:wald:gen} that applies when $\widehat{\Phi}$ satisfies \nameref{remark:local-ADL}, rather than \hyperref[ADL]{ADL}. To obtain this result, we further modify the above adjustments as follows. Suppose that $\widehat{\Phi}$ satisfies \nameref{remark:local-ADL} and \hyperref[DUI]{DUI}. Let $C$ have rich domain, be \nameref{defi:ups}, and satisfy $C = \widehat{\Phi}(C')$ for some  \hyperref[axiom:prior:invariant]{Prior Invariant} $C' \in \C$ that is \nameref{defi:lq} and \nameref{defi:sp}. \nameref{remark:local-ADL} implies that $k_{C'}$ is an upper kernel of $C$ on $\Delta^\circ(\Theta)$. Since $\widehat{\Phi}$ satisfies \hyperref[DUI]{DUI} and $C'$ is \nameref{defi:sp}, \cref{cor:ker-SP} and \cref{thm:qk:gen} imply that $k_{C'}$ is a lower kernel of $C$ on $\Delta^\circ(\Theta)$. It follows that $C$ is \nameref{defi:lq} on $\Delta^\circ(\Theta)$ with kernel $k_C = k_{C'}$. Since (Step 1 in the proof of) \cref{lem:pi-w-implies-lpi-w} implies that $C'$ is \nameref{defi:lpi} on $\Delta^\circ(\Theta)$, applying \cref{lem:lpi-kernel,lem:ups-lpi-wald-local} as in the proof of \cref{thm:wald} then delivers the desired result.}

    For the ``$\leftharpoondown$'' direction, suppose that $\widehat{\Phi}$ satisfies \hyperref[AIE]{AIE} and \hyperref[DUI]{DUI}. The statement and proof of \cref{lem:wald-is-spi}(i) (see \cref{ssec:wald-is-spi}) apply verbatim (with $\widehat{\Phi}$ used in place of $\Phi$) and yield the desired conclusion after one minor adjustment. Specifically, we now use the ``$\rightharpoonup$'' direction of \cref{thm:flie:gen} in the final sentence of that proof to obtain $\widehat{\Phi}(\overline{C}_\text{Wald}) = C_\text{Wald}$.
\end{proof}

\begin{proof}[Proof of \cref{lem:cost-domain-restrict}]
    Points (i) and (ii) are trivial. For point (iii), let $W\subseteq \Delta(\Theta)$ be convex. 
    
    To begin, suppose that $C \in C$ is \hyperref[axiom:mono]{Monotone}. Let $\pi,\pi' \in \Ex$ such that $\pi' \geq_\text{mps} \pi$ be given. There are three cases. First, if $\pi' \in \Ex^\varnothing$, then $\pi \in \Ex^\varnothing$ and therefore $C|_W(\pi') = C|_W(\pi) = 0$. Second, if $\pi' \in \Delta(W)\backslash \Ex^\varnothing$, then $\supp(\pi')\subseteq W$ (by definition) and therefore $\supp(\pi) \subseteq \text{conv}(\supp(\pi')) \subseteq W$, where the first inclusion is by $\pi' \geq_\text{mps} \pi$ and the second inclusion holds because $W$ is convex. 
    %\footnote{\awb{[Here, and elsewhere, we implicitly use Phelps' Prop 1.2 characterization of closed convex hull.]}} 
    It follows that $\pi \in \Delta(W)$, and hence that $C|_W(\pi') = C(\pi') \geq C(\pi) = C|_W(\pi)$, where the inequality holds because $C$ is \hyperref[axiom:mono]{Monotone}. Third, if $\pi' \notin \Delta(W)\cup\Ex^\varnothing$, then $  C|_W(\pi') = +\infty \geq C|_W (\pi)$. Overall, we conclude that $C|_W$ is \hyperref[axiom:mono]{Monotone}.

    Next, suppose that $C \in \C$ is \hyperref[axiom:POSL]{Subadditive}. Fix any $\Pi \in \Delta^\dag(\Ex)$. There are two cases:

    \textbf{\emph{Case 1: Let $\E_\Pi[\pi_2] \in  \Delta(W) \cup \Ex^\varnothing$.}} Then we have
    \[
    C|_W (\E_\Pi[\pi_2]) \, = \, C (\E_\Pi[\pi_2]) \, \leq \,  C(\pi_1) + \E_\Pi[C(\pi_2)] \, \leq \, C|_W(\pi_1) + \E_\Pi[C|_W(\pi_2)],
    \]
    where the first equality is by the supposition and definition of $C|_W$, the second inequality holds because $C$ is \hyperref[axiom:POSL]{Subadditive}, and the final inequality is by point (i).

    \textbf{\emph{Case 2: Let $\E_\Pi[\pi_2] \notin  \Delta(W) \cup \Ex^\varnothing$.}} Then $C|_W(\E_\Pi[\pi_2]) = +\infty$, so it suffices to show that 
    \begin{equation}\label{eqn:restrict-subadditive-infinite}
    C|_W(\pi_1) + \E_\Pi[C|_W(\pi_2)] = +\infty.
    \end{equation}
    There are three sub-cases to consider, depending on $\pi_1 \in \Ex$. 
    
    First, suppose that $\pi_1 \notin \Delta(W) \cup \Ex^\varnothing$. Then $C|_W(\pi_1) = +\infty$, which directly implies \eqref{eqn:restrict-subadditive-infinite}.

    Second, suppose that $\pi_1 \in \Delta(W)$. If $\supp(\Pi)\backslash\Ex^\varnothing = \emptyset$, then $\E_\Pi[\pi_2] = \pi_1 \in \Delta(W)$, which contradicts the hypothesis that $\E_\Pi[\pi_2] \notin \Delta(W)\cup\Ex^\varnothing$. Thus, we have $\supp(\Pi)\backslash\Ex^\varnothing \neq \emptyset$. Since $\Pi \in \Delta^\dag(\Ex)$, there exists an $n \in \mathbb{N}$ and an enumeration $\supp(\Pi)\backslash\Ex^\varnothing = \{\pi_2^i\}_{i=1}^n$, where $\Pi(\{\pi^i_2\}) >0$ for all $i \in \{1,\dots,n\}$. We claim that $\pi_2^i \notin \Delta(W)$ for some $i \in \{1,\dots, n\}$. Note that this claim implies that $\E_\Pi[C|_W(\pi_2)] = +\infty$, and hence that \eqref{eqn:restrict-subadditive-infinite} holds, as desired. Suppose, towards a contradiction, that $ \{\pi_2^i\}_{i=1}^n \subseteq \Delta(W)$. Define the Borel measure $\mu_1$ on $\Delta(\Theta)$ as $\mu_1(B) := \Pi(\{\pi_2 \in \Ex \mid p_{\pi_2} \in B\} \cap \Ex^\varnothing)$ for all Borel $B \subseteq \Delta(\Theta)$. By construction, we have $\mu_1(B) \leq \pi_1(B)$ for all Borel $B \subseteq \Delta(\Theta)$, which implies  $\supp(\mu_1) \subseteq \supp(\pi_1)\subseteq W$. Moreover,  
    \[
    \E_\Pi[\pi_2] \, =  \, \sum_{i=1}^n \Pi (\{\pi_2^i\}) \cdot \pi_2^i \, + \, \int_{\Ex^\varnothing} \pi_2 \dd \Pi(\pi_2) \, =  \, \sum_{i=1}^n \Pi (\{\pi_2^i\}) \cdot \pi_2^i \, + \, \mu_1,
    \] 
    where the first equality is by definition and the second equality is by a change of variables. Since $\supp(\mu_1)\cup \big[ \bigcup_{i=1}^n \supp(\pi^i_2)\big] \subseteq W$ by supposition, it follows that $\supp(\E_\Pi[\pi_2]) \subseteq W$. This contradicts the hypothesis that $\E_\Pi[\pi_2] \notin \Delta(W)\cup\Ex^\varnothing$, and thereby proves the claim.

    Third, suppose that $\pi_1 \in \Ex^\varnothing \backslash \Delta(W)$. Then, by definition, $\pi_1 = \delta_p$ for some $p \in \Delta(\Theta)\backslash W$. If $\supp(\Pi)\backslash\Ex^\varnothing = \emptyset$, then $\E_\Pi[\pi_2] = \pi_1 = \delta_p \in \Ex^\varnothing$, which contradicts the hypothesis that $\E_\Pi[\pi_2] \notin \Delta(W)\cup\Ex^\varnothing$. Thus, we have $\supp(\Pi)\backslash\Ex^\varnothing \neq \emptyset$. As in the previous sub-case, we consider the enumeration $\supp(\Pi)\backslash\Ex^\varnothing = \{\pi_2^i\}_{i=1}^n$ and claim that $\pi_2^i \notin \Delta(W)$ for some $i \in \{1,\dots, n\}$, which then implies \eqref{eqn:restrict-subadditive-infinite}. Suppose, towards a contradiction, that $ \{\pi_2^i\}_{i=1}^n \subseteq \Delta(W)$. Then, by definition, $\supp(\pi^i_2) \subseteq W$ and $p_{\pi^i_2} \in \text{conv}(\supp(\pi^i_2))$ for all  $i \in \{1,\dots,n\}$. Since $W$ is convex, it follows that $p_{\pi^i_2} \in \text{conv}(\supp(\pi^i_2)) \subseteq W$ for all  $i \in \{1,\dots,n\}$. But since $\pi_1 = \delta_p$, we also have $p_{\pi^i_2} = p \notin W$ for all $i \in \{1,\dots,n\}$. This yields the desired contradiction.

    Since these three sub-cases are exhaustive, we conclude that \eqref{eqn:restrict-subadditive-infinite} holds. 

    \textbf{\emph{Wrapping Up.}} Together, Cases 1 and 2 imply that $ C|_W (\E_\Pi[\pi_2]) \leq C|_W(\pi_1) + \E_\Pi[C|_W(\pi_2)]$. Since the fixed $\Pi \in \Delta^\dag (\Ex)$ was arbitrary, we conclude that $C|_W$ is \hyperref[axiom:POSL]{Subadditive}. 
\end{proof}

\end{document}